\documentclass[twoside, openright]{thesis}

\usepackage[T1]{fontenc}
\usepackage[latin1]{inputenc}

\usepackage{color}
\usepackage{url}
\usepackage{epsfig}
\usepackage{makeidx}
\usepackage{amssymb,latexsym, amsmath}
\usepackage{xspace}
\usepackage{tabularx}
\usepackage[authoryear]{natbib}
\usepackage[french, english, italian]{babel}
\usepackage{afterpage}

\newcommand{\authstyle}[1]{{\scshape #1}}

\newcommand{\textcmb}{cosmic microwave background~}
\newcommand{\FRW}{Friedmann-Robertson-Walker}
\newcommand{\eg}{e.g.\@\xspace }
\newcommand{\ie}{i.e.\@\xspace}
\newcommand{\eq}{\textrm{\tiny eq}}
\newcommand{\sn}{supernov\ae\ }
\newcommand{\vc}{{\textrm{\tiny (v)}}}
\newcommand{\ts}{{\textrm{\tiny (t)}}}
\newcommand{\pert}{{\textrm{\tiny pert}}}
\newcommand{\CF}{cf.\@\xspace}
\newcommand{\SEC}[1]{\S~\ref{#1}}
\newcommand{\FIG}[1]{Fig.~\ref{#1}}
\newcommand{\FIGPAG}[1]{\FIG{#1} on page \pageref{#1}}
\newcommand{\refp}[1]{\ref{#1}, page \pageref{#1}}
\newcommand{\TAB}[1]{Table~\ref{#1}}
\newcommand{\CHAP}[1]{Chapter~\ref{#1}}
\newcommand{\EQ}{\text{eq}}
\newcommand{\RD}{0}
\newcommand{\MD}{\text{MD}}
\newcommand{\dec}{\text{dec}}

\newcommand{\Hbl}{\mathcal{H}}
\newcommand{\DEG}{^\circ}
\newcommand{\arcmin}{^\prime}
\newcommand{\piv}{\text{P}}
\newcommand{\reion}{\text{re}}
\newcommand{\chisq}{chi-square}
\newcommand{\obs}{^\text{obs}}
\newcommand{\sky}{_\text{sky}}
\newcommand{\Rshift}{\mathcal{R}^\text{shift}}
\newcommand{\dbfk}{\delta_{\bs{k}}}
\newcommand{\dirac}{\delta^{\text{\tiny{(D)}}}}
\newcommand{\OSW}{\text{\scriptsize{(OSW)}}}
\newcommand{\ISW}{\text{\scriptsize{(ISW)}}}
\newcommand{\Doppler}{\text{\scriptsize{(Dpl)}}}

\newcommand{\eqdot}{\, .}
\newcommand{\eqcomma}{\, ,}
\newcommand{\tocsection}[1]{\section*{#1}
\addcontentsline{toc}{section}{\numberline{}#1}}

\newcommand{\curv}{\mathcal{K}}
\newcommand{\dr}{\textrm{d}}
\newcommand{\ol}[1]{\Bar {#1}}
\newcommand{\crit}{\textrm{\tiny crit}}
\newcommand{\pnorm}{\vert {\bf p} \vert}
\newcommand{\Lie}{{\cal L}}
\newcommand{\ra}{\quad \rightarrow \quad}
\newcommand{\comp}[1]{#1}
\newcommand{\varep}{\boldsymbol{\varepsilon}}

\newcommand{\taudot}{\dot{\tau}}
\newcommand{\DS}{\displaystyle}
\newcommand{\EX}[1]{\langle #1 \rangle}
\newcommand{\bxz}{{\bf{x}_0}}
\newcommand{\bn}{{\bf{n}}}
\newcommand{\bk}{{\bf{k}}}
\newcommand{\bkh}{{\hat{\bf{k}}}}
\newcommand{\TQ}{\Theta^{Q}}
\newcommand{\bs}[1]{\mathbf{#1}}
\newcommand{\IN}{{\text{in}}}
\newcommand{\OUT}{{\text{out}}}
\newcommand{\EIN}{\bs{\call{E}}^{\IN}}
\newcommand{\EOUT}{\bs{\call{E}}^{\OUT}}
\newcommand{\pdf}{\mathcal{P}}
\newcommand{\alm}{a_{\ell m}}
\newcommand{\halm}{\hat{a}_{\ell m}}
\newcommand{\almp}{a_{\ell' m'}}

\newcommand{\Cl}{C_\ell}
\newcommand{\hCl}{\hat{C}_\ell}
\newcommand{\bfr}{{\bf r}}
\newcommand{\data}{{\bf d}}
\newcommand{\params}{\boldsymbol{\theta}}
\newcommand{\mean}{\boldsymbol{\mu}}
\newcommand{\like}{L}
\newcommand{\lnlike}{\mathcal{L}}
\newcommand{\ML}{^*}
\newcommand{\noise}{w_b^{-1} e^{\ell(\ell+1)/ \ell_b^2}}
\newcommand{\evalML}{{\Big\arrowvert_{\params\ML}}}
\newcommand{\R}{\Rshift}
\newcommand{\kA}{\mathcal{A}}
\newcommand{\kB}{\mathcal{B}}
\newcommand{\kV}{\mathcal{V}}
\newcommand{\kR}{\mathcal{R}}
\newcommand{\kM}{\mathcal{M}}
\newcommand{\rZ}{\mathcal{T}}

\newcommand{\fres}{f_e^{\textrm{res}}}
\newcommand{\tres}{\tau^{\textrm{res}}}
\newcommand{\zdec}{z_{\textrm{dec}}}
\newcommand{\treion}{\tau_{\textrm{re}}}
\newcommand{\zreion}{z_{\rm re}}
\newcommand{\step}[1]{&\quad & (#1)}
\newcommand{\rn}{r_\nu}
\newcommand{\Pin}{\Pi_\nu}
\newcommand{\Piz}{\Pi_0}

\newcommand{\cdm}{\text{cdm}}
\newcommand{\PD}{\mathcal{P}_n}
\newcommand{\On}{\mathcal{SO}_n}
\newcommand{\TOT}{{\rm tot}}
\newcommand{\CRIT}{{\rm crit}}
\newcommand{\MAT}{{\rm m}}
\newcommand{\CDM}{{\cdm}}
\newcommand{\BAR}{{\rm b}}

\newcommand{\SCAL}{{\rm s}}
\newcommand{\UUNIT}[2]{
{\;\mathrm{#1}^{#2}} }
\newcommand{\MIX}{{\rm MIX}}
\newcommand{\AD}{{\rm AD}}

\newcommand{\CI}{{\rm CI}}
\newcommand{\NID}{{\rm ND}}
\newcommand{\NIV}{{\rm NV}}
\newcommand{\OLa}{\Om_\La}
\newcommand{\lims}[2]{_{#1}^{#2}}
\newcommand{\EFF}{{\rm eff}}
\newcommand{\GAL}{{\rm g}}
\newcommand{\tensud}[3]{{#1}^{#2}_{\;\:#3}}
\newcommand{\enter}{\text{\tiny{ent}}}
\newcommand{\ENTR}{e}
\newcommand{\CORR}{c}

\newcommand{\tabomegab}{baryon density& $\omega_b$}
\newcommand{\tabomegam}{matter density& $\omega_m$}
\newcommand{\tabomegal}{cosmological constant density& $\omega_\Lambda$}
\newcommand{\tabomegals}{$\La$ density& $\omega_\Lambda$}
\newcommand{\tabns}{spectral index &$n_\SCAL$}
\newcommand{\tabnorm}{normalization &$Q$}
\newcommand{\tabshift}{shift parameter &$\R$}
\newcommand{\tabal}{fine structure constant &$\alpha_\dec$}
\newcommand{\tabreion}{reionization optical depth &$\tau_\reion$}

\newcommand{\onefigwidth}{8.5cm}
\newcommand{\twofigswidth}{0.50\linewidth}
\newcommand{\separator}[1]{\multicolumn{6}{|c|}{#1} \\\hline}
\newcommand{\dline}{\hline \hline}

\newcommand{\al}{\alpha}
\newcommand{\ba}{\beta}

\newcommand{\ep}{\epsilon}
\newcommand{\ga}{\gamma}
\newcommand{\Ga}{\Gamma}

\newcommand{\La}{\Lambda}
\newcommand{\la}{\lambda}
\newcommand{\Om}{\Omega}
\newcommand{\om}{\omega}
\newcommand{\si}{\sigma}

\newcommand{\te}{\theta}

\newcommand{\tr}{\mbox{tr}}

\newcommand{\eff}{{\rm eff}}

\newcommand{\ev} {\mbox{eV}}
\newcommand{\MeV}{\mbox{MeV}}

\newcommand{\mpc}{\mbox{Mpc}}

\newcommand{\yr} {\mbox{yr}}

\newcommand{\be}{\begin{equation}}
\newcommand{\ee}{\end{equation}}
\newcommand{\bes}{\begin{equation*}}
\newcommand{\ees}{\end{equation*}}

\newcommand{\lsim}{\,\raise 0.4ex\hbox{$<$}\kern -0.8em\lower 0.62ex\hbox{$\sim$}\,}
\newcommand{\gsim}{\,\raise 0.4ex\hbox{$>$}\kern -0.7em\lower 0.62ex\hbox{$\sim$}\,}
\newcommand{\bsea}{\begin{subequations} \begin{align}}
\newcommand{\esea}{\end{align} \end{subequations}}
\newcommand{\bean}{\begin{eqnarray*}}
\newcommand{\eean}{\end{eqnarray*}}
\newcommand{\rr}[1]{Eq.~\eqref{#1}}
\newcommand{\rrp}[1]{Eq.~(\ref{#1}, page \pageref{#1})}

\newcommand{\call}[1]{\mathcal{#1}}

\newcommand{\calo}{\mathcal{O}}

\title{Cosmic microwave background anisotropies: Beyond standard parameters}
\author{Roberto Trotta}
\qualification{Doctor of Philosophy}
\date{{\bf DRAFT VERSION 1.0 \\ April, 26th 2004}}

\begin{document}
\selectlanguage{english}

%
%
%
%

\begin{titlepage}
\hsize=16truecm
\hoffset=-0.5truecm \voffset=-0.5truecm

\thispagestyle{empty}

\begin{tabular}{lr}
\makebox[7.25cm][l]{UNIVERSITY OF GENEVA}&
\makebox[7.25cm][r]{Department of Theoretical Physics}
\\ \hline
\end{tabular}


\vspace{6\baselineskip}

\begin{center}
 {\LARGE \bf
    \noindent Cosmic Microwave Background Anisotropies:
 \vspace{0.5\baselineskip} \\
    \noindent Beyond Standard Parameters
 }
\end{center}

\vspace*{1.5cm}

\begin{center}
{\Large \noindent Ph.D. Thesis}
\end{center}

\vspace*{1.5cm}

\begin{center}
{\large \noindent submitted at the Department of Theoretical
Physics \vspace{0.5\baselineskip} \\
of the \vspace{0.5\baselineskip} \\
UNIVERSITY OF GENEVA
\vspace{0.5\baselineskip} \\
to obtain the degree of
\vspace{0.5\baselineskip} \\
{\it Docteur ès Sciences, mention Physique} \noindent}
\end{center}

\begin{center}
\large \noindent by
\end{center}

\vspace*{\stretch{0.6}}

\begin{center}
{\Large \bf \noindent Roberto Trotta}
\end{center}

\vspace*{\stretch{1}}

\begin{center}
\noindent \large Thesis N.~3534
\end{center}

\vspace*{\stretch{1}}

\begin{center}
2004
\end{center}

\pagebreak

\end{titlepage}

\thispagestyle{empty} \newpage
\thispagestyle{empty}

\noindent This thesis was presented on June, 18th 2004 at the
Department of Theoretical Physics of the University of Geneva,
Geneva, Switzerland.

\vspace{3\baselineskip}
\begin{tabbing}
Members of the Jury extraspace:    \= \kill
{\bf Supervisor:} \> Prof.~\authstyle{Ruth Durrer} \\
\phantom{a} \\
{\bf Members of the Jury:}  \>  Prof.~\authstyle{Joseph Silk}  \\
 \>  Prof.~\authstyle{Thierry Courvoisier} \\
  \>  Dr.~\authstyle{Pedro G.~Ferreira}

\end{tabbing}
\newpage

\newcommand{\citsp}{\quad}

\vspace*{2cm}

\begin{tabbing}
\setlength{\tabbingsep}{0cm} \hspace{7cm}\=nobody,not even the
rain,has such small hands\= \kill
 \> {\it \Large  To Elisa} \\
 \> \\
 \> nobody,not even the rain,has such small hands\\
 \> \\
 \> \> \authstyle{E.~E.~Cummings} \' \\
\end{tabbing}
\clearpage

\thispagestyle{empty} \cleardoublepage

\begin{acknowledgments}
It is a pleasure to thank the people who have contributed to the
realization of this work and who have accompanied me along the
way. During these three years, I have enjoyed working with and
learning from all the members of the Geneva Cosmology Group and of
the group of Michele Maggiore. I have had the chance to
collaborate with several people, to whom I would like to express
my appreciation: Pedro P.~Avelino, Rachel Bean, Rebecca Bowen,
Ruth Durrer, Steen H.~Hansen, Carlos J.~A.~Martins, Alessandro
Melchiorri, Alain Riazuelo, Graça Rocha, Joseph Silk and Pedro
T.~P.~Viana.

Ruth Durrer has been a wonderful supervisor, never counting the
hours she spent on my questions, constantly stimulating my
interests while allowing me to pursue my research in all freedom.
I thank her for her teaching and for her example. The help of
Alain Riazuelo was precious during my first year and gave me a
swift start into real research work. I am grateful to Alessandro
Melchiorri for involving me in many collaborations which
constitute a major part of this thesis. I enjoyed having various
interesting and promising discussion with Filippo Vernizzi, Martin
Kunz and Céline Boehm, which I am sure one day we will be able to
finalize. Sam Leach is guilty to have converted me to the Bayesian
school during many restless and humorous discussions. I am
indebted to Christophe Ringeval and Thierry Baertschiger for their
help in solving my various computer problems, and to Andreas
Malaspinas for his Sisyphus work of troubleshooting our computer
network. I shall join my thanks to the ones of the cosmology
community to Anthony Lewis for developing, supporting and making
publicly available the {\sc camb} and {\sc cosmomc} codes. I would
like to express my gratitude to Prof.\ Silk, Prof.\ Courvoisier
and Dr.\ Ferreira for accepting to be part of the jury.

The European Network CMBNet and the Schmidheiny Foundation have
provided generous support for some of the collaborations I was
involved with, in the form of travel grants. I am indebted to
Oxford Astrophysics and to Prof.\ Silk for their kind hospitality
in many occasions, and to Profs.\ Spergel and Kosowsky and to
Princeton University for financial support of my visit.

\vspace{\baselineskip}

But there is not only science, even to the life of a PhD student.
My time in Geneva would not have been the same without the
friendship of Timon ``Lincoln'' Boehm: I wish him all the best on
his new path. I have had the pleasure of sharing many refreshing
moments with Stefano Foffa, Marj Tonini, Yasmin Friedmann, Simone
Lelli, Anna Rissone, Martin Zimmermann, Davide ``Dutturi''
Lazzati. To my parents, my affectionate thoughts for their
encouraging presence. To Elisa, my fiancée, my deepest gratitude
for having been at my side in all marvellous and all difficult
moments, and for the ones which are still to come.

This work was much improved both in form and contents by the
careful reading of Christophe Ringeval and Elisa Cunial (for the
french part), Sam Leach and Ruth Durrer for the English part. I
thank them for their time and competence. I alone bear the
responsibility for any mistake which might still be present.

\end{acknowledgments}

 \tableofcontents
 \listoffigures
 \listoftables

\selectlanguage{italian} \thispagestyle{empty} \cleardoublepage
\thispagestyle{empty}

\vspace*{2cm}

\begin{tabbing}
 \hspace{8cm}\= \citsp  che porta 'l ciel, per un pertugio tondo; \= \kill
 \setlength{\tabbingsep}{0cm}
 \> Lo duca e io per quel cammino ascoso \\
 \> \citsp intrammo a ritornar nel chiaro mondo \\
 \> \citsp  e sanza cura aver d'alcun riposo \\
 \> salimmo su, el primo e io secondo, \\
 \> \citsp  tanto ch'i' vidi de le cose belle \\
 \> \citsp  che porta 'l ciel, per un pertugio tondo; \\
 \> e quindi uscimmo a riveder le stelle.
 \> \\
 \vspace{\baselineskip} \\
 \> \> \authstyle{Dante}, La divina commedia \' \\
 \> \> Inferno XXXIV, 133-139. \' \\

\end{tabbing}

\clearpage

\thispagestyle{empty} \cleardoublepage

\starttext



\selectlanguage{english}
\chapter*{Overview and conclusions}
\label{chap:conclusion}
\tocsection{Towards a cosmological standard model}

The study of cosmic microwave background anisotropies is one of
the pillars of modern cosmology. The cosmic microwave background
(hereafter CMB) consists of photons left over by the hot phase
after the Big-Bang and is very homogeneous and isotropic. Its
existence was predicted by \cite{Gamov:1946}, and accidentally
discovered only much later by Penzias and Wilson
\citep{Penzias:1965wn}, but it was only in 1992 that the COBE
satellite \citep{Smoot:1992td} detected the presence of tiny
temperature fluctuations (1 part in 100'000), which are thought to
have been generated by quantum fluctuations in the very early
universe. The observational study of these temperature
fluctuations, known as anisotropies, has been a great
technological achievement. Over the last ten years, there has been
a spectacular advancement in the accuracy of measurements, using
ground-based, balloon-born and orbital instruments. The WMAP
satellite \citep{Bennett:2003bz} has recently measured the
anisotropies with a precision which, on certain scales, is close
to a fundamental statistical limit, called ``cosmic variance''.

The importance of such a wealth of data for theoretical cosmology
cannot be overstated. In a few seconds on a desktop computer, it
is nowadays possible to produce accurate numerical predictions of
the statistical distribution of the anisotropies on the sky for
any cosmological model of interest, \ie of the CMB angular power
spectrum. If the primordial fluctuations are Gaussian distributed,
then the power spectrum encodes all of the statistical
information: its computation is based on linear perturbation
theory and the underlying physics is well understood. The detailed
shape of the power spectrum carries characteristic signatures
depending on the value of the late Universe cosmological
parameters and on the initial conditions for the perturbations. By
``late Universe cosmological parameters'' we mean the quantities
controlling the expansion history of the Universe, \ie its matter
budget, complemented by some description of the reionization
history. In the former category, an incomplete list would include
the Hubble parameter, the energy density in baryons, cold dark
matter and dark energy, the dark energy equation of state
parameter (possibly including a description of its time
evolution), the neutrino masses and the number of massless
families plus the density parameters and effective equation of
state of any other exotic form of matter one might wish to
include; specifying how the Universe was reionized in the context
of stellar evolution theory might require three or four additional
parameters, which however usually reduce to the optical depth to
reionization or equivalently to the redshift of reionization, as
far as the CMB is concerned. Specifying the initial conditions
requires the value of ``primordial parameters'' for the amplitudes
of the primordial fluctuations in each of the matter components
and their scale dependence.

The fact that CMB anisotropies are sensitive both to the late
Universe cosmological parameters and to primordial parameters
means that CMB observations only constrain a (degenerate)
combination of both: until now, disentangling the former required
rather strong assumptions about the nature of initial conditions.
Some guidance is offered by the inflationary paradigm: in its
simplest incarnation, the decay of the inflaton field produces
adiabatic initial conditions, in which there is no fluctuation in
the relative number density of the species, hence no entropy
perturbations (``adiabatic''). The presence of entropy
fluctuations can excite up to four other non-decaying modes for
the perturbations. Those are collectively termed ``isocurvature'',
because in three cases the total matter density is unperturbed and
hence there is no curvature perturbation in the spatial sections
either. The observation of the first acoustic peak in the CMB
power spectrum \citep{Page:2003fa} at $\ell = 220.1 \pm 0.8$ has
substantially confirmed the predominance of the adiabatic mode.
However, a subdominant isocurvature contribution to the prevalent
adiabatic mode cannot be excluded: after all, there is no
compelling reason why the physics of the early universe should
boil down to only one degree of freedom.

Even though in principle the number of late Universe parameters
can be very large, easily exceeding a dozen, only an handful of
them seems to be required by the currently available observational
evidence \citep{Spergel:2003cb,Tegmark:2003ud,Liddle:2004nh}:
 \begin{itemize}
 \item the Hubble parameter $h \sim 0.7$;
 \item the density parameter for baryons $\Om_b \sim 0.05$;
 \item the density parameter for cold dark matter (CDM) $\Om_\cdm \sim
 0.25$;
 \item the density parameter for a cosmological constant $\OLa \sim 0.7$;
 \item the optical depth to reionization $\tau_\reion \sim 0.15$.
 \end{itemize}
Summed together, $\Om_\cdm + \Om_b + \Om_\La \sim 1$ imply a flat
Universe. The crucial point is that for the CMB these results only
hold once we make the rather strong assumption of purely adiabatic
initial conditions. In that case, the primordial parameters reduce
to the spectral index for the fluctuations, $n_\SCAL \sim 1$, and
an overall adiabatic amplitude $A_\text{AD}$. These two quantities
together with the above five late Universe parameters are what we
call ``standard CMB parameters'', because they build the basis of
the ``concordance model'' of present-day cosmology\footnote{We do
not discuss the possibility of gravitational waves, which are
indeed predicted by any inflationary scenario; presently there are
merely upper limits to their contribution, which could be small
enough to be very difficult to detect in the CMB. Our discussion
here and in the following focuses on the scalar sector only.}.

By combining CMB data with other cosmological and astrophysical
measurements -- such as galaxy distribution statistics,
supernov\ae~ luminosity distance measurements, gravitational
lensing statistics, Lyman $\alpha$ absorption lines, local
determination of the Hubble parameter, light elements abundance --
we have reached an unprecedented precision in determining the
standard cosmological parameters, which are now known with an
accuracy of a few percent. This is even more astonishing if we
think that only ten years ago it was only possible for most
parameters to estimate their order of magnitude. Most importantly,
various independent observations -- which probe very different
epochs of the cosmic history and are based on totally different
physical processes -- seem to be converging to the same answer.

We are now in a position where we can move on from parameter
fitting to model testing: in other words, in order to establish a
``cosmological standard model'' we need to assess the consistency
and completeness of our theoretical framework. In order to be sure
that we can trust the error-bars on the standard parameters beyond
the quoted statistical error, we have to confront ourselves with
the question of possible systematic errors in the measurements on
one side, and of hidden flaws in our theoretical interpretation of
the data on the other. Given the intrinsic difficulty of many
cosmological observations, an assessment of systematic errors for
a certain data-set can come from the combination with other,
independent measurements of the same quantity. Discrepancies in
the results will indicate a flaw in the underlying theory, or in
the data, or in both. This is one of the reasons why the
comparison of many data-sets is so important, the other being that
often the combined data have a superior constraining power due to
the breaking of degenerate directions in parameter space. From the
point of view of model-building, it is now becoming possible to
relax some assumptions which were before necessary in order to
extract from the data any information at all, and thereby check
whether our results are robust or else whether they critically
depend on our prejudices. If it is found that our conclusions
depend strongly on the underlying model assumptions, then we need
to critically review our theoretical paradigm and open our mind to
alternative explicative models.

\tocsection{Testing the concordance model with the CMB}

The CMB is an excellent testing ground to carry out this program:
our theoretical understanding is based on General Relativity and
linear perturbation theory, which suffices to describe almost all
of the relevant physical processes. This makes us confident that
we understand quite well CMB anisotropies, and we can exploit them
to go beyond the standard cosmological parameters in two different
ways: the first path leads directly to the primordial Universe,
via the dependence of the CMB on the nature of initial conditions;
the second approach makes use of the high quality of recent CMB
data to look for effects which were previously ignored because
thought to be irrelevant, but which are now within the
constraining power of the observations. In both cases, the
microwave background plays the role of a Universe-sized laboratory
for the study of fundamental physics which is often unaccessible
to any particle physics laboratory. This work pursues both those
aspects, as we detail in the following.

In the first part, we introduce in \CHAP{chap:introduction} the
homogeneous and isotropic Friedmann-Robertson-Walker universe,
which is the background on which perturbation theory is built, and
we briefly present a few other observations which we later compare
and combine with the CMB. We then give the derivation of all the
relevant perturbation equations needed to describe the CMB in
\CHAP{chap:perturbation}. Those are applied to the temperature
fluctuations in the cosmic photons in the second part: in
\CHAP{chap:cmb} we obtain under various approximations analytical
expressions for the growth of perturbations in an Universe
containing photons, cold dark matter, massless neutrinos, baryons
and a cosmological constant; in \CHAP{chap:params} we present a
thorough account of the main features of the CMB temperature and
polarization angular power spectra. In particular, we are
concerned with characteristic signatures on the angular power
spectra of the standard cosmological parameters, which constitute
the basis for their determination using CMB data. We also
introduce the most general type of initial conditions, which
consist of one adiabatic and four isocurvature modes. The third
part focuses on the interplay between theoretical modelling and
observational data. The comparison of theoretical models with
actual data needs some basis in probability theory and statistics,
which we give in \CHAP{chap:data}, emphasizing their application
to the problem of parameter estimation from CMB observations. The
last two chapters contain most of the original research work,
which is developed along the two lines sketched above:
\CHAP{chap:beyondsp} deals with the observational consequences and
constraints when we add to the standard cosmological parameters
new quantities describing possible departures from known physics,
while \CHAP{chap:genic} explores the consequences of relaxing the
fundamental assumption of adiabaticity.

In \SEC{chap:beyondsp;sec:rel} we focus on the effective number of
massless neutrino families, $N_\eff$ \citep*{Bowen:2001in}.
Although in the standard model of particle physics $N_\eff = 3$,
there are several mechanism which would give $N_\eff \neq 3$ as
measured by the two cosmological probes we discuss, namely
Big-Bang Nucleosynthesis (BBN) combined with observations of the
light elements abundances, and CMB. This is because both of them
are sensitive not only to the number of weakly interacting
neutrinos, but rather to the total energy density of relativistic
particles which sets the expansion rate at early times, and
therefore can constrain \eg the existence of sterile neutrinos
unobservable in Z-decay experiments. Using pre-WMAP CMB data
alone, we obtain fairly broad bounds on $N_\eff$, $0.04 < N_\eff <
13.37$ with $2\si$ likelihood content, which are reduced by
including prior information coming from supernov\ae~luminosity
distance measurements and large scale structure observations.  We
show that $N_\eff$, or equivalently $\om_\text{rel}\equiv
\Om_{\text{rel}}h^2$, the energy density parameter in relativistic
particles, is nearly degenerate with the amount of energy in
matter, $\omega_m\equiv \Om_m h^2$, and that its inclusion in CMB
parameter estimation also affects the constraints on other
parameters such as the curvature or the scalar spectral index of
primordial fluctuations. However, even though this degeneracy has
the effect of limiting the accuracy of parameter estimation from
the WMAP satellite, we find that it can be broken by measurements
on smaller scales such as those provided by the Planck satellite
mission. We forecast that Planck will be able to constrain
$N_\eff$ within $0.24$ ($1\si$).

The primordial $^4$He mass fraction, $Y_p$, is predicted by BBN
along with the abundances of the other light elements as a
function of two free parameters, namely the baryon density $\om_b$
and the relativistic energy density $\om_\text{rel}$. If we fix
$N_\eff = 3$ and thereby $\om_\text{rel}$ as motivated by the
particle physics standard model, then in standard BBN the
abundances of D, $^3$He, $^4$He and $^7$Li depend on the baryon
density alone: comparison with the observed values in
astrophysical systems indicates a slight discrepancy, which
however presently cannot clearly be ascribed to systematical
errors or to deviations from the standard BBN scenario. We explore
in \SEC{chap:bspII;sec:helium} the potentiality of using the CMB
as a totally independent way of measuring $Y_p$ via its impact on
the reionization history, thereby possibly allowing to
discriminate between the various hypothesis \citep{Trotta:2003xg}.
We find that WMAP data give only a marginal detection, $ 0.160 <
Y_p < 0.501$ at 68\% likelihood content. We estimate that the
Planck satellite will determine the helium mass fraction within
$5\%$ (or $\Delta Y_p \sim 0.01$), which however will only allow a
marginal discrimination between different astrophysical
measurements. Equally important, we identify degeneracies between
$Y_p$ and other cosmological parameters, most notably the baryon
abundance, the redshift and optical depth of reionization and the
spectral index; we conclude that even though present-day CMB data
accuracy does not require the inclusion of $Y_p$ as a free
parameter, the uncertainty of the helium fraction will have to be
taken into account in order to correctly estimate the errors on
the baryon density from Planck.

The search for observational evidence for time or space variations
of the ``fundamental'' constants that can be measured in our
four-dimensional world is an extremely exciting area of current
research, with several independent claims of detections in
different contexts emerging in the last few years, together with
other improved constraints. Most efforts have been concentrating
on the fine-structure constant, $\alpha$, both due to its
obviously fundamental role and to the availability of a series of
independent methods of measurement. Of particular interest is the
result of Webb and collaborators, who claim a $4\sigma$ detection
of a fine-structure constant that was smaller in the past
\citep{Murphy:2002ve,Webb:2002vd}. Noteworthy among the
possibilities of independently check those results is the CMB,
which probes $\alpha_\dec$, the value of $\alpha$ at decoupling,
$z \sim 1100$ \citep{Martins:2002iv,Martins:2003pe,Rocha:2004}. As
we show in \SEC{chap:bspIII;sec:alpha}, by analyzing the first
year WMAP data for time-variations of $\al$ we obtain the
constrain $0.95 < \al_\dec/\al_0 < 1.02$ with $95\%$ likelihood
content, where $\al_0$ denotes the present value. We clarify the
issue of degeneracies between $\al$ and other standard parameters,
and give exhaustive forecasts of the expected performance of the
full four year WMAP data, of the Planck satellite and of an ideal
CMB experiment. We emphasize the role of polarization measurements
to lift flat directions (\ie, degeneracies) in parameter space,
and discuss the role of reionization in the determination of
$\al_\dec$.

In \CHAP{chap:genic} we relax the assumption of adiabaticity by
allowing for the most general initial conditions
\citep{Bucher:1999re} and we investigate two complementary
aspects: the first is the degradation in the accuracy of the late
Universe standard parameters as a consequence of the introduction
of new degrees of freedom in the primordial Universe
\citep{Trotta:2001yw}; the second is the robustness of the
measurement of a non-zero cosmological constant, $\Om_\La \neq 0$,
when different statistical approaches (frequentist rather then
Bayesian) are applied to the data, or when general isocurvature
modes are included in the analysis \citep{Trotta:2002iz}. We also
explicitly test the paradigm of adiabaticity by using CMB
observations to put constraints on the isocurvature contribution.

For the first point, the results in \SEC{chap:genic;sec:precision}
demonstrate that the determination of the Hubble parameter and the
baryon density from pre-WMAP CMB data is essentially impossible
without strong assumptions about the nature of initial conditions.
Conversely, it becomes very difficult to put limits on the type of
the initial conditions without using external, non-CMB priors on
the late Universe parameters. Indeed, the CMB is perhaps the most
effective way to directly probe the very early Universe, and
thereby constrain or falsify the models for the generation of
perturbations. It is therefore very important to extract the most
information about the conditions in the early Universe. Adding
polarization information greatly enhances the power of the CMB to
simultaneously constrain the late Universe parameters and the
primordial ones: we show in \SEC{chap:genic;sec:future} that the
full four year WMAP data will measure orthogonal combinations of
the late Universe parameters with an accuracy of the order
$10\%-30\%$ for most parameters even in the general initial
conditions case. The Planck mission will have a better
polarization resolution and will be able to do precision cosmology
almost independently on the type of initial conditions
\citep{Trotta:2004}. As for the possibility of mitigating the
cosmological constant problem by introducing isocurvature modes,
our findings in \SEC{chap:genic;sec:lambda} indicate that $\Om_\La
\neq 0$, as obtained from a combination of CMB and large scale
structure data, is indeed robust even in the presence of
isocurvature contributions. The more conservative frequentist
statistics -- as compared to the usual Bayesian approach --
excludes $\Om_\La = 0$ only at the $2\si$ confidence level for
pre-WMAP CMB data combined with the 2dF Galaxy Redshift Survey,
but this only if we admit a rather low value for the Hubble
constant, $h \sim 0.5$, which would be in contradiction with the
result of the Hubble Space Telescope, $h=0.72\pm0.08$
\citep{Freedman:2000cf}.

\tocsection{Outlook and conclusion}

The CMB has become a well established tool for the study of our
Universe, and an unavoidable testing ground for any theoretical
model. The ever improving quality of the data permits on one side
to look for new physics in the early Universe, as shown in our
study of time variations of $\al$, on the presence of extra
relativistic particles and on the existence of non-adiabatic
modes; on the other hand, it also requires an upgrade of our
modelling, so to properly treat subtle effects such as the
uncertainty coming from our unprecise knowledge of the primordial
Helium fraction, or from our ignorance on the correct model for
the generation of fluctuations. For this reasons, it is important
to look ahead, to the goals for the next generation of
experiments, and to their potential to constrain or falsify the
theoretical models.

 More than ever, the central issue is becoming how
to efficiently and reliably extract the most information from
upcoming high-quality data: there are about 2000 observable
independent multipoles for each of the three angular power
spectra, namely temperature, E-polarization and
temperature-polarization cross-correlation, which however are
highly redundant due to the smooth oscillatory nature of the
spectra. The amount of information which can be extracted is much
less, and can be condensed in maybe a dozen of well-chosen
parameters. The best choice for those quantities is the one which
takes into account the physics and selects orthogonal directions
in parameters space on the basis of fundamental degeneracies. This
idea has been a {\it leitmotiv} of the works presented here, and
there is probably still space to apply it further, especially in
connection with the primordial parameters.

Despite this encouraging picture, there are still open challenges
for our understanding of the Universe: the nature of dark energy
and dark matter, the details of the initial conditions and the
epoch of reionization, for example. The CMB will provide key
advancements on all these issues over the next years. The
polarization of the anisotropies has been detected by the
experiments DASI \citep{Kovac:2002fg} and WMAP and will be
precisely mapped by the forthcoming experiments PolarBear, Bicep,
SPOrt, AMiBA and QUEST, opening up a new line of research and
allowing to reconstruct the cosmological parameters with still
higher precision. This process will culminate with the European
Space Agency satellite Planck \citep{Planck:Website}, which
starting in 2007 will observe the temperature spectrum with the
ultimate possible precision and provide accurate mapping of the
polarization as well. In view of this wealth of data, and in order
to fully exploit its potential, it is of fundamental importance
that theoretical research on the subject advances accordingly.
There is a need of more powerful and efficient computational and
statistical techniques which can handle the considerably larger
amount of data expected. Also, our theoretical understanding of
model-building has to be refined and in particular we need to
further develop the interdisciplinary link between models coming
from high energy physics, string theory, astrophysics and their
observational signature on the CMB. This approach will strengthen
the role of the CMB as a universe-size laboratory for
investigating the most elusive domains of fundamental physics.

\part{BASICS}{
All men, Socrates, who have any degree of right feeling, at the
beginning of every enterprise, whether small or great, always call
upon God. And we, too, who are going to discourse of the nature of
the universe, how created or how existing without creation, if we
be not altogether out of our wits, must invoke the aid of Gods and
Goddesses and pray that our words may be acceptable to them and
consistent with themselves.} {\authstyle{Plato}}
 {Timaeus}

\chapter{Introduction}
\label{chap:introduction}
\section{Notation and conventions}

We begin by introducing the notation and conventions which are
used throughout this work.
\begin{itemize}
 \item The metric signature is $- + + +$.
 \item The spacetime metric is denoted by ${g}_{\mu
 \nu}$, where the spacetime coordinate are $x^\mu, \mu = 0,1,2,3$.
 Greek indexes always run from 0 to 3.
 \item The 3-space of constant curvature has metric
 ${\gamma}_{ij}$. Latin indexes always run from 1 to 3.
 \item When we discuss perturbations, the background, unperturbed
 quantities are denoted by an overline. Therefore for instance
 $\rho = \ol{\rho} + \delta \rho$, where $\ol{\rho}$ denotes the
 background energy density and $\rho$ the perturbed (background
 plus linear perturbation) energy density.
 \item The overdot ``$\;\;\dot{}\;\;$'' denotes the derivative with respect to conformal
 time, $\eta$.
 \item Bold character denote the $i=1,2,3$ components of the
 corresponding 4-vector.
 \item Unless otherwise stated we use natural units, in which the speed of light, the
 Boltzmann constant and the Planck constant are unity, $c = k_B =
 \hbar = 1$.
 \item The Hubble parameter today is written as $H_0 \equiv 100\,h
\UUNIT{km}{} \UUNIT{s}{-1} \UUNIT{Mpc}{-1}$.
 \item The symbol $\Om_X$ denotes the density parameter in the component $X$
 (where $X$ can stand for baryons, photons, cold dark matter, etc.), expressed
 in units of the critical energy density. In
general, $\Om_X = \Om_X(\eta)$, but whenever we omit the explicit
time dependence, it is understood that the quantity is evaluated
today, \ie $\Om_X \equiv \Om_X(\eta_0)$, where $\eta_0$ is the
present value of conformal time.
 \item The critical energy density today is
 $\rho_{\crit}(\eta_0) \approx 1.88 \cdot 10^{-29} \,h^2 \text{
g/cm}^3$, and the present energy density of component $X$ is
written $\rho_X(\eta_0) = \om_X \,1.88 \cdot 10^{-29} \text{
g/cm}^3$, where we have defined $\om_X \equiv \Om_X(\eta_0) h^2$.
\end{itemize}

\section{Friedmann-Robertson-Walker cosmology}
\label{chap:intro;sec:FRW_cosmology}

In this section, we briefly review the standard treatment of an
homogeneous and isotropic universe. We present the background
Einstein and conservation equations for perfect fluids, along with
the unperturbed Boltzmann equation describing relativistic
particles.

\subsection{Einstein equations}
\label{chap:intro;sec:Einstein}

 The \textcmb is homogeneous and isotropic to better
than one part in 100'000. This justifies the assumption that the
universe, on large enough scale, can be treated as being
homogeneous and isotropic. We then consider a 4-dimensional
manifold $\mathcal{M}$ endowed with a metric ${g}_{\mu \nu}$, so
that constant-time hypersurfaces are constant-curvature, maximally
symmetric 3-spaces. The Friedmann-Robertson-Walker (FRW) metric
reads
 \be \label{eq:FRW_metric_t}
  {g}_{\mu\nu} \dr x^{\mu} \dr x^{\nu} = -\dr t^2 + a(t) {\gamma}_{ij} \dr x^i \dr
  x^j \eqcomma
 \ee

 with the 3-space metric of curvature $\curv = \{ 0, +1, -1 \}$ given by
 \be \label{eq:metric_with_chi}
  {\gamma}_{ij} \dr x^i \dr x^j = \dr r^2 + \chi^2(r)(\dr \theta^2 +
  \sin(\theta)^2 \dr \phi^2) \eqdot
 \ee
Here the {\it scale factor} $a(t)$ depends only on time, and
 \be \label{eq:define_chi_function}
 \chi(r) = \left\{
 \begin{array}{l}
 \begin{aligned}
 & r  & &\quad \text{for } \curv = 0 \text{ (flat universe)} \\
 & \sin (r)  & &\quad \text{for } \curv =  +1 \text{ (closed
 universe)}\\
 & \sinh (r)  & &\quad \text{for } \curv = -1 \text{ (open universe)}
\end{aligned}
 \end{array}
 \right. \eqdot
 \ee

We will mostly work in {\it conformal time} $\eta$, defined
through $\dr \eta \equiv a^{-1}(t)\dr t$, so that the FRW metric
reads
 \be \label{eq:FRW_metric_tau}
  {g}_{\mu\nu} \dr x^{\mu} \dr x^{\nu} = a(\eta) (-\dr \eta^2 +  {\gamma}_{ij} \dr x^i \dr
  x^j) \eqdot
 \ee
Following the assumptions of homogeneity and isotropy, the
background energy-momentum tensor, ${T}_{\mu \nu}$ is bound to be
of the perfect fluid form
 \be \label{eq:background_energy_tsr}
  {T}_{\mu \nu} = (\rho + P) u_\mu u_\nu + P
  {g}_{\mu \nu} \eqcomma
 \ee
where $\rho, P$ are functions of the conformal time $\eta$ only,
and represent the fluid energy density and pressure, respectively.
The fluid 4-velocity is the timelike 4-vector $\bf{u}$, with
 \be
 u^{\mu} = \left(\frac{1}{a}, 0 , 0 , 0 \right) \qquad \textrm{and} \qquad
 u_\mu u^\mu = -1 \eqdot
 \ee
We suppose that the equation of state of the fluid is of the form
 \be \label{eq:eq_of_state}
 P = w(\rho) \rho \eqcomma
 \ee
where the enthalpy $w(\rho)$ depends only on the local energy
density. In many cases of interest, the enthalpy is simply a
constant, in which case it is termed {\it equation of state
parameter}: for cold, non-relativistic, pressureless matter $w_m =
0$ (dust), for relativistic particles $w_r = 1/3$ (radiation) and
$w_\Lambda = -1$ for a cosmological constant (vacuum energy). The
energy density of a cosmological constant is contained in
${T}_{\mu\nu}$, and is of the form $\rho_\Lambda = \Lambda/(8\pi
G)$. Another relevant quantity is the {\it adiabatic sound speed}
of the fluid, defined as
 \be \label{eq:adiabatic_sound_speed}
 c_s^2 \equiv \dot{P}/\dot{\rho} \eqdot
 \ee
The Einstein equations
 \be \label{eq:background_Einstein}
 {G}_{\mu \nu} = 8 \pi G {T}_{\mu \nu}
 \ee
 with the FRW metric (\ref{eq:FRW_metric_tau}) and the
 energy-momentum tensor (\ref{eq:background_energy_tsr}) yield the
 two {\it Friedmann equations}. The first Friedmann equation is a first order
 differential equation for the {\it conformal Hubble parameter} $\Hbl(\eta)
 \equiv \dot{a}/a$
 \be
 \label{eq:background_Friedmann_1}
 \dot{\Hbl}  = -\frac{4 \pi G}{3}a^2(\rho + 3 P) \eqdot
 \ee
 The second one is a constraint equation,
\be
 \label{eq:background_Friedmann_2}
 \Hbl^2 + \curv  = \frac{8 \pi G}{3} a^2 \rho \eqdot
\ee

An evolution equation for the fluid energy density follows from
the 0 component of the energy-momentum conservation equation,
$\nabla_\mu {T}^{\mu\nu} = 0$:

 \be \label{eq:background_energy_conservation}
 \dot{\rho} + 3 \Hbl(\rho + P) = 0 \eqcomma
 \ee
supplemented with the fluid equation of state,
\rr{eq:eq_of_state}. If the universe contains (or is dominated by)
only one fluid with $w = \text{const}$, it follows from
\rr{eq:background_energy_conservation} that its energy density
behaves as
 \be \label{eq:background_energy_density}
 \rho \propto a^{-3(1+w)} \eqcomma
 \ee
hence from \rr{eq:background_Friedmann_1} the scale factor of a
flat universe ($\curv = 0$) is
 \be
 \DS a = \Big \vert \frac{2A}{1+3w} \eta \Big \vert ^{\frac{2}{1+3w}} \quad \text{for } w
\ne -1/3\eqdot
 \ee
with $A^2 = 8\pi G / 3 \rho a^{3(1+w)} = \textrm{const}$. In
particular, in the radiation dominated universe ($w=1/3$) we have
$a \propto \eta$, while in the matter dominated universe ($w
\approx 0$) $a\propto\eta^2$.

In the standard cosmological picture, the universe contains
non-relativistic, pressureless matter (baryons and cold dark
matter), photons, massless neutrinos and a vacuum energy
component. In this case, the stress-energy tensor is the sum of
the fluid components
 \be
 {T}^{\mu \nu} = \sum_\al {T}_{\comp{\al}}^{\mu \nu}
 \eqdot
 \ee
The Friedmann equations (\ref{eq:background_Friedmann_1},
\ref{eq:background_Friedmann_2}) apply to the total energy density
and pressure, which are just the sum of the contributions from
each fluid. The energy conservation equation,
\rr{eq:background_energy_conservation}, still applies to the total
variables, while in general for each component we have
 \be \label{eq:background_energy_conservation_multifluid}
 \nabla_\mu {T}^{\mu \nu}_\al = {Q}_{\comp{\al}}^{\nu}
 \eqcomma
 \ee
where the 4 vector ${Q}_{\comp{\al}}^{\mu \nu}$ describe the
energy-momentum transfer from the component $\alpha$. The
conservation of total energy requires
 \be
 \sum_\al {Q}_{\comp{\al}}^{\nu} = 0 \eqdot
 \ee

In the general case, the Friedmann equations have to be solved
numerically. However, we can easily write down solutions of simple
cases. From \rr{eq:background_energy_density} it follows that for
radiation $\rho_r \propto a^{-4}$ while for matter $\rho_m \propto
a^{-3}$. Physically, the energy density of matter is diluted by
the growth of the physical volume of the 3-space, while for
radiation an extra $a^{-1}$ factor comes in from the redshifting
of the particles energy. Hence, since $a$ is growing, at early
enough time the universe is radiation dominated. The {\it equality
time} is defined as the time at which the two contributions are
equal, \ie $\rho_r = \rho_m$, after which the universe becomes
matter dominated. Therefore
 \be
 \dfrac{a_{\eq}}{a_0} = \dfrac{\rho_r}{\rho_m}\bigg \vert_{\eta_0}
 \approx 3\cdot 10^{-3} \eqcomma
  \ee
or in terms of the {\it redshift } $z \equiv a_0/a -1 $ we have
 \be
 z_{\eq} \approx 3000 \eqdot
 \ee
The subscript $0$ indicates that the quantity is evaluated today.
The numerical estimate comes from the measurement of the present
day radiation density in the cosmic microwave background, which
together with the assumption of three massless neutrino families
yields
 \be
 \rho_r = 7.94 \cdot 10^{-34}
 \left( \dfrac{T_\text{CMB}}{2.737\;\text{K}}\right)^4
  \quad \text{g/cm}^3 \eqdot
  \ee
The matter content of the Universe is obtained from the
combination of CMB, large scale structure and supernov\ae~ type IA
measurements. We shall see in \SEC{chap:params:sec:normal} that
the CMB itself is a good probe to determine the redshift of
equality.

Since for a cosmological constant $w_\Lambda =-1$, $\rho_\Lambda =
\textrm{const}$, its contribution is negligible in the early
universe, and indeed for a redshift
 \be
 z \gg \left( \dfrac{\Om_m}{\Om_\La}\right)^3 -1 \approx 0.5
 \eqdot
 \ee
However, if $\Lambda \ne 0$, the late universe will be dominated
by the vacuum energy term. In that case, $a(t) \propto
\exp\left[(\Lambda/3)^{1/2} t\right]$ and the expansion becomes
exponential (in physical time).

It is customary to introduce the {\it critical energy density} as
the energy density for which the universe is flat
 \be \label{eq:critical_energy_density}
 \rho_{\crit} \equiv \frac{3 \Hbl^2}{8\pi G a^2} \eqdot
 \ee
We also define the {\it Hubble parameter} $H_0 \equiv \Hbl/a_0$
and the {\it fudge factor} $h$
 \be
 H_0 \equiv 100\,h \UUNIT{km}{} \UUNIT{s}{-1} \UUNIT{Mpc}{-1} \eqdot
 \ee
The critical energy density today then evaluates to
 \be
 \rho_{\crit}(\eta_0) \approx
 1.88 \cdot 10^{-29} \,h^2 \text{  g/cm}^3\eqdot
 \ee
At all times, the {\it density parameters} $\Om_X$ give the
contribution of the component $X$ in units of the critical energy
density:
 \begin{align}
 \Om_r (\eta) & \equiv  \frac{\rho_r}{\rho_\crit} \eqcomma \\
 \Om_m (\eta) & \equiv  \frac{\rho_m}{\rho_\crit} \eqcomma\\
 \Om_\La (\eta) & \equiv  \frac{\rho_\La}{\rho_\crit} = \frac{\La}{8\pi G \rho_\crit} \eqcomma\\
 \Om_\curv (\eta) & \equiv  \frac{-3\curv}{8\pi G a^2 \rho_\crit} \label{eq:cosmological_parameters_last}\eqdot
 \end{align}
By definition the sum of the density parameters has to be unity
 \be
 \Om_r(\eta) + \Om_m(\eta) + \Om_\La(\eta) + \Om_\curv(\eta) = 1
 \eqdot
 \ee
The physical energy density of the component $X$ is then given by
 \be
 \rho_X(\eta) = \Om_X(\eta) \rho_\crit(\eta) \eqcomma
 \ee
and in particular when evaluating this quantity at the present
time we define $\om_X \equiv \Om_X(\eta_0) h^2$ and write
 \be
 \rho_X(\eta_0) = \om_X \,1.88 \cdot 10^{-29} \text{  g/cm}^3\eqdot
 \ee

The definition (\ref{eq:cosmological_parameters_last}) expresses
the energy density due to the curvature of the spatial sections
for $\curv = \pm 1$. Since $\Om_\curv \propto \Hbl^{-2} \propto
\eta^2$, the curvature is always negligible in the early universe.
Various cosmological observations indicate that today $\Om_\curv
\approx 0$. However, if the universe is not exactly flat, this
would imply that at Planck time $\vert \Om_\curv \vert \approx
\mathcal{O}(10^{-60})$. The smallness of this number is the
essence of the ``flatness problem''. The inflationary mechanism
indeed naturally provides a solution for this fine tuning problem:
as the universe inflates quasi-exponentially, its curvature is
driven to 0.

A key quantity is the {\it angular diameter distance} $D_A(z)$:
consider an object of physical length $d$ sitting at a redshift
$z_1$ (corresponding to conformal time $\eta_1$ and radial
distance $r_1$), which is observed at our present position ($z_0 =
0, r_0 = 0$) under an angle $\theta$. Then the angular diameter
distance is defined as
 \be \label{eq:angular_diameter_distance}
 D_A(\eta_1) \equiv \frac{d}{\theta} = a(\eta_1) \chi(\eta_0 - \eta_1)
 \eqcomma
 \ee
where in the second equality we have used $d = \lambda a(\eta_1)$,
with $\lambda$ the comoving length of the object, and $\theta =
\lambda/\chi(r_1)$, noting that $r_1 = \eta_0 - \eta_1$ since
light travels on null geodesics. We can now integrate
\rr{eq:background_Friedmann_2} to find
 \be \label{eq:Delta_tau}
 \Delta \eta \equiv \eta_0 - \eta_1 = \frac{1}{H_0 a_0^2} \int_{a_1}^{a_0} \frac{\dr a}
 {\DS \left[ \Om_r  + \Om_m \frac{a}{a_0} + \Om_\curv \frac{a^2}{a_0^2}
 + \Om_\La \frac{a^4}{a_0^4}   \right]^{1/2}} \eqcomma
 \ee
This equation is more conveniently written in redshift space
 \be \label{eq:delta_tau_in_redshift}
 \Delta \eta  = \frac{1}{H_0 a_0} \int_{0}^{z_1} \frac{\dr z}
 {\left[ \Om_r (1+z)^4  + \Om_m (1+z)^3 + \Om_\curv (1+z)^2
 + \Om_\La \right]^{1/2}} \eqdot
 \ee
Recall that the quantities $\Om_X$ above are evaluated at the
present time. So if we know the physical length of an object at a
given redshift, and we measure the angle subtended by it on the
sky, we are in principle able to extract the value of the
cosmological parameters using \rr{eq:delta_tau_in_redshift}. The
CMB provides exactly such a standard rod on the sky: the acoustic
oscillations of the photon fluid just before recombination have a
characteristic length scale, which shows up as the first peak in
the angular power spectrum, see \SEC{chap:params;sec:acoustic}.
The redshift of recombination is also known with good accuracy,
hence the CMB measures with high precision the angular diameter
distance to the last scattering surface. This piece of information
alone is however insufficient to reconstruct completely the
matter-energy content of the Universe: this problem is known as
{\it geometrical degeneracy}, and it is explained in
\SEC{chap:params;sec:acoustic}.

\subsection{Boltzmann equation} \label{chap:intro;sec:Boltzmann}

At early time, the energy density of the universe is dominated by
the relativistic species, and to leading order we can neglect in
the contribution of non-relativistic components to the total
energy. As long as photons are in local thermodynamical
equilibrium, the photon temperature $T$ is related to the energy
density of radiation by
 \be \label{eq:temperature_of_the_photons}
 \rho_r = \frac{\pi^2}{30}g_\star T^4 \eqcomma
 \ee
where $g_\star$ counts the total number of relativistic degrees of
freedom
 \be
 g_\star \equiv \sum_b g_b \frac{T_b^4}{T^4} + \sum_f g_f \frac{T_f^4}{T^4}
 \ee
and $b$ and $f$ run over the bosonic and fermionic species
respectively. The factors $T_b$ and $T_f$ take into account
possible temperature differences between the photons and the other
relativistic particles. From \rr{eq:temperature_of_the_photons}
and $\rho_r \propto a^{-4}$ it follows that while the photons are
in thermodynamical equilibrium, $T \propto 1/a$.

For $T > 4000 \textrm{K} \approx 0.4 \ev$ hydrogen nuclei are
ionized, and photons are coupled to baryons via non-relativistic
Thomson scattering off free electrons, see
\SEC{chap:perts;sec:thomson}. As the temperature drops below $0.30
\ev$, corresponding to $z_{\dec} \approx 1100$, almost all the
hydrogen nuclei quickly recombine, the mean free path of photons
becomes larger than the Hubble length $1/\Hbl$: the universe
becomes transparent. This event is called {\it last scattering} or
{\it decoupling}.

After recombination, the photon distribution function
 \be
 \label{eq:photon_distribution_function}
 f(\eta, E) = \frac{1}{\exp(E/T) - 1}
 \ee
evolves according to the {\it collisionless Boltzmann equation},
which can be derived by requiring that the total derivative of $f$
with respect to the affine parameter $\lambda$ vanishes
 \be \label{eq:df_over_dlambda}
 \frac{\dr f}{\dr \lambda} = 0 \eqdot
 \ee
In general $f = f(\eta, x^i, E, n^i)$, where the momentum 4-vector
$p^\mu = (p^0, {\bf p})$ is written as
 \be
   p^\mu  = \frac{E}{a}(1, {\bf n}) \eqcomma
 \ee
with
 \begin{alignat}{2} \label{eq:def_photons_directions}
  & p^i   = \frac{\pnorm}{a}n^i \eqcomma \quad &
  & p^0   = \frac{E}{a} = \frac{\pnorm}{a} \eqcomma \\
  & \sqrt{p_i p^i}  \equiv \pnorm \eqcomma \quad &
  & n^i n^j {\ga}_{ij}  =  1 \eqdot \label{eq:def_photons_directions_2}
 \end{alignat}
From  \rr{eq:df_over_dlambda} we have
 \be \label{eq:df_dlambda_expandend}
 \frac{\partial f}{\partial \eta}
 +   \frac{\partial f}{\partial x^i}n^i
 +   \frac{\partial f}{\partial E}\dot{E}
 +   \frac{\partial f}{\partial n^i} \dot{n}^i = 0 \eqdot
 \ee
Because of isotropy, $\partial f/\partial n^i = 0$, while
homogeneity implies $\partial f/\partial x^i = 0$. Using the 0
component of the geodesics equation
 \be
 \frac{\dr p^\al}{\dr \la} + \Gamma^\al_{\mu\nu} p^\mu p^\nu = 0
 \eqcomma
 \ee
which in the FRW universe reads
 \be
 \dot{E} + \Hbl E = 0
 \ee
we obtain from \rr{eq:df_dlambda_expandend} the background
Boltzmann equation
 \be \label{eq:background_Bolzmann_equation}
 \frac{\partial f}{\partial \eta} - \Hbl E \frac{\partial
 f}{\partial E} = 0 \eqdot
 \ee
This equation is satisfied by any $f$ of the form $f = f(aE)$. We
conclude that after decoupling the energy of the cosmic photons is
redshifted by the expansion as $E \propto a^{-1}$. The black body
distribution, \rr{eq:photon_distribution_function}, retains its
spectrum. The spectrum of the \textcmb photons has been measured
very accurately by the FIRAS spectrometer onboard the COBE
satellite \citep{Fixsen:1996}, and was found to be exceedingly
close to thermal. Deviations from a perfect black body spectrum
can be measured by the Comptonization parameter $y$, the chemical
potential $\mu$ and the parameter $Y_{ff}$ describing
contamination by free-free emission. The $95\%$ confidence limits
on those parameters are
 \be
 \vert \mu \vert < 9 \cdot 10^{-5}\eqcomma\quad
 \vert y \vert < 1.2 \cdot 10^{-5}\eqcomma\quad
 \vert Y_{ff} \vert < 1.9 \cdot 10^{-5}\eqdot
 \ee

After decoupling, $T$ is no longer a temperature in the
thermodynamical sense, rather a parameter in the distribution
function, which drops as $T \propto a^{-1}$.

\section{Cosmological observations}
\label{chap:intro;sec:observations}

It is only in comparatively recent times that cosmology has become
a data driven science, in which theoretical hypothesis can be
falsified or validated against observational data. It is amazing
that only 15 years ago the total energy density of the universe
was known with order-of-magnitude accuracy only. Nowadays, most
cosmological parameters are constrained within a few percent. The
discovery and accurate mapping of CMB fluctuations has constituted
a major pillar in this evolution and represents a fundamental
cornerstone of modern cosmology, see \SEC{chap:data;sec:obs} for
an overview.

It is nevertheless of equal importance that many other
cosmological probes have been developed in parallel, and this for
at least two good reasons. Firstly, all observation suffers in one
form or in another from the {\it degeneracy problem}: only a
certain combination of cosmological parameters can be measured
accurately. Since degeneracy directions are different for
different observations, combining two or more measurements leads
to tighter constrains on the parameters we are interested in. The
second reason is that cosmologically relevant measurements are
intrinsically difficult. One obvious obstacle is that there is
only one universe for which the experimental conditions cannot be
manipulated at will. Very often the interesting physics is hidden
behind foreground emissions, poor statistical sampling, faint
signals and non-linearities. It is common to try and extract
cosmological information by using objects whose physical
properties are poorly understood, and in general systematics are
very difficult to assess in cosmology. Hence a cosmological
measurement is usually considered as valid only if confirmed by
one or more independent pieces of evidence.

The so-called $\Lambda$CDM {\it concordance model} is strongly
supported by several independent observational data. It is
generally accepted that our universe is very close to flat
($\Om_\curv \approx 0$); that it is dominated by ``dark energy''
($\Om_\Lambda \approx 0.7$), perhaps in form of vacuum energy, or
quintessence or a tracking scalar field; that around 25\% is
non-interacting cold dark matter, and that only the remaining 5\%
is constituted of baryons. If the three neutrino families of the
Standard Model of particle physics are not massless (as the large
mixing angle solution to the solar neutrino problem seems to
suggest), than their mass is bounded from above to be $m_\nu \lsim
\mathcal{O}(1) \ev$. Structure formation proceeded by
gravitational instability from quantum fluctuations stretched to
super-horizon scale by a period of superluminal expansion
(inflation). The simplest inflationary model, in which inflation
is driven by one single slow-rolling scalar field, successfully
predicts the absence of non-Gaussianity, the (predominantly)
adiabatic nature of the fluctuations and the almost scale
invariant spectral index ($n_\SCAL \sim 1$) for the perturbations.
The age of the universe, around $13$ Gyrs, easily accommodates the
oldest observed objects. For definiteness, in
\TAB{table:concordance_model} we give the parameters of what we
believe is a currently widely accepted ``concordance model'', to
which we will refer throughout this work for illustrative and
comparative purposes.
\begin{table}[tb]
 \centering
\begin{tabular}{|l  l l l|}
\hline Quantity      &  & Value & Observations \\ \hline
Baryon density       & $ \om_b $      & $0.024$  &CMB, BBN, light elements abundance\\
Cold dark matter density & $ \om_\cdm$ & $0.116$ & CMB+LSS+SN, clusters\\
$\La$ density        & $\om_\La$      &  $0.378$ & CMB+LSS+SN+weak lensing\\
Hubble constant      & $ h$           & $0.72$   & HST, SZ, strong lensing\\
Optical depth        &  $\treion$     & $0.17$   & CMB \\
Spectral index & $n_\SCAL$            & $1.00$   & CMB, LSS, Lyman-$\al$, clusters\\
\dline
Baryons              & $\Omega_b$      & $0.046$  &\\
Cold dark matter     & $\Omega_\cdm$      & $0.224$  &\\
Cosmological constant& $\Om_\La    $   & $0.73$  &\\
Radiation            & $\Omega_{\textrm{rad}}$  & $7.95 \cdot 10^{-5}$ & CMB \\
Massless $\nu$ families & $N_\nu $     & 3.04 & CMB+LSS\\
Curvature            & $\Omega_{\curv}$  & $0.00$  & CMB+LSS+SN+weak lensing\\
\dline
 Initial conditions   & \multicolumn{2}{l}{purely adiabatic} & CMB
\\ \hline
\end{tabular}
\caption[Parameters of present-day ``$\La$CDM cosmological
concordance model''.]{\label{table:concordance_model} Parameters
of today's ``$\La$CDM cosmological concordance model'', which is
in good agreement with most of the current observational evidence
coming from CMB \citep{Spergel:2003cb}, large scale structures
(LSS) \citep{Tegmark:2003ud}, Big-Bang Nucleosynthesis (BBN)
\citep{Sarkar04}, supernov\ae~ type Ia (SN) \citep{Tonry:2003zg},
strong \citep{Kochanek:2003pi} and weak lensing
\citep{Contaldi:2003hi}, Lyman-$\al$ absorption systems
\citep{Seljak:2003jg} and galaxy clusters \citep{Bahcall:2002wx}
observations.}
\end{table}

Apart from CMB anisotropies, which we will discuss in depth in the
rest of this work, we briefly present some of the pieces of
observational evidence which corroborate the (presently) standard
$\La$CDM scenario.

\subsection{Big-Bang Nucleosynthesis}

Big-Bang Nucleosynthesis is based on the Standard Model of
particle physics, and gives predictions for the abundance of light
elements D, $^3$He, $^4$He and $^7$Li synthesized in the early
Universe, which are in good overall agreement with the observed
abundances, see \cite{Olive:1999ij} for a review and
\cite{Sarkar04} for more recent results.

Below a temperature $T \sim 1 ~\MeV$ the neutron-proton conversion
rate falls below the expansion rate, and the neutron to proton
ratio freezes out at the value $n/p = \exp{(- Q / T)} \approx
1/6$, where $Q= 1.293~\MeV$ is the neutron-proton mass difference.
The light elements production starts slightly afterwards, at a
temperature $T \sim 0.1 ~\MeV$, which is well below the binding
energy of deuterium, $B_D = 2.23 ~\MeV$ because photo-dissociation
prevents the formation of deuterium and other nuclei until then.
By this time, $\beta$-decay has further reduced the
neutron-to-proton ratio to $n/p \approx 1/7$. The surviving
neutrons end up almost completely in $^4$He, while the abundance
of the other elements is sensitively dependent on the nuclear
reactions rates, which in turn depend on the baryon density,
usually expressed with respect to the photon density by defining
the parameter $\eta_{10}$ as
 \be \label{eq:def_eta10}
 \eta_{10} \equiv \dfrac{n_b}{n_\ga} \times 10^{10} \approx 274
 \cdot
 \om_b(\eta_0) \eqcomma
 \ee
where $\eta_0$ is the conformal time today. A simple counting
argument, see \rrp{eq:Yp_estimate}, yields that the primordial
$^4$He mass fraction is about $25\%$, while the number densities
of the other elements relative to hydrogen turn out to be of the
order D/H $\sim$ $^3\text{He/H} \sim 10^{-5}$ and $^7\text{Li/H}
\sim 10^{-10}$ . The predictions are very reliable and accurate,
with a residual numerical uncertainty which depends on the
experimentally determined reaction rates; interestingly, it turns
out that most of this uncertainty is associated with our only
approximative knowledge of the neutron lifetime
\citep{Cuoco:2003cu}. The other free parameter of BBN is the
radiation density in the early Universe, which sets the Hubble
expansion rate and therefore determines the freeze-out temperature
for the weak reactions and is usually parameterized with the
equivalent number of (massless) neutrino families. We comment on
the possibility of a non-standard number of neutrino families and
discuss BBN-related issues in \SEC{chap:beyondsp;sec:Neff}.

In summary, agreement between the abundance of the light elements
as inferred from astrophysical measurement and the corresponding
prediction of BBN is a powerful tool to verify the Standard Model
of particle physics. In \SEC{chap:bspII;sec:bbn} we present in
detail the determination of light elements, discuss the slight
discrepancies between them and the BBN predictions and give some
possible interpretations. However, the overall agreement is
satisfactory, and (for a standard number of neutrino families) the
light elements abundances can be explained by a baryon density
compatible with the one independently inferred from CMB, namely
$\eta_{10} \sim 5.5$ or $\om_b \sim 0.02$.

\subsection{Matter distribution}

Structure formation proceeds from small inhomogeneities in the
matter distribution which grow by gravitational instability,
eventually giving rise to the large scale structures like galaxies
and clusters observed today. From the determination of the
statistical distribution of matter one tries to reconstruct the
properties of the primeval fluctuations, and to validate the
structure formation model.

In \SEC{chap:cmb;sec:mattpower} we introduce the linear matter
power spectrum $P_m(k)$, which represents the Fourier transform of
the 2-point correlation function for the matter density contrast.
Observations of the distribution of galaxies out to a redshift $z
\sim 0.1$ probe the galaxy-galaxy power spectrum, $P_{gg}$; the
Sloan Digital Sky Survey, for example, currently contains
approximately $2 \times 10^5$ galaxies \citep{Tegmark:2003uf}, and
upon completion will achieve $10^6$ galaxies. The problem is then
to relate $P_{gg}(k)$, which probes the luminous matter
distribution, with the underlying $P_m(k)$ describing (mostly) the
dark matter distribution. This is the issue of {\it bias},
introduced by Kaiser to explain the different amplitudes of the
correlation function for galaxies and for clusters
\citep{Kaiser:1984sw,Kaiser:1987qv}: the basic idea is that
galaxies represent peaks of the matter distribution, and therefore
our observations of $P_{gg}$ actually select only the regions of
the underlying matter distribution above some threshold. This
concept has been extended to various kinds of bias:
luminosity-dependent, morphology-dependent, color-dependent bias,
scale-dependent bias, anti-bias, and others. The simplest form is
to assume a scale-independent bias, which seems to be justified on
large (linear) scales, setting
 \be
 P_{gg}(k) = b^2 P_m(k) \quad \text{ for } k < k_{\text{NL}}
 \approx 0.3\; h \mpc^{-1}
 \ee
with the bias parameter $b$ which is just an unknown constant
factor (\citealp[see however \eg ][]{Durrer:2002zu} for a critical
discussion). In practice, this prescription amounts to introducing
a free parameter which controls the amplitude of the matter power
spectrum. There are methods which allow to determine the bias from
the higher-order n-point function of the distribution: for
instance \cite{Verde:2001sf} found $b = 1.04 \pm 0.11$ from the
data of the 2dF Galaxy Redshift survey \citep{Colless:2001gk},
which plans to measure $2.5 \times 10^5$ galaxies.

One can also consider the distribution of galaxy clusters as a
function of redshift, which in principle one should be able to
predict by using hydro-dynamical simulations. Comparison with the
observed distribution would then allow to constrain the
cosmological parameters. This simple sounding program is in
practice complicated by the need of accurately simulating all the
relevant physics, and despite the great amount of computational
power nowadays available, recent works in the field still involve
many approximations. As a result, cluster data mainly constrain a
combination of the matter power spectrum at clusters scales and
the value of $\Om_m$, see \eg \cite{Bahcall:2002wx}.

 Another way to probe the mass distribution is offered
by the Lyman $\al$ forest, the absorption lines in the spectra of
distant quasars produced by the neutral hydrogen in regions of
overdense intergalactic gas along the line of sight at a redshift
$2-4$ \citep{Croft:2000hs}. Since the overdensities probed at
these redshifts are still close to the linear regime, one hopes to
be able to connect the observations to the matter power spectrum
by modelling numerically the relevant physics
\citep{Mandelbaum:2003km,Seljak:2003jg}.

Weak gravitational lensing is very promising as a tool to
constrain cosmological parameters, and in particular the matter
distribution. It uses the distortion in the images of distant
galaxies induced by inhomogeneities in the intervening matter
distribution \citep{Kaiser:1993ps}, and reconstructs with a
statistical analysis the so-called ``cosmic shear''
\citep{Wittman:2000tc,Bartelmann:1999yn}. The technique is now
rapidly becoming mature to help constrain the matter budget
\citep{Contaldi:2003hi}.

One of the most important aspects is that all of the above
observations can be combined to achieve superior constraining
power on the CDM model parameters, while testing the consistency
of the theory itself, or the soundness of each data-set. A
technique to merge galaxy surveys, cluster distribution, weak
lensing and Lyman $\al$ data with the CMB to probe a larger
portion of the matter power spectrum is presented in
\cite{Tegmark:2002cy}. There is presently a general agreement that
the matter content of the Universe is low, around $\Om_m \sim
0.3$.

\subsection{Type Ia supernov\ae}

Supernov\ae\ (SN) are classified according to their spectrum: the
type Ia is characterized by the absence of hydrogen (the ``I''),
and by strong silicon features (the ``a''). The standard picture
is a progenitor binary system, with a white dwarf which accretes
matter from its companion until it reaches the Chandrasekhar
limit, and the gravitational infall triggers a thermonuclear
explosion which we observe as a supernova. At the peak of its
brightness, a SN can easily exceed the luminosity of its host
galaxy, making it a promising candidate to measure distances out
to very high ($z \sim 1 - 2$) redshifts.

Their most important property is the remarkable homogeneity in
their spectra, in the shape of their light-curve and in their peak
absolute magnitude, which makes them {\it nearly} ``standard
candles''. In fact, it was  discovered that intrinsically brighter
SNIa decline more slowly than dim ones \citep{Hamuy:1996st}. By
exploiting an empirical correlation between the shape of the light
curve and the intrinsic luminosity, and correcting for extinction
effects via measurements at different wavelengths, it is
nevertheless possible to produce a ``calibrated candle'', with a
very narrow peak magnitude dispersion \citep{Riess:1996pa}. For a
review of the cosmological applications, see \eg
\cite{Filippenko:2003pr}.

\begin{figure}[tb]
 \centering
 \includegraphics[width=\onefigwidth]{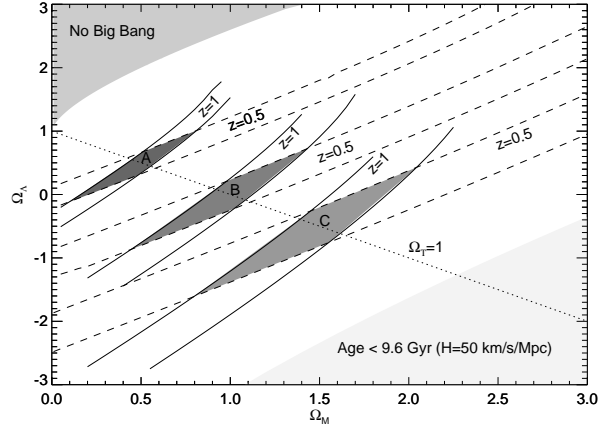}
\caption[Illustration of the determination of ($\Om_m, \OLa)$
using supernov\ae\ data.]{Illustration of the determination of
($\Om_m, \OLa)$ using supernov\ae\ data: the dashed (solid) curves
are lines of constant $D_L$ for the given measured apparent
magnitude of a standard candle at a redshift $z=0.5$ ($z=1.0$). If
the apparent magnitude $m$ can be measured with  accuracy $\Delta
m = 0.05$ combining the two observations gives the dark shaded
allowed region for ($\Om_m, \OLa$). Figure reprinted from
\cite{Goobar:1995af}.} \label{fig:sn_plane}
 \end{figure}

The measured apparent magnitude $m$ is related to the absolute
magnitude $M$ via the {\it luminosity distance} $D_L$
 \be \label{eq:magitude_redshift}
 m = M + 5 \log \left[ H_0 D_L(z, \Om_m, \OLa) \right] + K
 \ee
where the ``$K$-correction'' compensates for the difference in
wavelength of the emitted and received photons due to the
expansion, and the luminosity distance of an object at redshift
$z$ is defined in terms of the intrinsic luminosity $L$ and of the
measured flux $\ell$ as
 \be \label{eq:luminosity_distance}
 D_L(z) \equiv \left( \frac{L}{4\pi \ell} \right)^{1/2} \eqdot
 \ee
The luminosity distance is related to the angular diameter
distance by $D_L(z) = (1+z)^2 D_A(z)$. Supernov\ae\ essentially
measure the angular diameter distance over a redshift range of $z
\sim 0.5 - 2$, much lower than range probed by the CMB. At such
low redshift, the radiation content is negligible, and with
$\Om_\curv = 1 - \Om_m - \Om_\La$ we obtain from
(\ref{eq:angular_diameter_distance}) and
(\refp{eq:delta_tau_in_redshift}) \be
 \begin{split}
 H_0 D_L (z_1, \Om_m, \Om_\La) = &
 \dfrac{1+z_1}{\sqrt{\vert \Om_\curv} \vert} \times \\
 & \chi  \left(
   \dfrac{1+z_1}{\sqrt{\vert \Om_\curv} \vert} \int_0^{z_1}
   \left[ (1+z)^2(1 + z \Om_m) - \Om_\La z(2+z) \right]^{-1/2} \dr z
   \right) \eqcomma
 \end{split}
 \ee
where the function $\chi$ is defined in
\rrp{eq:define_chi_function}. Notice that magnitude-redshift
relation (\ref{eq:magitude_redshift}) does not depend on the
Hubble parameter. Therefore, assuming that we are able to reliably
reconstruct the intrinsic luminosity $M$, from the measurement of
one SN \rr{eq:magitude_redshift} yields one degeneracy line for
the possible values of $(\Om_m, \OLa)$. By measuring a second
standard candle at $z_2 \neq z_1$ we are able to determine the
intersection of the degenerate luminosity distance lines in the
$(\Om_m,\OLa)$ plane, and thus to measure separately the matter
and cosmological constant content. When we add the measurements
error, both lines widen to two strips, and we obtain a region of
confidence for the two parameters, independently on the Hubble
parameter, see \FIG{fig:sn_plane}.

In practice, of course, a larger number of measurements is
necessary, and it turns out that the approximate combination
$\Om_m - \OLa$ is well constrained, as it is intuitively clear
from \FIG{fig:sn_plane}. For instance, \cite{Tonry:2003zg} found
 \be
 \OLa - 1.4 \Om_m = 0.35 \pm 0.14 \quad \text{(at 1$\si$).}
 \ee
This degeneracy direction is almost orthogonal to the one in
inferred from the angular diameter distance at $z \sim 1100$
measured by the CMB, \CF \FIG{fig:shift_param2D}. Combination of
\sn and CMB data is thus a very effective way to break the angular
diameter distance degeneracy and to constraint the matter and
vacuum energy contents separately. As we have seen, observations
of the matter distribution on large scales independently constrain
the matter density parameter: it is a remarkable achievement of
modern cosmology that this ``cosmic complementarity'' seems to be
pointing toward the same value, namely $\Om_m \sim 0.3$ and $\OLa
\sim 0.7$, see \eg \cite{Spergel:2003cb}. At the same time, the
puzzle of the nature of dark matter and dark energy remains
unsolved, and we offer some further remarks regarding the
cosmological constant in \SEC{chap:genic;sec:lambda}.

\chapter{Cosmological perturbation theory}
\label{chap:perturbation}
In order to understand the physical origin of CMB anisotropies, we
are interested in studying the evolution of perturbations in the
photon distribution function, by perturbing at linear order around
the ``background'' solution for the homogeneous and isotropic \FRW
~(FRW) universe of \SEC{chap:intro;sec:FRW_cosmology}. That linear
perturbation theory is sufficient to describe almost all aspects
of CMB physics is a consequence of the smallness of the
fluctuations.

In \SEC{chap:pert;sec:variables} we introduce the relevant
perturbation variables, discuss the issues of gauge
transformations and gauge invariant formalism, extend the
treatment to multiple fluids and define entropy perturbations. We
then present the perturbed Einstein
(\SEC{chap:perts;sec:einstein}) and conservation equations
(\SEC{chap:perts;sec:conservation}) for an Universe filled with
four different particle species: baryons, cold dark matter (CDM),
photons and massless neutrinos. The Bardeen equation is presented
in \SEC{chap:perts;sec:bardeen}, while
\SEC{chap:perts;sec:boltzmann} is devoted to the derivation of the
collisionless Boltzmann equation, which describes massless
neutrinos and photons after decoupling. The last section
\SEC{chap:perts;sec:thomson} concerns the Thomson scattering
process which couples photons and baryons before recombination,
and explains the origin of CMB polarization.

Cosmological perturbation theory in the four-dimensional FRW
universe is a well studied subject, see \eg \cite{Kodama:1985bj,
Mukhanov:1992me,Ma:1995,Durrer:1993db}. More recently, the
formalism has been extended to higher-dimensional manifolds,
involving extra dimensions \cite[see \eg][]{Riazuelo:2002mi}, in
view of the recent interest in string theory motivated braneworlds
scenarios.

\section{Perturbation variables}
\label{chap:pert;sec:variables}

In this section, background (unperturbed) quantities are denoted
by an overline, so that the perturbed energy density, \eg, is
denoted by $\rho = \ol{\rho} + \delta \rho$. The background
quantities depend on time only, while the linear perturbations are
functions of time and of the 3-space coordinate, \ie $\delta \rho
= \delta \rho (\eta, {\bf x})$.

\subsection{Metric perturbations}
\label{chap:pert;sec:metric_perts}

We perturb to linear order the FRW metric of
\rrp{eq:FRW_metric_tau} by setting
 \be \label{eq:pertd_metric}
 g_{\mu\nu} \dr x^{\mu} \dr x^{\nu} = \ol{g}_{\mu\nu} \dr x^{\mu} \dr x^{\nu}
 + a^2 h_{\mu\nu} \dr x^{\mu} \dr x^{\nu}
 \ee
where the perturbation $ h_{\mu\nu} $ is given by
 \be \label{eq:pertd_metric_expanded}
 h_{\mu\nu} \dr x^{\mu} \dr x^{\nu} =
 -2A\dr \eta^2 + 2 B_i \dr x^i \dr \eta + 2H_{ij} \dr x^i \dr x^j
 \eqdot
 \ee
The perturbation variables $A, B_i, H_{ij}$ are arbitrary
functions of the 4-coordinate vector \mbox{$x^\mu = (\eta, {\bf
x})$}.

It is convenient to split them into components which transform
irreducibly under the rotation group $SO(3)$. The vector field
$B_i$ can thus be written as the sum of a gradient of a scalar and
a divergenceless component (vector)
 \be
 B_i = B_{\vert i} + B^\vc_i \eqcomma \quad B_i^{\vc \vert i} =
0 \eqdot
 \ee
We split $H_{ij}$ into an isotropic and an anisotropic part
 \be
 H_{ij} = C \ol{\gamma}_{ij} + E_{ij} \eqcomma
 \ee
and $E_{ij}$ is further decomposed in irreducible scalar (spin 0),
vector (spin 1) and tensor (spin 2) components as
 \be
 E_{ij} = E_{\vert ij}
 + \frac{1}{2}(E_{j\vert i}^{\vc} + E_{i\vert j}^{\vc} )
 + E_{ij}^\ts \eqcomma
 \ee
 where
 \begin{align}
  E_{\vert j}^{\vc j} = E_{\vert j}^{\ts ij} = 0 \quad
&  \textrm{(divergenceless)}\eqcomma \\
 E_j^{\ts j} = 0  \quad  & \textrm{(traceless)} \eqdot
 \end{align}
Note that at this stage we are still working in real space and we
do not perform an harmonic analysis of the perturbation variables
(\citealp[see][]{Kodama:1985bj,Durrer:1993db} instead). At linear
order, the different spin components do not mix, and we can treat
them separately.

\subsection{Perturbations of the energy-momentum tensor}
\label{chap:pert;sec:energy_impulse_perts}

The perturbed energy-momentum tensor is obtained by perturbing in
\rr{eq:background_energy_tsr} the energy density
 \be
 \rho = \ol{\rho} + \delta \rho = \ol{\rho}(1 + \delta) \eqcomma
 \quad \textrm{with} \quad
 \delta \equiv \delta \rho / \ol{\rho} \eqcomma
 \ee
the pressure
 \be
 P = \ol{P} + \delta P \equiv \ol{P}(1 + \pi_L)  \eqcomma
 \quad \textrm{with} \quad
 \pi_L \equiv \delta P / \ol{P} \eqcomma
 \ee
and the space components of the observer's 4-velocity
 \begin{align}
 u^i & =  \delta u ^i \equiv - \frac{v^i}{a} = -\frac{1}{a}(v^{\vert i} + v^{\vc i})
 \eqcomma \label{eq:pertd_u^i} \\
 u^0 & =  \ol{u}^0 + \delta u^0 = \frac{1}{a}(1 - A) \eqcomma
 \end{align}
and the second line follows from the norm of the 4-velocity $u^\mu
u_\mu = -1$.

The perturbation of the energy-momentum tensor is then written as
 \be
 \delta T_{\mu \nu} = \left( \ol{\rho}\delta +
 \ol{P}\pi_L\right)\ol{u}_\mu \ol{u}_\nu
 +\left(\ol{\rho} + \ol{P}\right)\left(\delta u_\mu \ol{u}_\nu + \delta u_\nu
 \ol{u}_\mu\right) + \ol{P}\left(\pi_L \ol{g}_{\mu \nu} + a^2 h_{\mu
 \nu} + a^2 \Pi_{\mu \nu} \right) \eqcomma
 \ee
where we have introduced the {\it anisotropic stress perturbation}
$\Pi_{\mu \nu}$, which is a traceless tensor and orthogonal to the
4-velocity, $u^\mu \Pi_{\mu \nu} = 0$. It describes off-diagonal,
space-space perturbations in the stress-energy tensor, and can be
split into a scalar $\Pi$, a divergenceless vector $\Pi^\vc_i$ and
a trace-free tensor part $\Pi_{ij}^\ts$, according to:
 \be
 \Pi_{ij} = (\nabla_i \nabla_j - \frac{1}{3} \ol{\ga}_{ij} \nabla_k \nabla^k ) \Pi
 + \frac{1}{2}(\Pi_{i\vert j}^\vc + \Pi_{j\vert i}^\vc )
 + \Pi_{ij}^\ts \eqcomma
 \ee
The perturbation components of the stress-energy tensor therefore
take the form
 \begin{subequations} \label{eq:perturbed_energy_tsr}
 \begin{align}
 \delta \tensud{T}{0}{0} &= - \ol{\rho} \delta \eqcomma \\
 \delta \tensud{T}{0}{i} &= (\ol{\rho} + \ol{P})(B_i - v_i) \eqcomma\\
 \delta \tensud{T}{i}{0} &= (\ol{\rho} + \ol{P}) v^i \eqcomma\\
 \delta \tensud{T}{i}{j} &= \ol{P}(\tensud{\ol{\gamma}}{i}{j} \pi_L + \tensud{\Pi}{i}{j}) \eqdot
 \end{align}
 \end{subequations}

\subsection{Gauge transformations}
\label{chap:pert;sec:gauge_trafo}

By choosing the background spacetime manifold and metric to be of
the FRW form, we implicitly assume that for all quantity of
interest $Q$ we are able to define a spatially averaged $\ol{Q}$,
which represents the background, homogeneous and isotropic value
of $Q$ on $(\ol{\mathcal{M}}, \ol{g})$. Consider now a slightly
perturbed manifold, $\mathcal{M}_{\mbox{\pert}}$, endowed with a
coordinate system $x^\mu$. The value of $Q$ on
$\mathcal{M}_{\mbox{\pert}}$ depends on the choice of the
coordinate system, $Q_{\mbox{\pert}} = \ol{Q} + \delta Q(x^\mu)$.
Along with $x^\mu$, any other coordinate system which leaves
$\ol{g}$ invariant is admissible, \ie we can arbitrarily transform
the coordinates by an infinitesimal amount
 \be \label{eq:gauge_trafo}
 x^\mu \ra y^\mu = x^\mu + \delta x^\mu
 \ee
thereby obtaining for $Q$ in this newly defined coordinates
 \be \label{eq:Stewart_Walker_Lemma}
 Q_{\pert}(x^\mu) \ra  Q_{\pert}(y^\mu) =  Q_{\pert}(x^\mu) + \Lie_{\delta x}(\ol{Q})
 \eqcomma
 \ee
where $\Lie_{X}(\ol{Q})$ is the Lie derivative of $Q$ with respect
to the vector field $X$, see \eg \cite{Straumann}. Such
infinitesimal coordinate transformations are called {\it gauge
transformations}, and the above result is known as Stewart--Walker
Lemma. Fixing the coordinate system on
$\mathcal{M}_{\mbox{\pert}}$ is called a {\it gauge choice}.
Clearly, physical observables are geometrical quantities, and are
therefore independent of the coordinate system in which they are
calculated. The form of the equations, however, can be very
different according to the gauge choice. It is often convenient to
fix the gauge in the way which is best suited for the problem at
hand.

The gauge transformation \rr{eq:gauge_trafo} can be written in all
generality as
 \be \label{eq:gauge_trafo_expandend}
 \delta x^0 = T \eqcomma \quad \delta x^i = L^{\vert i} + L^{\vc
 i} \eqdot
 \ee
By applying the transformation law (\ref{eq:Stewart_Walker_Lemma})
to the perturbed metric (\ref{eq:pertd_metric}) under a gauge
transformation of the type (\ref{eq:gauge_trafo_expandend}), we
obtain the following transformation properties for the metric
variables:
\begin{subequations} \label{eq:gauge_trafo_metric}
 \begin{align}
 A  & \ra A + \Hbl T + \dot{T} \eqcomma \label{eq:trafo_for_A}\\
 B  & \ra B - T + \dot{L} \eqcomma\\
 C  & \ra C + \Hbl T \eqcomma \\
 E  & \ra E + L \eqcomma \\
 B^{\vc i}   & \ra B^{\vc i} + \dot{L}^{\vc i} \eqcomma \\
 E^{\vc i}   & \ra E^{\vc i} + L^{\vc i} \eqcomma \\
 E^{\ts ij}  & \ra E^{\ts ij} \eqdot
 \end{align}
\end{subequations}

The same procedure applied on the background stress-energy tensor
$\ol{T}_{\mu\nu}$ and 4-velocity $\ol{u}^{\mu}$ gives for the
matter perturbation variables:
\begin{subequations} \label{eq:gauge_trafo_matter}
 \begin{align}
  \delta & \ra \delta - 3 T \Hbl (1 + w) \eqcomma \\
  \pi_L  & \ra \pi_L - \frac{3 c_s^2}{w}(1+w) \Hbl T \eqcomma \\
  \Pi & \ra \Pi \eqcomma \\
  v & \ra v + \dot{L} \eqcomma \\
  v^{\vc}_i & \ra v^{\vc}_i + \dot{L}^{\vc}_i \eqcomma \\
  \Pi^{\vc}_i & \ra \Pi^{\vc}_i \eqcomma \\
  \Pi^{\ts}_{ij} & \ra \Pi^{\ts}_{ij} \eqdot
 \end{align}
 \end{subequations}

In order to completely fix the gauge, we need to specify in
\rr{eq:pertd_metric_expanded} the functional form of two scalar
functions, corresponding to a specific choice for  $(T, L)$, and
one vector, corresponding to a choice for $L^{\vc i}$. In the
following, we briefly summarize some popular gauge choices.

\subsubsection*{Longitudinal gauge}

Longitudinal gauge (also sometimes called ``Newtonian gauge'') is
defined by requiring $B = E = B^{\vc i} = 0$, so that the
perturbed metric element takes the form
 \be  \label{eq:long_gauge}
 \dr s^2 = a^2 \left[ - (1 + 2 \Psi) \dr \eta^2 + (1 - 2 \Phi) \ol{\ga}_{ij} \dr x^i \dr x^j
 \right]\eqcomma
 \ee
and we have defined the {\it Bardeen potentials} $\Psi = A$ and
$\Phi = -C$ \citep{Bardeen:1980kt}, which represent the
gravitational time dilation and the perturbation to the 3-space
curvature, respectively. From any other gauge, the transformation
$T = B - \dot{E}$, $L = -E$ and $\dot{L}^{\vc i} = - B^{\vc i}$
leads to the longitudinal gauge.

\subsubsection*{Flat slicing gauge}

This gauge owns its name to the choice $E = C = E^{\vc i} = 0$,
which makes the spatial hypersurfaces unperturbed, and the metric
element is
 \be  \label{eq:flat_slicing_gauge}
 \dr s^2 = a^2 \left[ - (1 + 2 A) \dr \eta^2 + 2 B_i \dr x^i \dr \eta + \ol{\ga}_{ij} \dr x^i \dr x^j
 \right]\eqdot
 \ee
The coordinate transformation which leads to flat slicing gauge is
$T = -C/\Hbl$, $L = - E$ and $L^{\vc i} = - E^{\vc i}$.

\subsubsection*{Synchronous gauge}

In synchronous gauge, constant time hypersurfaces are orthogonal
to the 3-space (hence the name), \ie $(\eta, x^i)$ are Gaussian
coordinates. This can be obtained by imposing $A = B = B^{\vc i} =
0$. Thus the metric presents perturbations in the space-space part
only, and it is often written as
\begin{subequations} \label{eq:synchronous_gauge}
\begin{align}
 \dr s^2 & = a^2 \left[ - \dr \eta^2 +  (\ol{\ga}_{ij} + h_{ij})\dr x^i \dr
 x^j \right] \eqcomma\\
 h_{ij} & \equiv h_{\vert ij} (\eta, {\bf x})
 + ( \nabla_i \nabla_j - \frac{1}{3} \ol{\ga}_{ij} \nabla_k \nabla^k)6 \eta(\eta, {\bf x}) \eqdot
\end{align}
\end{subequations}

The above choice does not fix completely the gauge: in fact, the
gauge transformation which leads to synchronous gauge is
 \begin{subequations}
 \begin{align}
 T & = - \dfrac{1}{a}\int a A \dr \eta + \dfrac{\al}{a} \\
 L & = \int(T-B)\dr \eta + \beta \\
 L^{\vc i} & = -\int B^{\vc i}\dr \eta + \beta^{\vc i} \eqcomma
 \end{align}
 \end{subequations}
which presents a residual gauge freedom in the four arbitrary
integration constants $\al$ and $\beta^i = \beta^{\vert i} +
\beta^{\vc i}$ (where $\beta^{\vc i}$ must be divergenceless). The
four constants correspond to different choices of the constant
time hypersurface and of the spatial coordinates on it. This leads
to the presence of fictitious ``gauge modes'' in the perturbation
equations, which must be removed because they are just an artifact
of the choice of the coordinate. Despite this difficulty,
synchronous gauge is quite popular in the literature.

\subsubsection*{Comoving gauge}
\label{par:comoving_gauge}

In the comoving gauge the total bulk velocity vanishes, $\delta
\tensud{T}{0}{i} = 0$, which translates into the condition $B_i =
v_i$. In order to completely fix the gauge one further requires
$E=0$ and $E^{\vc i}=0$. This is achieved with the transformation
$T = B - v - \dot{E}$, $L = -E$ and ${L}^{\vc i} = - E^{\vc i}$.
This gauge is the one which resembles most the gauge invariant
formalism (defined below), since for the variables in comoving
gauge we have
 \begin{alignat}{2}
 C      & = - \zeta & & \quad \text{ see \rr{eq:def_curvature_pert}}\notag \\
 \delta & = D       & & \quad \text{ see \rr{eq:def_gi_D}} \notag\\
 \delta_\al & = \Delta_\al  & & \quad \text{ see \rr{eq:def_Delta}}\notag \\
 v      & = V       & & \quad \text{ see \rr{eq:def_gi_V}} \eqdot
 \end{alignat}

\subsection{Gauge invariance}

General covariance guarantees that all equations in general
relativity can be written in a form which is independent of the
gauge choice \citep{Bardeen:1980kt,Kodama:1985bj,Durrer:1993db}.
From (\ref{eq:Stewart_Walker_Lemma}) it follows that for all
tensor fields with vanishing or constant background contribution,
so that $\Lie_X(\ol{Q}) = 0 \quad \forall X$, we can construct
{\it gauge invariant perturbation equations}. Such perturbation
variables are invariant under a gauge transformation of the type
\rr{eq:gauge_trafo}. Since we can cast all general relativistic
equations in the form $Q = 0$, it is always possible to construct
gauge invariant perturbation equations \citep{Stewart:1977}.

This approach has the advantage of leading to equations which are
independent of the coordinate choice, and which are often easier
to interpret physically. Furthermore, gauge independent equations
are free from spurious gauge modes. In order to write down the
relevant gauge invariant perturbation equations, we make use of
the transformation properties of the metric and matter variables
under a change of gauge, Eqs.~(\ref{eq:gauge_trafo_metric}) and
(\ref{eq:gauge_trafo_matter}).

\subsubsection*{Metric variables}

From \rr{eq:gauge_trafo_metric} we can construct the following 4
gauge invariant metric variables:
 \begin{subequations}
  \begin{align}
 \Phi & \equiv -C - \Hbl(B-\dot{E}) \eqcomma \\
 \Psi & \equiv A + \Hbl(B - \dot{E}) + (\dot{B} - \ddot{E})
         \eqcomma \label{eq:def_gi_Psi}\\
 \Sigma^{\vc}_i & \equiv \dot{E}^{\vc}_i - B^{\vc}_i
 \label{eq:def_Sigmavec} \eqcomma \\
 H^{\ts}_{ij}   & \equiv E^{\ts}_{ij} \eqdot
  \end{align}
  \end{subequations}
The two scalar variables $\Phi$ and $\Phi$ are called Bardeen
potentials \citep{Bardeen:1980kt}. Another very useful variable is
the {\it gauge invariant curvature perturbation} $\zeta$, which is
defined as
 \be \label{eq:def_curvature_pert}
 \zeta \equiv -C + \Hbl (v-B) \eqcomma
 \ee
where $v$ is defined in \rr{eq:pertd_u^i}. From the constraint
equation (\ref{eq:constraint_V}), it follows that for a flat
universe, $\curv = 0$, the gauge invariant curvature perturbation
is related to the Bardeen potentials by
 \be \label{eq:curvature_pert_related_to_potentials}
 \zeta = \Phi + \frac{\Hbl}{\Hbl^2 - \dot{\Hbl}}(\Hbl \Psi +
 \dot{\Phi}) \eqdot
 \ee

There is only one gauge invariant vector perturbation constructed
out of metric variables, \rr{eq:def_Sigmavec}. Tensor variables
are automatically gauge invariant, since there is no spin-2
coordinate transformation.

\subsubsection*{Matter variables}

Because of the Stewart--Walker Lemma
(\ref{eq:Stewart_Walker_Lemma}), the variables $\Pi$, $\Pi^\vc_i$
and $\Pi^\ts_i$ are already gauge invariant, since the background
anisotropic stress vanishes.

From scalar matter variables alone we can construct the gauge
invariant variable
 \be \label{eq:def_gi_Gamma}
 \Gamma \equiv \pi_L - \frac{c_s^2}{w}\delta \eqcomma
 \ee
which measures the intrinsic non-adiabaticity of the matter
content. More precisely, as we shall see below, $\Gamma$ is
related to the entropy production rate. If the pressure is a
function of the local energy density only, $P = P(\rho)$, then we
can write
 \be
 \frac{\delta P}{\delta \rho} =  \frac{\dot{P}}{\dot{\rho}}
 \ee
and since by definition $\delta \rho = \delta \cdot \rho$, $\delta
P = \pi_L \cdot P$, it follows that $\Gamma = 0$. In the case of a
perfect fluid, $P = w \rho$ and $\Gamma$ vanishes. Non-zero
contributions to $\Gamma$ arise from the relative entropy of a
mixture of several fluid components, which is discussed in
\SEC{sec:multiple_fluids}.

The choice of a gauge invariant density contrast is not unique,
and requires the use of metric variables. Meaningful combinations
are
\begin{subequations} \label{eq:def_gi_D}
  \begin{alignat}{2}
  D_s & \equiv \delta - 3(1+w)\Hbl(B - \dot{E}) \quad &&\text{(longitudinal),} \label{eq:def_gi_Ds} \\
  D_g & \equiv \delta + 3(1+w)C \quad &&\text{(flat slicing),} \label{eq:def_gi_Dg}\\
  D   & \equiv \delta - 3(1 + w)\Hbl(B - v) \quad &&\text{(comoving). \label{eq:def_gi_D_comoving}}
  \end{alignat}
 \end{subequations}
On super-horizon scales, $D_s$ corresponds to the density contrast
in the longitudinal gauge; $D_g$ is the density contrast on
homogeneous 3-space hypersurfaces (flat slicing); $D$ reduces to
the density contrast in the comoving gauge. The distinction is
only important on super-horizon scales, since on small
(sub-horizon) scales, all the above variables reduce to the same
\citep{Durrer:2001gq}.

The remaining velocity perturbation can be written in gauge
invariant form as
 \begin{subequations} \label{eq:def_gi_V}
  \begin{align}
   V & \equiv v - \dot{E} \eqcomma \\
   V^{\vc}_i & \equiv v^{\vc}_i - \dot{E}^{\vc}_i \eqdot
  \end{align}
  \end{subequations}

Useful relations between those gauge invariant variables are
\begin{subequations}  \label{eq:relations_between_Ds}
 \begin{align}
 D_g & =   D_s - 3(1+w)\Phi \eqcomma \\
 D   & =   D_s + 3(1+w)\Hbl V \eqcomma \\
 D   & =   D_g + 3(1 + w) \zeta \eqcomma \\
 \zeta & =  \Phi + \Hbl V \label{eq:useful_relation_zeta}\eqdot
 \end{align}
 \end{subequations}

\subsection{Multiple fluids}
\label{sec:multiple_fluids}

 The above definitions assume that the universe is filled
with, or dominated by, only one fluid component. In a more
realistic modelling, we must account for the presence of several
matter components. We will usually consider four of them, namely
photons (subscript $\ga$), massless neutrinos (subscript $\nu$),
non-interacting cold dark matter (CDM, subscript $c$) and baryons
(subscript $b$). The subscripts $r$ (radiation) and $m$ (matter)
will refer generically to a relativistic ($w_r = 1/3$) and a
non-relativistic, dust-like ($w_m = 0$) fluid, respectively.
Variables without subscript designate the total perturbation.

If multiple matter components are present, the total perturbation
variables are the weighted sum of the variables for each
component:
\begin{subequations} \label{eq:def_multiple_comps_variables}
 \begin{align}
 \delta & = \sum_\al \frac{\ol{\rho}_\comp{\al}}{\ol{\rho}}
 \delta_\comp{\al} \eqcomma \\
 v^j    & = \sum_\al \frac{\ol{\rho}_\comp{\al} + \ol{P}_\comp{\al}}{\ol{\rho} + \ol{P}}
 v^j_\comp{\al}\eqcomma \\
 \Pi^{ij} & = \sum_\al \frac{\ol{P}_\comp{\al}}{\ol{P}}
 \Pi^{ij}_\comp{\al} \eqdot
 \end{align}
 \end{subequations}
The equation of state and the adiabatic sound speed are defined
for each component
 \be
 w_\comp{\al} \equiv
 \frac{\ol{P}_\comp{\al}}{\ol{\rho}_\comp{\al}}
 \quad \textrm{and} \quad
 c^2_\comp{\al} \equiv
 \frac{\dot{\ol{P}}_\comp{\al}}{\dot{\ol{\rho}}_\comp{\al}}
 \eqcomma
 \ee
and for the mixture we have
 \be
 w \equiv \frac{\ol{P}}{\ol{\rho}}
 \quad \textrm{and} \quad
 c^2_s \equiv \frac{\dot{\ol{P}}}{\dot{\ol{\rho}}}
 \eqdot
 \ee

The transformation properties of the variables for each components
are the same as for the total variables,
Eqs.~(\ref{eq:gauge_trafo_matter}). Hence for each matter
component we can define gauge invariant variables as in
Eqs.~(\ref{eq:def_gi_Gamma}, \ref{eq:def_gi_D},
\ref{eq:def_gi_V}), yielding for the scalar part:
\begin{subequations}
  \begin{align}
  \Gamma_\comp{\al} & \equiv \pi_{\comp{\al}, L} - \frac{c_\comp{\al}^2}{w_\comp{\al}}\delta_\comp{\al}
  \eqcomma\\
  V_\comp{\al}      & \equiv v_\comp{\al} - \dot{E} \eqcomma \\
  D_{\comp{\al},s}  & \equiv \delta_\comp{\al} - 3(1+w_\comp{\al})\Hbl(B - \dot{E}) \eqcomma \\
  D_{\comp{\al},g} & \equiv \delta_\comp{\al} + 3(1+w_\comp{\al})C \eqcomma\\
  D_\comp{\al}      & \equiv \delta_\comp{\al} - 3(1 + w_\comp{\al})\Hbl(B - v_\comp{\al})
  \eqdot \label{eq:def_gi_D_al}
  \end{align}
 \end{subequations}

In the presence of multiple matter components, it is often useful
to work with the gauge invariant density contrast
 \be \label{eq:def_Delta}
 \Delta_{\comp{\al}} \equiv \delta_\comp{\al} - 3(1 + w_\comp{\al})
 \Hbl (B - v ) \eqcomma
 \ee
which corresponds to the density contrast in the gauge where the
total matter is at rest, \ie the comoving gauge introduced on page
\pageref{par:comoving_gauge}. Notice that on the right hand side
it appears the total velocity $v$, rather then the velocity of the
$\al$ component as in (\ref{eq:def_gi_D_al}). This new variable is
related to the density contrast in the flat slicing gauge by
 \be \label{eq:relation_Dg_Delta}
 \Delta_{\comp{\al}} = D_{g, \comp{\al}} + 3(1+w_\comp{\al})\left(\Phi + \Hbl
 V\right)\eqdot
 \ee

\subsection{Entropy perturbations}
\label{chap:pert;sec:entropy}

When more than one component is present, entropy perturbations can
arise even for a mixture of perfect fluids. The total
non-adiabaticity of the mixture is given by
(\ref{eq:def_gi_Gamma}), where the quantities appearing on the
right hand side have to be interpreted as total variables. Using
the definitions (\ref{eq:def_multiple_comps_variables}), we obtain
 \be \label{eq:Gamma_multifluids_1}
 \ol{P} \Gamma = \ol{P}\Ga_{\textrm{int}} +
 \sum_\al \delta_\comp{\al} \ol{\rho}_\comp{\al} (
 c_\comp{\al}^2 - c_s^2) \eqcomma = \ol{P}\left( \Ga_{\textrm{int}} +
 \Ga_{\textrm{rel}} \right) \eqdot
 \ee
We have introduced the total intrinsic entropy perturbation
 \be
 \Ga_{\textrm{int}} = \sum_\al \frac{\ol{P}_\comp{\al}}{\ol{P}}
 \Gamma_\comp{\al}
 \ee
and the relative entropy perturbation $\Ga_{\textrm{rel}}$, which
using the background energy conservation,
\rrp{eq:background_energy_conservation_multifluid}, can be recast
as
 \be \label{eq:Gamma_rel}
 \ol{P} \Ga_{\textrm{rel}} =
 \frac{1}{2} \sum_{\al, \ba}
 \frac{(1+w_\comp{\al}) (1 + w_\comp{\ba})
 \ol{\rho}_\comp{\al} \ol{\rho}_\comp{\ba}}{(1 + w)\ol{\rho}}
 (c^2_\comp{\al} - c^2_\comp{\ba})
 \left( \frac{\delta_\comp{\al}}{1 + w_\comp{\al}} -
  \frac{\delta_\comp{\ba}}{1 + w_\comp{\ba}} \right) \eqdot
 \ee
Here we have assumed that the components are decoupled from each
other, \ie that $\ol{Q}_\comp{\al}^\nu = 0$ in
(\refp{eq:background_energy_conservation_multifluid}), see
\citep{Malik:2002jb} for a generalization to the case of
interacting fluids.

The quantity $\Ga_{\textrm{rel}}$ represents {\it relative entropy
perturbations} which are produced by the different dynamical
behavior of the matter components with different sound speed. The
{\it entropy perturbation} between the components $\al$ and $\ba$
is defined as
 \be \label{eq:def_entropy_perturbation}
 S_{\al \ba} \equiv
 \frac{\delta_\comp{\al}}{1 + w_\comp{\al}} -
 \frac{\delta_\comp{\ba}}{1 + w_\comp{\ba}} \eqdot
 \ee
It is easy to see that the entropy perturbations are gauge
invariant quantities by substituting the gauge dependent density
contrasts on the right hand side with the gauge invariant density
variables defined in (\ref{eq:def_Delta}), obtaining
 \be
 S_{\al \ba} =
 \frac{\Delta_\comp{\al}}{1 + w_\comp{\al}} -
 \frac{\Delta_\comp{\ba}}{1 + w_\comp{\ba}} \eqdot
 \ee

In order to clarify the physical meaning of $S_{\al \ba}$,
consider a mixture of radiation and dust-like matter. We are
interested in fluctuations of the number density (per physical
volume) ratio of the two species:
 \be
 \delta \left( \frac{n_r}{n_m} \right ) /(n_r n_m) =
 \frac{\delta n_r}{n_r} -  \frac{\delta n_m}{n_m} \eqdot
 \ee
Recall that \cite[see \eg][]{Inflation_Kolb} $n_r \propto s
\propto T^3$, with $s$ the radiation entropy per volume, hence
 \be
 \frac{\delta n_r}{n_r} = \frac{\delta s}{s} =
 3 \frac{\delta T}{T} =  \frac{3}{4} \frac{\delta \rho_r}{\ol{\rho}_r}
 \ee
For matter we have
 \be
 \frac{\delta n_m}{n_m} = \frac{\delta \rho_m}{\ol{\rho}_m}
 \eqcomma
 \ee
and therefore
 \be
 \frac{\delta n_r}{n_r} -  \frac{\delta n_m}{n_m} =
 \frac{\delta_\comp{r}}{(1 + w_\comp{r})} -   \frac{\delta_\comp{m}}{(1 + w_\comp{m})}
 = S_{rm} \eqdot
 \ee
Thus a non vanishing relative entropy perturbation means that
there are spatial inhomogeneities in the relative number density
of the the two fluids, which can be understood as a spatial
variation in the equation of state. The above results are
generalized in \SEC{chap:params;sec:ic}.

\section{Perturbation equations}

In this section, we write down the first order perturbation
equations using the gauge invariant formalism and variables
defined above. For completeness, we also give the vector and
tensor equations, but in the rest of this work we will concentrate
exclusively on the scalar sector.

\subsection{Einstein equations}
 \label{chap:perts;sec:einstein}

The perturbed Einstein equations
 \be
 \delta G_{\mu \nu} = 8 \pi G \delta T_{\mu \nu}
 \ee
are split in their scalar, vector and tensor parts.

\subsubsection*{Scalar equations}

There are 4 scalar equations for the 4 gauge invariant quantities
$\Phi, \Psi, V$ and $D$:
 \begin{alignat}{2}
 (\triangle + 3 \curv) \Phi & =  4 \pi G a^2 \ol{\rho} D
 &&  \text{(Poisson),}  \label{eq:poisson}\\
\Hbl \Psi + \dot{\Phi} & =  4 \pi G a^2 \ol{\rho}(1 + w) V
 &&  \text{(constraint),} \label{eq:constraint_V}\\
 \Phi - \Psi & =  8 \pi G a^2 \ol{\rho} w \Pi
 && \text{(anisotropic stress),}
 \label{eq:anisotropic_stress} \\
 \Hbl \dot{U} + (\Hbl^2 + 2 \dot{\Hbl}) U & =  4 \pi G a^2 \ol{\rho}
 \left(c_s^2 D_g  + w \Gamma + \frac{2}{3} w \triangle \Pi  \right)
 && \eqcomma
 \end{alignat}
where
 \be
 U \equiv \Psi + \frac{\Hbl^2 - \dot{\Hbl}}{\Hbl^2} \Phi
 + \frac{\dot{\Phi}}{\Hbl} \eqdot
 \ee

Recall that $D_g$ is related to $D$, $V$ and $\Phi$ via
Eqs.~(\refp{eq:relations_between_Ds}), and we have assumed an
equation of state of the form (\refp{eq:eq_of_state}).
Eq.~(\ref{eq:poisson}) is the general relativistic analogue of the
Poisson equation. In order to close this system, we need to
specify the matter content by giving $w$, $c_s^2$, $\Gamma$ and
$\Pi$. For a single perfect fluid, $\Gamma = \Pi = 0$, hence from
the anisotropic stress equation (\ref{eq:anisotropic_stress}) it
follows that $\Psi = \Phi$.

We shall see below that an evolution equation for $\Pi$ follows
\eg from the kinetic description provided by the Boltzmann
equation, see \rrp{eq:collisional_Boltzmann_hierarchy_trunc}. For
multiple fluids, we will also rewrite $\Gamma$ in terms of the
relative entropy perturbations, as in \rr{eq:Gamma_rel}.

\subsubsection*{Vector equations}

The vector part yields a constraint and an evolution equation for
$V^{\vc}_i$ and $\Sigma^{\vc}_i$:
 \begin{align}
 \left( 2\curv + \triangle + 4 (\dot{\Hbl} - \Hbl^2) \right) \Sigma^{\vc}_i & =
 16 \pi G \ol{\rho} a^2 (1 + w) V^{\vc}_i \eqcomma\\
 \dot{\Sigma}^{\vc}_i + 2 \Hbl \Sigma^{\vc}_i & =
 8 \pi G \ol{\rho} a^2 w \Pi^{\vc}_i \eqdot
\end{align}

For a perfect fluid, $\Pi^{\vc}_i = 0$, the above equations give
in a flat universe on large scales (such that gradients can be
neglected)
\begin{equation}
 \Sigma^{\vc}_i = - V^{\vc}_i \propto \frac{1}{a^2} \eqdot
\end{equation}
Therefore in the absence of active seeds, vector perturbations are
always decaying on large scales.

\subsubsection*{Tensor equation}

The tensor part yields an equation describing the gravitational
waves. It is the equation of a forced harmonic oscillator, with a
damping term due to the expansion of the universe:
 \be \label{eq:gravitational_waves}
 \ddot{E}_{ij}^{\ts} + 2 \Hbl \dot{E}_{ij}^{\ts} + (2
 \curv - \triangle)
 E_{ij}^{\ts} = 8 \pi G \ol{\rho}a^2 \Pi_{ij}^{\ts} \eqdot
 \ee
On super-horizon scales and for zero curvature, the term $\propto
E_{ij}^{\ts}$ is negligible. The homogeneous equation in the
radiation era, when $\Hbl = \eta^{-1}$, has a decaying solution $
{E}_{ij}^{\ts} \propto \eta^{-1}$ and a constant solution,
${E}_{ij}^{\ts} = \text{const}$. As a mode enters the horizon, the
oscillatory behavior takes over, and the wave propagates with a
frequency $k^2 + 2\curv$ and is damped as $a^{-1}$. In the absence
of anisotropic stress and in a flat universe, $\curv = 0$, the
general solution of \eqref{eq:gravitational_waves} for $\Pi = 0$,
writing $E_{ij}^{\ts} = h(\bs{x}, \eta) \varepsilon_{ij}(\bs{x})$
and going to Fourier space in a flat universe, is given by
 \be
 h = (k \eta)^{1-q}\left[A j_{q-1}(k\eta) + B n_{q-1}(k\eta)
 \right] \eqcomma
 \ee
where $j_\nu(x)$ and $n_\nu(x)$ are the Bessel and von Neumann
functions of order $\nu$, respectively (see
Eqs.~\refp{eq:asymptotic_bessel_neumann}) and $a \propto \eta^q$.

\subsection{Conservation equations}
 \label{chap:perts;sec:conservation}

The conservation equations, which follow from the contracted
Bianchi identity, offer evolution equations which are sometimes of
a simpler form and are handy to manipulate. From the perturbed
energy conservation equation
 \be
 \delta(\nabla_\mu \ol{T}^{\mu \nu}) = 0
 \ee
we obtain the following equations for a mixture of non-interacting
fluids.

\subsubsection*{Scalar equations}

There are two scalar conservation equations, one for the density
contrast and the second for the velocity perturbation. In terms of
$D_{g, \comp{\al}}$ the conservation equations read:
 \be \label{eq:pertd_conservation_eq_1_Dg}
 \dot{D}_{g, \comp{\al}} + 3\Hbl(c_\comp{\al}^2 -w_\comp{\al}) D_{g,
 \comp{\al}} =
 - 3\Hbl \Gamma_\comp{\al} w_\comp{\al} + (1 + w_\comp{\al}) \triangle
 V_\comp{\al}  \eqcomma
 \ee
 \be \label{eq:pertd_conservation_eq_2_Dg}
 \dot{V}_\comp{\al} + (1 - 3c_\comp{\al}^2) \Hbl V_\comp{\al} =
 \Psi + 3c_\comp{\al}^2\Phi +
 \frac{w_\comp{\al}}{1 + w_\comp{\al}} \left( \Gamma_\comp{\al} +
 \frac{c_\comp{\al}^2}{w_\comp{\al}} D_{g, \comp{\al}} + \frac{2}{3}
 (\triangle+ 3\curv) \Pi_\comp{\al} \right) \eqdot
 \ee
Is is sometimes convenient to express the above in terms of the
density contrast $D_\comp{\al}$:
 \be \label{eq:pertd_conservation_eq_1_D}
 \dot{D}_{\comp{\al}} -3 w_\comp{\al} \Hbl  D_{ \comp{\al}} =
 (\triangle + 3 \curv) \left[(1+ w_\comp{\al})V_\comp{\al} + 2 \Hbl w_\comp{\al} \Pi_\comp{\al} \right]
 + 3 \frac{1 + w_\comp{\al}}{1 + w}(\Hbl^2 + \curv)(V -
 V_\comp{\al}) \eqcomma
  \ee
 \be \label{eq:pertd_conservation_eq_2_D}
 \dot{V}_\comp{\al} + \Hbl V_\comp{\al} =
 \Psi + \frac{c_\comp{\al}^2}{1 + w_\comp{\al}}D_\comp{\al} +
 \frac{w_\comp{\al}}{1 + w_\comp{\al}} \left( \Gamma_\comp{\al}+ \frac{2}{3}
 (\triangle+3 \curv) \Pi_\comp{\al} \right) \eqdot
 \ee

\subsubsection*{Vector equation}

We obtain one evolution equation for the {\it vorticity} $\Om_{i
\comp{\al}}^\vc \equiv \Sigma_{i \comp{\al}}^\vc + V_{i
\comp{\al}}^\vc$:
 \be
 \dot{\Om}_{i,\comp{\al}}^\vc + \Hbl \Om_{i,
\comp{\al}}^\vc (1 - 3c_{\comp{\al}}^2) =
\frac{1}{2}\frac{w_\comp{\al}}{1 + w_\comp{\al}} \triangle \Pi_{i,
\comp{\al}}^\vc \eqdot
 \ee
If the anisotropic stress source term is absent, we can rewrite
the above equation as
 \be
 \frac{\dr}{\dr \eta} ( \Om_{i,
\comp{\al}}^\vc a^{1 - 3c_\comp{\al}^2}) = 0 \eqcomma
 \ee
hence
 \be
 \Om_{i, \comp{\al}}^\vc \propto a^{3 c_\comp{\al}^2 - 1} \eqdot
 \ee

\subsection{The Bardeen equation}
 \label{chap:perts;sec:bardeen}

It is often convenient to have an evolution equation for the
Bardeen potential in terms of the total matter content. By
combining the conservation equation
\rr{eq:pertd_conservation_eq_1_Dg} with the Einstein equations
(\ref{eq:poisson}--\ref{eq:anisotropic_stress}) we obtain a second
order equation, called the Bardeen equation, for $\Phi$:
 \be \label{eq:Bardeen_eq}
 \ddot{\Phi} + 3\Hbl(1 + c_s^2) \dot{\Phi} +
 \left[ 3(c_s^2-w)\Hbl^2 - (1+3c_s^2)\curv - c_s^2 \triangle \right] \Phi =
 g_\Phi \eqcomma
 \ee
where the source term $g_\Phi$ is generated by the matter
anisotropic stress and entropy perturbation:
 \be
 g_\Phi = 8 \pi G a^2 P \left[
 \Hbl \dot{\Pi} + [2\dot{\Hbl} + 3 \Hbl^2(1 - c_s^2/w) ]\Pi
 + \tfrac{1}{2}\triangle \Pi + \tfrac{1}{2} \Gamma \right] \eqdot
 \ee
The above equation can be recast in an evolution equation for the
gauge invariant curvature perturbation,
\rr{eq:curvature_pert_related_to_potentials}. For hydrodynamical
matter, \ie setting $\Pi = 0$ and for a flat universe ($\curv =
0$) we find
 \be \label{eq:evolution_of_zeta}
 \dot{\zeta} = \frac{\Hbl}{\Hbl^2 - \dot{\Hbl}}
 \left[ c_s^2\triangle \Phi + \tfrac{3}{2} \Hbl^2 w \Gamma \right]
 \eqdot
 \ee
This expression will be used when discussing the evolution of
curvature and entropy perturbations.

\subsection{Collisionless Boltzmann equation}
 \label{chap:perts;sec:boltzmann}

We briefly recall in the following the basics of relativistic
kinetic theory, for more details see \eg \cite{Book:DeGroot:80}.
Consider the phase space given by the the tangent bundle
 \be
 \mathcal{T} \equiv \{(x^\mu, p^\mu) \vert x^\mu \in \mathcal{M}, p^\mu \in \mathcal{T}_x \}
 \ee
where $\call{M}$ is the spacetime manifolds and $\call{T}_x$ its
tangent space at the point $x^\mu$. For a particle of mass $m$,
its distribution function $f(x^\mu, p^\mu)$ is defined on the
mass-shell
 \be
 \call{P}_m(x^\mu) \equiv \{p^\mu \in \call{T}_x \vert p_\mu p^\mu = - m^2 \}
 \ee
The Liouville operator $\call{L}$ is defined on $\call{T}$, and it
gives the evolution of $f(x^\mu, p^\mu)$ along the particle world
lines, according to the {\it Boltzmann equation}
 \be \label{eq:symbolic_Bolzmann_eq}
 \call{L} \left[ f \right] = C \left[ f \right] \eqcomma
 \ee
which states that the rate of change of $f$ is due to the {\it
collision term} $C \left[ f \right]$. For the purpose of studying
relativistic particles such as photons and massless neutrinos, we
will treat the case $m = 0$ only. The hereby derived equations
will then be applied to the description of neutrinos and of
photons after recombination. Further details and the general case
for massive particles can be found in \eg
\cite{Durrer:1993db,Uzan:1998mc}.

We now proceed with perturbing the left hand side of
\rr{eq:symbolic_Bolzmann_eq}. Its background solution was
presented in \SEC{chap:intro;sec:Boltzmann}, and was shown to be
of the form $ \ol{f} = \ol{f}(ap)$, see
\rr{eq:background_Bolzmann_equation}, where $E^2 = p^2 \equiv
p_\mu p_\nu g^{\mu \nu}$. By splitting the distribution function
into a background and a perturbed part,
 \be
 f(\eta, x^i, p, n^i) = \ol{f}(\eta, p) + F(\eta, x^i, p, n^i)
 \ee
we move to a phase space which differs to linear order from the
one of $\ol{f}$. Therefore the choice of $F$ and its
transformation properties depend on the isomorphism relating the
``background'' and the ``perturbed'' phase space. By an opportune
choice of the isomorphism, it can be shown \citep{Durrer:1993db}
that under a gauge transformation $F$ transforms as
 \be
 F \ra F + p \frac{\partial \ol{f}}{\partial p} \left[ \Hbl T + n^i  T_i \right]
 \eqdot
 \ee
It follows that the following variable
 \be
 \call{F} \equiv F - p  \frac{\partial \ol{f}}{\partial p} \left[
 C + n^i (\dot{E}_i - B_i) \right] \eqcomma
 \ee
is gauge invariant. In terms of $\call{F}$, the collisionless
Boltzmann equation reads
 \be \label{eq:pertd_Boltzmann_for_F}
 \frac{\partial \call{F}}{\partial \eta} +
 \frac{\partial \call{F}}{\partial x^i}n^i -
 p \Hbl  \frac{\partial \call{F}}{\partial p} -
 \phantom{}^{(3)}\Gamma^i_{jk}n^j n^k  \frac{\partial \call{F}}{\partial n^i}
 =
 p \frac{\partial \ol{f}}{\partial p}
 \left[ n^i \partial_i (\Psi + \Phi) \right] \eqcomma
 \ee
and $^{(3)}\Gamma^i_{jk}$ are the Christoffel symbols of the
background 3-space. The above equation is in manifestly gauge
invariant form, and we notice that spatial variations in the
Bardeen potential act as source for perturbations in the
distribution function.

By integrating this equation over the particle energies, we obtain
a differential equation for the {\it brightness perturbation}
$\call{I}$, defined as
 \be
 I = \ol{I}(\eta) +  \call{I}(\eta, x^i, n^i) \equiv
 4 \pi \int_0^\infty \ol{f}p^3 \dr p +
 4 \pi \int_0^\infty \call{F}p^3 \dr p \eqdot
 \ee
The brightness represents the energy per unit solid angle as
measured by an observer at position $x^i$. The photon energy is
just the monopole of the brightness, \ie
 \be
 \rho_\comp{\ga} = \int \frac{\dr \Om}{4 \pi} I \eqcomma
 \ee
and therefore $\ol{\rho}_\comp{\ga} = \ol{I}$. From
\rr{eq:pertd_Boltzmann_for_F} we obtain
 \be \label{eq:pertd_Boltzmann_for_I}
 \dot{\call{I}} +
 \left( n^i \partial_i + 4 \Hbl  -
 \phantom{}^{(3)}\Gamma^i_{jk}n^j n^k \frac{\partial}{\partial n^i} \right)
 \call{I} =
 - 4 \ol{I} \left[ n^i \partial_i (\Psi + \Phi) \right] \eqdot
 \ee

The above can be rewritten in terms of the {\it temperature
contrast}
 \be
 \Theta(\eta, x^i, n^i) \equiv \frac{\delta T}{T} =
 \frac{1}{4}\frac{\call{I}}{\ol{I}}
 \ee
and using the background energy conservation equation we obtain
 \be \label{eq:pertd_Boltzmann_for_Theta}
 \dot{\Theta} +
 \left( n^i \partial_i -
 \phantom{}^{(3)}\Gamma^i_{jk}n^j n^k \frac{\partial}{\partial
 n^i} \right )
 \Theta =
 - n^i \partial_i (\Psi + \Phi) \eqdot
 \ee
This is the Boltzmann equation for relativistic, collisionless
particles, which relates gravitational perturbations to
temperature fluctuations of their distribution function.

\subsubsection*{The Boltzmann hierarchy}

We now go to Fourier space, and we restrict ourselves to the
spatially flat case, $\curv = 0$, so that the eigenfunctions of
the Laplacian are just plane waves and
$\phantom{}^{(3)}\Gamma^i_{jk} = 0$ (an harmonic decomposition for
non-flat spaces can be found \eg in
\citealp{Vilenkin:1964,Kodama:1985bj}), so that for any scalar $f$
 \be \label{eq:def_FT}
 f(\eta, {\bf x}) = \frac{1}{(2\pi)^{3/2}}
 \int \dr ^3 {\bf k}  f(\eta, {\bf k}) e^{\imath {\bf k x}}
 \eqcomma
 \ee
and in general we denote the real space $f$ and its harmonic
transform with the same symbol. Defining $\mu \equiv n^j k_j/k$
and $k \equiv \sqrt{k_i k^i}$ we obtain from
\rr{eq:pertd_Boltzmann_for_Theta}
 \be \label{eq:pertd_Boltzmann_for_Theta_FT}
 \dot{\Theta} + \imath \mu k \Theta =
 - \imath \mu k (\Psi + \Phi) \eqdot
 \ee
Assuming that $\Theta$ does not depend explicitly on $k_i$, then
the dependence on the photons momentum direction comes in only via
$\mu$. In that case $\Theta = \Theta(\eta, k, \mu)$, and we will
suppress the explicit time dependence. We now perform an expansion
in Legendre polynomials\footnote{Different normalizations for the
expansion coefficient are commonly used in the literature and
their relation with the one used here is: in \cite{Hu:1995jd}
$\Theta^{\text{HS}} = \imath^\ell (2\ell+1) \Theta_\ell$ (notice
that in this work the Bardeen potentials are such that
$\Psi^\text{HS} = \Psi$ but $\Phi^\text{HS} = -\Phi$); in
\cite{Ma:1995} $\Theta$ is denoted by $\Psi$ and
$\Psi_\ell^\text{MB} = \imath^\ell \Theta_\ell$, which is the same
convention used by \cite{Seljak:1996is}; in \cite{Durrer:1993db}
$\Theta$ is denoted by $\call{M}$ and $\call{M}_\ell =
\Theta_\ell/2$.}
 \begin{align} \label{eq:Theta_expansion_in_P}
 \Theta(\mu, k) & = \sum_\ell (2\ell+1) P_\ell \Theta_\ell
 \eqcomma
 \\
 \Theta_\ell(k) & \equiv
 \dfrac{1}{2} \int_{-1}^{1} \dr \mu \Theta (\mu, k) P_\ell(\mu)
 \eqcomma \label{eq:Legendre_expansion}
 \end{align}
where $P_\ell(x)$ is the Legendre polynomial of order $\ell$,
which satisfy
 \begin{align}
 P_0(x)  = & 1 \eqcomma \\
 P_1(x)  = & x \eqcomma \\
 P_2(x)  = & \frac{1}{2}(3x^2 -1) \eqcomma\\
 (\ell+1)P_{\ell+1}(x)  = &
 (2 \ell + 1) x P_\ell(x)- \ell P_{\ell-1} (x) \eqdot
 \end{align}
From \rr{eq:pertd_Boltzmann_for_Theta_FT} follows an infinite
hierarchy of equations for the moments of the Boltzmann equation:
  \begin{align}\label{eq:pertd_Boltzmann_hierarchy}
 \dot{\Theta}_0 + \imath k \Theta_1 & = 0 \eqcomma \\
 \dot{\Theta}_1 + \frac{1}{3} \imath k \Theta_0 +
 \frac{2}{3} \imath k \Theta_2 & = - \frac{1}{3} \imath k (\Phi + \Psi) \eqcomma \\
 \dot{\Theta}_\ell + \frac{\ell}{2\ell+1} \imath  k
 \Theta_{\ell-1} + \frac{\ell+1}{2 \ell+1} \imath  k
 \Theta_{\ell+1} & = 0 \quad (\ell \geq 2) \eqdot
  \end{align}

Gradients of the Bardeen potentials act as a source for the first
moment. Because of the recursion relation, each multipole moment
$\ell$ is coupled to the preceding and the following moment.
Therefore, power is transferred to higher moments, and in
principle we need to solve an infinite number of coupled
differential equations. Simply truncating the hierarchy by
imposing $\Theta_{\ell^{max}} = 0$ is not an optimal solution,
since the error due to the truncation will reflect back to lower
moments via the coupling. A more effective truncation scheme is
discussed in \cite{Ma:1995}. We notice that at early times and
super-horizon scales (\ie $k\eta \ll 1$) higher moments are
suppressed by successive powers of $k\eta$, $\Theta_\ell \sim
\call{O}(\Theta_{\ell -1} k \eta)$, and hence the first few
moments are sufficient to accurately describe the temperature
fluctuation.

\subsubsection*{Relations with macroscopic quantities}

From the definition of the stress-energy tensor
\citep{Book:DeGroot:80}
 \be
 T^{\mu \nu}(x^\al) = \int \frac{\dr^3p}{p^0} p^\mu p^\nu f(x^\al, p^\mu)
 \ee
and comparing with \rrp{eq:perturbed_energy_tsr}, we can establish
the hydrodynamical gauge invariant variables as integrals over
momenta of the gauge invariant brightness perturbation:
 \begin{subequations} \label{eq:def_macroscopic_quantities}
 \begin{align}
 D_{g,\comp{\ga}} & = \frac{1}{\ol{\rho}_\comp{\ga}}
 \int \frac{\dr \Om}{4 \pi} \call{I} \eqcomma \\
 V^j_\comp{\ga} & = - \frac{1}{(1 + w_\comp{\ga}) \ol{\rho}_\comp{\ga}}
 \int \frac{\dr \Om}{4 \pi} n^j \call{I} \eqcomma \\
 \Pi^{ij}       & = \frac{1}{w_\comp{\ga}\ol{\rho}_\comp{\ga}}
 \int \frac{\dr \Om}{4 \pi} n^{ij} \call{I} \eqdot
 \label{eq:def_Pi_macroscopic}
 \end{align}
 \end{subequations}
Rewriting the above in terms of multipole moments of the
temperature perturbation, we have the identities in harmonic
space\footnote{Notice that the monopole of our $\call{F}$
corresponds (up to multiplicative constants) to the density
perturbation in the comoving gauge; in the literature the
temperature perturbation in Newtonian gauge is often employed (as
in \citep{Hu:1995jd}), in which case an extra term $\propto \Phi$
appears along with $\Theta_0^{\text{N}}$. With the normalization
convention of \citep{Hu:1995jd}, the relation between our monopole
and the one in Newtonian gauge is $\Theta_0 = \Theta_0^{\text{N}}
- \Phi$. All other multipoles $\ell> 0$ do not suffer from this
ambiguity and are gauge independent.}
 \begin{subequations}
  \begin{align}
  \Theta_0 & = \dfrac{1}{4} D_{g,\comp{\ga}} \eqcomma \\
  \Theta_1 & = - \frac{1}{3}  \imath k V_\comp{\ga} \eqcomma \\
  \Theta_2 & = -\frac{1}{12} k^2 \Pi_\comp{\ga} \eqdot
 \end{align}
 \end{subequations}

Truncating the Boltzmann hierarchy at the third moment by setting
$\Theta_\ell = 0$ for $\ell \ge 3$, we obtain
  \begin{align} \label{eq:pertd_Boltzmann_hierarchy_trunc}
  \dot{D}_{g, \comp{\ga}} + \frac{4}{3}k^2 V_\comp{\ga} &= 0 \eqcomma \\
  \dot{V}_\comp{\ga} - \frac{1}{4}D_{g, \comp{\ga}} &= -
  \frac{1}{6}k^2 \Pi_\comp{\ga} + \Phi + \Psi \eqcomma \label{eq:ell1_collisionless}\\
  \dot{\Pi}_\comp{\ga} - \frac{8}{5} V_\comp{\ga} &= 0 \label{eq:ell2_collisionless}\eqdot
  \end{align}

Unsurprisingly, we recover the two conservation equations of
(\ref{eq:pertd_conservation_eq_1_Dg}-\refp{eq:pertd_conservation_eq_2_Dg})
for radiation (with $w_\comp{\ga} = c_\comp{\ga}^2 = 1/3$ and
$\Gamma = 0$), supplemented with an evolution equation for
$\Pi_\comp{\ga}$. These equations are appropriate for
relativistic, collisionless and massless particles such as
neutrinos. At later times, however, higher order moments need to
be taken into account. Photons are scattered by electrons, and to
describe their evolution we now turn to the appropriate collision
term.

\subsection{Thomson scattering} \label{chap:pert;sec:collision_term}
 \label{chap:perts;sec:thomson}

We now consider the case of elastic Thomson scattering between
photons and non-relativistic electrons. We give some elements of
the derivation for the collision term for the total photon
intensity, while we just outline the polarization treatment. A
detailed derivation can be found in
\cite{Kosowsky:1996cy,Durrer:2001gq}.

Thomson scattering of unpolarized light generates linear
polarization if the incident intensity has a quadrupolar
anisotropy. In the tight coupling regime, collisions make the
photons distribution function uniform in the electrons rest frame,
and therefore no polarization can arise. However, during the weak
coupling regime just before last scattering, the mean free path of
photons grows and a sizable temperature quadrupole is generated,
which acts as a source for polarization, as we briefly describe in
this section. After decoupling, free streaming conserves the
polarization state, which can only be changed by further
rescattering due to reionization, see \SEC{chap:params:sec:reion}.

\subsubsection{Stokes parameters}

The polarization state of light is usually described in terms of
Stokes parameters, see \eg \cite{Book:Jackson}. The electric field
of a plane monochromatic electromagnetic wave propagating in the
$z$ direction is
 \be
 \bs{E}({\bf x}, t) =
 \bs{\call{E}} e^{\imath{( \omega t - k z) }} \eqcomma
 \ee
where the complex vector $\bs{\call{E}}$ describing the
polarization state of the wave is given by \be
 \bs{\call{E}} = \left(
 \begin{array}{l}
 a_x e^{\imath \te_x} \\
 a_y e^{\imath \te_y} \\
 0
 \end{array}
 \right) \eqdot
 \ee

Instead of using the four numbers $(a_x, a_y, \te_x, \te_y)$, it
is convenient to introduce the {\it Stokes parameters}
 \begin{align}
 I & \equiv a_x^2 + a_y^2 \eqcomma\\
 Q & \equiv a_x^2 - a_y^2  \eqcomma\\
 U & \equiv 2 a_x a_y \cos(\te_x - \te_y) \eqcomma \\
 V & \equiv 2 a_x a_y \sin(\te_x - \te_y) \eqcomma
 \end{align}
which can be directly measured with a linear polarizer and a
quarter-wave plate. Their physical interpretation is
straightforward: $I$ gives the total intensity, $Q$ measures the
difference between $x$ and $y$ polarization, $U$ gives phase
information for the two linear polarizations, and $V$ determines
the difference between positive and negative circular
polarization. $I$ and $V$ are physical observables independent of
the coordinate system, but $Q$ and $U$ mix under a rotation by an
angle $\phi$ of the $x-y$ plane:
 \begin{subequations} \label{eq:QU_rotation}
 \begin{align}
 Q' &= Q \cos(2\phi) + U \sin(2\phi) \\
 U' &= - Q \sin(2\phi) + U \cos(2\phi)\eqcomma
 \end{align}
 \end{subequations}
from which it is easy to derive that the physically observable
quantity is the polarization vector $\bs{P}$, lying in the $x-y$
plane, with magnitude $(Q^2 + U^2)^{1/2}$ and with polar angle
$\alpha = \frac{1}{2}\tan^{-1}\frac{U}{Q}$.

Finally, the four stokes parameters are not independent, but
satisfy the relation
 \be \label{eq:not_independent_Stokes}
 I^2 = Q^2 + U^2 + V^2 \eqdot
 \ee

\subsubsection{Scattering cross section}
\label{chap:perts;sec:crosssec}

\begin{figure}[tb]
 \centering
 \unitlength1cm
  \begin{picture}(8,8)
  \put(0,1){\includegraphics[width=\onefigwidth]{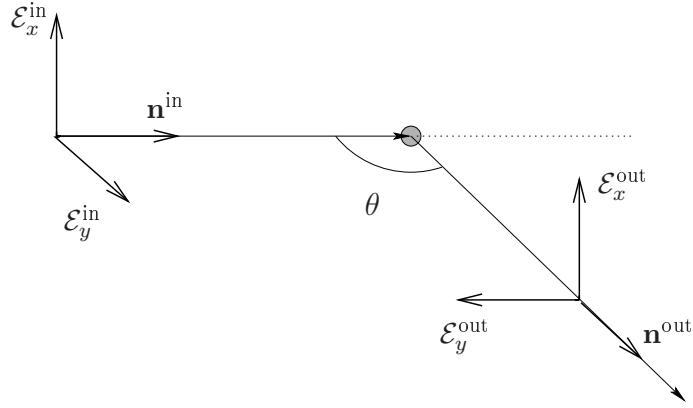}}
  \put(4.2,3.5){\large $\theta$}
  \put(0.2, 3.3){$\EIN_y$} \put(-0.5, 6.0){$\EIN_x$}
  \put(5.2, 1.8){$\EOUT_y$} \put(7.3, 3.8){$\EOUT_x$}
  \put(1.3, 4.8){$\bs{n}^{\text{in}}$} \put(7.9, 1.8){$\bs{n}^{\text{out}}$}
  \end{picture}
 \caption[Geometry of the the Thomson scattering process.]
{Geometry of the the Thomson scattering process in the rest frame
of the electron, represented by the sphere in the center. A photon
beam is incoming from the left and is scattered off with an angle
$\theta$.} \label{fig:pol_geometry}
 \end{figure}

We now consider the scattering process in the rest frame of the
electron, with the geometry of \FIG{fig:pol_geometry}. The Thomson
scattering cross section for an incident wave with linear
polarization $\EIN$ into a scattered wave with polarization
$\EOUT$ is
 \be
 \frac{\dr \si}{\dr \Om} = \frac{3 \sigma_T}{8 \pi} \vert \EIN
 \cdot \EOUT \vert ^2 \eqcomma
 \ee
with $\sigma_T$ the Thomson scattering cross section. It is
convenient to work with the partial intensities $I_x$ and $I_y$,
defined as
 \be
 I_x \equiv \frac{I+Q}{2} \quad \text{and} \quad
 I_y \equiv \frac{I-Q}{2} \eqdot
 \ee
The incoming wave is unpolarized by assumption, so $I_x^\IN =
I_y^\IN = I^\IN/2$, and for the outgoing wave we find \be
 I_x^\OUT  = \frac{3 \si_T}{16 \pi} I^\IN  \quad \text{and} \quad
 I_y^\OUT  = \frac{3 \si_T}{16 \pi} I^\IN  \cos^2(\theta)
 \ee
or, in terms of the outgoing Stokes parameters
\begin{align}
 I^\OUT & = \frac{3 \si_T}{16 \pi}I^\IN(1 + \cos^2(\theta))
 \eqcomma \label{eq:I_OUT}\\
 Q^\OUT & = \frac{3 \si_T}{16 \pi}I^\IN \sin^2(\theta) \eqcomma \\
 U^\OUT & = 0 \label{eq:Q_OUT} \eqdot
 \end{align}

The value of $U^\OUT$ has been found by recalculating $Q$ in an
outgoing basis which has been rotated by $\pi/4$. Thomson
scattering does not generate circular polarization, so $V=0$ and
we will not consider it further. Since from
(\ref{eq:not_independent_Stokes}) there are only three independent
Stokes parameters, and $V=0$ all the time, the description in
terms of $I$ and $Q$ is sufficient, and we wont use $U$ any
further.

The total outgoing intensities are obtained by integrating over
all incoming directions, and rotating the result into a common
coordinate system using (\ref{eq:QU_rotation}):
 \begin{align}
 I^\OUT & = \frac{3 \si_T}{16 \pi} \int \dr \Om (1 + \cos^2(\theta))
 I^\IN(\theta, \phi) \eqcomma \label{eq:scattered_intensity}\\
 Q^\OUT & = \frac{3 \si_T}{16 \pi} \int \dr \Om
 \sin^2(\theta)\cos(2\phi)
 I^\IN(\theta, \phi) \label{eq:scattered_Q} \eqdot
 \end{align}

\subsubsection{Temperature hierarchy}

We are now in the position of deriving the collision term due to
Thomson scattering for the intensity distribution function $f$,
which is of the form
 \be
 C \left[ f \right] =
 \frac{\dr f^+}{\dr \eta} - \frac{\dr f^-}{\dr \eta}
 \eqdot
 \ee
where $f^+(x^\mu, p^\mu)$ ($f^-$) denotes the distribution of
particles within $(\Delta x^\mu, \Delta p^i/p^0)$ of $(x^\mu,
p^\mu)$ gained (lost) in the scattering process. According to the
hypothesis of molecular chaos \citep{Book:DeGroot:80}, the
contribution lost is just proportional to the electron density
times the photon distribution, hence with the definitions
(\ref{eq:def_photons_directions}--\refp{eq:def_photons_directions_2})
 \be
 \frac{\dr f^-}{\dr \eta}(x^\mu, p, n^i) =  \taudot
 f(x^\mu, p,  n^i) \eqcomma
 \ee
where
 \be \label{eq:differential_Thomson_optical_depth}
 \taudot \equiv a \sigma_T n_e
 \ee
 is the {\it differential
Thomson optical depth},  and $n_e$ is the free electron density.
The contribution scattered into $p^i = p n^i$ is most easily
evaluated in the electron's rest frame, which we denote by a
tilde. After averaging over incoming and summing over outgoing
polarization states, we obtain
 \be
 \frac{\dr \tilde{f}^+}{\dr \tilde{t}}(x^\mu, \tilde{p}, {\bf \tilde{n}})  =  \sigma_T n_e
 \int \frac{\dr \tilde{\Om}_{\bf \varep}}{4 \pi} \tilde{f}(\tilde{p}, {\bf \tilde{n}})
 \om ({\bf \tilde{n}}, {\bf \varep}) \eqcomma
\ee where the angular dependence of the scattered intensity is,
from (\ref{eq:scattered_intensity})
 \be
 \om ({\bf \varep}, {\bf \varep'}) = \frac{3}{4}[1 - ({\bf \varep \cdot \varep'})^2] =
 1 + \frac{3}{4}\varepsilon_{ij}\varepsilon^{'ij}
 \ee
with $\varepsilon_{ij} \equiv \varepsilon_i \varepsilon_j -
\frac{1}{3}\delta_{ij}$. We now transform into the coordinate
system, in which the photon distribution function $f$ is defined.
To first order we have the relations
 \begin{align}
 \tilde{p}  & =  p\left(1 + n_i(v^i_\comp{b} - B^i)
 \right) \eqcomma
 \\
 {\bf \tilde{n}} & =  \bf{n} \eqcomma
 \end{align}
since aberration appears only at second order. We have used the
baryon 3-velocity $v^i_\comp{b}$, since electrons and baryons are
electromagnetically coupled and their velocities are the same.
Note that the above transformation assumes $v_b \ll 1$, \ie that
the electrons are non-relativistic, consistent with the fact that
we consider $v_b$ as a perturbation. Splitting the distribution
function in an isotropic part and a (gauge dependent)
perturbation, $f = \ol{f}(\eta, p) + \delta f(x^i, p^i)$, we then
compute the energy integrated collision term
 \be \label{eq:collision_term_gauge_dep}
 4\pi \int p^{3} \dr p C \left[ f \right] =
 a \sigma_T n_e \left[
 -4 n_i( v^i_\comp{b} - B^i) \ol{\rho}_\comp{\gamma}
 + \delta \rho_\comp{\gamma}
 - \delta I({\bf n}) +
 \frac{3}{4}n^{ij} \delta I_{ij} \right]  \eqcomma
 \ee
and we have introduced the gauge dependent brightness perturbation
$\delta I \equiv 4\pi \int \dr p p^3 \delta f $ and its second
moment
 \be
 \delta I_{ij} \equiv \int \frac{\dr \Om_{\varep}}{4\pi}  \varepsilon_{ij} \delta I (\varep)
\eqdot
 \ee
The expression \rr{eq:collision_term_gauge_dep} can be brought in
explicit gauge invariant form by substituting the gauge dependent
variables with the corresponding gauge independent counterparts.
After some manipulations we obtain
 \be
  4 \pi \int p^{3} \dr p C \left[ f \right] =
  4 \taudot \ol{\rho}_\comp{\gamma} \left[
  \Theta_0 -  n_i V_\comp{b}^i
  - \Theta
  + \frac{1}{16} n_{ij} \Pi_\comp{\ga}^{ij} \right] \eqcomma
 \ee
where we have used the identity (\refp{eq:def_Pi_macroscopic}). In
view of adding the collision term on the right hand side of the
hierarchy (\refp{eq:pertd_Boltzmann_hierarchy}), it is convenient
to rewrite it in terms of multipoles of the temperature
fluctuation $\Theta$ and transform to Fourier space
 \be
 4\pi  \int p^{3} \dr p C \left[ f \right] =
  - 4 \taudot \ol{\rho}_\comp{\gamma} \left[
   (\imath k V_\comp{b} + 3\Theta_1) P_1
   + \frac{9}{2}\Theta_2 P_2
   + \sum_{\ell \geq 3} (2 \ell + 1) \Theta_\ell P_\ell \right]
   \eqdot
 \ee
 A few remarks are in order at this point: as a
consequence of the conservation of energy in the elastic
collision, non-relativistic Thomson scattering does not contain a
monopole, while the dipole corresponds to a velocity mismatch
between photons and baryons, as is apparent from the first term on
the right hand side with $3 \Theta_1 = -\imath k V_\comp{\ga}$.
The angular dependence of the scattering generates a quadrupole
moment. In the limit of very many collisions, $\taudot \gg \Hbl$,
all multipoles $\ell > 1$ are driven to zero, therefore in the
{\it strong coupling regime}, the photons and baryons velocity
coincide and higher order moments are suppressed: thus the
tight-coupled photons-baryons system can be described as an
hydrodynamical fluid in term of the zeroth and first moments only.

The Boltzmann hierarchy, \rrp{eq:pertd_Boltzmann_hierarchy},
supplemented with the above collision term for photons-electrons
Thomson scattering, now becomes:
 \begin{subequations} \label{eq:collisional_Boltzmann_hierarchy}
  \begin{align}
 \dot{\Theta}_0 + \imath k \Theta_1 & = 0 \eqcomma \\
 \dot{\Theta}_1 + \frac{1}{3} \imath k (\Theta_0 + \Phi + \Psi) +
 \frac{2}{3} \imath k \Theta_2  & =
 - \taudot ( \frac{1}{3} \imath k V_\comp{b} + \Theta_1)
 \eqcomma \\
 \dot{\Theta}_2 + \frac{2}{5} \imath k \Theta_1 + \frac{3}{5}
 \imath k \Theta_3 & =  - \taudot \frac{9}{10}\Theta_2 \\
 \dot{\Theta}_\ell + \frac{\ell}{2\ell+1} \imath  k
 \Theta_{\ell-1} + \frac{\ell+1}{2 \ell+1} \imath  k
 \Theta_{\ell+1} & = - \taudot \Theta_\ell \quad (\ell \geq 3) \eqdot
  \end{align}
  \end{subequations}

Rewriting the above in terms of macroscopic quantities and cutting
the hierarchy at $\ell = 2$ gives instead of
\rrp{eq:pertd_Boltzmann_hierarchy_trunc}
\begin{subequations} \label{eq:collisional_Boltzmann_hierarchy_trunc}
  \begin{align}
  \dot{D}_{g, \comp{\ga}} + \frac{4}{3}k^2 V_\comp{\ga} &  = 0 \eqcomma  \label{eq:ell_0_hierarchy}\\
  \dot{V}_\comp{\ga} - \frac{1}{4}D_{g, \comp{\ga}} +
  \frac{1}{6}k^2 \Pi_\comp{\ga} - \Phi - \Psi  & =
  - \taudot (V_\comp{\ga} - V_\comp{b})\eqcomma  \label{eq:ell_1_hierarchy} \\
  \dot{\Pi}_\comp{\ga} - \frac{8}{5} V_\comp{\ga} & = - \taudot \frac{9}{10}\Pi_\comp{\ga} \label{eq:ell_2_hierarchy}\eqdot
  \end{align}
  \end{subequations}

\subsubsection{Polarization hierarchy}

As discussed in \SEC{chap:perts;sec:crosssec}, photons scattered
at a right angle are are preferentially polarized along the
direction orthogonal to the scattering plane (\ie in the $\EOUT_x$
direction in \FIG{fig:pol_geometry} when $\theta = \pi/2$).
Expanding the incoming intensity in spherical harmonics according
to
 \be
 I^\IN(\theta, \phi) = \sum_{\ell} \sum_{m} I_{\ell m} Y_{\ell m} (\theta, \phi)
 \eqcomma
 \ee
then the resulting $Q^\OUT$, from (\ref{eq:scattered_Q}) is
 \be
 Q^\OUT = \frac{3 \si_T}{4\pi} \sqrt{\frac{2\pi}{15}} \text{Re }  I_{22}
 \eqcomma
 \ee
which shows that if the incoming photon intensity as a function of
direction has a non-zero component of  $Y_{22}$, associated with
an $\ell = 2$ quadrupolar moment, then there will be a net linear
polarization of the outgoing distribution.

In analogy with the intensity distribution function $f$, we denote
by $f^Q=\ol{f}^Q(\eta, p) + F^Q(\eta, x^i, p, n^i)$ the perturbed
distribution function in phase space and by $\Theta^Q$ the
brightness perturbation for the Stokes parameter Q,
 \be
 \Theta^Q = \dfrac{1}{4} \dfrac{\int_0^\infty \ol{f}^Q p^3 \dr p}{\int_0^\infty F^Q p^3 \dr
 p} \eqdot
 \ee
Then the collisional Boltzmann equation for the brightness
perturbation $f^Q$ in Fourier space is
\citep{Bond:1984fp,Kosowsky:1996cy}
 \be \label{eq:Q_collisional_Boltzmann}
 \dot{\Theta}^Q + \imath k \mu \Theta^Q
 = -
 \dot{\tau} \left[ \TQ + \dfrac{1}{2}(1 - P_2)
 \left( \Theta_2 + \TQ_2 - \TQ_0 \right) \right]  \eqdot
 \ee
Expanding the equation in Legendre polynomials as in
\rrp{eq:Legendre_expansion}, we obtain the Boltzmann polarization
hierarchy:
\begin{align}\label{eq:Qpol_Boltzmann_hierarchy}
 \dot{\Theta}^Q_0 + \imath k \TQ_1 & =
 -\dfrac{\taudot}{2}\left[\Theta_2 + \TQ_0 + \TQ_2 \right] \eqcomma \\
 \dot{\Theta}^Q_1 +
 \frac{1}{3} \imath k \left[ \TQ_0 + 2 \TQ_2 \right]
 & = - \taudot \TQ_1 \eqcomma \\
 \dot{\Theta}^Q_2 +
 \frac{2}{5} \imath k \TQ_1 + \frac{3}{5} \imath k \TQ_3
 & =  - \dfrac{\taudot}{10}
 \left[9 \TQ_2 - \Theta_2 + \TQ_0 \right] \eqcomma \\
 \dot{\Theta}^Q_\ell + \frac{\ell}{2\ell+1} \imath  k
 \TQ_{\ell-1} + \frac{\ell+1}{2 \ell+1} \imath  k
 \TQ_{\ell+1} & = - \taudot \TQ_\ell \quad (\ell \geq 3) \eqdot \label{eq:Qpol_Boltzmann_hierarchy_end}
  \end{align}

Polarization effects also feed back into the temperature collision
term, modifying the $\ell = 2$ equation in the temperature
hierarchy (\ref{eq:collisional_Boltzmann_hierarchy}) as follows:
 \be \label{eq:Qell2_temp_including}
 \dot{\Theta}_2 + \frac{2}{5} \imath k \Theta_1 + \frac{3}{5}
 \imath k \Theta_3  =  - \dfrac{\taudot}{10}
 \left[9 \TQ_2 - \Theta_2 + \TQ_0 \right] \eqdot
 \ee

\subsubsection{E and B polarization}
\label{chap:perts;sec:EB}

From the the hierarchy of equations
(\ref{eq:Qpol_Boltzmann_hierarchy}) it is possible to determine
the brightness perturbation for Q today, and define the
corresponding power spectrum. However, the approach using Stokes
parameters is limited by the fact that $U$ and $Q$ are not
rotationally invariant, but are defined with respect to a fixed
coordinate system on the sky. Not only the superposition of
different modes is cumbersome because of the behavior of $Q$ and
$U$ under rotation, but the coordinate system becomes ambiguous
and ill-defined on the whole sky, since it is impossible to define
a rotationally invariant orthogonal basis on the two-sphere.

The solution is to construct two spin 2 quantities from $Q$ and
$U$, which one then expands in the appropriate spin-weighted basis
on the two-sphere \citep{Zaldarriaga:1997xe},
 and reduces to scalar
quantities by acting on them with spin raising and lowering
operators. This manipulations yield two scalar quantities which
are rotationally invariant, and therefore well defined on the
whole sky. Furthermore, one can expand these quantities in terms
of usual spherical harmonics and build two linear combinations
which behave differently under parity transformation: the
combination labelled $E$, in analogy with the electric field, is
invariant under a parity change, while the $B$-type combination
changes it sign, analogous to the magnetic field. Another
terminology, sometimes found in the literature, is $C$ mode for
``curl'' (corresponding to the $B$-type) and $G$ for ``gradient``
(corresponding to the $E$-type).

Another advantage of this decomposition is that only the
cross-correlation between E-polarization and temperature is
needed, since the cross-correlation between $B$ and $E$ or $T$
vanishes since $B$ has opposite parity. Furthermore, scalar modes
do not generate $B$ polarization, due to the peculiar $\mu$
dependence of Thomson scattering, while tensor modes do.
Therefore, the separation of the polarization signal in $E$ and
$B$ modes is useful to separate scalar from tensor contribution,
and to identify foreground contamination or a lensing signal,
which can convert $E$ polarization into $B$ polarization for
scalar modes.

We do not give explicit expressions here, which are rather
technical and are not needed in the following, but refer the
reader to \cite{Zaldarriaga:1997xe} instead. A similar
decomposition, but with a different normalization has been
proposed by \cite{Kamionkowski:1997ks}.

\part{COSMIC MICROWAVE BACKGROUND}
{\selectlanguage{french}
 Nous partons des faits, pour composer des
 théories, et nous tâchons toujours de nous éloigner le moins
 possible de ces faits. Nous ignorons ce qu'est {\em l'essence}
 des choses, et n'en avons cure, parce qu'une telle étude sort de
 notre domaine.}
 {\authstyle{Vilfredo Pareto}}
 {Traité de sociologie générale}
\selectlanguage{english}

\chapter{Fundamental equations}
\label{chap:cmb}
The all sky picture of CMB anisotropy delivered by COBE and more
recently and with 30 times more resolution by WMAP can be
considered as a fingerprint of the early Universe. More precisely,
it is an accurate reproduction of the fluctuations in the
radiation-matter mixture at the epoch of recombination.

In this section we succinctly explain the origin of this picture,
by starting with the behavior of scalar perturbations in a
Universe containing one perfect fluid,
\SEC{chap:cmb;sec:one_fluid}; many of the fundamental features of
the anisotropies can be understood in a simple model with a
mixture of radiation and matter which are coupled only
gravitationally, as demonstrated in
\SEC{chap:cmb;sec:matter_radiation} where the concepts of
adiabatic and CDM isocurvature initial conditions are introduced;
adding a massless neutrino component yields two new growing modes,
the neutrino entropy/density and velocity isocurvature solutions,
derived in \SEC{chap:cmb;sec:neutrinos}. Although the results of
those two sections are already known in the literature, the
derivation presented in this work is original. We then refine the
picture of acoustic oscillations by including baryons in
\SEC{chap:cmb;sec:baryons}, and sketch the origin of damping in
\SEC{chap:cmb;sec:damping}. Finally we derive the line of sight
solution for the observed temperature fluctuations today and
introduce the CMB angular power spectra in
\SEC{chap:cmb;sec:observables}. The understanding and tools
developed in the following will build the basis for the next
chapters, where parameter extraction techniques will be discussed
(\CHAP{chap:data}) and applications presented (Chapters
\ref{chap:beyondsp} and \ref{chap:genic}).

There is a rich literature on the cosmic microwave background but
unfortunately an updated work which encompasses both and
introduction to the field and more advanced material, covering the
rapid evolution of the last few years, is presently lacking.
Throughout this and the next chapter we give ample references to
the classic and more recent research papers; as background
material, \cite{NatoProc96} is a valuable source which presents an
introduction to the CMB theory as well as some observational
issues; \cite{Durrer:2001gq} is built on a gauge invariant
formalism similar to the one used here; \cite{Book:Partridge} is a
good introductory overview written at the onset of the recent
data-driven epoch. A rather complete review of both theory and
data analysis is offered by \cite{Hu:2001bc}.

\section{One perfect fluid}
\label{chap:cmb;sec:one_fluid}

 We begin by examining the behavior
of scalar perturbations in a flat ($\curv = 0$) universe which
contains a single perfect fluid, described by $w = c_s^2 =
\textrm{const}$, and $\Gamma = \Pi = 0$.

Since the anisotropic stress vanishes, from
\rrp{eq:anisotropic_stress} it follows $\Psi = \Phi$. The
evolution of the perturbations is given by the two conservation
equations
(\ref{eq:pertd_conservation_eq_1_D}--\refp{eq:pertd_conservation_eq_2_D})
supplemented by the Poisson equation (\refp{eq:poisson}), which in
Fourier space read:
 \begin{align} \label{eq:1_fluid_system}
  \dot{D} - 3w \Hbl D & = -(1+w)k^2 V \eqcomma\\
  \dot{V} + \Hbl V & = \Psi + \frac{c_s^2}{1+w}D \eqcomma\\
  -k^2 \Psi & = \frac{3}{2}\Hbl^2 D \eqdot
 \end{align}
These equations can be combined into a second order equations for
the density contrast:
 \be
 \ddot{D} + (1-3w)\Hbl \dot{D}
 - \frac{3}{2}\Hbl^2(1 + 2w - 3w^2) D + c_s^2k^2 D = 0
 \ee
By defining a new variable $x \equiv k\eta$ and the parameter $\nu
\equiv 2/(1+3w)$, we obtain the following equation for $\call{D}
\equiv D x^{\nu -2}$
 \be \label{eq:second_order_D}
 \frac{\dr^2}{\dr x^2} \call{D} + \frac{2}{x} \frac{\dr}{\dr x}  \call{D} +
 \left[c_s^2 - \frac{\nu(\nu+1)}{x^2} \right]
 \frac{\call{D}}{x^2} = 0 \eqcomma
 \ee

For $c_s^2 \ne 0$ the solution is a linear combination of
spherical bessel ($j_\nu$) and von Neumann ($n_\nu$) functions of
order $\nu$ \citep{Book:Abramowitz:70}
 \be
 \call{D} = C_1 j_\nu(c_s x) + C_2 n_\nu(c_s x) \equiv Z_\nu(c_s x)
 \eqdot
 \ee
Therefore the general solution of Eqs.~(\ref{eq:1_fluid_system})
is
 \begin{align}
 D & =  x^{2 -\nu} Z_\nu(c_s x) \eqcomma \\
 V & =  \frac{3}{2} \nu \left[
 Z_\nu(c_s x) x^{1-\nu} + \frac{2-\nu}{3\nu(1+\nu)}
 x^{2-\nu}Z_{\nu-1}(c_sx) \right] \eqcomma \\
 \Psi & =  - \frac{3}{2} \nu^2 x^{-\nu} Z_\nu(c_s x) \eqdot
 \end{align}

The asymptotic behavior of the Bessel and von Neumann functions is
\begin{subequations} \label{eq:asymptotic_bessel_neumann}
 \begin{alignat}{2}
 j_\nu & \propto  x^\nu \quad \textrm{for $c_sx \ll 1$,} & \qquad
 j_\nu &  \propto  \frac{1}{x}\cos(c_s x - \ga_\nu) \quad \textrm{for $c_sx \gg 1$,}
 \quad\\
 n_\nu & \propto  x^{-(\nu+1)} \quad \textrm{for $c_sx \ll 1$,} & \qquad
 n_\nu & \propto  \frac{1}{x}\sin(c_s x - \ga_\nu) \quad \textrm{for $c_sx \gg 1$.}
 \end{alignat}
 \end{subequations}
with $\ga_\nu \equiv \pi(\nu + 1)/2$. For an expanding universe
($x>0$) and $\nu > -1$ (\ie $w < -1$ or $w > -1/3$) $n_\nu$ is
divergent at early times, $c_s x \ll 1$. Therefore we set $C_2 =
0$ and we obtain the asymptotic solutions (for $w > -1/3$)
 \be
 \left\{
 \begin{array}{l}
 \begin{aligned}
 \DS \Psi &= \Psi_0 \\
 \DS D & = -\frac{2}{3}\frac{\Psi_0}{\nu^2}x^2 \\
 \DS k V & = \frac{2}{(1+\nu)\nu^2}\Psi_0 x
 \end{aligned}
 \end{array}
 \right. \quad \text{for } c_s x \ll 1
 \ee
and
 \be
 \left\{
 \begin{array}{l}
 \begin{aligned}
 \DS \Psi &= \Psi_0 x^{-(1+\nu)}\cos(c_sx + \ga_\nu)\\
 \DS D &= -\frac{2}{3}\frac{\Psi_0}{\nu^2}x^{1-\nu}\cos(c_sx + \ga_\nu) \\
 \DS k V &= \dfrac{(\nu-2)\Psi_0}{3(1+\nu)} x^{1-\nu}\cos(c_sx + \ga_{\nu-1})
 \end{aligned}
 \end{array}
 \right. \quad \text{for } c_s x \gg 1.
 \ee
This solution was first discovered by \cite{Bardeen:1980kt}. The
Bardeen potential is constant on super-horizon scales, and decays
once inside the acoustic horizon. On scales smaller than the
acoustic horizon ($c_s x \ll 1$) density perturbations oscillate:
the gravitational attraction is resisted by the fluid pressure ($w
\ne 0$) and this sets up acoustic oscillations. The amplitude of
density and velocity fluctuations remains constant inside the
horizon in the case of radiation ($\nu= 1, w = 1/3$), while it
increases for $w > 1/3$ or $w > -1/3 $. The behavior of the
density and velocity perturbations on scales larger then the
horizon depends on the variable under consideration. While $D$,
corresponding to the density contrast in the comoving gauge, is
growing, the density contrast in the flat slicing gauge $D_g$
remains constant. Therefore there is no universal criterion to
establish the growth of perturbations outside the horizon: the
behavior depends on the chosen gauge. As we go to early times, $x
\rightarrow 0$, perturbation theory remains valid as long as it is
possible to find a gauge in which the largest perturbation
variable does not diverge. We come back to this point in
\SEC{chap:params;sec:ic}, where we derive the most general initial
conditions.

The case of dust $w=c_s^2 = 0$ has a power-law solution on all
scales. It suffices to remark that \rr{eq:second_order_D} reduces
to
 \be
 \frac{\dr^2}{\dr x^2} D + \frac{2}{x} \frac{\dr}{\dr x} D - \frac{6}{x^2}D = 0 \eqcomma
 \ee
whose general solution is $D = Ax^2 + Bx^{-3}$. The growing exact
solution is therefore
 \be
 \left\{
 \begin{array}{l}
 \begin{aligned}
 \DS \Psi & = \Psi_0 \label{eq:Psi_dust}\\
 \DS D & = -\frac{1}{6}\Psi_0 x^2 \propto a \\
 \DS k V & = \frac{1}{3} \Psi_0 x \propto a^{1/2} 
 \end{aligned}
 \end{array}
 \right. \quad \text{for dust, } w=0.
\ee
 Clearly, in a dust universe perturbations always grow on
sub-horizon scales, since there is no pressure to counterbalance
the gravitational attraction.

\section{Cold dark matter and radiation}
\label{chap:cmb;sec:matter_radiation}

In this section we investigate the evolution of perturbations in a
flat universe containing only radiation and a pressureless matter
component which is decoupled from radiation. Thus the matter has
only a gravitational effect and represents a cold dark matter
component. In the next section we include massless decoupled
neutrinos in the picture, while the role of baryons, which are
coupled to photons via Thomson scattering, is investigated in
\SEC{chap:cmb;sec:baryons}.

\subsection{Adiabatic and isocurvature modes}

In this section we use as density variable the density contrast in
the total comoving gauge $\Delta_\comp{\al}$, defined in
\rrp{eq:def_Delta}. We identify the radiation with photons
(subscript $\gamma$), and we have $w_\comp{\ga} = c_\comp{\ga}^2 =
1/3$, while for matter $w_\comp{m} = c_\comp{m}^2 = 0$. We
normalize the scale factor at the matter-radiation equality, so
that \be
 \ol{\rho}_\comp{m}(a_\EQ) = \ol{\rho}_\comp{\ga}(a_\EQ)
 \quad \text{with}\quad a_\EQ \equiv 1
 \quad
 \textrm{hence} \quad \frac{\ol{\rho}_\comp{m}}{\ol{\rho}_\comp{\ga}} = a \eqdot
\ee
 The total equation of state parameter and sound velocity are
therefore
 \be
 w = \frac{1}{3}\frac{1}{a + 1} \quad \text{and} \quad
 c_s^2 = \frac{1}{3}\frac{4}{4+3a} \eqdot
 \ee
As long as we are considering times well before decoupling, the
photons form a tight coupled fluid with baryons, since Thomson
scattering prevents the generation of anisotropic stress (and
higher multipoles in the Boltzmann hierarchy) in the photons
component, $\Pi_\comp{\ga} = 0$, as we show in
\SEC{chap:cmb;sec:baryons}. Therefore, via the anisotropic stress
equation (\refp{eq:anisotropic_stress}), the Bardeen potentials
are equal, $\Psi = \Phi$. The Bardeen equation for $\Phi$
(\refp{eq:Bardeen_eq}) is then
 \be \label{eq:Bardeen_rm}
 \ddot{\Phi} + 3\Hbl(1 + c_s^2) \dot{\Phi} + 3(c_s^2-w)\Hbl^2 \Phi =
 c_s^2 \triangle \Phi + \tfrac{3}{2}\Hbl^2 w \Gamma \eqcomma
 \ee
where $\Gamma = \Gamma_{\textrm{rel}}$ is related to the relative
entropy perturbation $S \equiv S_{m\gamma} =  \Delta_\comp{m} -
\tfrac{3}{4} \Delta_\comp{\ga}$ by \rrp{eq:Gamma_rel}. By using
the Poisson equation we can rewrite the above as an equation for
the total density contrast,
 \be \label{eq:ddot_D}
 \begin{split}
 \Hbl^{-2} \ddot{D} + (1 - 6w + 3c_s^2) \Hbl^{-1}\dot{D}
& - \tfrac{3}{2}(1 + 8w - 3w^2 - 6c_s^2) D = \\
& -  c_s^2\left( \frac{k}{\Hbl} \right)^2
 \left[ D - 3  c_z^2(1 + w) S \right] \eqcomma
 \end{split}
 \ee
where we have introduced $c_z^2 \equiv \ol{\rho}_\comp{\ga}
\ol{\rho}_\comp{m}(c_\comp{\ga}^2 - c_\comp{m}^2)/\left[{(1 + w)
\ol{\rho}}\right] = a/(3a + 4)$.

The energy conservation equation
(\refp{eq:pertd_conservation_eq_1_Dg}) reads for the radiation and
matter components:
 \begin{alignat}{2}
 & \dot{D}_{g, \comp{\ga}} + \dfrac{4}{3}k^2 V_\comp{\ga} = 0 &&
 \quad \text{(radiation),} \label{eq:energy_cons_radiation} \\
 & \dot{D}_{g, \comp{m}} +   k^2 V_\comp{c} = 0 &&
 \quad \text{(matter).} \label{eq:energy_cons_matter}
 \end{alignat}
Subtracting (\ref{eq:energy_cons_matter}) from
(\ref{eq:energy_cons_radiation}) and using that
 \be
 \dfrac{D_{g, \comp{\al}}}{1+w_\comp{\al}} - \dfrac{D_{g,
 \comp{\ba}}}{1+w_\comp{\ba}}=
 \dfrac{\Delta_{\comp{\al}}}{1+w_\comp{\al}} -
 \dfrac{\Delta_{\comp{\ba}}}{1+w_\comp{\ba}}= S_{\al, \ba}
 \ee
 we obtain
 \be \label{eq:dot_S_and_delta_V}
 \dot{S} = -k^2(V_\comp{m} - V_\comp{\ga}) \eqdot
 \ee
In order to find an evolution equation for the entropy $S$, we
derive (\ref{eq:dot_S_and_delta_V}) and making use of the momentum
conservation equation (\refp{eq:pertd_conservation_eq_2_D}) after
a lengthy manipulation we arrive at
 \be \label{eq:ddot_S}
 \Hbl^{-2}\ddot{S} + (1 - 3c_z^2) \Hbl^{-1} \dot{S} =
  \left( \frac{k}{\Hbl} \right)^2
 \left[ \frac{1}{3(1+w)} D - c_z^2 S \right] \eqdot
 \ee

Together, Eqs.~(\ref{eq:ddot_D}) and (\ref{eq:ddot_S}) describe
the evolution of adiabatic (curvature) and isocurvature (dark
matter) perturbations in a flat universe containing only dark
matter and radiation.

We start by considering large scales ($k \ll \Hbl$) at early
times, $a \rightarrow 0$. Then the right hand side of
(\ref{eq:ddot_D}) and (\ref{eq:ddot_S}) is negligible, thus $D$
and $S$ are decoupled. Using the scale factor $a$ as variable, we
obtain an homogeneous system
 \be \label{eq:homogeneous_system_dm_rad}
 \left\{
 \begin{array}{l}
 \begin{aligned}
 \DS a^2 \frac{\dr^2}{\dr a^2}D - 2 D  &= 0 \\
 \DS a^2 \frac{\dr^2}{\dr a^2}S + a \frac{\dr}{\dr a}S & = 0
\end{aligned}
 \end{array}
 \right.
 \qquad\text{(Large scales, radiation epoch)}
 \ee
 whose general solution consists of four modes,
 \be
 \left\{
 \begin{array}{l}
 \begin{aligned}
 D &= D_0 a^2 + D_1 a^{-1} \\
 S &= S_0 + S_1 \ln a
 \end{aligned}
 \end{array}
 \right. \eqdot
 \ee
We will call the mode with $D_0 \ne 0, D_1=S_0=S_1= 0$ the {\it
growing adiabatic mode}, while the one with $S_0 \ne 0, D_0 =
D_1=S_1$ the {\it growing isocurvature mode} (notice that for $a <
1$ the $S_1$ mode is indeed decaying). As we show below, the
isocurvature mode at early times has vanishing total density
contrast, Bardeen potential curvature perturbation, $\zeta = 0$,
hence its name\footnote{The CDM isocurvature mode is sometimes
termed ``isothermal'' in the literature: this comes from the fact
that $D=0$ implies $ \frac{\delta T}{T} = -
\frac{\ol{\rho}_m}{\ol{\rho}_\ga}\Delta_m \approx 0$ at early
times. Intuitively, it takes only a small perturbation in the
radiation component to compensate for a fluctuation in the matter
at early times, because the Universe is radiation dominated.}.

Consider first the growing adiabatic mode: we can now restore the
solution for $D$ in the source term on the right hand side of
\rr{eq:ddot_S} to find the solution for $S$ up to second order in
$k/\Hbl$. The Bardeen potential is easily found from the Poisson
equation, and the result is
 \be \label{eq:ad_growing_mode}
 \left\{
 \begin{array}{l}
 \begin{aligned}
 \DS D &= D_0 a^2 \\
 \DS S &= \frac{D_0}{64} \left( \frac{k}{\Hbl} \right) ^2 a^2  \propto a^4 \\
 \DS \Phi &= - \frac{3D_0}{2}\left( \frac{\Hbl a}{k} \right)^2 = \text{ const} \\
 \DS k V &= \dfrac{1}{2} \frac{k}{\Hbl} \Phi \propto a \\
 \DS \zeta & = - \frac{9D_0}{4}\left( \frac{\Hbl a}{k} \right)^2 = \text{ const}
  \end{aligned}
 \end{array}
 \right. \quad \text{(adiabatic, radiation epoch).}
 \ee
Clearly, we recover the behavior already found in the single
radiation fluid case for the potential. We also discover that the
entropy perturbation grows as $a^4$, but remains negligible on
large scales, thus the adiabaticity condition $S \approx 0$ is
maintained on large scales.

For the growing isocurvature mode we find, to the same
approximation
  \be \label{eq:iso_growing_mode}
 \left\{
 \begin{array}{l}
 \begin{aligned}
 \DS D & =   \frac{S_0}{12} \left( \frac{k}{\Hbl} \right)  ^2 a \propto a^3 \\
 \DS S & = S_0 \\
 \DS \Phi & = - \frac{S_0}{8} a \\
 \DS k V & = -\frac{S_0}{8}\frac{k}{\Hbl}a  \propto a^2 \\
 \DS \zeta & = - \frac{3S_0}{16} a
 \end{aligned}
 \end{array}
 \right. \quad \text{(isocurvature, radiation epoch).}
 \ee
We see that there is no generation of entropy on large scales
($\dot{S} = 0$), however the isocurvature condition $\Phi \approx
0$ is maintained only as long as $a \ll 1$. Naively we would
expect that, as long as the scale considered is outside the
horizon, the term containing $S$ on the right hand side of
\rr{eq:ddot_D} is suppressed as $k^2/\Hbl^2$, thus $D$ (hence
$\Phi$) should not grow significantly. However, since $\Phi
\propto \Hbl^2/k^2$, effects of magnitude $k^2/\Hbl^2$ in $D$ are
significant for $\Phi$. This can be seen more directly by
rewriting the right hand side of \rr{eq:Bardeen_rm} as $-c_s^2
k^2/\Hbl^2 \Phi - 2(1+w)c_s^2 c_z^2 S$. Therefore even on
super-horizon scale the term $\propto S$ act as a source for
$\Phi$ whenever $c_s^2c_z^2$ is significantly non-zero. This is
the case during the transition from the radiation to the matter
dominated epoch.

Having established the behavior in the early epoch, we now turn
our attention to scales which enter the horizon when the universe
is well matter dominated, \ie to wavelengths such that
 \be
 k \ll k_\EQ \equiv \Hbl(a_\EQ) \eqdot
\ee
 The effects of the radiation-matter transition are easiest to discuss by looking
at the behavior of the curvature perturbation $\zeta$. To this end
we rewrite the evolution equation $(\refp{eq:evolution_of_zeta})$
as
 \be \label{eq:dotzeta}
 \dot{\zeta} = - c_s^2 \Hbl
 \left[ \frac{2}{3(1+w)} \left( \frac{k}{\Hbl} \right)^2 \Phi
 + 3 c_z^2 S \right] \eqdot
 \ee
The term $\propto \Phi$ on the right hand side is always
negligible on super-horizon scales ($k/\Hbl \ll 1$); for adiabatic
perturbations we also have $S=0$, and thus we obtain
 \be \label{eq:zeta_is_constant}
 \zeta = \text{ const} \qquad \text{(adiabatic, all times),}
 \ee
the usual conservation law for $\zeta$ in the adiabatic case. For
the isocurvature mode ($S = S_0 = \text{const}$) we find by
integration
 \be \label{eq:zeta_is_constant_isoc}
 \zeta = - 3S_0 \int_0^a \frac{\dr a}{a} c_s^2 c_z^2
  \underset{a \rightarrow \infty}{\longrightarrow}
 - \frac{1}{3} S_0 \quad  \text{(isocurvature, matter epoch).}
 \ee
The radiation-matter transition generates a curvature perturbation
from the initial isocurvature one, and this even on super-horizon
scales.

Since $\zeta = \text{ const }$ in the matter era independently on
the initial conditions, we can find the value of the Bardeen
potential in the matter epoch simply by integrating the definition
of the curvature perturbation, using that $w = \text{const}$ as
well. We then obtain the relation (valid only in the regime where
$\zeta = \text{const}$, $w = \text{const}$)
  \be \label{eq:Phi_with_w}
  \Phi = \frac{3(1+w)}{5+3w} \zeta + C a^{-\tfrac{5+3w}{2}}
  \eqcomma
  \ee
and we can drop the second term, which is decaying for $w >
 -5/3$. Therefore
  \be \label{eq:Phi_as_fct_of_Zeta}
  \Phi(a \gg a_\EQ) = \text{ const } =
  \tfrac{3}{5} \zeta \quad \text{(matter epoch, independent of IC).}
 \ee
For the adiabatic mode, $\zeta = \text{const}$ in the radiation
era as well, therefore we can apply (\ref{eq:Phi_with_w}) with $w
\approx \text{const} = 1/3$, getting
  \be \label{eq:Phi_as_fct_of_Zeta_RD}
  \Phi(a \ll a_\EQ) = \text{ const } =
  \tfrac{2}{3} \zeta \quad \text{(radiation epoch, adiabatic).}
 \ee

Let us denote by $\Phi_\RD$ the value of $\Phi$ at the moment when
the initial conditions for the perturbations are specified, deep
in the radiation era. The adiabatic mode corresponds to $S_0 = 0,
\Phi_\RD \neq 0$, while the isocurvature mode has $S_0 \neq 0,
\Phi_\RD = 0$. From (\ref{eq:Phi_as_fct_of_Zeta}) we know that
$\Phi$ is constant on super-horizon scales in the matter era,
independent of the type of initial conditions; we denote its value
by $\Phi_\MD$, and we wish to express it in terms of $S_0,
\Phi_\RD$. For adiabatic perturbations, $\zeta$ stays constant
through the transition, and therefore combining
(\ref{eq:Phi_as_fct_of_Zeta}) with
(\ref{eq:Phi_as_fct_of_Zeta_RD})
 \begin{align} \label{eq:Phi_MD_and_RD}
 \Phi_\MD & \approx \tfrac{9}{10} \Phi_\RD \quad
 \text{(adiabatic, large scales)}. \\
\intertext{For isocurvature perturbations, the growth of $\zeta$
through the transition gives a non-zero $\Phi$ in the matter
epoch, from (\ref{eq:Phi_as_fct_of_Zeta}) and
(\ref{eq:zeta_is_constant_isoc}) :}
 \Phi_\MD & \approx -\frac{1}{5} S_0  \quad
 \text{(isocurvature, large scales)}. \label{eq:Phi_MD_and_S_0}
\end{align}

In conclusion, we can summarize our results in terms of a transfer
matrix as
 \be \left(
 \begin{array}{l}
 \Phi \\
 S
 \end{array}
 \right)_{a \gg a_\EQ}
 =
 \left(
 \begin{array}{l l}
 9/10   & -1/5 \\
 0      & 1
 \end{array}
 \right)
 \left(
 \begin{array}{l}
 \Phi_0 \\
 S_0
 \end{array}
 \right) \eqdot
 \ee
It is often useful to use the curvature perturbation as a variable
describing the adiabatic mode, instead of $\Phi$. In terms of the
initial values of the curvature and entropy perturbations,
$(\zeta_0, S_0)$, the final values in the matter era are given by
a transfer matrix of the form
 \be \left(
 \begin{array}{l}
 \zeta \\
 S
 \end{array}
 \right)_{a \gg a_\EQ}
 =
 \left(
 \begin{array}{l l}
 T_{\zeta \zeta} & T_{\zeta S} \\
 0               & T_{S S}
 \end{array}
 \right)
 \left(
 \begin{array}{l}
 \zeta_0 \\
 S_0
 \end{array}
 \right) \eqdot
 \ee
From the above analysis, we conclude that for scales $k \ll k_\EQ$
the transfer coefficients are
 \be
 T_{\zeta \zeta} = 1 \eqcomma \quad  T_{\zeta S} = - \frac{1}{3}
 \eqcomma \quad
 T_{S S} = 1 \eqdot
 \ee
For smaller scales, which enter the horizon before the universe is
completely matter dominated, the coefficients have to be found
numerically.

\subsection{Acoustic oscillations}

We have seen in \SEC{chap:cmb;sec:one_fluid} that perturbations in
a fluid of photons oscillate on scales smaller than the horizon.
We now discuss the corresponding behavior in the presence of
matter, and link the phase of the oscillations to the adiabatic or
isocurvature initial conditions on large scales.

Neglecting the anisotropic stress, $\Pi_\comp{\ga} = 0$, the
conservation equations
(\ref{eq:pertd_conservation_eq_1_Dg}--\refp{eq:pertd_conservation_eq_2_Dg})
for photons read
 \begin{align}
 \dot{D}_{g, \comp{\ga}} + \tfrac{4}{3}k^2 V_\comp{\ga} & =   0\\
 \dot{V}_\comp{\ga} - \tfrac{1}{4}D_{g, \comp{\ga}} & =  2 \Phi
 \end{align}
where $\Phi$ can be considered as an external potential determined
by the Poisson equation. We can recast the above in a second order
equation for the density perturbation:
 \be \label{eq:HO_second_order}
 \ddot{D}_{g, \comp{\ga}} + c_\ga^2 k^2 D_{g, \comp{\ga}}  =   2
 \Phi \eqdot
 \ee

\subsubsection*{Adiabatic initial conditions}

Let's consider \rr{eq:HO_second_order} deep in the matter era,
when the driving force is just a constant set by the dominating
matter contribution in the adiabatic case. Then the general
solution of \rr{eq:HO_second_order} is
 \begin{align}
 D_{g,\comp{\ga}} & = C_1 \cos(c_\comp{\ga} k \eta) + C_2 \sin(c_\comp{\ga} k \eta)
 - 8 \Phi  \label{eq:HO_solution_D_AD} \\
 k V_{\comp{\ga}}   & = \frac{1}{ 4 c_\comp{\ga}} \left[
 C_1 \sin(c_\comp{\ga} k \eta) - C_2 \cos(c_\comp{\ga} k \eta)
 \right] \label{eq:HO_solution_V_AD} \eqdot
 \end{align}

For small scales, where all choices of density perturbation are
equivalent, we recover the oscillatory behavior already found in
\SEC{chap:cmb;sec:one_fluid}. The density perturbations perform
harmonic oscillations around a zero point displaced by a constant
factor.

The constants $C_1$ and $C_2$ are fixed by the initial conditions,
adiabatic or isocurvature, established by matching the above
solution on large scales with the results of the previous section.
To this end, we shall use the following relation between $D_{g,
\comp{\ga}}$ and $\Delta_\comp{\ga}$, which follows from the
definitions of the variables:
 \be \label{eq:D_g_relation}
 \tfrac{1}{4} D_{g, \comp{\ga}} =  \tfrac{1}{3} \Delta_\comp{m}
 - \tfrac{1}{3} S - \Hbl V  - \Phi \eqdot
 \ee
From the momentum conservation equation
(\refp{eq:pertd_conservation_eq_2_D}) we obtain for the total
velocity perturbation in the matter era
 \be
 \dot{V} + \Hbl V = \Phi \eqcomma
 \ee
with solution
 \be \label{eq:V_in_MD}
 V = V_1 a^{-1} + \frac{2}{3} \Hbl^{-1} \Phi \eqdot
 \ee
The term $\propto a^{-1}$ is decaying, therefore we retain $V \sim
\tfrac{2}{3} \Hbl^{-1} \Phi$. Inserting this into
\rr{eq:D_g_relation} and using that in the matter era $\Phi = 9/10
\Phi_0 - S_0/5$ we obtain on large scales, where $\Delta_\comp{m}
\sim (k/\Hbl)^2 \Phi \ll \Phi$,
 \be \label{eq:D_g_ga_superhorizon}
 \tfrac{1}{4} D_{g, \comp{\ga}}(a \gg a_\EQ) \approx \text{const } = - \tfrac{3}{2} \Phi_\RD \eqdot
 \ee
Thus on large scales and in the matter epoch, $D_{g, \comp{\ga}}$
is independent of the entropy perturbation, and is simply related
to the primordial Bardeen potential.

The adiabatic mode stays decoupled from the isocurvature mode on
super-horizon scales, therefore we can set the initial conditions
for the solution
(\ref{eq:HO_solution_D_AD}--\ref{eq:HO_solution_V_AD}) by taking
its constant-time super-horizon limit, \ie $k\rightarrow 0$, $\eta
= \text{const} \gg \eta_\EQ$. This gives, with $S_0=0$
 \be
 \tfrac{1}{4} D_{g, \comp{\ga}} = \tfrac{1}{4} C_1 - 2 \Phi_\MD
 \ee
and comparing with \rr{eq:D_g_ga_superhorizon} and using again
(\ref{eq:Phi_MD_and_RD}) we obtain
 \be \label{eq:C1_AD}
C_1 = \tfrac{4}{3} \Phi_\MD
 \eqdot
 \ee

The constant $C_2$ is set by noting that the adiabatic condition
$S = 0$ is preserved on super-horizon scales, and that, because of
energy-momentum conservation for matter and radiation, this
implies
 \be
 V_\comp{\ga} = V_\comp{m} \eqdot
 \ee
Since
 \be \label{eq:V_as_Vr_plus_Vm}
 V = \frac{4}{4 + 3a} V_\comp{\ga} + \frac{3a}{4+ 3a}V_\comp{m}
 \ee
we have that $V \approx V_\comp{m}$ for $a \gg a_\EQ$, and with
(\ref{eq:V_in_MD}) it follows that
 \be \label{eq:Vgamma_in_MD}
 V_\comp{\ga} = V_\comp{m} \approx \tfrac{2}{3} \Phi \Hbl^{-1}
 \eqdot
 \ee
Comparing this with the large scale limit of
\rr{eq:HO_solution_V_AD},
 \be
 \lim_{k \rightarrow 0 , \eta = \text{const}} V_\comp{\ga}
 =
 \frac{\eta}{4}
 \left[C_1 - C_2 \lim_{y \rightarrow 0}\frac{\cos y}{y} \right]
 \eqcomma
 \ee
we see that we need to impose $C_2 = 0$, otherwise $V_\comp{\ga}$
would diverge in the large-scale limit $y \rightarrow 0$, and we
recover again (\ref{eq:C1_AD}) by using $\Hbl = 2/\eta$:
 \be
 C_1 = \tfrac{4}{3}\Phi_\MD
 \quad \text{and}\quad
 C_2 = 0 \eqdot
 \ee

In conclusion, the adiabatic solution is
 \be
  \left\{
 \begin{array}{l} \label{eq:HO_adiabatic}
 \begin{aligned}
 \DS D_{g,\comp{\ga}} & = \frac{4}{3} \Phi \cos(c_\comp{\ga} k \eta)  - 8 \Phi
 \\
 \DS k V_{\comp{\ga}}   & = c_\comp{\ga} \Phi \sin(c_\comp{\ga} k \eta)
 \end{aligned}
 \end{array}
  \right. \quad \text{(adiabatic).}
 \ee
\subsubsection*{Isocurvature initial conditions}

As we have seen in the previous section, $\Phi = 0$ is no longer
maintained in the matter era for isocurvature initial conditions.
It is therefore convenient to solve (\ref{eq:HO_second_order}) at
early times in the radiation regime, where we know that the
driving term on the right hand side is $\Phi \propto \eta$ (\CF
\rrp{eq:iso_growing_mode}):
 \begin{align}
 D_{g,\comp{\ga}} & = C_1 \cos(c_\comp{\ga} k \eta) + C_2 \sin(c_\comp{\ga} k \eta)
 - \frac{3}{4} k^{-2} \eta_\EQ^{-1} S_0 \eta  \eqcomma \label{eq:HO_solution_D_iso} \\
 k V_{\comp{\ga}}   & = \frac{1}{ 4 c_\comp{\ga}}
 \left[ C_1 \sin(c_\comp{\ga} k \eta) - C_2 \cos(c_\comp{\ga} k \eta)
 \right] + \frac{9}{16}k^{-3}\eta_\EQ^{-1} S_0
 \label{eq:HO_solution_V_iso} \eqdot
 \end{align}

The constants $C_1$ and $C_2$ are determined by looking at the
early time limit on super-horizon scales, $\eta \rightarrow 0$, $k
= \text{const} \ll k_\EQ$. From the early-times solution
(\ref{eq:iso_growing_mode}) we have that $D_{g, \comp{\ga}}
\rightarrow 0$ for $\eta \rightarrow 0$, and therefore we need to
set $C_1 = 0$. The early time limit for \rr{eq:HO_solution_V_iso}
gives
 \be \label{eq:Vlimit_ISO}
 \lim_{\eta \rightarrow 0, k = \text{const}} k V_\comp{\ga} =
 - \frac{C_2}{ 4 c_\comp{\ga}} + \frac{9}{16}k^{-3}\eta_\EQ^{-1} S_0
 \eqcomma
 \ee
while from the isocurvature solution (\ref{eq:iso_growing_mode})
combined with (\ref{eq:V_as_Vr_plus_Vm}) we have for $a \ll a_\EQ$
 \be
  \lim_{\eta \rightarrow 0, k = \text{const}} k V_\comp{\ga} =
  k V \propto \eta^2 \rightarrow 0 \eqdot
 \ee
By requiring that the left hand side of (\ref{eq:Vlimit_ISO})
vanishes we conclude that
 \be
 C_2 = \frac{3}{4c_\comp{\ga}} k^{-3} \eta_\EQ^{-1} S_0 \eqdot
 \ee

In conclusion, isocurvature initial conditions excite a sine
oscillation in the radiation density:
 \be \label{eq:HO_isocurvature}
 \left\{
 \begin{array}{l}
 \begin{aligned}
 \DS  D_{g,\comp{\ga}} & = \frac{3}{4} k^{-2} \eta_\EQ^{-1} S_0 \left[
 \sqrt{3}{k} \sin(c_\comp{\ga} k \eta)  - \eta \right] \\
 \DS k V_{\comp{\ga}}  & = - \frac{3\sqrt{3}}{16} k^{-2} \eta_\EQ^{-1} S_0 \left[
 \sqrt{3}{k} \cos(c_\comp{\ga} k \eta)  - 1 \right]
 \end{aligned}
 \end{array}
  \right. \quad \text{(isocurvature).}
 \ee

An heuristic argument \citep{Hu:1995jd} explains why adiabatic
initial conditions excite the cosine mode while isocurvature
initial conditions produce the sine mode: at early times, the
potential acting as a driving force on the right hand side of
\rr{eq:HO_second_order} is constant for adiabatic initial
conditions, while it is $\propto \eta$ in the isocurvature case.
This mimics a cosine and a sine forcing term, respectively, and
thus the corresponding modes get excited. An approximated
analytical solution valid until recombination and through the
radiation-matter transition can be found in \cite{Hu:1995uz}.

\section{Neutrinos and initial conditions}
\label{chap:cmb;sec:neutrinos}

In this section we extend the above treatment to include massless
neutrinos. They are described as an additional relativistic
component, which is decoupled from the others below a temperature
of a few MeV, and therefore their distribution function obeys the
collisionless Boltzmann equation. We shall see in the following
that the anisotropic stress created by free streaming of neutrinos
considerably complicates the simple picture of the previous
section.

By including one more component in the mixture, we generally
expect two additional modes to arise, which we will be able to
identify with the so-called ``neutrino isocurvature density''
(NID) and ``neutrino isocurvature velocity'' (NIV) modes. In the
following, we shall refer to both of them as to ``neutrino
isocurvature modes''\footnote{The term ``isocurvature'' is
somewhat abused for the neutrino density mode, see the remark
after \rr{eq:zeta_for_nid} on page \pageref{eq:zeta_for_nid}. We
nevertheless employ this terminology for simplicity and
consistency with the literature.}, and we will sometimes call the
neutrino density mode ``neutrino entropy'', which is a more
appropriate definition in our view. These two modes were first
found by \cite{Bucher:1999re}, who solved a formal expansion in
powers of $\eta$ of the Einstein and conservation equations at
early times and on large scales (\ie for $\eta k \rightarrow 0$)
in synchronous gauge, an analysis repeated in the gauge invariant
formalism in \cite{Trotta:2001tesi}. The approach we propose here
offers a more physical understanding and the approximations we
employ could be extended to a refined analytical model of the
sub-horizon structure of the neutrino modes angular power spectra.
We explicitly give some details of the derivation, since to our
knowledge this calculation is new.

We argue in \SEC{chap:cmb;sec:anisoic} that an ``anisotropic
stress mode'', which is characterized by a non-vanishing
$\Pi_\comp{\nu}$ at early times, is non-physical, since it leads
to incurable divergences in the perturbation variables.

\subsection{Evolution equations for a three components model}

In the presence of neutrinos, the background radiation energy
density is written as
 \be
 \ol{\rho}_\comp{r} = \ol{\rho}_\comp{\ga} + \ol{\rho}_\comp{\nu}
 = \ol{\rho}_\comp{\ga}(1 + \rn) \eqcomma
 \ee
where we have defined the constant $\rn \equiv (7
N_\nu/8)(4/11)^{4/3} \approx 0.68$ for $N_\nu = 3$ neutrino
families. As before, the scale factor is normalized to
matter-radiation equality, the conformal Hubble parameter is
 \be
 \Hbl = \dfrac{1 + \eta/2}{\eta + \eta^2/4} = \dfrac{(1 +
 7a)^{1/2}}{7a} \eqcomma
 \ee
and the cosmological parameters as a function of the scale factor
are of the form
 \begin{align}
 \Om_\nu(a) & = \dfrac{\rn}{(1+\rn)(1+a)} \eqcomma \\
 \Om_\ga(a) & = \dfrac{1}{(1+\rn)(1+a)} \eqcomma \\
 \Om_m(a) & = \dfrac{a}{(1+a)} \eqdot
 \end{align}
We still neglect the dynamical effect of baryons, which to lowest
order is unimportant, but continue to assume that Thomson
scattering drives to zero all multipoles $\ell \geq 2$ in the
Boltzmann hierarchy for photons, which are then described as a
relativistic perfect fluid. Neutrinos become collisionless after
neutrino decoupling, therefore the fluid approximation is
insufficient. A neutrino anisotropic stress is generated by free
streaming and to lowest order we cut the Boltzmann hierarchy for
neutrinos, \rrp{eq:pertd_Boltzmann_hierarchy_trunc}, by setting to
zero all moments $\geq 3$. The goal is to derive second order
evolution equations for the three relevant and physical
quantities: the total density contrast $D$, the entropy
perturbations in the dark matter, $S_{m \ga}$, and in the
neutrinos, $S_{\nu \ga}$, supplemented by an evolution equation
for the neutrino anisotropic stress.

The source term in the Bardeen equation is modified in two ways:
there is an additional entropy contribution coming from the
neutrino entropy perturbation $S_{\nu \ga}$, and we have to take
into account the anisotropic stress term. This gives for the
evolution equation of the total density contrast $D$ (compare with
(\refp{eq:ddot_D}))
 \be \label{eq:ddot_D_with_nu}
 \begin{split}
 \Hbl^{-2} \ddot{D} + (1 - 6w + 3c_s^2) \Hbl^{-1}\dot{D}
& - \tfrac{3}{2}(1 + 8w - 3w^2 - 6c_s^2) D = \\
 -     \left( \frac{k}{\Hbl} \right)^2 &\Bigg\{
 \left[c_s^2 D - 3 c_s^2 c_z^2(1 + w)
 \left( S_{m \ga} - \dfrac{\rn}{1 + \rn} S_{\nu \ga}  \right) \right]   \\
& + \dfrac{2\rn}{3(1+\rn)(1+a)} \left[ \Hbl \dot{\Pi}_\comp{\nu} -
\left[(1+3w) - 3c_z^2 \right]\Hbl^2 \Pin - \frac{1}{2}k^2 \Pin
\right] \Bigg\}.
 \end{split}
 \ee
Equation (\refp{eq:ddot_S}) acquires extra terms coming from
$S_{\nu \ga}$, reading
 \be \label{eq:ddot_S_with_nu}
 \Hbl^{-2}\ddot{S}_{m \ga} + (1 - 3c_z^2) \Hbl^{-1} \left[ \dot{S}_{m \ga} - \dot{S}_{\nu\ga}\right]  =
  \left( \frac{k}{\Hbl} \right)^2
 \left[ \frac{1}{3(1+w)} D - c_z^2 S_{m \ga} - \dfrac{4 w \rn}{3 (1+w)(1+\rn)} S_{\nu \ga}
 \right].
 \ee
In deriving the above equations we have made use of
(\refp{eq:relation_Dg_Delta}) and (\refp{eq:dot_S_and_delta_V})
together with the following useful relations:
 \begin{align}
 \dfrac{1}{4} \Delta_\comp{\ga} & = \dfrac{1}{3(1+w)} D -
 \dfrac{4\rn}{3(4+3a)(1+\rn)} S_{\nu \ga} - \dfrac{a}{4+3a} S_{m
 \ga} \eqcomma\\
 k V_\comp{\ga} & = k V - \dfrac{4 \rn}{(4+3a)(1+\rn)}(V_\comp{\nu}
 - V_\comp{\ga})  - \dfrac{3a}{4+3a}(V_\comp{m}
 - V_\comp{\ga}) \eqdot
 \end{align}

We obtain an equation for the neutrino entropy perturbation by
deriving the difference of the momentum conservation equation for
neutrinos (\rrp{eq:ell1_collisionless} written for $\nu$ instead
of $\ga$) and the momentum conservation for the photon fluid,
(\refp{eq:energy_cons_radiation}), with the result
 \be \label{eq:ddot_Snuga}
 \ddot{S}_{\nu \ga} + \dfrac{k^2}{3} S_{\nu \ga} = \dfrac{k^4}{6} \Pin \eqdot
 \ee

The coupled system (\ref{eq:ddot_D_with_nu}),
(\ref{eq:ddot_S_with_nu}) and (\ref{eq:ddot_Snuga}) describes the
evolution of adiabatic and entropy perturbations in a mixture of
photons, dark matter and radiation, once we specify $\Pi_\nu$.
However, on super-horizon scales and for early times, $k/\Hbl\ll
1$, the anisotropic stress is unimportant, since from
(\refp{eq:ell2_collisionless}) written for $\nu$ instead than for
$\ga$, it obeys
 \be \label{eq:aniso_stress_evolution}
 a \dfrac{\dr}{\dr a} k^2 \Pin = \dfrac{8}{5} \dfrac{k}{\Hbl} k
 V_\nu \approx 0 \eqcomma
 \ee
which shows that on super-horizon scales there is no generation of
anisotropic stress, a result expected on the grounds of causality
arguments. At earlier times, the neutrinos were coupled to
electrons via weak interaction processes, which isotropized the
neutrino distribution function suppressing any appreciable
anisotropic stress; hence we can assume that at the time under
consideration (just after neutrino decoupling) there is no
anisotropic stress to zeroth order in powers of $a$, \ie $\Pin =
\calo(a)$ at least.

In the above approximation and for $a \ll 1$ we thus obtain the
simple system
 \be
 \left\{
 \begin{array}{l}
 \begin{aligned}
 \DS a^2 \frac{\dr^2}{\dr a^2}D - 2 D  &= 0 \eqcomma\\
 \DS a^2 \frac{\dr^2}{\dr a^2}S_{m \ga} + a \frac{\dr}{\dr a}S_{m \ga} &= a \frac{\dr}{\dr a}S_{\nu \ga} \eqcomma\\
 \DS a^2 \frac{\dr^2}{\dr a^2}S_{\nu \ga} & = 0 \eqcomma
\end{aligned}
 \end{array}
 \right.
 \ee
 whose general solution consists of six modes,
 \be \label{eq:large_scale_six_modes}
 \left\{
 \begin{array}{l}
 \begin{aligned}
 D &= D_0 a^2 + D_1 a^{-1} \eqcomma\\
 S_{m \ga} &= S_0 + S_1 \ln a + N_v a \eqcomma \\
 S_{\nu \ga} &= N_d + N_v a \eqdot
 \end{aligned}
 \end{array}
 \right.
 \ee
We recognize the growing and decaying adiabatic (the $D_0$ and
$D_1$ terms, respectively) and isocurvature dark matter ($S_0$ and
$S_1$ terms, respectively) modes, and we also find two new
non-decaying modes, a constant neutrino entropy mode $N_d$, and a
neutrino velocity mode $N_v a$ (the reason for this terminology is
explained below).

In order to go beyond this large scales solution, we need to
include the effect of the anisotropic stress. To this end, we
recast \rr{eq:aniso_stress_evolution} by substituting $k
V_\comp{\nu}$ with
 \be
 k V_\comp{\nu} = k V - \dfrac{a \Hbl}{k}
 \left[\dfrac{1}{1+\rn} \dfrac{\dr}{\dr a}S_{\nu \ga} + \frac{3(1+\rn)a }{4}\frac{\dr}{\dr a}S_{m \ga} \right]
 \eqdot
 \ee
From now on we drop the last term on the right hand side, which is
always suppressed by a power of $a$ except in the dark matter
isocurvature case, which we do not investigate further here. For
the total velocity, the constraint equation
(\refp{eq:constraint_V}), combined with the the anisotropic stress
equation (\refp{eq:anisotropic_stress}) and the Poisson equation
(\refp{eq:poisson}) yield, in the early time $a \ll 1$ limit
 \be
 k V = \dfrac{\Hbl}{k}
 \left( \dfrac{3}{4} D - \dfrac{3a}{4} \dfrac{\dr}{\dr a} D
 - \dfrac{\rn}{1+\rn} k^2 \Pin \right) \eqdot
 \ee

The evolution equation (\ref{eq:aniso_stress_evolution}) for the
anisotropic stress then reads, for $a \ll 1$
 \begin{align}
 & a \dfrac{\dr}{\dr a} k^2\Pin + \dfrac{4}{5} \dfrac{\rn}{1+\rn}
 k^2 \Pin  =
 \dfrac{6}{5} D - \dfrac{6a}{5} \dfrac{\dr}{\dr a} D -
 \dfrac{8a}{5(1+\rn)} \dfrac{\dr}{\dr a} S_{\nu \ga} \eqdot\label{eq:aniso_stress_wrt_D_Snuga}\\
\intertext{In the same limit and in terms of the scale factor $a$,
the equations for $D$ and $S_{\nu \ga}$ become (dropping the last
term $\propto k^2\Pin$ on the right hand side of
(\ref{eq:ddot_D_with_nu}) which is always negligible compared to
the others):}
 & a^2 \dfrac{\dr^2}{\dr^2 a}D - 2D  =
  - \left( \dfrac{k}{\Hbl} \right)^2 \dfrac{\rn}{3(1+\rn)} S_{\nu
  \ga}
  - \dfrac{2\rn}{3(1+\rn)} \left[a \dfrac{\dr}{\dr a} k^2 \Pin - 2 k^2 \Pin
  \right] \eqcomma \label{eq:D_wrt_aniso_Snuga}\\
 & a^2 \dfrac{\dr^2}{\dr^2 a}S_{\nu \ga} + \dfrac{1}{3}\left( \dfrac{k}{\Hbl}
  \right)^2 S_{\nu \ga}  = \dfrac{1}{6} \left( \dfrac{k}{\Hbl}
  \right)^2k^2 \Pin \label{eq:Snuga_wrt_D_aniso}\eqdot
\end{align}
The system of coupled differential equations
(\ref{eq:aniso_stress_wrt_D_Snuga}), (\ref{eq:D_wrt_aniso_Snuga})
and (\ref{eq:Snuga_wrt_D_aniso}) is too difficult to solve
analytically. To find an approximate solution valid to leading
order in powers of $a$ for early times, we treat the anisotropic
stress iteratively as a perturbation to the large scale solution,
\rr{eq:large_scale_six_modes}, in analogy with the procedure in
\cite{Hu:1995uz}. More specifically, we use the large scale
solution for $D$ and $S_{\nu \ga}$ as a source on the right hand
side of \rr{eq:aniso_stress_wrt_D_Snuga} to determine the
anisotropic stress, then we re-insert the solution for $\Pin$ on
the right hand side of (\ref{eq:D_wrt_aniso_Snuga}) and
(\ref{eq:Snuga_wrt_D_aniso})
 to find self-consistent corrections
to the large scale behavior.

As an illustration, let us first consider the adiabatic growing
mode, $D = D_0 a^2, D_1 = S_0 = S_1 = N_d = N_v = 0$. In that
case, the right hand side of (\ref{eq:aniso_stress_wrt_D_Snuga})
is dominated by the terms in $D$, giving
 \be
  a \dfrac{\dr}{\dr a} k^2\Pin + \dfrac{4}{5} \dfrac{\rn}{1+\rn}
 k^2 \Pin  =
 - \dfrac{6}{5} D_0 a^2 \eqcomma
 \ee
 which has the particular solution
 \be
 k^2 \Pin = - \dfrac{3(1+\rn) D_0}{7\rn + 5} a^2 \eqdot
 \ee
Notice that, although the above form of $\Pin \propto a^2$ is of
the same order as the adiabatic solution $D \propto a^2$, its
contribution on the right hand side of
(\ref{eq:D_wrt_aniso_Snuga}) cancels out because of the factor 2
in the exponent. Thus it is consistent to have neglected the
anisotropic stress in the first place when deriving the large
scale solution.

With the above approximation for $\Pin$, from
(\ref{eq:Snuga_wrt_D_aniso}) we can determine the growth of
neutrino entropy perturbations in the adiabatic mode, finding to
leading order in powers of $a$
 \be
 S_{\nu \ga} = - \dfrac{(1+\rn) D_0}{48(7\rn + 5)} \left( \dfrac{k}{\Hbl}
  \right)^2 a^2 \propto a^4 \ll D\eqdot
 \ee
The growth of the dark matter entropy perturbation is also
modified by the coupling to the neutrino entropy perturbations on
the left hand side of (\refp{eq:ddot_S_with_nu}), but the term
$\propto \dot{S}_{\nu\ga} \propto a^4$ has the same scaling as the
term $\propto D$ on the right hand side, and the approximate
solution is
 \be
 S_{m\ga} = \dfrac{1}{64} \left[1 - \dfrac{1+\rn}{3(7\rn + 5)}
 \right]
 D_0  \left( \dfrac{k}{\Hbl}
  \right)^2 a^2 \propto a^4 \ll D \eqdot
 \ee

In conclusion, the growing adiabatic mode at early times in the
presence of neutrinos and anisotropic stress has the approximate
solution (compare with the solution (\refp{eq:ad_growing_mode})):
 \be \label{eq:ad_growing_mode_with_nu}
 \left\{
 \begin{array}{l}
 \begin{aligned}
 \DS D &= D_0 a^2 \\
 \DS S_{m \ga}    & \propto  \left( \frac{k}{\Hbl} \right) ^2 a^2  \propto a^4 \\
 \DS S_{\nu \ga}  & \propto  \left( \frac{k}{\Hbl} \right) ^2 a^2  \propto a^4 \\
 \DS k^2 \Pin     & \propto   a^2  \\
 \DS \Phi &= -\frac{3}{2} \left( \frac{k a}{\Hbl} \right) ^2 D_0 = \Phi_0 = \text{ const} \\
 \DS \Psi &= \Phi_0 + \dfrac{3\rn}{7(1+\rn)} \left( \frac{k a}{\Hbl} \right)
 ^2 \equiv \Psi_0 = \text{ const} \\
 \DS k V &= \dfrac{1}{2} \frac{k}{\Hbl} \Phi_0 \propto a \\
 \DS \zeta & = - \frac{9D_0}{4}\left( \frac{\Hbl a}{k} \right)^2 = \text{ const}
  \end{aligned}
 \end{array}
 \right. \quad \text{(adiabatic).}
 \ee
The Bardeen potentials are no longer equal due to the anisotropic
stress, the fractional correction being
 \be
 \Bigg\vert \dfrac{\Phi_0 - \Psi_0}{\Phi_0} \Bigg\vert = \dfrac{2}{7} \dfrac{\rn}{1+\rn}  \approx 0.1
 \eqcomma
  \ee
of order $10\%$, in good agreement with \cite{Hu:1995uz}.

\subsection{Neutrino entropy mode}

Let us now turn our attention to the $N_d \neq 0$ mode, with $N_v
= D_0 = D_1 = S_0 = S_1 = 0$: this is clearly a neutrino entropy
mode, since $S_{\nu \ga} = \text{const}$ for $a \rightarrow 0$.

To determine the growth of perturbations in the total density $D$
beyond the large scale solution $D = 0$, consider the right hand
side of \rr{eq:D_wrt_aniso_Snuga}: if the anisotropic stress goes
at least as $a^2$, then the part containing $\Pin$ cancels (for
$\Pin \propto a^2$) or is subdominant with respect to the $S_{\nu
\ga}$ term (for $\Pin = \calo(a^3)$ or higher). In any case, we
can neglect the anisotropic stress term as a source for $D$ with
respect to the neutrino entropy perturbation, with the caveat that
at the end of our calculation we have to check that this
assumption is satisfied - indeed, \CF \rr{eq:result_for_Pin}. By
this argument, we look for a particular solution of
 \be
 a^2 \dfrac{\dr^2}{\dr^2 a}D - 2D  =
  - \left( \dfrac{k}{\Hbl} \right)^2 \dfrac{\rn}{3(1+\rn)} N_d
  \eqcomma
 \ee
which is given by
 \be
 D = - \dfrac{\rn}{9(1+\rn)} N_d \left( \dfrac{k}{\Hbl} \right)^2
 \ln(a) \propto a^2 \ln(a) \eqdot
 \ee
The logarithmic dependence can be neglected if we do not apply
this solution over a too large time range (say, less than a few
orders of magnitude), and replaced by the value of $\ln(a)$
evaluated at the typical value of the scale factor in the range
considered, $a_*$, which we reabsorb in the overall normalization
by defining a new constant $N_d^* \equiv N_d \ln(a_*)$.

We can now solve for $\Pin$ by inserting the above expression for
$D$ in \rrp{eq:aniso_stress_wrt_D_Snuga}, and observing that on
the right hand side  $\frac{\dr S_{\nu \ga}}{\dr a} = 0$, thus
obtaining
 \be \label{eq:result_for_Pin}
 k^2 \Pin = N_d^* \left( \dfrac{k}{\Hbl} \right)^2 \dfrac{\rn}{3(7\rn +
 5)}\propto a^2 \eqcomma
 \ee
 which is consistent with our initial assumption for $\Pin$.

 Finally, the Bardeen potentials follow from the Poisson equation
 and the anisotropic stress equation, yielding
  \begin{align}
  \Phi & = \dfrac{\rn N_d^*}{6(1+\rn)} = \text{ const} \eqcomma \\
  \Psi & = \Phi \left(1 - \dfrac{2 \rn}{7\rn + 5}\right) = \text{ const}  \eqdot
  \end{align}
The gauge invariant curvature perturbation $\zeta$ is given by
(\refp{eq:curvature_pert_related_to_potentials}) and it can be
rewritten as
 \be \label{eq:zeta_rewritten}
 \zeta = \dfrac{3}{2}\Phi + \dfrac{a}{2}\dfrac{\dr}{\dr a}\Phi -
 \dfrac{\rn}{2(1+\rn)}\left( \dfrac{\Hbl}{k} \right)^2 k^2 \Pin
 \eqdot
 \ee
yielding for the neutrino entropy mode
 \be \label{eq:zeta_for_nid}
 \zeta = \dfrac{\rn N_d^*}{1+\rn}\left( \dfrac{1}{4} - \dfrac{\rn}{6(7\rn + 5)}
 \right)= \text{ const} \eqdot
 \ee

This results agree with the power law solution found by
\cite{Bucher:1999re}, which they called ``neutrino isocurvature
density'' mode; we prefer however to term this mode ``neutrino
entropy'', since the initial curvature perturbation does not
vanish, and indeed is of the same order as the entropy
perturbation.

\subsection{Neutrino velocity mode}

The mode with $N_v \neq 0$ has vanishing entropy at early times,
since $S_{\nu \ga} \rightarrow 0$ for $a \rightarrow 0$, but the
bulk velocity difference between neutrinos and photons in
non-zero,
 \be
 k(V_\comp{\nu} - V_\comp{\ga}) = - \dfrac{\dot{S}_{\nu \ga}}{k} =
 \text{const}
 \ee
hence its name.

From the power-law solution for this mode
\cite[see][]{Bucher:1999re,Trotta:2001tesi} we expect that the
anisotropic stress goes to leading order as $\Pin \propto a$.
Indeed, by replacing the large-scale solution $D=0, S_{\nu\ga} =
N_v a$ on the right hand side of
(\ref{eq:aniso_stress_wrt_D_Snuga}) we find the particular
solution
 \be
 k^2 \Pin = - \dfrac{8N_v}{9\rn + 5} a \eqdot
 \ee
 We now use this expression as a source on the right hand side of
(\ref{eq:D_wrt_aniso_Snuga}) to determine the corrections to $D$,
and we can ignore the contribution of the term $\propto S_{\nu
\ga}$ which goes as $a^3$ compared to the part containing $\Pin$,
which is dominant, being proportional to $a$. We thus have to
solve
 \be
  a^2 \frac{\dr^2}{\dr a^2}D - 2 D = - \dfrac{16 \rn
  N_v}{3(1+\rn)(9\rn +5)}a \eqcomma
 \ee
and we find the particular solution
 \be
 D = \dfrac{8 \rn N_v}{3(1+\rn)(9\rn +5)} a \eqdot
 \ee

As already noticed in \cite{Bucher:1999re}, the Bardeen potentials
are decaying
 \begin{align}
 \Phi & = - \dfrac{4 \rn N_v}{(1+\rn)(9\rn +5)} \left( \dfrac{\Hbl}{k} \right)^2
 a\propto a^{-1} \eqcomma \\
 \Psi &= - \Phi \eqcomma
 \end{align}
but this does not necessarily mean that perturbation theory breaks
down for $a \rightarrow 0$. In general, a solution is considered
non divergent if it is possible to find a gauge in which {\it all}
the perturbation variables do no diverge in the limit $a
\rightarrow 0$. The synchronous gauge potentials for the neutrino
velocity mode are indeed non-singular at early times
\citep{Bucher:1999re}. In fact, even though the Bardeen potential
diverge, the gauge invariant curvature perturbation $\zeta$
vanishes to leading order. This is most easily seen by making use
of \rrp{eq:useful_relation_zeta}, finding
 \be
 \zeta = \dfrac{1}{2}\left(\Psi + \Phi \right) = 0 \eqcomma
 \ee
and thus the velocity mode is indeed an isocurvature mode.

The leading order corrections to $S_{m \ga} = 0$ induced by the
neutrino modes can be obtained as particular solutions to
\rrp{eq:ddot_S_with_nu}, which for early times reads
 \be
 a^2 \dfrac{\dr}{\dr a^2}S_{m\ga} +
 a \left[ \dfrac{\dr}{\dr a}S_{m\ga} - \dfrac{\dr}{\dr
 a}S_{\nu\ga}\right] = -
 \left(\dfrac{k}{\Hbl}\right)^2\dfrac{\rn}{3(1+\rn)} S_{\nu \ga}
 \eqdot
 \ee
 Summarizing, the early time solutions
for neutrino entropy ($N_d \neq 0$) and neutrino isocurvature
velocity ($N_v \neq 0$) initial conditions are:
 \begin{alignat}{2}
 \notag & \text{\bf Neutrino entropy} & & \text{\bf Neutrino velocity}\\
 \DS
 &   S_{\nu \ga}   = N_d
 &&  S_{\nu \ga}   = N_v a \notag  \\
 \DS
 &  D  =  - \left( \frac{k}{\Hbl} \right)^2\dfrac{\rn N_d^*}{9(1+\rn)}   \propto a^2
 && D  =  \dfrac{8 \rn N_v}{3(1+\rn)(9\rn +5)} a \notag \\
 \DS
 &  S_{m \ga}  = - \left(\dfrac{k}{\Hbl}\right)^2 \dfrac{\rn N_d}{12(1+\rn)} \propto
 a^2
 && S_{m \ga} = a N_v \notag\\
 \DS
 &  k V  = \dfrac{1}{2}\dfrac{k}{\Hbl}\Psi \propto a
 && k V  = \frac{k}{\Hbl} \Psi = \text{ const} \\
 \DS
 &  k^2 \Pin = \left( \dfrac{k}{\Hbl} \right)^2 \dfrac{\rn N_d^* }{3(7\rn +
 5)}\propto a^2
 && k^2 \Pin = - \dfrac{8N_v}{9\rn + 5} a \notag \\
 \DS
 &  \Phi = \dfrac{\rn N_d^*}{6(1+\rn)} = \text{ const}
 && \Phi = - \dfrac{4 \rn N_v}{(1+\rn)(9\rn +5)} \left( \dfrac{\Hbl}{k} \right)^2
 a\propto a^{-1} \notag\\
 \DS
 &  \Psi = \Phi \left(1 - \dfrac{2 \rn}{7\rn + 5}\right) = \text{ const}
 && \Psi = - \Phi \notag\\
 \DS
 &  \zeta = \dfrac{\rn N_d^*}{1+\rn}\left( \dfrac{1}{4} - \dfrac{\rn}{6(7\rn + 5)}
 \right) = \text{ const} \quad
 && \zeta = 0 \eqdot \notag
 \end{alignat}

\subsection{The divergent nature of the anisotropic stress mode}
\label{chap:cmb;sec:anisoic}

One could ask whether it would be possible to excite a growing
``neutrino anisotropic stress mode'', characterized by initial
conditions $D= S_{\nu \ga} = S_{m\ga} = V_{\nu \ga} = V_{m \ga} =
0$ and $\Pin \neq 0$ for $a \rightarrow 0$. Even though highly
exotic, such a mode, if it existed, should be included if we want
to consider the most general type of perturbations. We now show
that this mode is divergent in all gauges, and therefore is
non-physical, since it would lead to the breakdown of perturbation
theory for $a \rightarrow 0$. Alternatively, we can see it as a
decaying mode, which therefore does not need to be considered
since it quickly disappears.

Consider the anisotropic stress equation
(\refp{eq:anisotropic_stress}) with $\Pin = \Piz = \text{const}$
on the right hand side,
 \be
 \Psi = \Phi - \dfrac{\rn}{(1+\rn)(1+a)} \Hbl^2 \Piz \eqdot
 \ee
Since $\Hbl = \eta^{-1}$ to leading order for $a \ll a_\eq$, it
follows that $\Psi \propto \eta^{-2}$. The fact that the Bardeen
potential diverges at early times is not by itself sufficient to
discard the corresponding mode, as we have seen in the example of
the neutrino velocity mode. A necessary condition, however, is the
existence of a gauge in which all of the perturbation variables
constructed out of $A, B, C, E, \delta, v, \pi_L$ are
non-divergent. For the neutrino velocity mode, this gauge is the
synchronous gauge. Clearly, since $\Psi$ is a gauge invariant
variable, by construction it does not change under a gauge
transformation but the variables $A, B, C, E$ do, according to the
transformation laws (\refp{eq:gauge_trafo_metric}). If we expand
in a Laurent series around $\eta=0$ the definition of $\Psi$,
\rrp{eq:def_gi_Psi}, and we allow terms $\eta^n$ with exponent $n
\geq -2$, because of $\Hbl = 1/\eta$ we obtain to leading order
 \be
 A = \Psi \propto \eta^{-2} \eqdot
 \ee
In other words, in the radiation dominated universe a metric
perturbation of the form $A \propto \eta^{-2}$ is gauge invariant.
This can also be seen directly from the transformation law for
$A$, \rrp{eq:trafo_for_A}: the part $ \Hbl T + \dot{T} $ does not
contain terms $\propto \eta^{-2}$ if $T$ is written as a Laurent
series in $\eta$. We conclude that $\Piz \neq 0$ induces a
divergence of $A$ for early times, which does not disappear in any
gauge. One could conceive to combine $A$ with other diverging
variables to construct via cancellation a non-diverging metric
variable: this however would unavoidably produce divergent terms
in the matter variables. Therefore a neutrino anisotropic stress
mode is always decaying in all gauges.

In principle, there is a whole hierarchy of modes coming from
setting $\Theta_\nu^\ell \neq 0$ for $\ell \geq 3$ as initial
conditions in the neutrino Boltzmann hierarchy. As we noticed in
\SEC{chap:perts;sec:boltzmann}, higher order moments are coupled
to the potentials and to the velocity and density perturbations by
successive powers of $k \eta$. By reversing the argument, we see
that $\Theta_{\ell-1}^\nu = \mathcal{O}\left(
\Theta_\ell^\nu/k\eta\right)$ implies that in the early Universe
and on super-horizon scales, $k \eta \ll 1$, choosing $\Theta_\ell
= \mathcal{O}(1)$ for $\ell \geq 3$ would produce divergent
behavior in the lower-order multipoles of the hierarchy. Since for
$\ell \geq 2$ the multipole moments are gauge invariant, it
follows that there is no gauge in which such a mode is growing. In
summary, the adiabatic and the general isocurvature modes
presented above constitute the most general type of perturbation.

\section{The role of baryons}
\label{chap:cmb;sec:baryons}

In this section, we go back to the model of a Universe containing
dark matter and photons, and refine the treatment given in
\SEC{chap:cmb;sec:matter_radiation}  by taking into account the
role of baryons in the dynamic of the oscillations. For
simplicity, we neglect the corrections induced by the neutrinos
anisotropic stress, omitting neutrinos entirely.

Before recombination photons interact with electrons via Thomson
scattering (see section \ref{chap:pert;sec:collision_term}). The
time-scale for the scattering process is set by the Compton
scattering time $\taudot^{-1}$, which represents the typical time
between two collisions. Tight coupling is an expansion in powers
of $\taudot^{-1}$, assuming that the scattering rate is rapid
enough to equilibrate changes in the photon-baryons fluid, and in
this limit moments $\ell \geq 2$ in the photon distribution
function are suppressed by successive powers of $\taudot^{-1}$.
Therefore to lowest order the photon distribution function is
described by its zeroth and first multipoles only, and we can set
$\Pi_\comp{\ga} = \Theta_{\ell \geq 3} =0$, which justifies the
approximation taken in the previous section. Therefore the
truncated Boltzmann hierarchy
(\refp{eq:collisional_Boltzmann_hierarchy_trunc}) gives for
photons
  \begin{align}
  \dot{D}_{g, \comp{\ga}} + \frac{4}{3}k^2 V &  =  0 \eqcomma
  \label{eq:photons_1} \\
  \dot{V}_\comp{\ga} - \frac{1}{4}D_{g, \comp{\ga}} -
  2  \Phi  & =
  - a \sigma_T n_e (V_\comp{\ga} - V_\comp{b})
  \label{eq:photons_2} \eqdot
  \end{align}

To ensure conservation of the total momentum, we need to
supplement the conservation equation for baryons with the {\it
Thomson drag force} term coming from the scattering process,
obtained as the first moment of the collision term
 \be \label{eq:Thomson_drag_force}
 F_j^{\textrm{drag}} = a \sigma_T n_e \rho_\ga
 \int \frac{\dr \Om}{4 \pi} n_j C \left[ f \right] \eqdot
 \ee
The momentum conservation for baryons,
\rrp{eq:pertd_conservation_eq_1_Dg}, therefore gives
 \begin{align}
 \dot{D}_{g, \comp{b}} + k^2 V_b & =  0
 \label{eq:baryons_1}\\
 \dot{V}_b + \Hbl V_b - \Phi & =
 - \frac{1}{R} a \sigma_T n_e (V_b - V_\ga) \label{eq:baryons_2}
 \eqcomma
 \end{align}
and we have defined $R \equiv 3 \ol{\rho}_b / (4 \ol{\rho}_\ga) $,
which can easily be estimated
 \be \label{eq:R_estimate}
 R \approx \left( \frac{670}{1+z} \right)
 \left( \frac{\Om_\comp{b} h^2}{0.022} \right) \eqdot
 \ee

The set of Eqs.~(\ref{eq:photons_1}--\ref{eq:photons_2}) and
(\ref{eq:baryons_1}--\ref{eq:baryons_2}) describes the evolution
of perturbations for the tight-coupled photon-baryon fluid, while
the dark matter component enters via its influence on the
gravitational potential $\Phi$. To lowest order in $1/\taudot$,
collisions force the baryons and photons velocities to coincide,
 $V_\comp{\ga} = V_\comp{b}$,
which via \rrp{eq:dot_S_and_delta_V} implies
 $\dot{S}_{b\ga} = 0$, hence the entropy per baryon
is conserved.

Equations (\ref{eq:photons_1}, \ref{eq:photons_2} and
\ref{eq:baryons_2}) can now be combined into the equation of a
damped, forced harmonic oscillator:
 \be \label{eq:HO_with_baryons}
 \frac{\dr}{\dr \eta} \left[ (1+R) \dot{D}_{g, \comp{\ga}}\right]
 + \frac{k^2}{3} D_{g, \comp{\ga}} =
 - \frac{4}{3}(2+R) k^2 \Phi \eqdot
 \ee

By comparing with \rrp{eq:HO_second_order}, we see that baryons
have two effects: they change the effective mass of the system
(factor $(1+R)$ on the left hand side) and they displace the zero
point of the oscillation by adding to the potential $\Phi$. Both
modifications are a consequence of the fact that baryon add to the
mass of the system but not to the restoring pressure, which is
still given by the photons alone.

The time dependence of $R$ is of the order of the Hubble time,
hence large compared to the time scale of one oscillation. For
illustrative purpose, we can then neglect the time dependence of
$R$ and obtain from \rr{eq:HO_with_baryons}
 \be
 \label{eq:HO_with_baryons_R_const}
 \ddot{D}_{g, \comp{\ga}}
 + c_s^2 k^2 D_{g, \comp{\ga}} =
 - 4 (2+R) c_s^2 k^2 \Phi \eqcomma
 \ee
where the sound speed of the coupled fluid is $c_s^2 =
1/(3(1+R))$. At early times, $c_s^2 \rightarrow 1/3$, as
appropriate for radiation, while at late times $c_s^2 \approx 0$,
when the universe is dominated by matter. The homogeneous solution
is still a superposition of sine and cosine oscillations, but
adding the baryons slows down the period by decreasing $c_s^2$
with respect to the pure photons fluid. This is responsible for a
shift in the acoustic peak positions and for a larger distance
between the peaks in the CMB power spectrum, see the explanations
regarding the role of the shift parameter on page
\pageref{eq:def_kB}.

The adiabatic solution (\ref{eq:HO_adiabatic}) becomes
 \begin{align} \label{eq:HO_with_baryons_solution}
 D_{g,\comp{\ga}} & =  \frac{4}{3}(1+R) \Phi \cos (c_s k \eta)
 - 4 (2 + R) \Phi \eqcomma \\
 k V_\comp{\ga} & =  \left( \frac{1+R}{3} \right)^{1/2}\Phi \sin (c_s k \eta)
  \eqdot  \label{eq:HO_with_baryons_solution_V}
 \end{align}
The amplitude of the cosine oscillation has increased by a factor
$(1+R)$, and the potential well has deepened by an extra factor
$(1+R/2)$. This displacement of the zero point of the oscillations
induces a boost (decrease) of the odd (even) peaks in the power
spectrum sometimes denotes as ``baryon driving'', which is
discussed in \SEC{chap:params;sec:barsig} and shown in
\FIGPAG{fig:kB}. Finally, the amplitude of the velocity
oscillation becomes smaller, since it is suppressed by a factor
$c_s$ with respect to the density and $c_s$ is smaller in the
presence of baryons. This leads to a suppression of the Doppler
contribution to the acoustic peak structure. From
\rr{eq:R_estimate} we obtain that at the moment of decoupling,
$z_\dec \approx 1100$, we have $R \approx 0.6$.

The solution to (\ref{eq:HO_with_baryons}) for time-dependent R
can be found in the WKB approximation \citep{Hu:1995uz}, in which
case the qualitative picture sketched above slightly changes: the
sound speed becomes $k \int c_s \dr \eta$, while the amplitude of
the oscillations grows in time as $c_s^{1/2}$. This can be seen
simply by considering the quantity $m\omega A^2$, which for an
harmonic oscillator is an adiabatic invariant: since in our case
the effective mass $m = (1+R)^{1/2}$ decreases in time, it follows
that the amplitude $A \sim (1+R)^{-1/4} \sim c_s^{1/2}$.

\section{Damping}
\label{chap:cmb;sec:damping}

In the above discussion, we have neglected the fact that
recombination takes a finite time to complete, and the acoustic
oscillations are not frozen instantly. This ``finite thickness''
of the last scattering surface has a twofold effect: photon
diffusion and cancellation. Diffusion damping arises because of
the imperfect coupling between photons and baryons, so that
photons diffuse out of over-dense into under-dense regions and
erase fine scale anisotropies; cancellation occurs for scales
which have the time to oscillate through recombination, so that
the effect of photons that last scattered on a crest of the
oscillation is cancelled by the contribution of the photons coming
from a trough. Cancellation produces a power law damping of the
fluctuations \citep{Hu:1995uz}, while diffusion damping is
exponential and is by far the dominant effect, and the one to
which we now turn our attention. It is often referred to as ``Silk
damping'' \citep{Silk:1968kq}.

In view of obtaining a dispersion relation $\om (k)$ for photons
accurate to first order in $\taudot^{-1}$, we look for solutions
of the form $V_\comp{\ga} \propto \exp \imath \int \om \dr \eta$.
At this order we need to include the photon anisotropic stress,
which to first order in $\taudot^{-1}$ from
\rrp{eq:ell_2_hierarchy} is given by (neglecting polarization
effects)
 \be
 \Pi_\comp{\ga} = \taudot^{-1} \frac{16}{9} V_\comp{\ga} \eqdot
 \ee

Using the anisotropic stress equation
(\refp{eq:anisotropic_stress}) we can substitute in the dipole
equation (\refp{eq:ell_1_hierarchy}) $\Phi = \Psi + \Hbl^2
\Pi_\comp{\ga}$. However, we assume that the oscillation time
scale is much shorter that the expansion time scale, \ie
$\om^{-1}\ll \Hbl^{-1}$, so that we can neglect the term $ \Hbl^2
\Pi_\comp{\ga}$ in the photon dipole. By the same token, in the
following we also neglect all time dependencies of the potentials
and of $R$ compared with the oscillation time scale.

We now expand the baryon momentum conservation equation
(\ref{eq:baryons_2}) up to {\it second order} in $\taudot^{-1}$,
and find, under the above assumptions
 \be
 V_\comp{b} = V_\comp{\ga} - \taudot^{-1} R(\imath \om V_\ga - \Phi) -
 \taudot^{-2} (R \om)^2 V_\ga + \calo (\taudot^{-3})\eqdot
 \ee

Inserting this into \rrp{eq:ell_1_hierarchy} we obtain
 \be
 \imath \om (1+R) V_\comp{\ga} =
 \frac{1}{4}D_{g,\comp{\ga}} + (2+R)\Phi
 - \taudot^{-1} V_\comp{\ga}
 \left[ (R\om)^2  - \frac{8}{27}k^2 \right] \eqdot
 \ee
To lowest order in $\taudot^{-1}$ we have found in
\SEC{chap:cmb;sec:baryons} that the quantity
$\frac{1}{4}D_{g,\comp{\ga}} + (2+R)\Phi$ oscillates with the same
frequency as $V_\ga$, see \rr{eq:HO_with_baryons_R_const}.
Therefore we set $\frac{1}{4}D_{g,\comp{\ga}} + (2+R)\Phi \propto
\exp \imath \int \om \dr \eta$, and using the photon monopole
equation (\ref{eq:photons_1}) we arrive at
 \be
 \om^2 = \dfrac{k^2}{3(1+R)}
 + \imath \taudot^{-1} \frac{\om}{1+R}
 \left[R^2 \om^2 + \frac{8}{27}k^2
 \right] \eqdot
 \ee

To zeroth order we find as before $\om^2 = k^2/[3(1+R)]$, which we
can use to obtain the first order solution
 \be
 \om = \dfrac{k}{\sqrt{3(1+R)}} + \imath \taudot^{-1} \frac{k^2}{6(1+R)}
 \left[ \dfrac{R^2}{(1+R)} + \dfrac{8}{9} \right] \eqdot
 \ee

The imaginary term in the frequency induces an exponential damping
of the oscillatory solutions of the form $\exp
(-k^2/k^2_{\text{D}})$, with the characteristic damping scale
given by
 \be \label{eq:damping_scale}
 k^{-1}_\text{D} = \int
 \frac{1}{6 \taudot}
 \left[ \dfrac{R^2}{(1+R)^2} + \dfrac{8}{9(1+R)} \right] \dr \eta \eqdot
 \ee

Including polarization effects via
Eqs.~(\refp{eq:Qpol_Boltzmann_hierarchy}) and
(\refp{eq:Qell2_temp_including}) would increase the damping, by
changing the numerical factor $8/9$ in the above equation to
$16/15$.

\section{Observable quantities}
 \label{chap:cmb;sec:observables}

\subsection{Temperature fluctuations}
\label{chap:cmb;sec:line_of_sight}

 We now calculate the
fluctuations in the CMB photon temperature on the sky. When the
photon mean free path becomes larger than the horizon scale,
$1/\taudot \gg 1/\Hbl$, the Universe becomes transparent and
photons propagate along null geodesics ({\it free streaming
regime}).

In this section we calculate the photon temperature today with the
line of sight method: we formally integrate the Boltzmann equation
along the photon path, and obtain the temperature measured today
as an integral over a time dependent source term. This approach
includes in principle all the effects due to imperfect
photons-electrons coupling and reionization as well, and it is the
core of the fast numerical algorithms for the integration of the
photon Boltzmann equation, such as CMBfast \citep{Seljak:1996is}.
Another derivation of the same result based on a more physical
understanding of the free streaming regime can be found in
\cite{Durrer:1990mk}.

Consider the collisional Boltzmann equation for the photons
temperature $\Theta(\eta, k, \mu=\bf{\hat{k}}\cdot\bf{n})$ (were
we neglect polarization)
 \be \label{eq:collisional_Boltzmann}
 \dot{\Theta} + \imath k \mu \Theta
 + \imath k \mu (\Psi + \Phi) = -
 \dot{\tau} \left[\Theta + \imath \mu k V_\comp{b} -
 \Theta_0 - \dfrac{1}{2} P_2 \Theta_2 \right] \eqcomma
 \ee
and denote with
 \be
  \tau(\eta) \equiv \int_{\eta}^{\eta_0} \taudot
  \dr \tilde{\eta}
 \ee
the total opacity from the time $\eta$ until today. Using the
equality
 \be
 \dfrac{\dr}{\dr \eta}
 \left(\Theta e^{\imath k \mu \eta} e^{-\tau} \right)
 = e^{\imath k \mu \eta} e^{-\tau}
 \left [\dot{\Theta}
 + \imath k \mu \Theta
 + \taudot \Theta \right]
 \ee
we obtain from (\ref{eq:collisional_Boltzmann})
 \be
 \Theta = - \int_0^{\eta_0} e^{\imath k \mu(\eta - \eta_0)}
 e^{-\tau}
 \left[ \taudot \left( \imath \mu k V_\comp{b} -
 \Theta_0 - \dfrac{1}{2} P_2 \Theta_2 \right)
 + \imath k \mu (\Psi + \Phi) \right] \eqdot
 \ee

The second term on the right hand side can be integrated by parts
and we drop the boundary term, which contributes only to the
monopole and is thus unobservable, obtaining
 \be \label{eq:integral_solution}
 \begin{split}
 \Theta(\eta_0, k, \mu) = \int_0^{\eta_0} \dr \eta e^{\imath k \mu(\eta - \eta_0)}
 g(\eta)
 & \left[- \imath \mu k V_\comp{b} +
 \Theta_0 + \tfrac{1}{2} P_2 \Theta_2 + \Psi + \Phi
 \right]\\
 & + \int_0^{\eta_0} \dr \eta  e^{\imath k \mu(\eta - \eta_0)} e^{-\tau}
 (\dot{\Psi} + \dot{\Phi}) \eqcomma
 \end{split}
 \ee
and we have defined the {\it visibility function}
 \be
 g(\eta) \equiv
 \taudot e^{-\tau} \eqdot \label{eq:def_visibility_function}
 \ee

Equation (\ref{eq:integral_solution}) is an integral system of
equations, since moments $\ell < 3$ of the photons temperature
appear on both sides. However, the left hand side can be
determined given the time evolution of an handful of quantities
which act as a source on the right hand side: the photons moments
$\ell < 3$ are calculated from the Boltzmann hierarchy
(\refp{eq:collisional_Boltzmann_hierarchy}), the baryon and CDM
velocity and density perturbation from the fluid conservation
equations
(\ref{eq:pertd_conservation_eq_1_D}--\refp{eq:pertd_conservation_eq_2_D}),
while the Bardeen potentials follow from the Poisson equation
(\ref{eq:poisson}) and either the constraint equation
(\ref{eq:constraint_V}) or the anisotropic stress equation
(\refp{eq:anisotropic_stress}). Neutrinos can be included via a
collisionless Boltzmann hierarchy,
\rrp{eq:pertd_Boltzmann_hierarchy}. The great advantage is that
only the first few moments of the collisional Boltzmann hierarchy
for photons need to be computed accurately in order to obtain the
sources of (\ref{eq:integral_solution}), reducing the number of
coupled differential equations which needs solving from several
thousands to a few dozens. This line of sight integration approach
is the core algorithm of all modern codes for the computation of
the CMB power spectrum \citep{Seljak:1996is}.

The visibility function $g(\eta)\dr \eta$ in
(\ref{eq:integral_solution}) encodes the information regarding the
ionization history of the Universe, and can be interpreted as the
probability that a given CMB photon was last scattered between
$\eta$ and $\eta+\dr \eta$. The sharp drop of the free electron
density $n_e$ at decoupling makes the visibility function sharply
peaked around $\eta_\dec$, \CF the solid line in
\FIG{fig:visibility_function}. When the Universe is reionized at
later time, the visibility function becomes non-zero again, and
the free streaming regime goes once again over in a collisional
regime (\SEC{chap:params:sec:reion}).

In the limit of instantaneous recombination, the LSS becomes
infinitely thin and the visibility function a delta function
peaked at $\eta_\dec$, while we can approximate $e^{-\tau}$ with
the Heaviside step function $u(\eta-\eta_\dec)$. In this limit,
the tight coupled fluid approximation for photons goes over
directly to the free streaming regime, and there is no generation
of photons anisotropic stress nor polarization. Performing the
time integral of (\ref{eq:integral_solution}) and setting to
zeroth order $V_\comp{b} = V_\comp{\ga}$ we find
 \be \label{eq:temperature_fluctuation}
 \Theta(\eta_0, k, \mu)  =
 e^{\imath k \mu (\eta_\dec - \eta_0)}
 \left[ \Theta^\OSW + \Theta^\Doppler +
 \Theta^\ISW \right] \eqcomma
 \ee
 where
 \begin{align}
 \begin{split}
 \label{eq:OSW_term}
 \Theta^\OSW & \equiv
 \left[ \Theta_0 + \Psi + \Phi \right](\eta_\dec, k)
  = \left[ \frac{1}{4} D_{g, \comp{\ga}} + \Phi + \Psi \right](\eta_\dec, k)\\
 & = \left[ \frac{1}{4} D_{s, \comp{\ga}} + \Psi  \right](\eta_\dec, k)
 \end{split} \\
 \Theta^\Doppler & \equiv
 - \imath \mu k V_\comp{\ga}(\eta_\dec, k)  \label{eq:Doppler_term} \\
  \Theta^\ISW & \equiv
  \int_{\eta_\dec}^{\eta_0}  \dr \eta  e^{\imath k \mu (\eta - \eta_0)}
 (\dot{\Psi} + \dot{\Phi})(\eta,
 k) \label{eq:ISW_term}
 \end{align}

The temperature fluctuation consists of three terms:
 \begin{itemize}
 \item {\bf The ordinary Sachs-Wolfe (OSW)} part, $\Theta^\OSW$.
The photons temperature monopole $\Theta_0$ on the last scattering
surface, together with the potential terms $\Phi$ and $\Psi$,
reflect intrinsic inhomogeneities in the radiation fluid and in
the metric at the moment of decoupling. On large scales, the
ordinary SW effect is responsible for the SW plateau in the
temperature power spectrum, while on intermediate scales the
oscillations of $D_{g,\comp{\ga}}$ produce the familiar peak
structure.
 \item {\bf The Doppler term} $\Theta^\Doppler \propto k V_\comp{b}$ arises because of
 the relative velocity of observer and emitter. Its contribution shows up
 on the acoustic peak scale.
 \item {\bf The integrated Sachs-Wolfe (ISW)}
 effect produces the term $\Theta^\ISW$, and it is induced by a time dependence of the Bardeen
potentials along the path of the photons. The {\it early} ISW
effect is due to the fact the the universe is not completely
matter dominated at recombination and therefore the potentials are
not exactly constant; the {\it late} ISW is generated when the
late universe becomes dominated by the curvature or a cosmological
constant term, both of which induce a time dependence in the
potentials.
 \end{itemize}

The dependence of the anisotropies on the cosmological parameters
is presented in \SEC{chap:params;sec:cosmo}.

\subsection{Angular power spectra}

The relevant quantities for the comparison of theoretical models
and observations are the temperature and polarization angular
power spectra, which we introduce in this section. We refer the
reader to \SEC{chap:data;sec:stattools} for precise definitions of
the terminology. We denote by $\EX{\cdot}$ the theoretical
ensemble average over realizations.

\subsubsection*{Temperature power spectrum}

The temperature fluctuation in direction $\bf{n}$ on the sky
measured by an observer today ($\eta_0$) and here ($\bf{x}_0$) is
a superposition of plane wave contributions (in a flat Universe)
 \be
 \Theta(\eta_0, \bxz, \bn) = \frac{1}{(2\pi)^{3/2}}
 \int \dr^3k \; \Theta(\eta_0, \bk, \bn) e^{\imath \bxz \bk}
 \ee
and each Fourier mode can be expanded in spherical harmonics on
the 2-sphere as
 \be \label{eq:harmonic_expansion}
 \Theta(\eta_0, \bk, \bn) = \sum_{\ell=0}^{\infty} \sum_{m=-\ell}^{\ell}
 a_{\ell m}(\bk, \eta_0)Y_{\ell m}(\bn) \eqcomma
 \ee
where the expansion coefficients $a_{\ell m}(\bk)$ are given by
 \begin{align}
 a_{\ell m}(\bk) & = \int \dr \Omega_{\bn} \Theta(\bk, \bn) Y_{\ell m}(\bn) \\
 &= 4\pi  \Theta_\ell(\bs{k}) Y_{\ell m}(\bkh) \label{eq:expansionalms}\eqdot
 \end{align}
In deriving the last expression we have expanded the temperature
fluctuation in Legendre polynomials as in
(\refp{eq:Theta_expansion_in_P}) and used the addition theorem and
orthogonality relation for spherical harmonics:
 \begin{align} \label{eq:useful_properties_of_Ylm}
 \sum_{m=-\ell}^{\ell} Y_{\ell m}(\bn) Y_{\ell m}^*(\bn') &=
 \dfrac{2\ell+1}{4\pi} P_\ell(\bn \cdot \bn') \eqcomma \\
 \int \dr \Om_\bn Y_{\ell m}(\bn) Y_{\ell' m'}^*(\bn) & = \delta_{\ell
 \ell'} \delta_{m m'} \eqdot
 \end{align}
We can perform the harmonic expansion
(\ref{eq:harmonic_expansion}) directly in real space rather than
in Fourier space, with coefficients $\alm(\bxz)$ (for which we
will neglect the argument $\bxz$ from now on), obviously related
to $\alm(\bk)$ by
 \be \label{eq:real_space_alm}
 a_{\ell m} = \dfrac{1}{(2\pi)^{3/2}} \int \dr^3 \; \bk a_{\ell m}(\bk)e^{\imath \bk \bxz} \eqdot
 \ee

We are interested in the 2-point temperature correlation function
$C$ on the sky between two directions $\bn$ and $\bn'$. By
choosing our coordinate system in such a way that the direction
$\bn$ corresponds to the $z$-axis, and introducing spherical
coordinates we can write $\bn' = (\phi, \vartheta)$ and $\bn \cdot
\bn' = \cos(\vartheta)$. If we assume statistical homogeneity and
isotropy for the random field $\Theta$, see
\SEC{chap:data;sec:stattools}, then the correlation function does
not depend on the observer's position (homogeneity) nor on the
azimutal angle $\phi$ (isotropy). Therefore
 \be \label{eq:def_angular_corr_fct}
 \begin{split}
 C(\vartheta) & \equiv \EX{ \Theta(\eta_0, \bxz, \bn) \cdot \Theta(\eta_0, \bxz, \bn') } \\
              & = \dfrac{1}{4\pi} \sum_\ell (2\ell + 1) C_\ell P_\ell(\bn \cdot
 \bn')\eqcomma
\end{split}
 \ee
where we have defined the {\it CMB angular power spectrum} by
 \be \label{eq:def_C_ell}
 \EX{a_{\ell m} \cdot a_{\ell' m'}^* } = \delta_{\ell \ell'}
 \delta_{m m'} C_\ell \eqdot
 \ee
The fact that $C_\ell$ does not depend on $\bxz$ is a consequence
of the assumption of homogeneity, while isotropy requires that it
does not depend on the index $m$, which would introduce an
azimutal dependence. It is also customary to assume that the
$\alm$'s are Gaussian random fields, as motivated by inflation,
but this is not strictly necessary at this stage.
Eq.~(\ref{eq:def_angular_corr_fct}) shows that the angular power
spectrum is the harmonic transform of the correlation function on
the 2-sphere and for Gaussian variables it contains the full
statistical information. If the $\alm$'s are Gaussian distributed,
then the Fourier coefficients $\alm(\bk)$ are Gaussian random
variables as well. From the assumption of homogeneity it follows
that $\EX{\alm(\bk)} = \dirac(\bk)$, where $\dirac$ denotes the
Dirac delta function. Homogeneity and isotropy together imply that
 \be \label{eq:almsquared}
 \EX{\vert a_{\ell m} \vert^2} = \dfrac{1}{(2\pi)^3} \int \dr^3\bk \EX{\vert a_{\ell m}(\bk)
 \vert^2} \eqdot
 \ee

We now relate the angular power spectrum to the temperature
multipoles: this is done by observing that the evolution equations
(\refp{eq:collisional_Boltzmann_hierarchy}) for $\Theta_\ell$ are
independent of $\bkh$, and therefore we can write
 \be \label{eq:thetaellandchi}
 \Theta_\ell(\eta, \bk) = \Theta_\ell(\eta, k) \chi(\bk) \eqcomma
 \ee
where we assume that $\chi(\bk)$ are the Fourier components of a
Gaussian, isotropic and homogeneous random field. As a consequence
 \be
 \EX{\chi(\bk) \cdot \chi^*(\bk') } =
  \dirac(\bk - \bk')\; \EX{\vert \chi(k) \vert^2} \eqdot
 \ee
Now from (\ref{eq:def_C_ell}) and using
Eqs.~\eqref{eq:thetaellandchi}, \eqref{eq:almsquared} and
\eqref{eq:expansionalms} we obtain
 \be \label{eq:C_ell_and_source}
 C_\ell = 4\pi \int \dfrac{\dr k}{k}\: P_\chi(k)
 \: \vert \Theta_\ell (\eta_0, k) \vert ^2 \eqdot
 \ee
We shall later identify $\chi$ with the primordial curvature or
entropy perturbation, see \rrp{eq:source_for_mixed_AD_ISO}, and
call
 \be
 P_\chi(k) \equiv \frac{k^3}{2\pi^2}\EX{\vert \chi
\vert^2}
 \ee
the curvature (or entropy) power spectrum: this quantity gives the
contribution to $\Cl$ per logarithmic $k$-interval of the
primordial fluctuation.

\begin{figure}[tb]
\centering
\includegraphics[width=\twofigswidth]{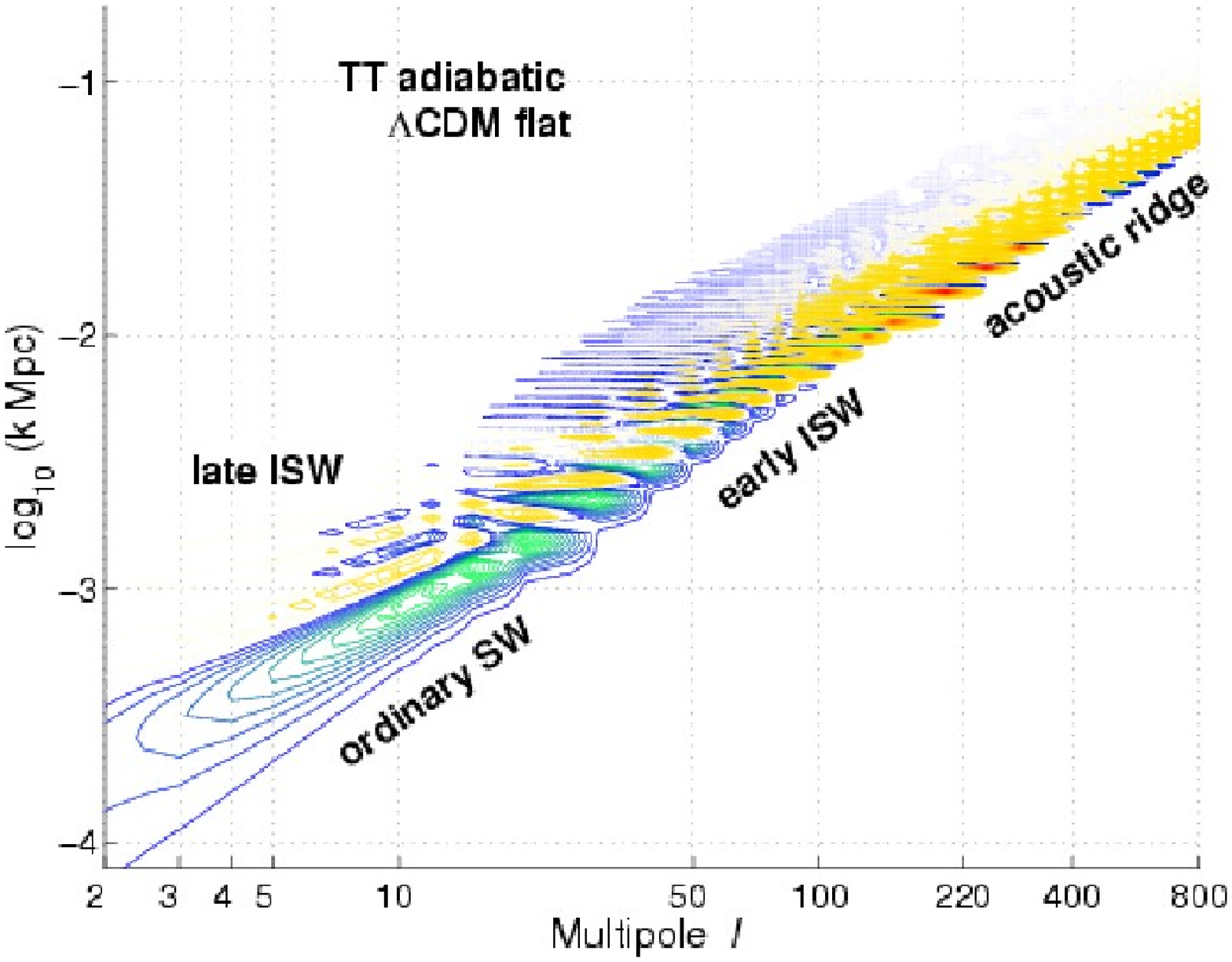}\hfill%
\includegraphics[width=\twofigswidth]{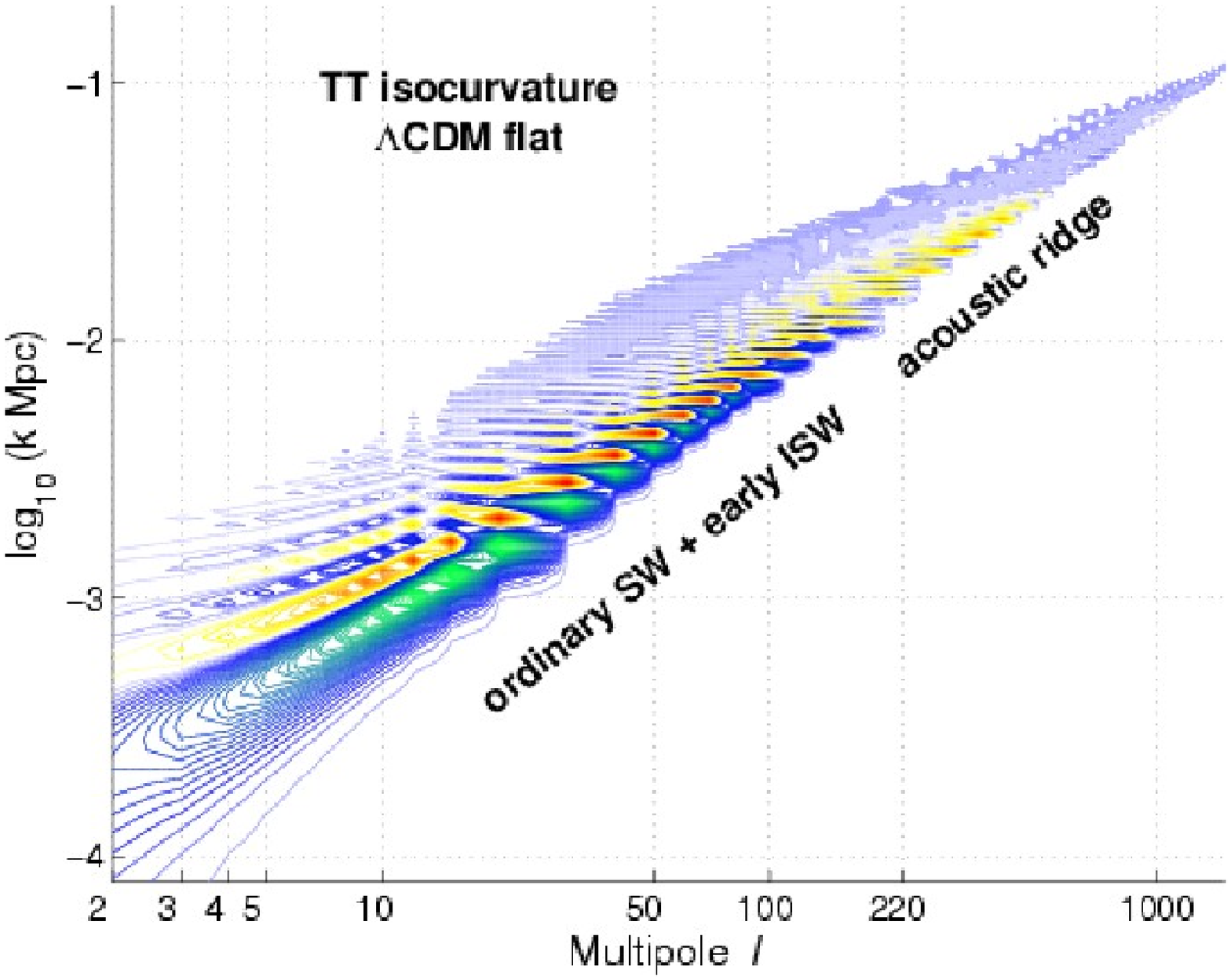}
\includegraphics[width=\twofigswidth]{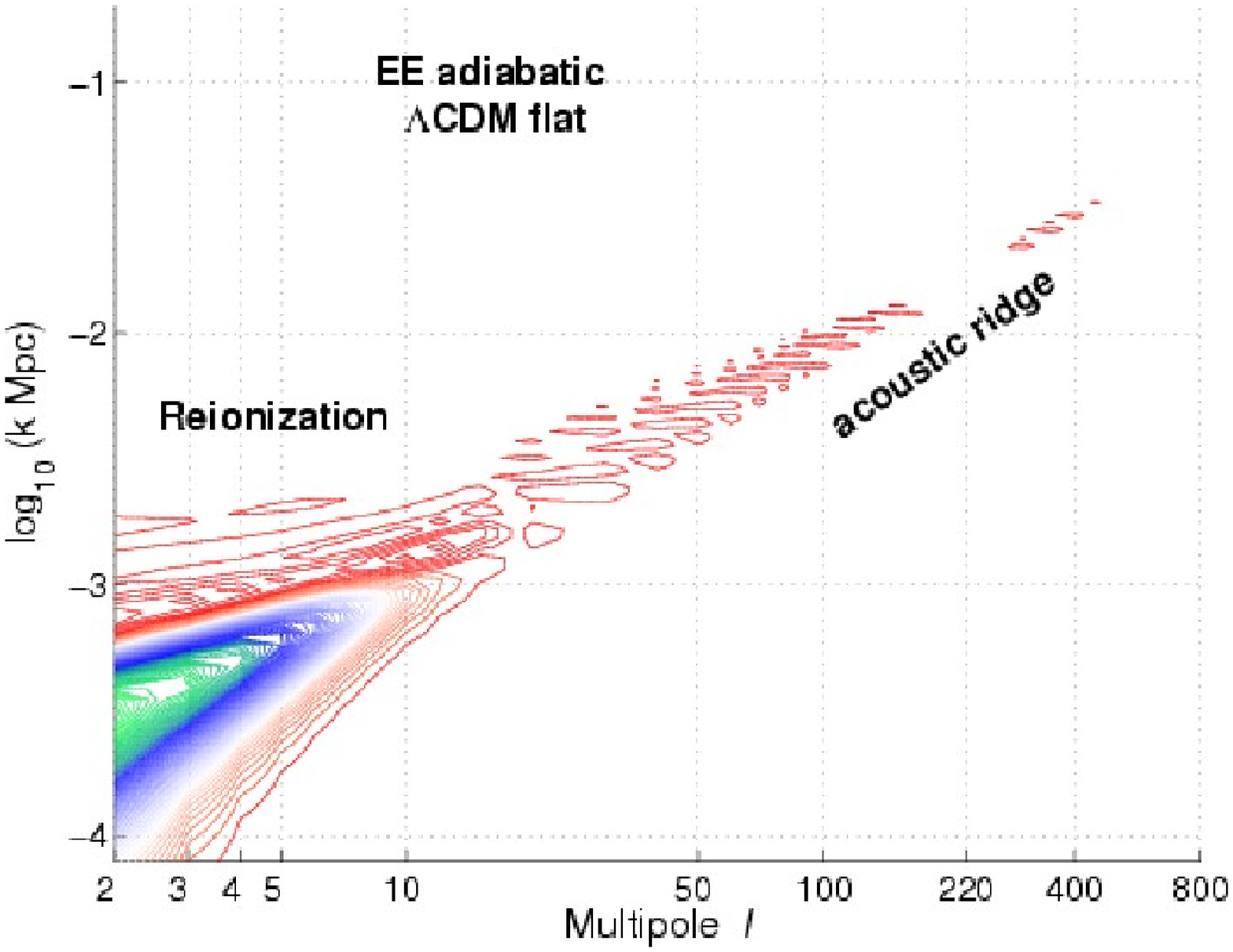}\hfill%
\includegraphics[width=\twofigswidth]{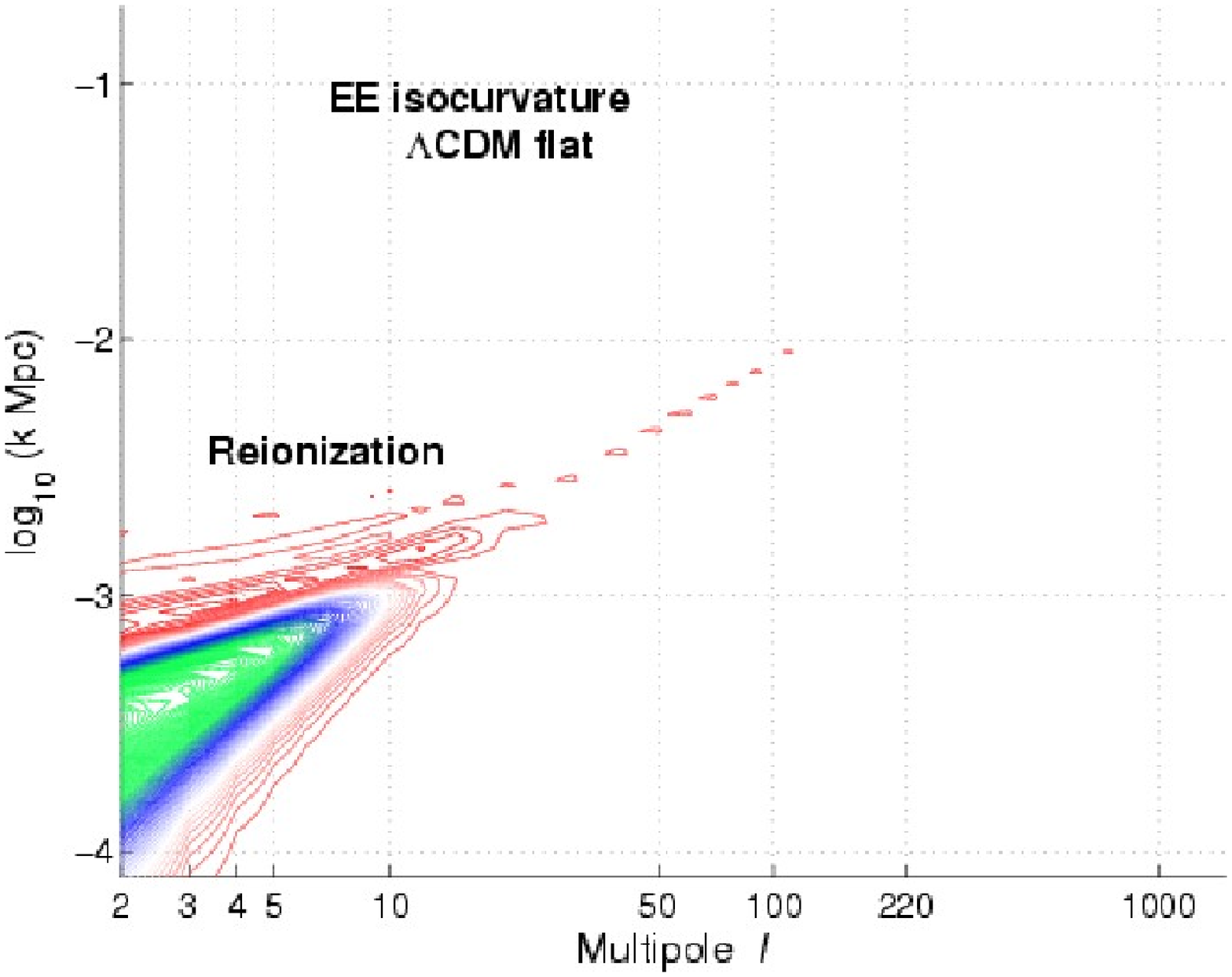}
\caption[CMB transfer functions for adiabatic and isocurvature
initial conditions.]{Temperature (top) and polarization (bottom)
transfer function $\Theta_\ell(\eta_0, k)$ and
$\Delta^{E}_\ell(\eta_0, k)$ for adiabatic (left panels) and
isocurvature CDM (right panels) initial conditions. The color
scales are arbitrary, and have been chose as too highlight the
features of the transfer functions. In particular, the color
coding is not in scale between the different plots.
\label{fig:transfer_fct}}
\end{figure}

The {\it photons transfer function} $\Theta_\ell (\eta_0, k)$ in
\rr{eq:C_ell_and_source} above is an intrinsically 2-dimensional
quantity which gives information about how the initial power is
mapped onto the angular power spectrum. It can be evaluated from
\rrp{eq:integral_solution}, by observing that the angle $\mu =
\bkh \cdot \bn$ in the integrand can be eliminated by replacing
 \be
 e^{\imath k \mu(\eta - \eta_0)} g \imath k \mu  V_\comp{b}
 =
 \frac{\dr}{\dr \eta} \left(e^{\imath k \mu(\eta - \eta_0)} g
 V_\comp{b} \right)
 - e^{\imath k \mu(\eta - \eta_0)} \dot{g}  V_\comp{b}
 - e^{\imath k \mu(\eta - \eta_0)}  g  \dot{V}_\comp{b}
 \ee
 and dropping the total derivative which only gives an
 unobservable monopole term. Therefore we can rewrite
 (\refp{eq:integral_solution}) as
  \be
  \Theta(\eta_0, k, \mu) =
  \int_0^{\eta_0} \dr \eta e^{\imath k \mu(\eta - \eta_0)}
  \mathcal{S}(\eta, k)
  \ee
 with the source term of the form
 \be
 \begin{split} \label{eq:general_form_for_Source}
 \mathcal{S}(\eta, k) & =
 g \left[\dfrac{\dot{V}_\comp{b}}{k} +
 \Theta_0 - \frac{\Theta_2}{4}  + \Psi + \Phi
 \right]
  - \dot{g} \left[\dfrac{V_\comp{b}}{k} + \dfrac{3}{4}
 \dfrac{\Theta_2}{k^2}\right]
 - \ddot{g}\dfrac{3}{4} \dfrac{\dot{\Theta}_2}{k^2} \\
 & + e^{-\tau}(\dot{\Psi} + \dot{\Phi}) \eqdot
\end{split}
\ee
 Now we expand the plane wave in radial and angular
eigenfunctions, Bessel functions and Legendre polynomials
respectively, using the Rayleigh formula
 \be
 e^{\imath k \mu(\eta - \eta_0)} = \sum_\ell \imath^\ell
 (2\ell+1) j_\ell(k (\eta_0 - \eta)) P_\ell(\mu) \eqcomma
 \ee
and we obtain for the temperature transfer function
 \be
\label{eq:Theta_ell_today}
 \Theta_\ell(\eta_0, k) = \imath^{\ell} \int_0^{\eta_0} \dr \eta
 \mathcal{S}(\eta, k) j_\ell(k (\eta_0 - \eta)) \eqdot
 \ee
This is shown in the top panels of \FIG{fig:transfer_fct} for
adiabatic and isocurvature CDM initial conditions.

 Together,
Eqs.~(\ref{eq:Theta_ell_today}) and (\ref{eq:C_ell_and_source})
allow the computation of the CMB angular power spectrum and neatly
split the geometric effects from the physics: all of the dynamical
evolution is encoded in the source function $\mathcal{S}(\eta,
k)$, while the Bessel function accounts for the projection from
3-dimensional $k$-space on the 2-sphere. The generalization of
this result for the $\curv \neq 0$ case can be found in
\cite{Zaldarriaga:1997va,Zaldarriaga:1999ep,Lewis:1999bs}.
 The temperature and E-polarization spectra of a concordance model for adiabatic and
isocurvature CDM initial conditions are displayed in the top left
panel of \FIG{fig:general_ic1} on page \pageref{fig:general_ic1}.

\subsubsection*{Polarization power spectrum}

As mentioned in \SEC{chap:perts;sec:EB}, polarization of scalar
modes is conveniently described by the $E$ polarization mode,
supplemented by the cross-correlator between $E$ and $T$
(temperature). As for temperature, we can formally integrate the
Boltzmann equation for the Stokes parameter $Q$,
\rrp{eq:Q_collisional_Boltzmann}, along the line of sight and
obtain
 \be \label{eq:Q_integral_solution}
 \Theta^Q(\eta_0, k, \mu) = - \dfrac{1}{2}\int_0^{\eta_0} e^{\imath k \mu(\eta - \eta_0)}
 g(\eta)
 \left( 1 - P_2 \right)
 \left( \Theta_2 + \TQ_2 - \TQ_0 \right) \eqdot
 \ee

The E-polarization power spectrum and the ET-correlator
(superscript $C$) are defined as
 \begin{align}
 \label{eq:def_E_and_C_C_ell}
 \EX{a^E_{\ell m} \cdot {a^{*E}_{\ell' m'}}} & = \delta_{\ell \ell'}
 \delta_{m m'} C^E_\ell \eqcomma \\
 \EX{a^T_{\ell m} \cdot {a^{*E}_{\ell' m'}}} & = \delta_{\ell \ell'}
 \delta_{m m'} C^C_\ell \eqcomma
 \end{align}
and in analogy with the treatment for the temperature spectrum
they can be computed as a superposition of $k$ modes of a source
function integrated over time:
 \begin{align}
 C^E_\ell & = 4\pi \int \dfrac{\dr k}{k} \: P_\chi(k) \:
 \vert \Delta^E_\ell (\eta_0, k) \vert ^2 \eqcomma
 \label{eq:E_C_ell_and_source} \\
 \Delta^E_\ell (\eta_0, k) & = \sqrt{\frac{(\ell + 2)!}{(\ell-2)!}}
 \int_0^{\eta_0} \dr \eta \call{S}^E(\eta,
 k) j_\ell(k(\eta_0-\eta)) \eqcomma\\
 \call{S}^E(\eta,k) & = \frac{3 g(\eta)}{4 k^2(\eta_0-\eta)^2}
  \left( \Theta_2 + \TQ_2 - \TQ_0 \right) \label{eq:E_source}\eqdot
 \end{align}

The cross-correlator spectrum is computed using
(\ref{eq:Theta_ell_today}) as
 \be
 C^C_\ell  = 4\pi \int \dfrac{\dr k}{k} \: P_\chi(k) \:
 \Theta_\ell^*(\eta_0, k) \Delta^E_\ell (\eta_0, k)  \eqdot
\ee
 The polarization transfer function $\Delta^E_\ell (\eta_0, k)$ is
plotted in \FIG{fig:transfer_fct} for adiabatic and isocurvature
CDM initial conditions.

The degree of polarization is proportional to the magnitude of the
temperature quadrupole at last scattering. Since during the tight
coupling regime the temperature quadrupole cannot grow,
polarization is generated in the relatively short transition
between the strong coupling and the free streaming regime. To
first order in $\taudot^{-1}$, the temperature quadrupole is
proportional to the temperature dipole, see
(\refp{eq:quadrupole_propto_dipole}). The polarization amplitude
is thus proportional to the temperature dipole at recombination
times the width of the last scattering surface
\citep{Zaldarriaga:1995gi}, resulting in a polarization signal two
orders of magnitude lower than the temperature signal.

\subsection{Matter power spectrum}
\label{chap:cmb;sec:mattpower}

Let $\delta(\eta, \bs{x})$ denote the real-space density contrast
in the matter component in the comoving gauge; hence $\delta$
corresponds to the gauge invariant variable $\Delta_m$ defined in
\rrp{eq:def_Delta}. We will drop the time dependence when not
needed, and write $\delta$ instead of $\Delta_m$ to simplify the
notation. For clarity, the Fourier transform of the variables is
denoted by a subscript ``$\bk$'', in this section only.

The real space correlation function is defined as
 \be \label{eq:correlation_fct}
 \xi(\bs{r}) \equiv \EX{\delta(\bs{x}) \cdot \delta(\bs{x} +
 \bs{r})}\eqcomma
 \ee
where $\EX{\cdot}$ denotes an average over realizations, see
\SEC{chap:data;sec:stattools} for precise definitions. It is the
expectation value of $\delta_2 = \delta(\bs{x} + \bs{r})$ and
$\delta_1 = \delta(\bs{x})$ under the 2-point probability
distribution function for $\delta_1, \delta_2$. We write
$\delta(\bs{x})$ as
 \be \label{eq:delta_k}
 \delta({\bf x}) = \frac{1}{(2\pi)^{3/2}}
 \int \dr ^3 {\bf k}  \dbfk  e^{\imath \bk \bs{x}}
 \ee
where we denote by $\dbfk$ the Fourier transform (in flat space)
of $\delta(\bs{x})$. We postulate that $\delta(\bs{x})$ is a
Gaussian distributed, isotropic and homogeneous random field, see
\SEC{chap:data;sec:stattools}, and therefore the quantity
$\EX{\delta_\bk^* \cdot \delta_{\bs{k'}}}$ vanishes for $\bk \neq
\bs{k'}$ (homogeneity) and it only depends on the modulus, not the
direction of $\bk$ (isotropy):
 \be
 \EX{\delta_\bk^* \cdot \delta_{\bs{k'}}} =
 (2\pi)^{3/2} \dirac (\bk - \bs{k'}) \; P_m(k)
 \ee
where $\dirac$ denotes the Dirac delta function. We call $P_m(k)$
the {\it matter power spectrum}. Replacing (\ref{eq:delta_k}) in
(\ref{eq:correlation_fct}) we obtain
 \be \label{eq:corr_funct_res}
 \xi(r) = \dfrac{1}{(2\pi)^{3/2}} \int \dr^3 \bk P_m(k)
 e^{\imath \bk \bfr} =
 \dfrac{2}{\sqrt{2\pi}} \int \dr k k^2 \dfrac{\sin r k}{r k} P_m(k)
  \eqcomma
 \ee
showing that the correlation function is the Fourier transform of
the matter power spectrum.

Our aim is to compute the power spectrum today as a function of
the spectral distribution in the early Universe in the adiabatic
CDM scenario. To this end, we make use of the results of linear
perturbation theory presented in the previous sections for the
growth of matter perturbations in a Universe containing CDM and
photons only. Clearly, these computations are valid only as long
as the scale considered is in the linear regime, \ie $\dbfk \ll
1$. We only sketch the elements which are needed in the following,
referring the reader to \eg
\cite{Structure_Peebles,Structure_Padmanabhan,Pert_Liddle} for a
full account.

Perturbations $\dbfk$ over a comoving length $\la \sim k^{-1}$
behave differently depending whether they are outside ($k < \Hbl$)
or inside ($k > \Hbl$) the Hubble length. For a given scale $k$,
we denote by $\eta_\enter$ the time at which that scale crosses
inside the horizon, \ie $\Hbl(\eta_\enter) = k$ and by $k_\eq$ the
wavelength which enters the horizon at the time of
matter-radiation equality, \ie $k_\eq = \Hbl(\eta_\eq)$. We thus
need to distinguish two cases: scales $k
> k_\eq$ enter the horizon in the radiation dominated epoch, while
$k < k_\eq$ enter the horizon after matter domination. We shall
restrict ourselves to length scales $\la$ which are large enough
not to be wiped out by free streaming, \ie $\la > \la_\text{FS}$,
see \cite{Structure_Padmanabhan} for details.

For $k > k_\eq$ and $\eta_\enter < \eta < \eta_\eq$, $\dbfk(\eta)$
stays approximately constant after horizon crossing because the
radiation dominated epoch suppresses the growth of perturbation in
a dust-like component; this is called the Meszaros effect
\citep{Meszaros:1974}. For $\eta > \eta_\eq$ the Universe is
matter dominated and the situation is analogous to the single
fluid case examined in \SEC{chap:cmb;sec:one_fluid}, and the
perturbation grows as $\dbfk \propto a$, see \rrp{eq:Psi_dust}.
Wavelengths which enter the horizon in the matter dominated epoch,
$k < k_\eq$, start growing as soon as they cross the horizon,
$\dbfk(\eta) \propto a$ for $\eta > \eta_\enter$, by the same
argument given above. Summarizing, we have that
 \be \label{eq:dbfk_1}
 \dbfk(\eta > \eta_\enter) \propto \left\{
 \begin{array}{l}
 \begin{aligned}
 &\dbfk(\eta_\enter) \dfrac{a}{a_\eq} & \quad \text{for } k >
 k_\eq \\
 &\dbfk(\eta_\enter) \dfrac{a_\eq}{a_\enter}\dfrac{a}{a_\eq} & \quad \text{for } k
 < k_\eq
 \end{aligned}
 \end{array}
 \right. \eqcomma
 \ee
and therefore we know $\dbfk$ for all subsequent times once we
specify $\dbfk(\eta_\enter)$, the value of the density contrast
for the wavelength $k$ {\it at the moment when that wavelength
crossed inside the horizon}. Since for a given wavelength
$\eta_\enter \propto 1/k$, horizon crossing happens at a different
time for each scale. We notice that in the second line of
\rr{eq:dbfk_1} we can rewrite the factor $a_\eq/a_\enter$ as
 \be
 \dfrac{a_\eq}{a_\enter} = \left( \dfrac{\eta_\eq}{\eta_\enter} \right)^2
 = \left( \dfrac{k}{k_\eq} \right)^2 \propto k^2 \eqcomma
  \ee
 where in the first equality we have used the fact that $a \propto \eta^2$
 in the matter dominated universe.

Given that the range of scales of cosmological interest is not too
wide, we can make the following power law Ansatz for the scale
dependence of the perturbation at horizon crossing
 \be
 \dbfk(\eta_\enter) = A k^{-\al} \eqdot
 \ee
An important quantity is $k^3/(2\pi)^{3/2}P_m(k)$, which from
(\ref{eq:corr_funct_res}) gives the contribution per logarithmic
$k$-interval to the real space correlation function, and which
with the above Ansatz evaluates to
 \be
 \dfrac{k^3}{(2\pi)^{3/2}}P_m(k)\Big \vert_{\eta_\enter} \propto k^{3-2\al} = \text{
 const for } \al = 3/2 \eqdot
 \ee
This quantity can also be interpreted as the variance of the mass
contained in spheres of diameter $\la \sim 1/k$ at horizon
crossing, see \eg \cite{Structure_Padmanabhan}; for the value $\al
= 3/2$ the variance is the same on all scales.

We might prefer to specify our Ansatz not at horizon crossing, but
rather for some fixed initial time (the same for all scales)
$\eta_i$. In order to relate $\dbfk(\eta_i)$ with
$\dbfk(\eta_\enter)$, we notice that on super-horizon scales $k <
\Hbl$ and for times $\eta_\enter > \eta > \eta_\eq$ we have $\dbfk
\propto a \propto \eta^2$ from \rrp{eq:Psi_dust}. For the case $k
< \Hbl$ in the radiation epoch, $\eta < \eta_\enter < \eta_\eq$ we
can use the adiabatic solution (\refp{eq:ad_growing_mode}) and the
relation
 \be
 \dbfk \equiv \Delta_m = \dfrac{3 + 3a}{4+3a} D + \dfrac{4}{4+3a}
 S \approx \dfrac{3}{4}D + S \propto a^2 \propto \eta^2 \eqcomma
 \ee
and the approximation is valid for $a < a_\eq$. In conclusion, the
comoving dark matter density contrast grows as $\eta^2$ at all
epochs while outside the horizon. Therefore we obtain (with
$\eta_\enter > \eta_i$ for all scales of interest)
 \be \label{eq:dbfk_enter_and_dbfk_i}
 \dbfk(\eta_\enter) = \left( \dfrac{\eta_\enter}{\eta_i} \right)^2
 \dbfk(\eta_i) \propto k^{-2} \dbfk(\eta_i) \eqdot
 \ee
It is customary to make a power law Ansatz for the matter power
spectrum at the time $\eta_i$ of the form
 \be
 P_m(k, \eta_i) = B k^n
 \ee
and by the relation (\ref{eq:dbfk_enter_and_dbfk_i}) the index $n$
is related to $\al$ by
 \be
 n = -2 \al + 4 \eqdot
 \ee
The value $\al = 3/2$ which yields a constant-mass-variance on all
scales at horizon crossing corresponds to $n=1$, the so-called
``scale invariant spectral index'', also known as
Harrison-Zel'dovich spectrum \citep{Harrison:1970,
Zel'dovich:1972}. The power spectrum today then becomes in terms
of $n$, from (\ref{eq:dbfk_1})
 \be \label{eq:dbfk_2}
 \dbfk(\eta_0) \propto \left\{
 \begin{array}{l}
 \begin{aligned}
 & k^{n-4} & \quad \text{for } k > k_\eq \\
 & k^n & \quad \text{for } k < k_\eq
 \end{aligned}
 \end{array}
 \right. \eqdot
 \ee
The length scale which crosses the horizon at equality, $\la_\eq
\approx 13/(\Om_m h^2)\;\mpc$ corresponds to a peak in the power
spectrum: fluctuations on larger scales, $k < k_\eq \sim
1/\la_\eq$ retain their primordial shape, while perturbations on
smaller scales have their spectrum multiplied by $k^{-4}$. The
above arguments only apply in the linear region, \ie for $k \lsim
0.3\; h/\mpc$, above which non-linear growth of the fluctuations
invalidate perturbation theory and a full numerical simulation is
required to follow the evolution.

Finally, we can easily relate the matter power spectrum to the
Bardeen potential by using the Poisson equation
(\refp{eq:poisson}). If we consider the value of
$\Psi_\bk(\eta_\enter)$, the Fourier transform of $\Psi$ evaluated
at horizon crossing, we have from the Poisson equation, noticing
that $\Hbl(\eta_\enter) = 1/k$, $\dbfk = \Delta_m \sim \Delta_\ga
\sim D$ by the adiabaticity condition, that $\Psi_\bk(\eta_\enter)
\sim - \dbfk(\eta_\enter)$. Therefore for the power spectrum of
the Bardeen potential, defined as
 \be
 P_\Psi \equiv \dfrac{k^3}{2\pi^2} \EX{ \vert \Psi_\bk \vert^2}
 \ee
we have that
 \be
 P_\Psi(k)\big\vert_{\eta_\enter} \propto k^3
 P_m(k)\big\vert_{\eta_\enter}\propto k^{n-1} \eqcomma
 \ee
and the $n=1$ scale invariant spectrum corresponds to
$P_\Psi(\eta_\enter) = \text{ const}$.  Or we can specify $P_\Psi$
at a fixed initial time $\eta_i$, in which case we obtain again
from the Poisson equation
 \be
 P_\Psi(k)\big\vert_{\eta_i} \propto k^{-1}
 P_m(k)\big\vert_{\eta_i}\propto k^{n-1} \eqdot
 \ee
The fact that there is no evolution in the power spectrum of
$\Psi$ until horizon crossing is of course a consequence of the
fact that $\Psi_{\bk} \approx \text{const}$ on super-horizon
scales, as shown in \SEC{chap:cmb;sec:matter_radiation}. The same
scaling applies for the power spectrum of the gauge invariant
curvature perturbation $\zeta$, which is constant on super-horizon
scales for adiabatic perturbations, and proportional to $\Psi$.

\chapter{Parameter dependence}
\label{chap:params}
This chapter presents a brief review of the dependence of the CMB
power spectra on the standard cosmological parameters and on
general initial conditions, building on the results of the
previous sections. Understanding the impact of the parameters on
the observable spectra builds the framework for parameter
extraction from data, which is the subject of Part III.

In \SEC{chap:params;sec:cosmo} we concisely review the origin and
main parameters dependencies of well known features of the power
spectrum: the large scale Sachs-Wolfe plateau, the acoustic
oscillations, and the damping tail. Introductory reviews on this
topic can be found in \eg \cite{Kosowsky:2001ue} and
\cite{Hu:1995hf}. A detailed physical understanding in a fully
analytical approach is explained in
\cite{Hu:1995uz,Hu:1995jd,Hu:1996en}. In view of efficient and
accurate parameter estimation, fundamental degeneracies in the CMB
spectra are best understood by introducing a set of analytical
functions of the parameters which the CMB probes directly, and
upon which the spectra dependence is almost linear
\citep{Kosowsky:2002zt}. We call this new basis in parameter space
``normal parameters set'', and we illustrate it in
\SEC{chap:params:sec:normal}.

In \SEC{chap:params;sec:ic} the CMB angular power spectra for
general isocurvature initial conditions in a Universe containing
CDM, baryons, photons and neutrinos are presented. The four modes
adiabatic, CDM isocurvature, neutrino density and neutrino
velocity -- along with a baryon isocurvature mode which is equal
to the CDM mode up to a rescaling constant -- span the whole space
of non-diverging solutions of Einstein's equations at early times
\citep{Bucher:1999re}, and thus their superposition constitutes
the most general type of initial conditions for CMB anisotropy.

\section{Standard parameters}
\label{chap:params;sec:cosmo}

The detailed shape of the CMB temperature and polarization spectra
depends on the value of the cosmological parameters and on the
type of initial conditions in characteristic ways. However,
certain combination of parameters lead to very similar spectra:
this causes degeneracies among some parameters, which cannot be
reconstructed with CMB alone, but require the inclusion of
external data-sets.

Polarization information helps breaking temperature degeneracies
because of two characteristic features: the first is that after
decoupling the polarization state is preserved by free streaming,
and the polarization spectrum is only modified by rescattering due
to reionization (\SEC{chap:params:sec:reion}). Therefore in a
sense polarization is a more clean probe of the decoupling than
temperature. The second reason is that while the acoustic peaks in
temperature are dominated by the monopole of the temperature
fluctuation on the LSS, the peaks in E-polarization reflect the
dipole component at decoupling, \ie the photon bulk velocity
(\SEC{chap:params;sec:peaksloc}).

In the following we revisit the main parameter dependence of the
CMB spectra: for the sake of illustrating the physical effects
involved, we divide the CMB power spectrum in three distinct
regions, corresponding to different angular separations on the sky
with the approximate relation $\vartheta \sim \pi/\ell$.

 \begin{itemize}
 \item {\bf Large scales:} on scales larger than the Hubble radius at
 decoupling, $k\eta_\dec \ll 1$, perturbations are dominated by the ordinary
 Sachs-Wolfe effect, given by the combination of the intrinsic
 temperature fluctuations on the LSS and the gravitational
 redshift induced by climbing out of the potential well. In non-flat
 cosmologies, or models with a considerable value of the
 cosmological constant, the late ISW effect also contributes.
 This
 region corresponds roughly to the COBE scale, $\ell \lsim 30$ and
 $\vartheta \gsim 7\DEG$.

 Reionization produces a a characteristic increase of
 E-polarization on large scales, the so-called ``polarization
 bump''.
 \item {\bf Acoustic region:} inside the sound horizon photon pressure
cannot be neglected, and scales within the sound horizon $k \int
c_s \eta \gsim 1$ oscillate, while gravitational infall becomes
negligible because of potential decay inside the horizon. On
intermediate scales $50 \lsim \ell \lsim 600$ the CMB power
spectrum displays a rich peak structure, reflecting the
contributions of density oscillations and Doppler term on the LSS.
The early ISW effect contributes at roughly the $20\%$ level up to
the first acoustic peak (for adiabatic models). Those scales have
a typical angular separation on the sky ranging from about
$10\DEG$ down to a few $10\arcmin$.
 \item {\bf Damping tail:} wavelengths smaller than the diffusion damping scale
 $1/k_\text{D}$ given in (\refp{eq:damping_scale}) are
 exponentially suppressed and this causes a drop in power above
 $\ell \sim 800$ or $\vartheta \lsim 1\arcmin$. This effect combines with rescattering due to reionization,
 which also erases fine-scale anisotropies.
 \end{itemize}

\subsection{Large scales}
\label{chap:cmb;sec:large_scales}

We wish to investigate the expected temperature fluctuations on
very large scales in the general case of a superposition of
primordial adiabatic and isocurvature CDM initial conditions. We
look at wavelength $k \ll k_\dec$ which at decoupling where still
outside the horizon and we consider a zeroth order approximation
which neglects any anisotropic stress and the baryon influence
(\ie set $R=0$). If we take decoupling to happen well into matter
domination, we can also neglect the ISW contribution since the
potentials are equal and constant -- see \rrp{eq:Psi_dust} -- and
to this level of approximation we can set $V_\comp{b} =
V_\comp{\ga}$. With this approximations we have for each Fourier
mode from Eqs.~(\refp{eq:OSW_term}) and (\refp{eq:Doppler_term})
 \be \label{eq:Theta_OSW_and_Doppler}
 \Theta(\eta_0, k, \mu)  =
 e^{\imath k \mu (\eta_\dec - \eta_0)}
 \left[\frac{1}{4}
 D_{g, \comp{\ga}} + 2\Phi  - \imath k \mu V_\comp{\ga} \right] (\eta_\dec,
 k)\eqdot
 \ee
In the adiabatic case, we can neglect the contribution of the
Doppler term which behaves as a sine and hence disappears on large
scales, $k \eta_\dec \ll 1$, while the cosine oscillation of the
density perturbation $D_{g, \comp{\ga}}$ becomes constant, see
(\refp{eq:HO_adiabatic}). Therefore for adiabatic initial
conditions, from the solution (\ref{eq:HO_adiabatic}) it follows
 \be \label{eq:cancelling_potentials}
 \Theta(\eta_0, k, \mu)  \approx
 e^{\imath k \mu (\eta_\dec - \eta_0)}
 \left[ \left( \dfrac{1}{3}\Phi_\MD -2\Phi_\MD \right) + 2\Phi_\MD \right]
 \quad \text{(adiabatic),}
 \ee
 where $\Phi_\MD$ denotes the value of $\Phi$ at decoupling well
within matter domination. On the right hand side, the term
$-2\Phi_\MD$ comes from the solution (\ref{eq:HO_adiabatic}), and
its negative sign reflects the fact that the temperature is larger
inside potential wells ($\Phi < 0$), so that photons are
blushifted when they fall {\it into} the well. The term
$2\Phi_\MD$ represents the gravitational redshift which photons
experience when they climb {\it out} of the potential as they free
stream after decoupling, which exactly cancels the gravitational
blueshift term in the absence of baryons. In conclusion we have
 \be \label{eq:dT_AD}
 \Theta(\eta_0, k, \mu)  \approx e^{\imath k \mu (\eta_\dec - \eta_0)}
 \dfrac{1}{3} \Phi_\MD \quad
 \text{(adiabatic).}
 \ee

For isocurvature initial conditions, we have that $D_{g,
\comp{\ga}}(\eta_\dec) = 0$, which follows from
(\refp{eq:D_g_ga_superhorizon}) with the isocurvature condition
$\Phi_0 = 0$. The Doppler term can again be neglected with respect
to the potential, because from (\refp{eq:Vgamma_in_MD}) we have
that $k V_\ga \sim k/\Hbl \Phi \ll \Phi$ and
(\ref{eq:Theta_OSW_and_Doppler}) reduces to
 \be \label{eq:dT_ISO}
 \Theta(\eta_0, k, \mu)
 \approx e^{\imath k \mu (\eta_\dec - \eta_0)}
 2  \Phi_\MD \quad
 \text{(isocurvature),}
 \ee
the well-known result that isocurvature initial conditions produce
large scale fluctuations six times larger than in the adiabatic
case {\it for the same value of the Bardeen potential} on the last
scattering surface.

More interestingly, we can relate the large-scale temperature
fluctuations to the amplitude of the primordial curvature and
entropy spectra. Rewriting (\ref{eq:dT_AD}--\ref{eq:dT_ISO}) in
terms of the curvature and entropy perturbations in the radiation
era via
Eqs.~(\ref{eq:Phi_as_fct_of_Zeta}--\refp{eq:Phi_MD_and_S_0}),
yields for the source term (\refp{eq:general_form_for_Source})
 \be \label{eq:source_for_mixed_AD_ISO}
 \mathcal{S}(\eta, k) = \delta(\eta - \eta_\dec)
 \left[ \dfrac{\zeta_0}{5}  \psi (k)
 - \dfrac{2}{5} S_0 \phi(k) \right] \eqcomma
 \ee
where $\psi(k)$ and $\phi(k)$ are the Fourier components of random
fields which we assume are Gaussian distributed, isotropic and
homogeneous, see \SEC{chap:data;sec:stattools}, evaluated at some
initial time $\eta_i$ deep in the radiation epoch. For their {\it
power spectrum} we make a power low Ansatz
 \begin{align}
 P_\psi(k)\big\vert_{\eta_i} &\equiv
 \frac{k^3}{2\pi^2}\EX{\vert \psi(k) \vert^2 }  =
 \zeta_0^2 \left( \frac{k}{k_\piv} \right)^{n_\SCAL -1} \eqcomma \label{eq:PS_for_zeta}\\
 P_\phi(k)\big\vert_{\eta_i} &\equiv \frac{k^3}{2\pi^2}\EX{\vert \phi(k) \vert^2 }  =
 S_0^2 \left( \frac{k}{k_\piv} \right)^{n_\ENTR -1} \eqcomma \\
 P_{\CORR}(k)\big\vert_{\eta_i} &\equiv \frac{k^3}{2\pi^2}\EX{\psi(k) \cdot \phi^*(k)}  =
  \zeta_0 S_0
  \left( \frac{k}{k_\piv} \right)^{n_{\CORR} -1} \cos(\Delta_{\CORR}) \eqdot
\end{align}
The constants $\zeta_0$ and $S_0$ are dimensionless and positive,
while the angle $\Delta_{\CORR}$ parameterizes the correlation
between entropy and isocurvature perturbations; the constant
$k_\piv$ is a pivot scale, for which a popular choice is $k_\piv =
0.05~\mpc^{-1}$, and we have defined $n_{\CORR} \equiv (n_\SCAL +
n_\ENTR)/2$.

The power law index $n_\SCAL$ is the {\it scalar spectral index}:
$n_\SCAL \approx 1$ is a generic prediction of inflation, almost
independently of the particular model, and is called
``scale-invariant'' or Harrison-Zel'dovich \citep{Harrison:1970,
Zel'dovich:1972} spectral index. The reason for the name is
explained in \SEC{chap:cmb;sec:mattpower}. Since $\Psi \propto
\zeta$ up to constant factors, $\Psi$ and $\zeta$ have the same
spectrum.

From (\refp{eq:C_ell_and_source}) the angular power spectrum on
large scales ($\ell \lsim 20$) is then given by
 \be
 \begin{split}
 C_\ell = & 4\pi
 \int \frac{\dr k}{k}
 \left[ \frac{\zeta_0^2}{25} \left( \frac{k}{k_\piv} \right)^{n_\SCAL
 -1}+
 \frac{4 S_0^2}{25} \left( \frac{k}{k_\piv} \right)^{n_\ENTR -1} -
 \frac{4}{25}\zeta_0 S_0 \cos(\Delta_{\CORR}) \left( \frac{k}{k_\piv}
 \right)
 ^{n_{\CORR -1}} \right] \times \\
  & \times j_\ell^2\left(k(\eta_0 - \eta_\dec) \right) \eqdot
  \end{split}
 \ee
The integral can be performed analytically provided all the
indexes are within the range $-3 < n_X < 3$ and in the
approximation $k(\eta_0 - \eta_\dec) \approx k\eta_0$
\citep{Gradshtein}. The result is
 \be
 C_\ell = 2\pi^2 \left[
 \dfrac{\zeta_0^2}{25} f(n_\SCAL, \ell)
 + \dfrac{4 S_0^2}{25} f(n_\ENTR, \ell)
 - \dfrac{4}{25}\zeta_0 S_0 \cos(\Delta_{\CORR}) f(n_\CORR, \ell)
  \right] \eqdot
 \ee
 The function $f$ contains the dependence on the spectral indexes, and it is given by
 \be
 f(n, \ell) \equiv  (\eta_0 k_\text{P})^{1 - n}
\frac{\Ga(3-n)\Ga(\ell-\frac{1}{2}+\frac{n}{2})}{
2^{3-n}\Ga^2(2-\frac{n}{2})\Ga(\ell+\frac{5}{2}-\frac{n}{2})}
\eqcomma
 \ee
where $\Ga$ is the gamma function, which for a scale invariant
spectrum, $n = 1$, evaluates to
 \be
 f(n=1, \ell) = \frac{1}{\pi(\ell(\ell+1))} \eqdot
 \ee

If both the curvature and entropy spectral indexes are close to
scale invariant ($n_\SCAL = n_\ENTR = 1$), we find that the
so-called {\it Sachs-Wolfe (SW) plateau} for $\ell \lsim 20$ is
constant:
 \be \label{eq:COBEplateau}
 \frac{\ell(\ell+1)}{2\pi} C_\ell = \dfrac{1}{25}\zeta_0^2 +
 \dfrac{4}{25} S_0^2 - \dfrac{4}{25}\cos(\Delta_{\CORR}) \zeta_0 S_0
 \approx 10^{-10} \eqcomma
 \ee
and the numerical value is the measurement of the DMR instrument
aboard the COBE satellite averaged on scales $\lsim 7\DEG$
\citep{Smoot:1992td}. Clearly, uncorrelated entropy and curvature
perturbations (\ie with $\cos(\Delta_{\CORR})= 0$) both add to the
SW plateau, but a positive correlation (defined by
$\cos(\Delta_{\CORR}) > 0$) {\it reduces} the power on large
scales, while a negative correlation increases it, as shown in the
top left panel of \FIGPAG{fig:general_ic1}. If there is no
correlation, the isocurvature Sachs-Wolfe plateau from
(\ref{eq:dT_AD}) and (\ref{eq:dT_ISO}) is 36 times larger than the
adiabatic one for the same value of $\Psi$ at last scattering, and
4 times larger for the same amplitude of the primordial curvature
and entropy perturbations, \rr{eq:COBEplateau}. In the pure
adiabatic case, $S_0 = 0$, we obtain from (\ref{eq:COBEplateau})
an estimate of the primordial amplitude of the curvature
perturbation:
 \be
 \zeta_0 \approx 5 \cdot 10^{-5} \eqdot
 \ee

For models with a non-zero cosmological constant, the Universe
becomes $\La$ dominated  for $a/a_0 \geq (\Om_m/\Om_\La)^{1/3}$,
and the potentials start again to decay. This produces a {\it
late} time ISW which contributes on large scales, where it is
dominant with respect to the ordinary SW part described above,
producing a rise of the SW plateau at low multipoles. The details
differ considerably for adiabatic and isocurvature models, and
also depend on the spectral index, see \cite{Hu:1995jd} for a
detailed explanation.

\subsection{Acoustic region}
\label{chap:params;sec:acoustic}

The structure of the power spectrum on intermediate scales is the
result of several physical effects, sometimes with contrasting
impacts. The most distinctive features are acoustic oscillations
and projection.

\subsubsection{Peak locations}
 \label{chap:params;sec:peaksloc}

 Scales $k r_s  = k \int_0^{\eta_\dec} c_s \dr
\eta > 1$ enter the horizon before decoupling and thus
$D_{g,\comp{\ga}}$ oscillates as $\cos(r_s k)$ -- \CF
(\refp{eq:HO_adiabatic}) -- for adiabatic perturbations or as
$\sin(r_s k)$ -- \CF (\refp{eq:HO_isocurvature}) -- in the
isocurvature mode. Thus scales which at the moment of decoupling
have reached an extremum of their oscillation will yield
corresponding peaks in the temperature power spectrum. Notice that
since the power spectrum is a quadratic quantity, both maxima and
minima of the oscillations give peaks. The $k$ modes which at
recombination are at maximum compression or expansion are
 \begin{alignat}{2}
 k_{\text{ad}}^{(m)} & =  \dfrac{m \pi}{r_s(\eta_\dec)}, \quad m=1,2,3,\dots
 &&  \quad \text{(adiabatic),} \label{eq:enumerate_ad}\\
 k_{\text{is}}^{(m)} & = \dfrac{m \pi + 1/2}{r_s(\eta_\dec)}, \quad m=0,1,2,\dots
 &&  \quad \text{(isocurvature).\label{eq:enumerate_ci}}
\end{alignat}
The corresponding physical scale $\la^{\text{phys}}=a_\dec \pi/ k$
subtends an angle $\vartheta$ on the sky given by the angular
diameter distance relation (\refp{eq:angular_diameter_distance}),
and the peaks in the angular power spectrum show up at $\ell \sim
\pi/\vartheta$ or
 \begin{alignat}{2}
 \ell^{(m)} & \sim m \pi \frac{D_A}{a r_s}(\eta_\dec)
 && \quad \text{(adiabatic),} \label{eq:ad_peak_position}\\
 \ell^{(m)} & \sim (\tfrac{1}{2} + m) \pi \frac{D_A}{a r_s}(\eta_\dec)
 && \quad \text{(isocurvature).}
 \end{alignat}

Since $D_{g, \comp{\ga}}(k=k_{\text{ad}}^{(1)}) < 0$, the first
adiabatic peak corresponds to a compression maximum, while the
first ``isocurvature hump'' is an expansion maximum, $D_{g,
\comp{\ga}}(k=k_{\text{is}}^{(0)})
> 0$. In the literature, ``first acoustic peak'' usually
designates the compression peaks, \ie the first adiabatic extremum
and the second isocurvature one, which in the notation of
(\ref{eq:enumerate_ad}--\ref{eq:enumerate_ci}) correspond both to
the index $m=1$. For a flat universe ($\curv = 0$) without
cosmological constant ($\Om_\Lambda = 0$) and a baryon content as
inferred from BBN ($\Om_b h^2 \approx 0.02$), the location of the
first acoustic peak is approximately
 \begin{align}
  \ell^{(1)} & \sim 220 \quad \text{(adiabatic)} \quad \text{and}
  \\
  \ell^{(1)} & \sim 330 \quad \text{(isocurvature).}
\end{align}

The WMAP data allow a very precise determination of the position
of the first peak, $\ell^{(1)} = 220.1 \pm 0.8$
\citep{Page:2003fa}, thereby confirming that the adiabatic mode is
the dominant one. However, subdominant isocurvature contributions
cannot be ruled out, see \CHAP{chap:genic}.

The location of the peaks depends on the of initial conditions,
but the inter-peaks distance is independent on the type of
perturbations, and in the above estimate is $\Delta \ell \approx
220$. The peak spacing depends on the baryon content, which sets
$r_s$, and on the spatial geometry which enters in $D_A$. A larger
baryon content slows down the oscillations, thus decreasing the
sound horizon and the spacing between peaks grows larger. The
dependence of $D_A$ is primarily on the curvature of the universe:
in a crude approximation we neglect $\Om_\curv \ll \Om_m$ and
$\Om_\La$ when integrating (\refp{eq:delta_tau_in_redshift}) up to
$z_\dec \approx 1100 \gg 1$ and neglect $\Om_r$ as well (which is
not a good approximation for a large redshift) and we obtain
 \be
 D_A(z_\dec) \approx \frac{2a_\dec}{H_0 a_0} \Om_m^{-1/2} \eqdot
\ee

Therefore the peak position scales as $\Om_m^{-1/2}$, which means
that the peaks are shifted to larger $\ell$ values in an open
universe. Introducing a non-zero cosmological constant complicates
matters, since it is then possible to obtain the same value of the
angular diameter distance, and hence the same peak location, by
compensating a change in $\Om_m$ with a different value of
$\Om_\La$, an effect which goes under the name of angular diameter
distance degeneracy \citep{Efstathiou:1998xx,Melchiorri:2000px}.
The angular diameter distance test is no longer sufficient to
determine alone the curvature of the universe, but an independent
measurement of $\Om_m$ or $\Om_\La$ is necessary.

To illustrate this fundamental degeneracy, let us introduce the
{\it shift parameter} $\Rshift$, which gives the first peak's
position (in an adiabatic model) with respect to its location in a
flat reference model with $\Om_m = 1$:
 \be \label{eq:def_shift_parameter}
 \ell^{(1)} = \ell^{(1)}_\textrm{ref}/ \Rshift \eqcomma
 \ee
which can be evaluated from (\ref{eq:ad_peak_position}). To this
end, we need the explicit expression for the sound horizon at
decoupling, which is given by
 \be
 \begin{split}
 r_s(a_\dec) & = \int_0^{a_\dec} c_s \dfrac{\dr \eta}{\dr \tilde{a}}\dr \tilde{a} \\
 &= \dfrac{1}{H_0 a_0 \sqrt{3}} \int_0^{a_\dec/a_0} \dfrac{\dr x}
 {\left[ \left( 1 + \frac{3\Om_b}{4 \Om_\ga}x \right)
 \left(\Om_m x + \Om_r + \Om_\curv x^2 + \Om_\La x^4 \right)
 \right]^{1/2}}
 \end{split}
 \ee
(where all the $\Om_X$'s are evaluated today). Neglecting the
curvature and cosmological constant term in the early universe
($a_\dec/a_0 \ll 1$) yields the approximate result
 \be \label{eq:sound_horizon}
 \begin{split}
 r_s(a_\dec) \approx & \dfrac{1}{\sqrt{3} H_0 a_0 \Om_m^{1/2}}
 \left( \dfrac{a_\eq/a_0}{R_\eq} \right)^{1/2} \times \\
 & \times \ln\left[ \dfrac{1 + R_\eq + 2R_\dec + 2 \sqrt{(1+R_\dec)(R_\eq + R)}}
 {1 + R_\eq + 2 \sqrt{R_\eq}} \right] \eqcomma
 \end{split}
 \ee
 where
 \be
  R(a) \equiv \dfrac{3\Om_b}{4 \Om_\ga} \dfrac{a}{a_0} \quad
  \text{and} \quad R_\eq \equiv R(a_\eq), R_\dec \equiv R(a_\dec) \eqdot
  \ee

\begin{figure}[bt]
 \includegraphics[width=0.31\linewidth]{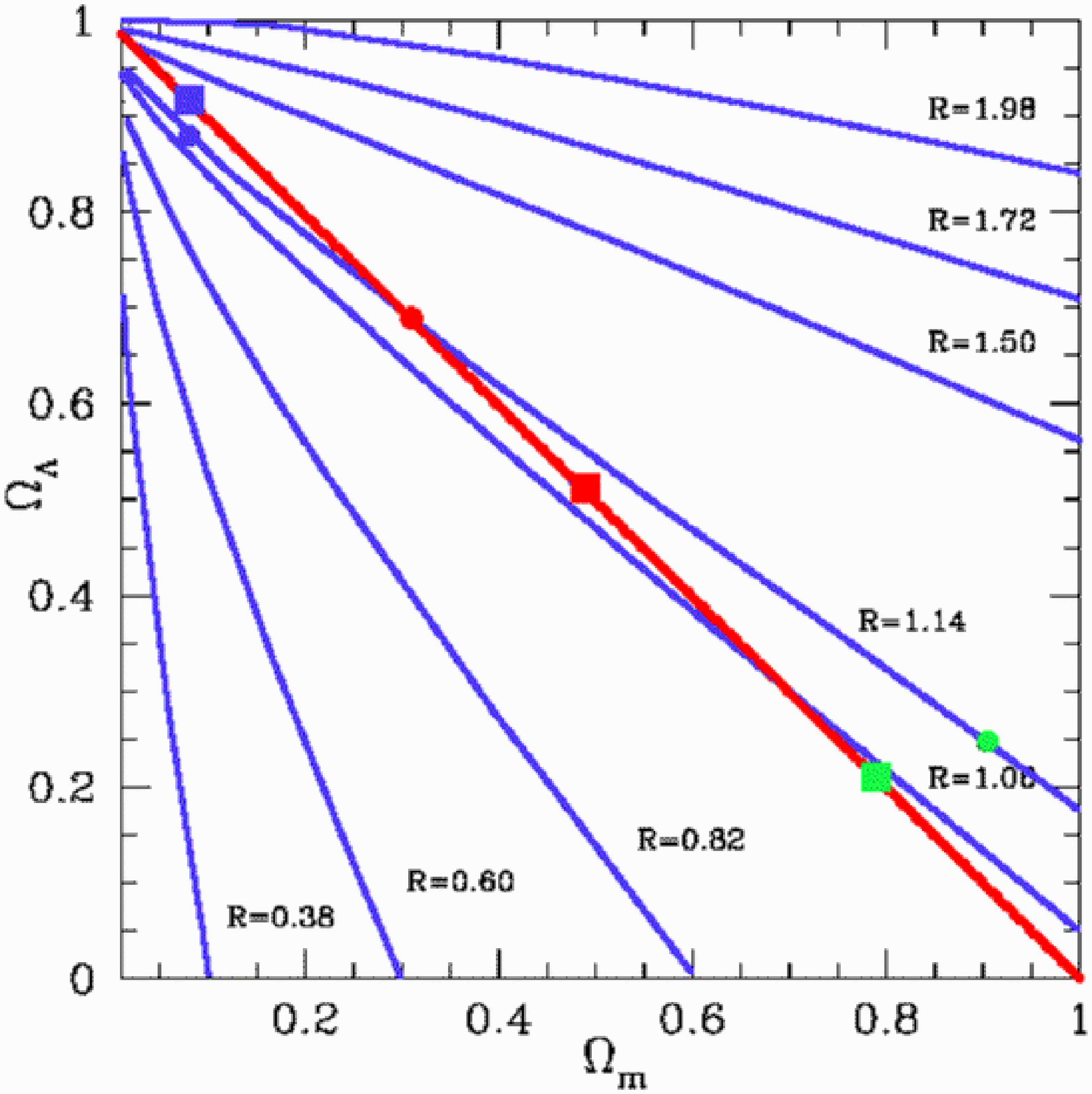}
 \includegraphics[width=0.31\linewidth]{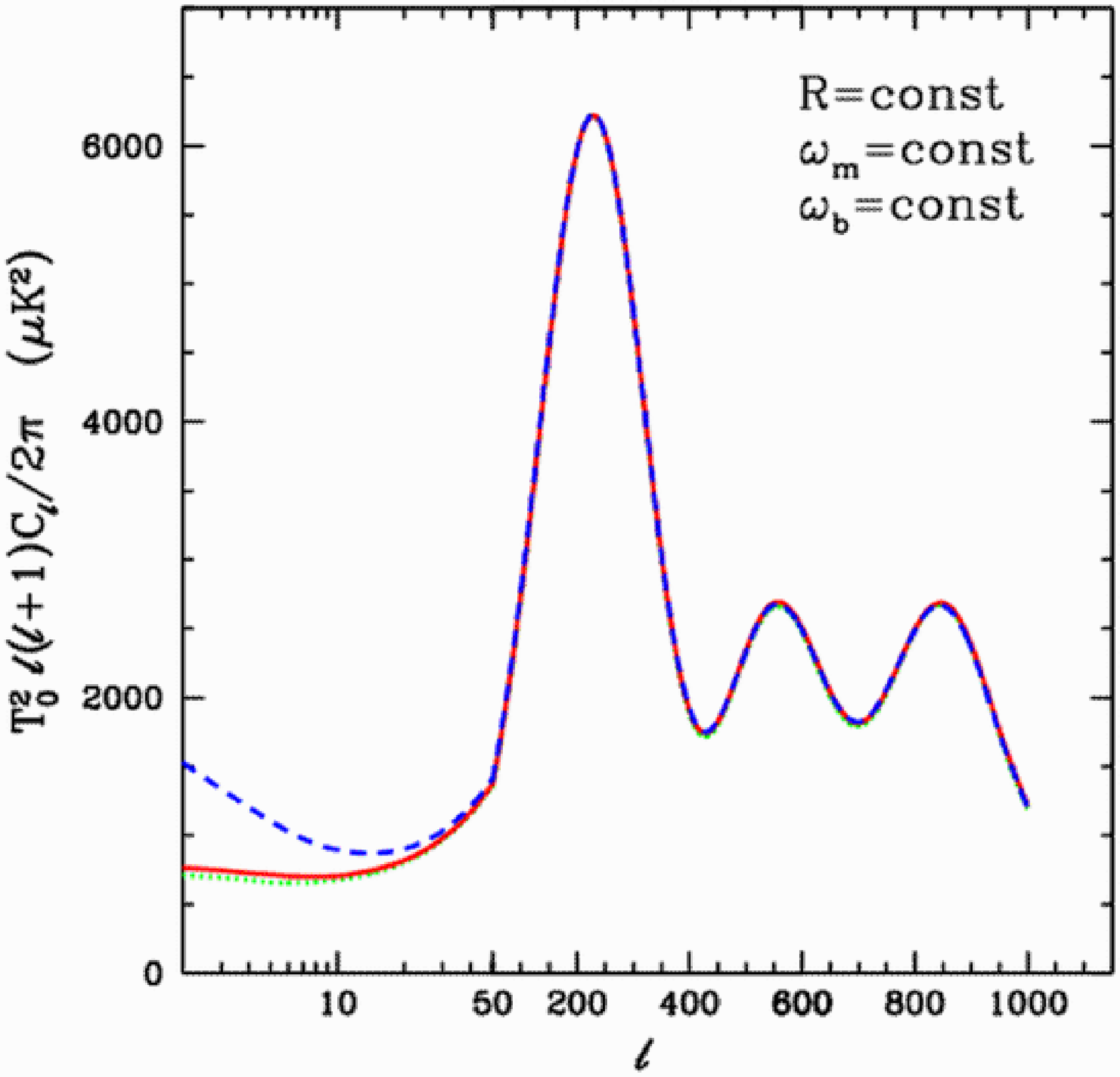}
 \includegraphics[width=0.31\linewidth]{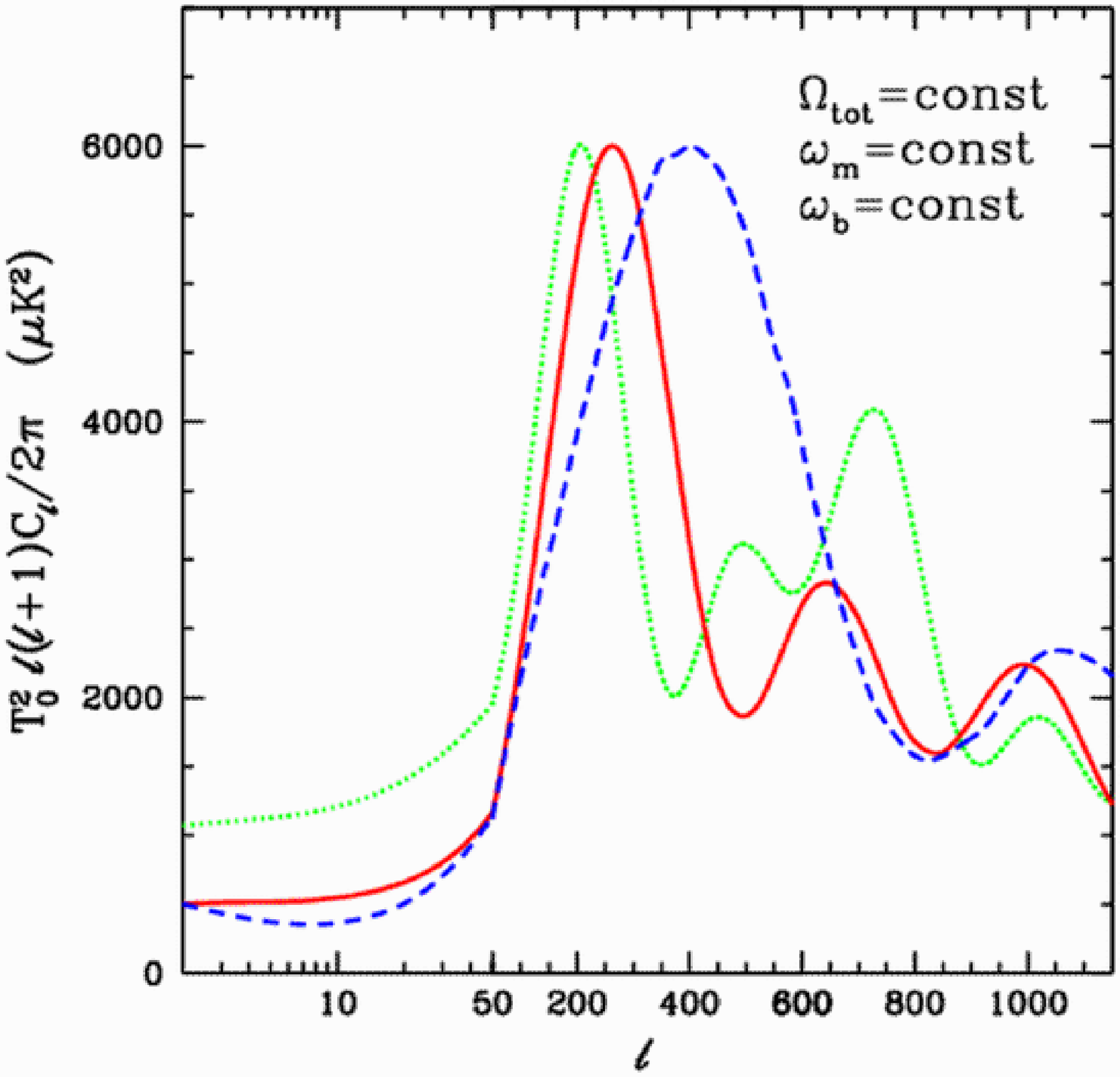}
 \centering \caption[Illustration of the geometrical degeneracy.]
 {Left panel: lines of constant shift parameter (\ref{eq:approximate_Rshift}) in the $(\Om_m,
\Om_\La)$ plane (in blue) correspond to models in which the
acoustic peaks are in the same position; those lines are not
parallel to lines of constant curvature (in red, the line of
$\Om_\text{tot} = 1$ is the locus of flat models). Middle panel: a
closed (blue, long-dashed), a flat (solid, red) and an open model
(dotted green) with parameters corresponding to the three colored
dots in the left panel on the $\Rshift = 1.14$ line are almost
completely degenerate. Right panel: three flat models with
different shift parameters (and values corresponding to the three
colored squares in the left panel) exhibit a very different peak
structure. In particular, measuring the position of the first peak
alone is not enough to determine the curvature of the Universe.
\label{fig:shift_param2D}}
\end{figure}

In order to find a simple approximate expression for $\Rshift$,
let us ignore the logarithmic dependence on the parameters of
$r_s$, and neglect the parameter dependence of the factor
$(a_\eq/a_0)^{1/2}/R_\eq^{1/2}$ as well; we shall relax those
approximations in \SEC{chap:params:sec:normal}. Then the sound
horizon for $\curv \ne 0$ models scales as
 \be \label{eq:rs_scaling}
 r_s(a_\dec) \approx \alpha \sqrt{\dfrac{\vert \Om_\curv \vert}{\Om_m}}
 \eqcomma
 \ee
while for the reference model with $(\Om_m, \Om_\La) = (1, 0)$ we
have
 \be
 \dfrac{D_A(a_\dec)}{a_\dec r_s(a_\dec)} = 2 \alpha \eqcomma
 \ee
 with $\alpha$ being approximately the same factor as in
(\ref{eq:rs_scaling}). For the shift parameter
(\ref{eq:def_shift_parameter}) of a model with arbitrary $(\Om_m,
\Om_\La)$ we then obtain the simple expression
  \be \label{eq:approximate_Rshift}
   \Rshift \approx \dfrac{2}{\chi(\Delta \eta)} \sqrt{\dfrac{\vert \Om_\curv
   \vert}{\Om_m}}\eqcomma
  \ee
where $\Delta \eta$ is given in \rrp{eq:delta_tau_in_redshift} and
$\chi$ in \rrp{eq:define_chi_function}. This handy expression
gives the approximate position of the first peak as a function of
$\Om_m$ and $\Om_\La$, with $\Om_\curv$ obtained from the
constrain $1 = \Om_m + \Om_\La + \Om_\curv$. Here we have ignored
the dependence on the radiation content of the model, which is
explicitly included in (\rrp{eq:def_r}). In the left panel of
\FIG{fig:shift_param2D} we plot lines of $\Rshift = \text{const}$
in the $(\Om_m, \Om_\La)$ plane, which are not parallel to lines
of constant curvature (diagonal lines).

Along with $\Rshift$, two other physical quantities determine the
structure of the peaks: the baryon density $\Om_b h^2$ controls
the relative height of the peaks, see
\SEC{chap:params;sec:barsig}, while the amount of matter $\Om_m
h^2$  sets the redshift of equality, for a fixed relativistic
energy content. Therefore by fixing the three quantities $\Rshift,
\Om_mh^2, \Om_bh^2$ we obtain models with almost indistinguishable
power spectra in the acoustic region. This is illustrated in the
middle panel of \FIG{fig:shift_param2D}, where a flat, a closed
and an open model result completely degenerate, with the only
difference showing up on large scales because of the different
amount of late ISW effect. The right panel shows that conversely
the first peak's position in three flat models can be very
different if the shift parameters differ, and therefore the
statement that the first peak position alone can determine the
curvature of the Universe is imprecise.

Polarization peaks are displaced by $\pi/2$ with respect to
temperature peaks, hence polarization maxima occur at temperature
minima. This can be seen by expanding to first order in
$\taudot^{-1}$ the polarization hierarchy
(\ref{eq:Qpol_Boltzmann_hierarchy}--\refp{eq:Qpol_Boltzmann_hierarchy_end}),
finding for the polarization monopole and quadrupole
 \be
 \TQ_0 = - \dfrac{5}{4}\Theta_2 \quad \text{and} \quad
 \TQ_2 = - \dfrac{1}{4}\Theta_2 \eqdot
 \ee
The temperature quadrupole is found to the same order from the
temperature hierarchy, including the polarization feedback as in
(\refp{eq:Qell2_temp_including}), giving
 \be \label{eq:quadrupole_propto_dipole}
 \Theta_2 = - \taudot^{-1} \dfrac{8}{15}\imath k \Theta_1 \eqdot
 \ee
The E-polarization source term (\refp{eq:E_source}) becomes in the
instantaneous decoupling approximation
 \be
 \call{S}^E = - \taudot^{-1} (\eta_0 - \eta_\dec)^{-2}
 \dfrac{\imath}{k} \Theta_1(\eta_\dec) \eqcomma
 \ee
showing that E-polarization probes the temperature dipole, \ie the
bulk velocity of the photons-baryons fluid, at decoupling. Since
$\Theta_1 \propto V_\comp{\ga} \propto \dot{D}_{g, \comp{ \ga}}$
we see that polarization oscillations are out of phase of $\pi/2$,
as visible in the top left panel of \FIGPAG{fig:general_ic1}.

\subsubsection{Baryon signature}
\label{chap:params;sec:barsig}

Let us now examine in more detail the role of baryons in the
adiabatic scenario. The relevant quantity for the final
temperature fluctuations is, from Eqs.~(\ref{eq:OSW_term}) and
(\refp{eq:Doppler_term}) with $\Phi=\Psi$
 \be
 \begin{split}
 \dfrac{1}{4}D_{g, \comp{\ga}}  +  2\Phi - \imath \mu k V_\comp{\ga}
 =&
 \dfrac{1}{3}(1+R) \Phi \cos(c_s k \eta) - (2+R)\Phi \\
 & + 2\Phi
 - \imath \mu \frac{\sqrt{1+R}}{\sqrt{3}} \Phi \sin (c_s k \eta)
 \eqcomma
 \end{split}
 \ee
where we have inserted the adiabatic solution
(\ref{eq:HO_with_baryons_solution}--\refp{eq:HO_with_baryons_solution_V})
and explicitly restored the Doppler contribution. The effect of
baryons, $R > 0$, is twofold: the amplitude of the cosine
oscillation is larger and the zero point is now displaced to
$-R\Phi$, \ie the gravitational effects of falling into and
climbing out of the potential at decoupling no longer exactly
cancel as in \rr{eq:cancelling_potentials}, where we had taken
$R=0$. Therefore a larger baryon content enhances compression
peaks, which correspond to negative extrema of the
cosine\footnote{Note that $\Phi < 0$ inside potential wells, thus
$\cos(c_sk\eta) < 0$ indeed gives $D_{g, \comp{\ga}} > 0$,
according to \rrp{eq:HO_with_baryons_solution}, \ie it corresponds
to an overdensity with $\delta T / T > 0$.}, while it suppresses
expansion peaks. This leads to a distinctive signature of the
baryon density on the CMB spectrum: a larger baryon content boosts
odd peaks and reduces the even ones, hence a precise measurement
of the first three peaks leads to an accurate measurement of the
baryon content, as is evident from \FIGPAG{fig:kB}.

\begin{figure}[tb]
 \centering
 \includegraphics[width=\onefigwidth]{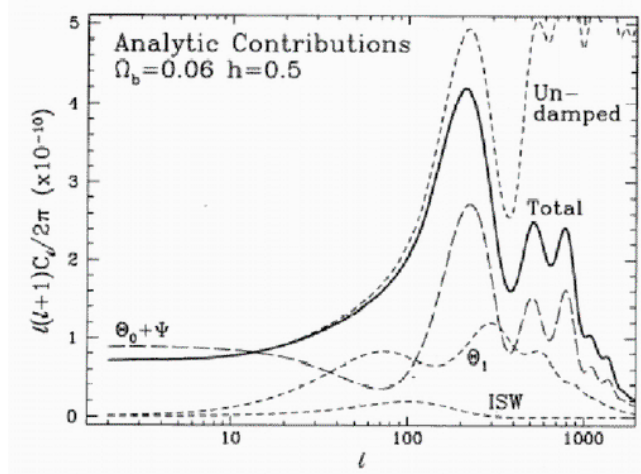}
 \caption[Individual contributions to the adiabatic temperature spectrum.]
 {Contributions to the adiabatic temperature spectrum (solid) from the temperature monopole
 (long-dashed), the temperature dipole (Doppler term, short dashed with label $\Theta_1$),
 and ISW effect \cite[reprinted from][]{Hu:1995uz}.
 \label{fig:reprinted_peaks_contrib}}
 \end{figure}

Up to now we have put aside the Doppler term $V_\comp{\ga} \propto
\sin(c_s k \eta)$: the sine is out of phase of $\pi/2$ with
respect to the density oscillation, and its maxima fill in the
zeros of the cosine. In the absence of baryons, this would lead to
an exact cancellation and to the disappearance of the acoustic
peaks: adding the density and velocity term incoherently in
quadrature for $R=0$ gives a constant. However, $R>0$ suppresses
the Doppler term by a factor $(1+R)$ (in quadrature) with respect
to the density term, and the net effect is that the velocity
contribution partially fills in the minima of the density
oscillation without erasing the peak structure, as shown in
\FIG{fig:reprinted_peaks_contrib}. Also the peak structure for the
velocity contribution gets more washed out by the free streaming
conversion than for the density, a consequence of the fact that
the velocity term is multiplied by $\mu$ \citep{Hu:1995uz}.

\subsubsection{Early ISW effect}
 \label{chap:params;sec:isw}

At recombination, the Universe is not completely matter dominated,
since $a_\dec \approx 3 a_\eq$ and thus the Bardeen potentials are
not exactly constant. This gives an early ISW contribution to the
anisotropy, which is spread out over a large multipole range,
adding in particular to the rise from the large scale plateau to
the first acoustic peak for the adiabatic scenario, \CF
\FIG{fig:reprinted_peaks_contrib}. Since most of the contribution
comes from early times, when $\eta \ll \eta_0$, we can
approximatively set $j_\ell(k(\eta_0 - \eta)) \approx j_\ell(k
\eta_0)$ and write for the ISW contribution to
(\refp{eq:Theta_ell_today})
 \be
 \Theta_\ell^{\text{(ISW)}} =
 \imath^\ell \int_{\eta_\dec}^{\eta_0}(\dot{\Psi} + \dot{\Phi})
 j_\ell(k(\eta_0 - \eta))
 \approx \imath^\ell \left[\dot{\Psi} +
 \dot{\Phi}\right]^{\eta_0}_{\eta_\dec} j_\ell(k \eta_0) \eqdot
 \ee

The early ISW is more prominent if the epoch of equality is
delayed due to a smaller matter content or to a larger radiation
content, for instance in the presence of extra relativistic
particles, as shown in \SEC{chap:beyondsp;sec:rel}.

\subsection{Damping tail}
\label{chap:params;sec:dampingtail}

\subsubsection{Recombination}

Temperature fluctuations on small angular scales are exponentially
suppressed by diffusion damping due to the breakdown of tight
coupling at recombination, as discussed in
\SEC{chap:cmb;sec:damping}. The effect can be roughly incorporated
into the undamped solution (\refp{eq:temperature_fluctuation}) by
multiplying it with the damping factor
 \be \label{eq:recombination_damping}
 \call{D}(k) \equiv \int \dr \eta g(\eta) e^{- \left[ k/k_\text{D}(\eta) \right]^2}
 \approx e^{- \left[ k/k_\text{D}(\eta_\dec) \right]^2}
 \eqcomma
 \ee
using the damping length scale $k_\text{D}^{-1}$ of
\rrp{eq:damping_scale}.

The main parameter dependence of the damping scale is easy to
understand physically: the matter content sets the horizon scale
at decoupling, while the baryon density controls the Compton
scattering time $\sim \taudot^{-1}$. Before recombination, photons
diffuse by a random walk over a typical length $\lambda_\text{D} =
\sqrt{N}/\taudot$, where $N$ is the number of collisions, $N \sim
\eta \taudot$. Hence the damping length scales as
 \be
 \lambda_\text{D} \sim \sqrt{\eta_\dec / \taudot} \propto
 \om_m^{-1/4} \om_b^{-1/2} \eqcomma
 \ee
where the last proportion takes advantage of the fact that $n_e
\propto \omega_b$ (see \rrp{eq:ombproptone}) and $\eta_\dec
\propto \om_m^{-1/2}$ if decoupling happens in the matter
dominated era. A more detailed estimate is given in
\rrp{eq:damping_scale_with_Yp}, which also includes the effect of
the helium fraction, which we have ignored here.

Clearly, when recombination occurs the mean free path goes to
infinity very rapidly, and therefore the above argument no longer
applies, and one has to use a more sophisticated analysis. More
details and precise fitting formulas for
(\ref{eq:recombination_damping}) can be found in \cite{Hu:1997mn},
while useful fitting formulas for many relevant recombination
quantities are detailed in \citealp[Appendix E]{Hu:1996en}.

\subsubsection{Reionization}
\label{chap:params:sec:reion}

When the Universe is reionized, the free electron fraction becomes
unity again and CMB photons can be rescattered. Fairly little is
known about the details of the reionization mechanism and its
redshift dependence \citep[for a review see][]{Haiman:2003he} but
the null detection of Gunn-Peterson troughs indicates that the
Universe was completely ionized after redshift $\approx 6$
\citep{Becker:2001ee}, possibly for the second time
\citep{Cen:2002zc}. The recent WMAP results \citep{Spergel:2003cb}
seem to indicate that reionization happened quite early, at a
redshift $z_\reion \approx 17$, corresponding to an optical depth
of $\tau_\reion \approx 0.16$ for a standard $\La$CDM model.

Reionization has two effects on the power spectrum: temperature
anisotropies on scales below the angle subtended by the horizon at
recombination get washed out, and on the same scale there is a
generation of polarized power. Let us take for simplicity a model
in which all the hydrogen is suddenly reionized at a redshift
$z_\reion$, and ignore helium reionization which happens around $z
\approx 3$ which only contributes a few percent. Then the
corresponding optical depth to reionization, $\tau_\reion$, is
given by
\be
 \begin{aligned}
 \label{eq:taureionvszreion}
 \tau_\reion & = \int_{t_0}^{t_\reion} c \si_T n_e  \dr t \\
 & = \dfrac{c \si_T}{H_0} \int_0^{z_\reion} \dfrac{n_e(z)}{(1+z)}
 \dfrac{\dr z} {\left[ \Om_r (1+z)^4  + \Om_m (1+z)^3 + \Om_\curv (1+z)^2
 + \Om_\La \right]^{1/2}} \eqdot
 \end{aligned}
 \ee
The free electron density (per cm$^3$) can be expressed as (see
 \rrp{eq:ombproptone})
 \be
 n_e(z) = 11.3\cdot10^{-6}(1- Y_p)\om_b(1+z)^3\eqcomma
 \ee
 where we have included the Helium mass fraction $Y_p$ for future
reference (see \SEC{chap:bspII;sec:cmb}). For a flat Universe
($\Om_\curv = 0$) and neglecting the contribution of radiation,
which is a good approximation if $z_\reion \ll 100$, the integral
in \eqref{eq:taureionvszreion} can be performed analytically,
giving \citep{Hu:1997mn}
 \be \label{eq:tau_reion_approx}
 \tau_\reion =
  4.6 \cdot 10^{-2} (1 - Y_p) \frac{\Om_b h}{\Om_m}
\left[ \sqrt{\Om_\La + \Om_m(1+z_\reion)^3} - 1 \right] \eqdot
 \ee

From the definition of the visibility function $g$, the
probability that a photon last scattered between today and
redshift $z$ is
 \be
 P(z) = \int_0^{z} g(\tilde{z}) \dr \tilde{z} = 1 - e^{-\tau(z)}
 \eqcomma
 \ee
and therefore the fraction of photons which arrive to us directly
from the recombination epoch is $1 - P(z_\reion) =
\exp(-\tau_\reion)$. Above the horizon scale at reionization, all
photons contribute to the anisotropy, while below that scale only
the fraction $\exp(-\tau_\reion)$ which did not rescatter
contribute. Thus power on small scales will be suppressed by a
factor $\exp(-2\tau_\reion)$ and the reionization damping factor
is given by
 \be
 \call{D}_\reion(k) = \left\{
 \begin{array}{l}
 \begin{aligned}
 & 1 & \quad \text{for } k \tau_\reion \ll 1 \\
 & e^{-2 \tau_\reion} & \quad \text{for } k \tau_\reion \gg 1
 \end{aligned}
 \end{array}
 \right.
 \eqdot
 \ee

The angular scale subtended by the horizon at reionization can be
found using (\ref{eq:angular_diameter_distance}), yielding the
approximate scaling \citep{Tegmark:1995dy}
 \be \label{eq:reion_bump_position}
 \vartheta  \propto \sqrt{\frac{\Om_m}{z}}
 \eqdot
 \ee

 Without polarization information, reionization is highly
degenerate with the spectral tilt and a tensor or isocurvature
contribution which would add power only on large scales: a larger
reionization optical depth can easily be accommodated by adding
tensors or an isocurvature component an reducing at the same time
the overall normalization, thereby exactly compensating the
reionization power suppression. This degeneracy can be expressed
by introducing a suitable combination of $\tau_\reion$ and the
overall normalization, see \rr{eq:def_rZ} and compare
\FIG{fig:rZ}. However, the characteristic signature of
reionization is the generation of polarized power on the horizon
scale of reionization, and the corresponding ``polarization
bump'', clearly visible in the bottom right panel of
\FIGPAG{fig:alpha_peaks}, around $\ell \approx 20$ in the
E-polarization spectrum can be used to break the degeneracies with
other parameters.

The position and scaling of this bump can easily be understood
physically \citep{Zaldarriaga:1997ke}: the temperature quadrupole
at reionization, which determines the reionization induced
polarization, is given by the free stream of the temperature
monopole at decoupling:
 \be
 \Theta_2 (\eta_\reion) = (\Theta_0 + 2 \Phi)(\eta_\dec)
 j_2 \left( k(\eta_\reion - \eta_\dec) \right) \eqdot
 \ee
Given that the $k$-oscillation of the monopole is much slower than
the one of the Bessel function, $r_s \ll \eta_\reion - \eta_\dec$,
the first peak corresponds approximately to the maximum of the
Bessel function, which occurs for $k \approx 2/(\eta_\reion -
\eta_\dec)$. This translates into $\ell \approx k (\eta_0 -
\eta_\reion) \approx 2 (\eta_0 - \eta_\reion)/(\eta_\reion -
\eta_\dec) \approx 2 \sqrt{z_\reion}$. This peculiar scaling of
the position of the reionization bump in the E-spectrum could
potentially be used to distinguish the effect of a possible time
variation of the fine-structure constant, see
\SEC{chap:bspIII;sec:alphareion}.

Only one parameter is sufficient to characterize the simple model
of sudden reionization presented above, namely the reionization
redshift $z_\reion$ or equivalently $\tau_\reion$; but it has been
shown that there are up to five principal reionization modes which
could be extracted from CMB measurements \citep{Hu:2003gh}.
Furthermore, it is possible to link the reionization history to
specific stellar models and try to constrain the parameters of
star formation and evolution modelling using CMB data
\citep{Bruscoli:2002kv,Holder:2003eb,Kaplinghat:2002vt}.

\section{Normal parameters}
\label{chap:params:sec:normal}

The physical understanding of the characteristic signature of the
cosmological parameters can be exploited to build a set of
analytical functions which describe quantities directly probed by
the CMB. We call such a set a ``normal parameter basis'', because
the effect of the new parameters is almost orthogonal, in the
sense that correlations among the parameters should be very small.
The normal parameter set has the advantage of taking into account
the most severe CMB degeneracies, such as the geometrical
degeneracy described above, a feature which improves the
efficiency of parameter space exploration (see
\SEC{chap:data;sec:mcmc}). The dependence of the CMB spectrum on
the normal parameters is almost linear over a wide range of
values, a very important property which makes them ideal as a
basis set for the Fisher matrix analysis, see the explanations in
\SEC{chap:data;sec:fma} and \SEC{chap:bspII;sec:fma} for an
application. In terms of the normal parameters, it is easy to
disentangle and understand the physical effects on the CMB power
spectra of each parameter while keeping the other constant, to the
contrary of what happens for cosmological parameters.

We have seen in \SEC{chap:params;sec:acoustic} that the shift
parameter $\Rshift$, the baryon and matter density determine the
location and relative height of the acoustic peaks. We now expand
those considerations by introducing a normal parameter set, based
on the discussion of \cite{Kosowsky:2002zt}, to which the reader
is referred for further details. See also \cite{Sandvik:2003ii}
for an application to parameter estimation techniques and
\cite{Jimenez:2004ct} for recent improvements including the
polarization spectrum.
\begin{itemize}
 \item The position of the peaks is set by the ratio between the
angular diameter distance relation
(\refp{eq:angular_diameter_distance}) and the physical size of the
acoustic horizon at decoupling, \rrp{eq:sound_horizon}. Hence a
first normal parameter which determines the overall angular scale
is
 \be
 \kA \equiv \dfrac{D_A(a_\dec)}{a_\dec r_s(a_\dec)} \eqcomma
 \label{eq:def_kA}
 \ee
\CF \rr{eq:ad_peak_position}, which is just a general expression
for the shift parameter. The scale factor at decoupling $a_\dec$,
or equivalently the redshift of decoupling, depends upon $\Om_b
h^2$ and the $\Om_m/\Om_r$, for which \cite{Hu:1996en} provide an
accurate analytical fitting formula. The effect of a change in
$\kA$ while keeping the other normal parameters fixed is displayed
in \FIG{fig:kA}.
 \begin{figure}[tb]
\centering
\includegraphics[width=\twofigswidth]{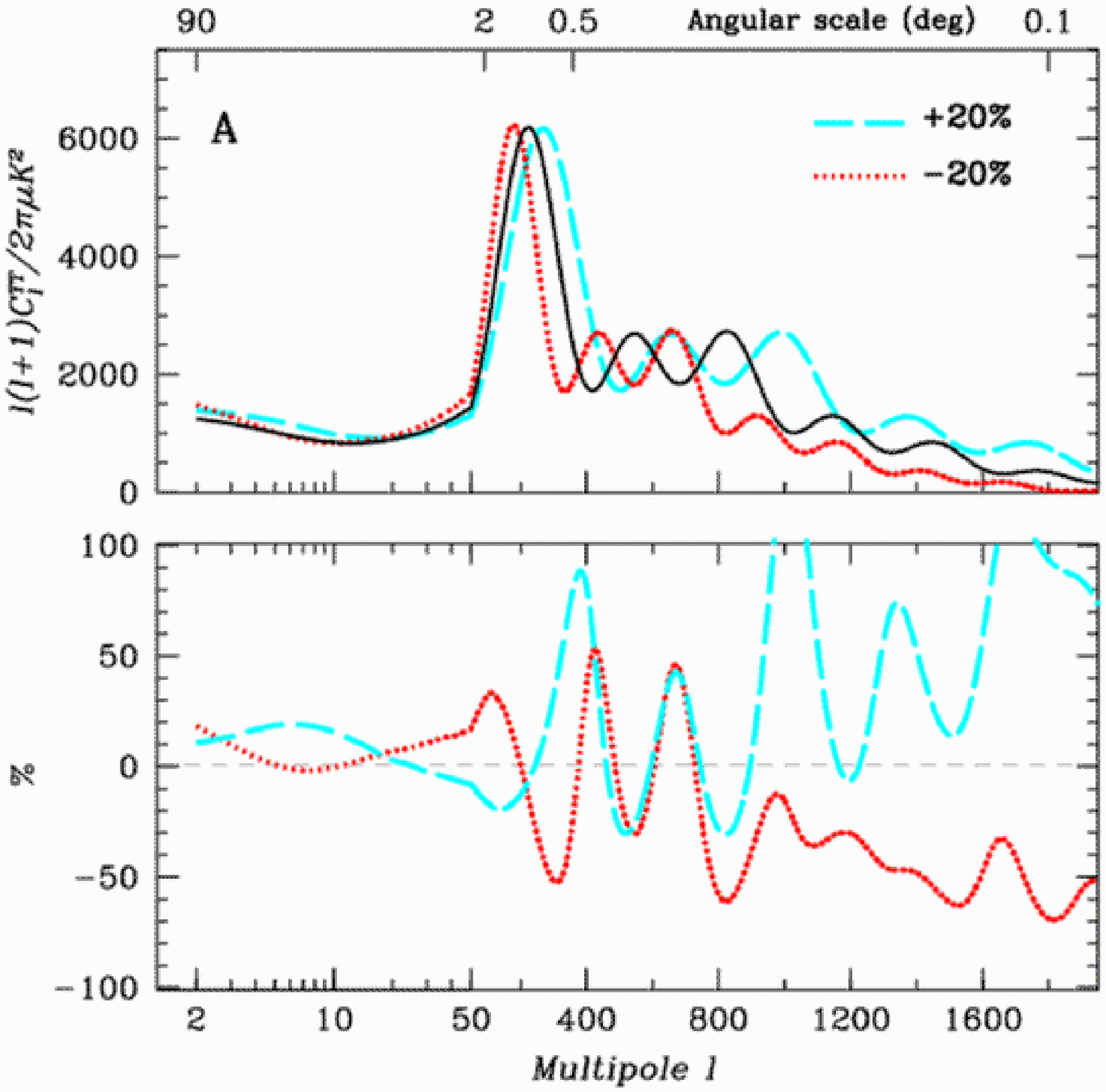}\hfill%
\includegraphics[width=\twofigswidth]{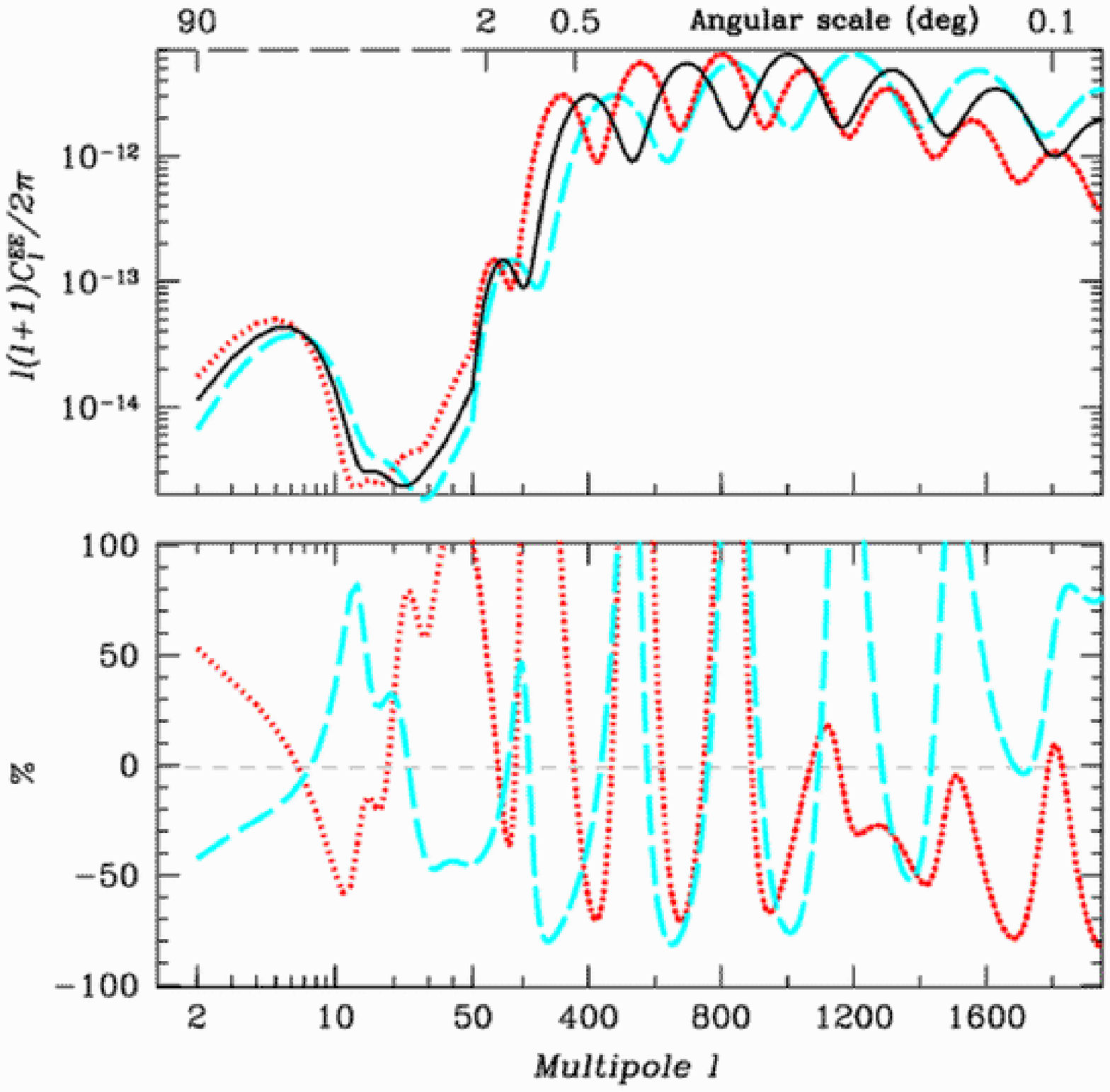}\hfill%
\caption[Impact of the shift parameter on the temperature and
polarization spectra.]{Impact of the shift parameter
(\ref{eq:def_kA}) on the CMB temperature (left) and polarization
(right) spectra, all other normal parameters kept fixed. The
geometrical projection effect affects temperature and polarization
in the same way. In the bottom panel, we plot the percent
difference with respect to the reference model (black).
\label{fig:kA}}
\end{figure}
 \item The
radiation/matter ratio sets the epoch of equality, which in turn
determines the amount of early ISW, thus we introduce the
parameter
 \be
 \kR \equiv \dfrac{\Om_m}{\Om_r}\dfrac{a_\dec}{a_0} \eqcomma \label{eq:def_kR}
 \ee
which gives the matter to radiation density ratio at the time of
decoupling. The boost of the first acoustic peak due to the early
ISW is visible in \FIG{fig:kR}.
 \begin{figure}[!tb]
\centering
\includegraphics[width=\twofigswidth]{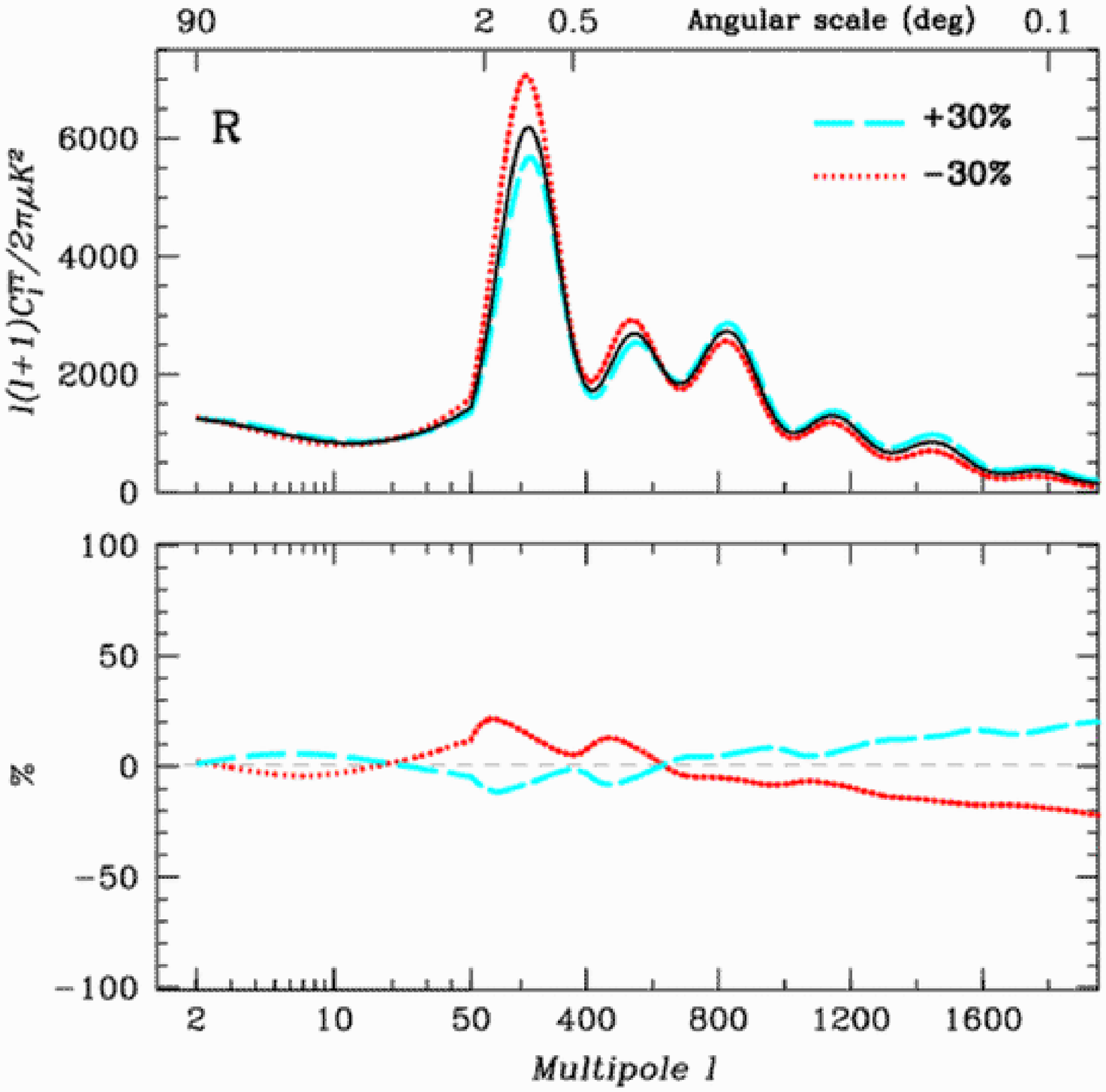}\hfill%
\includegraphics[width=\twofigswidth]{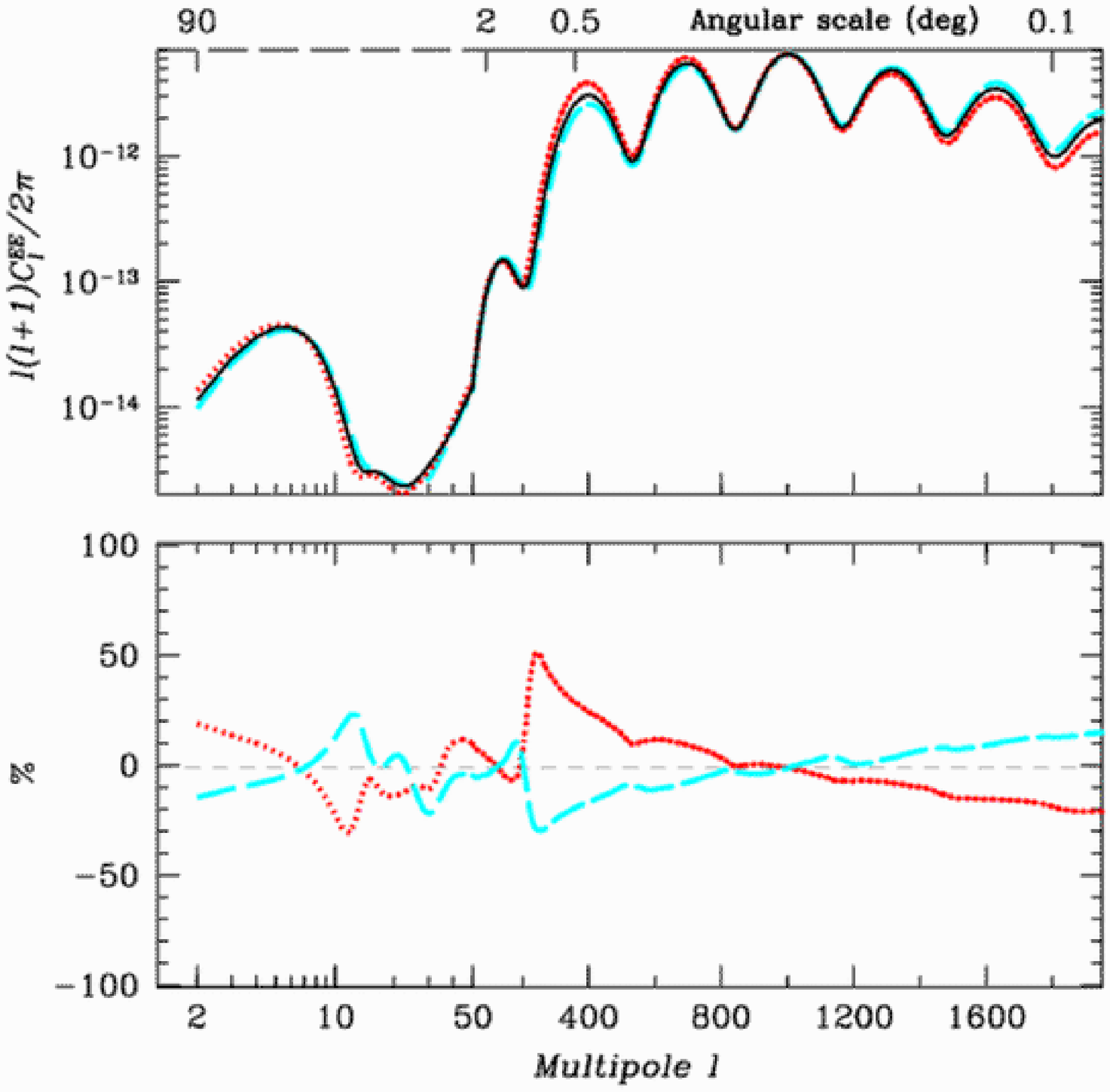}\hfill%
\caption[Impact of a change in the epoch of equality on the
temperature and polarization spectra.]{Impact of a change in the
radiation to matter energy density ratio at decoupling
(\ref{eq:def_kR}) on the temperature (left) and polarization
(right) spectra, all other normal parameters kept fixed. This can
more easily be interpreted as a shift in the epoch of
matter-radiation equality, which changes the amount of early ISW
effect contribution around the first acoustic peak.
\label{fig:kR}}
\end{figure}
 \item The geometrical degeneracy is along the energy density in
 the cosmological constant, which also gives the amount of late
 ISW effect. Thus we use the parameter
  \be
  \kV \equiv \Om_\La h^2 \eqdot \label{eq:def_kV}
  \ee
As shown in \FIG{fig:kV}, the impact is quite small in magnitude
and solely on large angular scales, where cosmic variance limits
our ability to constrain this parameter, making of the
cosmological constant one of the worst determinable parameters
with CMB data alone.
 \begin{figure}[!tb]
\centering
\includegraphics[width=\twofigswidth]{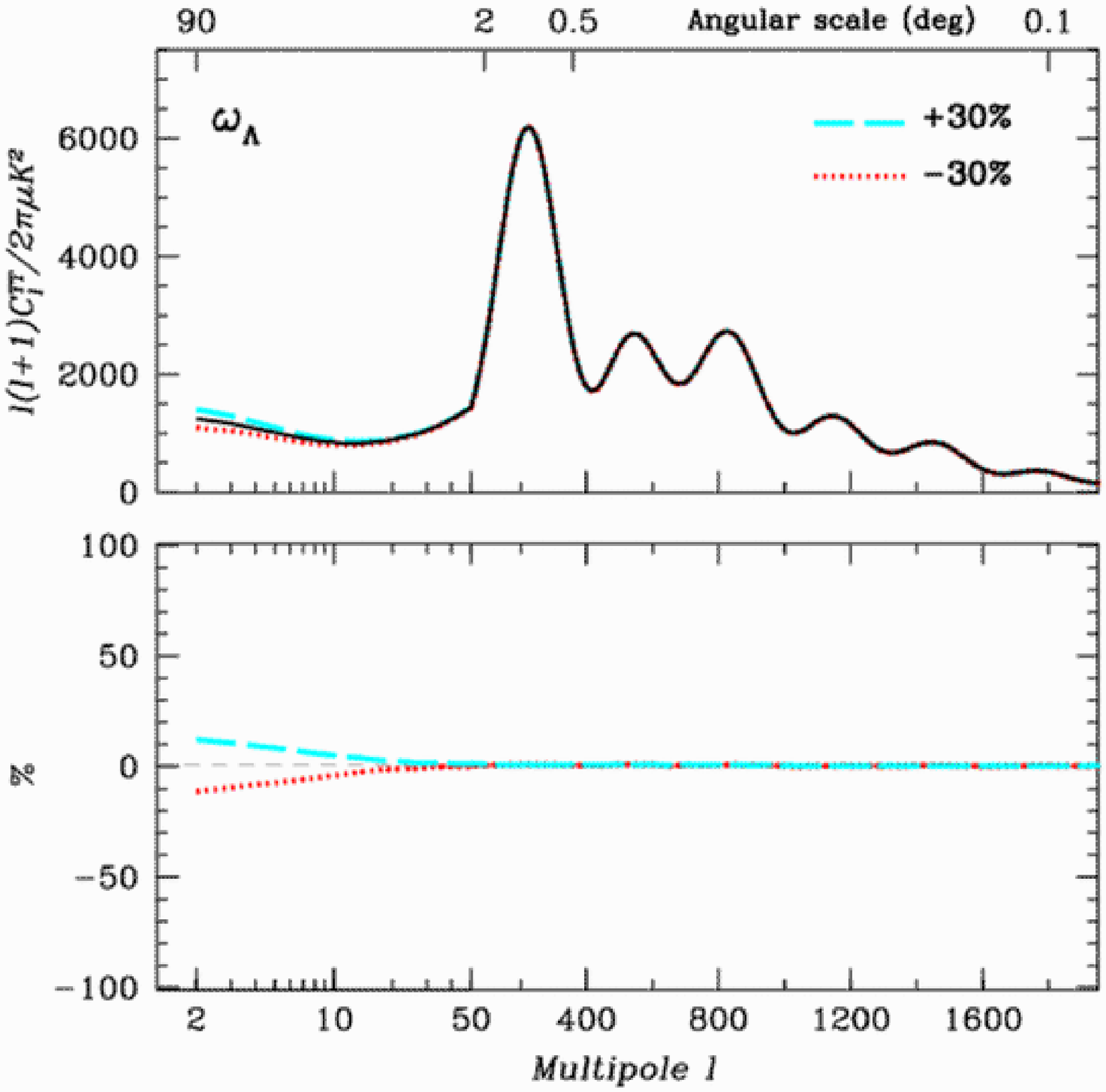}\hfill%
\includegraphics[width=\twofigswidth]{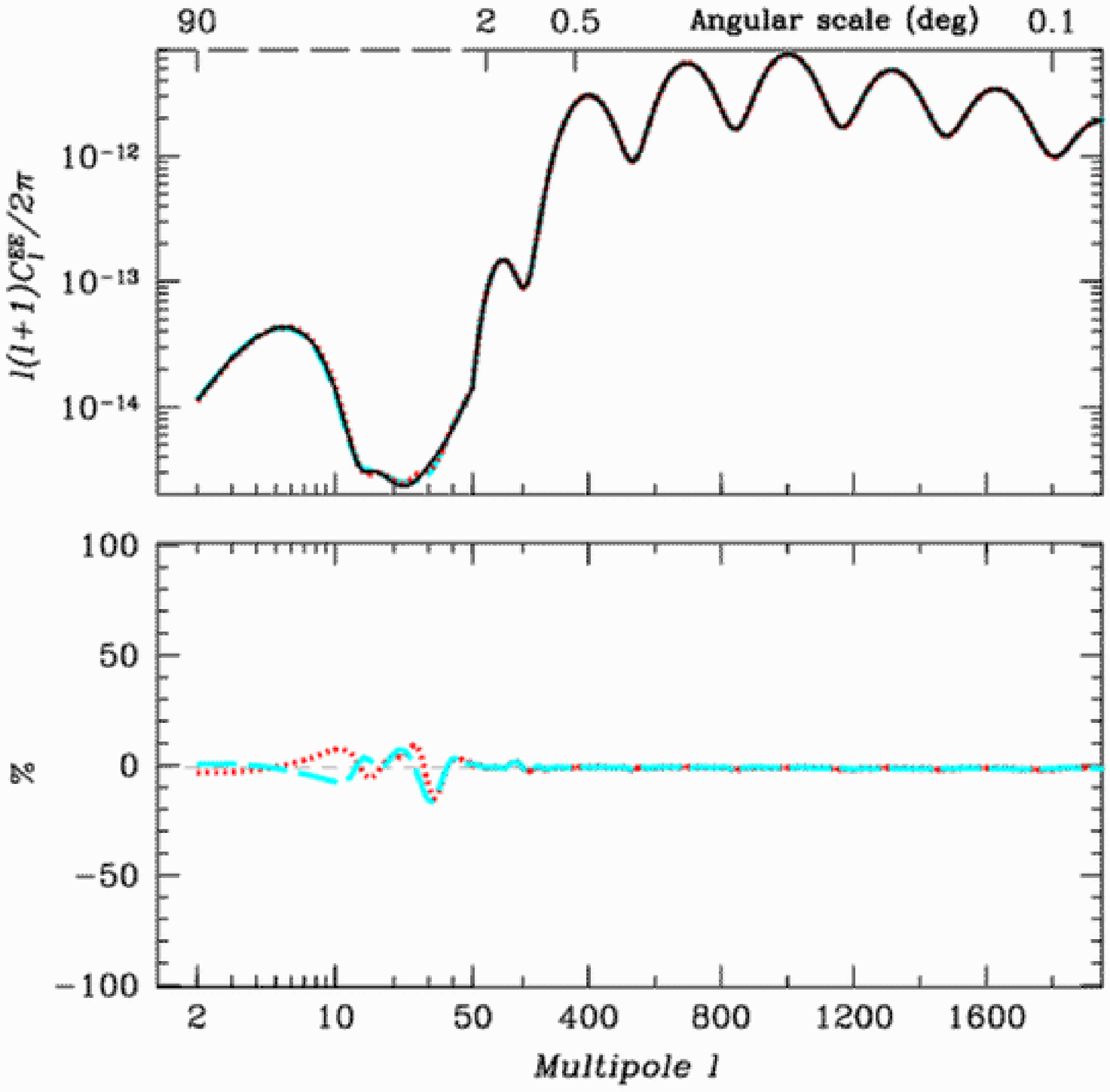}\hfill%
\caption[Impact of the energy density in the cosmological constant
on the CMB temperature and polarization spectra.]{Impact of the
energy density in the cosmological constant (\ref{eq:def_kV}) on
the CMB temperature (left) and polarization (right) spectra, all
other normal parameters kept fixed.  The impact is only on large
angular scales due to the late ISW effect, where measurements are
limited by cosmic variance and therefore cannot constraint much
this parameter.\label{fig:kV}}
\end{figure}
 \item The parameter $\kA$ already includes the effect of
the baryon density on the spacing and location of the peaks, which
is produced by the dependence of the sound horizon on the baryon
content. Therefore keeping the other normal parameters and in
particular $\kA$ fixed while varying
  \be
  \kB \equiv \Om_b h^2  \label{eq:def_kB}
  \ee
 isolates the baryon driving
effect on the acoustic oscillations, which sets the relative
height of the peaks. Since the polarization amplitude is
proportional to the temperature dipole at recombination, which in
turn is suppressed by a factor $(1+R)^{1/2}$ with $R \propto \Om_b
h^2$, a larger baryon density reduces the height of polarization
peaks (\FIG{fig:kB}).
\begin{figure}[tb]
\centering
\includegraphics[width=\twofigswidth]{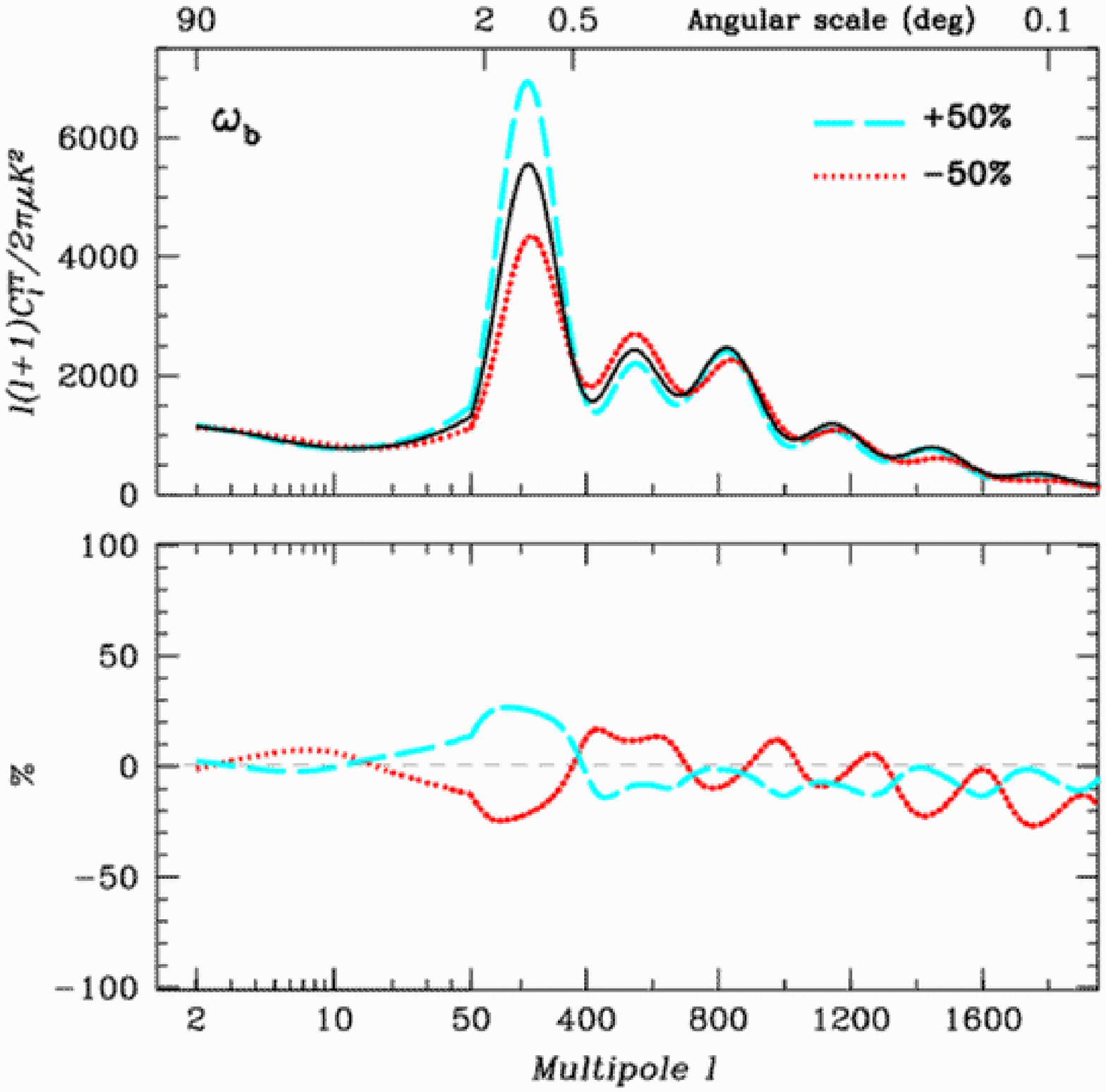}\hfill%
\includegraphics[width=\twofigswidth]{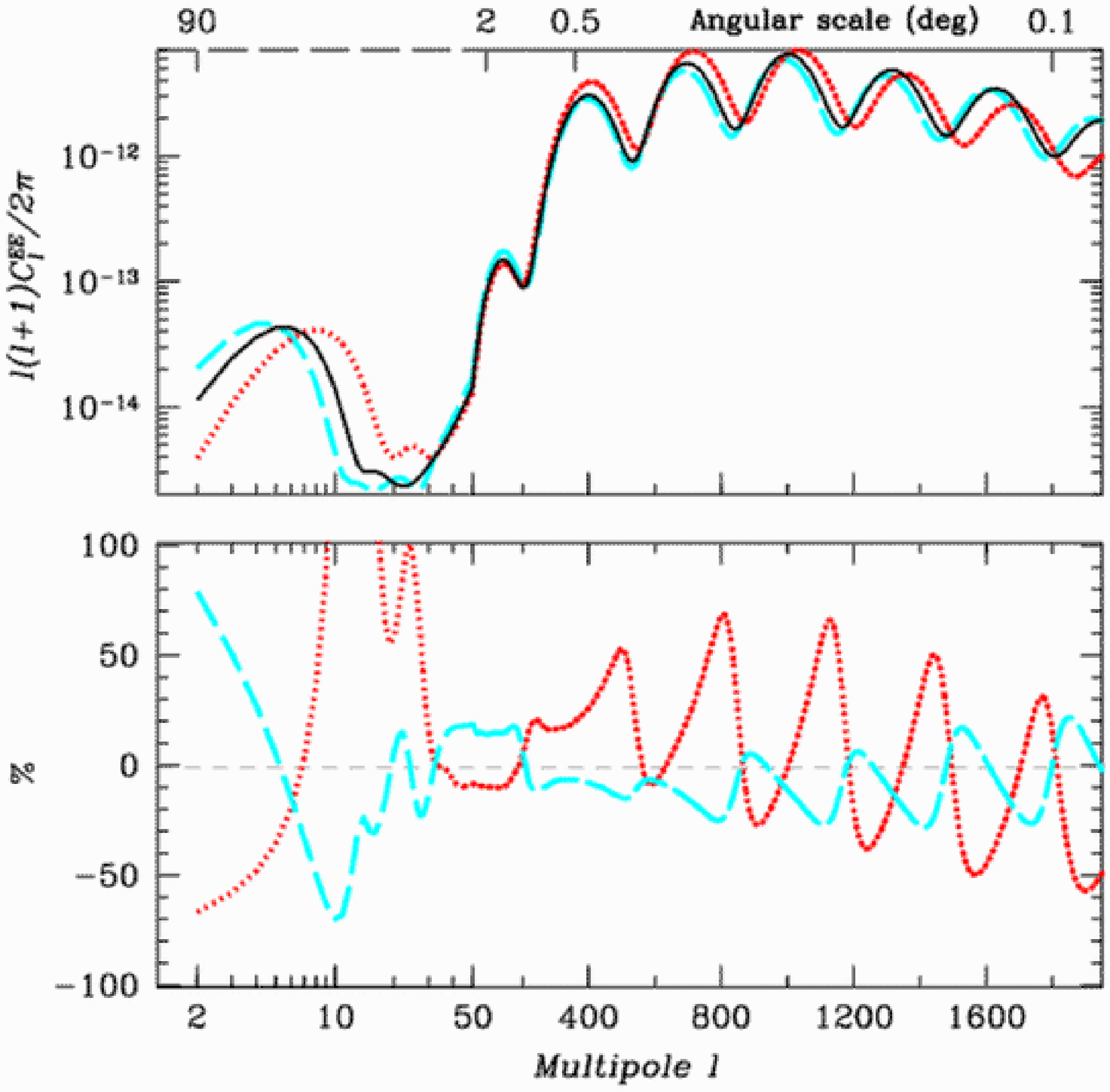}\hfill%
\caption[Impact of the baryon density on the temperature and
polarization spectra.]{Impact of the baryon density
(\ref{eq:def_kB}) on the CMB temperature (left) and polarization
(right) spectra, all other normal parameters kept fixed. A larger
baryon content boosts odd peaks and suppresses even ones, see
\SEC{chap:params;sec:barsig}. The height of the polarization peaks
is reduced by a larger baryon content. \label{fig:kB}}
\end{figure}
 \item The CMB spectrum turns out to be almost linear in the
 combination
   \be
   \kM \equiv \Om_m h^2 \left(1 + \dfrac{\Om_r^2}{a_\dec^2 \Om_m^2}
   \right)^{1/2} = \Om_m h^2 \left(1 + \dfrac{1}{\kR}
   \right)^{1/2}\eqcomma \label{eq:def_kM}
 \ee
 which is a refinement of our previous approach of taking simply
 $\Om_m h^2$ as a determining parameter, see
 \cite{Kosowsky:2002zt} for more details.
  \item A good way of taking into account the degeneracy between
  the optical depth to reionization and the scalar normalization
  described in \SEC{chap:params:sec:reion} is to adopt the
  parameter
   \be
   \rZ \equiv A_s \exp(-2\tau_\reion) \eqcomma \label{eq:def_rZ}
   \ee
where for the adiabatic model considered here $A_s \equiv
\zeta_0^2$ is the scalar amplitude of the power spectrum of the
gauge invariant curvature perturbation, \CF \rrp{eq:PS_for_zeta}.
When adopting a change in $\tau_\reion$, the normalization $A_s$
is also changed as to keep the power above the third peak
unchanged, thus avoiding artificial degeneracies with the other
normal parameters, which would disappear if one adopted a
different normalization convention \citep{Kosowsky:2002zt}, see
\FIG{fig:rZ}.
 \begin{figure}[tb]
\centering
\includegraphics[width=\twofigswidth]{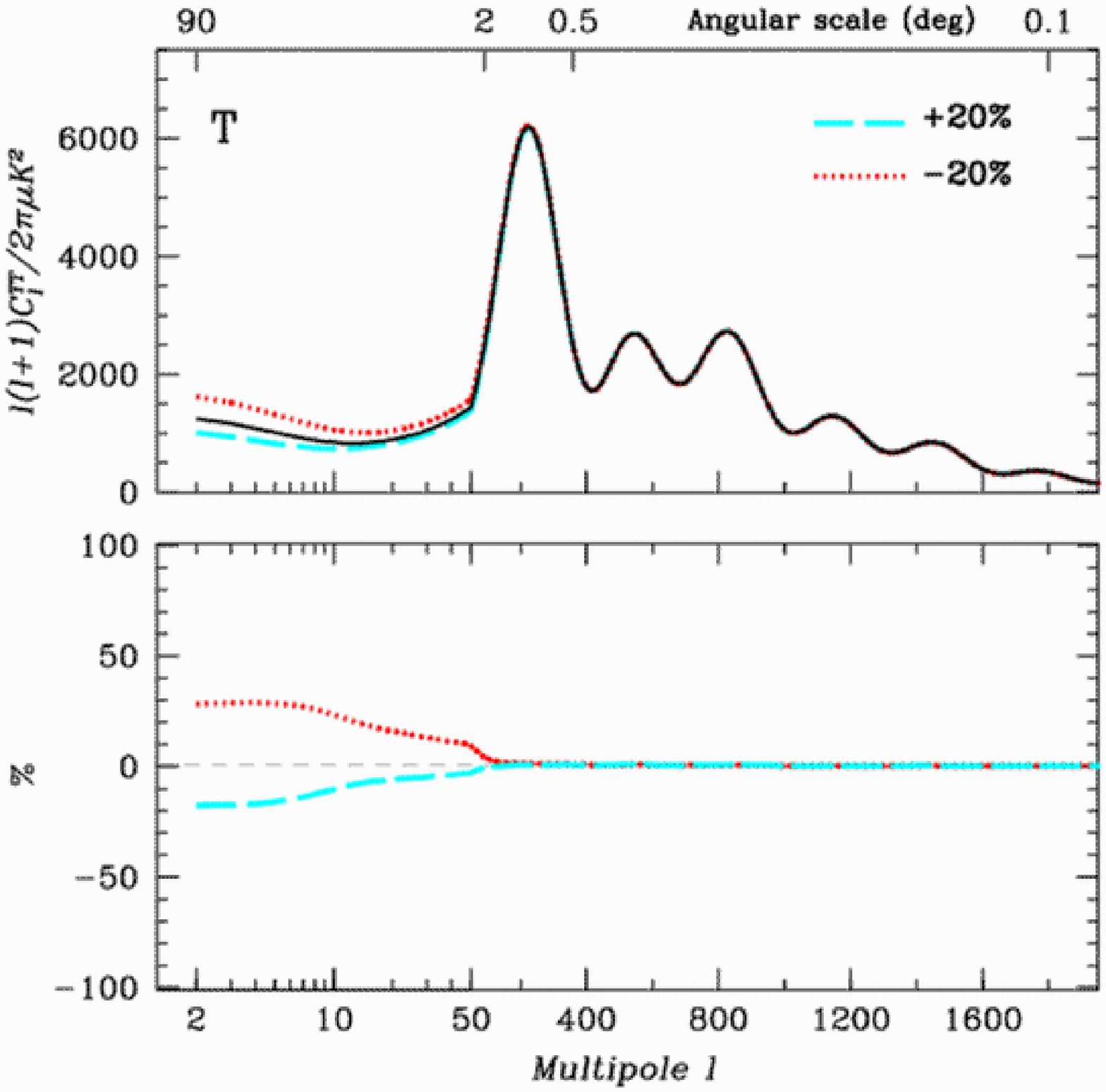}\hfill%
\includegraphics[width=\twofigswidth]{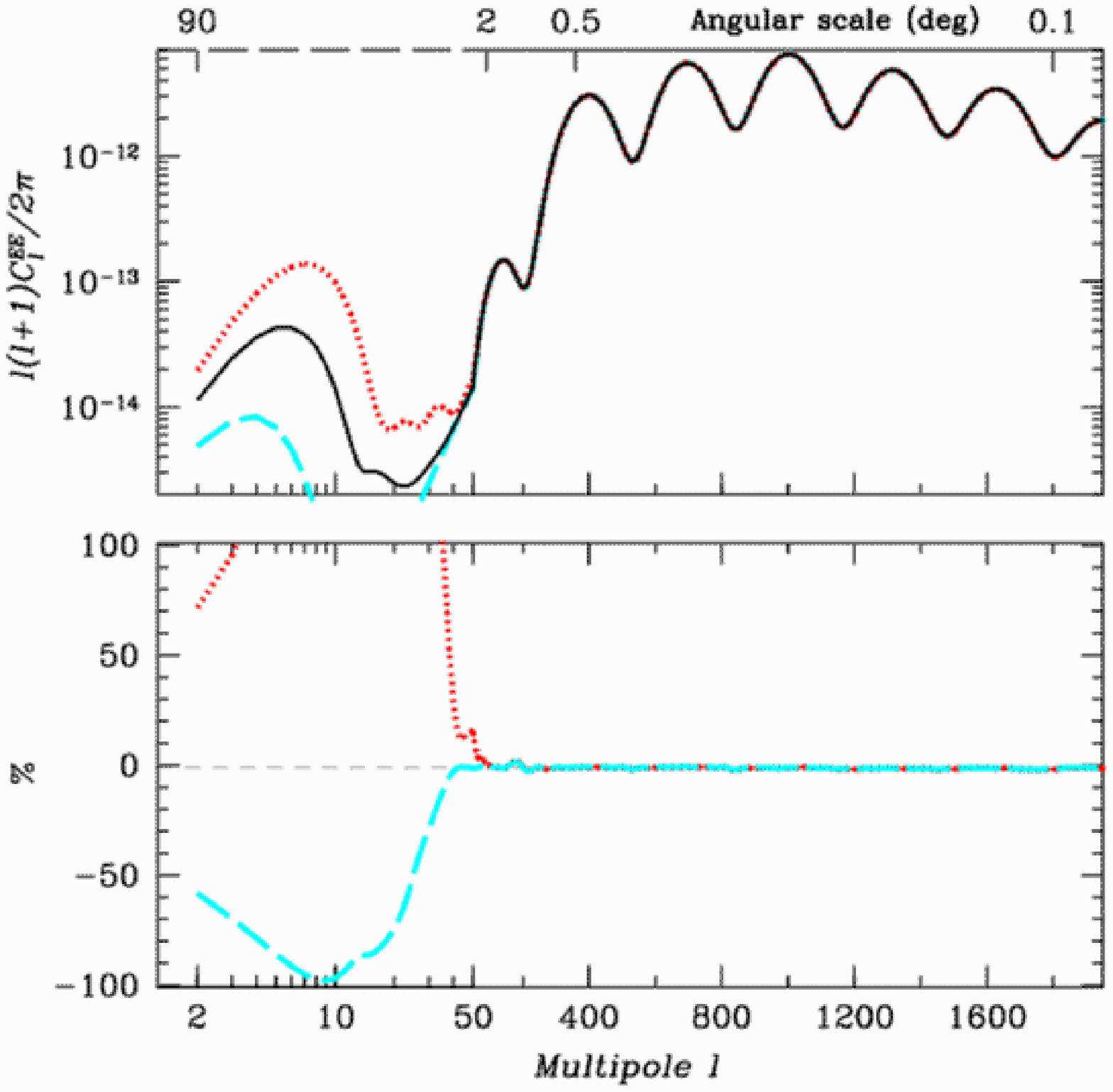}\hfill%
\caption[Impact of a degenerate combination of the normalization
and the reionization optical depth on the temperature and
polarization spectra.]{Impact of the parameter $\rZ$ defined in
(\ref{eq:def_rZ}) on the CMB temperature (left) and polarization
(right) spectra, all other normal parameters kept fixed.
Increasing $\tau_\reion$ and the overall normalization at the same
time as to keep the power above the third peak unchanged reveals
the degeneracy between normalization and reionization. The only
measurable effect is at large scales, where the temperature signal
is {\it enhanced} for smaller $\rZ$ (and hence larger
$\tau_\reion$) as well as the reionization bump in the
polarization spectrum. \label{fig:rZ}}
\end{figure}
  \item The scale dependence of the initial power spectrum is
 described by the scalar spectral index $n_s$, as in
(\ref{eq:PS_for_zeta}). A value $n_s > 1$ (``blue index'')
increases the power for wavevectors larger than the pivot scale,
and thus yields more power for large multipoles; the converse is
true for $n_s < 1$ (``red index''), see \FIG{fig:ns}. Therefore
the impact on the CMB spectrum can be approximately modelled as
 \be
 C_{\ell T,E} (n_s) \approx C_{\ell T,E} (n_s = 1)
 \left(\dfrac{\ell}{\ell_0}\right)^{n_s - 1}
 \ee
 with $\ell_0$ a pivot point which should be chosen as to match
 $k_P$ (even though a different choice will only correspond to a
 change in overall normalization).
\end{itemize}
 \begin{figure}[!tb]
\centering
\includegraphics[width=\twofigswidth]{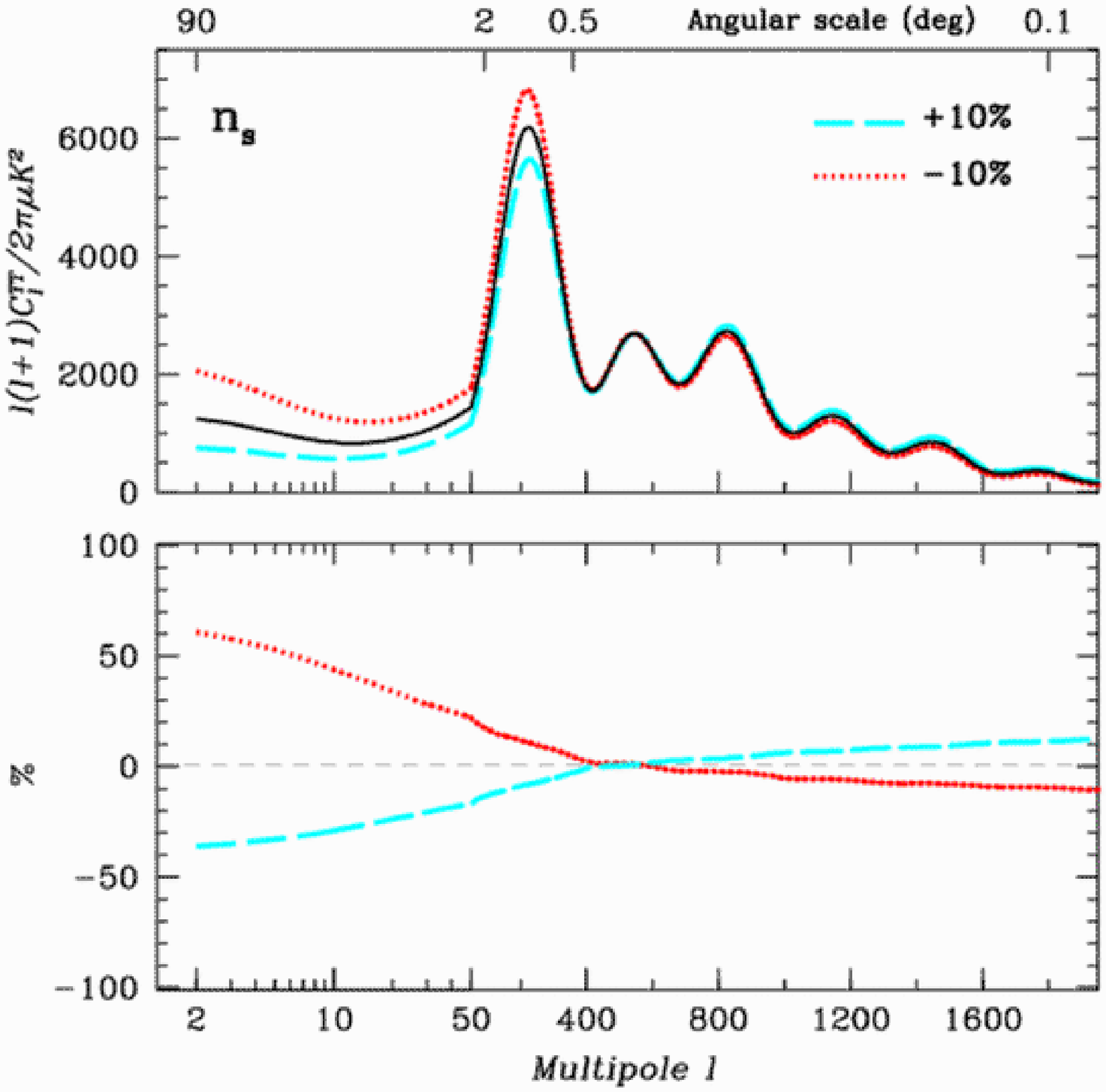}\hfill%
\includegraphics[width=\twofigswidth]{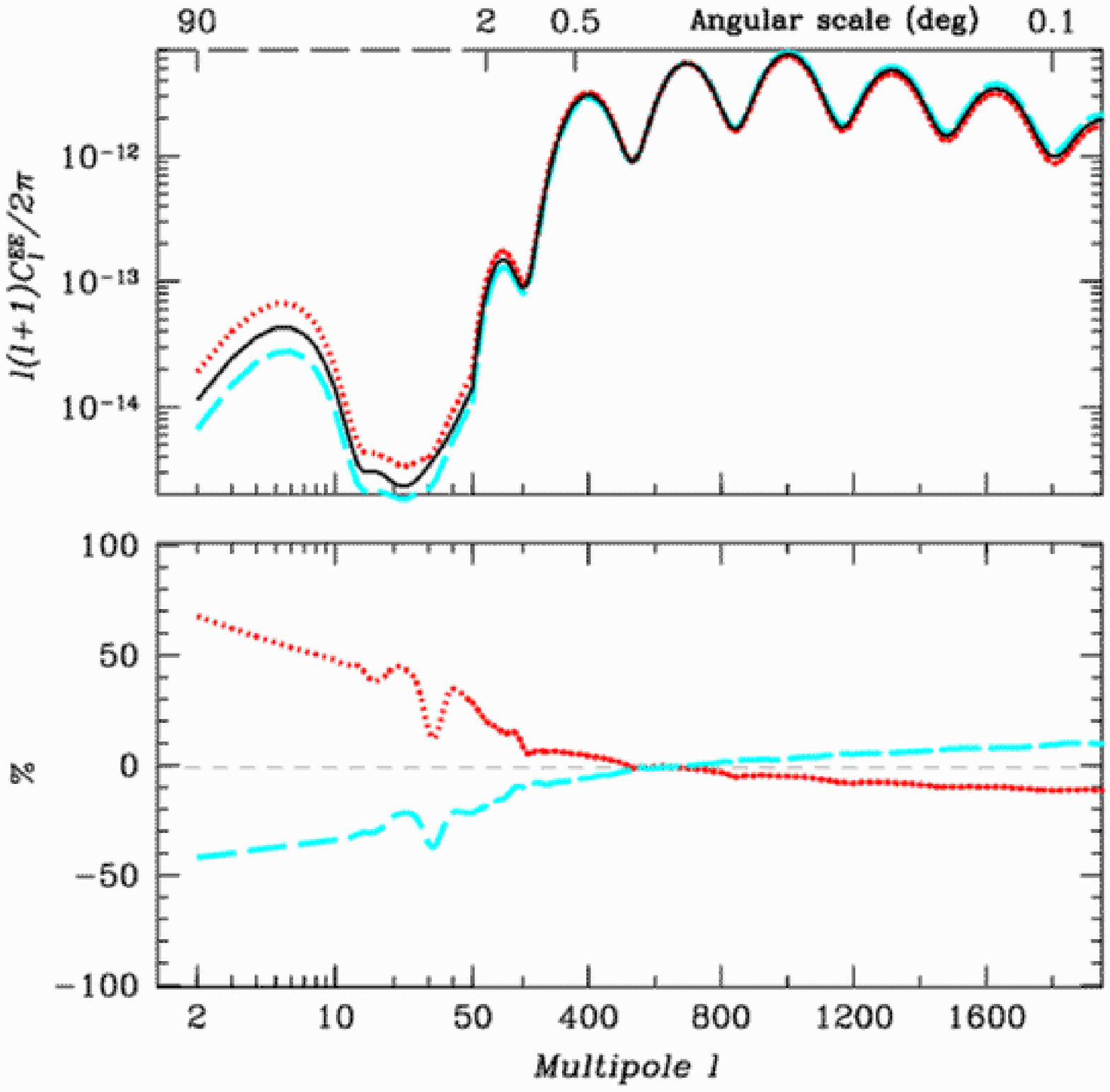}\hfill%
\caption[Impact of the scalar spectral index on the temperature
and polarization spectra.]{Impact of the scalar spectral index on
the CMB temperature (left) and polarization (right) spectra, all
other normal parameters kept fixed. A blue spectrum ($n_s > 1$)
gives more power at larger multipoles. The glitches are numerical
artifacts. \label{fig:ns}}
\end{figure}

\enlargethispage{\baselineskip}

 Given the above correspondences,
we can transform from the cosmological parameter set $(\Om_m,
\Om_\La, \Om_b, \Om_r, h)$ into the normal basis $(\kA, \kR, \kV,
\kB, \kM$) and vice-versa by numerically inverting the relations
(\ref{eq:def_kA}--\ref{eq:def_kM}).

\clearpage

\section{General initial conditions}
\label{chap:params;sec:ic}

As we have seen in \SEC{chap:cmb;sec:matter_radiation} and
\SEC{chap:cmb;sec:neutrinos}, a Universe containing photons,
massless neutrinos, cold dark matter and photons coupled to
baryons admits four growing modes for the perturbations. To this
set, one should add a baryon isocurvature entropy mode, which we
have not described, but which behaves exactly as the cold dark
matter mode, only rescaled by an overall constant
$\Om_\comp{b}/\Om_\comp{\cdm}$ \citep{Gordon:2002gv}. Thus without
loss of generality, we can treat the CDM and baryon isocurvature
modes as one single mode, and restrict our considerations to the
four modes: adiabatic, CDM isocurvature, neutrino entropy and
neutrino velocity.

\subsection{Angular power spectra for all modes}

\begin{figure}[!tb]
\centering
\includegraphics[width=\twofigswidth]{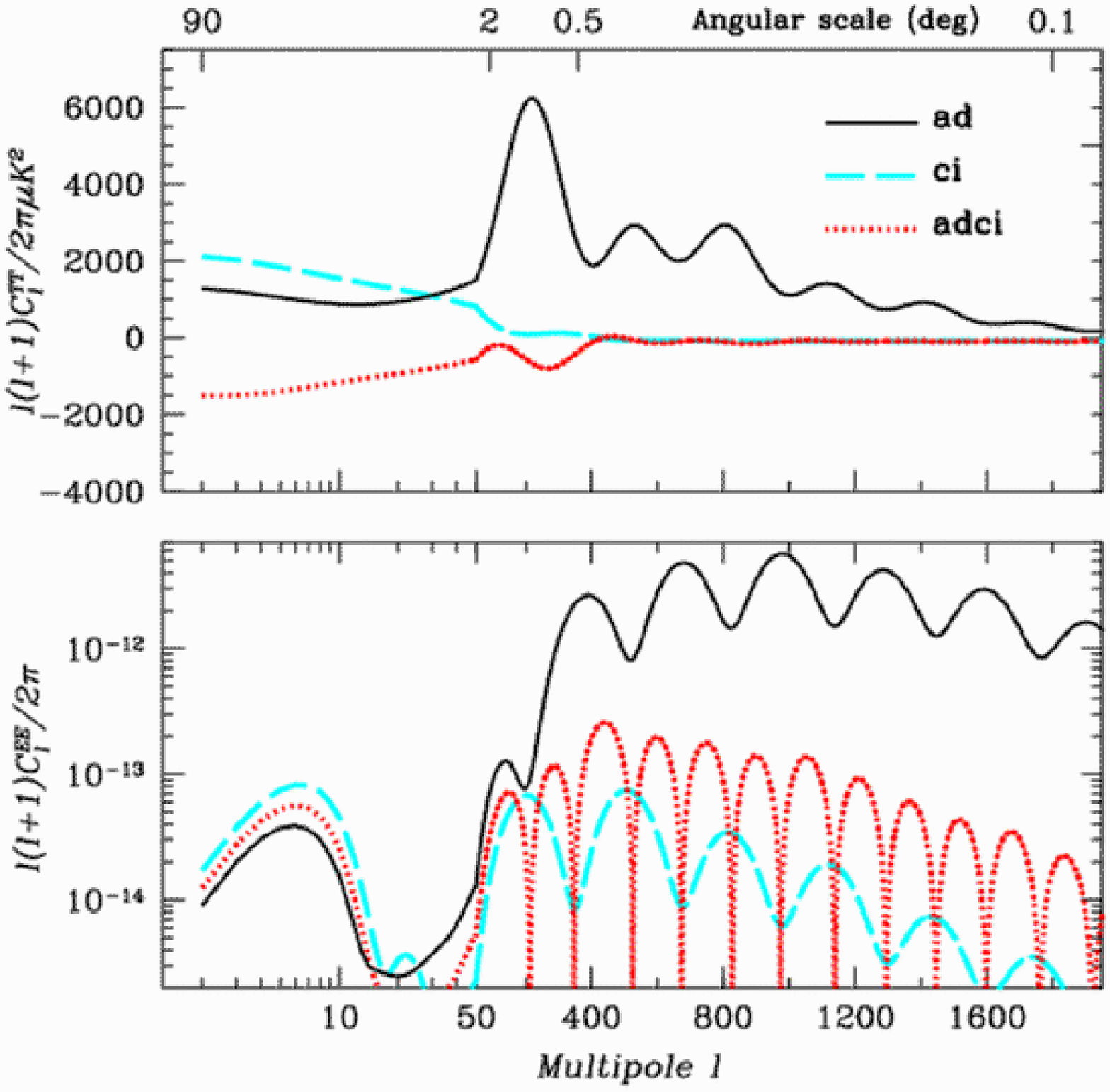}\hfill%
\includegraphics[width=\twofigswidth]{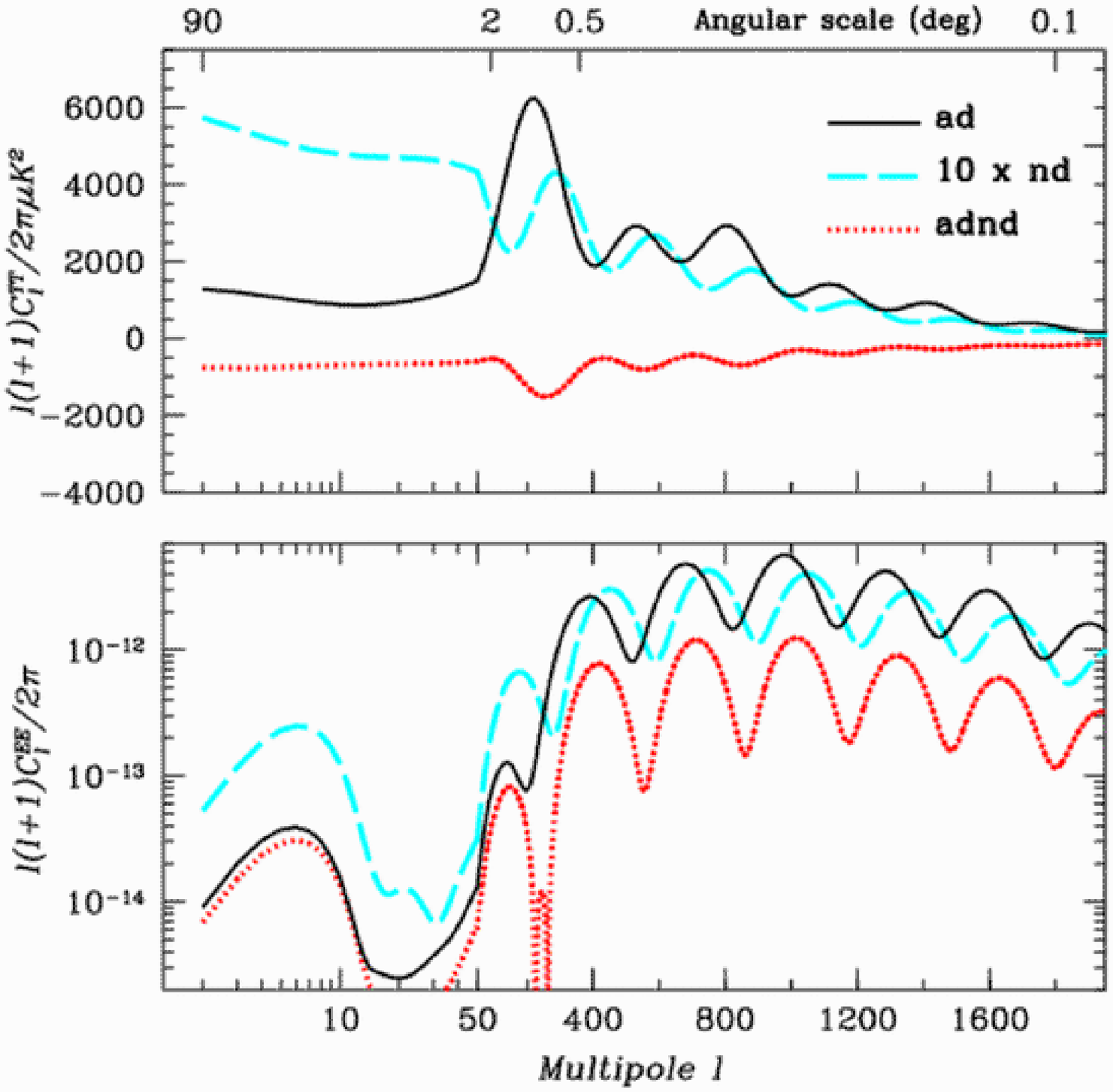}
\includegraphics[width=\twofigswidth]{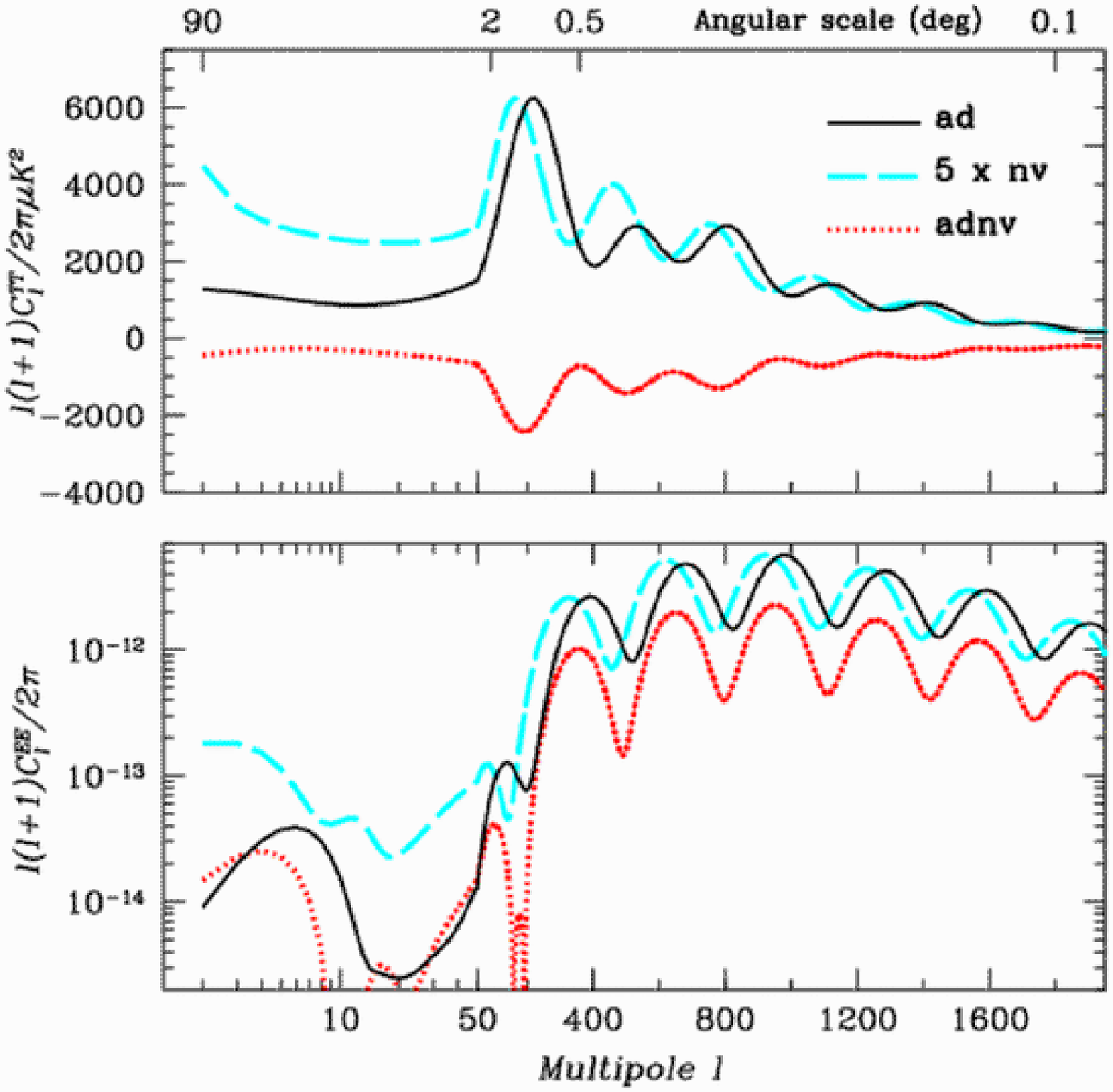}\hfill%
\includegraphics[width=\twofigswidth]{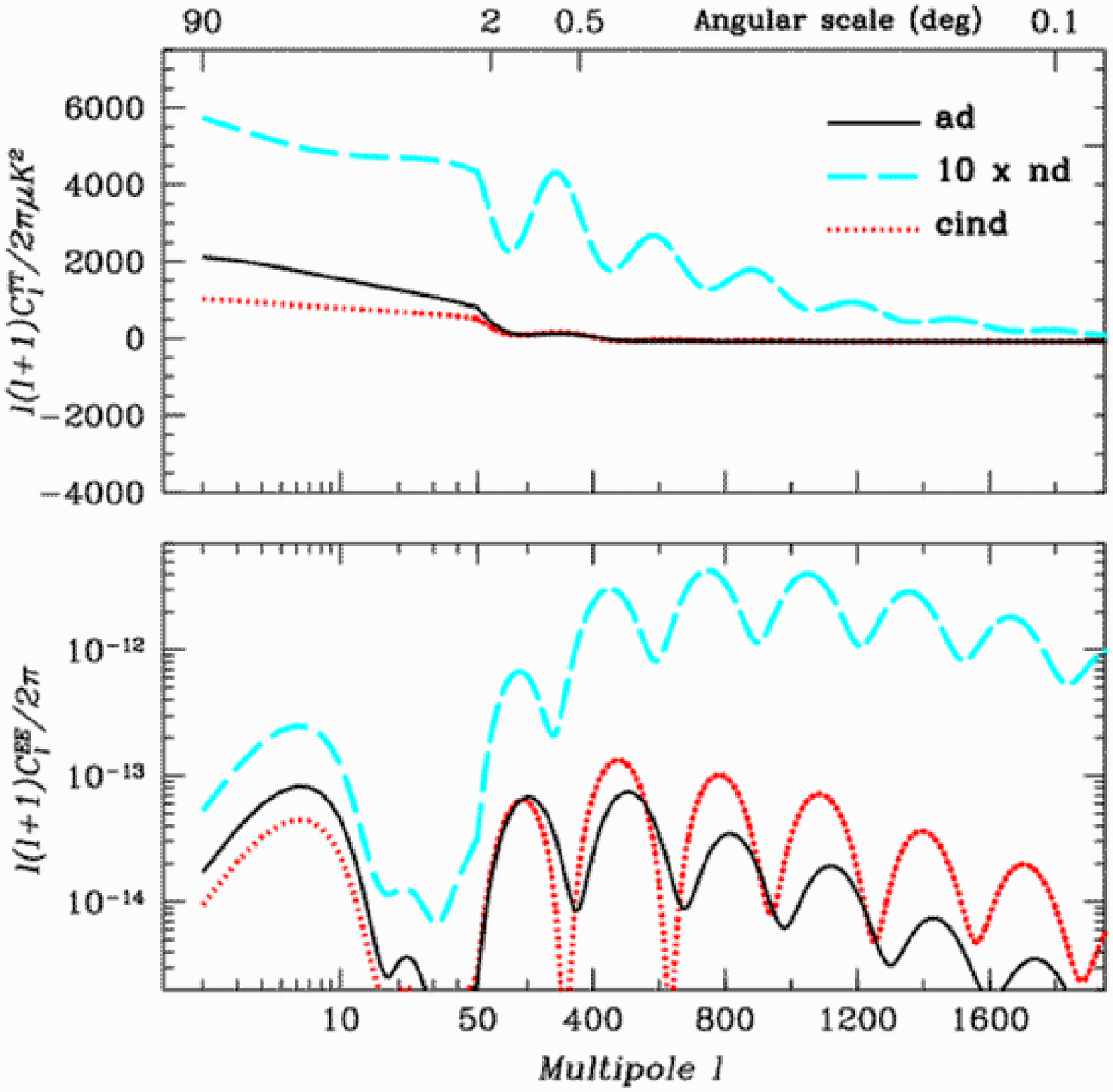}
\caption[Temperature and polarization spectra for general initial
conditions (I).]{Temperature and E-polarization angular power
spectra for the four modes constituting the most general initial
conditions for CMB anisotropies, Figure 1 of 2. The correlators
are for positive total correlation between the modes, and we take
all spectral indexes to be unity. The remaining cosmological
parameters are fixed to a concordance, flat $\La$CDM model. In the
lower panel, the correlators are plotted in absolute value. The
four modes are: ad (adiabatic), ci (CDM isocurvature), nd
(neutrino density/entropy), nv (neutrino velocity).}
\label{fig:general_ic1}
\end{figure}

\begin{figure}[tb]
\centering
\includegraphics[width=\twofigswidth]{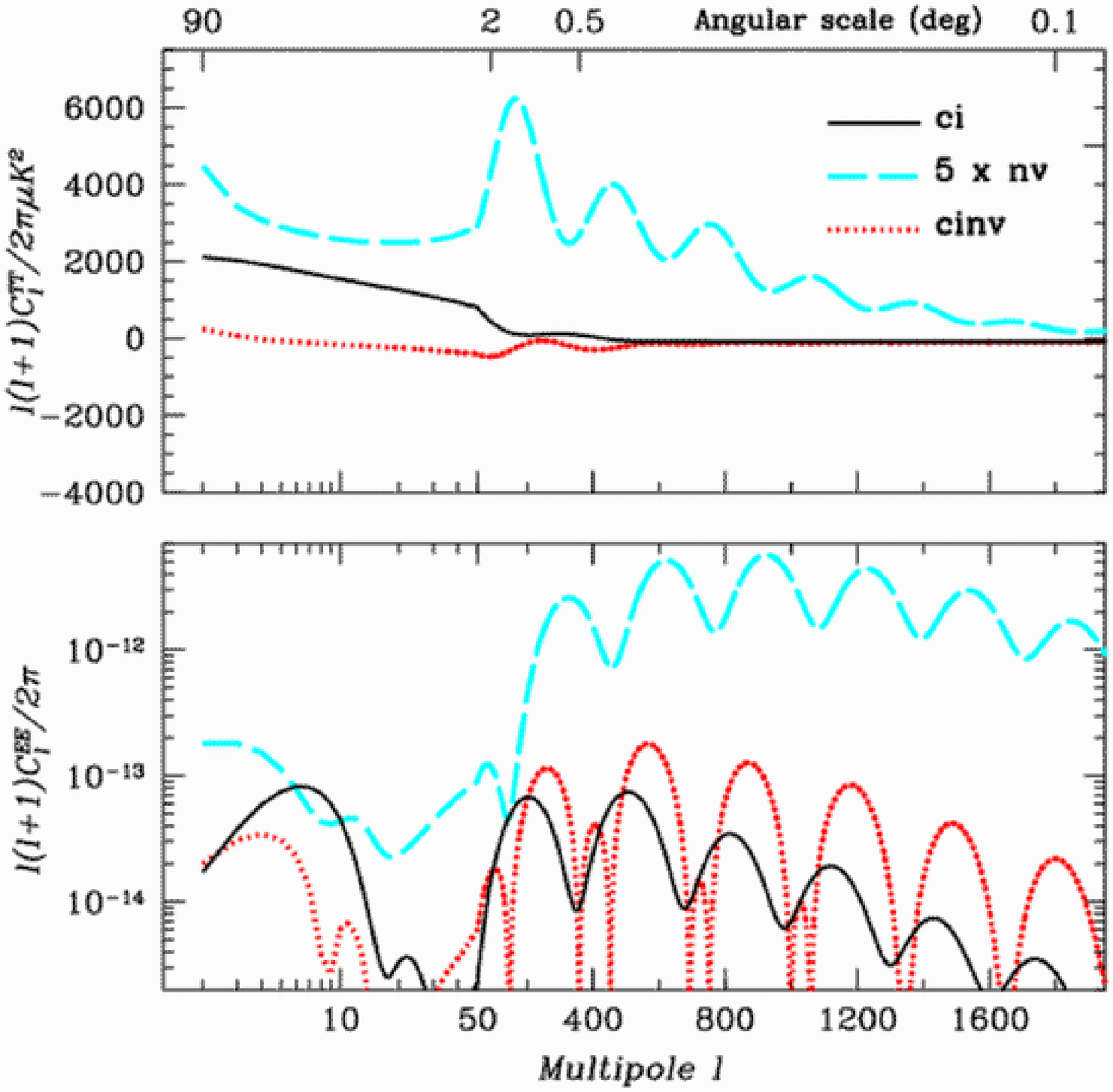}\hfill%
\includegraphics[width=\twofigswidth]{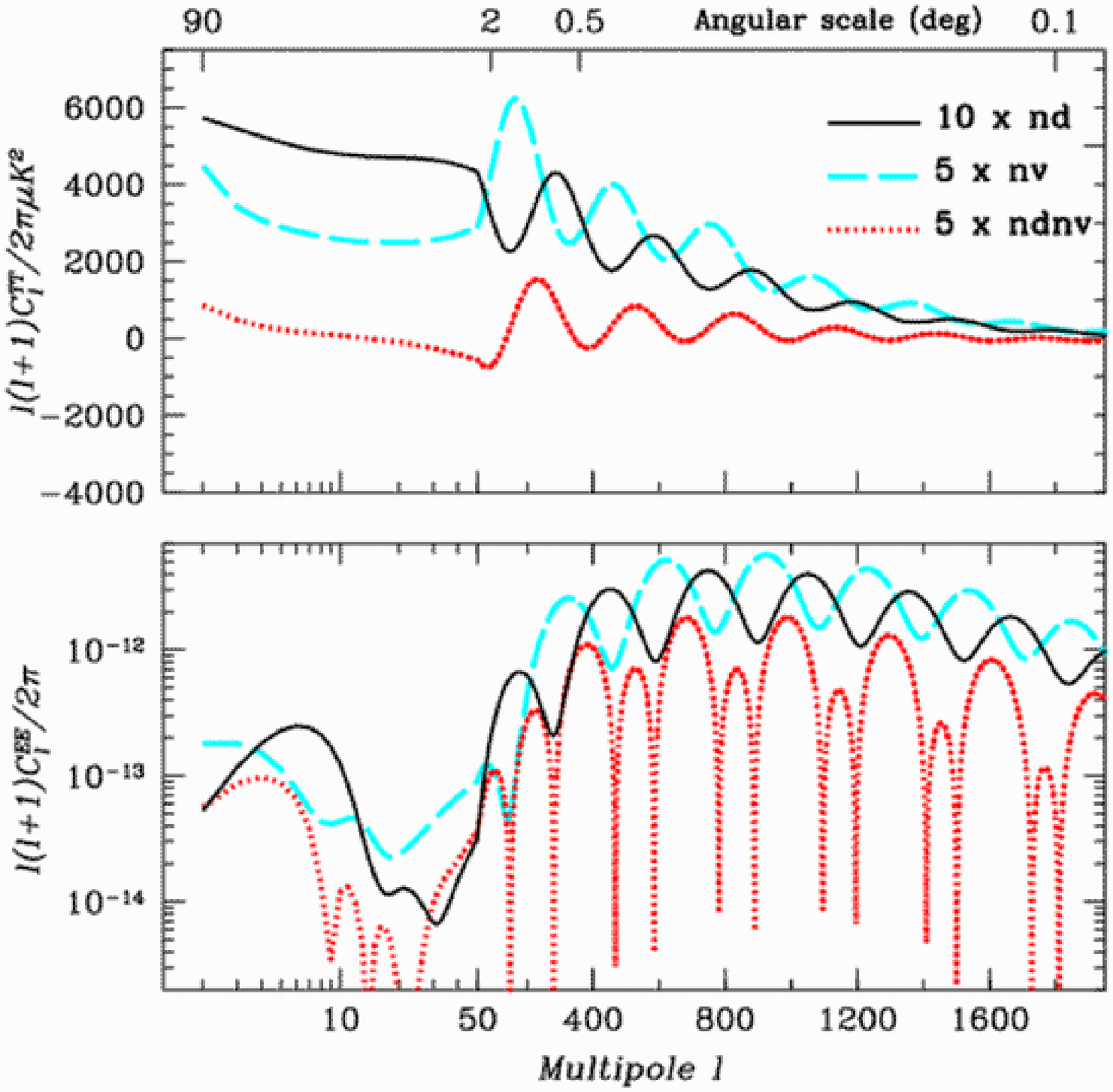}
\caption[Temperature and polarization spectra for general initial
conditions (II).]{Temperature and E-polarization angular power
spectra for the four modes constituting the most general initial
conditions for CMB anisotropies, Figure 2 of 2.}
\label{fig:general_ic2}
\end{figure}

The numerical integration of the evolution equations is necessary
to go beyond the early time approximative solutions derived
earlier and obtain the full angular power spectra for the
different types of initial conditions. Recent versions of
$\textsc{camb}$ include the possibility of specifying neutrino
entropy and velocity initial conditions, along with the adiabatic
and isocurvature CDM ones. The resulting temperature and
E-polarization spectra are displayed in Figures
\ref{fig:general_ic1} and \ref{fig:general_ic2}.
Analogously to the adiabatic-CDM isocurvature case discussed in
\SEC{chap:cmb;sec:large_scales}, in the most general case the
modes are arbitrarily correlated with each other, and each of them
possesses its own spectral index. In the figures we plot the
correlators for total positive correlation between the modes, take
scale invariant spectral indexes for all modes, $n = 1$ and we fix
the other cosmological parameters to a flat, concordance $\La$CDM
model with early reionization, as emerged from the WMAP data for
the pure adiabatic case.

The collection of modes presents a wide variety of oscillatory
structures, and very different amplitude ratios between the
large-scale plateau and the peaks. Since the perturbation
equations are linear, the most general CMB power spectrum is a
positive definit superposition of all the modes. From a
phenomenological point of view, we expect that widening the
initial condition space to include all of the four possible modes,
will lead to large degeneracies between initial conditions and
cosmological parameters. We dedicate
\SEC{chap:genic;sec:precision} to a thorough investigation of this
issue. On the other hand, if the neutrino isocurvature modes were
non-zero, their contribution could conceivably allow to fit the
CMB data without the need for a cosmological constant, a
possibility which we analyze and reject in
\SEC{chap:genic;sec:lambda}.

\subsection{Modes superposition}
\label{chap:params;sec:modes_superposition}

In the purely adiabatic scenario, initial conditions for scalar
perturbations are described by two parameters, namely the overall
normalization and the spectral index of the curvature perturbation
power spectrum, as in \rrp{eq:PS_for_zeta}. By enlargening 
the initial conditions space to include all of the four possible
modes, we add nine amplitudes (three for the CDM isocurvature,
neutrino density and velocity modes, and six for the correlators
between the four modes) and three spectral indexes, for a total of
14 parameters describing the most general initial conditions.

Although the dependence of the modes on the amplitudes is trivial,
the numerical search in the initial conditions parameter space is
complicated by the positive definiteness conditions on the total
spectrum. The total temperature (or polarization) angular power
spectrum obtained by superposing the modes must be positive
 \be \label{eq:modes_superposition}
 \DS \Cl = \sum_{i,j=1}^{4} M_{ij} \Cl^{ij} \geq 0 \quad \forall \quad
 \ell\eqcomma
 \ee
with the {\it modes correlation matrix} $\bs{M}\in \PD$, where
$\PD$ denotes the space of $n \times n$ real, positive
semi-definite, symmetric matrices with in our case $n=4$, and the
$\Cl^{ij}$ are computed for a fixed choice of cosmological
parameters when only the corresponding element of the correlation
matrix is non-zero, \ie for $M_{ij} = 1$, all others vanishing.
The elements of the correlation matrix are arranged so that the
amplitudes of the pure modes are along the diagonal (so that
$M_{ii} \geq 0$ for $i=1,\dots, 4$) while the off-diagonal
elements are the correlators amplitudes. Each correlator amplitude
must satisfy Schwartz' inequality
 \be \label{eq:Schwartz_inequality}
 M_{ij}^2 \leq  M_{ii}M_{jj} \quad i,j=1, \dots, 4
 \ee
because of the positive definiteness condition \cite[see][Appendix
A for a proof]{Trotta:2001tesi}, but in general the correlators
amplitudes can of course be negative. Finally, Schwartz'
inequality between all pairs $i \neq j$ of $\bs{M}$ is a necessary
but not sufficient condition for the positive definiteness of the
correlation matrix. A sufficient condition is that all
sub-determinants of $\bs{M}$ are positive or zero \citep[see
\eg][proposition 172.5]{Book:Heuser:93}, giving the four
sufficient conditions on the elements of $\bs{M}$:
\begin{subequations} \label{eq:positive_definit_conditions}
 \begin{align}
 & M_{11}  \geq 0 \eqcomma \\
 &M_{11}M_{22} - M_{12}^2  \geq 0 \eqcomma\\
 &M_{11}M_{22}M_{33} + 2 M_{12}M_{23}M_{13}^2M_{22} -
 M_{13}^2M_{33} - M_{12}^2M_{33} - M_{23}^2M_{11}  \geq 0 \eqcomma \\
 & \det \bs{M}  \geq 0 \eqdot
 \end{align}
\end{subequations}
When numerically searching the initial conditions parameter space,
the conditions (\ref{eq:positive_definit_conditions}) must be
imposed by hand to avoid regions which would lead to non-physical
(\ie negative) angular power spectra. This approach is used in
\cite{Trotta:2001yw} and some related issues are discussed in
\SEC{chap:genic;sec:precision}.

A more convenient parametrization of the correlation matrix is
employed in \cite{Trotta:2002iz}, where the matrix $\bs{M} \in
\PD$ is written as
\begin{equation}
\mathbf{M} = \mathbf{U} \mathbf{D} \mathbf{U} ^{T} ,
\label{eq:M_matrix_diagonalized}
\end{equation}
 $\mathbf{U} \in \On$, $\mathbf{D} =
\textrm{diag}(d_1, d_2, \dots, d_n)$ and $d_i \geq 0$, $i \in
\{1,2,\dots,n\}$. Here $\On$ is the space of $n \times n$ real,
orthogonal matrices with $\det = 1$ and $n = 4$. We can write
$\mathbf{U}$ as an exponentiated linear combination of generators
$\mathbf{H}_i$ of $\On$:
\begin{equation}
\mathbf{U} = \exp\left(\sum_{i = 1}^{(n^2-n)/2}
                         \alpha_i{\mathbf{H}_i} \right) ,
\label{eq:matrix_U1}
\end{equation}
with
\begin{equation}
\mathbf{H}_1
 = \left( \begin{array}{cccc}
          0 & 1 & 0 & \ldots \\
          -1 & 0 & 0 & \ldots \\
          0 & 0 & 0 & \ldots \\
          \vdots & \vdots &  \vdots & \ddots
          \end{array} \right) ,
\label{eq:matrix_H1}
\end{equation}
and so on, with $- \pi / 2 < \alpha_i < \pi / 2$, $i \in
\{1,2,\dots, (n^2 - n) / 2 \}$. In analogy to the Euler angles in
three dimensions, we can re-parameterize $U$ in the form
\begin{equation}
\mathbf{U}
 = \prod_{i = 1}^{(n^2-n)/2} \exp \left( \psi_i{\mathbf{H}_i} \right) ,
\label{eq:matrix_U2}
\end{equation}
with some other coefficients $- \pi / 2 < \psi_i < \pi / 2$, $i
\in \{1,2,\dots, (n^2 - n) / 2 \}$, whose functional relation with
the $\alpha_i$'s does not matter.  The diagonal matrix
$\mathbf{D}$ can be written as
\begin{equation}
\mathbf{D}
 = \textrm{diag}\left( \tan(\theta_1), \ldots, \tan(\theta_n) \right) ,
\label{eq:matrix_D1}
\end{equation}
with $0 \leq \theta_i < \pi / 2$, for $i \in \{1,2,\dots,n\}$.  In
this way, the space of initial conditions for $n$ modes is
efficiently parameterized by the $(n^2 + n) / 2$ angles $\theta_i,
\psi_j$. In our case, $n = 4$ and the initial conditions are
described by the ten dimensional hypercube in the variables
$(\theta_1, \ldots, \theta_4, \psi_1, \ldots, \psi_6)$. This is of
particular importance for the numerical search in the parameter
space. One can then go back to the explicit form of $\mathbf{M}$
using Eqs.~(\ref{eq:matrix_U2}),~(\ref{eq:matrix_D1}) and
(\ref{eq:M_matrix_diagonalized}). This more efficient
parametrization is employed in \SEC{chap:genic;sec:lambda}.
\label{sec:hypercube_params}

There is no optimal solution for an efficient and physically
motivated parametrization of the initial amplitudes; another
possibility, based on a ten-dimensional hypersphere, is employed
in the analysis of \cite{Bucher:2004an}.

\part{PARAMETER EXTRACTION}{
  The fundamental problem of scientific progress, and
  a fundamental one of everyday life, is that of learning
  from experience. Knowledge obtained in this way is partly
  merely description of what we have already observed, but
  part consists of making inferences from past experience
  to predict future experience.}
  {\authstyle{Harold Jeffreys}}
  {Theory of probability}
\selectlanguage{english}

\chapter{Statistics and data analysis}
\label{chap:data}
We are now in a position to attack the task of actually
determining the values of cosmological parameters from the
observed CMB anisotropy. To this end, we need several statistical
tools, which we introduce in \SEC{chap:data;sec:stattools}. The
emphasis is on their application to the CMB: we work out the
cosmic variance limit from first principles in
\SEC{chap:data;sec:cosmicvar}, and we present the Maximum
Likelihood principle and its application to data analysis in
\SEC{chap:data;sec:ML}; we focus on the differences between the
frequentist (\SEC{chap:data;sec:frequentist}) and Bayesian
approach (\SEC{chap:data;sec:Bayesian}) to statistics, explaining
the procedures to assess likelihood and confidence intervals and
their interpretation; we then discuss the implementation of two
popular methods to sample the parameters space, the traditional
gridding method (\SEC{chap:data;sec:numest}) and the more
efficient Monte Carlo sampling (\SEC{chap:data;sec:mcmc}). In
\SEC{chap:data;sec:fma} we explain the details of the Fisher
matrix analysis, an handy and accurate technique to produce
forecasts for the expected capabilities in terms of parameters
extraction of future CMB observations. In the last section,
\SEC{chap:data;sec:obs}, we offer a brief historical review of the
last decade of CMB observation, presenting the data-sets which are
then exploited in Chapters \ref{chap:beyondsp} and
\ref{chap:genic}.

\section{Elements of probability and statistics}

\subsection{Some concepts of probability theory}
\label{chap:data;sec:stattools}

We work in real, three-dimensional space, and we consider a field
$X$ which is defined in all points $\bfr \in \bs{R}^3$ in such a
way that the probability of obtaining the value $X$ at the point
$\bfr$ is $\pdf(X,\bfr)$. We call $X$ an {\it infinite dimensional
random field} and $\pdf$ its 1-point probability distribution
function (pdf). In order to fully describe the random field $X$,
we need to specify not only $\pdf$, but also the 2-point pdf,
denoted by $\pdf_2(X_1, \bfr_1, X_2, \bfr_2)$, which describes the
probability of getting the value $X_1$ at the point $\bfr_1$ and
the value $X_2$ at the point $\bfr_2$; then the probability
distribution for all triples of points, $\pdf_3$, and so on for an
arbitrarily large number of points.

From the definition of probability, the n-point pdf's are not all
independent, obeying the relations
 \be \label{eq:pdf_relations}
 \pdf_n(X_1, \dots, X_n) = \int \pdf_{n+1} (X_1, \dots, X_n, X_{n+1})
 \dr X_{n+1} \eqdot
 \ee
The field $X$ is said to be {\it statistically homogeneous} if its
1-point pdf is the same in all points of space:
 \be
 \pdf(X, \bfr) = \pdf(X) \quad \text{(statistical homogeneity),}
 \ee
and {\it statistically isotropic} if the 2-point pdf depends only
on the distance between the points but not on the direction of the
vector joining them:
 \be
 \pdf_2(X_1, \bfr_1, X_2, \bfr_2) = \pdf_2(X_1, X_2, r)
 \quad \text{(statistical isotropy),}
 \ee
with $r \equiv \vert \bfr_1 - \bfr_2 \vert$. In cosmology, all
random fields are assumed to be homogeneous and isotropic. From
now on we will always make this assumption. We denote with
$\EX{\cdot}$ the ensemble average over realizations of the field
$X$ (expectation value). For a function $f(X)$, its expectation
value is
 \be
 \EX{f(X)} \equiv \int_\Om f(X) \pdf(X) \dr X \eqcomma
 \ee
where the integration goes over all possible realizations of $X$
defining the sample space $\Om$. The expectation value of $f(X) =
X$ is called the mean of $X$. Under the assumption of isotropy,
$\EX{X}$ is a constant independent on $\bf{r}$. Therefore in
cosmological perturbation theory we can always take the
perturbations to have zero mean, since a constant offset can
always be reabsorbed in a redefinition of the corresponding
background quantity.

Consider $X(\bk)$, the harmonic transform of $X$ with respect to
the eigenfunctions of the Laplace operator; in ${\bf R}^3$ this is
the usual Fourier transform. Then as a consequence of homogeneity
and isotropy, $X(\bf{k})$ has the following properties:
\begin{align}
\EX{X(\bk)} & = \dirac(\bk) \EX{X} \\
\EX{X(\bk) \cdot X(\bs{k'})} & = \dirac(\bk - \bs{k'}) g(k)
\end{align}

The {\it real space correlation function} is defined as
 \be
 \xi(\bfr) \equiv \EX{X(\bfr_1) \cdot X(\bfr_1 + \bfr)} \eqdot
 \ee
It is the expectation value of $X_1 \equiv X(\bfr_1)$ and $X_2 =
X(\bfr_1 + \bfr)$ under the 2-point pdf,
 \be \label{eq:corr_fct_and_pdf_2}
 \xi(r)  =
 \int \dr X_1 \int \dr X_2 \; \pdf_2(X_1,X_2,r) X_1 X_2  \eqcomma
 \ee
where in writing $\xi(r)$ instead of $\xi(\bfr)$ we have assumed
statistical isotropy.

The field $X$ is called {\it space ergodic} if we can perform a
spatial average instead of an ensemble average and obtain the same
result:
 \be
 \lim_{R \rightarrow \infty} \left( \frac{4}{3}\pi R^3
 \right)^{-1}
 \int_{\vert \bfr \vert < R} f\left[X(\bfr)\right] \dr^3 \bfr
 = \EX{f\left[X\right]} \eqdot
 \ee
Notice that ergodicity requires that the field is defined over an
infinite space, such as ${\bf R}^3$. The temperature field of the
CMB however lives on the two-sphere $S^2$, which is a compact
manifolds and therefore not ergodic. Therefore even if we could
measure the anisotropies with no experimental error, we still
would not be able to perform the ensemble average with perfect
accuracy, see \SEC{chap:data;sec:cosmicvar}.

We denote by $\hat{f}$ the estimator for $f(X)$, \ie a procedure
applied to a random sample of $X$ to produce a numerical value for
$f$, which is called the estimate. When applied to a set of
observations $X_1\obs, X_2\obs, \dots X_n\obs$ which constitute a
random sample, the estimator $\hat{f}$ produces a distribution of
estimates, and as such it too is a random variable.

An important particular case is the Gaussian random field, for
which all the n-point pdf's are Gaussian. The 1-point pdf is then
 \be
 \pdf(X) = \dfrac{1}{\sqrt{2\pi}\si} \exp \left(
 -\dfrac{X^2}{2\si^2} \right) \eqcomma
 \ee
while the 2-point pdf is given in terms of the field's correlation
function $\xi$ as
 \be
 \pdf_2(X_1, X_2, r) = \dfrac{1}{2\pi \si^2 \sqrt{1 - \xi^2(r)}}
 \exp \left( - \dfrac{X_1^2 + X_2^2 - 2 \xi(r)X_1 X_2}{2\si^2\left[1-\xi^2(r)\right]}\right)
 \ee
and the 2-point pdf (or equivalently, the correlation function)
contains the full statistical information.

The statement that the correlation function determines the 2-point
pdf completely is true only for a Gaussian field; in general, from
(\ref{eq:corr_fct_and_pdf_2}) it is clear that after the
integration $\xi(r)$ only contains part of the information encoded
in $\pdf_2$. For instance, \cite{Jones1996} gives an interesting
counter-example of a Gaussian and a non-Gaussian distribution with
the same correlation function and yet with two different 2-point
pdf's.

\subsection{The origin of cosmic variance}
\label{chap:data;sec:cosmicvar}

It is instructive to compute explicitly the variance of the
observed $\Cl$ starting from basic principles. If we assume that
the temperature fluctuation $\Theta$ is an isotropic and
homogeneous random field, then the coefficients of the harmonic
expansion on the 2-sphere, the $\alm$'s, have zero mean and
variance given by the true $\Cl$'s:
\begin{align}
\EX{\alm} & = 0 \\
\EX{\alm^* \cdot \almp} & = \delta_{\ell \ell'} \delta_{m m'} \Cl
\label{eq:Cl_expectation} \eqdot
\end{align}
Inflation predicts that the $\alm$'s are very close to Gaussian
variables, so we make the assumption of Gaussianity and for the
pdf of $\alm$ we take
 \be
 \pdf(\alm) = \dfrac{1}{\sqrt{2 \pi} \Cl} e^{- \dfrac{\alm^2}{2 \Cl}}
 \eqdot
 \ee
The true $\alm$'s are of course inaccessible to us, but from the
measured temperature fluctuation we obtain an estimate which we
denote by $\halm$. As an estimator for the power spectrum we
define
 \be \label{eq:estimator_for_Cl}
 \hCl \equiv \dfrac{1}{2\ell + 1}\sum_{m = -\ell}^\ell
 \vert \halm^2 \vert =
 \dfrac{\Cl}{2\ell + 1} V \eqcomma
 \ee
where we have introduced the variable
 \be
 V \equiv \sum_{m = -\ell}^\ell
 \dfrac{\vert \halm^2 \vert}{\Cl^2} \eqdot
 \ee
Eq.~(\ref{eq:estimator_for_Cl}) implies an ergodic hypothesis,
since in the estimator we replaced the expectation value in
(\ref{eq:Cl_expectation}) by an average over independent azimutal
directions by summing over $m$.

The variable $V$ is a sum of $2\ell + 1$ squared Gaussian
variables with unit variance, and therefore \citep{Book:Kendall}
its pdf is the \chisq\ pdf with $2\ell+1 = l$ degrees of freedom
(dof):
 \be \label{eq:chi_pdf}
 \pdf_{\chi^2_l} (V) = \dfrac{V^\frac{l-2}{l}}{2^{l/2}
 \Gamma(l/2)}
 e^{-V/2} \eqdot
 \ee

From this we can write down the pdf for the estimator $\hCl$,
which is
 \be \label{eq:pdf_for_hCl}
 \pdf(\hCl) = \dfrac{l}{\Cl} \pdf_{\chi^2_l}
 \left( \frac{l \hCl}{\Cl} \right)
 \ee
which shows that our estimator is distributed according to a
\chisq\ pdf. For $l \rightarrow \infty$ the Central Limit Theorem
guarantees that the distribution will become Gaussian, hence
 \be
 \lim_{\ell \rightarrow \infty} \hCl = \Cl
 \ee
and the estimator is said to be {\it consistent}. From
(\ref{eq:pdf_for_hCl}) we can calculate the expectation value of
$\hCl$, finding
 \be
 \EX{\hCl} = \Cl \quad \text{(unbiasedness),}
 \ee
and its variance
 \be
 \EX{\hCl^2} - \EX{\hCl}^2 = \dfrac{2}{2\ell + 1} \Cl^2
 \quad \text{(efficiency).}
 \ee
We conclude that the fact that there are only $2 \ell + 1$
independent directions on the sky for a given multipole $\ell$
limits the efficiency of our estimator for the power spectrum with
variance
 \be
 \dfrac{\EX{\hCl^2} - \EX{\hCl}^2}{\Cl} = \dfrac{2}{2\ell + 1}
 \quad \text{(cosmic variance).}
 \ee

Despite the fact that cosmic variance is a fundamental statistical
limit, an ingenious method to circumvent it and to measure the
temperature quadrupole with better than cosmic variance precision
has recently been proposed by \cite{Skordis:2004xr}.

\subsection{The principle of Maximum Likelihood}
\label{chap:data;sec:ML}

The estimation problem can be generally stated as follows:
starting from a limited number of observations, which constitute a
random sample, one wants to reconstruct some properties of the
underlying pdf. It is simpler to think of the properties of the
pdf as unknown parameters, which we seek to determine. Consider a
set of $n$ observations $\data= \left\{ d_1\obs, d_2\obs, \dots,
d_n\obs \right\}$ of the variable $X$ and a set of $p$ parameters
$\params=\left\{\te_1\obs, \te_2\obs, \dots, \te_p\obs \right\}$.
The measurements have a conditional probability $\pdf(d_i \vert
\params)$ to be observed given the value $\params$ for the parameters.
The problem at hand is to estimate the joint conditional
probability
 \be \label{eq:def_LF}
 \like(\data \vert \params) \equiv \prod_{i=1}^n \pdf(d_i \vert \params)
 \ee
from the observations $\data$. In the above definition, we thought
of $\like$ as a function of the random variable $X$; however, once
the observations have been done, we can think of $\like$ rather as
a function of the unknown parameters $\params$ for a given value
of $\data$ and call it the {\it likelihood function} (LF).

The maximum likelihood (ML) principle affirms that as an estimate
for $\params$ we should choose the value $\params\ML$ which makes
the probability of the actual result obtained, $\data$, as large
as it can be, \ie
 \be \label{eq:ML_definition}
 \like(\data \vert \params\ML) \ge \like(\data \vert \params)
 \quad \text{(Maximum Likelihood)}
 \ee
for all possible values of $\params$.

Instead of maximizing the LF, one can minimize the quantity
 \be \label{eq:lognormal}
 \lnlike \equiv -2 \ln \like \eqcomma
 \ee
which we will call lognormal LF.

If the pdf is Gaussian, then the ML estimation reduces to the
usual least square fit: suppose that the measured $d_i\obs$ are
independent from each other and Gaussian distributed around their
(unknown) true values $d_i(\params)$, with variance given by the
experimental error $\sigma_i\obs$. Then minimizing $\lnlike$ is
equivalent to minimization of the quantity
 \be \label{eq:chi_definition}
 \chi^2(\params) \equiv \sum_{i=1}^n
 \left(\dfrac{d_i\obs - d_i(\params)}{\sigma_i\obs}\right)^2
 \eqcomma
 \ee
which is called the \chisq.

Applied to the problem of parameter extraction from CMB data, the
ML prescription means that, given the measured power spectrum,
$\Cl\obs$, with errors $\sigma_\ell$, we have to minimize the
value of the chi-square by varying the cosmological parameters of
interest. This procedure only gives information about the set of
parameters which are the ``most probable'' to have generated the
measurements at hand. However, quantifying the error on our
estimate for the parameters is a more subtle business, since it
involves dwelling into the exact definition of what probability
means. There is a long dispute going on among specialists about
the correct interpretation of probability, and some fundamental
issues are still unresolved. One can take fundamentally two
different point of views on the subject, the orthodox
(frequentist) approach or the Bayesian point of view, as we now
explain. A good introduction to Bayesian methods and a comparison
with the sampling theory approach can be found in \cite{Book:Box},
while \cite{Book:Kendall} give full details about frequentist
theory calculations. \cite{Book:Jaynes} is a very enjoyable book,
which provides a wider perspective on the logic of science and
probability theory. A useful textbook with many stimulating
examples of Bayesian inference is \cite{Book:MacKay}.
\cite{Book:Frodesen} -- written by experimentalists who have used
on the field the methods described -- is more praxis-oriented, and
explains in a practical way the statistical mambo-jumbo.

\subsection{Orthodox probabilities -- Confidence intervals}
\label{chap:data;sec:frequentist}

The orthodox definition of probability -- also known as ``sampling
theory'' approach -- is based on the empirical repeatability of
the experiment, see \eg \cite{Book:Jaynes}. If an experiment is
performed $N$ times and the outcome $A$ occurs in $M$ of this
cases, then the probability of the outcome $A$ is
  \be \label{eq:prob_limit}
  P(A) \equiv \lim_{N \rightarrow \infty} \dfrac{M}{N} \eqdot
  \ee
In the case of continuous variables, the concept of probability is
defined as the limiting process \eqref{eq:prob_limit} reached from
a finite subdivision in $N$ equiprobable intervals of the sample
space \cite[Section 7.11, Vol. 1]{Book:Kendall}. The frequentist
approach allows the definition and interpretation of {\it
exclusion regions} or {\it confidence intervals} for the
parameters, see below. It is the point of view usually adopted in
particle physics, where an experiment can be repeated many times
under the same circumstances. It is not very popular in cosmology
though, where there is only one particular realization to observe.

\subsubsection*{Confidence intervals -- frequentist}

Confidence intervals in the frequentist approach have a
straightforward interpretation: consider a random variable $X$
whose pdf depends on the parameter $\theta$ which we wish to
estimate from a random sample $\left\{x_1\obs, x_2\obs, \dots,
x_N\obs\right\}$ with an estimator $\hat{\te}$. For instance, one
can think of $\te$ as the true mean $\mu$ of a normal
distribution, and the estimator as the sample mean, $\hat{\mu} =
N^{-1} \sum_i x_i\obs$.

The estimates are distributed according to some pdf, which we
denote by $\pdf_e$. Then a $100 \ga \%$ {\it confidence interval}
for the estimated parameter $\hat{\te}$ is the range
$\left[\te_1;\te_2\right]$ such that the probability content for
the estimator is $\ga$, \ie
 \be
 P(\te_1 < \hat{\te} < \te_2 ) \equiv \int_{\te_1}^{\te_2}
 \pdf_e \dr \te =
 \ga \eqdot
 \ee

Notice that this is a statement about the probability of our {\it
estimate} $\hat{\te}$ to lie in a certain range, with the
interpretation that, if we would draw the $N$ samples $L$ times
under identical circumstances, then the estimates produced by
$\hat{\theta}$ fall in the range $\left[\te_1;\te_2\right]$
$\gamma L$ times. Therefore at this stage we are merely making a
statement of the distribution of our estimator. If we want to
convert this into a confidence statement for the true value $\te$,
we can say that there is a probability $\ga$ that the random
interval $\left[\te_1;\te_2\right]$ will cover the true value
$\te$. In other words, in the long run the limits $\te_1$ and
$\te_2$ are such that the statement
 \be
 \te_1 < \te < \te_2
 \ee
will be true in $100\ga \%$ of the cases.

Unfortunately, the above interpretation is unapplicable to
cosmology, where we cannot draw new samples at will from the
underlying distribution, but we have to content ourselves with the
only realization we happen to observe. However, we can still use
as an estimator the least-square fit to the observed value, and
interpret the result in frequentist's terms.

Consider the least-square fit of (\ref{eq:chi_definition}), which
applied to the CMB power spectrum is
 \be \label{eq:chi_for_Cl}
 \chi^2(\params) \equiv \sum_{\ell}
 \left(\dfrac{
 \Cl\obs - \Cl(\params)}{\sigma_\ell\obs}\right)^2
 \eqcomma
 \ee
where the observed $\Cl\obs$ are estimated using the estimator
(\ref{eq:estimator_for_Cl}): since each term is a sum of $2\ell +
1$ Gaussian variables squared (the $\halm$'s), its distribution
becomes Gaussian by virtue of the Central Limit Theorem only for
large $\ell$. The $\sigma_\ell\obs$ are the estimated errors from
the observations for each multipole, and $\params$ is the vector
containing the $p$ cosmological parameters of interest. The
functional dependence of $\Cl(\params)$ is given by the underlying
theory, which we try to falsify by comparing its predictions with
the actual observations.

The least-square estimate for $\params$ -- which is equivalent to
the ML estimator for Gaussian variables -- is the value
$\params\ML$ for which the $\chi^2$ reaches the minimum value
${\chi^2}\ML$, which is called {\it least square estimate}. Until
this point, the least-square estimation makes no assumptions about
the underlying pdf for the variables. To the extent to which the
$\hCl$'s can be considered as independent Gaussian variables, then
the quantity ${\chi^2}\ML$ is distributed as a \chisq\ pdf with
$f=n-p$ dof, denoted by $\pdf_{\chi^2_f}$, see (\ref{eq:chi_pdf}).
Here $n$ is the number of multipoles observed and $p$ the number
of fitted parameters.

Under these assumptions, the distribution $\pdf_{\chi^2_f}$
provides a measure of the {\it goodness of fit}: assume that a
given parameter set $\params_0$ is the correct one, and that the
measured \chisq\ in our Universe for $\params_0$ is $\chi^2_0$;
then if the measurement would be repeated many times in different
realizations, the probability that the outcome will be equal or
larger than the true value $\chi^2_0$ is
 \be \label{eq:chi_probability_content}
 P(\chi^2 > \chi^2_0) = \int_{\chi^2_0}^\infty \pdf_{\chi^2_f}(u)
 \dr u \equiv 1 - \ga_0 \eqdot
 \ee
The interpretation in frequentist terms is straightforward: if
some other parameters $\params_1$ have $\chi^2(\params_1)
=\chi^2_1 \gg \chi^2_0$, the chance that $\params_1$ is the
correct set and we are actually seeing a realization far out in
the tail of the distribution is very small.

It now remains to define {\it confidence intervals} for the
parameters basing on the above frequentist interpretation: a
$100\ga \%$ confidence interval encompasses parameters whose
measured $\chi^2$ is smaller than the value of corresponding to
the quantile\footnote{Given the pdf $\pdf$, $x$ is said to be the
quantile of $q$ if it satisfies $\int_x^\infty \pdf(u) \dr u =
q$.} of $1-\ga$ for the distribution $\pdf_{\chi^2_f}$. In other
words, if the measurements could be repeated many times, in the
long run the above confidence interval would include the true
value of the parameters $100\ga\%$ of the time. Thus the parameter
space outside the estimated confidence interval is a proper {\it
exclusion region} at the given confidence level. Notice that the
frequentist confidence levels depend both on the {\it total}
number of parameters fitted and on the number of independent data
points we are using.

We conclude this section with two remarks: firstly, the above
assumptions of Gaussianity and independency are only partially
fulfilled by the $\hCl$'s, therefore the outcome of such a
frequentist analysis is only approximative (see
\citealp{Abroe:2001de} for a strictly correct frequentist
parameter estimation, which involves the numerical sampling of the
pdf which we simply took as a \chisq); and second, the clean
interpretation of the frequentist approach is somewhat weakened by
the fact that we are compelled to invoke measurements in other
realizations which cannot take place, not even in principle.
Bayesian statistics takes instead a more pragmatic approach, by
dealing only with actual observations.

\subsection{Statistical inference -- Likelihood intervals}
\label{chap:data;sec:Bayesian}

Bayesian statistics does not consider possible outcomes of
measurements which are never performed. Instead, it exploits the
actual data to update our knowledge about the probability of a
certain statement, starting from our prior degree of belief.
Criticism has been raised against this approach because the final
inference depends on the prior information available, and
therefore seems to suffer from a certain degree of subjectivity.
However, Bayesian inference can be applied to theories which are
not repeatable and are unscientific in the frequentist point of
view (\eg the probability that it will rain tomorrow). It is based
on Bayes' Theorem\footnote{Rev. Thomas Bayes, 1763.}, which is
nothing more than rewriting the definitions of conditional
probability:
  \be
  \pdf(A \vert B) = \dfrac{\pdf(B \vert A) \pdf(A) }{\pdf(B)}
  \quad
  \text{(Bayes' Theorem).}
  \ee
In order to clarify the meaning of this relation, let us write
$\params$ for $A$ and $\data$ for $B$, obtaining
  \be \label{eq:Bayes_Theorem}
  \pdf(\params \vert \data) = \dfrac{\like(\data \vert \params)
  \pdf(\params)}{\int \dr \params \pdf(\data \vert \params) \pdf(\params) }
  =
  \dfrac{\like(\data \vert \params) \pdf(\params)}{\pdf(\data)}
  \eqcomma
  \ee
which relates the {\it posterior probability} $\pdf(\params \vert
\data)$ for the parameters $\params$ given the data $\data$ to the
likelihood function $\like(\data \vert \params)$ if the {\it prior
pdf} $\pdf(\params)$ for the parameters is known. The quantity in
the denominator is independent of $\params$ and it is called the
{\it evidence} of the data for a certain model
\citep{Book:MacKay}. It is important for model comparison, but
here we shall regard it just as a normalization constant. In short
 \be
 \text{posterior} =
 \dfrac{\text{likelihood} \times\text{ prior}}{\text{evidence}} \eqdot
 \ee
The prior distribution contains all the (subjective) knowledge
about the parameters before observing the data: our physical
understanding of the model, our insight into the experimental
setup and its performance, in short the amount of all our prior
scientific experience. This information is then updated via Bayes
theorem to the posterior distribution, by multiplying the prior
with the LF which contains the information coming from the data.
The posterior probability is the base for inference about
$\params$: the most probable value for the parameters is the one
for which the posterior probability is largest.

Bayes' postulate\footnote{Bayes' postulate is also known --
perhaps with an hint of sarcasm -- as the Postulate of
Equidistribution of Ignorance.} states that in absence of other
arguments, the prior probability should be assumed to be equal for
all values of the parameters over a certain range,
$\params_\text{min} \leq \params \leq \params_\text{max}$. This is
called a ``flat prior'', \ie \be
 \pdf(\params) = \left[ H(\params -
  \params_\text{min}) H(\params_\text{max} - \params)\right]
  \prod_{i=1}^p \left[\theta_{\text{max},i} -
 \theta_{\text{min},i}\right]^{-1} \eqcomma
 \ee
 where $H$
is the Heaviside step function and $\params_{\text{max},i} >
\params_{\text{min},i}\; \forall \; i$. This is one of the principal
conceptual difficulties of Bayesian inference: a flat prior on
$\params$ does not correspond to a flat prior on some other set
$f(\params)$, obtained via a non-linear transformation $f$.
Therefore the result of Bayesian inference do depend on the choice
of priors, even though this usually does not constitue a major
obstacle in practical problems -- see however \cite{Bucher:2004an}
for an instructive example of the role of priors.

We see from \rr{eq:Bayes_Theorem} that the Maximum Likelihood
principle is equivalent to Bayesian inference in the case of flat
priors. We will always work with flat, top-hat priors unless
otherwise stated. There is however an important conceptual
difference. By writing the posterior distribution as
 \be
 \pdf(\params\vert\data) = \dfrac{\pdf(\params, \data)}{\pdf(\data)} \eqcomma
 \ee
it follows that Bayes' Theorem imposes to maximise the {\it joint
probability} $\pdf(\params, \data)$ of $\params, \data$, while
Maximum Likelihood requires that the {\it conditional probability}
$\like(\data\vert\theta)$ should be maximised.

\subsubsection*{Likelihood intervals -- Bayesian}

Bayesian statistics use the LF to perform an interval estimation
for $\params$: basing on Bayes' Theorem, \rr{eq:Bayes_Theorem}, we
not only consider the ML point in parameter space as the ``most
likely'' value of the unknown parameter; we shall also interpret
values further and further away as less and less likely to have
generated the particular measurement at hand. Hence {\it
likelihood intervals} drawn from the LF measure our  ``degree of
belief'' that the particular set of observations was generated by
a parameter belonging to the estimated interval. This is radically
different from the frequentist interpretation sketched above.

Let us simplify the notation by writing $\like(\params)$ instead
of $\like(\data \vert\params)$, since now we consider the LF as a
function of the parameters given a data set $\data$. Assume
further that the LF is a multivariate Gaussian distribution in the
$p$ parameters $\params$, \ie
 \begin{align}
 \like(\params ) & = (\det \bs{C})^{-1/2} (2\pi)^{-p/2}  \exp({-
 \lnlike/2)}
 \eqcomma \label{eq:like_MVG}\\
 \lnlike & = - 2 \ln \like = (\params - \mean)^T \bs{C}^{-1} (\params-\mean) \label{eq:Gaussian_lnlike}
 \end{align}
where $T$ denotes transposition, $\mean$ is the expectation value
of the parameters $\mean \equiv \EX{\params}$ and $\bs{C}$ is the
{\it covariance matrix}
 \be
 C_{ij} \equiv \EX{(\te_i - \mu_i)(\te_j - \mu_j)} \eqdot
 \ee
 From the likelihood one
can then obtain the posterior distribution via
(\ref{eq:Bayes_Theorem}), once the prior is specified. For the
prior distribution $\pdf(\bs{\te})$ a simple choice are so-called
``flat'' priors, a multidimensional top-hat function over some
range which is supposed to encompass all the values of interest.
Usually, in grid-based method the prior coincides with the
extension of the grid, so that the prior is just a multiplicative
constant and we can identify the likelihood with the posterior. As
mentioned, this choice is somewhat arbitrary, since it depends on
the basis chosen for the parameters.

We can Taylor expand a general LF around its maximum which is
given by our ML estimate $\params\ML$ of $\mean$, which on average
coincides with the true mean for a normal distribution,
$\EX{\params\ML} = \mean$. By definition of the ML point the first
derivatives vanish, $\partial \lnlike/\partial \te_i (\params\ML)
= 0$, and we obtain
 \be \label{eq:like_second_order_expansion}
 \lnlike(\params) \approx \lnlike(\params\ML)
 + \dfrac{1}{2}\sum_{ij} (\te_i-\te_i\ML) \dfrac{\partial^2 \lnlike}{\partial \te_i \partial
 \te_j}(\te_j-\te_j\ML) \eqdot
 \ee
If the LF is sharply peaked around $\params\ML$, \ie the errors on
the parameters are small enough, then third order terms are
unimportant and the above Gaussian form is a good enough
approximation everywhere in parameter space. By comparing with
(\ref{eq:Gaussian_lnlike}) we find that the covariance matrix can
thus be estimated as
 \be \label{eq:define_Fisher_matrix}
 \hat{\bs{C}} = \bs{F}^{-1} \quad \text{where} \quad
 F_{ij} \equiv  \Big \langle \dfrac{1}{2} \dfrac{\partial^2 \lnlike}{\partial \te_i \partial
 \te_j}\Big \rangle {\Big\arrowvert_{\params\ML}}
 \ee
is called {\it Fisher information matrix} \cite[Chap.15,
Vol.1]{Book:Kendall}.

According to our understanding of the LF as a measure of our
degree of belief for the possible values of $\params$, the
probability that parameters within a certain region from the ML
point have generated the observations should be proportional to
the likelihood content of the region. The probability content
depends on whether we are estimating all parameters jointly, or
keeping some of them fixed to their ML value, or rather
disregarding a certain subset by integrating over them
(marginalization). We consider each case in turn.

\subsubsection*{Estimation of all $p$ parameters jointly.}

Without loss of generality we can take in the following $\mean =
\bs{0}$ in \rr{eq:Gaussian_lnlike}, which can always be achieved
by shifting the origin of the coordinate system in parameter
space. Contours of constant likelihood define hyperellipses in
parameter space with some probability content we wish to
determine. To this aim we consider the quadratic form
 \be
 Q(\params) \equiv \params ^T \bs{C}^{-1} \params
 \ee
and for the LF (\ref{eq:like_MVG}) the condition $Q(\params) =
Q_\ga^s$ for some constant $Q_\ga^s$ gives the contours of
constant likelihood. We write $Q_\ga^s$ to indicate that the
numerical value of the constant depends on the number of
parameters under consideration, $s$, and on the desired
probability content of the hyperellipse, $\ga$. It can be shown
\cite[Chap.8, Vol.1]{Book:Kendall} that the quadratic form $Q$ is
\chisq\ distributed with $s$ dof, which allows us to relate
$Q_\ga^s$ with the probability content of the ellipse.

If we want a confidence region containing $100\ga\%$ of the {\it
joint} probability for all $p$ parameters, then $s = p$ and
$Q_\ga^p$ is determined by solving
 \be
 \int_0^{Q_\ga^p} \pdf_{\chi^2_p}(u) \dr u = \gamma \eqdot
 \ee
The projection ({\it not} the intersection) of the hyperellipse
$Q(\params) = Q_\ga^p$ onto each of the parameter axis gives the
corresponding likelihood interval for each parameter when all
parameter are estimated simultaneously (which we will call ``joint
likelihood interval'').

It is a simple geometrical problem to find an analytical
expression for the joint likelihood interval for each parameter:
for the parameter $1 \le d \le p$, the intersection of the
hyperellipse with the hyperplane defined by $\theta_d = c$, with
$c$ a constant, gives either an hyperellipse in $p-1$ dimensions,
or a point or else an empty set. The extrema of the joint
likelihood interval for the parameter $d$ are given by the values
of $c$ for which the $p-1$ dimensional ellipse reduces to a point.

To find the equation of the $p-1$ dimensional ellipse we proceed
as follows: define $\bs{C}^{-1}  \equiv \bs{M}$ and write
$Q(\params) = Q_\ga^p$ in the form
 \be \label{eq:subellipse_eq_1}
 \tilde{\params}^T \bs{\tilde{M}} \tilde{\params} + 2c\sum_{j \ne
 d} m_{dj} \tilde{\te}_j = Q_\ga^p - m_{dd}c^2 \eqcomma
 \ee
where we have defined
 \begin{align}
 \tilde{\params} & \equiv
 \left(\te_1, \dots, \te_{d-1}, \te_{d+1}, \dots , \te_p \right)
 \in \bs{R}^{p-1}
 \\
 \bs{\tilde{M}} & \equiv
 \left(
 \begin{matrix}
 m_{11} & \dots & m_{1,d-1} & m_{1,d+1} & \dots & m_{1p} \\
 \vdots &       &           &           &       & \vdots \\
 m_{d-1, 1}&    &  \dots    &           &       & m_{d-1,1} \\
 m_{d+1, 1}&    &   \dots   &           &       & m_{d+1,1} \\
 \vdots&        &           &           &       & \vdots \\
 m_{p1}&        &  \dots    &           &       & m_{pp}
 \end{matrix}
 \right) \in \bs{R}^{(p-1 \times p-1)}.
 \end{align}

Now we diagonalize the submatrix $\bs{\tilde{M}}$,
 \be
 \text{diag}\left(\la_1, \dots, \la_{p-1} \right) \equiv
 \bs{\La} = \bs{U}^T \bs{\tilde{M}} \bs{U}
 \ee
finding the eigenvalues $\la_i, i \le 1 \le p-1$ and eigenvectors
$\left( u_1, \dots, u_{p-1} \right)$, and after some algebraic
manipulations of (\ref{eq:subellipse_eq_1}) we arrive at the
equation of the $p-1$ dimensional hyperellipse
 \be
 \sum_{i=1}^{p-1}\la_i z_i^2 =
 Q_\ga^p - m_{dd}c^2 + \sum_{i=1}^{p-1} \dfrac{c^2}{\la_i}
 \left(\sum_{j \ne d} m_{dj} u_{ji} \right)^2 \eqcomma
 \ee
where we have defined the new variables
 \be
 z_i \equiv (\bs{\tilde{\te}} \bs{\tilde{U}})_i + \dfrac{c}{\la_i}
 \sum_{j \ne d}m_{dj} u_{ji} \eqcomma \quad
 0 \le i \le p-1 \eqdot
 \ee
  The above hyperellipse becomes degenerate if
 \be
 \sum_{i=1}^{p-1}\la_i z_i^2 = 0
 \ee
from which we obtain a quadratic equation for $c$ with solutions
 \be
 c_{\text{min, max}} =
 \dfrac{\pm  \sqrt{Q_\ga^p}}
 {\left[ m_{dd} - \sum_{i=1}^{p-1} \la_i^{-1} \left(\sum_{j \ne d} m_{dj} u_{ji} \right)^2 \right]^{1/2}}
 \eqdot
 \ee
It is easy to show that the positive definiteness condition for
the Fisher matrix guarantees that the quantity under the square
root in the denominator is always $\ge 0$. In conclusion, the
joint likelihood interval for the parameter $\te_d$ with
likelihood content $\ga$ is given by
 \be
 c_{\text{min}} \le \te_d \le c_{\text{max}} \eqdot
 \ee

\subsubsection*{Estimation of $k < p$ parameters, the others fixed.}

We are sometimes interested in giving confidence intervals for
some subset $k < p$ of the parameters, while assuming the other
$p-k$ parameters as (exactly) known. Without loss of generality we
shall take the first $k$ parameters as the one we are interested
in, and we split the parameter vector as
 \be \label{eq:params_split}
 \bs{\te} = \left(
 \begin{matrix}
 \bs{t} \\
 \bs{u}
 \end{matrix}
 \right)
 \ee
with $\bs{t} \in \bs{R}^{k}$ and $\bs{u} \in \bs{R}^{p-k}$.
Correspondingly we write the covariance matrix in
(\ref{eq:Gaussian_lnlike}) as the Fisher matrix estimate of
(\ref{eq:define_Fisher_matrix}),
 \be \label{eq:Fisher_split}
 \bs{C}^{-1} = \bs{F} = \left(
 \begin{matrix}
 \bs{A} & \bs{G} \\
 \bs{G}^T & \bs{B}
 \end{matrix}
 \right)
 \ee
where $\bs{A} \in \bs{R}^{k \times k}$, $\bs{B} \in \bs{R}^{p-k
\times p-k}$ and $\bs{G} \in \bs{R}^{p-k \times k}$.

If the known parameters $\bs{u}$ are held fixed at their ML value,
the LF for the parameters of interests $\bs{t}$ is simply the full
LF restricted to the $k$ subspace,
 \be
 \like \left(\bs{t} \vert
 \bs{u}\ML\right) \propto \exp(- \dfrac{1}{2}\bs{t}^T
 \bs{A} \bs{t} )
 \eqcomma
 \ee
with an appropriate normalization constant, and the new covariance
matrix $\bs{V} \in \bs{R}^{k\times k}$ for the $k$ parameters of
interest is
 \be
 \bs{V} = \bs{A}^{-1}  \quad
 \text{(conditional).}
 \ee

In particular, we often consider the best case scenario in which
all parameters but one are supposed to be known exactly, say from
independent observations or theoretical prejudice, and therefore
$k = 1$. Then the $1\si$ likelihood interval for the first
parameter only is the square root of the covariance matrix
element, and it is given by (all others fixed to their ML value)
 \be
 \si_{1} = \dfrac{1}{\sqrt{f_{11}}} \eqdot
 \ee

\subsubsection*{Estimation of $k < p$ parameters, the others
marginalized.}

Instead of fixing some parameters, we may prefer to disregard them
completely, by integrating over them in order to obtain the {\it
marginalized} likelihood in the $k$ parameter of interest:
 \be \label{eq:marginalization_integral}
 \like(\bs{t})
 \propto \int_{\Om_{\bs{u}}} \like(\bs{t}, \bs{u})\dr \bs{u} \eqcomma
 \ee
with a suitable normalization constant so that the probability
content of the marginalized LF is equal to unity.

 The marginal LF for $\bs{t}$
is still a multivariate Gaussian, with the same covariance matrix
as the full LF, only with the last $p-k$ rows and columns deleted:
  \be
  V_{ij} = \left[ \bs{F} ^{-1} \right]_{ij} \quad 0 \le i,j \le k
  \quad
 \text{(marginalized).}
 \ee
This result can be obtained by performing explicitly the
integration (\ref{eq:marginalization_integral}) or more elegantly
by using the properties of the characteristic function
\cite[Chap.4, Vol.1]{Book:Kendall}. In terms of the splitting
(\ref{eq:Fisher_split}), the covariance matrix for the
marginalized distribution is
 \be \label{eq:marginalized_covariance}
 \bs{V} = \left[ \bs{A} - \bs{G} \bs{B}^{-1} \bs{G}^T \right]^{-1}
 \eqdot
 \ee

Very often one quotes marginalized likelihood intervals for one
parameter alone, $k=1$ with all other parameters marginalized, in
which case the $1\si$ error is given by
 \be
 \si_{1} =  \sqrt{\left( \bs{F} ^{-1} \right)_{11}} \eqdot
 \ee
If the parameters are uncorrelated, then $\bs{F}$ is diagonal, and
fixing $\bs{u}$ or marginalizing over them is equivalent,
otherwise the resulting likelihood intervals for the parameter(s)
of interest are in general different, with the marginalized
interval being broader.

\subsection{Gridding method}
 \label{chap:data;sec:numest}

In the numerical fit to the data, the shape of the LF is
determined by evaluating the least-square estimator
(\refp{eq:chi_definition}) at each point on a grid in the $p$
dimensional parameter space and the minimization of the \chisq\ in
the desired range of parameters gives the ML estimate.

Assuming that the measurements are normally distributed around
their true value we have
 \be \label{eq:LF_from_chi}
 L(\data \vert \params) = L_\text{max} \exp\left[-
 \chi^2(\params)/2\right] \eqdot
 \ee
From this we can use the above prescriptions to determine
likelihood or confidence intervals from real data.

In the frequentist analysis, the boundaries of the confidence
regions represent exclusion plots at the given confidence level:
they are found as the contours of constant $\chi^2$ using the
relation (\refp{eq:chi_probability_content}), independently of the
value of the \chisq\ at the ML point. In Bayesian statistics, the
likelihood intervals are instead drawn {\it around} the ML point,
hence their extension depends on the best fit value. This applies
only to the gridding method, not to the Monte Carlo sampling
described below in \SEC{chap:data;sec:mcmc}. It is customarily to
quote marginalized likelihood intervals for one parameter only or
to plot two-dimensional likelihood contours to show degenerate
direction between two parameters (also see below the paragraph
discussing the maximization approach instead of marginalization);
for these two cases, the cook-book prescription for Bayesian
(Maximum Likelihood) statistics on a grid of samples in parameter
space is:
 \begin{itemize}
 \item find the ML point $\like_{\max}$ in the grid of parameters by
minimizing the $\chi^2$ of \rrp{eq:chi_for_Cl} and mark this point
as $\chi^2_\text{min}$, your least-square estimate of the best
fit;
 \item determine the boundaries of the region containing $100\ga\%$
 of likelihood as the values of the parameters for which the $\chi^2$ has increased
by an amount $\Delta \chi^2 = Q_\ga^{k}$ ($k = 1,2$ the number of
parameters considered) with respect to $\chi^2_{\text{min}}$.
 \item  The values of $Q_\ga^{k}$ can be found for every
 desired likelihood content using the relation, \CF (\refp{eq:chi_probability_content})
  \be
  \ga = \int_0^{Q_\ga^{k}} \pdf_{\chi^2_k}(u) \dr u \eqdot
  \ee
 Table \ref{table:chi_content} displays the values of $\Delta \chi^2$ for $k=1,2$
 and for some popular choices of likelihood content.
 \end{itemize}

\begin{table}
\centering
\begin{tabularx}{\linewidth}{|p{4cm}|XXXXX|}
  \hline
  $100\ga\%$            &  $68.3\% $ & $95\%$     & $95.4\%$ & $99\% $    & $99.7\% $\\
  Likelihood content    &  $(1\si)$  & $(1.96\si)$& $(2\si)$ & $(2.58\si)$&
        $(3\si)$ \\
    \hline \hline
  1 parameter, $Q_\ga^1$  & 1.00 & 3.84 & 4.00 & 6.63 & 9.00  \\
  2 parameters, $Q_\ga^2$ & 2.30 & 5.99 & 6.17 & 9.21 & 11.80 \\
  \hline
\end{tabularx}
\caption[Chi-square difference for one- and two-dimensional
marginalized likelihood plots.]{$\Delta \chi^2 = Q_\ga^{k}$ for
marginalized likelihood intervals in one parameter ($k=1$) or
marginalized likelihood contours in two parameters ($k=2$) for the
given joint likelihood content.} \label{table:chi_content}
\end{table}

In a real situation, the LF computed using (\ref{eq:LF_from_chi})
will not be exactly a multivariate Gaussian, and the likelihood
intervals obtained with this method will only approximatively
encompasses the stated probability content. There are methods
which improve on the assumption of a normal distribution presented
here, see for instance
\cite{Bond:1998qg,Bartlett:1999fw,Wandelt:1998qd,Jaffe:2003iv}.

Finally, notice that likelihood (Bayesian) contours are usually
much tighter than the confidence contours drawn from the
frequentist point of view. This is a consequence of the ML point
having often a $\chi^2 / f$ much smaller than $1$, because the
data-sets are highly consistent with each other and also because
usually not all points are completely independent. For the CMB,
this was the case when one considered a combination of several
data-sets before WMAP, as we discuss in
\SEC{chap:genic;sec:precision}. If we consider the usual situation
in which likelihood contours are drawn in a two dimensional plane
with all other parameters marginalized over, the frequentist
approach is more conservative than Bayesian statistics: the region
corresponding to the desired confidence level (frequentist) or
likelihood content (Bayesian) $\ga$, has bounds given by
$\chi^2(\params) = Q_k^\ga$, with $k=2$ for Bayesian statistics
and two-dimensional plots, and $k=f$ for frequentist statistics
independently on the number of parameters considered. Since in
general and for reasonably good ML values $\chi_{\rm{min}}^2 \lsim
\calo(f)$ and $f > 2$, we have that the probability/likelihood
content is the same, \ie
\begin{equation}
\int_{Q_f^\ga}^{\infty} \pdf_{\chi_f^2}(u) \dr u
 = \int_{Q_2^\ga}^{\infty} \pdf_{\chi_2^2}(u) \dr u
\end{equation}
only for $Q_f^\ga > Q_2^\ga$. When looking at Bayesian likelihood
contours one should thus keep in mind that a point more than, say,
$3 \si$ away from the ML point is not necessarily ruled out by
data. In order to establish this, one has to look at confidence
contours, \ie ask the frequentist's question. This is pointed out
in a penetrating way by \cite{Gawiser:2001yt}.

\subsubsection*{\bf Maximization instead of marginalization}

In practical applications, involving up to a dozen parameters, it
is an exceptionally demanding task to perform the multidimensional
integral of \rr{eq:marginalization_integral}. A computationally
more feasible alternative which avoids the time consuming
integration is to maximize the parameters we are not interested
in, $\bs{u}$, for each value of the parameters of interest,
$\bs{t}$, obtaining
 \be
 \like(\bs{t}) \propto \max_{\bs{u}} \like(\bs{t}, \bs{u}) \eqdot
 \ee

If the distribution is Gaussian, then the two procedures give the
same result: maximizing $\like(\bs{t}, \bs{u})$ corresponds to
minimization over $\bs{u}$ of the quadratic form $\params^T
\bs{C}^{-1}
\params$, with the notations of (\ref{eq:params_split}) and
(\ref{eq:Fisher_split}). Differentiating with respect to $\bs{u}$,
we find that the minimum of the quadratic form lies at
 \be
 \bs{u} = - \bs{B}^{-1} \bs{G}^T \bs{t} \eqcomma
 \ee
and therefore
 \be
 \like(\bs{t}) \propto \exp{-\dfrac{1}{2}\bs{t}^T
 \left[\bs{A} - \bs{G}\bs{B}^{-1} \bs{G}^T  \right]\bs{t}}\eqcomma
 \ee
which is the same result we found by marginalizing over $\bs{u}$,
\rr{eq:marginalized_covariance}. Numerical investigations have
found that maximization tends to underestimate errors when the
assumption of a Gaussian distribution is not accurately fulfilled
\citep{Efstathiou:1998qr}.

\subsection{Markov chain Monte Carlo}
 \label{chap:data;sec:mcmc}

A big practical limitation to grid based parameter extraction
techniques is that the number of CMB spectra needed scales
exponentially with the dimensionality of the parameter space
considered. Even with fast parallel computing, the required
computational time quickly becomes very large, even for a moderate
number of points in each dimension. Interpolation algorithms and
other optimization techniques have been employed to circumvent
this fundamental limitation, allowing the handling of up to a
dozen parameters \citep{Tegmark:2000qy}. Nevertheless, this method
shows a lack of flexibility if one wants to add new data-sets or
incorporate new parameters or theoretical priors. At the latest
with the coming of WMAP data, the days of grid-based parameter
extraction seem to be over, since the accuracy of WMAP-like data
cannot be exploited with the insufficient resolution and
flexibility offered by this technique.

Markov chain Monte Carlo (hereafter MCMC) methods are now becoming
the standard tool to determine parameters from CMB data, combine
it with large scale structure constraints or investigate the
effect of different priors. As advocated \eg by
\cite{Christensen:2001gj}, MCMC is a method to generate a sequence
of (correlated) samples, called a Markov chain, from the posterior
pdf of the parameters given the data, (\refp{eq:Bayes_Theorem}).
The great advantages are that the computational time scales
approximately linearly with the number of dimensions of the
parameter space, and that once the chain has properly converged
(see below for more details), the marginalized posterior
distribution for the parameter(s) of interest can be simply
recovered by plotting histograms of the sample list, thus avoiding
completely the costly integration. It is easy to adjust the prior
information or to include new data-sets into an existing chain
without having to recompute it, with a procedure called
``importance sampling''.

One can think of the MCMC algorithm as an efficient integration
technique to evaluate the posterior distribution in Bayes'
Theorem, \rrp{eq:Bayes_Theorem}. The Monte Carlo sampling does not
rely on the assumption of Gaussian pdf's: indeed, the direct
sampling of the posterior permits to reveal features due to its
non-Gaussian distribution, and therefore vastly improves on the
methods based on \chisq\ goodness-of-fit described above. Besides
those undeniable advantages over the grid method, the popularity
of MCMC in the cosmology community has been boosted by the timely
public release of the {\textsc{cosmomc}} package
\citep{Lewis:2002ah}, which integrates the code {\textsc{camb}}
for the computation of the CMB power spectra\footnote{Both codes
are available at: {\tt http://cosmologist.info}.} and several
useful tools for the generation and interpretation of Markov
chains using CMB and other cosmological data-sets. Further details
about MCMC methods can be found \eg in
\cite{Book:MCMCinpractice,Book:MacKay}.

The Metropolis-Hastings algorithm
\citep{Metropolis:1953am,Hastings1970} is the core of the sample
generation, and produces a Markov chain whose equilibrium
distribution is the target probability density, here the posterior
$\pdf(\params \vert \data)$. The chain is started from a random
point in parameter space, $\params_0$, and a new point $\params_1$
is proposed with an arbitrarily {\it proposal density
distribution} $q(\params_n,
\params_{n+1})$. The {\it transition kernel}
$T(\params_n,\params_{n+1})$ gives the conditional probability for
the chain to move from $\params_n$ to $\params_{n+1}$, and it must
satisfy the ``detailed balance''
 \be
 \pdf(\params_{n+1} \vert \data)T(\params_{n+1},\params_{n}) =
 \pdf(\params_{n} \vert \data)T(\params_n,\params_{n+1})
 \ee
so that the posterior $\pdf(\params \vert \data)$ is the
stationary distribution of the chain. This is achieved by defining
the transition kernel as
 \begin{align}
  T(\params_n,\params_{n+1}) & \equiv q(\params_n, \params_{n+1})
 \alpha (\params_n, \params_{n+1}) \eqcomma\\
 \alpha (\params_n, \params_{n+1}) & \equiv
 \min \left\{1, \dfrac{\pdf(\params_{n+1} \vert \data) q(\params_{n+1}, \params_{n})}
 {\pdf(\params_{n} \vert \data) q(\params_{n}, \params_{n+1})}
 \right\} \eqcomma
 \end{align}
 where $\alpha (\params_n, \params_{n+1})$ gives the probability
 that the new point is accepted. Since $\pdf(\params \vert \data)
 \propto \like(\data \vert \params) \pdf(\params)$ and for the
 usual case of a symmetric proposal density, $q(\params_n,
 \params_{n+1})= q(\params_{n+1}, \params_{n})$, the new step is
 always accepted if it improves on the posterior, otherwise it is
 accepted with probability $\like(\data \vert \params_{n+1})
 \pdf(\params_{n+1})/\like(\data \vert \params_{n})
 \pdf(\params_{n})$.

The result is a sample list from the target distribution, from
which all the statistical quantities of interest can readily be
evaluated. The samples are correlated with each other, a fact
which does not constitute a problem for the statistical inference
on the parameters; however, importance sampling does require
uncorrelated samples, which can be obtained from the original
chain by suitably ``thinning'' the chain, \ie by retaining only
one sample every $N$, with $N$ of the order of a few thousands.
Other important practical issues in working with MCMC methods
involve:
 \begin{itemize}
 \item {\bf Burn in period:} the initial samples need to be
 discarded, since the chain is not yet sampling from the equilibrium
 distribution. The burn in can roughly be assessed by looking at
 the evolution of the posterior and at the position of the chain
 in parameter space as a function of the step number.
 When the chain is started at a random point of the parameter space,
 the logarithm of the posterior pdf is large
 (and thus the posterior probability is small), and becomes smaller at
 every step as the chain approaches the region where the fit to the data is better. Only
 when the chain has moved in the neighborhood of the ML point the
 curve of the log posterior as a function of the step
 number
 flattens around the best fit value. This is illustrated in
 the left panel of \FIG{fig:chains_evolution}. Another useful diagnostic is the
 evolution in parameter space of multiple chains, which are started from
 different points. In a well-behaved situation
 all of the chains converge after the burn-in period to the same region
 around the ML point, see the right panel of \FIG{fig:chains_evolution} for an illustration.
 \begin{figure}[tb]
\centering
\includegraphics[width=\twofigswidth]{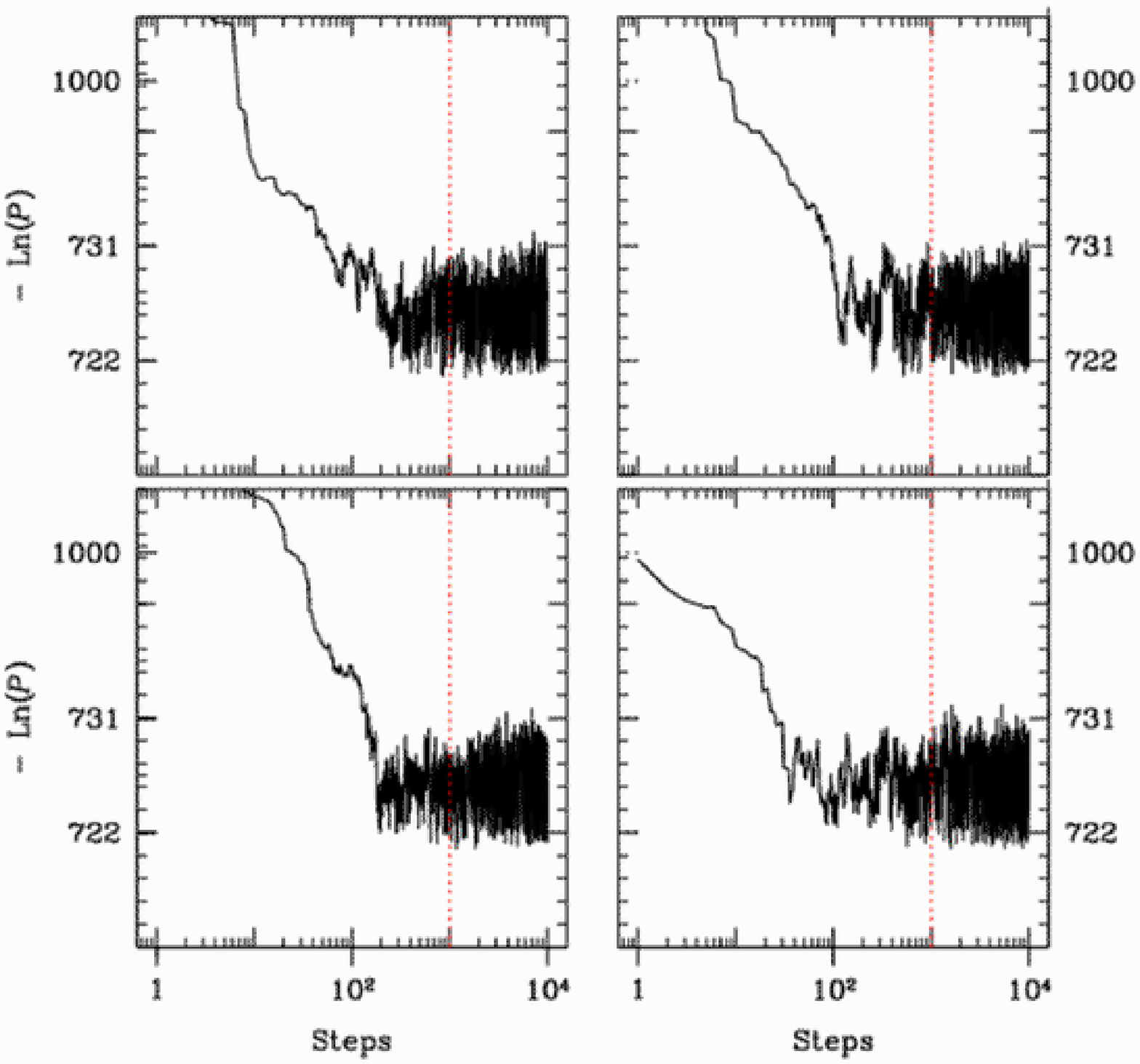}%
\includegraphics[width=\twofigswidth]{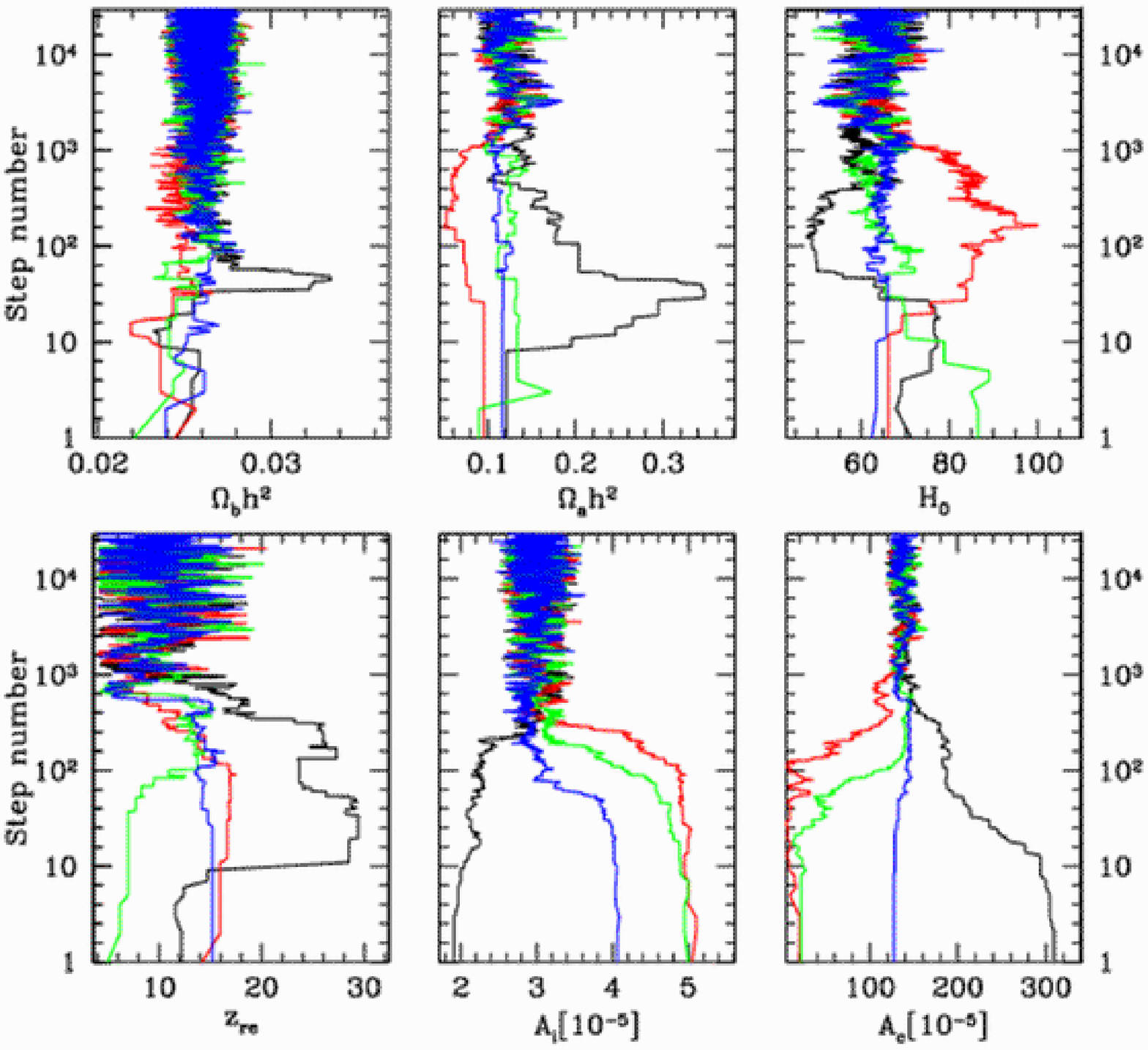}\hfill
\caption[Illustration of the burn-in period for Monte Carlo Markov
chains.]{Illustration of the burn-in period. Left panel: the
logarithm of the (non-normalized) posterior, $ - \ln \pdf(\params
\vert \data)$, as a function of the step number for four Monte
Carlo chains. After the burn-in period (dotted, vertical lines),
the value flattens and the chains are sampling from the target
distribution. Right panel: the four chains (in different colors)
are started in different points of a 6-dimensional parameter space
and all converge to the same region after the burn-in. The
vertical axis gives the number of steps.}
\label{fig:chains_evolution}
\end{figure}
 \item {\bf Convergence:} assessing convergence of the chain essentially means to know when we can stop,
having gathered a number of samples large enough to correctly
derive the statistical quantities of interest. This is in general
a difficult question, see \eg \cite{Cowles96,StatProc} and
references therein. The \textsc{cosmomc} package offers several
useful diagnostic tools, including the \cite{rafterylewis:195}
statistics and the \cite{Gelman92b} criterion.
 \item {\bf Multiple chains:} there is a debate among experts
 about the best strategy between having one long chain or rather
 several shorter ones running in parallel, see \eg
 \cite{Gelman92a,Gelman92b,rafterylewis:195}. Multiple independent chains
 offer the advantage of being computed in parallel, and can be
 started in different points of the parameter space to ensure good
 mixing, \ie an adequate exploration of the whole parameter space.
 \item {\bf Starting points:} after the burn in period, the
 converged chains do not depend on the initial starting points.
 However, it is convenient to start the chains in the proximity of
 the parameter region where the best fit is supposedly located, so
 that convergence will be quickly achieved, and the sophisticated choice of the starting points
 proposed by \cite{Gelman92b} is usually not necessary in cosmological applications. Also one has to take
 into account the fact that the MCMC is a local algorithm, which
 can be trapped inside local minima far away from the global
 minimum of the posterior, an issue which is intimately related
 with the choice of
 the proposal density. The use of simulated annealing algorithm
 via the introduction of a finite temperature for the MC can sometimes
 help in achieving convergence in a weird-shaped parameter space.
 \item {\bf Proposal density:} the optimal choice of the proposal
 density is the key parameter for an efficient implementation of the
 MCMC method. A simple possibility for the proposal density $q(\params_n, \params_{n+1})$
 is a Gaussian with step size $s_i$ along the
 parameter direction $i$, independently on the chain position.
 Finding the optimal value of $s_i$ is a trade-off between a large
 step size, which will result in almost all step being rejected
 and therefore in low efficiency, and a too small value, for which
 the chain performs a random walk and the tails of the distribution will not be adequately sampled,
 giving serious underestimate of the likelihood intervals for the
 parameters. One can also roughly sample the distribution with a
 short chain, construct from the samples the covariance matrix of
 the posterior distribution and use this information to construct
 a new parameter basis approximately aligned with the degeneracy
 directions \citep{Lewis:2002ah}, which ensures a more efficient
 exploration. A sampling method which exploits the known
 degeneracies of the CMB and uses normal parameters as basis
 has been proposed by \cite{Slosar:2003da}, and it can dramatically enhance the
 efficiency of the MCMC algorithm, especially for large data-sets
 as the one expected for the Planck satellite.
 \end{itemize}

\section{Fisher matrix forecasts}
\label{chap:data;sec:fma}

An important issue is to assess quantitatively the expected
performance of future CMB experiments in terms of the precision
reached in the determination of cosmological parameters: this
helps in understanding whether an observed degeneracy is a
consequence of the lack of precision in the data, or else it is of
fundamental nature and will not be lifted by upcoming or even
ideal (\ie cosmic variance limited) measurements; it also gives
estimates of the necessary instrumental characteristics to achieve
a certain precision, and on the optimal observing strategies, \eg
full sky coverage versus high resolution mapping of a patch only.

It is possible and indeed necessary at the development stage of a
CMB experiment to investigate in detail the above questions by
producing mock realizations of the CMB sky and run Monte Carlo
simulations of the observations. From the theorist's point of
view, however, it is often sufficient and preferable to resort to
a simpler alternative, which gives quantitative and accurate
results with very small computational requirements: a Fisher
matrix analysis (FMA)
\citep{Knox:1995dq,Kosowsky:1996mp,Tegmark:1996bz,
Zaldarriaga:1997ch,Bond:1997wr,Eisenstein:1998hr,Efstathiou:1998xx,Tegmark:1999ke}.

\subsection{Experimental parameters}
 \label{chap:data;sec:fma_expepars}

As explained in \SEC{chap:data;sec:Bayesian}, if the LF is a
multivariate Gaussian then the Fisher information matrix defined
in \rr{eq:define_Fisher_matrix} is an estimate of the inverse of
the covariance matrix for the parameters under scrutiny. Since any
LF can be expanded up to second order in the vicinity of the ML
point as in (\ref{eq:like_second_order_expansion}), the goal is to
compute the Fisher matrix for the CMB power spectrum, including
the noise of the future experiment, and estimate from it the
covariance matrix using the results for Bayesian statistics
presented in \SEC{chap:data;sec:Bayesian}.

The estimator (\ref{eq:estimator_for_Cl}) for the CMB temperature
power spectrum (below we generalize the result to include
polarization information as well,
\SEC{chap:data;sec:generalization}) needs to be modified to
subtract off the noise contribution and correct for the fact that
the measured $\alm$'s are a smeared out version of the true ones,
resulting from the convolution of the signal with the experimental
beam, giving \citep{Knox:1995dq,Bond:1997wr}
 \be \label{eq:estimator_with_noise}
 \hCl \equiv
 \left( \dfrac{1}{2\ell + 1}\sum_{m = -\ell}^\ell
 \vert \halm^2 \vert - w_b^{-1} \right) e^{\ell(\ell+1)/ \ell_b^2}
 \eqdot
 \ee

In the above expression, the two experimental parameters are the
{\it inverse weight per solid angle} $w_b$, which accounts for the
experimental noise, and the {\it beam width} $\ell_b$, which
corrects the smoothing due to the Gaussian profile of the beam.
These two parameters are written in terms of the fundamental
specifications of the experiments, namely the rms pixel noise (or
sensitivity per resolution element) $\si_b$ and the angular
resolution $\te_b$ (FWHM) expressed in degrees as
 \be \label{eq:expe_params}
 w_b^{-1} = (\si_b \te_b)^{2} \quad \text{and} \quad
 \ell_b = \sqrt{8\ln2} / \te_b \eqdot
 \ee
In the limit of infinite resolution, $\te_b \rightarrow 0$, and no
experimental noise, $\si_b \rightarrow 0$, we recover the cosmic
variance limited estimator (\ref{eq:estimator_for_Cl}).

As in \SEC{chap:data;sec:cosmicvar}, we can now find the pdf for
(\ref{eq:estimator_with_noise}),
  \be \label{eq:pdf_for_estimator_with_noise}
 \pdf(\hCl) = \dfrac{l}{\Cl +  \noise} \pdf_{\chi^2_l}
 \left( l \dfrac{\hCl+\noise}{\Cl + \noise} \right) \eqcomma
 \ee
recalling $l \equiv 2\ell+1$ and the \chisq\ distribution
displayed in \rr{eq:chi_pdf}. The correction for the noise and the
beam size makes this estimator biased, \ie
 \be \label{eq:estimator_with_noise_EX}
 \EX{\hCl} = \Cl + \noise \eqcomma
 \ee
which is exactly what we need to compensate for the experimental
noise and beam width. From this it follows from (\ref{eq:def_LF})
and (\ref{eq:Gaussian_lnlike}) that the log-normal LF has the form
 \be
 \lnlike(\params) = \sum_{\ell} l
 \left[ \ln \left( \Cl(\params) + \noise \right)
 + \dfrac{\hCl}{\Cl(\params) + \noise}
 \right]
 \ee
and we have dropped several normalization factors which do not
depend on $\params$. Using (\ref{eq:estimator_with_noise_EX}) we
then obtain for the Fisher information matrix defined in
(\ref{eq:define_Fisher_matrix})
 \be \label{eq:Fisher_matrix_expression}
 F_{ij} = \sum_{\ell = \ell_{\text{min}}}^{\ell_{\text{max}}} \dfrac{1}{(\Delta \Cl)^2}
 \dfrac{\partial \Cl}{\partial \te_i}
 \dfrac{\partial \Cl}{\partial \te_j}\evalML
 \eqcomma \ee
where the quantity $(\Delta \Cl)^2$ is the standard deviation on
the estimate of $\Cl$, and takes into account both the cosmic
variance and the experimental error,
 \be \label{eq:Delta_Cl}
 (\Delta \Cl)^2 = \dfrac{2}{2\ell + 1} \left( \Cl + \noise \right)^2
 \eqdot
 \ee
The sum over multipoles runs over the multipole coverage of the
experiment, between $\ell_\text{min}$ and ${\ell_{\text{max}}}$.

Thus once the experimental parameters are specified, the
computation of the Fisher matrix only requires the knowledge of
the derivatives of the power spectrum with respect to the
cosmological parameters. The derivatives are determined
numerically as double sided derivatives, see
\SEC{chap:data;sec:accuracy}, and this requires the computation of
$2p + 1$ spectra only for $p$ parameters, which is a very small
computational effort compared with the full numerical exploration
of the likelihood surface.

\subsection{Generalizations}
\label{chap:data;sec:generalization}

In this section, we develop the necessary general machinery which
refines the above results including a more detailed experimental
parametrization and polarization information.

Most experiments present several frequency channels, each of them
characterized by its own sensitivity $\si^{T,P}_c$ and angular
resolution $\te^{T,P}_c$, both for temperature ($T$) and
E-polarization ($P$). Furthermore, even full-sky experiments only
cover a fraction of the sky, since point source subtraction,
foreground removal and galactic plane cuts have to be performed on
the full-sky maps. This can be approximately taken into account by
assigning a ``clean'' fraction $f\sky$ to the experimental
coverage. These factors are accounted for by generalizing the
expression (\ref{eq:Delta_Cl}) to \citep{Efstathiou:1998xx}
 \be \label{eq:Delta_Cl_general}
 (\Delta \Cl)^2 = \dfrac{2}{(2\ell + 1)f\sky} \left( \Cl + B_\ell^{-2} \right)^2
 \eqcomma
 \ee
where the {\it inverse noise term} $B_\ell$ is given by
 \be \label{eq:noise_term}
 B_\ell^2 \equiv \sum_c w_c e^{-\ell(\ell+1)/\ell_c^2}
 \ee
and $w_c, \ell_c$ are given by (\ref{eq:expe_params}) for each
channel $c$.

In the more general case, we also want to include $E$ polarization
and temperature-polarization correlation ($C$) along with
temperature information: then instead of a single derivative we
have a vector of three derivatives with the weighting given by the
the inverse of the covariance matrix of the spectra, and the
Fisher matrix is given by \citep{Zaldarriaga:1997xe},
\begin{equation}
F_{ij}=\sum_{\ell = \ell_{\text{min}}}^{\ell_{\text{max}}}
\sum_{X,Y}{\partial {C}_{X\ell} \over
\partial \theta_i} {\rm Cov}^{-1}({C}_{X\ell}{C}_{Y\ell}){\partial
{C}_{Y\ell} \over
\partial \theta_j} \evalML
\end{equation}
where $\rm Cov^{-1}$ is the inverse of the covariance matrix for
the spectra evaluated at the ML point $\params\ML$, $\theta_i$ are
the cosmological parameters we want to estimate and $X,Y$ stands
for $T$ (temperature), $E$ (polarization mode), or $C$
(cross-correlation of the power spectra for $T$ and $E$).

For each $\ell$ one has to invert the covariance matrix and sum
over $X$ and $Y$. The diagonal terms of the covariance matrix
between the different estimators are given by
\begin{eqnarray}
{\rm Cov }({C}_{T\ell}^2)&=&\frac{2}{(2 \ell + 1) f_{\rm sky}
}({C}_{Tl}+ {B}_{T\ell}^{-2})^2
 \\
{\rm Cov }({C}_{E\ell}^2)&=&\frac{2}{(2 \ell + 1) f_{\rm
sky}}({C}_{E\ell}+ {B}_{P\ell}^{-2})^2
 \\
{\rm Cov }({C}_{C\ell}^2)&=&\frac{1}{(2 \ell + 1) f_{\rm
sky}}\left[{C}_{C\ell}^2+ ({C}_{T\ell}+{B}_{T\ell}^{-2})
({C}_{E\ell}+{B}_{P\ell}^{-2})\right] \eqcomma
\end{eqnarray}
and the off diagonal terms are
\begin{eqnarray}
{\rm Cov }({C}_{T\ell}{C}_{E\ell})&=&\frac{2}{(2 \ell + 1) f_{\rm
sky} }{C}_{C\ell}^2
 \\
{\rm Cov }({C}_{T\ell}{C}_{C\ell})&=&\frac{2}{(2 \ell + 1) f_{\rm
sky}}{C}_{C\ell} ({C}_{T\ell}+{B}_{T\ell}^{-2})
 \\
{\rm Cov }({C}_{E\ell}{C}_{C\ell})&=&\frac{2}{(2 \ell + 1) f_{\rm
sky}}{C}_{C\ell} ({C}_{E\ell}+{B}_{P\ell}^{-2}) \eqcomma
\end{eqnarray}
where ${B}_{T\ell}^{-2}={B}_{\ell}^{-2}$ given in
\rr{eq:noise_term}  and ${B}_{P\ell}^2$ is obtained using a
similar expression but with the experimental specifications for
the polarization channels.

\subsection{Accuracy issues}
\label{chap:data;sec:accuracy}

The accuracy of the Fisher matrix predictions for the errors
depends on a number of issues:
 \begin{itemize}
 \item The FMA \textit{assumes}
that the true values of the parameters are in the vicinity of the
ML point $\params\ML$. The validity of the results therefore
depends on this assumption, as well as on the assumption that the
$a_{\ell m}$'s are independent Gaussian random variables.
 \item This is a local method based on a quadratic expansion of
 the LF. Only if the FMA predicted errors are small
enough, the method is self-consistent and we can expect the FMA
prediction to correctly reproduce the exact behavior, and in
particular the correlations between parameters, thus revealing the
degeneracy directions. The expansion up to second order is {\it
exact} if the dependence of the $\Cl$ on the parameters is linear,
therefore great importance is attached to the choice of the
parameter set with respect to the FMA is performed. As shown in
\cite{Kosowsky:2002zt}, employing the normal parameters set
discussed in \SEC{chap:params:sec:normal} as a base, the accuracy
of the FMA predictions is greatly enhanced. This is because the
spectra are almost linear in the normal parameters in the vicinity
of the best fit.
 \item Special care must be taken when computing the derivatives of the power
spectrum with respect to the cosmological parameters. This
differentiation strongly amplifies any numerical errors in the
spectra, leading to larger derivatives, which would artificially
break degeneracies among parameters. Double--sided derivatives
reduce the truncation error from second order to third order
terms, but the correct choice of the step size is a trade-off
between truncation error and numerical inaccuracy dominated cases
\citep{Book:Press92}.
 \end{itemize}

\section{CMB observations: a brief historical account}
 \label{chap:data;sec:obs}

The experimental status of CMB observations has made giant leaps
over the last ten years, thanks to spectacular advancements in
detector technology. As demonstrated in \CHAP{chap:beyondsp}, CMB
data nowadays provide stringent tests which severely constrain
cosmological model building, and call for more refined theoretical
and computational approaches which take into account subtle
physical effects which were so far ignored or thought to be
irrelevant. Here we provide a personal selection of a few
milestones of this development, in order to put the current and
future experimental achievements into a wider perspective.

The first detection of temperature anisotropy came in 1992 with
the Differential Microwave Radiometer (DMR) aboard the COBE
satellite after one year of observations on angular scales larger
than $7\DEG$ \citep{Smoot:1992td,Wright:1992tf} or multipoles
$\lsim 20$. The key results of the full four year DMR observations
are summarized in \citet[see references therein]{Bennett:1996ce}:
the quadrupole amplitude was measured for the first time, the
spectral tilt of the large scale spectrum was found to be
compatible with an Harrison-Zel'dovich spectrum and no evidence of
non-Gaussianity of the fluctuations was discovered in the data.
The FIRAS instrument was devoted to the study of the CMB spectrum
\citep{Fixsen:1996}, and obtained a precision measurement of its
temperature ($T=2.728 \pm 0.002$ K), while constraining deviations
from a perfect black body spectrum to be less than about one part
in $10^{5}$ with $95\%$ confidence.

The Saskatoon and Toco data provided the first hint for the
presence of the first adiabatic peak
\citep{Netterfield:1996nb,Miller:1999qz,Knox:2000dc}, but at the
turning of the millennium several groups independently reported
measurements of the temperature anisotropy with a resolution of a
few arcminutes, sufficient to unambiguously reveal the first peak
and start exploring the subsequent ones: BOOMERanG
\citep{deBernardis:2001gv,Netterfield:2001yq} and Maxima
\citep{Hanany:2000qf,Lee:2001yp}, both balloon-borne bolometric
experiments, mapped the multipole region $80 \lsim \ell \lsim
1000$; the CBI \citep{Padin:2000wh} and DASI
\citep{Halverson:2001yy} ground based interferometers covered a
similar multipole range but with a completely different
technology, which had the advantage of being free from the
calibration uncertainty of bolometric receivers. The Archeops
experiment \citep{Benoit:2002mk}, conceived as a balloon-borne
precursor of the HFI bolometric instrument for the Planck
satellite, observed a larger portion of the sky, and thus provided
an estimation of the temperature power spectrum which for the
first time encompassed the first peak region and also partially
overlapped with the COBE measurement, in the range $15 \leq \ell
\leq 350$. Given the experimental calibration uncertainty of the
bolometers, which is about $10-20\%$, this permits to test the
relative calibration between COBE and the other experiments with
data in the $\ell \gsim 50$ region, and perform a comparison of
the height of the first peak with respect to the large scale
plateau. All of this data generally agrees well on the position
and shape of the first peak, but their resolution is insufficient
to permit the reconstruction of the subsequent ones with high
confidence \citep{deBernardis:2001gv,Durrer:2001jz}.

\begin{figure}[!b]
\centering
\includegraphics[width=\onefigwidth]{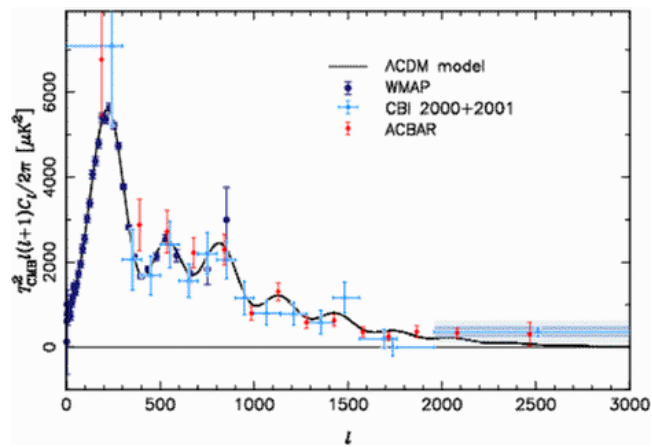}
\caption[The small scale temperature spectrum observed by the CBI
and ACBAR experiments.]{The small scale temperature angular power
spectrum observed by CBI ``mosaic'' during two years  and by
ACBAR. The shaded region shows the excess power at small scale,
compatible with the SZ effect. Reprinted from
\cite{Readhead:2004gy}.} \label{fig:cmb_high_ell}
\end{figure}

From the point of view of parameter extraction, each of the above
data sets by its own as well as their combination leads to a broad
agreement of an approximately flat $\Omega_\text{tot} \sim 1$
universe with scale invariant spectral index $n_s \sim 1$, with
the $1\si$ likelihood intervals being of the order of $10\%$ and
somewhat depending on the compilation of data and on the prior
assumed
\citep{Stompor:2001xf,Lange:2000iq,Pryke:2001yz,Netterfield:2001yq}.
The estimation of the baryon density proved to be more
controversial, because of discrepancies and a lack of resolution
at the level of the second and third peak: in particular, the
BOOMERanG 1998 and MAXIMA data seem to favor a baryon content
about $50\%$ larger than predicted by BBN, around $\Om_b h^2 \sim
0.03$ \citep{Tegmark:2000dr,Lange:2000iq,Stompor:2001xf}, a
discrepancy which disappears with the improved beam reconstruction
of the BOOMERanG 2000 observations \citep{Netterfield:2001yq}.
Inclusion of supernov\ae~ data or the Hubble Space Telescope prior
for the Hubble constant, together with the flatness determination,
points toward a universe dominated by a cosmological constant.

Before the WMAP satellite delivered its results, ground based
instruments pressed on and opened up two new observational
directions: very small scale observations ($4\arcmin-5 \arcmin$)
and E-polarization detection. The CBI interferometer, in two
different configurations called ``mosaic'' and ``deep field'',
obtained measurements of the temperature power spectrum up to
$\ell = 3500$ \citep{Sievers:2002tq,Mason:2002tm}, and it was
argued that the excess power observed at high multipoles could be
due to the SZ effect, from which a precise determination of
$\si_8$ could possibly be obtained \citep{Bond:2002tp}. The ACBAR
experiment, a bolometric instrument installed at the South Pole,
found small scale power consistent with the results of CBI,
without however being able to place tighter constraints on its
origin \citep{Goldstein:2002gf,Kuo:2002ua}. More recently, the
results of two years of observations with the CBI ``mosaic''
configuration, give smaller errors in the $\ell \sim 2000$ region,
due to the longer integration time and to an improved absolute
calibration derived from the WMAP data, see
\FIG{fig:cmb_high_ell}. Beside revealing effects due to secondary
anisotropies as the SZ effect, the small scale measurements are
helpful in better constraining $n_\SCAL$, $\tau_\reion$ and
possible features in the power spectrum (like a ``running'', \ie a
scale dependence of $n_\SCAL$) because of the larger lever arm
they offer when combined with WMAP and large scale structure data
\citep{Readhead:2004gy}.
\begin{figure}[tb]
\centering
\includegraphics[angle=90, width=\twofigswidth]{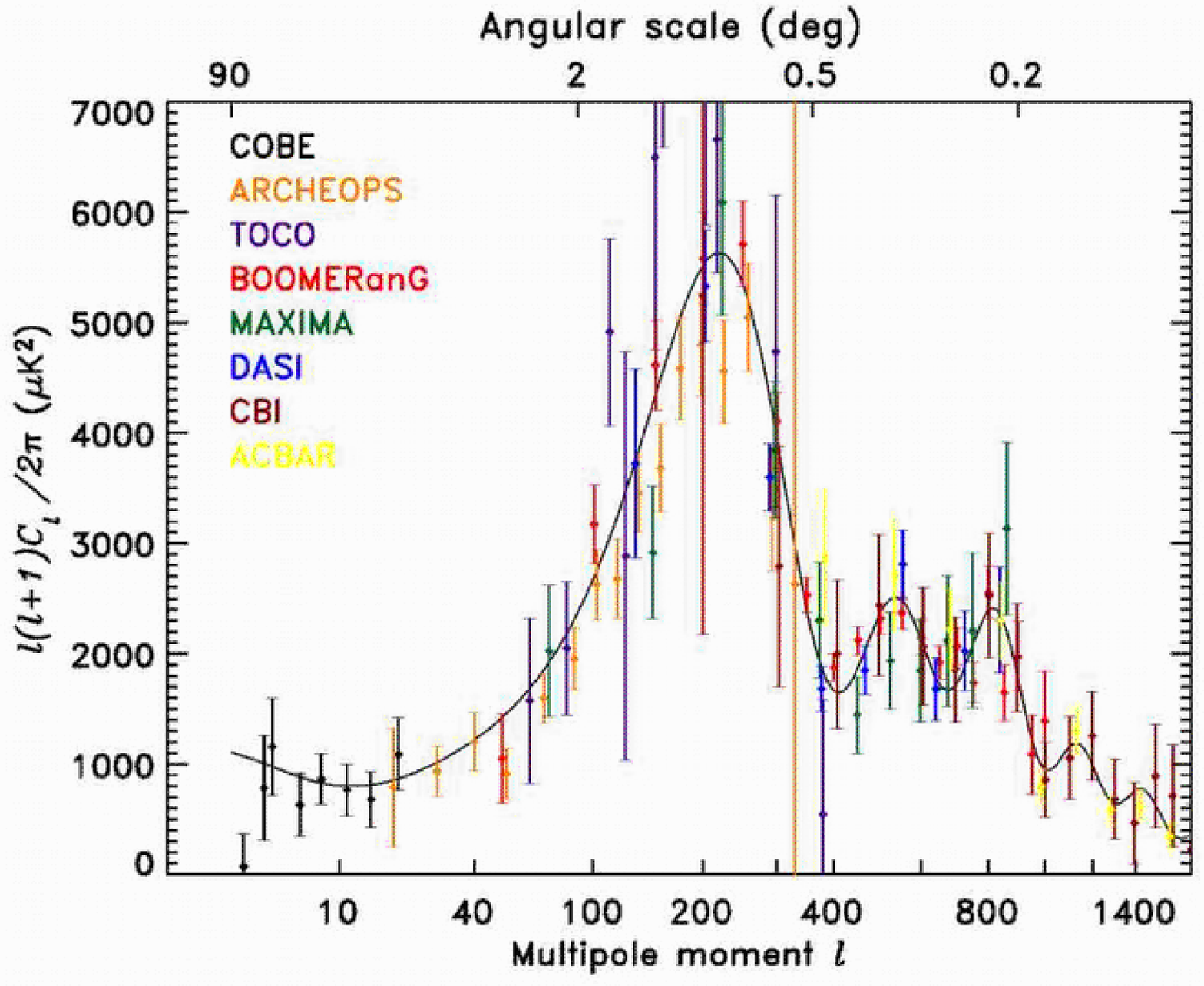}\hfill%
\includegraphics[angle=90, width=\twofigswidth]{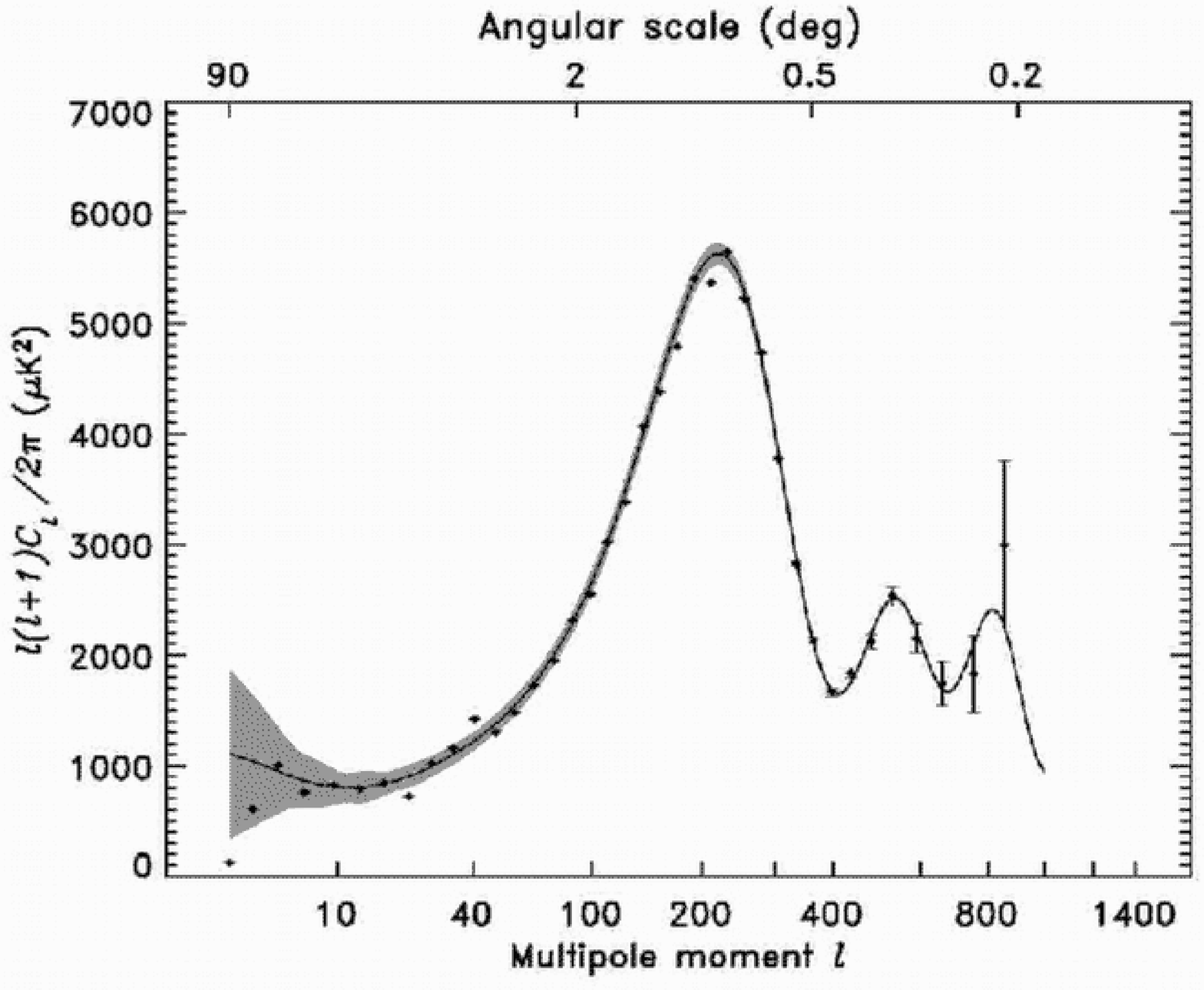}\hfill%
\caption[A compilation of pre-WMAP CMB temperature anisotropy data
compared with the WMAP temperature power spectrum.]{The
spectacular increase of the accuracy of CMB observations: in the
left panel, a compilation of pre-WMAP temperature power spectrum
measurements obtained between 1996 (COBE) and 2003 (CBI) is
compared with the WMAP first year data in the right panel,
released in February, 2003. The error-bars give the $1\si$
uncertainty due to the measurement errors, while the shaded region
represent the cosmic variance limit. Both figures reprinted from
\cite{Hinshaw:2003ex}.} \label{fig:cmb_data_overview}
\end{figure}

\afterpage{\clearpage}

 The DASI interferometer reported in the
second half of 2002 the first detection of E-polarization, which
was observed on degree angular scales with almost $5\si$
confidence \citep{Kovac:2002fg}, thereby opening the epoch of
polarization measurements.

The first year WMAP data, unveiled in February 2003
\citep{Bennett:2003bz,Hinshaw:2003ex}, essentially confirmed the
picture which had emerged from pre-WMAP observations, see
\FIG{fig:cmb_data_overview}: the height of the first peak was
corrected by about $10\%$, showing more power than in the previous
data, while the large scale spectrum confirmed the DMR results.
The second peak is now accurately outlined, while the full four
years data should allow to obtain good resolution up to $\ell \sim
1000$ in temperature. The low power of the quadrupole remains
troublesome, since it is still not clear whether it is pointing to
new physics or just a consequence of systematical errors. The
observation of the temperature-polarization correlation up to
$\ell \sim 500$ \citep{Kogut:2003et} has proved very useful in
order to better constrain parameters. The exquisite quality of the
power spectra has tightened the $1\si$ likelihood intervals to a
few percent for most cosmological parameters
\citep{Spergel:2003cb}, and the central value has remained in the
region preferred by earlier data, with two interesting exceptions:
the TE data favor a much larger reionization optical depth than
previously thought, and there seems to be a slight preference for
a ``running'' (\ie scale dependent) spectral index
\citep{Peiris:2003ff}.

 A complete overview of the evolution of data
and of the cosmological parameters derived from it can be found in
the review by \cite{Bond:2003ur}.

\chapter{Beyond standard parameters}
\label{chap:beyondsp}
This chapter is devoted to the investigation of three scenarios
involving non-standard cosmological parameters, and focuses on the
ability of constraining them using present and future CMB
observations: the existence of extra relativistic particles
(\SEC{chap:beyondsp;sec:rel}); the determination of the primordial
helium mass fraction (\SEC{chap:bspII;sec:helium}); and possible
time variations of the fine structure constant
(\SEC{chap:bspIII;sec:alpha}).

Until recently, the effects induced by these parameters on the CMB
where considered too small to be observable, or else irrelevant;
however, the era of precision cosmology that we are entering
requires on one hand that we check the consequences of our
assumptions on the standard results for other parameters (as in
the case of the neutrino families and the helium fraction); on the
other hand, it allows us to put under close scrutiny very subtle
effects which could previously be safely neglected because of the
less accuracy of the data sets.

\section{Extra relativistic particles}
\label{chap:beyondsp;sec:rel}

This section is based on the work published in
\citet*{Bowen:2001in}, which was carried out for the most part
during my stay in Oxford. We investigate one possible modification
to the standard scenario, namely variations in the parameter
$\omega_{\rm rel} = \Omega_{\rm rel} h^2$ which describes the
energy density of relativistic particles. The original work has
been performed in 2001, and therefore the results presented here
of the pre-WMAP data analysis are nowadays somewhat outdated.
However, the focus is on the degeneracies involving $\om_{\rm
rel}$ and as such the conclusions drawn are still valid.
Furthermore, the subsequent analysis by several groups of the
actual WMAP data permits a comparison between the forecasts
obtained with the Fisher matrix technique in 2001 and the real
case, showing a very satisfactory agreement and validating the
method used.

After offering the motivations for our study in
\SEC{chap:beyondsp;sec:motivation}, we review various physical
mechanisms that can lead to a change in $\omega_{\rm rel}$ with
respect to the standard value in \SEC{chap:beyondsp;sec:Neff}. In
\SEC{chap:beyondsp;sec:cmbomegar}, we illustrate how the CMB
angular power spectrum depends on this parameter and identify
possible degeneracies with other parameters, then present in
\SEC{chap:beyondsp;sec:cmbanalysis} a likelihood analysis from
pre-WMAP CMB data and show which of the constraints on the various
parameters are affected by variations in $\omega_{\rm rel}$.
Section \ref{chap:beyondsp;sec:fma} forecasts the precision in the
estimation of cosmological parameters for the space missions WMAP
and Planck, and then compares the predictions with actual data
analysis performed on the first year WMAP data.

\subsection{Motivation}
 \label{chap:beyondsp;sec:motivation}

CMB data analysis taking into account variations in the density of
relativistic particles has been previously undertaken by many
authors \citep{Hannestad:2000hc,Esposito:2000sv,
Kneller:2001cd,Hannestad:2001hn,Hansen:2001hi,Zentner:2001zr},
giving rather crude upper bounds, which are significantly improved
only by including priors on the age of the universe or by
including supernovae (SN) or large scale structure (LSS) data.  It
is worth emphasizing that there is little difference in the bounds
on $N_{\rm eff}$, the effective number of relativistic species,
obtained from old and recent CMB data because of the degeneracy
described in detail below. We focus here on the effects that the
inclusion of this parameter, $\omega_{\rm rel}$, has on the
constraints of the remaining parameters in the context of purely
adiabatic models.

As shown below -- and as observed previously, see \eg
\cite{Hu:1998tk} -- there is a strong degeneracy between
$\omega_{\rm rel}$ and the physical density of non-relativistic
matter, $\omega_m \equiv \Omega_m h^2$. This is important, because
an accurate determination of $\omega_m$ from CMB observations (and
of $\Omega_m$ by including the Hubble Space Telescope result
$h=0.72\pm0.08$) can be useful for a large number of reasons.
First of all, determining the cold dark matter content,
$\omega_{\rm cdm} = \omega_m-\omega_b$ can shed new light on the
nature of dark matter. The thermally averaged product of
cross-section and  thermal  velocity of the dark matter candidate
is related to $\omega_m$, and this relation can be used to analyze
the implications for the mass spectra in versions of the
Supersymmetric Standard Model, see \eg
\cite{Barger:2001yy,Djouadi:2001yk,Ellis:2001yu}. The value of
$\Omega_m$ can be determined in an independent way from the
mass-to-light ratios of clusters, and the present value is $0.1 <
\Omega_m < 0.2$ \citep{Carlberg:1996if,Bahcall:2000zu}.
Furthermore, a precise measurement of $\Omega_m$ will be a key
input for determining the redshift evolution of the equation of
state parameter $w(z)$ and thus discriminating between different
quintessential scenarios, see \eg \cite{Weller:2001gf}.

\subsection{Effective number of relativistic species}
 \label{chap:beyondsp;sec:Neff}

The energy density of relativistic particles can conveniently be
parameterized via the effective number of relativistic species,
$N_{\rm eff}$: in the standard model $\omega_{\rm rel}$ includes
photons and neutrinos, and it can be expressed as
\begin{equation}\label{eq:neff}
\omega_{\rm rel} = \omega_{\gamma} + N_{\rm eff} \cdot
\omega_{\nu}
\end{equation}
where $\omega_{\gamma}$ is the energy density in photons and $
\omega_{\nu}$ is the energy density in one active neutrino family.
In geometrical units, where $G=\hbar=c=1$, one has $\omega_x =
4\pi^3/45 \cdot g_x T_x^4$, where $g_x$ and $T_x$ are the
relativistic degrees of freedom and the temperature of species
$x=\gamma, \nu$, respectively. Measuring $\omega_{\rm rel}$ thus
gives a direct observation of the effective number of neutrinos,
$N_{\rm eff}$. Naturally there are only three active neutrinos,
and $N_{\rm eff}$ is simply a convenient parametrization for the
extra possible relativistic degrees of freedom
\begin{equation}
N_{\rm eff} = 3 + \Delta N  \eqdot
\end{equation}
Thus $\omega_{\rm rel}$ includes energy density from all the
relativistic particles: photons, neutrinos, and additional
hypothetical relativistic particles such as a light majoron or a
sterile neutrino. Such hypothetical relativistic particles are
strongly constrained from standard Big-Bang nucleosynthesis (BBN),
where the allowed extra relativistic degrees of freedom typically
are expressed through the effective number of neutrinos, $N_{\rm
eff} = 3 + \Delta N_{\rm BBN} $.  BBN bounds are typically about
$\Delta N_{\rm BBN}  < 0.2 - 1.0$
\citep{Burles:1999zt,Lisi:1999ng}.

One should, however, be careful when comparing the effective
number of neutrino degrees of freedom at the time of BBN (neutrino
decoupling) and at the formation of the CMBR (photon decoupling).
This is because the energy density in relativistic species may
change from the time of BBN ($T \sim$ MeV) to the time of last
rescattering ($T \sim$ eV), as explained in \cite{Hansen:2001hi}.
For instance, if one of the active neutrinos has a mass in the
range eV $< m <$ MeV and decays into sterile particles such as
other neutrinos, majorons etc. with lifetime $t(\mbox{BBN}) < \tau
< t(\mbox{CMBR})$, then the effective number of neutrinos at CMBR
would be substantially different from the number at BBN
\citep{White:1995as}. Such massive active neutrinos, however, do
not look very natural any longer in view of the recent
experimental results on neutrino oscillations
\citep{Fogli:2001xt,Gonzalez-Garcia:2000sq}, showing that all
active neutrinos are likely to have masses smaller than $0.1$ eV.
One could instead consider sterile neutrinos mixed with active
ones which could be produced in the early universe by scattering,
and subsequently decay.  The mixing angle must then be large
enough to thermalize the sterile neutrinos, and this can be
expressed through the sterile to active neutrino number density
ratio $n_s / n_\nu \approx 4 {\cdot} 10^4 \sin^2 2\theta \
(m/\mbox{keV}) (10.75/g^*)^{3/2}$ \citep{Dolgov:2000ew}, where
$\theta$ is the mixing angle, and $g^*$ counts the relativistic
degrees of freedom, such that $n_{s}/n_{\nu}=1$ or $\Delta
g^{*}=7/8$ increases $N_{\rm eff}$ by one unit. With $n_s/n_\nu$
of order unity we use the decay time, $\tau \approx 10^{20}
(\mbox{keV}/m)^5 /\sin^2 2\theta$ sec, and one finds, $\tau
\approx 10^{17} (\mbox{keV}/m)^4$~yr, which is much longer than
the age of the universe for $m \sim \mbox{keV}$, so they would
certainly not have decayed at $t(\mbox{CMBR})$. A sterile neutrino
with a mass of a few MeV would seem to have the right decay time,
$\tau \sim 10^5$~yr, but this is excluded by standard BBN
considerations \citep{Kolb:1991sn,Dolgov:1998it}. More inventive
models with particles decaying during last rescattering cannot
simply be treated with an $N_{\rm CMB} $ that is constant in time,
see \eg \cite{Kaplinghat:1998ry}, and we will not discuss such
possibilities further here.

Even though the simplest models predict that the relativistic
degrees of freedom are the same at BBN and CMB times, one could
construct models such as quintessence
\citep{Albrecht:1999rm,Skordis:2000dz} which effectively could
change $\Delta N$ between BBN and CMB \citep{Bean:2001wt}.
Naturally $\Delta N$ can be both positive and negative. For BBN,
$\Delta N$ can be negative if the electron neutrinos have a
non-zero chemical potential \citep{Kang:1992xa,Kneller:2001cd}, or
more generally with a non-equilibrium electron neutrino
distribution function \citep{Hansen:2000td}. To give an explicit
(but highly exotic) example of a different number of relativistic
degrees of freedom between BBN and CMB, one could consider the
following scenario. Imagine another two sterile neutrinos, one of
which is essentially massless and has a mixing angle with any of
the active neutrinos just big enough to bring it into equilibrium
in the early universe, and one with a mass of $m_{\nu_s}=3$ MeV
and decay time $\tau_{\nu_s}=0.1$ sec, in the decay channel $\nu_s
\rightarrow \nu_e + \phi$, with $\phi$ a light scalar. The
resulting non-equilibrium electron neutrinos happen to exactly
cancel the effect of the massless sterile state, and hence we have
$\Delta N_{\rm BBN}  = 0$. However, for CMB the picture is much
simpler, and we have just the stable sterile state and the
majoron, hence $\Delta N_{\rm CMB}  = 1.57$.  For CMB, one can
imagine a negative $\Delta N$ from decaying particles, where the
decay products are photons or electron/positrons which essentially
increases the photon temperature relative to the neutrino
temperature \citep{Kaplinghat:2000jj}. Such a scenario also
naturally dilutes the baryon density, and the agreement on
$\omega_b$ from BBN and CMB gives a bound on how negative $\Delta
N_{\rm CMB} $ can be. Considering all these possibilities, we will
therefore not make the usual assumption, $\Delta N_{\rm BBN} =
\Delta N_{\rm CMB} $, but instead consider $\Delta N_{\rm CMB} $
as a completely free parameter in the following analysis.

The standard model value for $N_{\rm eff}$ with three active
neutrinos is $3.044$.  This small correction arises from the
combination of two effects arising around the temperature $T \sim
\mbox{MeV}$. These effects are the finite temperature QED
correction to the energy density of the electromagnetic plasma
\citep{Heckler:1994tv}, which gives $\Delta N = 0.01$
\citep{Lopez:1998vk,Lopez:1998aq}. If there are more relativistic
species than active neutrinos, then this effect will be
correspondingly higher \citep{Steigman:2001px}.  The other effect
comes from neutrinos sharing in the energy density of the
annihilating electrons \citep{Dicus:1982bz}, which gives $\Delta N
= 0.034$ \citep{Dolgov:1997mb,Esposito:2000hi}. Thus one finds
$N_{\rm eff} = 3.044$. An accurate analysis which takes into
account both of this effects simultaneously has been performed by
\cite{Mangano:2001iu} and the result indicates that the combined
effect is slightly smaller, $N_{\rm eff} = 3.0395$.

\subsection{CMB theory and degeneracies}
\label{chap:beyondsp;sec:cmbomegar}

As explained in detail in \CHAP{chap:params}, the structure of the
$C_{\ell}$ spectrum depends on a restricted combination of
cosmological parameters, which are physically probed by the CMB;
simplifying somewhat the normal parameters set introduced in
\SEC{chap:params:sec:normal}, we focus here on the four
cosmological parameters
\begin{equation}
\omega_b \, \, \, , \, \, \, \omega_m \, \, \, , \, \, \,
\omega_{\rm rel} \, \, \, \mbox{and} \, \, \, \Rshift\, \, \, ,
\label{eq:4par}
\end{equation}
the physical baryonic density $\omega_b \equiv \Omega_b h^2$, the
energy density in matter $\omega_m \equiv (\Omega_{\rm cdm}
+\Omega_b)h^2$, the energy density in radiation $\omega_{\rm rel}$
and the shift parameter $\R \equiv \ell_{\rm ref}  / \ell$, which
gives the position of the acoustic peaks with respect to a flat,
$\Omega_\Lambda = 0$ reference model, see
\rrp{eq:def_shift_parameter}. In previous analysis
\citep{Efstathiou:1998xx,Melchiorri:2000px}, the parameter
$\omega_{\rm rel}$ has been kept fixed to the standard value,
while here we will allow it to vary.

It is therefore convenient to write
 \be
 \omega_{\rm rel} = 4.13 \cdot
10^{-5} (1 + 0.135 \cdot \Delta N_{\text{CMB}})
 \ee
 (taking $T_{\rm CMB}  = 2.726$
K), where $\Delta N_{\text{CMB}}$ is the excess number of
relativistic species with respect to the standard model, $N_{\rm
eff} = 3 + \Delta N_{\text{CMB}}$, and we drop the subscript CMB
from now on. The shift parameter $\R$ depends on $\Omega_m \equiv
\Omega_{\rm cdm} + \Omega_b$, on the curvature $\Omega_{\kappa}
\equiv 1 - \Omega_{\Lambda} - \Omega_m - \Omega_{\rm rel}$, and on
$\Omega_{\rm rel}=\omega_{\rm rel}/h^2$ through
 \be
 \label{eq:def_r}
 \R = \left( 1 - \frac{1}{\sqrt{1 + z_{\rm dec}}} \right)
 \frac{\sqrt{| \Omega_{\kappa}| }}{ \Omega_m}
       \frac{2}{\chi(\Delta \tau)}
    \left[ \sqrt{\Omega_{\rm rel} +
    \frac{\Omega_m}{1 + z_\dec} } - \sqrt{\Omega_{\rm rel}} \right]
    \eqcomma
 \ee
 where $z_{\rm dec} $ is a function of the physical baryon
density and $\chi(\Delta \tau)$ is given in \rrp{eq:Delta_tau}.
Eq.~(\ref{eq:def_r}) generalizes the expression for $\R$ given in
(\refp{eq:approximate_Rshift}) to the case of non-constant $
\Omega_{\rm rel}$.

\begin{figure}
\centering
\includegraphics[width=\twofigswidth]{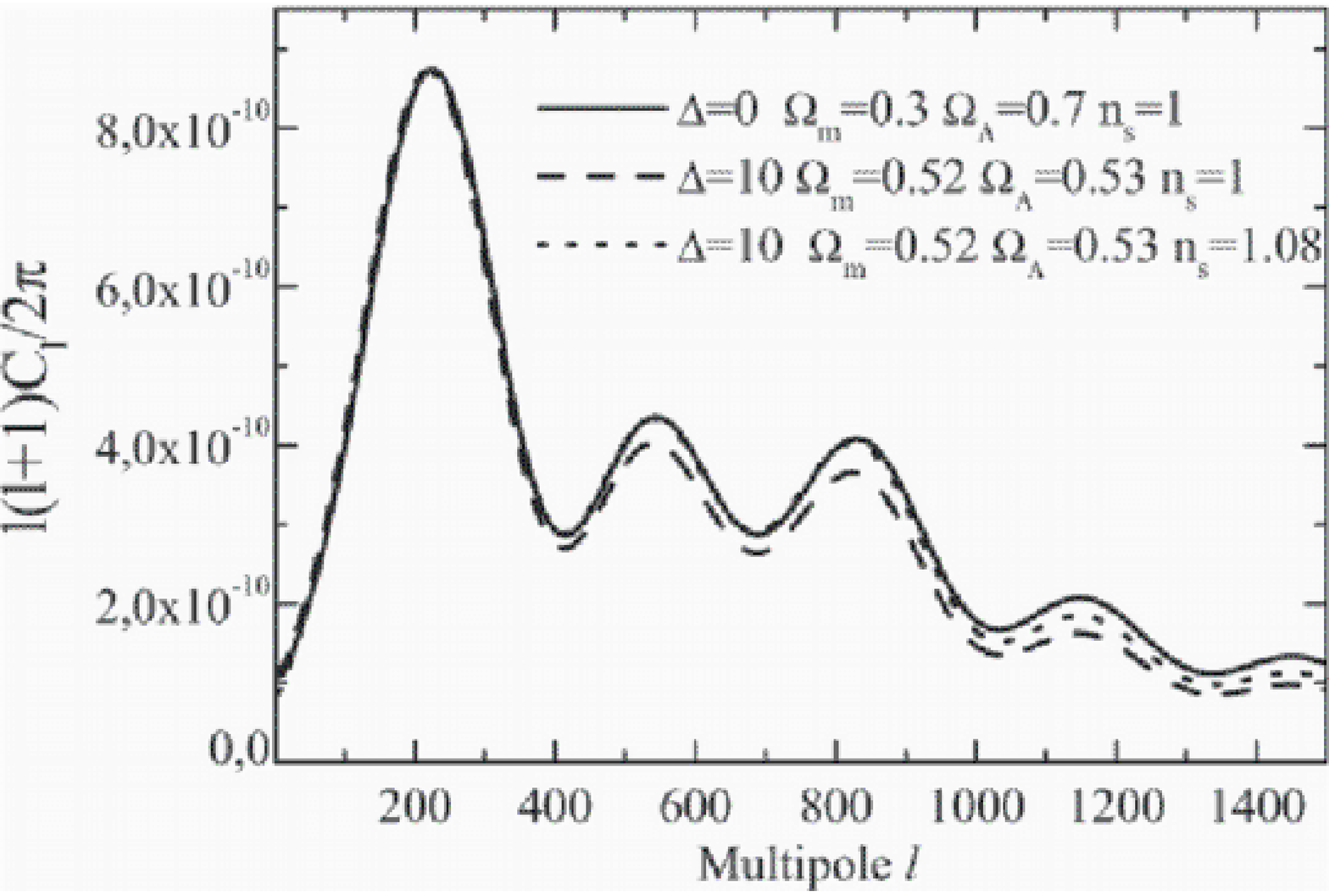}\hfill%
\includegraphics[width=\twofigswidth]{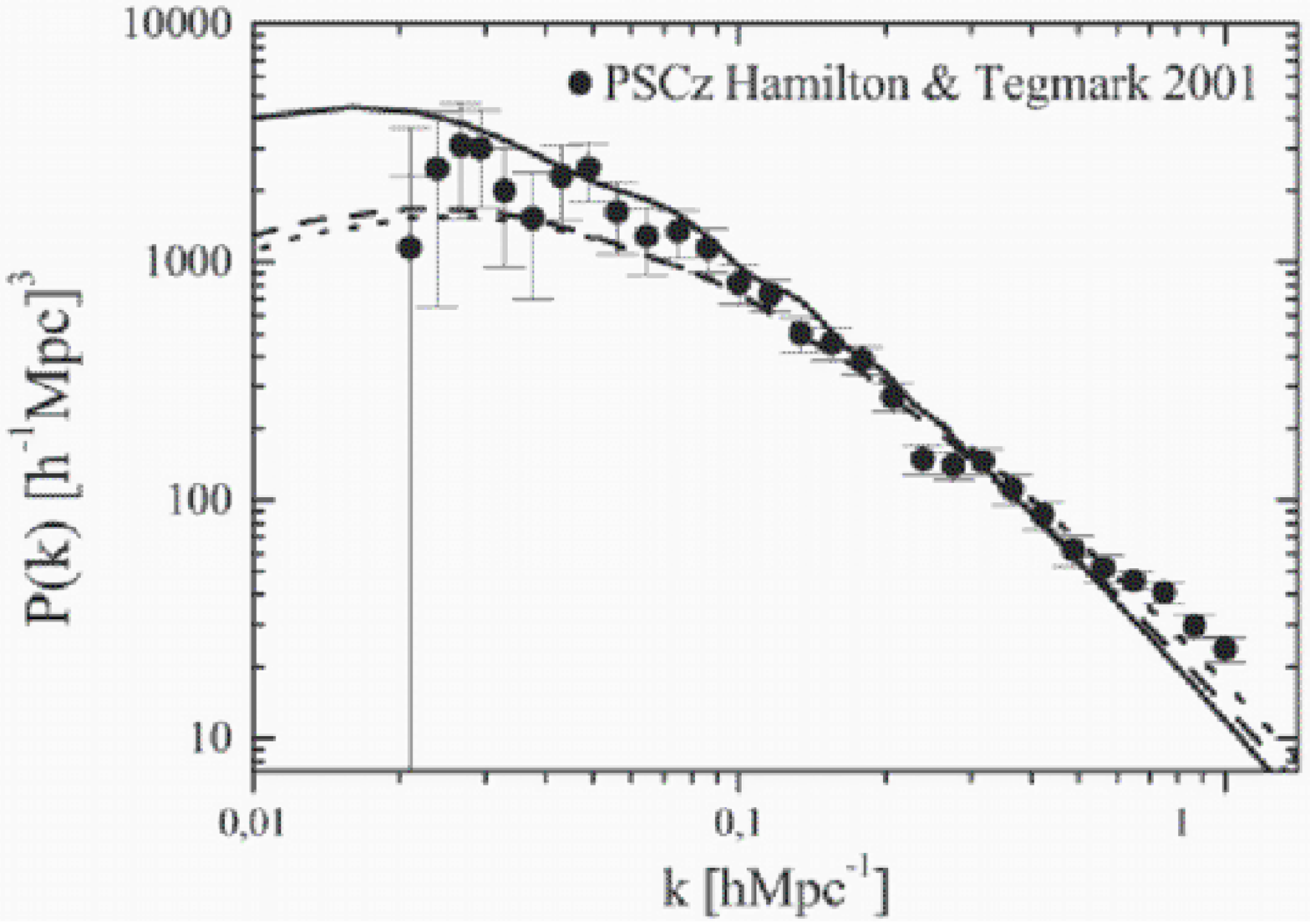}
\caption[CMB degeneracies including $\om_{\rm rel}$.]{Left panel:
CMB degeneracies between cosmological models. Keeping $z_{\rm eq},
\omega_b$ and ${\cal R}$ fixed while varying $\Delta N$ produces
nearly degenerate power spectra. The reference model (solid line)
has $\Delta N = 0$, $\Omega_{\rm tot} = 1.00$, $n_s = 1.00$; the
nearly degenerate model (dotted) has $\Delta N = 10$, $\Omega_{\rm
tot} = 1.05$, $n_s = 1.00$. The curves are normalized to the first
peak. The position of the peaks is perfectly matched, only the
relative height between the first and the other acoustic peaks is
somewhat different in this extreme example, due to the early ISW
effect. The degeneracy can be further improved, at least up to the
third peak, by raising the spectral index to $n_s = 1.08$
(dashed). Right panel: the matter power spectra of the models
plotted in the top panel together with the observed decorrelated
power spectrum from the PSCz survey \citep{Hamilton:2000du}. The
geometrical degeneracy is now lifted.} \label{fig:degg}
\end{figure}

By fixing the four parameters given in (\ref{eq:4par}), or
equivalently the set $\omega_b$, the redshift of equality $z_{\rm
eq}  \equiv \omega_m/\omega_{\rm rel}$, $\Delta N$ and $\R$, one
obtains a perfect degeneracy for the CMB anisotropy power spectra
on degree and sub-degree angular scales. On larger angular scales,
the degeneracy is broken by the late ISW effect because of the
different curvature and cosmological constant content of the
models. From the practical point of view, however, it is still
very difficult to break the degeneracy, since measurements are
limited by cosmic variance on those scales, and because of the
possible contribution of gravitational waves.

Allowing $\Delta N$ to vary, but keeping constant the other three
parameters $\omega_b$, $z_{\rm eq} $, and $\R$, we obtain nearly
degenerate power spectra which we plot in \FIG{fig:degg},
normalized to the first acoustic peak. The degeneracy in the
acoustic peaks region is now slightly spoiled by the variation of
the ratio $\Omega_{\gamma}/\Omega_{\rm rel}$: the different
radiation content at decoupling induces a larger (for $\Delta N >
0$) early ISW effect, which boosts the height of the first peak
with respect to the other acoustic peaks. Nevertheless, it is
still impossible to distinguish between the different models with
present (pre-WMAP) CMB measurements and without external priors.
Furthermore, a slight change in the scalar spectral index, $n_s$,
can reproduce a perfect degeneracy up to the third peak.

\begin{figure}[tb]
\centering
\includegraphics[width=10cm]{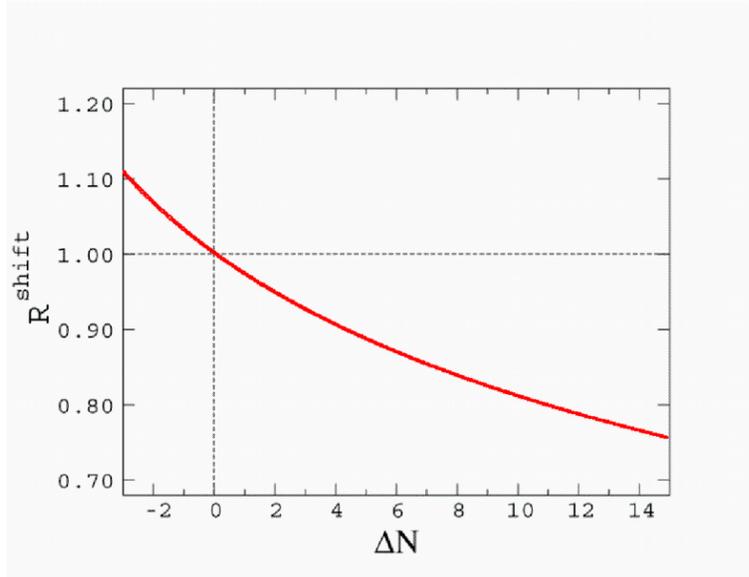}
\caption[The shift parameter as a function of the effective number
of relativistic species.]{The shift parameter $\Rshift$ as a
function of $\Delta N$ with $\Omega_{\Lambda}=0.7$ and
$\Omega_m=0.3$. The position of the peaks is only weakly affected
by $\Delta N$.} \label{fig:shift}
\end{figure}

The main result is that, even with a measurement of the first
three peaks in the angular spectrum, it is impossible to put
bounds on $\omega_{\rm rel}$ alone, even when fixing other
parameters such as $\omega_b$. Furthermore, since the degeneracy
is mainly in $z_{\rm eq} $, the constraints on $\omega_m$ from CMB
are also affected, see \SEC{chap:beyondsp;sec:cmbanalysis}.

In \FIG{fig:shift} we plot the shift parameter $\R$ as a function
of $\Delta N$, while fixing $\Omega_m=0.3$ and
$\Omega_{\Lambda}=0.7$. Increasing $\Delta N$ moves the peaks to
smaller angular scales, even though the dependence of the shift
parameter on $\Delta N$ is rather mild. In order to compensate
this effect, one has to change the curvature by increasing
$\Omega_m$ and $\Omega_{\Lambda}$. We therefore conclude that the
present bounds on the curvature of the universe are weakly
affected by $\Delta N$. Nevertheless, when a positive (negative)
$\Delta N$ is included in the analysis, the preferred models are
shifted toward closed (open) universes.

\subsection{Pre-WMAP constraints from CMB and other data-sets}
\label{chap:beyondsp;sec:cmbanalysis}

In this section, we compare pre-WMAP CMB observations with a set
of models with cosmological parameters sampled as follows: $0.1 <
\Omega_{m} < 1.0$, $0.1< \Omega_{\rm rel}/\Omega_{\rm rel}(\Delta
N=0) < 3$, $0.015 < \Omega_{b} < 0.2$; $0< \Omega_{\Lambda} < 1.0$
and $0.40 < h < 0.95$.  We vary the spectral index of the
primordial density perturbations within the range $n_s=0.50, ...,
1.50$ and we re-scale the fluctuation amplitude by a pre-factor
$C_{10}$, in units of $C_{10}^{\rm COBE}$.  We also restrict our
analysis to purely adiabatic, {\it flat} models ($\Omega_{\rm tot}
=1$) and we add an external Gaussian prior on the Hubble parameter
$h=0.65 \pm 0.2$.

\subsubsection*{Constraints from CMB only}

The theoretical models are computed using the publicly available
{\sc cmbfast} program \citep{Seljak:1996is} and are compared with
the BOOMERanG-98, DASI and MAXIMA-1 data. The power spectra from
these experiments were estimated in $19$, $9$ and $13$ bins
respectively, spanning the range $25 \le \ell \le 1100$. We
approximate the experimental signal $C_B^{ex}$ inside the bin to
be a Gaussian variable, and we compute the corresponding
theoretical value $C_B^{th}$ by convolving the spectra computed by
{\sc cmbfast} with the respective window functions. When the
window functions are not available, as in the case of
Boomerang-98, we use top-hat window functions.  The likelihood for
a given cosmological model is then given by
 \be
 \lnlike =(C_B^{th}-C_B^{ex})M_{BB'}(C_{B'}^{th}-C_{B'}^{ex})
 \ee
where $C_B^{th}$ ($C_B^{ex}$) is the theoretical (experimental)
band power and $M_{BB'}$ is the Gaussian curvature of the
likelihood matrix at the peak. This expression is a generalization
of \rrp{eq:chi_for_Cl} for the case of correlated experimental
points. We consider $10 \%$, $4 \%$ and $4 \%$ Gaussian
distributed calibration errors (in $\mu$K) for the BOOMERanG-98,
DASI and MAXIMA-1 experiments respectively. We also include the
COBE data using Lloyd Knox's {\sc RADPack} package
\citep{RADPack:Website}.

\begin{figure}[tb]
\centering
\includegraphics[width =
0.32\linewidth]{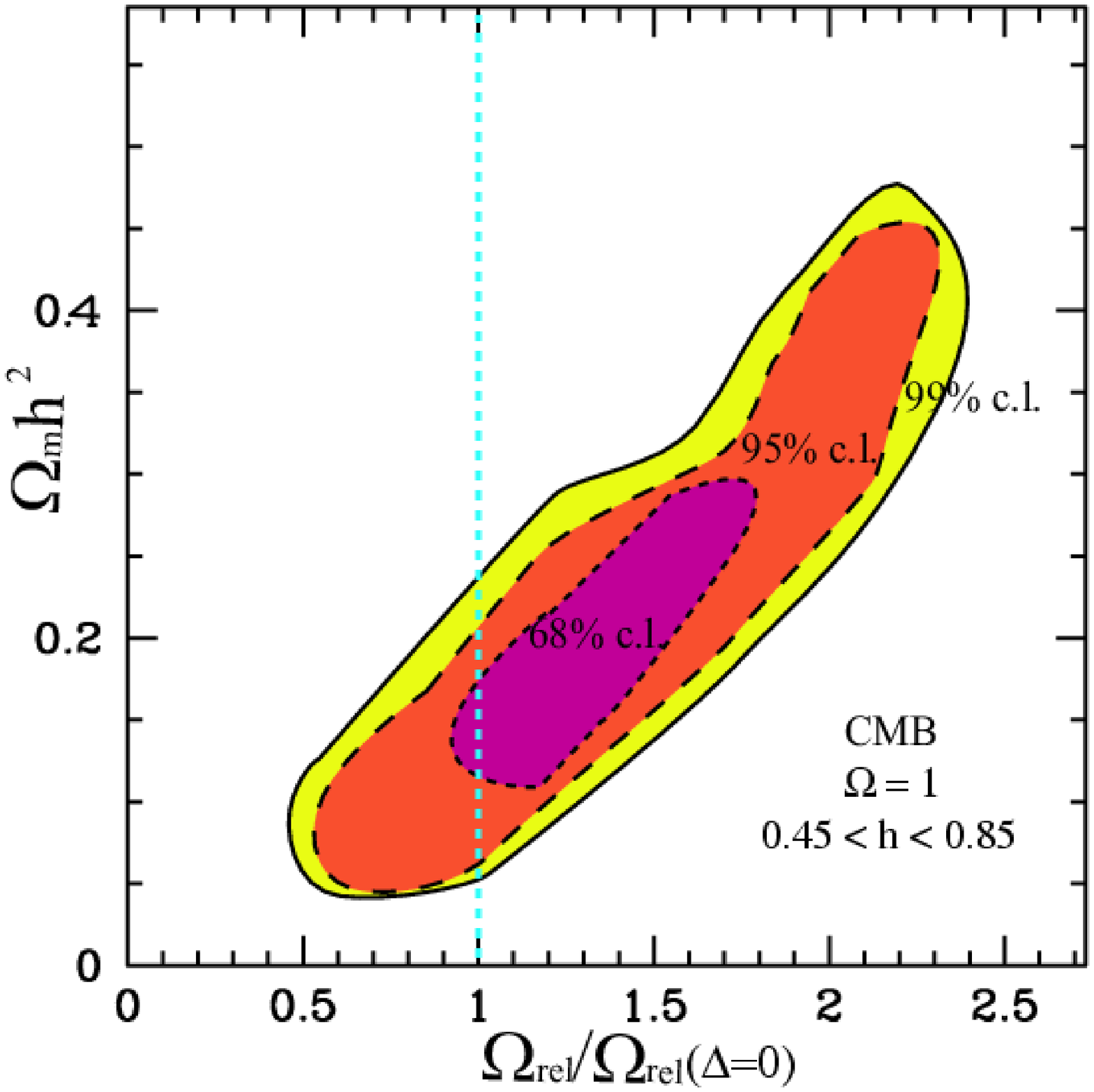}\hfill%
\includegraphics[width =
0.32\linewidth]{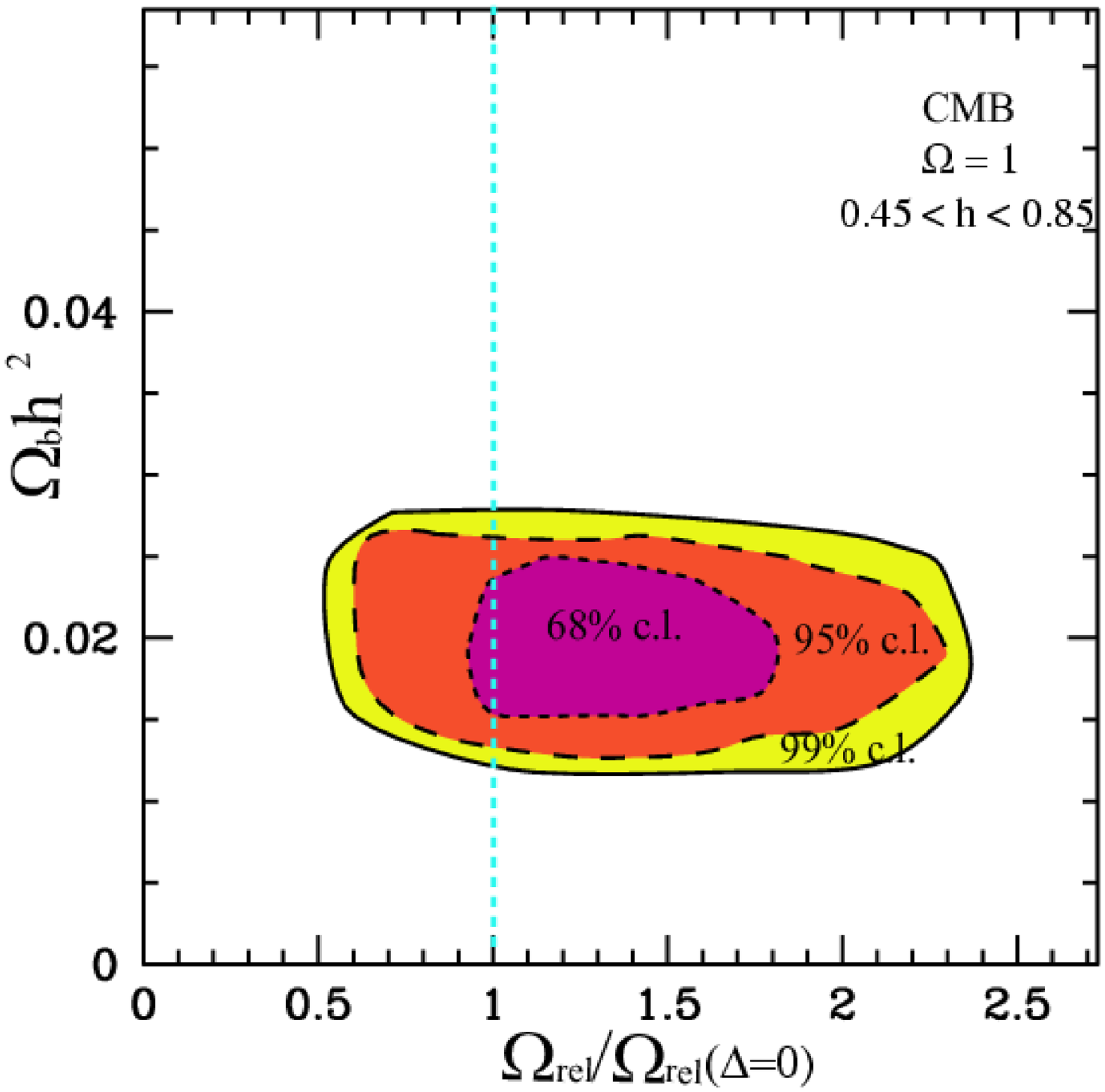}\hfill%
\includegraphics[width =
0.32\linewidth]{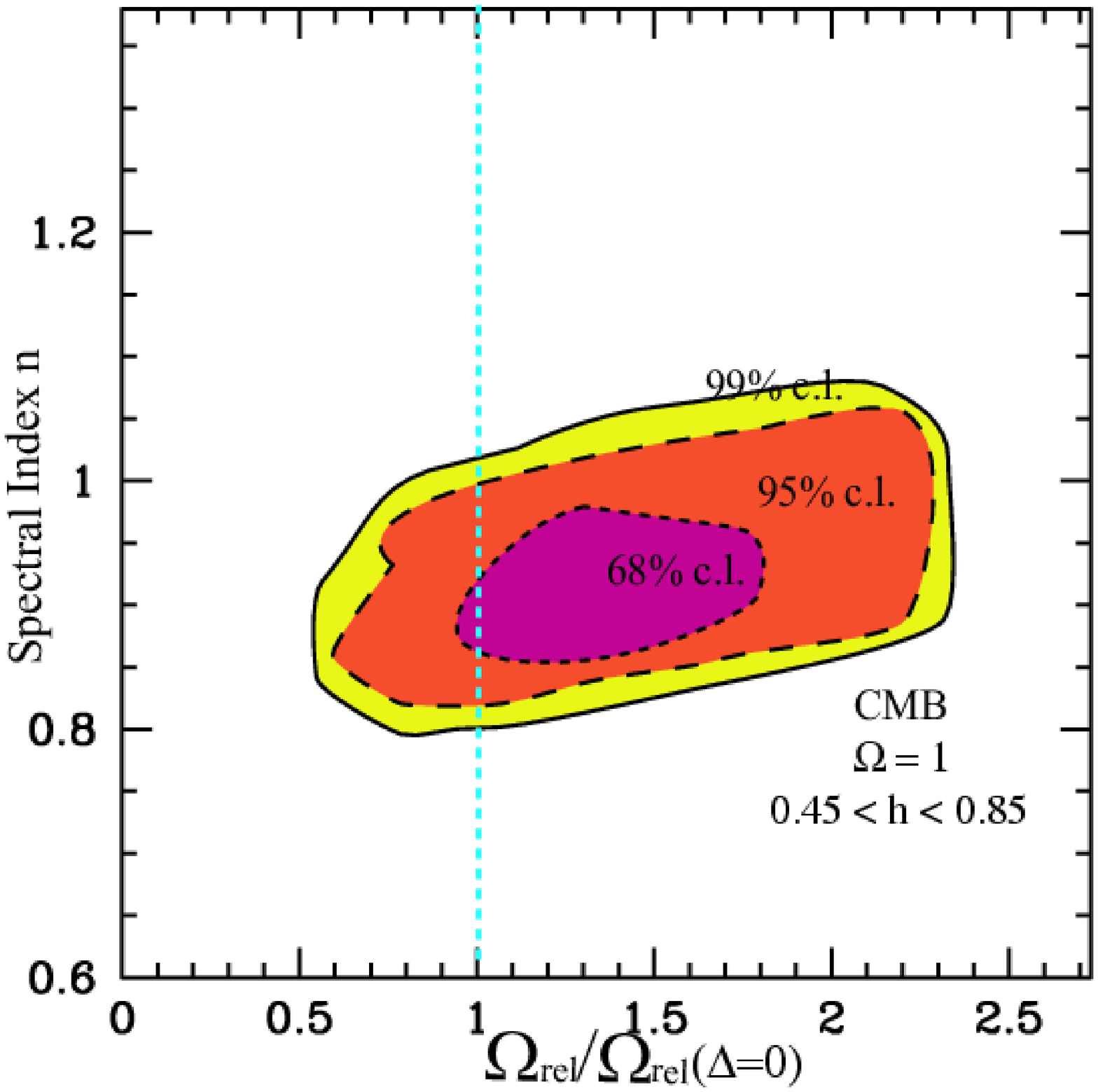}
 \caption[Two-dimensional likelihood plots for $\om_{\rm rel}$ and other parameters.]
 {Two-dimensional likelihood plots from analysis of CMB data.} \label{fig:plots}
\end{figure}

In order to show the effect of the inclusion of $\omega_{\rm rel}$
on the estimation of the other parameters, we plot likelihood
contours in the $\omega_{\rm rel}-\omega_{m}$, $\omega_{\rm
rel}-\omega_b$, $\omega_{\rm rel}-n_s$ planes. Proceeding as in
\cite{Melchiorri:1999br}, we calculate a likelihood contour in
those planes by maximizing the other parameters as explained in
\SEC{chap:data;sec:Bayesian}. In \FIG{fig:plots} we plot the
likelihood contours for $\omega_{\rm rel}$ vs $\omega_m, \omega_b$
and $n_s$.  As can be seen, $\omega_{\rm rel}$ is very weakly
constrained to be in the range $1 \le \omega_{\rm rel}/\omega_{\rm
rel}(\Delta N=0) \le 1.9$ at $1\sigma$ l.c. in all
plots\footnote{Here as in the following, the abbreviation ``l.c.''
stands for ``likelihood content'', in the Bayesian sense explained
in \SEC{chap:data;sec:Bayesian}.}. The degeneracy between
$\omega_{\rm rel}$ and $\omega_{m}$ is evident in the left panel
of \FIG{fig:plots}. Increasing $\omega_{\rm rel}$ shifts the epoch
of matter-radiation equality and this can be compensated only by a
corresponding increase in $\omega_m$. It is interesting to note
that even if we are restricting our analysis to flat models, the
degeneracy is still present and that the bounds on $\omega_m$ are
strongly affected.  We find $\omega_m=0.2\pm0.1$, to be compared
with $\omega_m=0.13\pm0.04$ when $\Delta N$ is kept to zero.  It
is important to realize that these bounds on $\omega_{\rm rel}$
appear because of our prior on $h$ and because we consider flat
models. When one allows $h$ and $\Omega_m$ to be free parameters,
then the degeneracy is almost complete and there are no bounds on
$\omega_{\rm rel}$.

In the central and right panel of \FIG{fig:plots} we plot the
likelihood contours for $\omega_b$ and $n_s$.  As we can see,
these parameters are not strongly affected by the inclusion of
$\omega_{\rm rel}$. The bound on $\omega_b$, in particular, is
completely unaffected by $\omega_{\rm rel}$. There is however, a
small correlation between $\omega_{\rm rel}$ and $n_s$: the boost
of the first peak induced by the ISW effect can be compensated (at
least up to the third peak) by a small change in $n_s$ (right
panel).

\begin{figure}[tb]
 \centering
 \includegraphics[width=\onefigwidth]{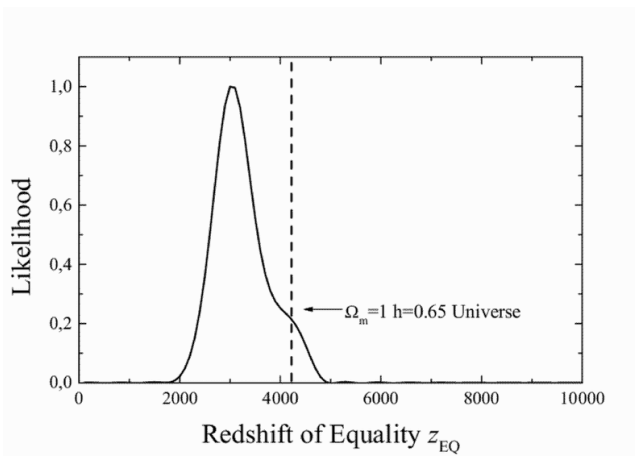}
 \caption{Likelihood probability distribution
function for the redshift of equality. \label{fig:likezeq}}
\end{figure}

Since the degeneracy is mainly in $z_{\rm eq} $, it is useful to
estimate the constraints we can put on this variable. In
\FIG{fig:likezeq} we plot the likelihood curve on $z_{\rm eq}$
alone obtained by maximizing over all other parameters. By
integration of this probability distribution function we obtain
 \be
 z_{\rm eq} = 3100_{-400}^{+600} \quad \text{at $68 \%$ l.c.}
 \ee

\subsubsection*{Adding other data-sets}

It is interesting to investigate how well the constraints from
CMB-independent data-sets can break the degeneracy between
$\omega_{\rm rel}$ and $\omega_m$.  The supernovae luminosity
distance is very weakly dependent on $\omega_{\rm rel}$ --  see
however \cite{Zentner:2001zr} -- and the bounds obtained on
$\Omega_m$ can be used to break the CMB degeneracy. Including the
SN-Ia constraints on the $\Omega_m-\Omega_{\Lambda}$ plane,
$0.8\Omega_m-0.6\Omega_{\Lambda}=-0.2\pm0.1$
\citep{Perlmutter:1998np}, we find
 \be
 \omega_{\rm rel}/\omega_{\rm
rel}(\Delta N=0) = 1.12_{-0.42}^{0.35} \quad \text{at $2\si \%$
l.c.}
 \ee

It is also worth including constraints from galaxy clustering and
local cluster abundances. The degeneracy between $\omega_m$ and
$\omega_{\rm rel}$ in the CMB cannot be broken trivially by
inclusion of large-scale structure (LSS) data, because a similar
degeneracy affects the LSS data as well \citep{Hu:1998tk}.
However, the geometrical degeneracy is lifted in the matter power
spectrum, and accurate measurements of galaxy clustering at very
large scales can distinguish between various models. This is
exemplified in the right panel of \FIG{fig:shift}, where we plot
three matter power spectra with the same cosmological parameters
as in the top panel, together with the decorrelated matter power
spectrum obtained from the PSCz survey.

The shape of the matter power spectrum in the linear regime for
galaxy clustering can be characterized by the shape parameter
 \be \Gamma \sim \dfrac{\Omega_mh}{\sqrt{1+0.135\Delta
N}}e^{-(\Omega_b(1+ \sqrt{2h}/\Omega_m)-0.06)} \eqdot
 \ee
From the observed data one has roughly \citep{Bond:1998qp} $0.15
\le \Gamma +(n_s-1)/2 \le 0.3$. The inclusion of this
(conservative) value on $\Gamma$ gives
 \be
 \omega_{\rm rel}/\omega_{\rm rel}(\Delta N=0) =
 1.40_{-0.56}^{0.49} \quad \text{at $2\si \%$
l.c.}
 \ee
 a bound which is less less restrictive than the one
obtained using the SN-Ia prior.

A better constraint can be obtained by including a prior on the
variance of matter perturbations over a sphere of size $8 h^{-1}$
Mpc, derived from cluster abundance observations. Comparing with
$\sigma_8=(0.55\pm0.05)\Omega_m^{-0.47}$, we obtain
 \be \omega_{\rm
rel}/\omega_{\rm rel}(\Delta N=0) = 1.27_{-0.43}^{0.35} \quad
\text{at $2\si \%$ l.c.}
 \ee

Our results are summarized in \TAB{table:rel_results}. Combination
of present day CMB data with SN and with LSS data yields a lower
bound $N_{\rm eff} > 0.8$ and $> 1.8$, respectively, with
$2\sigma$ likelihood content. Our result is in good agreement with
the analysis of \cite{Hannestad:2001hn}, which considered similar
data sets. It is worth emphasizing the fact that $N_{\rm eff} = 0$
is excluded at much more than $2\sigma$: this can be considered as
a strong cosmological evidence of the presence of a neutrino
background, as predicted by the Standard Model. The upper bounds
for the combined sets can be expressed as $N_{\rm eff} < 6.5$ for
CMB+SN and $N_{\rm eff} < 9.6$ for CMB+LSS, at $2\sigma$ l.c.

\begin{table}[tb]
\centering
\begin{tabular}{|l c c|} \hline
                       & $\omega_{\rm rel}/\omega_{\rm rel}(\Delta N=0)$
                                          & $N_{\rm eff}$ \\\hline
CMB only            & $1.50_{-0.90}^{+0.90}$  &  $0.04 \dots 13.37$ \\
CMB $+$ SN-Ia   & $1.12_{-0.42}^{+0.35}$  &  $0.78 \dots 6.48$ \\
CMB $+$ PSCz     & $1.40_{-0.56}^{+0.49}$  &  $1.81 \dots 9.59$ \\
CMB $+\;\sigma_8$ & $1.27_{-0.43}^{+0.35}$  &  $1.82 \dots 7.59$
\\\hline
\end{tabular}
\caption[$2\sigma$ likelihood intervals on the effective energy
density of relativistic particles from pre-WMAP data.]{Data
analysis results: $2\sigma$ likelihood intervals on the effective
energy density of relativistic particles, $\omega_{\rm
rel}/\omega_{\rm rel}(\Delta N=0)$, and on the corresponding
effective number of neutrino species, $N_{\rm eff}$, for different
data set combinations. Note that the bounds obtained with CMB data
only mainly reflect the priors used in the analysis.
\label{table:rel_results}}
\end{table}

\subsection{Fisher matrix forecast}
\label{chap:beyondsp;sec:fma}

 In this section we perform a Fisher
matrix analysis with the technique explained is
\SEC{chap:data;sec:fma} in order to estimate the precision with
which forthcoming satellite experiments will be able to constrain
the parameter $z_{\rm eq} $.

\TAB{table:expe_rel} summarizes the experimental parameters for
WMAP and Planck employed in the analysis, which considers
temperature information only. For both experiment we take a sky
coverage $f_{\rm sky} = 0.50 $. These values are indicative of the
expected performance of the experimental apparatus, but the actual
values may be somewhat different, especially for the Planck
satellite.

As base parameters for the Fisher matrix analysis, we use the
following nine dimensional parameter set:
 \be
 \params = \left\{ \omega_b, \omega_c,
\omega_\Lambda, \R, z_{\rm eq} , n_s, n_t, r, Q \right\} \eqdot
 \ee
 Here
$n_s, n_t$ are the scalar and tensor spectral indices respectively
and $r \equiv C_2^T/C_2^S$ is the tensor to scalar ratio at the
quadrupole. We adopt a phenomenological normalization parameter,
given by
 \be \label{eq:Q_normalization}
  Q \equiv \left( \sum_{\ell=2}^{\ell_\text{max}} \ell (\ell + 1) C_\ell
  \right)^{1/2} \eqcomma
 \ee
 so that $Q$ effectively measures the mean power seen by the experiment.
The shift parameter $\R$, including the radiation content as in
\rr{eq:def_r} takes into account the geometrical degeneracy.  Our
purely adiabatic reference model has parameters: $\omega_b =
0.0200$ ($\Omega_b = 0.0473$), $\omega_c = 0.1067$ ($\Omega_c =
0.2527$), $\omega_\Lambda = 0.2957$ ($\Omega_\Lambda = 0.7000$),
($h = 0.65$), $\R = 0.953$, $z_{\rm eq}  = 3045$, $n_s = 1.00$,
$n_t = 0.00$ , $r = 0.10$, $Q = 1.00$. This is a fiducial,
concordance model, which we believe to be in good agreement with
most recent determinations of the cosmological parameters (flat
universe, scale invariant spectral index, BBN compatible baryon
content, large cosmological constant). Furthermore, we allow for a
modest, $10 \%$ tensor contribution at the quadrupole in order to
be able to include tensor modes in the Fisher matrix analysis.
\begin{table}[tb]
\centering
\begin{tabular}{|l|ccc|cccc|}
\hline & \multicolumn{3}{|c|}{WMAP}& \multicolumn{4}{|c|}{Planck}
\\\hline $\nu$ (GHz) &  $40$  &  $60$  & $90$ &
              $100$  &  $150$  & $220$ & $350$ \\
$\theta_c$ (degrees)&  $0.46$ & $0.35$ & $0.21$ &
              $0.18$ & $0.13$ & $0.09$ & $0.08$ \\
$\sigma_c/10^{-6}$  & $6.6$  & $12.1$ & $25.5$ &
              $1.7$  & $2.0$  & $4.3$  & $14.4$ \\
$w^{-1}_c/10^{-15}$ & $2.9$  & $5.4$  & $6.8$  &
              $0.028$ & $0.022$ & $0.047$ & $0.44$  \\
$\ell_c$            & $289$ & $385$  & $642$  &
                  $757$ & $1012$ & $1472$ & $1619$ \\\hline
$\ell_{\rm max}$ & \multicolumn{3}{c|}{$1500$}
                 & \multicolumn{4}{c|}{$2000$} \\\hline
\end{tabular}
\caption[Experimental parameters used in the Fisher matrix
analysis.]{Experimental parameters used in the Fisher matrix
analysis for WMAP (first 3 channels) and Planck (last 4 channels).
\label{table:expe_rel} }
\end{table}

\begin{figure}[tb]
\centering
 \includegraphics[width=\twofigswidth]{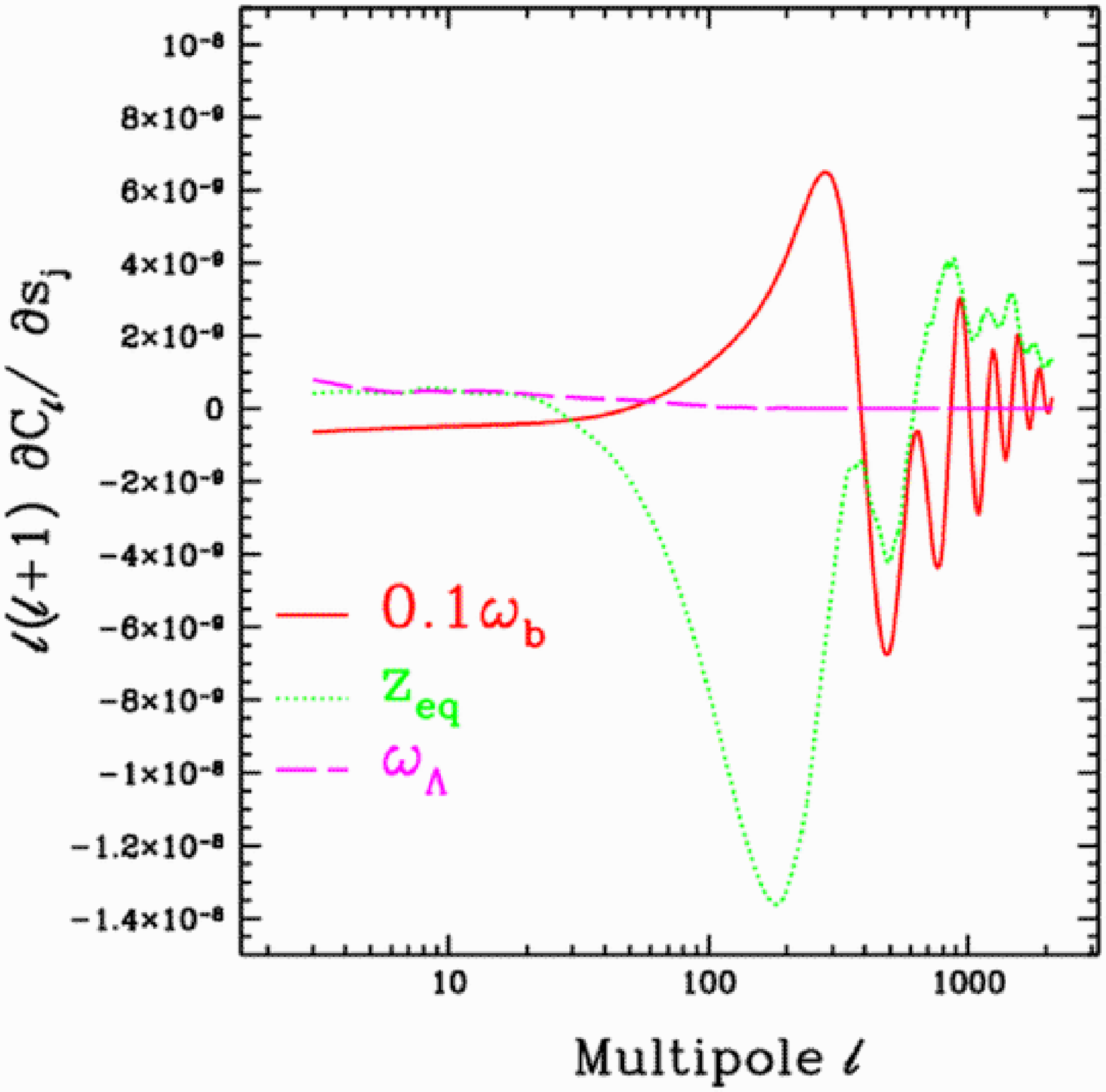}\hfill%
 \includegraphics[width=\twofigswidth]{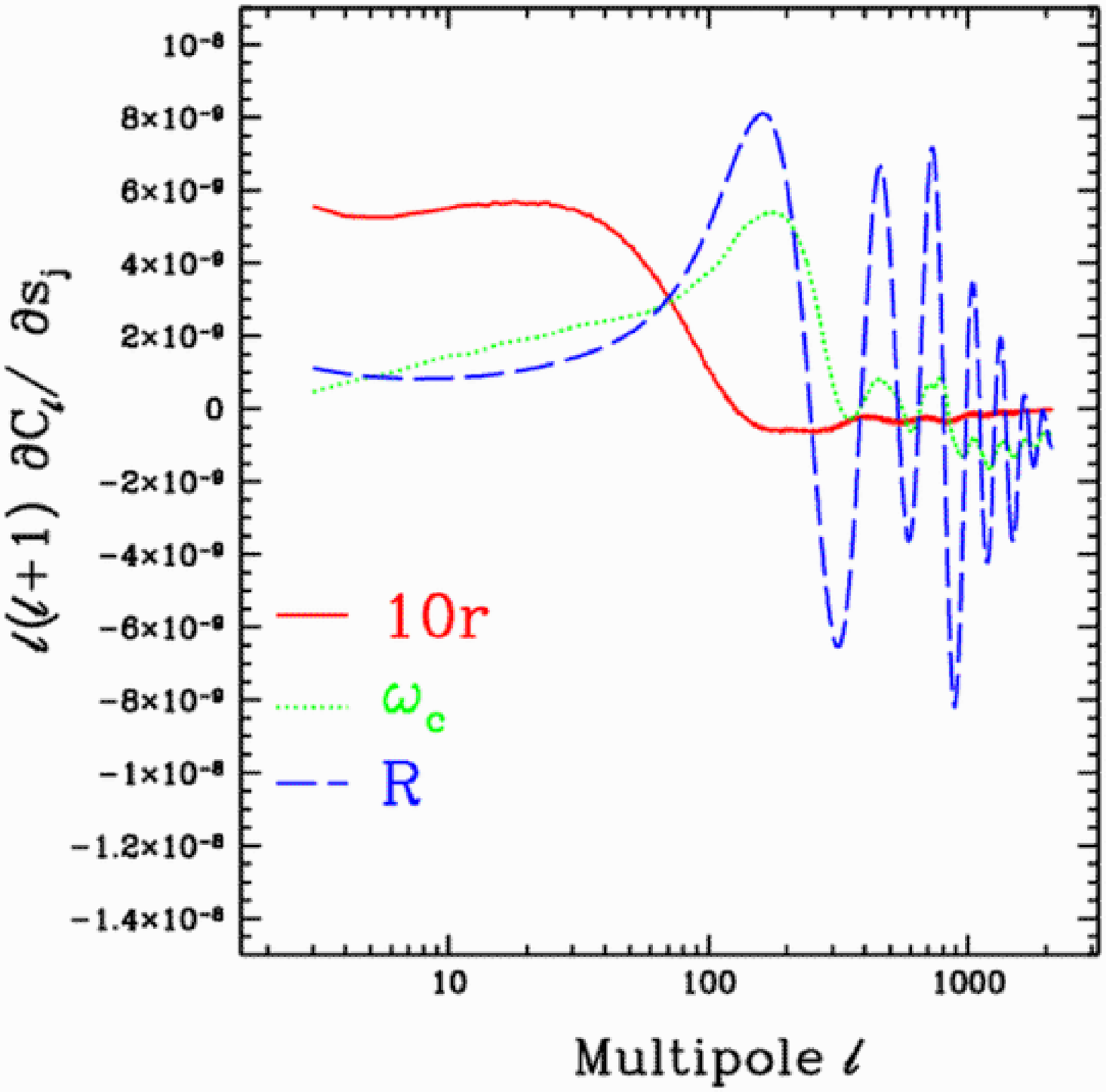}
 \includegraphics[width=\twofigswidth]{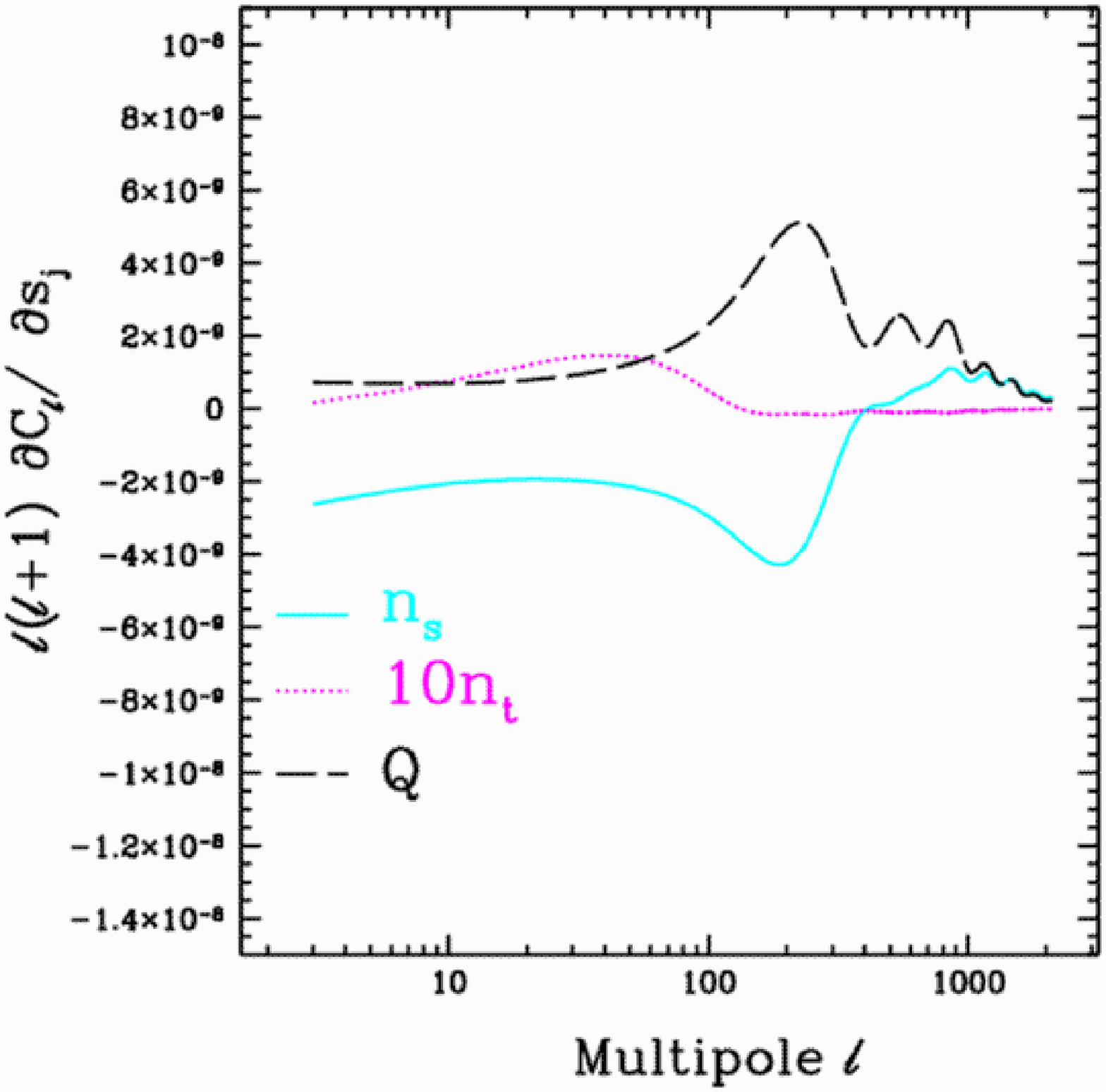}
 \caption[Derivatives of $C_\ell$ with respect to
the 9 parameters of the Fisher matrix analysis.]{Derivatives of
$C_\ell$ with respect to the 9 parameters evaluated at the
reference model described in the text. The numerical prefactor
indicates that the corresponding curve has been rescaled: thus
$0.1 \omega_b$ means that the displayed curve is $0.1 \cdot
\partial C_{\ell} /
\partial \omega_b$. The derivative $\partial C_\ell / \partial
\omega_\Lambda$ has been set to 0 for $\ell > 200$ in order to
suppress the effect of numerical errors, thus taking into account
the geometrical degeneracy.}
 \label{fig:derivatives_rel}
\end{figure}

We plot the derivatives of $C_\ell$ with respect to the different
parameters in \FIG{fig:derivatives_rel}. Generally, we note that
derivatives with respect to the combination of parameters
describing the matter content of the universe ($\omega_b$ and
$\omega_c$, $\R$, $z_{\rm eq} $) are large in the acoustic peaks
region, $\ell > 100$, while derivatives with respect to parameters
describing the tensor contribution ($n_t$, $r$) are important in
the large angular scale region. Since measurements in this region
are cosmic variance limited, we expect uncertainties in the latter
set of parameters to be large regardless of the details of the
experiment. The curve for $
\partial C_{\ell}/\partial Q $ is of course identical to the
$C_{\ell}$'s themselves. The cosmological constant is a notable
exception: variation in the value of $\omega_\Lambda$ keeping all
other parameters fixed produces a perfect degeneracy in the
acoustic peaks region. Therefore we expect the derivative
$\partial C_\ell / \partial \omega_\Lambda$ to be zero in this
region. Small numerical errors in the computation of the spectra,
however, artificially spoil this degeneracy, erroneously leading
to smaller predicted uncertainties. In order to suppress this
effect, we set $\partial C_\ell / \partial \omega_\Lambda = 0$ for
$\ell > 200$. From \rrp{eq:Fisher_matrix_expression} we see that a
large absolute value of $\partial C_\ell/\partial \te_i$ leads to
a large $F_{ii}$ and therefore to a smaller $1\sigma$ error
(roughly neglecting non-diagonal contributions). If the derivative
along $\te_i$ can be approximated as a linear combination of the
others, however, then the corresponding directions in parameter
space will be degenerate, and the expected error will be
important. This is the case for mild, featureless derivatives such
as $\partial C_\ell/\partial r$, while strongly varying
derivatives (such as $\partial C_\ell/\partial \R$) induce smaller
errors in the determination of the corresponding parameter.
Therefore the choice of the parameter set is very important in
order to correctly predict the standard errors of the experiment.

\subsubsection*{Error forecast}

The quantity $\epsilon_i \equiv 1/\sqrt{\lambda_i}$, where
$\lambda_i$ is the $i$-th eigenvalue of the Fisher matrix, is
sometimes used as a rough indication of the resolving power of an
experiment. It expresses the accuracy with which the $i$-th
eigenvector of the Fisher matrix can be determined. The principal
components describe to a good approximation which linear
combinations of the cosmological parameters can be directly
measured with the CMB. In fact, they represent linear
approximations to the orthogonal normal parameters introduced in
\SEC{chap:params:sec:normal}. For WMAP (Planck) the number of
eigenvectors with $\epsilon_i < 10^{-3}$ is 1 out of 9 (3 out of
9) and with $\epsilon_i < 10^{-2}$ is $3/9$ ($6/9$).

Table \ref{table:fma_rel} shows the results of our analysis for
the expected $1\sigma$ error on the physical parameters.
Determination of the redshift of equality can be achieved by WMAP
with $23 \%$ accuracy, while Planck will pinpoint it down to
within $2 \%$ relative error. From $\omega_{\rm rel} = (\omega_b +
\omega_c)/z_{\rm eq} $ it follows that the energy density of
relativistic particles, $\omega_{\rm rel}$, will be determined
within $43 \%$ by WMAP and $3 \%$ by Planck.  This translates into
an impossibility for WMAP alone of measuring the effective number
of relativistic species ($\Delta N_{\rm eff} \approx 3.17$ at
$1\sigma$), while Planck will be able to track it down to $\Delta
N_{\rm eff} \approx 0.24$.  As for the other parameters, while the
acoustic peak' positions (through the value of $\R$) and the
matter content of the universe can be determined by Planck with
high accuracy (of the order of or less than one percent), the
cosmological constant remains (with CMB data only) almost
undetermined, because of the effect of the geometrical degeneracy.
One could also see this as a consequence of an inappropriate
parameterization of the problem: we should in fact use the
parameters which the physics of the CMB measures best, \ie the
principal components. The scalar spectral index $n_s$ and the
overall normalization will be well constrained already by WMAP
(within $15 \%$ and $1 \%$, respectively), while because of the
reasons explained above the tensor spectral index $n_t$ and the
tensor contribution $r$ will remain largely unconstrained by both
experiments. Generally, an improvement of a factor ten is to be
expected between WMAP and Planck in the determination of most
cosmological parameters.

Our analysis considers temperature information only. Inclusion of
polarization measurements would tighten errors, especially for the
``primordial'' parameters $n_s, n_t$ and $r$
\citep{Zaldarriaga:1997ch,Bucher:2000hy}. This is especially
important for a WMAP-type experiment, since a precise
determination of $n_s$ and an higher accuracy in $\omega_m$ would
greatly improve the precision on $N_{\rm eff}$ which can be
obtained with temperature only. By the time Planck will obtain his
first results, polarization measurements will hopefully have been
performed. Combination of polarization information with the WMAP
temperature data would then considerably improve the precision of
the extracted parameter values.

\begin{table}
\centering
\begin{tabular}{|l|c c c|}
\hline
 Parameter     &                                & WMAP         &
Planck \\\hline
Redshift of equality & $\delta z_{\rm eq}  / z_{\rm eq} $                 & 0.23       & 0.02  \\
Relativistic energy  & $\delta \omega_{\rm rel} / \omega_{\rm rel}$     & 0.43       & 0.03  \\
Effective $\nu$ families & $\Delta N_{\rm eff}$ & 3.17 & 0.24
\\\hline
Baryons density       & $\delta \omega_b /\omega_b$              & 0.12        & $< 0.01$  \\
CDM density           & $\delta \omega_c /\omega_c$              & 0.50        & 0.04  \\
Cosmological constant & $\delta \omega_\Lambda / \omega_\Lambda$ & 3.40        & 1.71  \\
Shift parameter       & $\delta \R $                             &
$< 0.01$ & $< 0.01$
\\\hline
Scalar spectral index & $\delta n_s$                             & 0.15        & 0.01  \\
Tensor spectral index & $\delta n_t$                             & 1.96        & 1.08  \\
Scalar-to-tensor ratio& $\delta r / r$                           & 5.22        & 2.67  \\
Normalization         &$\delta Q$                               &
0.01 &$< 0.01$
\\\hline
\end{tabular}
 \caption[Fisher matrix forecasts for the errors on the energy density
 in relativistic particles.]{Fisher matrix analysis results: expected $1\sigma$ errors for
the WMAP and Planck satellites. See the text for details and
discussion. \label{table:fma_rel}}
\end{table}

A Fisher matrix analysis for $\Delta N_{\rm eff}$ was previously
performed by \cite{Lopez:1998aq} and repeated by
\cite{Kinney:1999pd} with the equivalent chemical potential $\xi$,
$\Delta N=15/7(2(\xi/\pi)^{2}+(\xi/\pi)^4$), and a strong
degeneracy was found between $N_{\rm eff}, h$ and
$\Omega_\Lambda$, and to lesser extent with $\Omega_b$. We have
seen here that the degeneracy really is between $\omega_{\rm rel},
\omega_m$ and $n_s$, and the degeneracy previously observed is
thus explained because they considered flat models, where a change
in $\Omega_\Lambda$ is equivalent to a change in $\omega_m$,
$\omega_m = (1 - \Omega_\Lambda -\Omega_b)h^2$. The results
regarding how precisely the future satellite missions can extract
the relativistic energy density, can be translated into
approximately $\Delta N_{\rm eff}=3.17$ ($\xi=2.4$) and $\Delta
N_{\rm eff}=0.24$ ($\xi=0.73$) for WMAP and Planck respectively.
However, including neutrino oscillation leads to equilibration of
the different chemical potentials, and hence BBN leads to the
stronger bound $| \xi | < 0.07$ for all neutrino species
\citep{Dolgov:2002ab}.

\subsubsection*{Comparison with WMAP data analysis}

After the release of the WMAP first year observations, several
groups have independently carried out an analysis similar to the
one presented above
\citep{Crotty:2003th,Hannestad:2003xv,Pierpaoli:2003kw}.
Unfortunately, none of these works includes tensor modes as in our
forecasts, and one has to keep in mind that the FMA assumed
temperature information only and experimental parameters as
appropriate for the original mission specifications, which may be
slightly different from the effective parameters for the first
year only. Despite the fact that the details of the data included
and the prior assumptions vary for each work, the overall
agreement of their findings with our forecasts for WMAP is
nonetheless very satisfactory. We briefly review their conclusions
and compare them with the above predictions.

In \cite{Crotty:2003th} the $1\si$ error on $N_{\rm eff}$ is found
to be $\Delta N_{\rm eff}=3.4$ using WMAP data only (but including
the TE-spectrum) and a weak top-hat prior on the Hubble parameter,
$0.5 < h < 0.9$, with the analysis limited to flat models only.
This result has to be contrasted with the prediction above, which
for the full WMAP data gives (at $1\si$) $\Delta N_{\rm
eff}=3.17$. As predicted, the WMAP observations improve
dramatically on the bounds for $N_{\rm eff}$ from CMB only, which
become with the above assumptions $-2.1 < \Delta N_{\rm eff} <
6.9$ (at $2\si$ likelihood content).

These findings are in good concordance with the more general
set-up of \cite{Pierpaoli:2003kw}, where curved models are
considered as well, the CBI data are used together with the WMAP
observations and constraints from the 2dF matter power spectrum
are also included. In this case the results do not compare
directly with our predictions because of the inclusion of external
constraints in the form of the matter power spectrum. The $95\%$
likelihood interval is then tighter because of the more powerful
observational data used, giving (without Hubble prior) $\Delta
N_{\rm eff} = 5.5$.

The quite complete investigation of \cite{Hannestad:2003xv} also
derives constraints on the neutrino masses, and considers the
effects of the inclusion of further observational constraints,
such as a prior on the Hubble parameter, a prior on $\Om_m$ from
supernov\ae\ data, a BBN prior on $\om_b$ and the 2dF matter power
spectrum. Where comparable, the findings are entirely compatible
with the other two works; in particular, for the case of massless
neutrinos and WMAP data only, the $95\%$ likelihood interval for
flat models only and a weak top-hat prior $0.5 \le h \le 0.85$ is
$\Delta N_{\rm eff} = 8.9$.

\clearpage

\section{The primordial helium fraction}
 \label{chap:bspII;sec:helium}

This section is based on the work \cite{Trotta:2003xg}, where the
first determination of the helium abundance from CMB data alone
was presented. After giving the motivation underlying this
investigation in \SEC{chap:bspII;sec:motivations}, we discuss in
\SEC{chap:bspII;sec:cmb} the role of the helium mass fraction for
CMB anisotropies, and in particular the details of the ionization
history of the Universe which are relevant for constraining the
helium abundance with the CMB. We then review the standard
Big-Bang Nucleosynthesis scenario for the abundance of light
elements and compare its predictions with current astrophysical
measurements in \SEC{chap:bspII;sec:bbn}; the present constraints
from CMB data are presented in \SEC{chap:bspII;sec:cmbresults},
while the future potential of using the CMB as an independent way
of determining the helium abundance is elucidated in
\SEC{chap:bspII;sec:fma}. There we also explore the impact of
helium for future accurate determination of the baryon abundance.

\subsection{Motivation}
\label{chap:bspII;sec:motivations}

Our understanding of the baryon abundance has increased
dramatically over the last few years, coming  from two independent
paths, namely BBN and CMB. Absorption features from high-redshift
quasars allow us to measure precisely the deuterium abundance,
D/H, which combined with BBN calculations provides a reliable
estimate of the baryon to photon ratio,
 \be
 \eta_{10} \equiv \dfrac{n_b}{n_\gamma} 10^{10}\eqdot
 \ee
An independent determination of the baryon content of the universe
from CMB anisotropies comes from the increasingly precise
measurements of the acoustic peaks, via the characteristic
signature of the photon-baryon fluid oscillations discussed in
\SEC{chap:params;sec:barsig}. The agreement between these two
completely different approaches is both remarkable and impressive
(see details below). The time is therefore ripe to proceed and
test the agreement between other light elements which are also
probed both by BBN and CMB.

Helium being the most abundant of the light elements, it is
natural to focus on this element by exploring the dependency of
CMB anisotropies on the value of the primordial helium mass
fraction $Y_p$, defined as
 \be
 \label{eq:define_Yp}
 Y_p \equiv 4 \dfrac{n_{\text{He}}}{n_b} \eqcomma
 \ee
where $n_{\text{He}}$ and $n_b$ denote the number densities of
$^4$He atoms and baryons, respectively. If we denote by $n_N$ and
$n_P$ the number densities of neutrons and protons, respectively,
and assume that all neutrons are in He nuclei, then a simple
counting argument gives the estimate
 \be \label{eq:Yp_estimate}
 Y_p = \dfrac{2 n_N/n_P}{1 + n_N/n_P} \approx 0.25 \eqcomma
 \ee
where the numerical value comes from a rough approximation to the
freeze-out value of the neutron to proton ratio $n_N/n_P \approx
1/7$, see \eg \cite{Inflation_Kolb}. The detailed value of $Y_p$
is predicted by BBN as a function of two parameters only, the
baryon abundance and the number of relativistic degrees of freedom
at BBN \citep{Sarkar04}.

The hope is that the CMB observations might provide an independent
measurement of $Y_p$, accurate enough to help clarify the
present-day discrepancies between direct observations of the
helium fraction derived from astrophysical systems, whose errors
are seemingly dominated by systematics which are hard to assess.
The latest CMB data are precise enough to allow taking this
further step, and in view of the emerging ``baryon tension''
between BBN predictions from observations of different light
elements \citep{Cyburt:2003fe} possibly requires taking such a
step. The advantage of using CMB anisotropies rather than the
traditional astrophysical measurements, is that the CMB provides a
clear measurement of the primordial helium fraction before it
could be changed by any astrophysical process. On the other hand
the dependence of the CMB power spectrum on the primordial helium
fraction is rather mild, a fact which makes it presently safe to
set the helium mass fraction to a constant for the purpose of CMB
data analysis of other cosmological parameters, but will have an
impact on the baryon abundance determination from Planck quality
data, as we show in \SEC{chap:bspII;sec:fma}.

\subsection{The impact of helium on the CMB: ionization history revisited}
\label{chap:bspII;sec:cmb}

We now resume our discussion of the recombination epoch and
reionization history of the Universe sketched in
\SEC{chap:params;sec:dampingtail}, and focus on the role of the
helium mass fraction, considered here as a free parameter. In a
second step, the aim will be to combine the CMB results with the
BBN predictions and compare the result with the independent
astrophysical determinations of the light elements abundance. We
thus have at our disposal three different tools, each of which
probes the same quantities at three vastly different epochs of the
cosmic history. It is important to stress that a good agreement
among the three is by no means trivial, and that testing their
concordance is a powerful way to check the consistency of the
standard cosmological scenario. On the other hand, significant
discrepancies would necessary imply the need for new physics.

The recent WMAP data allow us to determine with very high
precision the epoch of photon decoupling, $z_{\rm dec}$, \ie the
epoch at which the ionized electron fraction, $x_e(z) = n_e/n_H$,
has dropped from 1 to its residual value of order $10^{-4}$. Here
$n_e$ denotes the number density of free electrons, while $n_H$ is
the total number density of H atoms (both ionized and recombined).
The redshift of decoupling has been determined to be $z_{\rm
dec}=1088^{+1}_{-2}$ \citep{Spergel:2003cb}, which corresponds to
a temperature of about 0.25 eV. Helium recombines earlier than
hydrogen, roughly in two steps: around redshift $z=6000$ HeIII
recombines to HeII, while HeII to HeI recombination begins around
$z<2500$ and finishes just after the start of H
recombination~\citep{Liubarski83,Hu:1995fq,Seager:1999bc,Seager:1999km}.

\begin{figure}[tb]
\begin{center}
\includegraphics[width=\onefigwidth, angle=-90]{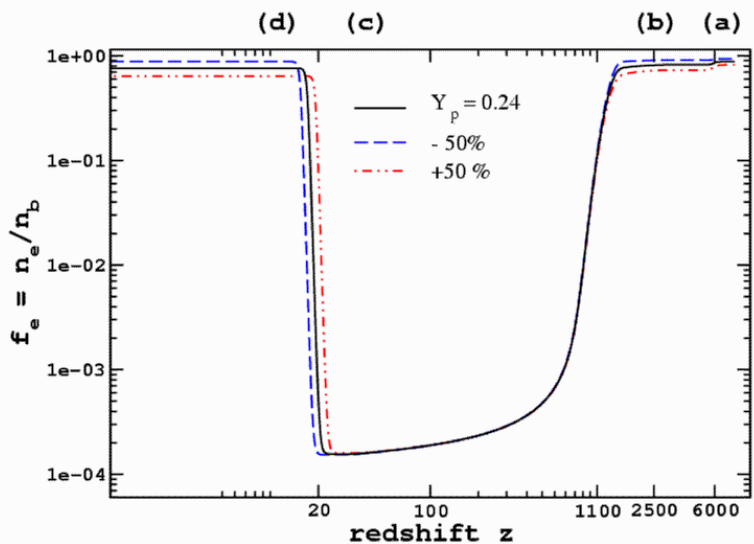}
\end{center}
\caption[Ionization history for different values of the helium
fraction.]{Evolution of the number density of electrons normalized
to the number density of baryons, $f_e=n_e/n_b$, as a function of
redshift for different values of the helium fraction $Y_p$. The
black-solid curve corresponds to the standard value $Y_p = 0.24$.
The labels (a) to (d) indicate the four different phases discussed
in the text.\label{fig:reionhist}}
\end{figure}

The baryon number density per m$^3$ $n_b(z)$ is related to the
baryon energy density today, $\omega_b$, by
 \be \label{eq:ombproptone}
 n_b = 11.3 (1+z)^3 \omega_b
  \ee
 and we have $n_H = n_b(1-Y_p)$.
 Usually, the ionization
history is described in terms of $x_e(z) = n_e/(n_b(1-Y_p))$.
However, for the purpose of discussing the role of $Y_p$, it is
more convenient to consider the quantity
 \be
 f_e(z) \equiv n_e/n_b
 \ee
instead, the ratio of free electrons to the total number of
baryons. For brevity, we will call $f_e$ the free electron
fraction. Once the baryon number density has been set by fixing
$\omega_b$, one can think of $Y_p$ as an additional parameter
which controls the number of free electrons available in the tight
coupling regime. The CMB power spectrum depends on the full
detailed evolution of the free electron fraction, but we can
qualitatively describe the role of helium in four different phases
of the ionization/recombination history, displayed in
\FIG{fig:reionhist}.
\begin{itemize}
 \item[(a)] Before HeIII recombination, all electrons are free,
therefore $f_e(z>6000) = 1 - Y_p/2$.
 \item[(b)] HeII progressively
recombines and just before H recombination begins, $f_e$ has
dropped to the value $f_e(z \approx 1100) = 1 - Y_p$.
 \item[(c)]
After decoupling, a residual fraction of free electrons freezes
out, giving $f_e(30 \lsim z \lsim 800) = \fres \approx 2.7 \cdot
10^{-5} \sqrt{\omega_m}/\omega_b$.
 \item[(d)] Reionization of all
the H atoms gives $f_e(z \lsim 20) = 1 - Y_p$.
\end{itemize}

During phase (a), the photon-baryons fluid is in the tight
coupling regime. However the presence of ionized He increases
diffusion damping, therefore having an impact on the damping scale
in the acoustic peaks region: the diffusion damping length
(\refp{eq:damping_scale}) including helium can be approximated as
\citep{Hu:1995uz}
 \be \label{eq:damping_scale_with_Yp}
 \la_\text{D}^{2} \approx 1.7 \times 10^7
 \left(1 - \dfrac{Y_p}{2}\right)^{-1}
 \om_b^{-1} \om_m^{-1/2} a^{5/2} \dfrac{1}{3\sqrt{a_\EQ/a} + 2}
 \quad \mpc^2 \eqdot
 \ee
As expected, a larger helium fraction implies an increased damping
length, and thus an extra power suppression on small scales.

When the detailed energy level structure of HeII is taken into
account~\citep{Seager:1999km}, the transition to phase (b) is
smoother than in the Saha equation approximation. Therefore the
plateau with $f_e = 1 - Y_p$ is not visible in
\FIG{fig:reionhist}. Before H recombination, He atoms remain
tightly coupled to H atoms through collisions, with the same
dynamical behavior. In particular, it is the total $\omega_b$
which determines the amount of gravitational pressure on the
photon-baryons fluid, and which sets the acoustic peak
enhancement/suppression, see \SEC{chap:params;sec:acoustic}. Hence
we do not expect the value of $Y_p$ to have any influence on the
boosting (suppression) of odd (even) peaks. The redshift of
decoupling (transition between (b) and (c)) depends mildly on
$Y_p$ in a correlated way with $\omega_b$, since the number
density of free electrons in the tight coupling regime (just
before H recombination) scales as $n_e = f_e n_b = n_b(1-Y_p)$.
Hence an increase in $\omega_b$ can be compensated by allowing for
a larger helium fraction. An analytical estimate along the same
lines as in e.g.~\cite{Inflation_Kolb} indicates that a $10\%$
change in $Y_p$ affects $\zdec$ by roughly $0.1\%$, which
corresponds to $\Delta \zdec \approx 1$. This is of the same order
as the current $1\sigma$ errors on $\zdec$, obtained by fixing
$Y_p = 0.24$.

After H recombination, the residual ionized electron fraction
$\fres$ does not depend on $Y_p$, but is inversely proportional to
the total baryon density (phase (c)). As the CMB photons
propagate, they are occasionally rescattered by the residual free
electrons. The corresponding optical depth, $\tres$ is given by
 \be
 \begin{split}
  \tres  & = \int_{t_0}^{t_{\textrm{dec}}}n_e^{\textrm{res}}
  c \sigma_T \dr t \\
  & \approx 1.86\cdot 10^{-6} \sqrt{\Omega_m}
  \int_0^{\zdec} \frac{(1+z)^2}{\left[ (1+z)^3 + \Om_\La/\Om_m \right]^{1/2}} dz
  \eqdot
  \end{split}
 \ee
Performing the integral we can safely neglect the contribution
of the cosmological constant at small redshift, since $\zdec \gg
\Om_\La/\Omega_m$. Retaining only the leading term, the
approximated optical depth from the residual ionization fraction
is estimated to be
\begin{equation}
\tres \approx 1.24\cdot10^{-6} (1+\zdec)^{3/2} \approx 0.045,
\end{equation}
{\it independent} of the cosmological parameters and of the helium
fraction. Therefore after last scattering we do not expect any
significant effect on CMB anisotropies coming from the primordial
helium fraction, until the reionization epoch, phase (d).

As pointed out in \SEC{chap:params:sec:reion}, CMB anisotropies
are sensitive only to the integrated reionized fraction if
temperature information only is available, while specific
signatures are imprinted on the E-polarization and ET-cross
correlation power spectra by the detailed shape of the
reionization history. There are several physically motivated
reionization scenarios, which however cannot be clearly
distinguished at present \citep{Haiman:2003ea,Hansen:2003yj}.
Therefore at the present level of accuracy it is safe for our
purpose to assume an abrupt reionization, \ie that at the
reionization redshift $\zreion$ all the hydrogen was quickly
reionized, thus producing a sharp rise of $n_e$ from its residual
value to $n_H$. More precisely, $\zreion$ is the redshift at which
$x_e(\zreion) = 0.5$. In our treatment we neglect HeII
reionization, for which there is evidence at a redshift $z \approx
3$ (see \citealp{Theuns:2002vm} and references therein). This
effect is small, since one extra electron released at $z \approx
3$ would change the reionization optical depth only by about
$1\%$. The effect of HeIII reionization, which happens still
later, is even smaller. We also neglect the increase of the helium
fraction due to non-primordial helium production, which has a
negligible effect on CMB anisotropies. Those approximations do not
affect the results at today's level of sensitivity of CMB data:
for WMAP noise levels, even inclusion of the polarization spectra
is not enough to distinguish between a sudden reionization
scenario and a more complex reionization history. At the level of
Planck a more refined modelling of the reionization mechanism will
be necessary~\citep{Holder:2003eb,Doroshkevich:2002wy}.

In the sudden reionization scenario adopted here, the relation
between reionization redshift and reionization optical depth,
$\treion$, is given by \rrp{eq:tau_reion_approx}. Once again,
since the number density of reionized electrons scales as
$\omega_b(1-Y_p)$, the redshift of reionization is positively
correlated with $Y_p$ (for fixed optical depth and baryon
density).

\begin{figure}[tb]
\begin{center}
\includegraphics[width=0.49\linewidth]{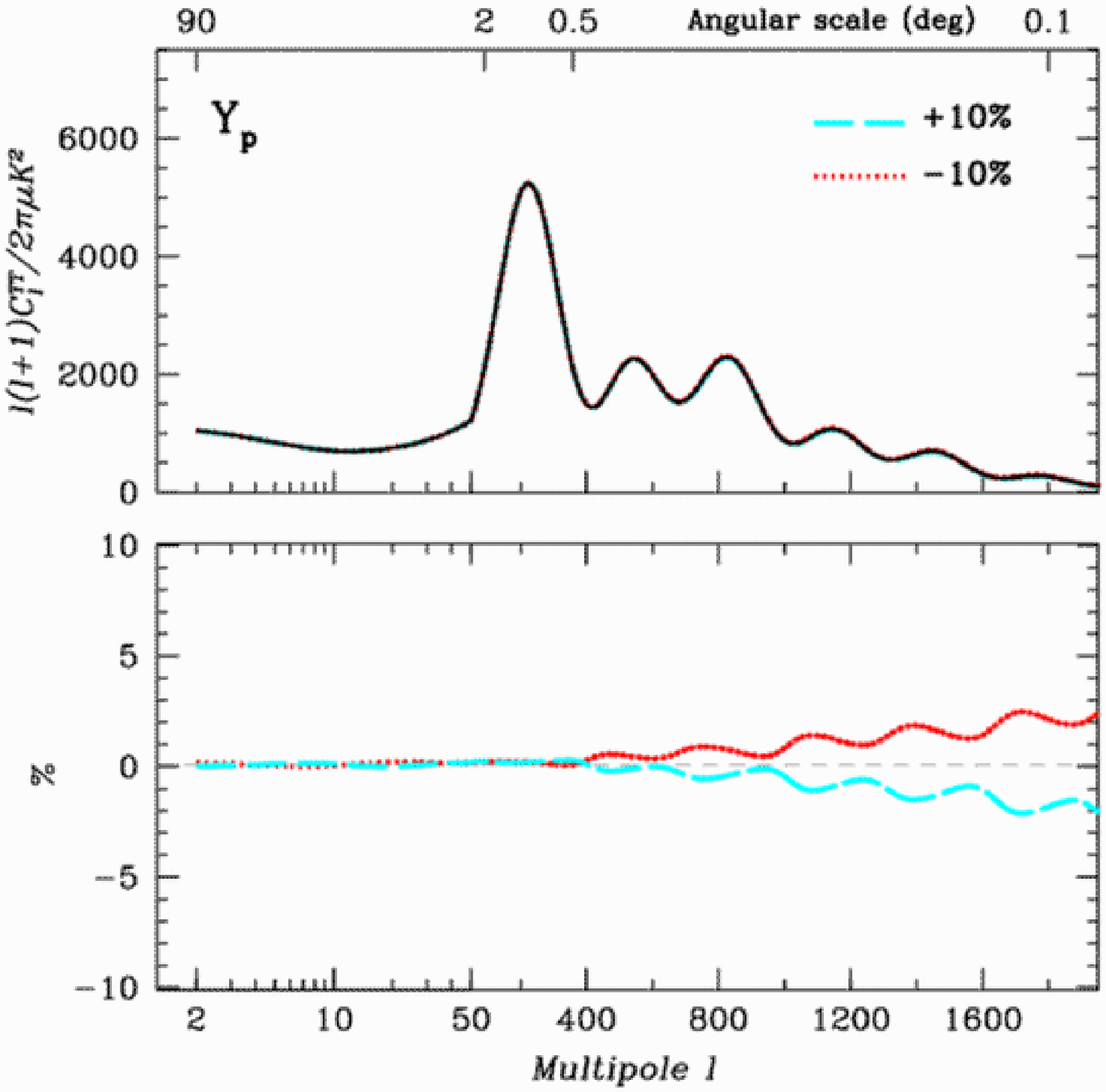}\hfill%
\includegraphics[width=0.49\linewidth]{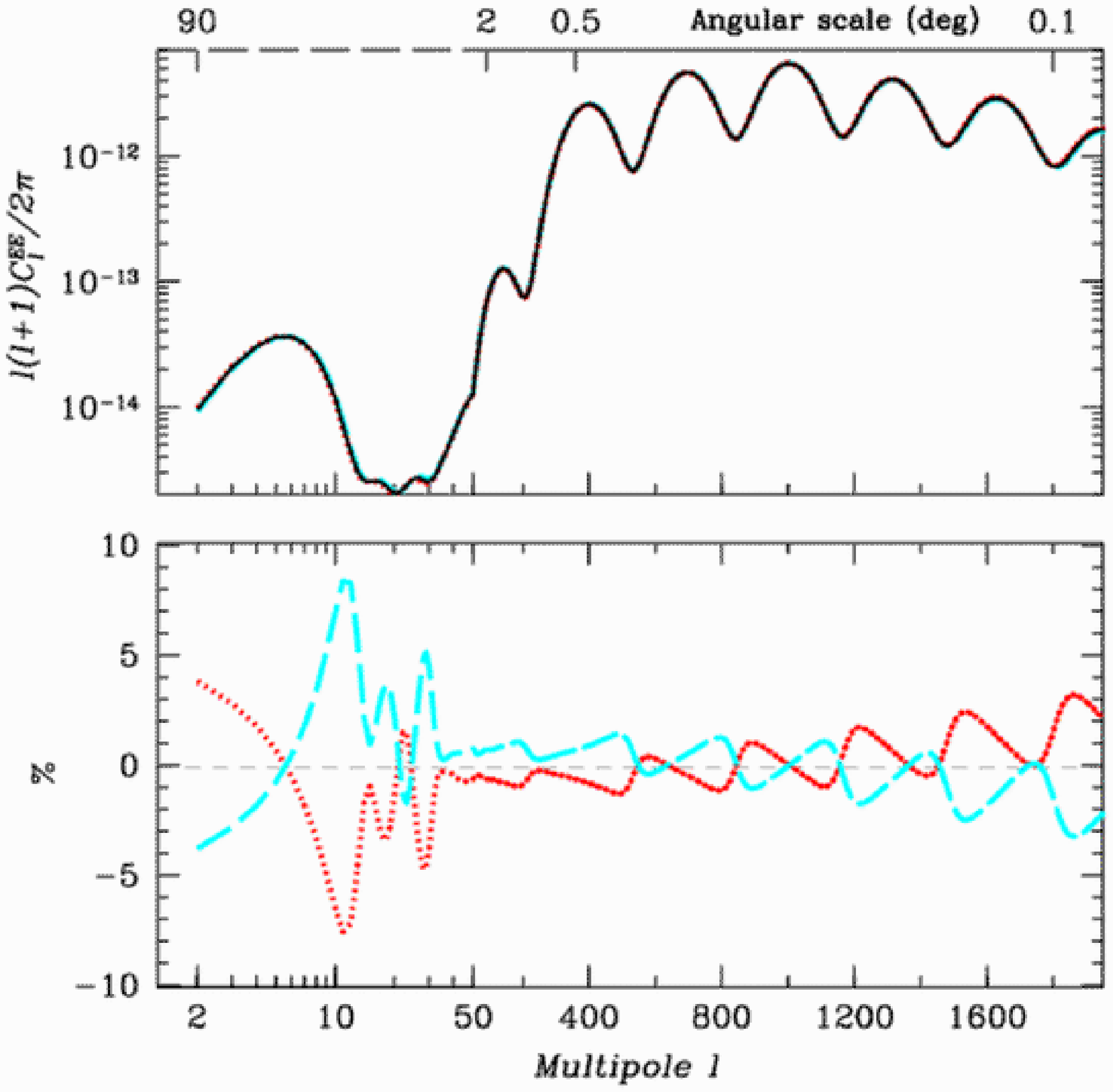}
\end{center}
\caption[Temperature and polarization
  power spectra for
  different values of the helium mass fraction.]{CMB temperature (left panel) and polarization
  (right panel) power spectra and percentage change (bottom panels) for
  a 10\% larger (smaller)
  value of the helium mass fraction, $Y_p$. The solid-black line in the top panels
  corresponds to a standard $\Lambda$CDM
  model, with $Y_p = 0.24$. The impact is at the percent level,
  and is almost indistinguishable in the top panels.
  All other parameters are fixed to the value of our fiducial model
  (\TAB{table:fiducial}),
  in particular, we have $\treion=0.166$.\label{fig:yheeffect}}
\end{figure}

As a result of the physical mechanism described above, a $10\%$
change in $Y_p$ has a net impact on the CMB power spectrum at the
percent level.  The impact on the CMB temperature and polarization
power spectra is highlighted in \FIG{fig:yheeffect}.
In the temperature panel, we notice that a larger helium fraction
slightly suppresses the peaks because of diffusion damping, while
it has no impact on large scales. Polarization is induced by the
temperature quadrupole component at last scattering and the
reionization bump induced in the polarization spectrum (see
\SEC{chap:params:sec:reion}) is clearly visible in the
polarization panel of \FIG{fig:yheeffect} in the $\ell \approx 15$
region. A change in the helium fraction implies a shift of the
redshift of reionization for a given (fixed) optical depth, and a
consequent shift of the position of the reionization bump via
\rrp{eq:reion_bump_position}. The value of $Y_p$ does not affect
the height of the bump, which is controlled by the optical depth
and is proportional to $\tau^2$. This effect is highlighted in the
polarization panel of \FIG{fig:yheeffect}: a 10\% change in $Y_p$
induces roughly a 10\% change in the position of the bump. The
subsequent two oscillatory features for $\ell \lsim 50$ reflect
the displacement of further secondary reionization induced
polarization oscillations. However, since the value of polarized
power is very low in that region, such secondary oscillations are
very hard to detect precisely.

In principle, given an accurate knowledge of the reionization
history, the effect of $Y_p$ on the polarization bump would assist
in the determination of the helium abundance. However, our
ignorance of the reionization history prevents us from recovering
useful information out of the measured reionization bump. The
displacement induced by $Y_p$ is in fact degenerate with a partial
reionization, or with other, more complex reionization mechanisms.
Hence constraints on $Y_p$ come effectively from the damping tail
in the $\ell \gsim 400$ region of the temperature spectrum, which
needs to be measured with very high accuracy. Other light elements
like deuterium and helium-3 are much less abundant, and will
therefore have even smaller effect on the CMB power spectrum, at
the order of $10^{-5}$.

\subsection{Astrophysical measurements and BBN predictions}
\label{chap:bspII;sec:bbn}

Once we fix the number of relativistic degrees of freedom by
specifying the number of massless neutrino families, the standard
model of Big-Bang Nucleo\-syn\-the\-sis (BBN) has only one free
parameter, namely the baryon to photon ratio $\eta_{10}$ defined
in (\refp{eq:def_eta10}), which for long has been known to be in
the range $1-10$~\citep{Inflation_Kolb}.  Thus by observing just
one primordial light element one can predict the abundances of all
the other light elements.

\subsubsection*{Astrophysical measurements}

The deuterium to hydrogen abundance, D/H, is observed by
Ly-$\alpha$ features in several quasar absorption systems at high
redshift, $D/H =
2.78^{+0.44}_{-0.38}\times10^{-5}$~\citep{Kirkman:2003uv}, which
in BBN translates into the baryon abundance, $\eta_{10} =
5.9\pm0.5$. Using BBN one thus predicts the helium mass fraction
to be in the range $0.2470 < Y_p < 0.2487$.  The dispersion in
various deuterium observations is, however, still rather large,
ranging from $D/H = 1.65\pm
0.35\times10^{-5}$~\citep{Pettini:2001yu} to $D/H =
3.98^{+0.59}_{-0.67}\times10^{-5}$~\citep{Kirkman:2003uv}, which
most probably indicates underestimated systematic errors.

\begin{figure}[tb]
\begin{center}
\includegraphics[width=\onefigwidth]{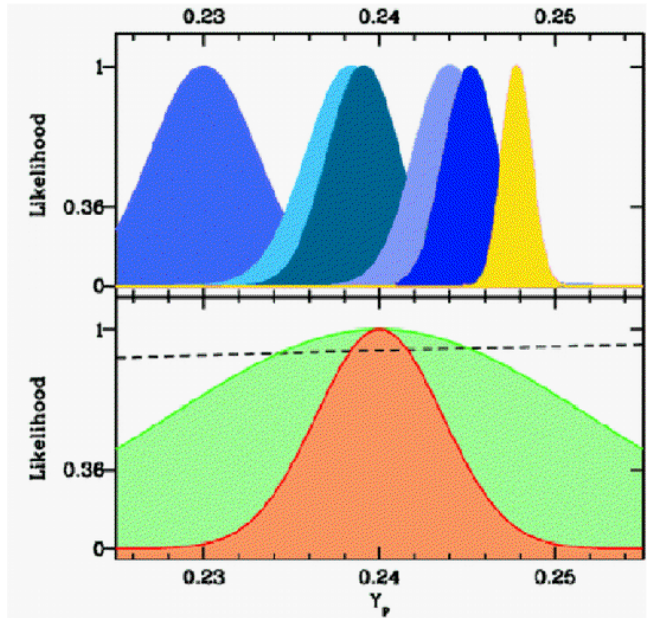}
\end{center}
\caption[Comparison between current astrophysical errors on the
helium fraction and the CMB potential.]{In the top panel we plot a
few current direct astrophysical measurements of the helium mass
fraction $Y_p$ as Gaussian likelihood curves with standard
deviation corresponding to the given $1\sigma$ (statistical) error
(blue/dark gray curves, on the left of the diagram), and the value
inferred from deuterium measurements combined with BBN
(yellow/light gray curve, on the far right), see the text for
references. In the bottom panel, a direct comparison with CMB
present-day accuracy (actual CMB data, black dashed line, this
work; the $1\si$ likelihood interval is $0.16 < Y_p < 0.50$) and
with its future potential (Fisher matrix forecast for Planck --
green/light gray curve -- and a Cosmic Variance Limited experiment
-- orange/dark gray curve).\label{fig:compilation}}
\end{figure}

The observed helium mass fraction comes from the study of
extragalactic HII regions in blue compact galaxies. A careful
study by \cite{Izotov:1998mj} gives the value $Y_P = 0.244 {\pm}
0.002$; however, also here there is a large scatter in the various
observed values, ranging from $Y_p = 0.230 \pm
0.003$~\citep{Olive:1997zu} over $Y_p = 0.2384 \pm
0.0025$~\citep{Peimbert:2001xf} and $Y_p = 0.2391 \pm
0.0020$~\citep{Luridiana:2003jy} to
$Y_p=0.2452\pm0.0015$~\citep{Izotov:1999wa}. Besides the large
scatter there is also the problem that the helium mass fraction
predicted from observations of deuterium combined with BBN,
$0.2470 < Y_p < 0.2487$, is larger than (and seems almost in
disagreement with) most of the observed helium abundances, which
probably points towards underestimated systematic errors, rather
than the need for new physics~\citep{Cyburt:2003fe,
Barger:2003zg}. Figure~\ref{fig:compilation} is a compilation of
the above measurements, and offers a direct comparison with the
current (large) errors from CMB observations, presented in
\SEC{chap:bspII;sec:cmbresults}, and with the potential of future
CMB measurements, discussed in \SEC{chap:bspII;sec:fma}.

The observed abundance of primordial $^7\textrm{Li}$ using the
Spite plateau is possibly spoiled by various systematic
effects~\citep{Ryan:1999vr,Salaris:2001vu}. Therefore it is more
appropriate to use the BBN predictions together with observations
to estimate the depletion factor
$f_7=\mbox{$^7\textrm{Li}_{\textrm{obs}}/^7\textrm{Li}_{\textrm{prim}}$}$
instead of using $^7\textrm{Li}_{\textrm{obs}}$ to infer the value
of $\eta_{10}$~\citep{Burles:2000zk,Hansen:2001hi}.

The numerical predictions of standard BBN (as well as various
non-standard scenarios) have reached a high level of
accuracy~\citep{Lopez:1998vk,Esposito:1999sz,Esposito:2000hh,Burles:2000zk},
and the precision of these codes is well beyond the systematic
errors discussed above.

\subsubsection*{BBN and the need for new physics}

If the CMB-determined helium mass fraction turns out to be as high
as suggested by BBN calculations together with the CMB observation
of $\Omega_bh^2$ (as discussed above), this could indicate a
systematic error in the present direct astrophysical helium
observations.

Alternatively, if the CMB could independently determine the helium
value with sufficient precision to confirm the present helium
observations, then this would be a smoking gun for new physics. In
fact, one could easily imagine non-standard BBN scenarios which
would agree with present observations of $\eta_{10}$, while having
a low helium mass fraction.  All what is needed is additional
non-equilibrium electron neutrinos produced at the time of
neutrino decoupling which would alter the $n-p$ reaction. This
could alter the resulting helium mass fraction while leaving the
deuterium abundance unchanged. One such possibility would be a
heavy sterile neutrino whose decay products include $\nu_e$. A
sterile neutrino with life-time of $1-5$ sec and with decay
channel $\nu_s \rightarrow \nu_e + \phi$ with $\phi$ a light
scalar (like a majoron), would leave the deuterium abundance
roughly untouched, but can change the helium mass fraction between
$\Delta Y_p = -0.025$ and $\Delta Y_p = 0.015$ if the sterile
neutrino mass is in the range $1-20$ MeV~\citep{Dolgov:1998st}. A
simpler model would be standard neutrino oscillation between a
sterile neutrino and the electron neutrino. The lifetime is about
1 sec when the sterile state has mass about 10 MeV, and the decay
channel is $\nu_s \rightarrow \nu_e +l + \bar l$ (with $l$ any
light lepton), and such masses and life-times are still
unconstrained for large mixing angle~\citep{Dolgov:2000jw}.
Related BBN issues are discussed by
\citet{Shi:1999kg,DiBari:2000wd,Kirilova:2001ab}. Such
possibilities are hard to constrain without an independent
measurement of the helium mass fraction.

Another much studied effect of neutrinos is the increased
expansion rate of the universe if additional degrees of freedom
are present (for BBN), and the degeneracy between the total
density in matter and relativistic particles (for CMB), which is
presented in detail in \SEC{chap:beyondsp;sec:rel}. The more
general set-up would then be to allow $N_\textrm{eff}$ as a
further free parameter both in the CMB and BBN analysis, but
because of the very weak dependence of the CMB on $Y_p$ this would
spoil any hope of being able to constrain the helium fraction with
the CMB; therefore we choose to fix $N_\textrm{eff} = 3.04$.

Also, an electron neutrino chemical potential could potentially
alter the BBN predictions~\citep{Kang:1992xa,Lesgourgues:1999wu},
however, with the observed neutrino oscillation parameters the
different neutrino chemical potentials would equilibrate before
the onset of
BBN~\citep{Dolgov:2002ab,Wong:2002fa,Abazajian:2002qx}, hence
virtually excluding this possibility \cite[see
however][]{Barger:2003rt}.

\subsection{WMAP Monte Carlo analysis}
 \label{chap:bspII;sec:cmbresults}

We use a modified version of the publicly available Markov Chain
Monte Carlo package \textsc{cosmomc} as described in
\cite{Lewis:2002ah} in order to construct Markov chains (see
\SEC{chap:data;sec:mcmc}) in our seven dimensional parameter
space. We sample over the following set of cosmological
parameters: the physical baryon and CDM densities, $\omega_b
\equiv \Omega_b h^2$ and $\omega_c \equiv \Omega_c h^2$, the
cosmological constant in units of the critical density,
$\Omega_{\Lambda}$, the scalar spectral index and the overall
normalization of the adiabatic power spectrum, $n_s$ and $A_s
\equiv \zeta_0^2$, \CF \rrp{eq:PS_for_zeta}, the redshift at which
the reionization fraction is a half, $\zreion$, and the primordial
helium mass fraction, $Y_p$. We restrict our analysis to flat
models, therefore the Hubble parameter is a derived parameter,
 \be
 h = [(\omega_c + \omega_b)/(1-\Omega_{\Lambda})]^{1/2} \eqdot
 \ee
 We consider purely adiabatic initial conditions and three massless neutrino
families for the reason given above. We do not consider either
gravitational waves or massive neutrinos. We include the WMAP data
from \cite{Kogut:2003et,Hinshaw:2003fc} (temperature and
polarization) with the routine for computing the likelihood
supplied by the WMAP team \citep{Verde:2003ey}. We make use of the
CBI \citep{Pearson:2002tr} and of the decorrelated
ACBAR~\citep{Kuo:2002ua} band powers above $\ell = 800$ to cover
the small angular scale region of the power spectrum.

Since $Y_p$ is a rather flat direction in parameter space with
present-day data, we find that a much larger number of samples is
needed in order to achieve good mixing and convergence of the
chains in the full 7D space. We use $M=4$ chains, each containing
approximately $N=3 \cdot 10^{5}$ samples. The mixing diagnostic is
carried out along the same lines as in \cite{Verde:2003ey}, by
means of the Gelman and Rubin criterion \citep{Gelman92b}.  The
burn-in of the chains also takes longer than in the case where
$Y_p$ is held fixed, and we discard 6000  samples per chain.

\subsubsection*{Results}

Marginalizing over all other parameters, we find that the helium
mass fraction from CMB alone is constrained to be
 \begin{align}
 & Y_p < 0.647 \quad \text{at 99\% l.c. (1 tail
limit)} \\
 \text{and} \quad  0.160 <  & Y_p < 0.501 \quad \text{at 68\% l.c. (2
 tails).}
 \end{align}
 Thus, for the first time the primordial
helium mass fraction has been observed using the cosmic microwave
background. However, present-day CMB data do not have by far
sufficient resolution to discriminate between the astrophysical
helium measurements, $Y_p \sim 0.244$, and the deuterium guided
BBN predictions, $Y_p \sim 0.248$, which would require percent
precision.

\begin{figure}[tb]
\begin{center}
\includegraphics[width=\onefigwidth]{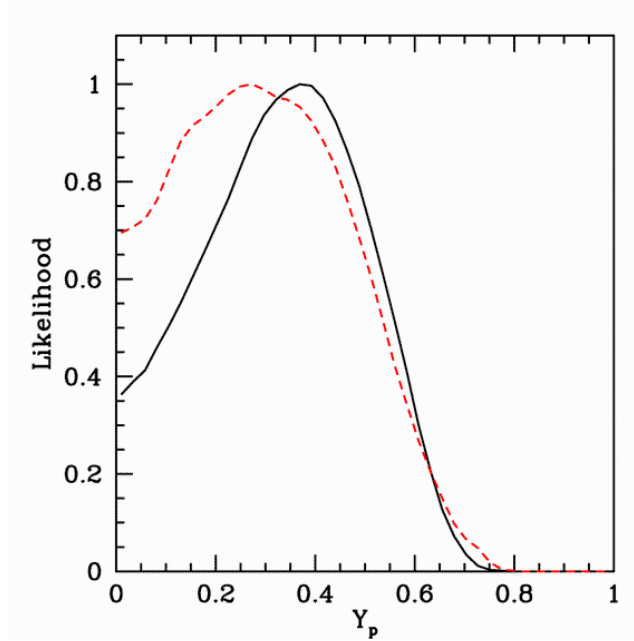}
\end{center}
\caption[Likelihood distribution for the helium mass fraction from
CMB data only.]{One-dimensional posterior likelihood distribution
for the helium mass fraction, $Y_p$, using CMB data only. The
solid-black line is for all other parameters marginalized, the
dashed-red line gives the mean likelihood.\label{fig:Ypmarg}}
\end{figure}

In \FIG{fig:Ypmarg} we plot the marginalized and the mean
likelihood of the Monte Carlo samples as a function of $Y_p$.  If
the likelihood distribution is Gaussian, then the 2 curves should
be indistinguishable. The difference between marginalized and mean
likelihood for $Y_p$ indicates that the marginalized parameters
are skewing the distribution, and therefore that correlations play
an important role. Although the mean of the 1D marginalized
likelihood is rather high, $\langle \like(Y_p) \rangle = 0.33$,
the mean likelihood peaks in the region indicated by astrophysical
measurements, $Y_p \sim 0.25$. In view of this difference, it is
important to understand the role of correlations with other
parameters, and we will turn to this issue now.

\begin{figure}[tb]
\begin{center}
\includegraphics[width=\onefigwidth]{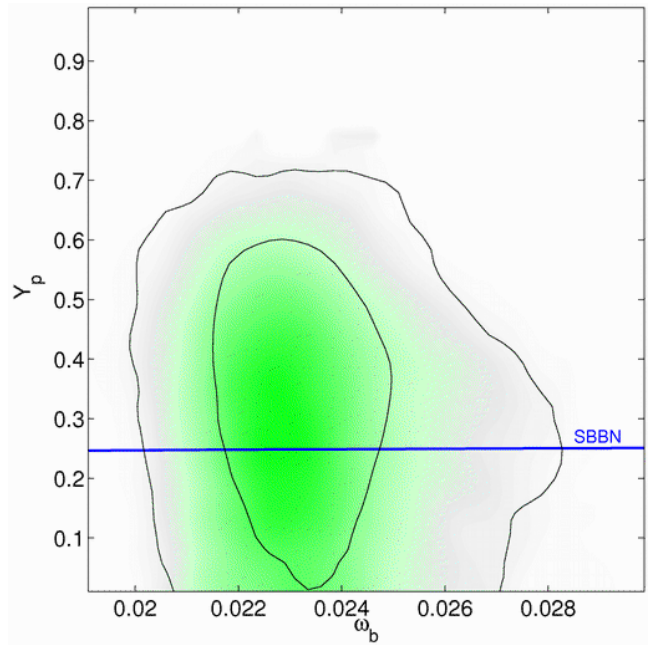}
\end{center}
\caption[Joint likelihood intervals in the ($\omega_b, Y_p$)-plane
from CMB data alone.]{Joint 68\% and 99\% likelihood contours in
the ($\omega_b, Y_p$)-plane from CMB data alone. The solid-blue
line gives the BBN prediction~\citep{Burles:1999zt}, which on this
figure almost looks like a straight line. \label{fig:obyp}}
\end{figure}

In \FIG{fig:obyp} we plot joint 68\% and 99\% confidence contours
in the ($\omega_b, Y_p$)-space. From the Monte Carlo samples we
obtain a small and negative correlation coefficient between the
two parameters, $\text{corr}(Y_p, \om_b )= -0.14$. Baryons and
helium appear to be anticorrelated simply because present-day WMAP
data do not map the peaks structure to sufficiently high $\ell$.
Precise measurements in the small angular scale region should
reveal the expected positive correlation between the baryon and
helium abundances, which is potentially important in order to
correctly combine BBN predictions and CMB measurements of the
baryon abundance. We turn to this question in more detail in the
next section. In BBN the baryon fraction and helium fraction are
correlated along a different direction, \CF \FIG{fig:obyp}.
However, this correlation is very weak, and the BBN relation gives
practically a flat line. Since the two parameters are not
independent from the CMB point of view, it is in fact not
completely accurate to perform the CMB analysis with fixed helium
mass fraction of $Y_p = 0.24$ to get the error-bars on the baryon
fraction, and then re-input this baryon fraction (and error-bars)
to predict the helium mass fraction from BBN. The most accurate
procedure is to analyse the CMB data leaving $Y_p$ as a free
parameter, thereby obtaining the correct (potentially larger)
error-bars on $\omega_b$ upon marginalization over $Y_p$.

In view of the emerging baryon tension between CMB and BBN, it is
important to check whether allowing helium as a free parameter can
significantly change the CMB determination of the baryon density
or its error. In order to evaluate in detail the impact of $Y_p$
on the error-bars for $\omega_b$, we consider the following three
cases.
\begin{itemize}
 \item[(a)] The usual case, when the helium fraction for the CMB
analysis is assumed to be known {\it a priori} and is fixed to the
canonical value $Y_p=0.24$.
 \item[(b)] A  case with a weak
astrophysical Gaussian prior on the helium fraction, which we take
to be $Y_p = 0.24 \pm 0.01$. As discussed above, the error-bars of
the astrophysical measurements are typically a factor 5 tighter
than this, but our prior is chosen to encompass the systematic
spread between the different observations.
 \item[(c)] The case in
which we assume a uniform prior for $Y_p$ in
  the range $0 \leq Y_p \leq 1$, \ie $Y_p$ is considered as a totally
  free parameter.
\end{itemize}

We do not find any significant change in the error-bars for
$\omega_b$ in the three different cases. The confidence intervals
on $\om_b$ alone are determined to be (case (c))  $0.0221 <
\omega_b < 0.0245$ at 68\% l.c. ($0.0204 < \omega_b < 0.0276$ at
99 \% l.c.). The standard deviation of $\omega_b$ as estimated
from the Monte Carlo samples is found to be $\hat{\sigma}_b = 1.3
\cdot 10^{-3}$. This is in complete agreement with the error-bars
on $\omega_b$ obtained by the WMAP team for the standard
$\Lambda$CDM case \citep{Spergel:2003cb}. We conclude that at the
level of precision of present-day CMB data, it is still safe to
treat the baryon abundance and the helium mass fraction as
independent parameters. This result is non-trivial, since the fact
that the damping tail is not yet precisely measured above the
second peak would a priori suggest that degeneracies between
$Y_p$, $\omega_b$ and $n_s$ could potentially play a role once the
assumption of zero uncertainty on $Y_p$ is relaxed. The impact of
$Y_p$ is small enough, and the error-bars on $\omega_b$ large
enough that a uniform prior on $Y_p$ can still be accommodated
within the uncertainty in the baryon abundance obtained for case
(a). However, the $Y_p-\omega_b$ correlation will have to be taken
into account to correctly analyze future CMB data, with a quality
such as Planck. We discuss this potential in the next section.

\begin{figure}[tb]
\begin{center}
\includegraphics[width=\onefigwidth]{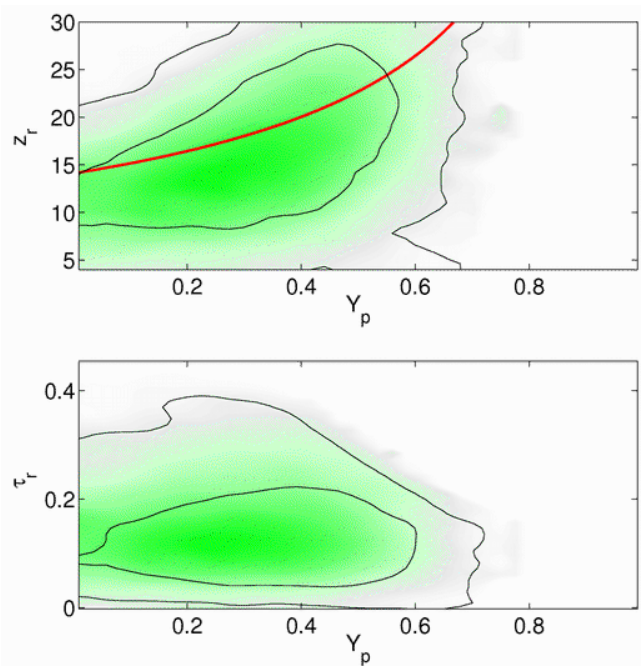}
\end{center}
\caption[Joint likelihood contours in the ($Y_p, \zreion$)-plane
and ($Y_p, \treion$)-plane from CMB data alone.]{Joint 68\% and
99\% likelihood
  contours in the ($Y_p, \zreion$)-plane (upper panel) and in the
  corresponding ($Y_p, \treion$)-plane (bottom panel) from CMB data
  alone.
  In the upper panel, the
  solid-red line is the relation $\zreion(Y_p)$ from \rrp{eq:tau_reion_approx},
  obtained by fixing the reionization optical depth to the value $\treion = 0.166$,
  while the other parameters are those of our fiducial $\Lambda$CDM
  model.  Although clearly the exact shape of $\zreion(Y_p)$ depends
  on the particular choice of cosmology, it is apparent that the
  $Y_p-\zreion$ degeneracy is along this direction. The correlation
  between $Y_p-\treion$ is almost negligible with present-day data
  (bottom panel).\label{fig:ypzr}}
\end{figure}

We observe the expected correlation between the redshift of
reionization and the helium fraction (\FIG{fig:ypzr}), which is
discussed above. The correlation coefficient between the two
parameters is found to be rather large and positive,
$\text{corr}(Y_p, \zreion )=0.40$. This correlation produces a
noticeable change in the marginalized 1D-likelihood distribution
for $\zreion$ as we go from case (a) to case (c). Marginalization
over the additional degree of freedom given by $Y_p$ broadens
considerably the error-bars on $\zreion$.  In fact, the 68\%
confidence interval for $\zreion$ increases by roughly 20\% (and
shifts  to somewhat higher values), from $10.2-20.9$ (case (a)) to
$10.6-23.3$ (case (c)).  Case (b) exhibits similar error-bars as
case (a). On the other hand, the determination of the reionization
optical depth is not affected by the inclusion of helium as a free
parameter, giving in all cases $0.08 < \treion < 0.23$.
Correspondingly, the correlation is less significant,
$\text{corr}(Y_p, \treion )=-0.11$. We therefore conclude that the
differences in the determination of $\zreion$ are due only to the
variation of the amount of electrons available for reionization as
$Y_p$ is changed.

\begin{figure}[tb]
\begin{center}
\includegraphics[width=\onefigwidth]{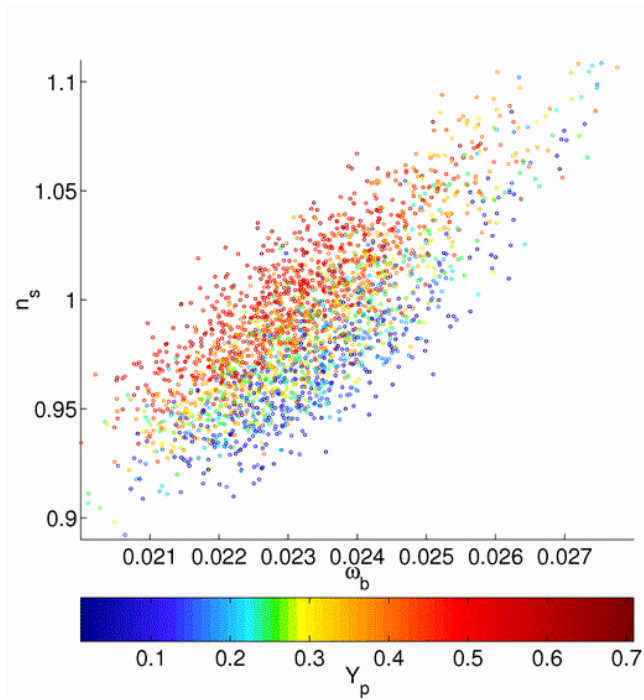}
\end{center}
\caption[Scatter plot in the $\om_b-n_s$ plane.] {Scatter plot in
the $\om_b-n_s$ plane, with the value of
  $Y_p$ rendered following the color scale. Green corresponds roughly
  to the BBN preferred value.\label{fig:3D}}
\end{figure}

Leaving $Y_p$ as a free parameter also has an impact on the
relation between $\om_b$ and the scalar spectral index, $n_s$. The
extra power suppression on small scales which is produced by a
larger $Y_p$ can be compensated by a blue spectral index, \CF
\FIG{fig:3D}.

\subsection{Potential of future CMB observations}
\label{chap:bspII;sec:fma}

In order to estimate the precision with which future satellite CMB
measurements  will be able to constrain the helium mass fraction
we perform a Fisher matrix analysis along the lines presented in
\SEC{chap:data;sec:fma}. As already emphasized, in order to obtain
a reliable prediction, it is extremely important to choose a
parameter set whose effect on the CMB power spectrum is as linear
and uncorrelated as possible. Here we improve upon the choice made
in \SEC{chap:beyondsp;sec:fma} by adopting the full set of normal
parameters introduced in \SEC{chap:params:sec:normal}. Our nine
dimensional basis parameter set is then
 \be
 \params = \left\{ \kA, \kB, \kV, \kR, \kM, \rZ, A_s, n_s, Y_p \right\} \eqcomma
 \ee
where the scalar power spectrum normalization constant is $A_s =
\zeta_0^2$, see (\refp{eq:PS_for_zeta}). The quantities $\kA, \kB,
\kV, \kR, \kM, \rZ$ are defined in
Eqs.~(\ref{eq:def_kA}--\refp{eq:def_kM}). It has been shown that
the normal parameter set is very well adapted to the FMA, which
give accurate predictions \citep{Kosowsky:2002zt}. Since here we
are interested in the predictions for $\kB = \Om_b h^2$ and $Y_p$,
we do not need to explicitly map the FMA forecasts in the normal
parameter space onto the cosmological parameter space.

The choice of the physical parameter set makes it easy to
implement in the FMA interesting theoretical priors. For instance,
we are interested in imposing flatness in our forecast, in order
to be able to directly compare present-day accuracy on $Y_p$ with
the potential of Planck and of and ideal CMB experiment (see
below). The prior on the curvature of the universe is imposed in
the FMA by fixing the value of the parameter $\kA$ to the one of
the fiducial model. In fact, the parameter  $\kA$ is a
generalization of the shift parameter, which describes the
sideways shift of the acoustic peak structure of the CMB power
spectrum as a function of the geometry of the universe and its
content in matter, radiation and cosmological constant. Although
imposing $\kA=$ const is not the same as having a constant spatial
curvature over the full range of cosmological parameters, for the
purpose of evaluating derivatives the two conditions reduce to the
same. The fact that our fiducial model is actually slightly open
(see below), does not make any substantial difference in the
results, apart from reducing the numerical inaccuracies which
would arise had we computed the derivatives around an exactly flat
model. We can also easily impose a prior knowledge of the helium
fraction, by fixing the value of $Y_p$, as is usually the case for
present CMB analyses, and investigate how this modifies the
expected error on the the baryon density.

\subsubsection*{Accuracy issues}

We numerically compute double sided derivative of the power
spectrum around the fiducial model with cosmological parameters
given in \TAB{table:fiducial}. We find it necessary to increase
the accuracy of {\textsc{camb}} by a factor of 3 in each of the
``accuracy boost'' values. As a fiducial model, we use the best
fit model to the WMAP data for the standard $\Lambda$CDM scenario,
as given in Table 1 of \cite{Spergel:2003cb}. However, in order to
avoid numerical inaccuracies which arise when differentiating
around a flat model, we reduce slightly the value of
$\Omega_\Lambda$ by imposing an open universe,
$\Omega_{\textrm{tot}} = 0.99$.

 We perform the FMA for the expected capabilities of Planck's High
Frequency Instrument (HFI) and for an ideal CMB measurement which
would be cosmic variance limited (CVL) both in temperature and in
E-polarization (and we do not consider the B-polarization
spectrum), and therefore represents the best possible parameter
measurement from CMB anisotropies alone. The complicated issues
coming from foreground removals, point source subtraction, etc.\
are assumed to be already (roughly) taken into account by the
experimental parameters, see \SEC{chap:data;sec:fma_expepars} for
definitions. These are the effective percentage sky coverage
$f\sky$, the number of channels, the sensitivity of each channel
$\sigma_c^{T,E}$ for temperature (T) and E-polarization (E) in
$\mu$K and the angular resolution $\theta_c^{T,E}$ (in arcmin).
For Planck HFI, we take the three channels with frequencies 100,
143 and 217 GHz, with respectively $\sigma_{c=1,2,3}^{T}= 5.4,
6.0, 13.1$ and $\sigma_{c=2,3}^{E}= 11.4, 26.7$ and we have
$f\sky=0.85$ \citep{Planck:Website} Since the CVL is an ideal
experiment, we put its noise to zero and assume perfect
foregrounds removal, so that $f\sky=1$. In order to test the
accuracy of our predictions and compare present-day results with
the forecasts, we also perform an FMA with WMAP first year
parameters, obtaining excellent agreement between the FMA results
and the error-bars from actual data. For the purpose of
comparison, we include forecasts for the full WMAP four year
mission, which will also measure E-polarization and reduce
present-day errors on the temperature spectrum by a factor of two.
We limit the range of multipoles to $\ell < 2000$, because at
smaller angular scales non-primary anisotropies begin to dominate
(Sunyaev-Zeldovich effect). \cite{Seljak:2003th} discuss the issue
of numerical precision of three different CMB codes and conclude
that they are accurate to within $0.1\%$. While this is
encouraging, it is not of direct relevance to this work, since
what matters in the computation of derivatives is not much the
absolute precision of the spectra, but rather their relative
accuracy.

\begin{table}[tb!]
 \centering
\begin{tabular}{|l  c c|}
\hline Parameter      &  & Value \\ \hline
Baryons        & $\Omega_b$      & $0.046$  \\
Matter         & $\Omega_m$      & $0.270$  \\
Dark Energy    & $\Om_\La    $ & $0.720$  \\
Radiation      & $\Omega_{\textrm{rad}}$  & $7.95 \cdot 10^{-5}$ \\
Massless $\nu$ families & $N_\nu $         & 3.04 \\
Total density  & $\Omega_{\textrm{tot}}$  & $0.990$  \\
Hubble constant & $ h$           & $0.72$   \\
Optical depth  & $\treion$          & $0.166$  \\
Spectral index & $n_s$           & $0.99$ \\
Normalization  & $A_s$           & $2\cdot10^{-9}$
\\ \hline
\end{tabular}
\caption[Fiducial model for the Fisher matrix
analysis.]{\label{table:fiducial} Cosmological parameters for the
fiducial $\Lambda$CDM model around which the FMA is performed. We
choose a slightly open model to avoid numerical inaccuracies in
the derivatives.}
\end{table}

\subsubsection*{Forecasts and discussion}

Table~\ref{table:fmares} summarizes our forecasts for the future
measurements and compares them with the results obtained from WMAP
actual data.

\begin{table}[tb]
\centering
\begin{tabular}{|l | c c |  c c | c |}
 \hline \separator{Temperature +   TE-cross  + E-polarization}
      &  \multicolumn{2}{c|}{No priors} &
      \multicolumn{2}{c|}{Flatness} & Flatness and\\
      &  \multicolumn{2}{c|}{}      &   \multicolumn{2}{c|}{} & $Y_p = 0.24$ \\
      & $\frac{\Delta Y_p}{Y_p}$ & $\frac{\Delta \om_b}{\om_b}$
      & $\frac{\Delta Y_p}{Y_p}$ & $\frac{\Delta \om_b}{\om_b}$
      & $\frac{\Delta \om_b}{\om_b}$ \\
WMAP 4yrs  \footnotemark[1] \rule{0pt}{4ex}
      & $\sim50$ & $2.92$                 & $\sim40  $ & $2.86$                      & $2.86$     \\
Planck& $7.60$& $1.31$                 & $ 4.96$ & $1.26$                      & $0.70$    \\
CVL   & $2.59$& $0.34$                 & $ 1.52$ & $0.32$ & $0.13$
\\\hline \separator{Temperature + TE-cross} WMAP 1st yr
\footnotemark[2] \rule{0pt}{4ex}
      &  N/A  & N/A                    & $71.25$ & $5.04$                      & $5.04$     \\
WMAP 4yrs  \footnotemark[1] \rule{0pt}{4ex}
      & $\sim75$& $4.10$                  & $\sim60  $ & $3.94$                      & $3.94$     \\
Planck& $8.91$& $1.74$                 & $ 6.60$ & $1.63$                      & $0.74$    \\
CVL   & $5.18$& $0.55$                 & $ 2.84$ & $0.55$                      & $0.19$      \\
\hline
\end{tabular}
\footnotetext[1]{~FMA forecast, four year mission including
E-polarization.} \footnotetext[2]{~Actual WMAP data and other CMB
experiments, this work.}
 \caption[Fisher
matrix forecasts and
  comparison with present-day results for the helium mass fraction.]{\label{table:fmares} Fisher
matrix forecasts and
  comparison with present-day results for different priors and
using different combinations of temperature and polarization CMB
spectra. Errors are in percent with respect to the values of the
fiducial model, $Y_p=0.24$ and $\omega_b = 0.0238$ ($1\sigma$ l.c.
all other marginalized).}
\end{table}

We notice that when the WMAP full four year data will be available
(including E-polarization), the error on the baryon density is
expected to decrease by a factor of two to about $3\%$, compared
to today's $5\%$ (assuming flatness). Nevertheless, inclusion of
$Y_p$ as a free parameter will still have no effect on the
determination of $\om_b$ for WMAP, \ie $Y_p$ will remain an
essentially flat direction when marginalized over. While the
determination of the helium fraction will improve, the FMA cannot
reliably assess quantitatively how much, since for such large
errors the likelihood distribution is not Gaussian and the
quadratic approximation breaks down. In the table we therefore
give the FMA estimation as an indication, with the caveat that the
Fisher approximation is likely to be inaccurate for the real
errors on $Y_p$ from WMAP's four year data-set.

\begin{figure}[tb]
\begin{center}
\includegraphics[width=\onefigwidth]{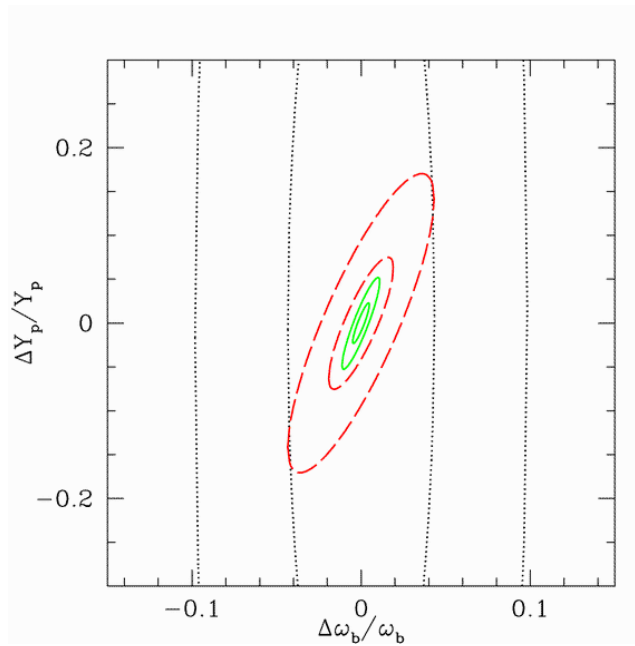}
\end{center}
\caption[FMA forecast for the expected errors on the helium
fraction.]{FMA forecast for the expected errors from WMAP four
year mission (dotted-black), Planck (dashed-red) and a CVL
experiment (solid-green). The ellipses encompass $1\sigma$ and
$3\sigma$ joint likelihood regions for $\omega_b-Y_p$ (all other
parameters marginalized). The axis values give the error in with
respect to the fiducial model values. This forecast is for the
full CMB information (Temperature, TE-cross, E-polarization) and
assumes flatness.\label{fig:fmaYpob}}
\end{figure}

It is interesting that for Planck, the effect of the helium
fraction can no longer be neglected. Inclusion of the helium
fraction increases the error on $\om_b$ by roughly $80\%$, from
$0.7\%$ to $1.3\%$. The correlation between the two parameters
will have to be taken into account, as is evident from
\FIG{fig:fmaYpob}. The expected correlation coefficient is
$\text{corr}(Y_p, \omega_b) = 0.84\quad(0.91)$ for Planck (for
CVL). The expected $1\sigma$ error on $Y_p$ is about $5\%$ for
Planck, or $\Delta Y_p \sim 0.01$. This is of the same order as
the spread in current astrophysical measurements. We conclude that
in Planck-accuracy data analysis, it will be necessary to include
the uncertainty in the determination of the helium mass fraction,
at least in the form of a Gaussian prior over $Y_p$ of the type we
used in the CMB data analysis presented above.

Finally, measuring CMB temperature and polarization with cosmic
variance accuracy would allow $Y_p$ to be constrained to within
$1.5\%$, or $\Delta Y_p \sim 0.0036$ (assuming flatness). Such an
ideal measurement would be able to discriminate between the
BBN-guided, deuterium based helium value and the current lowest,
direct helium observations (\CF \FIG{fig:compilation}).

Our forecasts for the uncertainty in the Helium mass fraction from
future observations are in excellent agreement with the findings
of \cite{Kaplinghat:2003bh}. There, the standard deviation on
$Y_p$ for Planck is estimated to be $\Delta Y_p = 0.012$.
\cite{Kaplinghat:2003bh} also consider an experiment (CMBPol) with
characteristics similar to our CVL, for which they forecast
$\Delta Y_p = 0.0039$, again in close agreement with our result.
In an earlier work, \cite{Eisenstein:1998tu} found for Planck
(temperature and polarization) $\Delta Y_p = 0.013$, also in
satisfactory concordance with our result. It should be noticed
that the forecast reported for MAP in Table 2 of
\cite{Eisenstein:1998tu}, namely $\Delta Y_p = 0.02$, is nothing
but the Gaussian prior $Y_p = 0.24 \pm 0.02$ which was assumed in
their analysis.

The main source of improvement for the determination of $Y_p$ will
be the better sampling of the temperature damping tail provided by
Planck and the CVL. Polarization measurements have mainly the
effect of reducing the errors on other parameters. In fact, we
have checked that excluding from our FMA the $2 \leq \ell \leq 50$
region of the E-polarization and ET-correlation spectra changes
the forecast precision on $Y_p$ less than about 10-15\% for Planck
and less than a few percent for CVL. This supports the conclusion
that the low-$\ell$ reionization bump is not very useful in
measuring the helium abundance, because of the degeneracy with
$\zreion$.

\clearpage
\section{Time variations of the fine-structure constant}
\label{chap:bspIII;sec:alpha}

The search for observational evidence for time or space variations
of the `fundamental' constants that can be measured in our
four-dimensional world is an extremely exciting area of current
research, with several independent claims of detections in
different contexts emerging in the past few years. In particular,
possible time variations of the fine-structure constant can be
tested with the CMB, and represent another line of investigation
going beyond the standard description of  cosmology. The contents
of this section summarize the latest result of a rather large
collaboration I have been involved with, aimed at constraining
time variations of the fine-structure constant using CMB
anisotropy. We thoroughly studied the issue of crucial
degeneracies with other cosmological parameters and discussing
what improvements can be expected with forthcoming data-sets
\citep*{Martins:2002iv,Martins:2003pe,Rocha:2004}.

We motivate the search for time variations of the fine-structure
constant in \SEC{chap:bspIII;sec:motivation}, and review the
current observational status of observations other than the CMB in
\SEC{chap:bspIII;sec:status}. After presenting the relevance of
the fine-structure constant for CMB anisotropies in
\SEC{chap:bspIII;sec:alphadec} and
\SEC{chap:bspIII;sec:alphareion}, in \SEC{chap:bspIII;sec:cmbres}
we provide up-to-date WMAP constraints on the value of $\alpha$ at
the epoch of decoupling; \SEC{chap:bspIII;sec:fma} is dedicated to
a detailed Fisher matrix analysis which encompasses the standard
parameters plus the fine-structure constant for the full WMAP four
year data, for the Planck satellite and for a cosmic variance
limited, ideal experiment.

\subsection{Motivation}
\label{chap:bspIII;sec:motivation}

Cosmology and astrophysics play an increasingly important role as
testing ground for our understanding of fundamental physics, since
they provide us with extreme conditions (that one has no hope of
reproducing in terrestrial laboratories) in which to carry out a
plethora of tests and search for new paradigms. Perhaps the more
illuminating example is that of multidimensional cosmology:
currently preferred unification theories
\citep{Polchinski1:1998,Damour:2002vu} predict the existence of
additional space-time dimensions, which will have a number of
possibly observable consequences, including modifications in the
gravitational laws on very large (or very small) scales
\citep{Will:2001mx} and space-time variations of the fundamental
constants of nature \citep{Martins:2002fm,Uzan:2002vq}.

The most promising case, and the one that has been the subject of
most recent work and speculation, is that of the fine-structure
constant
 \be
 \alpha = \dfrac{e^2}{\hbar c}
 \ee
where $e$ is the electron charge, $c$ the speed of light and
$\hbar$ Planck's constant.

There have been a number of recent reports of evidence for a time
variation of fundamental constants
\citep{Webb:2000mn,Webb:2002vd,Murphy:2000ns,Ivanchik:2002me},
which we review below. Apart from their obvious direct impact if
confirmed, they are also crucial in a different, indirect way,
since they provide us with an important (and possibly even unique)
opportunity to test a number of fundamental physics models, such
as string theory. Indeed here the issue is not \textit{if} such a
theory predicts such variations, but \textit{at what level} it
does so, and hence if there is any hope of detecting them in the
near future, or if we have done it already.

On the other hand, the theoretical expectation in the simplest,
best motivated model is that $\alpha$ should be a non-decreasing
function of time
\citep{Damour:1993id,Santiago:1998ae,Barrow:2001iw}. This is based
on rather general and simple assumptions, in particular that the
cosmological dynamics of the fine-structure constant is governed
by a scalar field whose behavior is akin to that of a dilaton. If
this is so, then it is particularly important to try to constrain
it at earlier epochs, where any variations relative to the
present-day value should be larger. However, one of the
interpretations of the Oklo results is that $\alpha$ was
\textit{larger} at an epoch corresponding to a redshift of about
$z\sim0.1$ than today, whereas the quasar results indicate that
$\alpha$ was {\it smaller} at $z\sim2-3$ than today, see below for
more details. If both results are validated by future experiments,
then the above theoretical expectation must clearly be wrong,
which would be a perfect example of using astrophysics to learn
about fundamental physics. Playing devil's advocate, one could
certainly conceive that cosmological observations of this kind
could one day prove string theory wrong. Indeed, it has been
argued \citep{Damour:2002vu,Damour:2003iz} that even the results
of Webb and collaborators may be hard to explain in the simplest,
best motivated models where the variation of the fine-structure
constant is driven by the spacetime variation of a very light
scalar field.

Cosmic microwave background anisotropies provide a tool to measure
the fine-structure constant at high redshift, being mostly
sensitive to the epoch of decoupling, $z \sim 1100$.

\subsection{The observational status}
\label{chap:bspIII;sec:status}

The recent explosion of interest in the study of varying constants
is mostly due to the results of Webb and collaborators
\citep{Murphy:2000pz,Webb:2000mn,Murphy:2000ns,Murphy:2001nu} of a
$4\sigma$ detection of a fine-structure constant that was smaller
in the past,
\begin{equation}
\frac{\Delta\alpha}{\alpha} = (-0.72 \pm 0.18)\times
10^{-5}\,,\quad z\sim0.5-3.5\,; \label{eq:webb_result}
\end{equation}
indeed, more recent work \citep{Murphy:2002ve,Webb:2002vd}
provides an even stronger detection. These results are obtained
through comparisons of various transitions (involving various
different atoms) in the laboratory and in quasar absorption
systems, using the fact that the size of the relativistic
corrections goes as $(\alpha Z)^{2}$. A number of tests for
possible systematic effects have been carried out, all of which
have been found either not to affect the results or to make the
detection even stronger if corrected for.

A somewhat analogous (though simpler) technique uses molecular
hydrogen transitions in damped Lyman-$\alpha$ systems to measure
the ratio of the proton and electron masses, $\mu=m_p/m_e$ (using
the fact that electron vibro-rotational lines depend on the
reduced mass of the molecule, and this dependence is different for
different transitions). The latest results \citep{Ivanchik:2001ji}
using two systems at redshifts $z\sim2.3$ and $z\sim3.0$ are
\begin{equation}
\frac{\Delta\mu}{\mu} = (5.7 \pm 3.8)\times 10^{-5}\eqcomma
\end{equation}
or
\begin{equation}
\frac{\Delta\mu}{\mu} = (12.5 \pm 4.5)\times 10^{-5}\eqcomma
\end{equation}
depending on which of the (two) available tables of ``standard''
laboratory wavelengths is used. This implies a $1.5 \sigma$
detection in the more conservative case, though it also casts some
doubts on the accuracy of the laboratory results, and on the
influence of systematic effects in general.

We should also mention a recent re-analysis \citep{Fujii:2002bi}
of the well-known Oklo bound \citep{Damour:1996zw}. Using new
Samarium samples collected deeper underground (aiming to minimize
contamination), these authors again provide two possible results
for both $\alpha$ and the analogous coupling for the strong
nuclear force, $\alpha_s$,
\begin{equation}
\frac{\dot\alpha}{\alpha} \sim \frac{\dot\alpha_s}{\alpha_s}= (0.4
\pm 0.5)\times 10^{-17} \yr^{-1} \label{eq:oklo1}
\end{equation}
or
\begin{equation}
\frac{\dot\alpha}{\alpha} \sim \frac{\dot\alpha_s}{\alpha_s}=
-(4.4 \pm 0.4)\times 10^{-17} \yr^{-1}\eqdot \label{eq:oklo2}
\end{equation}
Note that these are given as rates of variation, and effectively
probe timescales corresponding to a cosmological redshift of about
$z\sim 0.1$. Unlike the case above, these two values correspond to
two possible physical branches of the solution. See
\cite{Fujii:2002bi} for a discussion of why this method yields two
solutions (and also note that these results have opposite signs
relative to previously published ones, \citealp{Fujii:1998kn}).
While the first of these branches provides a null result,
(\ref{eq:oklo2}) is a strong detection of an $\alpha$ that was
{\it larger} at $z\sim0.1$, that is a relative variation that is
opposite to Webb's result (\ref{eq:webb_result}). Even though
there are some hints (coming from the analysis of other Gadolinium
samples) that the first branch is preferred, this is by no means
settled and further analysis is required to verify it.

Still we can speculate about the possibility that the second
branch turns out to be the correct one. Indeed this would
definitely be the most exciting possibility. While in itself this
wouldn't contradict Webb's results (since Oklo probes much smaller
redshift and the suggested magnitude of the variation is smaller
than that suggested by the quasar data), it would have striking
effects on the theoretical modelling of such variations. In fact,
proof that $\alpha$ was once larger than today's value would sound
the death knell for any theory which models the varying $\alpha$
through a scalar field whose behaviour is akin to that of a
dilaton. Examples include Bekenstein's theory
\citep{Bekenstein:1982eu} or simple variations thereof
\citep{Sandvik:2001rv,Olive:2001vz}. Indeed, one can quite easily
see \citep{Damour:1993id,Santiago:1998ae} that in any such model
having sensible cosmological parameters and obeying other standard
constraints, $\alpha$ must be a monotonically increasing function
of time. Since these dilatonic-type models are arguably the
simplest and best-motivated models for varying $\al$ from a
particle physics point of view, any evidence against them would be
extremely exciting, since it would point towards the presence of
significantly different, yet undiscovered physical mechanisms.

Finally, we also mention that there have been recent proposals
\citep{Braxmaier:2001ph} of more accurate laboratory tests of the
time independence of $\alpha$ and $\mu$ using monolithic
resonators, which could improve current bounds by an order of
magnitude or more.

However, given that there are both theoretical and experimental
reasons to expect that any recent variations will be small, it is
important to develop tools allowing us to measure $\alpha$ in the
early universe, as variations with respect to the present value
could be much larger then.

\subsection{Effects of $\alpha$ on the ionization history}
 \label{chap:bspIII;sec:alphadec}

The reason why the CMB can be used as a probe of variations of the
fine-structure constant is that these alter the ionization history
of the universe. Here we present the dominant effects, see
\cite{Hannestad:1998xp,Kaplinghat:1998ry} for a detailed
treatment.

The impact of the fine-structure constant on the CMB comes from
the dependence of the differential optical depth $\taudot$
(\refp{eq:differential_Thomson_optical_depth}) on the Thomson
scattering cross section, which is
 \be
 \si_T = \dfrac{8\pi \alpha^2 \hbar^2}{3m_e^2 c^2} \eqcomma
 \ee
where we have reintroduced the speed of light $c$ and the Planck
constant $\hbar$, and $m_e$ is the electron mass. Now the
equilibrium electron ionization fraction $x_e^{\text{eq}} \equiv
n_e/n_H$ goes approximately as
 \be
 x_e^{\text{eq}} \propto \left(\dfrac{m_e}{T}\right)^{3/2}
 \exp(-B/T) \eqcomma
 \ee
where $B$ is the Hydrogen binding energy
 \be
 B = \alpha^2m_e c^2/2
 \ee
\citep[see \eg][]{Inflation_Kolb}. If we ignore the fact that
$x_e(z)$ does not precisely track its equilibrium value, and since
the exponential factor dominates near recombination, we would
simply expect from $T \propto 1/a \propto z$ that the reionization
fraction be just a function of $z/\alpha^2$. This turns out to be
approximately correct, even if the effect of the factor
$(m_e/T)^{3/2}$ and the departure of $x_e$ from $x_e^\text{eq}$
need to be taken into account for a more precise estimation
\citep{Kaplinghat:1998ry}.

\begin{figure}[tb]
\centering
 \includegraphics[width=\twofigswidth]{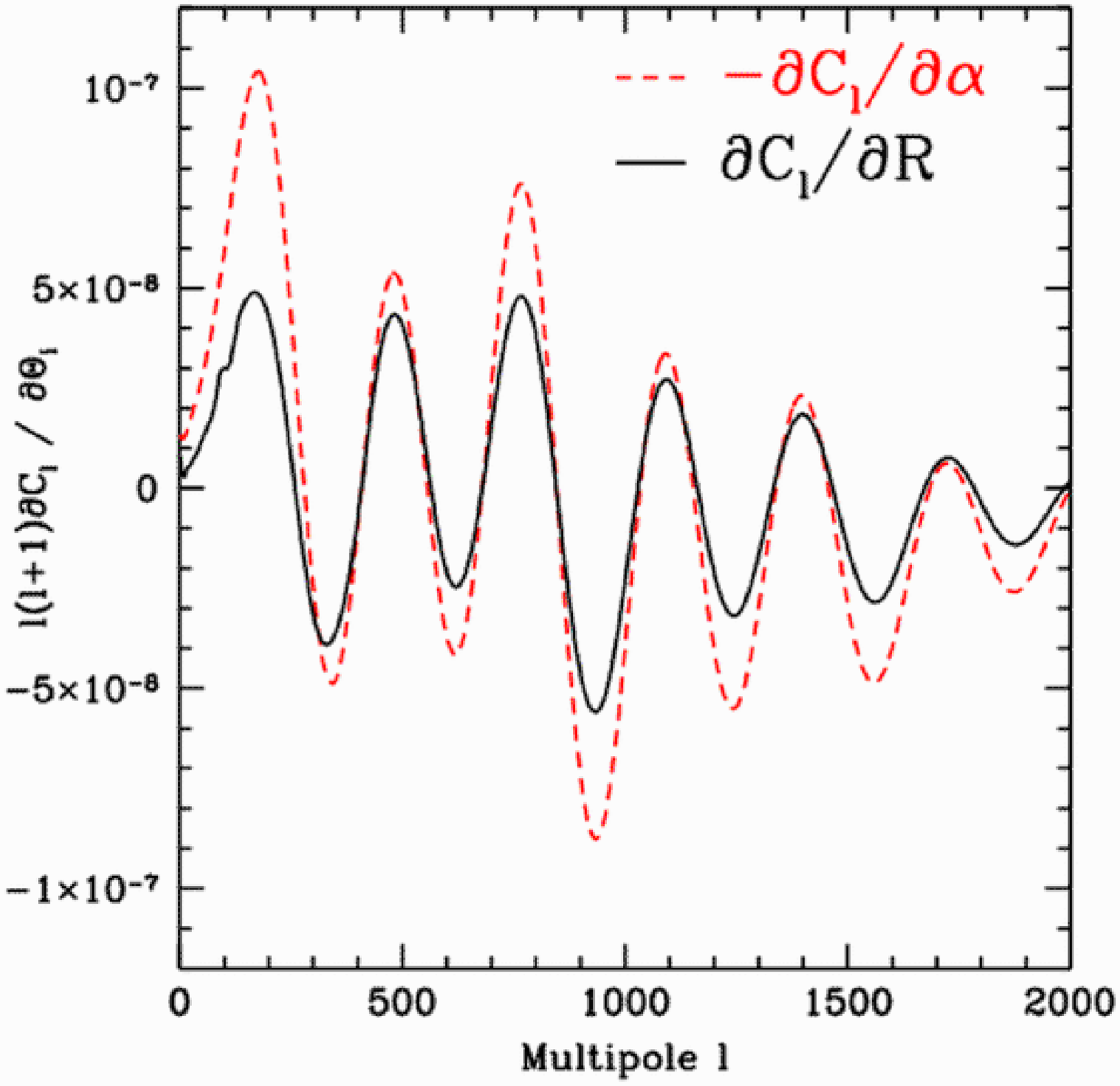}\hfill%
 \includegraphics[width=\twofigswidth]{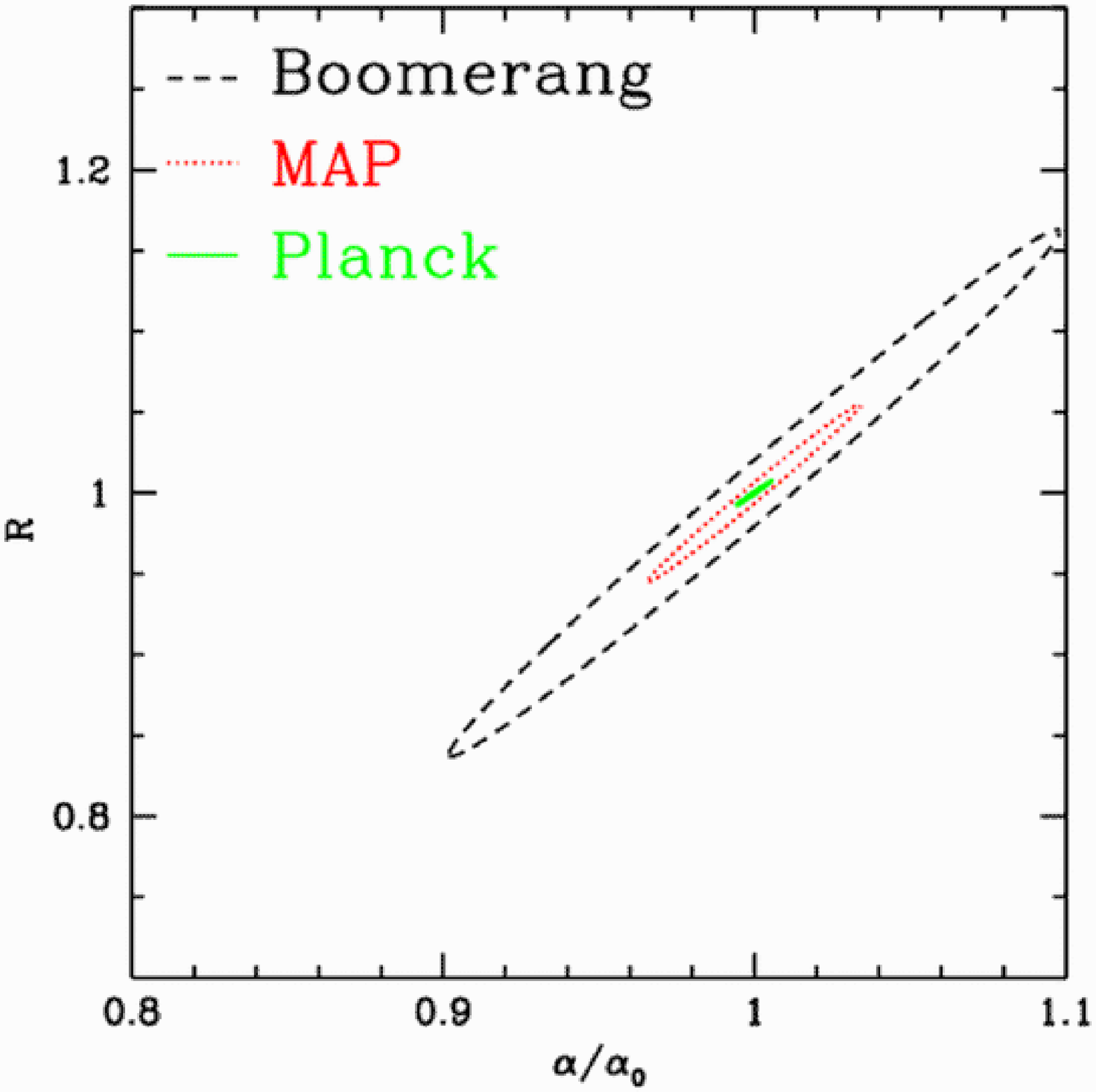}
 \caption[Degeneracy between the shift parameter and the fine-structure constant.]
 {Left panel: derivatives of the temperature spectrum with respect to $\al_\dec$ and the shift parameter
$\R$. We plot $-\partial C_\ell/ \partial \al_\dec$ to facilitate
the comparison with $\partial C_\ell/ \partial \R$. The two
derivatives are perfectly in phase: this is responsible for the
degeneracy between the corresponding parameters (right panel,
Fisher matrix analysis). Only the different amplitudes allow an
experiment which maps sufficiently high multipoles with high
accuracy to distinguish between them, in particular revealing the
change in the damping scale brought about by changes in
$\al_\dec$. In the right panel, the Fisher matrix results contain
$1\si$ of the likelihood (including temperature only), and clearly
indicate a strong correlation between the two parameters
\citep[see][]{Martins:2002iv}. \label{fig:alpha_shift_deg}}
\end{figure}

\begin{figure}[tb]
\centering
\includegraphics[width=\twofigswidth]{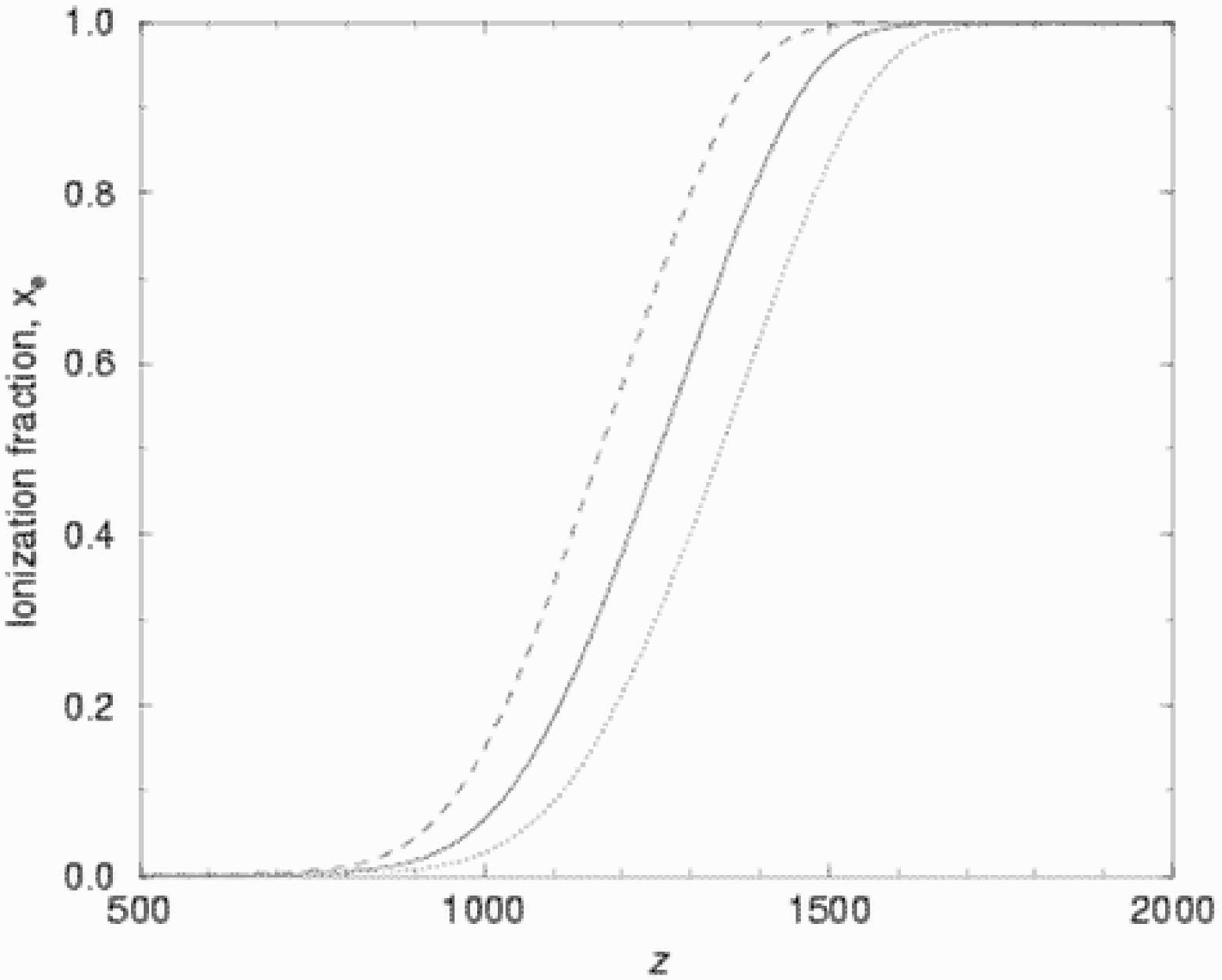}\hfill%
\includegraphics[width=\twofigswidth]{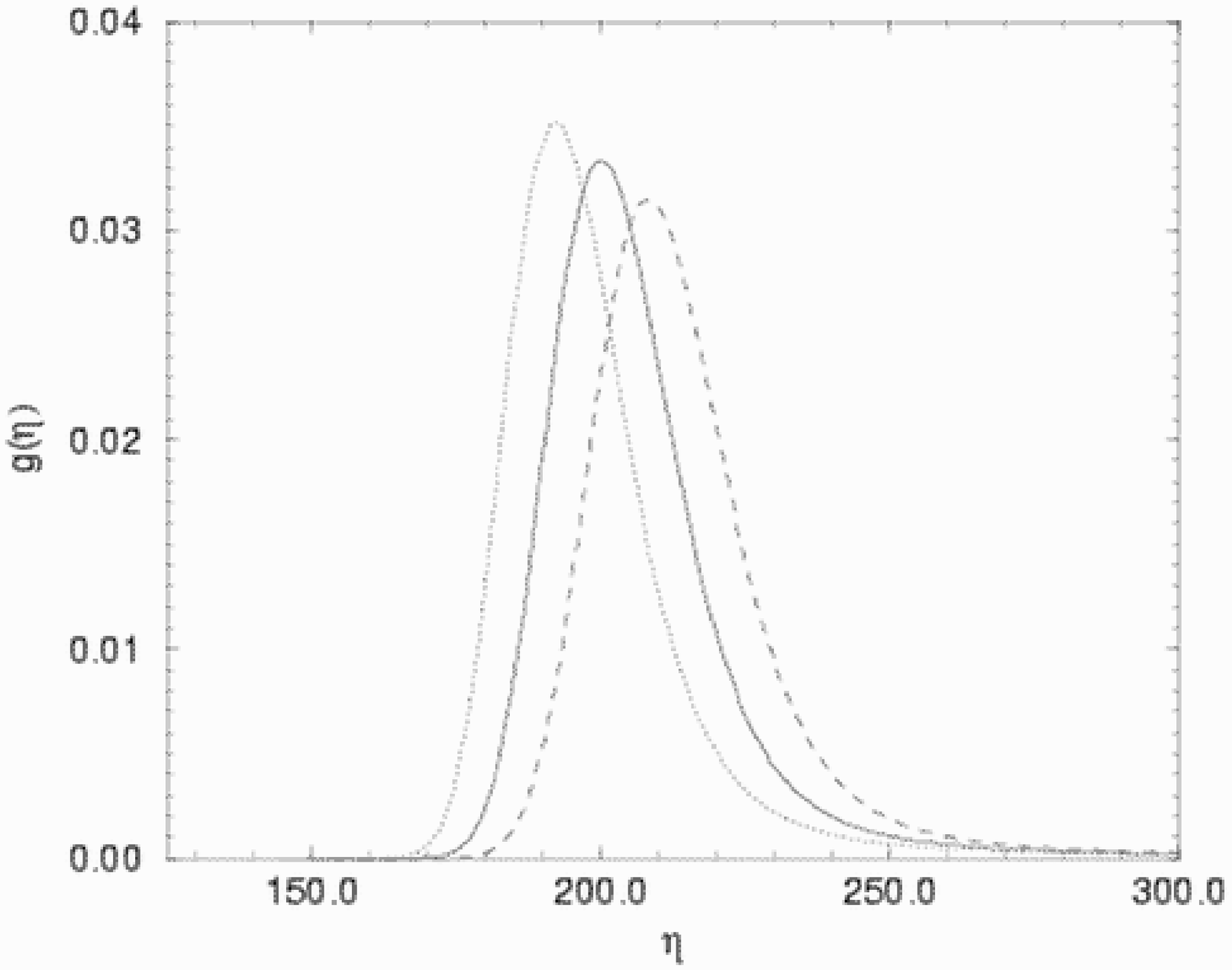}%
\caption[Ionization fraction and visibility function for different
values of the fine-structure constant at the epoch of decoupling.]
{Ionization fraction as a function of redshift (left panel) and
visibility function as a function of conformal time (right panel)
for different values of the fine-structure constant at decoupling:
$\alpha_\dec/\al_0 = 1$ (solid), $\alpha_\dec/\al_0 = 1.03$
(dotted), $\alpha_\dec/\al_0 = 0.97$ (dashed). Decoupling happens
earlier and the last scattering surface is narrower for
$\alpha_\dec/\al_0
> 1$. \label{fig:visibility_function}}
\end{figure}

In general, around the decoupling epoch relevant for the CMB, the
fine-structure constant can be expected to evolve with redshift,
$\alpha = \alpha(z)$, but we can take a constant value
$\alpha_\dec \equiv \alpha(z_\dec)$ instead and consider it as an
{\it effective} value averaged over the recombination process.
Summarizing, there are two important changes in the reionization
history brought about by a change in $\alpha_\dec$, the value of
$\alpha$ at the recombination epoch, which are best discussed in
terms of changes on the visibility function $g(z)$, defined in
\rrp{eq:def_visibility_function}. A larger value of $\alpha_\dec$
with respect to $\alpha_0$, its value today, implies:
 \begin{itemize}
 \item an increased redshift of last scattering: as estimated
 above, this follows from rescaling the reionization fractions as
 $z/\alpha_\dec^2$, hence decoupling happens earlier for a larger
 $\alpha_\dec$, which means that the sound horizon $r_s(z_\dec)$,
 see \rrp{eq:sound_horizon}, is smaller. As a consequence, we
 expect a shift of the peaks' structure to larger $\ell$ values,
 according to (\refp{eq:ad_peak_position}). This effect will be
 degenerate with the shift parameter $\Rshift$
 (\refp{eq:def_shift_parameter}) or equivalently with the normal
 parameter $\kA$, \rrp{eq:def_kA}, as shown in \FIG{fig:alpha_shift_deg}.
 There will also be a boost of
 the first acoustic peak due to the increased early ISW effect,
 see \SEC{chap:params;sec:isw}.
 \item A narrower peak of the visibility function: by increasing $\alpha_\dec$
the peak of the visibility function is moved to a larger redshift,
when the expansion rate is faster
 \be
 \dot{T} \propto - \Hbl \propto - (1+z)
 \ee
 and thus the temperature and therefore $x_e$ drop faster, which
makes $g(z)$ narrower, see \FIG{fig:visibility_function}. This
leads to a smaller damping scale, \CF
\rrp{eq:recombination_damping}, hence the small-scale power of the
CMB spectrum increases for $\alpha_\dec/\al_0 > 1$.
 \end{itemize}

\begin{figure}[tb]
\centering
\includegraphics[width=\twofigswidth,angle=0]{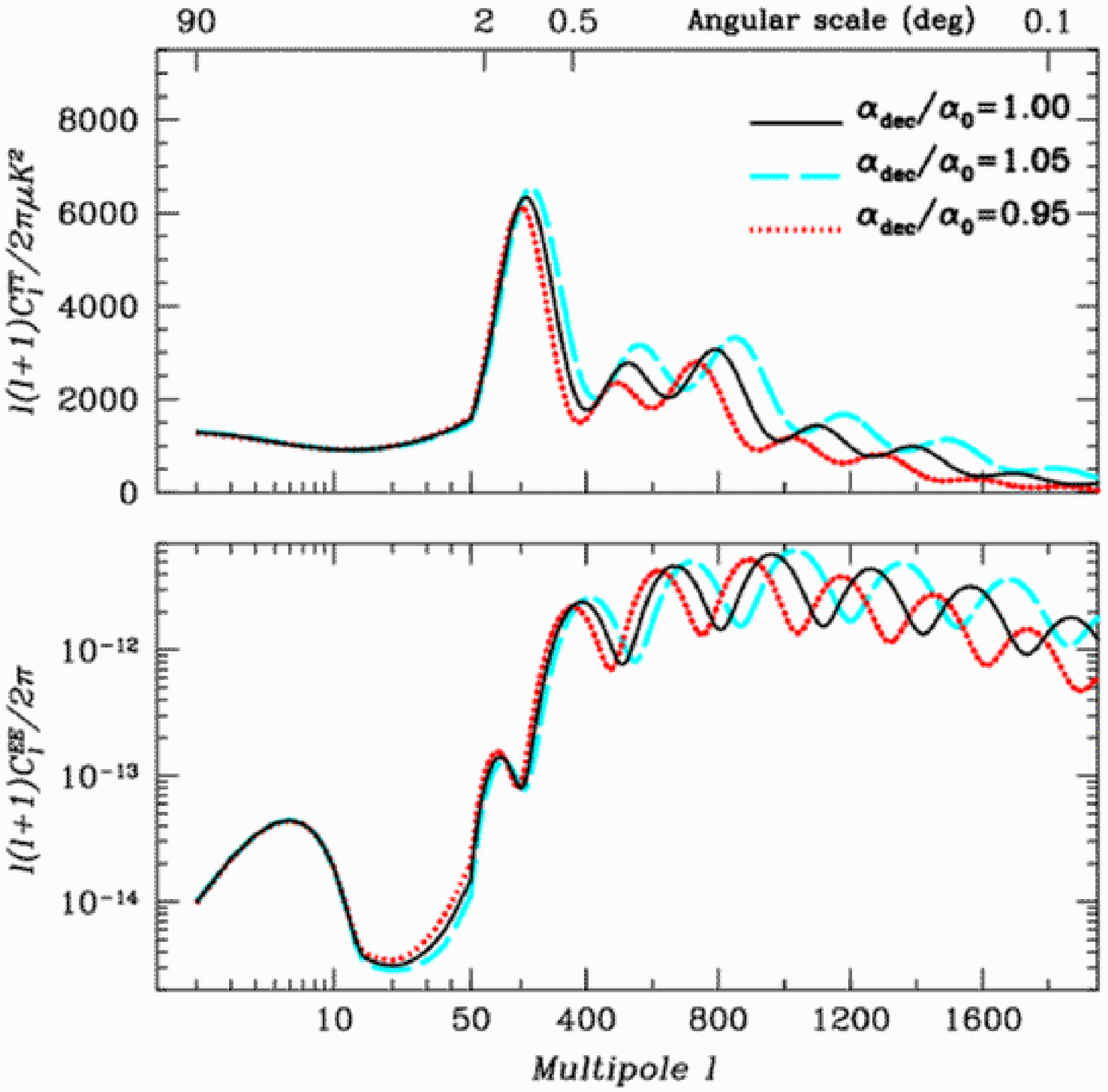}\hfill%
\includegraphics[width=\twofigswidth,angle=0]{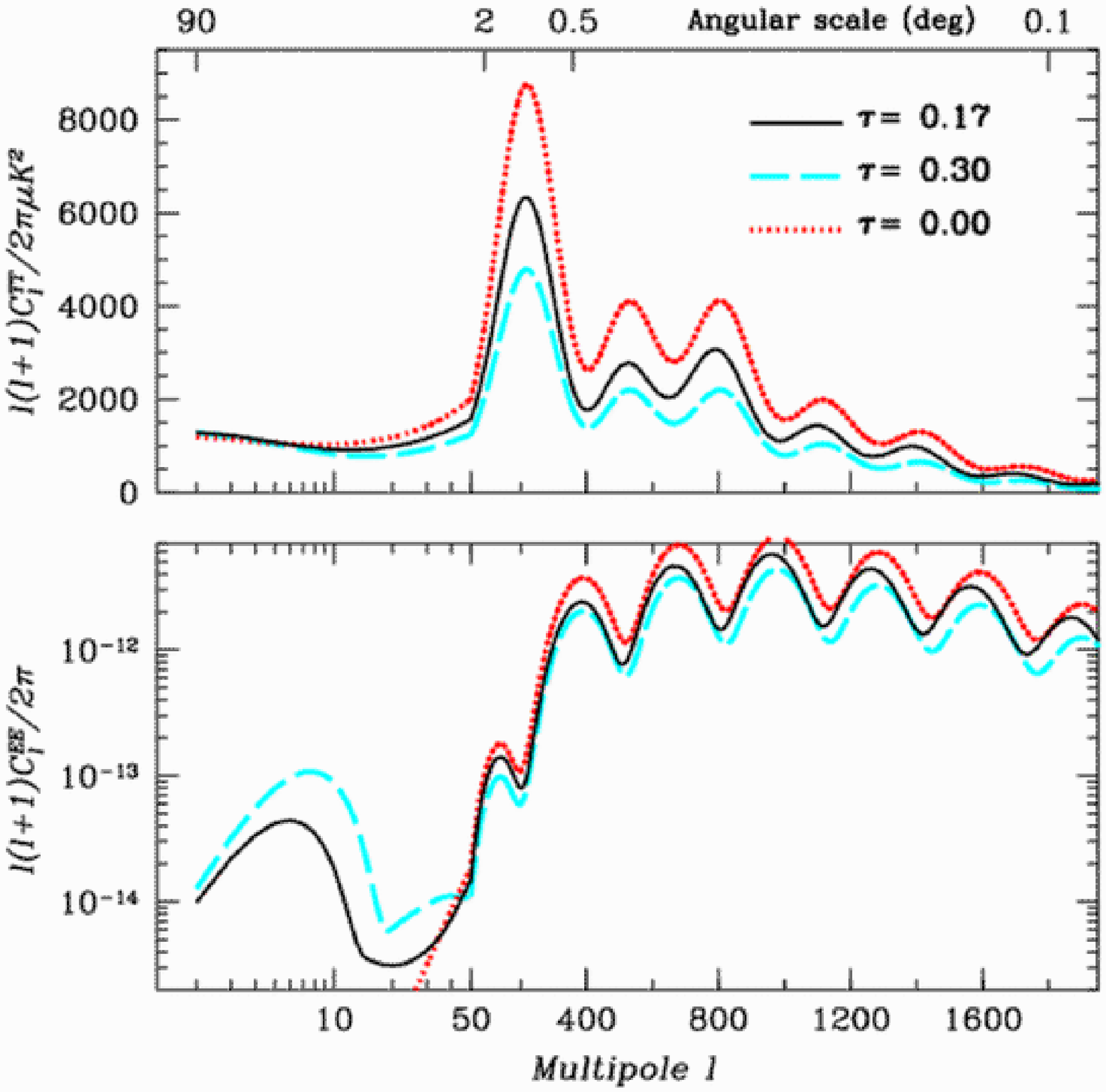}
\caption[Impact of variations of the fine-structure constant and
of the reionization optical depth on the CMB
spectra.]{\label{fig:alpha_peaks} Contrasting the effects of
varying $\al_\dec$ (left) and reionization optical depth
$\tau_\reion$ (right) on the CMB temperature (top) and
polarization (bottom). The reionization bump is not changed by
variations of $\al_\dec/\al_0$. The black lines are for the WMAP
best fit model, with $\al_\dec/\al_0 = 1$ and $\tau_\reion =
0.17$.}
\end{figure}

In \FIG{fig:alpha_peaks} we plot the resulting CMB temperature
spectrum, where the above mentioned changes are readily
distinguishable.

 \subsection{The role of reionization}
 \label{chap:bspIII;sec:alphareion}

After decoupling, the CMB is essentially insensitive to how
$\alpha$ varies, until the reionization epoch is reached, at which
point Thomson scattering becomes effective again. If the value of
$\alpha$ at reionization, $\alpha_\reion \equiv \alpha(z_\reion)$,
is different from its value today, it will affect the CMB spectrum
through a change in the reionization optical depth $\tau_\reion$.
However, $\tau_\reion$ is itself dependent on the cosmological
model and possibly on a number of relevant non-linear physical
processes related to the astrophysical mechanisms responsible for
the reionization. In general, this problem is solved by treating
$\tau_\reion$ as a free parameter, which accounts for the
relatively poor knowledge of the details of the reionization
history and in our case for the uncertainty about the exact value
of $\alpha$ during the reionization epoch. We conclude that
provided we treat $\tau_\reion$ as a free parameter the lack of a
precise knowledge of the value of $\alpha$ during the epoch of
reionization is unimportant, and we can take $\alpha_\reion =
\alpha_0$. On the more phenomenological side, the results of Webb
and collaborators for the value of $\alpha$ at a redshift of $2-3$
would suggest that at the epoch of reionization the possible
changes in $\alpha$ relative to the present day are already very
small. Therefore one can calculate the effect of a varying
$\alpha$ by simply assuming two values for the fine-structure
constant, one at low redshift, $z \lsim 20$, for which we take
today's value by the above argument, and one around the epoch of
decoupling, $\alpha_\dec$, which we want to determine.

\begin{figure}[tb]
\centering
\includegraphics[width=\onefigwidth]{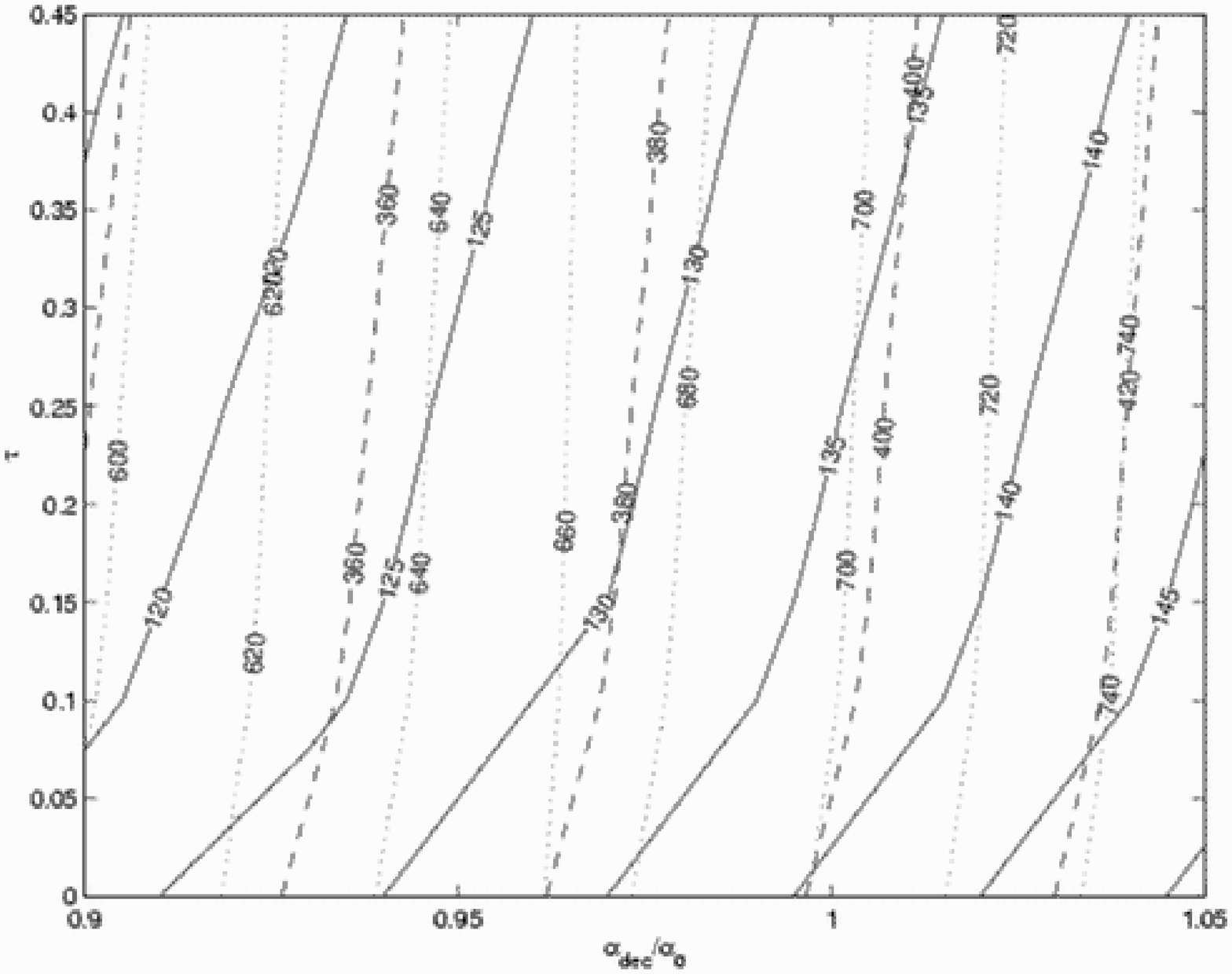}
\caption[Effect of variations of the fine-structure constant on
the CMB power spectra.]{\label{fig:combine_peaks} The separation
in $\ell$ between the reionization bump and the first (solid
lines), second (dashed) and third (dotted) peaks in the
polarization spectrum, as a function of $\alpha$ at decoupling and
$\tau$. A (somewhat idealized) description of how $\alpha$ and
$\tau_\reion$ can be measured using CMB polarization. }
\end{figure}

As shown in \SEC{chap:params:sec:reion}, reionization changes the
amplitude of the acoustic peaks in the temperature spectrum,
without affecting their position and spacing, while introducing
the reionization bump at low $\ell$ in the polarization spectrum.
If the value of $\alpha_\dec$ is different from the value today
(which corresponds to $\al_\reion$), then the peaks in the
polarization power spectrum at small angular scales will be
shifted sideways, while the reionization bump on large angular
scales will remain fixed. This is illustrated in
\FIG{fig:alpha_peaks} (lower right panel). It follows that by
measuring the separation between the acoustic peaks and the bump,
one could in principle measure both $\alpha$ and the reionization
optical depth $\tau_\reion$, as shown in \FIG{fig:combine_peaks}.
This holds true as long as one assumes a specific reionization
history, such as the sudden reionization scenario used here.
However, if we would allow for a more realistic reionization
modelling, the detailed dependence of the reionization bump on the
new reionization parameters is likely to wash out this effect.
Nevertheless, with present-day accuracy the CMB data are sensitive
only to the optical depth of reionization, as pointed out in
\SEC{chap:bspII;sec:cmb}, which justify the use of the simplest
reionization modelling. Within this framework, the fact that
$\tau_\reion$ unexpectedly turned out to be as large as $0.16$ as
derived from the WMAP data \citep{Spergel:2003cb} makes the
prospects of constraining $\alpha$ with the CMB much better
because of the above effect.

Finally, we point out that the modifications discussed above are
direct consequences of an $\alpha$ variation, and that indirect
effects are usually present as well since any variation of
$\alpha$ is necessarily coupled with the dynamics of the Universe
\citep{Mota:2003tc}. Here we take a pragmatic approach and say
that, since the CMB is insensitive to the details of $\alpha$
variations from decoupling to the present day, we do not in fact
need to specify a redshift dependence for this variation --
although we could have specified one if we so chose. At this
stage, we prefer to focus on model-independent constraints, and
hence do not attempt to include an explicit modelling for the
redshift dependence $\al(z)$. Nevertheless, given some
model-independent constraints one can always translate them into
constraints on the parameters of one's favorite model. Beside
possible time variations of $\alpha$, investigated here, one could
also envisage searching for spatial variations on the last
scattering surface \citep{Sigurdson:2003pd}.

\subsection{CMB constraints on $\alpha$ from WMAP alone}
\label{chap:bspIII;sec:cmbres}

We use a modified version of \textsc{cmbfast} which includes the
effects of varying $\alpha$ described above, to analyse the recent
WMAP temperature and cross-polarization data adopting the
likelihood estimator method described in \cite{Verde:2003ey}. The
models are sampled on an uniform grid in a 7 dimensional parameter
space as follows:
 \begin{alignat}{3}
  0.05  & < \Omega_c h^2              &< \quad & 0.20 \step{0.01} \notag\eqcomma \\
  0.010 &< \Omega_bh^2                &< \quad& 0.028 \step{0.001} \notag \eqcomma \\
  0.500 &< \Omega_{\Lambda}           &< \quad& 0.950 \step{0.025} \notag\eqcomma \\
  0.900 &< \alpha_\dec / \alpha_0     &< \quad& 1.050 \step{0.005} \eqcomma \\
  0.06  &< \tau_\reion                &< \quad& 0.30 \step{0.02} \notag\eqcomma \\
  0.880 &< n_s                        &< \quad& 1.08 \step{0.005} \notag\eqcomma \\
  -0.15 &< \dfrac{\dr n_s}{\dr \ln k} &< \quad & 0.05 \step{0.01} \notag \eqdot
 \end{alignat}
The numbers between parentheses give the step size along each
direction; $n_s$ is the scalar spectral index of the primordial
power spectrum, and $\dr n_s / \dr \ln k$ is the spectral index
running, \ie we introduce a scale dependence of the spectral index
of the form
 \be
 n_s(k) = n_s(k_\text{P}) + \dfrac{\dr n_s}{\dr \ln k}
 \ln\left(\dfrac{k}{k_\text{P}} \right) \eqcomma
 \ee
where $n_s \equiv  n_s(k_\text{P})$ is a constant and the pivot
scale $k_\text{P}$ is chosen to be $k_\text{P} = 0.002 \mpc^{-1}$.
We only include flat models, so that the Hubble parameter $H_0
\equiv 100h$ km s$^{-1}$ Mpc$^{-1}$ is a derived quantity. We
don't consider gravity waves or isocurvature modes since these
further modifications are not required by the WMAP data.

\begin{figure}[tb]
\centering
\includegraphics[width=\onefigwidth]{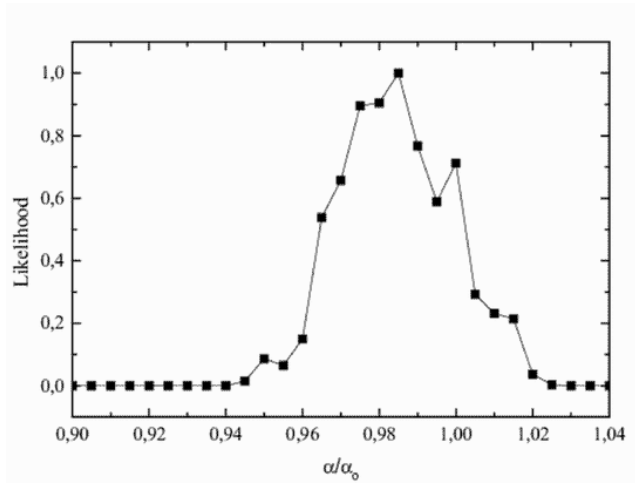}
\caption[Likelihood distribution function for variations in the
fine-structure constant from CMB alone.]{\label{figalpha}
Marginalized likelihood distribution function for variations in
the fine-structure constant at the time of decoupling obtained by
an analysis of the WMAP data (TT+TE, one-year).}
\end{figure}

The likelihood distribution function for $\alpha_{\text{dec}} /
\alpha_0$, obtained after marginalization over the remaining
parameters, see \SEC{chap:data;sec:Bayesian}, is plotted in
\FIG{figalpha}, and gives the marginalized confidence interval
 \be
 0.95 < \alpha_{\text{dec}} / \alpha_0 < 1.02 \quad \text{(at 95\% l.c.).}
 \ee
If we impose $\dr n_s / \dr \ln k = 0$ we obtain instead
 \be
 0.94 < \alpha_{\text{dec}} / \alpha_0 < 1.01 \quad \text{(at 95\% l.c.).}
 \ee

\begin{figure}[tb]
\centering
\includegraphics[width=\onefigwidth]{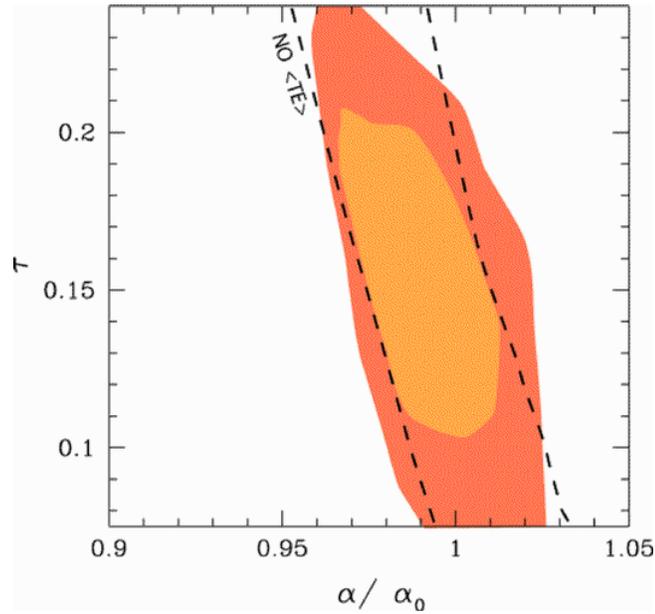}
\caption[Likelihood contour plot in the $\al_\dec / \alpha_0 -
\tau_\reion$ plane.]{\label{figalphavstau} Likelihood contour plot
in the $\alpha_\dec / \alpha_0 - \tau_\reion$ plane including
temperature information only (TT) and TT+TE together from WMAP
(68\% and 95\% l.c. from the inside out). The inclusion of
polarization data partially breaks the degeneracy between these
two parameters.}
\end{figure}

It is interesting to consider the correlations between a $\alpha /
\alpha_0$ and the other parameters in order to see how this
modification to the standard model can change our conclusions
about cosmology. In \FIG{figalphavstau} we plot the likelihood
contours in the $\alpha / \alpha_0 - \tau_\reion$ plane for two
cases: using the temperature only WMAP data and including the $TE$
cross correlation data. There is a clear degeneracy between these
two parameters if one uses only temperature information:
increasing the optical depth allows for an higher value of the
spectral index $n_S$ and a lower value of $\alpha / \alpha_0$.
Inclusion of the $TE$ data is already able to partially break this
degeneracy, but, as we explain below, more detailed measurements
of the polarization spectra are needed to constraint separately
the two parameters,

\begin{figure}\centering
\includegraphics[width=\onefigwidth]{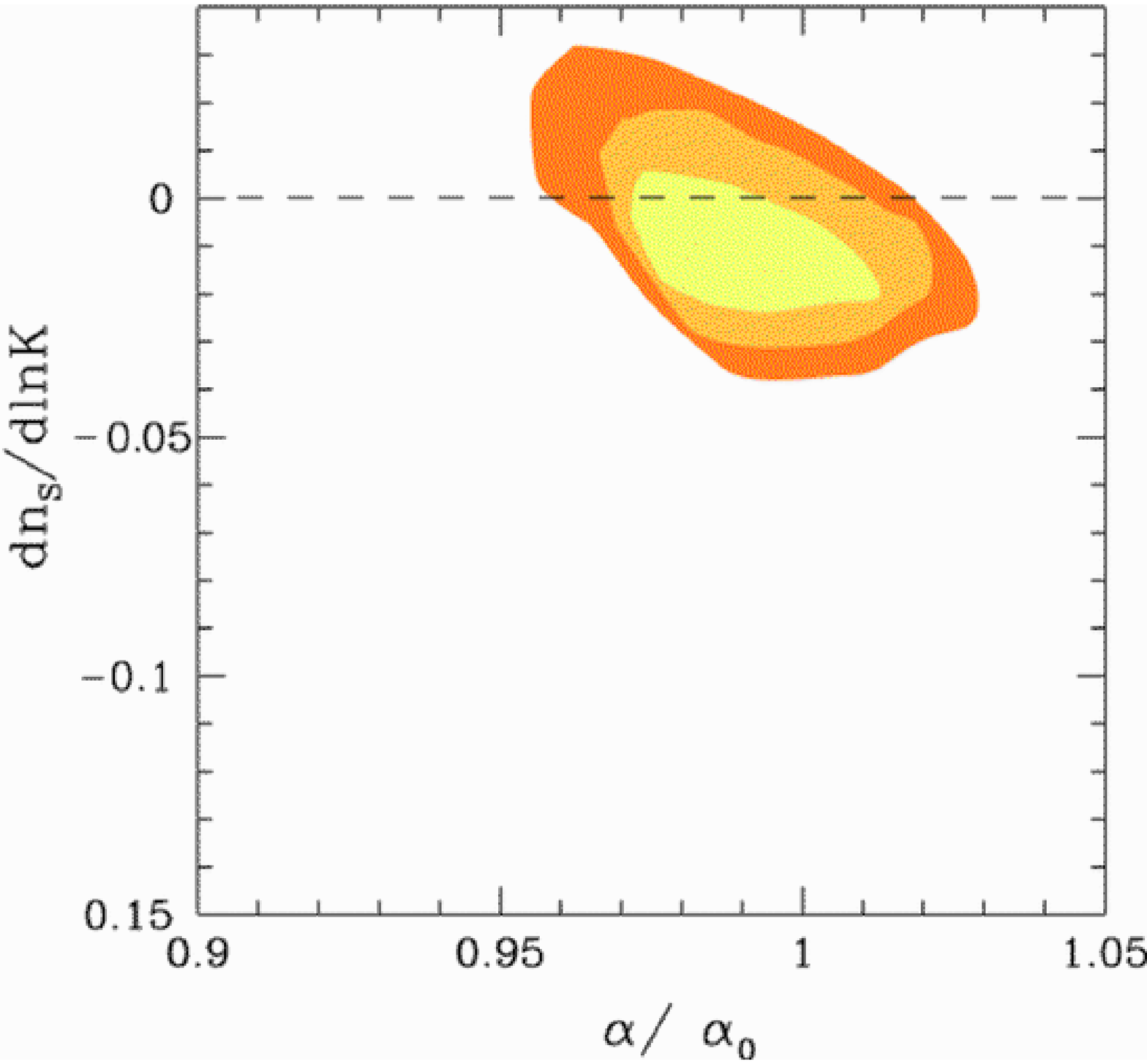}
\caption[Likelihood contour plot in the $\al_\dec / \alpha_0 - \dr
n_s / \dr \ln k$ plane.]{\label{figalphavsdn} Likelihood contour
plot in the $\al_\dec / \alpha_0-\dr n_s / \dr \ln k$ plane, from
WMAP temperature and ET correlation data (68\%, 95\% and 99\% l.c.
from the inside out). A zero scale dependence, as expected in most
of the inflationary models, seems to be more consistent with a
value of $\al_\dec / \alpha_0 < 1$.}
\end{figure}

One of the most unexpected results from the WMAP data is the hint
for a scale-dependence of the spectral index $n_s$ \citep[see
\eg][]{Peiris:2003ff,Kinney:2003uw}. Such a dependence should not
be detectable in most of the viable single field inflationary
models and, if confirmed, would have strong
 consequences on the possibilities of reconstructing the inflationary
potential. For this reason we included the running of the spectral
index in our parameter set. In \FIG{figalphavsdn} we plot
likelihood contours in the $\alpha / \alpha_0-\dr n_s / \dr \ln k$
plane, showing that a lower value of $\alpha / \alpha_0$ would
prefer the absence of running. As already pointed out in
\cite{Bean:2003kd}, a modification of the recombination scheme can
therefore provide a possible explanation for the large value of
$\dr n_s / \dr \ln k $ found from WMAP data.

In previous (pre-WMAP) work, CMB-based constraints on $\alpha$
were obtained with the help of additional cosmological data-sets
and priors, as in \cite{Martins:2002iv}. This procedure was
exposed to the criticism that different data-sets could possibly
have different systematic errors that are impossible to control
and could conceivably conspire to produce the results quoted. The
above results are obtain from WMAP only, and therefore eliminate
this possible uncertainty. For earlier works and pre-WMAP
constraints, see also
\cite{Avelino:2000ea,Avelino:2001nr,Battye:2000ds,Hannestad:1998xp}.

\subsection{Fisher matrix forecasts and degeneracies}
\label{chap:bspIII;sec:fma}

We apply the Fisher matrix analysis (FMA) technique explained in
\SEC{chap:data;sec:fma} to the problem of forecasting the expected
precision in the determination of $\alpha_\dec$ with CMB
anisotropy. For the accuracy reasons presented at length in
\SEC{chap:data;sec:fma}, \SEC{chap:beyondsp;sec:fma} and
\SEC{chap:bspII;sec:fma}, we choose to employ the following 8
dimensional base parameter set
 \be \label{eq:fma_params_alpha}
 \params = \left\{ \Om_b h^2, \Om_m h^2, \Om_\La h^2, \Rshift, n_s, Q, \tau_\reion, \alpha_\dec/\al_0
 \right\}
 \ee
which takes into account the severe geometrical degeneracy via the
shift parameter $\Rshift$, defined in \rr{eq:def_shift_parameter}.
The quantity $n_s$ is the scalar spectral index (without running)
and $Q$ a phenomenological normalization parameter as in
(\refp{eq:Q_normalization}). We restrict ourselves to scalar modes
and adiabatic initial conditions.
\begin{table}[tb]
\centering
\begin{tabular}{|l|ccc|ccc|}
\hline & \multicolumn{3}{c|}{WMAP}& \multicolumn{3}{c|}{Planck}
\\\hline $\nu$ (GHz) &  $40$  &  $60$  & $90$ &
               $100$ &  $143$ & $217$  \\
$\theta_c$ (arcmin)&
    $31.8$ & $21.0$  & $13.8$ &
    $10.7$ & $8.0$ & $5.5$ \\
$\sigma_cT$ ($\mu$K)  &
    $19.8$  & $30.0$ & $45.6$ &
    $5.4$  & $6.0$  & $13.1$  \\
$\sigma_{cE}$ ($\mu$K)  &
    $28.02$ & $42.43$ & $64.56$ &
    $n/a$  & $11.4$  & $26.7$  \\
$w^{-1}_c \cdot 10^{15}$ (K$^2$ ster) &
    $33.6$  & $33.6$  & $33.6$  &
    $0.215$ & $0.158$ & $0.350$  \\
$\ell_c$            & $254$ & $385$  & $586$  &
    $757$ & $1012$ & $1472$ \\\hline
$\ell_{\rm max}$ & \multicolumn{3}{c|}{$1000$} & \multicolumn{3}{c|}{$2000$} \\
$f_{\rm sky}$    & \multicolumn{3}{c|}{$0.80$} & \multicolumn{3}{c|}{$0.80$} \\
\hline \end{tabular}
 \caption[Experimental parameters for the Fisher matrix analysis.]{\label{exppar} Experimental
parameters for WMAP and Planck (nominal mission). Note that we
express the sensitivities in $\mu$K. See
\SEC{chap:data;sec:fma_expepars} for definitions.}
\end{table}

The maximum likelihood model around which the FMA for Planck and
the CVL is performed has parameters $\omega_b = 0.0200$, $\omega_m
= 0.1310$, $\omega_\Lambda = 0.2957$ (and $h = 0.65$), $\R =
0.9815$, $n_s = 1.00$, $Q = 1.00$, $\tau=0.20$ and
$\al/\al_0=1.00$. We differentiate around a slightly closed model
(as preferred by WMAP) with $\Omega_{\rm{tot}} = 1.01$ to avoid
extra sources of numerical inaccuracies, since open and closed
models are computed by \textsc{cmbfast} using different numerical
techniques which would introduce unwanted inaccuracies.

Regarding numerical accuracy issues in the computation of the
Fisher matrix, we implement in the present work double--sided
derivatives, which reduce the truncation error from second order
to third order terms. The choice of the step size is a trade-off
between truncation error and numerical inaccuracy dominated cases.
For an estimated numerical precision of the computed models of
order $10^{-4}$, the step size should be approximately 5\% of the
parameter value \citep{Book:Press92}, though it turns out that for
derivatives in direction of $\alpha$ and $n_s$ the step size can
be chosen to be as small as 0.1\%. After several tests, we have
chosen step sizes varying from 1\% to 5\% for $\omega_b, \omega_m,
\omega_\Lambda$ and $\R$. This choice gives derivatives with an
accuracy of about 0.5\%. The derivatives with respect to $Q$ are
exact, being the power spectrum itself.

\subsubsection*{Predictions for WMAP's four year data}

We present here the main results of the Fisher matrix forecasts;
the full tables and more detailed comments can be found in
\cite{Rocha:2004}. We first concentrate on the potential of the
WMAP four year data, and we compare in Tables
\ref{table:fma_tau_wmap4} and \ref{table:fma_al_wmap4} the
expected errors for two cases, for the base set of parameters
(\ref{eq:fma_params_alpha}) with and without inclusion of
$\al_\dec/\al_0$. In both cases, we take as reference model for
the Fisher matrix the WMAP best fit model of Table 1, in
\cite{Spergel:2003cb}, but with a slightly larger cosmological
constant which gives $\Om_\TOT = 1.01$, for the accuracy reasons
explained above.
\begin{table}[!tb]
\centering
\begin{tabularx}{\linewidth}{X l c c c | c c c}
Quantity &  & \multicolumn{6}{c}{$1\sigma$ errors (\%)} \\\dline
         &        & \multicolumn{6}{c}{WMAP four year}  \\
         &        & marg.  & fixed & joint  & marg.  & fixed & joint           \\
         \dline
         &  & \multicolumn{3}{c}{Polarization (EE)} &  \multicolumn{3}{c}{Temperature (TT)} \\
         \dline
\tabomegab   &    110.64   &    16.58  &    316.44   &        7.33 &    0.81  &     20.96  \\
\tabomegam   &    49.48   &    17.16  &    141.52   &        8.91 &    0.77  &     25.49  \\
\tabomegals  &    622.34   &    97.58  &   1779.93   &      113.30 &   83.39  &    324.06  \\
\tabns       &     69.43   &     4.89  &    198.58   &        6.68 &    0.53  &     19.11  \\
\tabnorm     &     79.22   &    13.51  &    226.58   &        0.90 &    0.32  &      2.58  \\
\tabshift    &     46.52   &    13.04  &    133.06   &        9.25 &    0.59  &     26.47  \\
\tabreion    &    100.84   &     8.21  &    288.40   & 102.72 &
16.70  &    293.79 \\
       \dline
        &   &\multicolumn{3}{c}{Temp+Pol (TT+EE)}               & \multicolumn{3}{c}{All (TT+EE+TE)} \\
         \dline
\tabomegab   &    2.14 &    0.80  &      6.11    &       2.13  &   0.80   &     6.08   \\
\tabomegam   &    3.09 &    0.77  &      8.85    &       3.08  &   0.77   &     8.81   \\
\tabomegals  &   90.70 &   63.84  &    259.41 &      86.97  &  62.69   &   248.75   \\
\tabns       &        1.46 &    0.52  &      4.18    &       1.45  &   0.52   &     4.15   \\
\tabnorm     &        0.52 &    0.32  &      1.48    &       0.52  &   0.32   &     1.48   \\
\tabshift    &        2.86 &    0.59  &      8.17    &       2.84  &   0.59   &     8.12   \\
\tabreion    &       10.52 &    7.45  &     30.08    & 10.41 &
7.44   &    29.78   \\ \hline
\end{tabularx}
\caption[Forecasts for the WMAP four year mission including
reionization.]{\label{table:fma_tau_wmap4}Fisher matrix analysis
results for a standard model with inclusion of reionization (for
the WMAP best fit model as the fisher analysis fiducial model,
with $\tau_\reion=0.17$): expected $1\sigma$ errors from the
WMAP-four year data. The column {\it marg.} gives the error with
all other parameters being marginalized over; in the column {\it
fixed} the other parameters are held fixed at their ML value; in
the column {\it joint} all parameters are being estimated
jointly.}
\end{table}
%
%
%
\begin{table}[!tb]
\centering
\begin{tabularx}{\linewidth}{X l c c c | c c c}
Quantity &  & \multicolumn{6}{c}{$1\sigma$ errors (\%)} \\\dline
         &        & \multicolumn{6}{c}{WMAP four year}  \\
         &        & marg.  & fixed & joint  & marg.  & fixed & joint           \\
         \dline
         &  & \multicolumn{3}{c}{Polarization (EE)} &  \multicolumn{3}{c}{Temperature (TT)} \\
         \dline
\tabomegab &   173.74   &    16.58  &    496.91    &      14.09   &     0.81   &    40.30 \\
\tabomegam &   260.62   &    17.16  &    745.40    &      13.76   &     0.77   &    39.36  \\
\tabomegals&   637.28   &    97.58  &   1822.66    &     133.73   &    83.39   &   382.47 \\
\tabns     &   108.18   &     4.89  &    309.41    &       7.86   &     0.53   &    22.47  \\
\tabnorm   &    96.60   &    13.51  &    276.30    &       2.33   &     0.32   &     6.67  \\
\tabshift  &   133.23   &    13.04  &    381.04    &      26.29   &     0.59   &    75.19  \\
\tabal     &    69.10   &     2.48  &    197.62    &       5.83   &     0.12   &    16.66  \\
\tabreion  &   228.69   &     8.21  &    654.07    &     103.86 &
16.70   &   297.05  \\
         \dline
        &   &\multicolumn{3}{c}{Temp+Pol (TT+EE)}               & \multicolumn{3}{c}{All (TT+EE+TE)} \\
         \dline
\tabomegab &   7.50     &      0.80  &     21.44   &        7.41 &       0.80 &      21.18 \\
\tabomegam &   5.48     &      0.77  &     15.66   &        5.46 &       0.77 &      15.62  \\
\tabomegals&  91.57     &     63.84  &    261.91   &       87.48 &      62.69 &     250.20 \\
\tabns     &      2.03  &      0.52  &      5.82   &        2.03 &       0.52 &       5.81   \\
\tabnorm   &      1.31  &      0.32  &      3.73   &        1.30 &       0.32 &       3.71   \\
\tabshift  &     14.34  &      0.59  &     41.01   &       14.17 &       0.59 &      40.53   \\
\tabal     &      3.08  &      0.11  &      8.80   &        3.05 &       0.11 &       8.71   \\
\tabreion  &     10.65  &      7.45  &     30.46   &       10.52 &
7.44 &      30.08   \\ \hline
\end{tabularx}
\caption[Forecasts for the WMAP four year mission including
fine-structure constant variations and
reionization.]{\label{table:fma_al_wmap4}Fisher matrix analysis
results for the model of \TAB{table:fma_tau_wmap4} with inclusion
of $\alpha_\dec$.}
\end{table}

Table \ref{table:fma_tau_wmap4} gives accurate predictions for the
errors on standard cosmological parameters, for models including
non-flat cosmologies. Clearly, with the WMAP sensitivity,
E-polarization alone will not constrain much the parameters, but
combining temperature information with the polarization channels
will reduce the errors on the baryon and matter density and on the
shift parameter by about a factor of three, with all other
parameters marginalized over. The error on the cosmological
constant will remain of order unity, since this is an expression
of the geometrical degeneracy which is fundamentally unbreakable
without external priors. The spectacular improvement of about a
factor 10 in determining $\tau_\reion$ with polarization
information is a consequence of the expected measurement of the
reionization induced polarization bump, which breaks the
degeneracy with normalization present with temperature alone. The
spectral index accuracy thus increases by a factor 4, because the
better determination of the reionization optical depth assists
into breaking the small scale degeneracy with $n_\SCAL$. The
column ``fixed'' gives the best case scenario in which all other
parameters are assumed to be known and fixed to their fiducial
model value. In this case, the errors obtained by combining all
channels are below $1\%$ for all parameters but the cosmological
constant.

Let us now compare this forecasts with the corresponding entries
in \TAB{table:fma_al_wmap4}, where the parameter $\al_\dec/\al_0$
has been added. The addition of a varying fine-structure constant
opens up new degeneracy directions, hence the marginalized and
joint error forecasts get worse (but not the errors with all other
parameters fixed, of course). The most degenerate direction is
with the shift parameter (marginalized errors larger by a factor 7
with all channels), as expected from the above considerations. Due
to its effect on the peak heights, the fine-structure constant is
largely degenerate with $\om_b$ up to the second acoustic peak; an
accurate mapping of the large multipole temperature spectrum can
nevertheless lift this degeneracy, also constraining better
$n_\SCAL$, see \cite{Martins:2002iv} for details. This explains
the larger errors on the baryon density and on the spectral index
as we include $\al$ in the parameter set. However, the optical
depth determination remains almost unaffected, as a consequence of
the simultaneous measurement of the reionization bump's position
and of the acoustic peaks angular scale, thereby validating our
method for the restricted class of sudden reionization models
considered here.

\subsubsection*{Predictions for Planck and an ideal experiment}

\begin{figure}[tb]
\centering
\includegraphics[width=\linewidth]{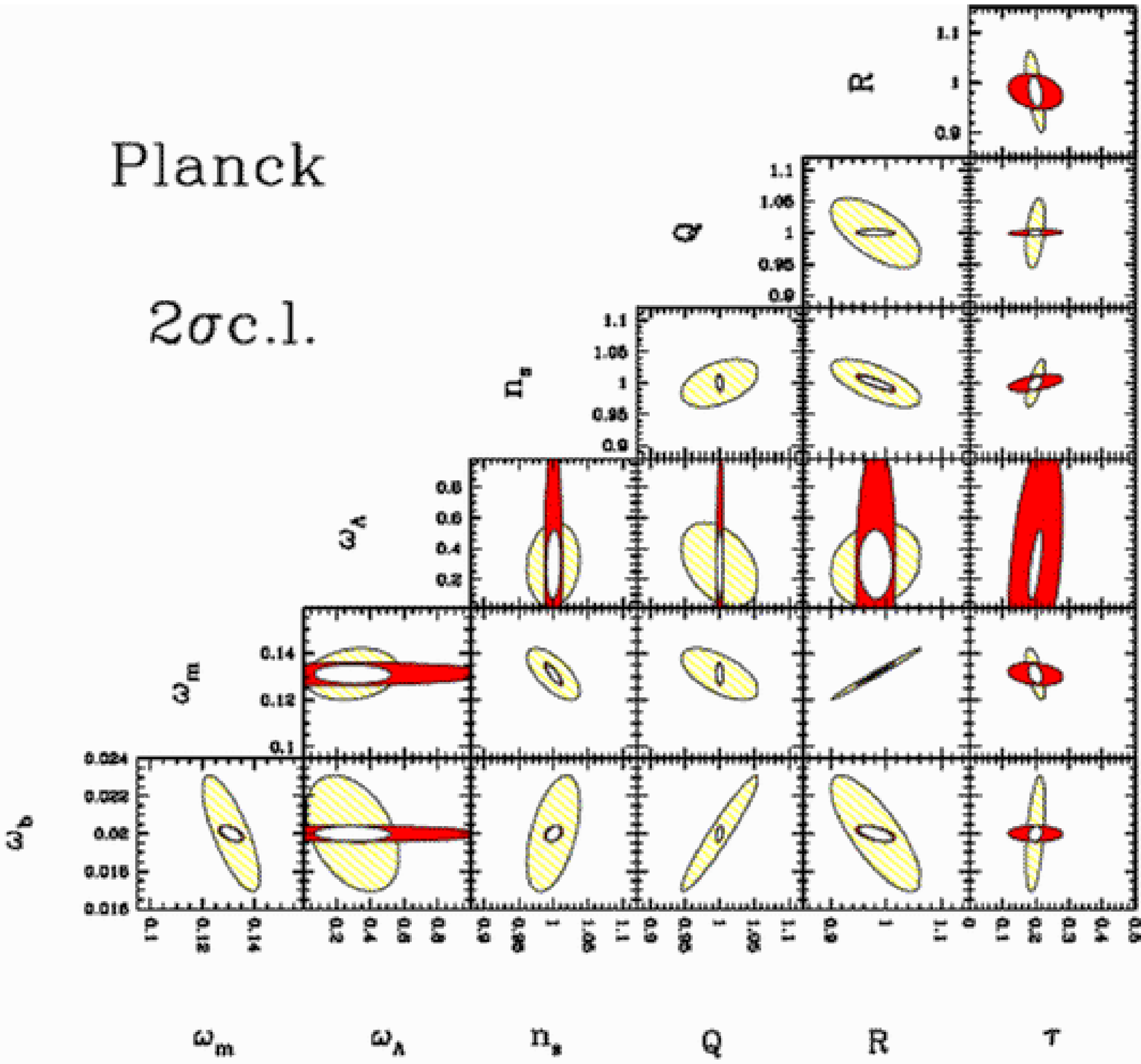}
\caption[Fisher matrix forecasts for Planck for all couples of
standard parameters.]{Ellipses containing $95.4\%$ ($2\sigma$) of
joint confidence (all other parameters marginalized) using
temperature alone (red), E-polarization alone (yellow), and both
jointly (white), for a standard model with inclusion of
reionization ($\tau_\reion=0.20$). Fisher matrix forecast for the
Planck HFI instrument.} \label{fig:fma_streion_Planck}
\end{figure}

\begin{figure}[tb]
\centering
\includegraphics[width=\linewidth]{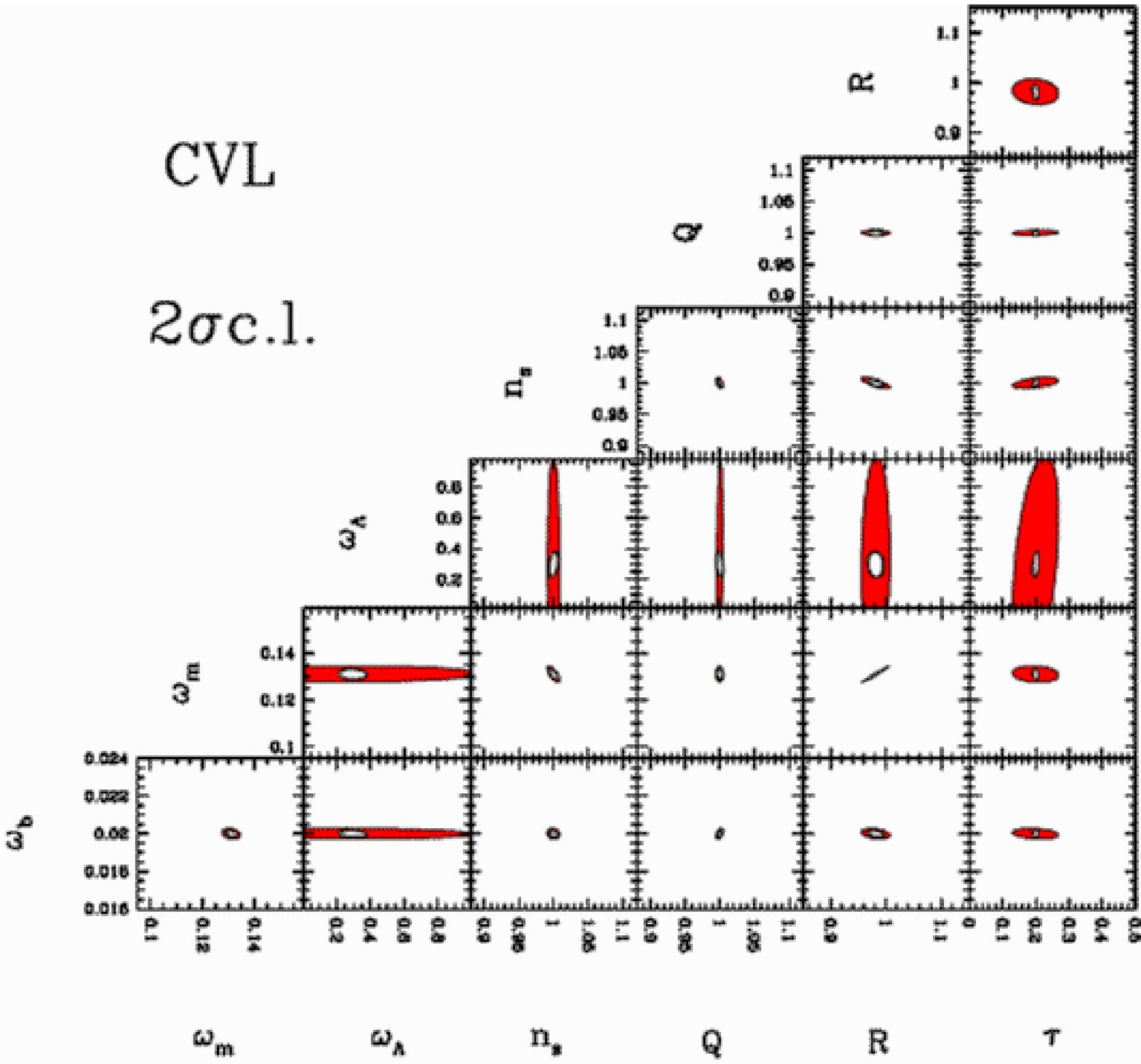}
\caption[Fisher matrix forecasts for an ideal CMB experiment for
all couples of standard parameters.]{Ellipses containing $95.4\%$
($2\sigma$) of joint confidence (all other parameters
marginalized) using temperature alone (red), E-polarization alone
(yellow), and both jointly (white), for a standard model with
inclusion of reionization ($\tau_\reion=0.20$). Fisher matrix
forecast for an ideal cosmic variance limited (CVL) experiment.}
\label{fig:fma_streion_CVL}
\end{figure}

\begin{figure}[tb]
\centering
\includegraphics[width=\linewidth]{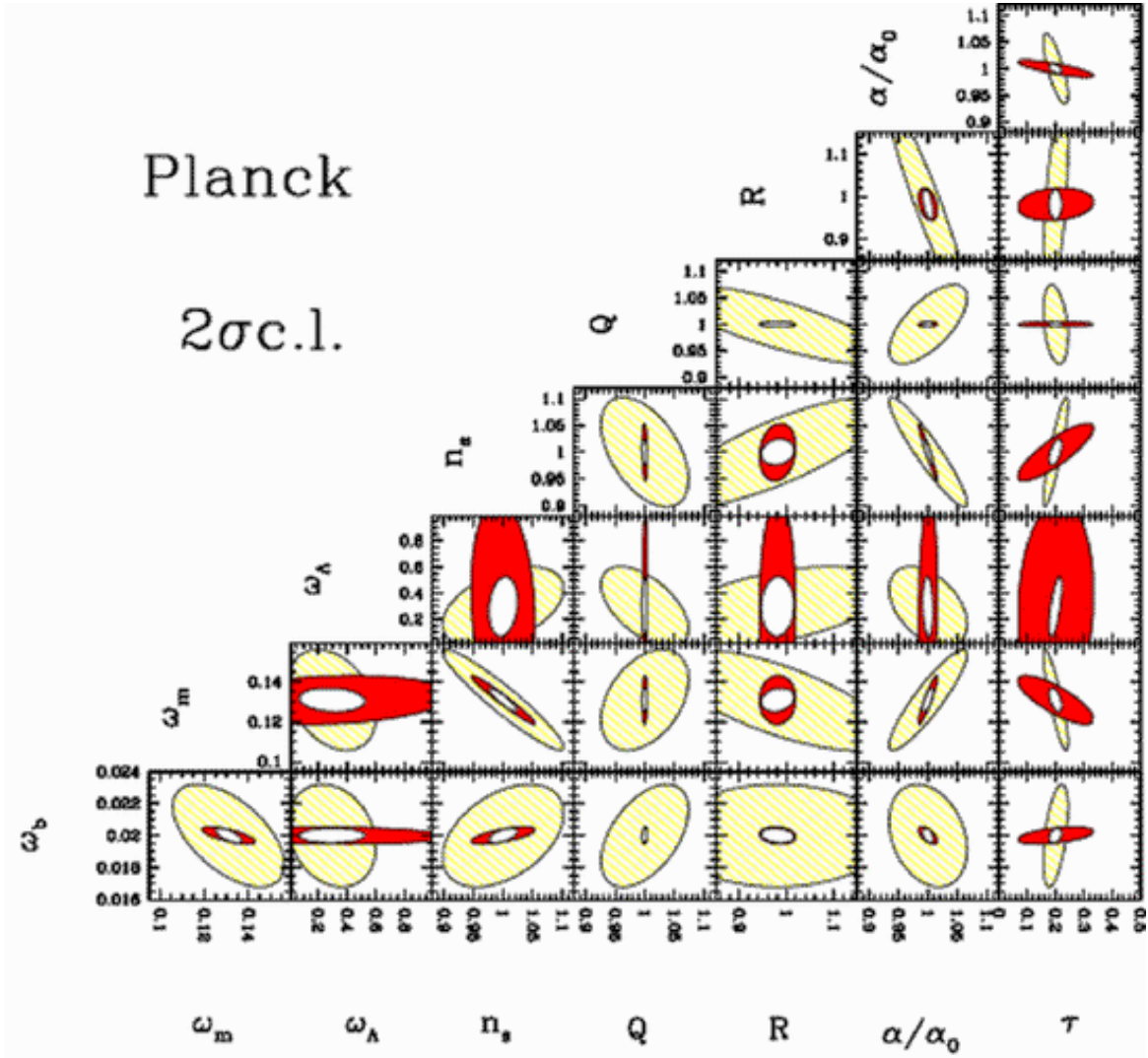}
\caption[Fisher matrix forecasts for Planck including variations
in the fine-structure constant.]{Ellipses containing $95.4\%$
($2\sigma$) of joint confidence (all other parameters
marginalized) using temperature alone (red), E-polarization alone
(yellow), and both jointly (white), for a standard model with
inclusion of reionization ($\tau_\reion=0.20$) and time variations
of the fine-structure constant. Fisher matrix forecast for the
Planck HFI instrument.} \label{fig:fma_alpha_Planck}
\end{figure}

\begin{figure}[tb]
\centering
\includegraphics[width=\linewidth]{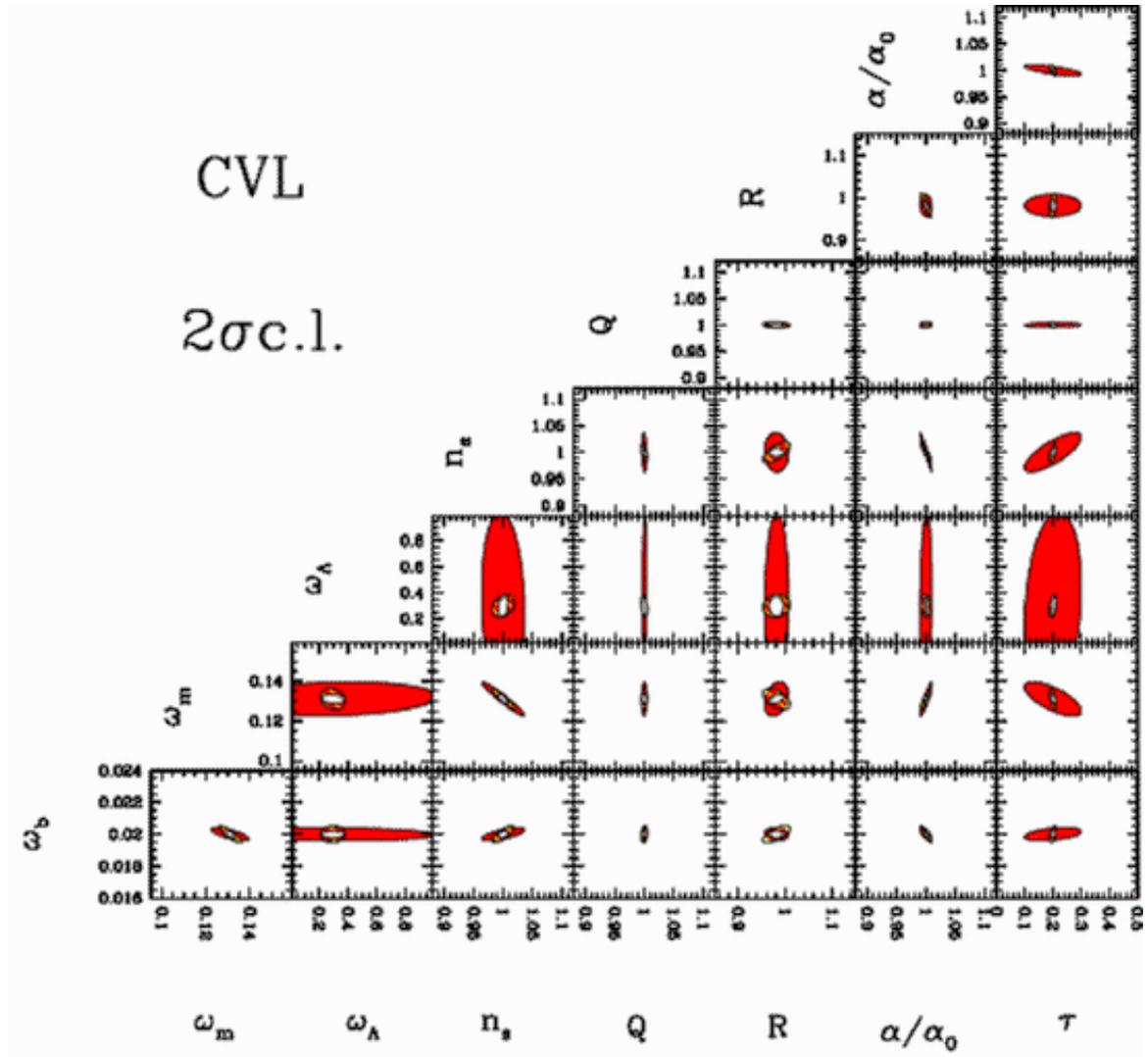}
\caption[Fisher matrix forecasts for an ideal CMB experiment
including variations in the fine-structure constant.]{Ellipses
containing $95.4\%$ ($2\sigma$) of joint confidence (all other
parameters marginalized) using temperature alone (red),
E-polarization alone (yellow), and both jointly (white), for a
standard model with inclusion of reionization ($\tau_\reion=0.20$)
and time variations of the fine-structure constant. Fisher matrix
forecast for an ideal cosmic variance limited (CVL) experiment.}
\label{fig:fma_alpha_CVL}
\end{figure}

We now focus on the Fisher matrix forecasts for the expected
performance of the Planck satellite, and compare them with the
results for an ideal CMB experiment, which would map both
temperature and E-polarization with cosmic variance limited (CVL)
accuracy up to $\ell = 2000$. Clearly, such a measurement is not
feasible in practice, because of foreground removal and limited
instrumental sensitivity, but it represents in principle the best
possible parameters determination using CMB alone. The full
results are tabulated in \TAB{table:fmast_tau} and
\TAB{table:fmaal_tau}. In order to clarify the role of
correlations between parameters, we plot in Figures
\ref{fig:fma_streion_Planck} and \ref{fig:fma_alpha_Planck} the
$2\si$ joint likelihood contours for all couples of parameters for
Planck, and in Figures \ref{fig:fma_streion_CVL} and
\ref{fig:fma_alpha_CVL} for the CVL experiment.

The first important fact is that E-polarization data alone from
Planck will constrain the standard parameters better than the four
year WMAP temperature data alone, compare
\TAB{table:fma_tau_wmap4} with \TAB{table:fmast_tau}. This follows
from the fact that the polarization spectrum is less plagued by
large scale degeneracies than the temperature spectrum.
Furthermore, as apparent from \FIG{fig:fma_streion_Planck},
degeneracy directions for the temperature spectrum are in many
cases almost orthogonal to the directions in the polarization
channel. This is especially the case for $\tau_\reion$, and in
fact combining temperature and polarization information reduces
its marginalized error from $16\%$ ($6\%$) for temperature
(polarization) alone to $4\%$. In general, the WMAP four year
error-bars will be approximately halved for all parameters by
Planck. Another significant aspect is that by comparing the
temperature only column for Planck to the one for the CVL
experiment, we conclude that Planck will be essentially cosmic
variance limited as far as the temperature spectrum is concerned.
This is not the case for the polarization channel, for which there
will still be room for a substantial improvement over Planck's
capabilities: the CVL experiment can do better than Planck by a
factor 5 or more on average. The comparison of Figures
\ref{fig:fma_streion_Planck} and \ref{fig:fma_streion_CVL}
immediately confirms this conclusion, which makes a strong case
for a post-Planck, polarization-dedicated experiment.

\begin{figure}[tb]
\centering
\includegraphics[width=\onefigwidth]{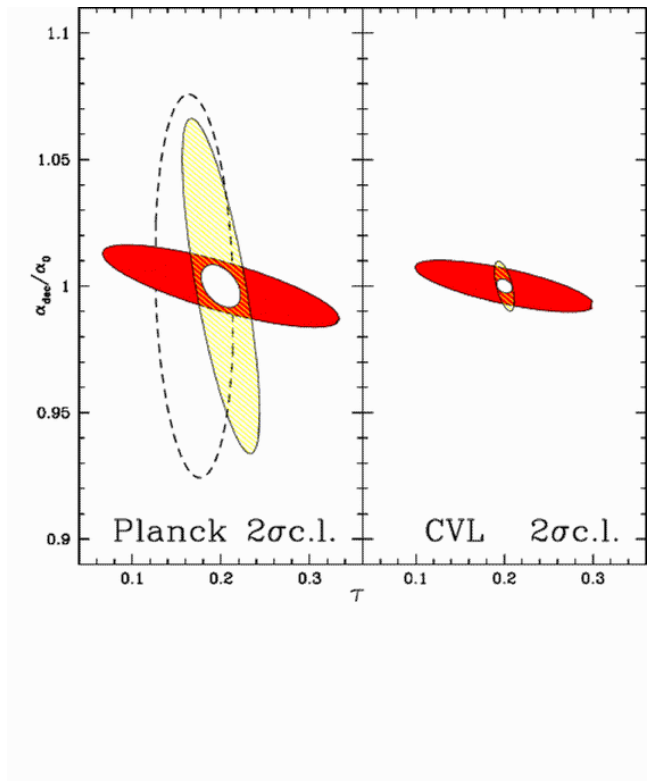}
\caption[Forecasts in the $\alpha_\dec/\al_0 - \tau_\reion$
plane.]{\label{fig:fma_ellipses_al_tau} Ellipses containing
$95.4\%$ ($2\sigma$) of joint likelihood in the $\alpha_\dec/\al_0
- \tau_\reion$ plane (all other parameters marginalized), for the
Planck and cosmic variance limited (CVL) experiments, using
temperature alone (red), E-polarization alone (yellow), and both
jointly (white). The dashed contour represents the WMAP - 4years
forecast using (TT+EE+TE) jointly. }
\end{figure}

When we add the fine-structure constant to the Planck parameter
set, the ellipses for temperature and polarization get larger for
all the couples of parameters involving degenerate directions with
$\al$, compare \FIG{fig:fma_alpha_Planck} with
\FIG{fig:fma_streion_Planck}. As before, this happens mostly for
the $\R$, $n_\SCAL$ and $\tau_\reion$ using temperature
information only. The degradation of the accuracy on those
parameters is less dramatic than for WMAP, because Planck will map
the spectrum to larger multipoles. It is remarkable that the {\it
combined} temperature and polarization error does not grow very
much when we add $\al$, because the degeneracies are in different
directions for the two channels. The fine-structure constant is
the only parameter which Planck will constrain better with
temperature only ($0.7\%$) than with polarization only ($2.7\%$,
all others marginalized), while the situation is opposite for
$\tau_\reion$, $27\%$ for temperature and $9\%$ for polarization.
Combining the two channels again lifts most of the degenerate
directions, and we conclude that Planck will achieve an accuracy
on $\al_\dec$ of order $0.3\%$ ($1\si$, all others marginalized),
thus improving by about a factor of 10 on the expected performance
of the four year WMAP mission and a factor of 5 on the current
upper bound (obtained however under the assumption of flatness).
At the same time, the reionization optical depth will be
constrained to about $4.5\%$. Our findings for $\al_\dec/\al_0$
and $\tau_\reion$ are summarized in \FIG{fig:fma_ellipses_al_tau},
where we compare degeneracy directions in the $\al_\dec/\al_0,
\tau_\reion$ plane for temperature alone, polarization alone and
the combined channels, for Planck and the CVL experiment. We also
superimpose the corresponding forecast for the WMAP four year
mission (all channels) in order to facilitate the comparison.

The columns in \TAB{table:fmaal_tau} regarding the CVL experiment
and the corresponding \FIG{fig:fma_alpha_CVL} give information
about further improvements on Planck's parameter accuracy. As
mentioned, a cosmic variance limited measurement of polarization
could further reduce Planck's error-bars by a factor 2 to 3,
reaching the highest possible accuracy from CMB alone. In
particular, our analysis indicate that CMB alone can constrain
variations of $\al$ up to $\calo(10^{-3})$ at $z \sim 1100$. Going
beyond will require additional priors on the other parameters.
%
%
\begin{table}[tb]
\centering
\begin{tabularx}{\linewidth}{X l c c c | c c c}
Quantity &  & \multicolumn{6}{c}{$1\sigma$ errors (\%)} \\\dline
         &        & \multicolumn{3}{c}{Planck HFI} & \multicolumn{3}{c}{CVL} \\
         &        & marg.  & fixed & joint  & marg.  & fixed & joint           \\\dline
         &  & \multicolumn{6}{c}{Polarization only (EE)} \\
         \dline
\tabomegab       &6.21       &1.11      &17.75    &0.48     &0.25    &1.38 \\
\tabomegam       &3.37       &0.39      &9.64     &0.70     &0.03    &1.99 \\
\tabomegal       &37.37      &22.87     &106.89   &11.40    &9.99    &32.61 \\
\tabns           &1.53       &0.96      &4.38     &0.30     &0.08    &0.86 \\
\tabnorm         &2.23       &0.51      &6.38     &0.24     &0.07    &0.67 \\
\tabshift        &3.33       &0.35      &9.52     &0.65     &0.03    &1.86 \\
\tabreion        &5.74       &2.78      &16.42    &1.81     &1.52
&5.18 \\
  \dline
  &   & \multicolumn{6}{c}{Temperature only (TT)} \\
  \dline
\tabomegab       &0.86       &0.60      &2.46     &0.57     &0.38    &1.64 \\
\tabomegam       &1.51       &0.13      &4.31     &1.10     &0.08    &3.14 \\
\tabomegal       &110.15     &96.15     &315.03   &98.15    &86.00   &280.72 \\
\tabns           &0.54       &0.13      &1.56     &0.36     &0.07    &1.04 \\
\tabnorm         &0.20       &0.11      &0.56     &0.17     &0.07    &0.50 \\
\tabshift        &1.47       &0.12      &4.21     &1.05     &0.07    &3.01 \\
\tabreion        &16.50      &8.28      &47.20    &14.02    &5.89
&40.09 \\
 \dline
  &   & \multicolumn{6}{c}{Temperature and Polarization (TT+EE)} \\
  \dline
\tabomegab      &0.80       &0.53      &2.30    &0.32      &0.21     &0.92 \\
\tabomegam      &1.24       &0.12      &3.55    &0.55      &0.03     &1.58  \\
\tabomegal      &30.58      &22.04     &87.46   &10.72     &9.85     &30.65 \\
\tabns          &0.43       &0.13      &1.23    &0.20      &0.05     &0.58  \\
\tabnorm        &0.19       &0.10      &0.53    &0.14      &0.05     &0.41  \\
\tabshift       &1.22       &0.11      &3.48    &0.52      &0.03     &1.49 \\
\tabreion       &4.04       &2.65      &11.56   &1.73      &1.48     &4.96 \\
\end{tabularx}
\caption[Fisher matrix forecast for the Planck satellite and and
ideal experiment.]{\label{table:fmast_tau}Fisher matrix analysis
results including reionization ($\tau_\reion=0.20$): expected
$1\sigma$ errors for the Planck satellite and for cosmic variance
limited (CVL) experiment.}
\end{table}

\begin{table}[tb]
\centering
\begin{tabularx}{\linewidth}{X l c c c | c c c}
Quantity &  & \multicolumn{6}{c}{$1\sigma$ errors (\%)} \\\dline
         &        & \multicolumn{3}{c}{Planck HFI} & \multicolumn{3}{c}{CVL} \\
         &        & marg.  & fixed & joint  & marg.  & fixed & joint           \\\dline
         &  & \multicolumn{6}{c}{Polarization only (EE)} \\
         \dline
\tabomegab       &6.46       &1.11       &18.47    &1.09        &0.25        &3.12 \\
\tabomegam       &7.75       &0.39       &22.17    &1.61        &0.03        &4.60 \\
\tabomegal       &41.61      &22.87      &119.01   &11.60       &9.99        &33.17 \\
\tabns           &4.14       &0.96       &11.85    &0.77        &0.08        &2.22 \\
\tabnorm         &2.99       &0.51       &8.55     &0.24        &0.07        &0.68 \\
\tabshift        &9.56       &0.35       &27.33    &1.19        &0.03        &3.40 \\
\tabal           &2.66       &0.06       &7.62     &0.40        &$<0.01$     &1.14 \\
\tabreion        &8.81       &2.78       &25.19    &2.26        &1.52        &6.45 \\
  \dline
  &   & \multicolumn{6}{c}{Temperature only (TT)} \\
  \dline
\tabomegab      &1.09       &0.60       &3.12     &0.83        &0.38        &2.37  \\
\tabomegam      &3.76       &0.13       &10.74    &2.64        &0.08        &7.55  \\
\tabomegal      &111.61     &96.15      &319.21   &98.97       &86.00       &283.05 \\
\tabns          &2.18       &0.13       &6.24     &1.49        &0.07        &4.26  \\
\tabnorm        &0.20       &0.11       &0.57     &0.18        &0.07        &0.50  \\
\tabshift       &1.58       &0.12       &4.53     &1.06        &0.07        &3.04  \\
\tabal          &0.66       &0.02       &1.88     &0.41        &0.01        &1.18  \\
\tabreion       &26.93      &8.28       &77.02    &20.32       &5.89        &58.11 \\
 \dline
  &   & \multicolumn{6}{c}{Temperature and Polarization (TT+EE)} \\
  \dline
\tabomegab      &0.91        &0.53      &2.61     &0.38        &0.21        &1.09  \\
\tabomegam      &1.81        &0.12      &5.17     &0.67        &0.03        &1.91 \\
\tabomegal      &30.89       &22.04     &88.36    &10.79       &9.85        &30.85\\
\tabns          &0.97        &0.13      &2.77     &0.33        &0.05        &0.93 \\
\tabnorm        &0.19        &0.10      &0.54     &0.14        &0.05        &0.41\\
\tabshift       &1.43        &0.11      &4.08     &0.60        &0.03        &1.72 \\
\tabal          &0.34        &0.02      &0.97     &0.11        &$<0.01$     &0.32 \\
\tabreion       &4.48        &2.65      &12.80    &1.80        &1.48        &5.15 \\
\end{tabularx}
\caption[Fisher matrix forecast for the Planck satellite and and
ideal experiment including variations of the fine-structure
constant.]{\label{table:fmaal_tau}Fisher matrix analysis results
as in \TAB{table:fmast_tau} but including $\al_\dec$.}
\end{table}


\chapter{Testing the paradigm of adiabaticity}
\label{chap:genic}

Combination of today's high quality CMB data with other
cosmological data sets allows us to constrain the eight parameters
  \be
 \params = \left\{\Om_{\cdm}, \Om_b, \Om_\La, N_\nu, h, \tau_\reion, n_s, A_s  \right\}
 \ee
with an accuracy of a few percent \citep{Tegmark:2003ud}, if we
assume flatness, \ie by imposing $\Om_\curv = 0$. This is a
spectacular achievement, even more so given the fact that many
completely independent measurements seem to be converging towards
the same values. In the previous sections we have discussed the
determination of most of the above parameters; here we highlight
that the accuracy of parameter extraction depends crucially on the
assumption that the initial conditions for the perturbations are
purely adiabatic, and explore the consequences of relaxing this
strong assumption by including the most general type of initial
conditions in the problem.

This chapter is organized as follows: we first present an
introductory survey on recent CMB analysis involving isocurvature
modes, \SEC{chap:genic;sec:intro}; we then investigate in a
specific example how the inclusion of isocurvature modes spoils
the precise determination of the baryon density from pre-WMAP CMB
data in \SEC{chap:genic;sec:precision}; in
\SEC{chap:genic;sec:lambda} we ask whether the presence of
non-adiabatic contribution can reproduce CMB and large scale
structure observations without the need for a cosmological
constant, and we conclude that $\OLa \neq 0$ is robust with
respect to the inclusion of isocurvature modes and to the use of a
frequentist (rather than Bayesian) approach; finally, in
\SEC{chap:genic;sec:future} we give the future prospects for the
determination by WMAP and Planck of cosmological parameters
independent of any assumption about the type of initial
conditions.

\section{Introductory survey} \label{chap:genic;sec:intro}

Until recently, most of the literature has focused on parameter
extraction assuming purely adiabatic initial conditions, because
the evidence for a first acoustic peak around $\ell \approx 220$
very soon ruled out the possibility of the simplest alternative,
namely purely isocurvature CDM initial conditions, see \eg
\cite{Enqvist:2000hp}. Nevertheless, subdominant CDM isocurvature
contributions cannot be excluded, and the constraints are even
less stringent if one allows for a correlated mixture, in which
case the correlator can cancel out most of the isocurvature
contribution on large scale
\citep{Langlois:2000ar,Amendola:2001ni}. This qualitative
conclusion holds even after the more precise measurements of WMAP
\citep{Valiviita:2003ty}.

In the works of \cite{Bucher:2000hy,Bucher:2000kb} the
consequences for parameter extraction are examined when the most
general initial conditions are allowed, with the conclusion that
only a precise measurement of polarization would allow for the
simultaneous reconstruction of cosmological parameters and of the
initial conditions correlation matrix. The first attempt of
including all the modes in a numerical parameter determination
from real data is performed in \cite{Trotta:2001yw}, as
illustrated in \SEC{chap:genic;sec:precision}, with the result
that the pre-WMAP CMB data can not constrain to any extent the
value of the baryon density and the Hubble parameter in the
general initial conditions case. After the release of the WMAP
first-year data, two groups have re-investigated the question of
the most general initial conditions in the wake of the improved
measurements: \cite{Crotty:2003rz} consider a correlated mixture
of the adiabatic mode with each of the isocurvature modes in turn,
finding that the pre-WMAP constraints on the isocurvature
contribution are significantly improved; \cite{Bucher:2004an}
refine the analysis of \cite{Trotta:2001yw} by using Monte Carlo
methods, and simultaneously including all the isocurvature modes
and six cosmological parameters, but the conclusions remained
qualitatively the same. The bottom line is that the relaxing the
assumption of adiabaticity spoils our ability to do precision
cosmology.

The phenomenological approach gives useful hints on the
``stiffness'' of current data, and indeed the possibility of
accommodating isocurvature modes has been considerably reduced by
WMAP. Although independent of any model for the generation of
perturbations, this approach has the disadvantage of introducing
many new free parameters in the description of the power spectrum.
To reduce this number somewhat, all analyses so far have assumed
the same spectral index for all modes, an assumption which is not
really motivated. Since the current CMB data are in excellent
agreement with purely adiabatic initial conditions, it is not
surprising however that there is no statistical evidence that such
extra parameters should be non-zero. Occam's razor would therefore
dictate to stick to the simplest adiabatic description, lacking
any evidence for a more complicated model. However, there is no
compelling reason why the physics of the early universe should
boil down to only one degree of freedom.

A second reason why model-independent constraints should be
regarded with care is that in any specific implementation, some of
the parameters will be correlated. For instance, in the curvaton
scenario
\citep{Moroi:2001ct,Lyth:2001nq,Enqvist:2001zp,Lyth:2002my}, the
adiabatic and residual isocurvature modes are always totally
correlated or anti-correlated. Therefore, not only the number of
extra degrees of freedom is reduced, but possibly the parameter
space of the model is a highly constrained subspace of the
model-independent parameter space. For this reason it is
interesting to derive model-specific constraints, which are more
stringent than those obtained with a general phenomenological
parametrization. For instance, WMAP constraints for the curvaton
model have been derived for the case of CDM and baryons
isocurvature fluctuations \citep{Gordon:2002gv,Lyth:2003ip}.  The
neutrino density mode can be generated from  perturbations of the
neutrino chemical potential \citep{Lyth:2002my}, and bounds have
recently been derived for this case \citep{Gordon:2003hw}. It
seems more difficult to produce a neutrino velocity mode: a
working model is at present still lacking.

Despite these difficulties, the CMB represents the most promising
data set to learn  about the type of initial conditions realized
in the observed Universe: it is our window to the very early
universe.

\section{Precision cosmology and general initial conditions}
\label{chap:genic;sec:precision}

In this section, based on the work published in
\cite{Trotta:2001yw}, we investigate the extent to which the
determination of cosmological parameters depends on the
assumptions about initial conditions. We show in a specific
example how the allowed parameter range is enlarged when the usual
requirement for purely adiabatic initial conditions is relaxed. In
order to limit the computational effort, we have chosen to vary
some cosmological parameters and keep the others fixed.  We
consider flat models only, and we fix the total density parameter,
the total matter density and the cosmological constant density
parameter as follows:
 \begin{align}
 \Omega_\TOT & \equiv \Omega_\Lambda + \Omega_\MAT = 1 \notag\eqcomma\\
 \Omega_\MAT & \equiv \Omega_\cdm + \Omega_\BAR = 0.3 \eqcomma \\
 \Omega_\Lambda & = 0.7 \notag\eqcomma
 \end{align}
where $\Omega_\CDM$ and $\Omega_\BAR$ are the density parameters
of cold dark matter (CDM) and baryons respectively, and
$\Omega_\Lambda$ denotes the density parameter due to a
cosmological constant, $\Omega_\Lambda \equiv \Lambda / 3 H_0^2$,
and $H_0 \equiv 100 h \UUNIT{km}{} \UUNIT{s}{-1} \UUNIT{Mpc}{-1}$
is the Hubble parameter today.  With $\Omega_\Lambda$ fixed to the
above values, we then vary the Hubble parameter $h$, the baryon
density $\omega_\BAR \equiv \Omega_\BAR h^2$ and the correlation
matrix $\bs{M}$ which describes the most general (i.e. mixed
adiabatic and isocurvature) initial conditions, as explained in
\SEC{chap:params;sec:ic}. We also fix to unity the scalar spectral
index, $n_\SCAL = 1$ for all modes and cross-correlators. Even by
varying only two cosmological parameters, our parameters space is
still 12-dimensional, since the initial condition correlation
matrix introduces ten free amplitudes.

We also investigate the following question: what is the preferred
isocurvature contribution to the perturbations? We shall see that,
with pre-WMAP CMB data, this question cannot be answered without
strong assumptions about the cosmological parameters.

\subsection{Pre-WMAP data analysis}

Our analysis uses the COBE \citep{Tegmark:1997jr} and
BOOMERanG~\citep{Netterfield:2001yq} data. For the latter, we
take into account the calibration and the beam size uncertainties
which treated just like two additional (normally distributed)
parameters of the problem (``nuisance parameters''). The two
cosmological parameters $h, \om_\BAR$ are sampled on a uniform
grid as follows (the number in parenthesis is the step size):
 \begin{alignat}{3}
  0.50  & < h                         &< \quad & 0.80 \step{0.05} \eqcomma \\
  0.015 &< \omega_\BAR                &< \quad& 0.085 \step{0.005}
  \eqdot
 \end{alignat}
For each grid point, we search the initial condition space by
minimizing the \chisq, as explained in
\SEC{chap:data;sec:Bayesian}. We look for the best fit point by
using a downhill simplex method~\citep{Book:Press92} initiated
after choosing a starting point randomly. The positive
semi-definiteness of the correlation matrix $\bs{M}$ is ensured by
penalty functions which guarantee that the conditions
(\refp{eq:positive_definit_conditions}) are satisfied \cite[more
details are given in][]{Trotta:2001tesi}. The best fit is then
estimated after $15,000$ minimization runs using this procedure.
It turns out that the topology of the $\chi^2$ surface on our
$14$-dimensional parameter space (including the two above nuisance
parameters) is quite complicated with many local minima and large
degeneracies, which considerably complicates the numerical search.
We assume that the likelihood function is Gaussian, and we
maximize instead of marginalize over the parameter we are not
interested in, see \SEC{chap:data;sec:Bayesian}.

\begin{figure}[tb]
\centering
\includegraphics[width=\twofigswidth]{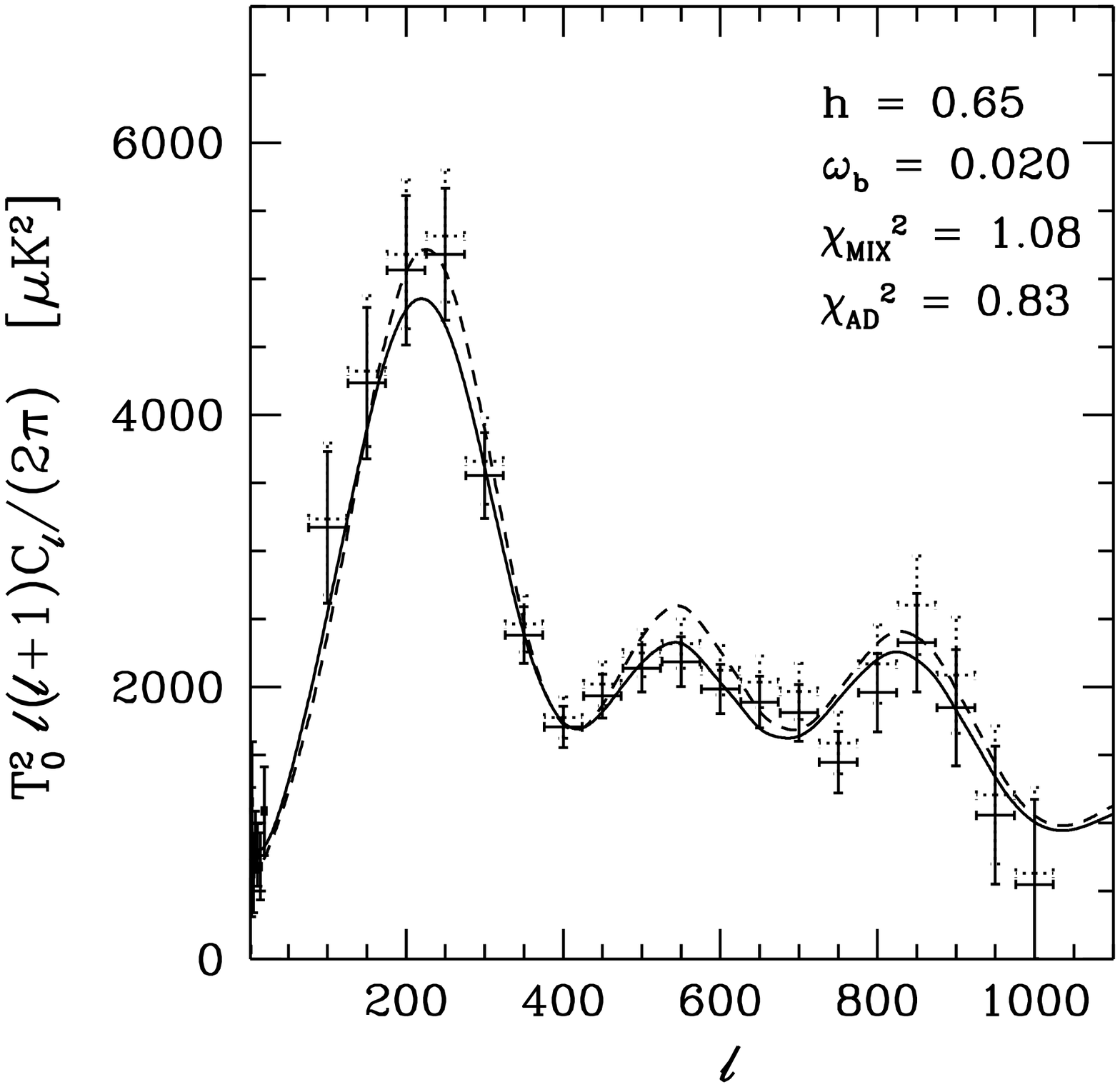}\hfill%
\includegraphics[width=\twofigswidth]{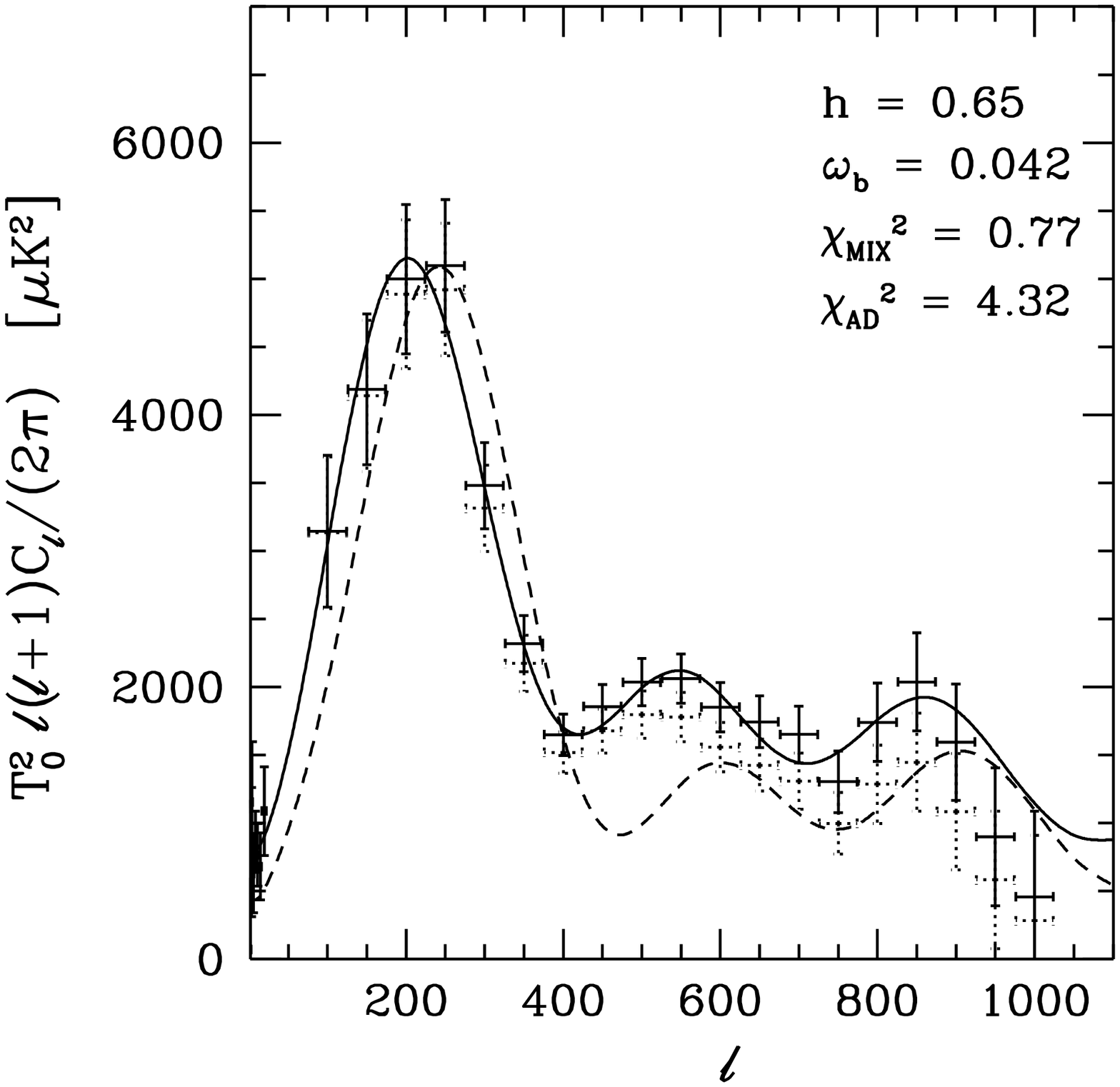}\hfill%
\caption[Best-fit models for purely adiabatic and mixed initial
conditions.]{CMB anisotropy temperature spectrum for different
values of the cosmological parameters $\omega_\BAR$ and $h$. We
plot the best-fit corresponding to a purely adiabatic case (dashed
line) and allowing general initial conditions, mixed models (solid
line). The calibration and the beam size of the BOOMERanG data
have been optimized to fit the mixed model (solid error bars) or
the adiabatic model (dotted error bars). The parameter choice in
the left panel ($\omega_\BAR = 0.02$, $h = 0.65$) can be fitted by
both models while the values $\omega_\BAR = 0.042$, $h = 0.65$
(right panel), can only be fitted by a mixed model.}
\label{fig:genic_bestfit}
\end{figure}

In \FIG{fig:genic_bestfit} we show the best-fit spectra for two
different choices of the cosmological parameters $\omega_\BAR$ and
$h$. Both of them are good fits if we allow for mixed initial
conditions. On the plot we have also indicated the reduced
$\chi^2$, \ie the value of $\chi^2/F$, where $F$ is the number of
degrees of freedom of the fit. For a fixed choice of
$\omega_\BAR$, $h$ the purely adiabatic model has only three
parameters (the amplitude of the adiabatic mode, and the two
nuisance parameters). With 26 data points (7 from COBE and 19 from
BOOMERanG) this leads to $F_\AD = 26 - 3 = 23$ degrees of freedom.
The mixed models have a symmetric $4 \times 4$ matrix determining
the initial amplitude, leading to a total of $12$ parameters and
hence $F_\MIX = 14$ degrees of freedom. If we also vary
$\omega_\BAR$ and $h$, the number of degrees of freedom is lowered
by two. It is not surprising that for fixed values $h = 0.65$,
$\omega_\BAR = 0.02$, which are well fitted by the adiabatic
model, the {\em reduced} $\chi^2$ of the adiabatic model is
somewhat lower than the one of the mixed model, since
$F_\MIX<F_\AD$ (as an example, see top panel of
\FIG{fig:genic_bestfit}). For the mixed model, the {\em absolute}
$\chi^2$ is always lower.

\begin{figure}[tb]
\centering
\includegraphics[width=\twofigswidth]{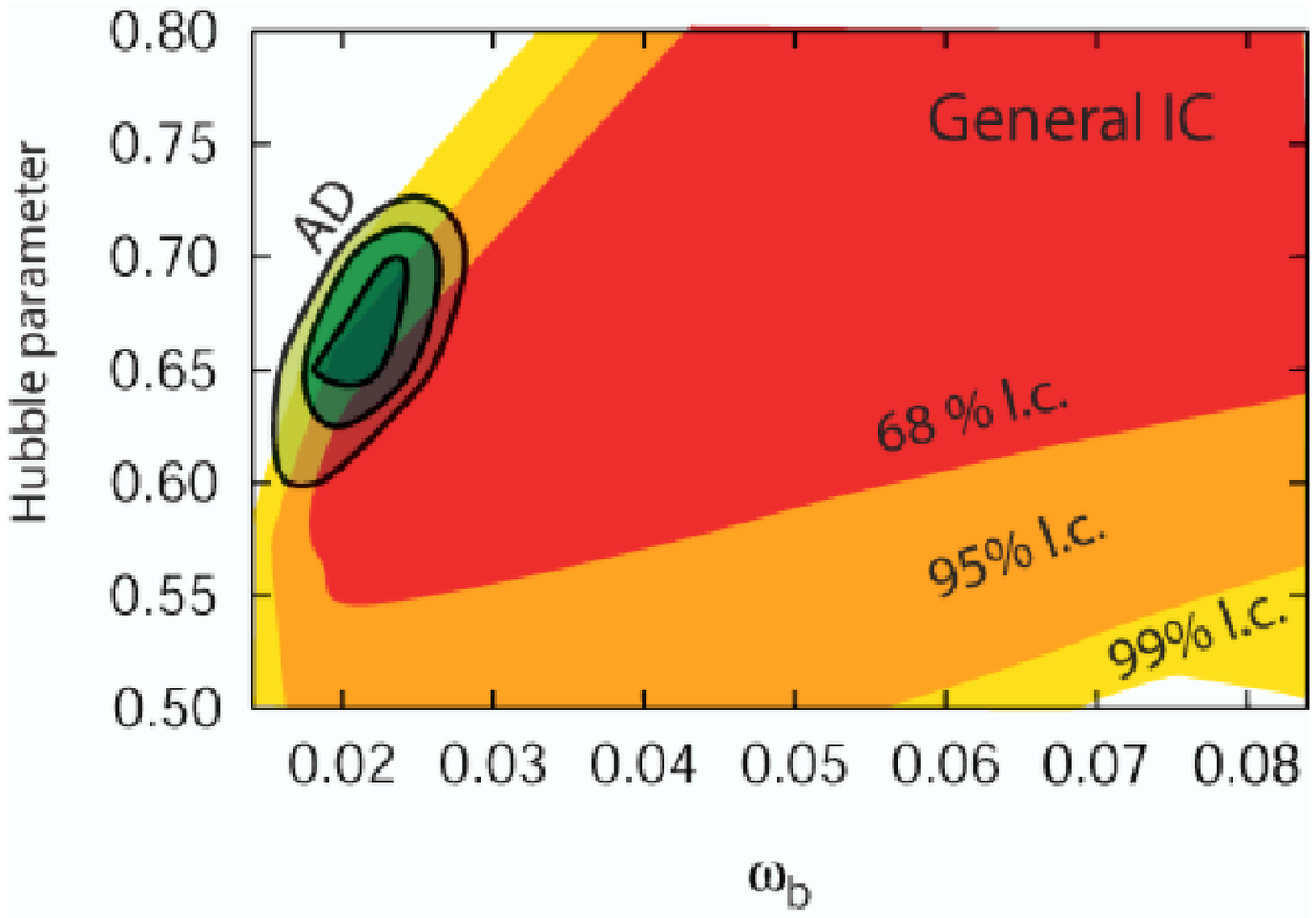}%
\includegraphics[width=\twofigswidth]{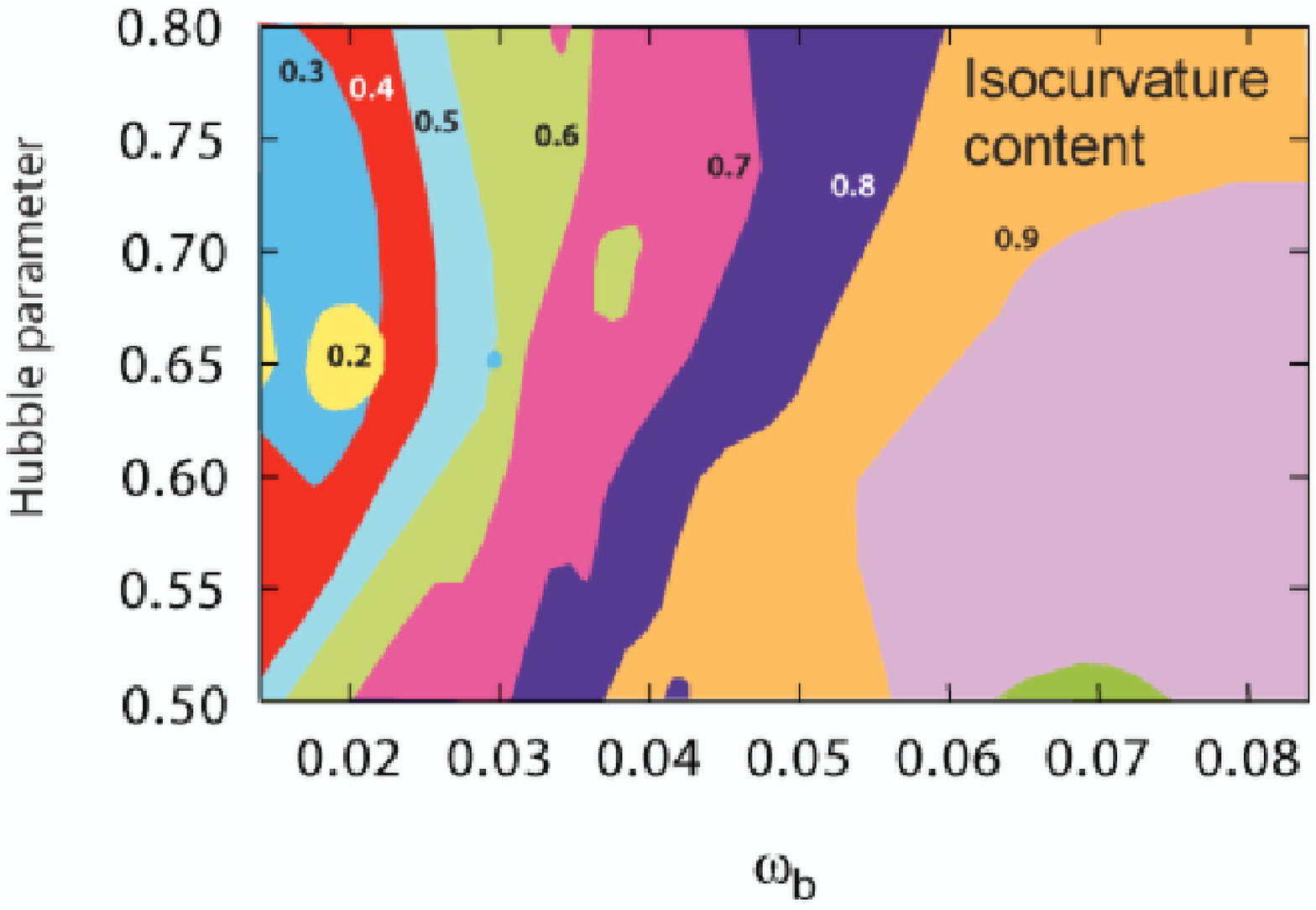}\hfill
\caption[Likelihood contours for purely adiabatic and mixed
isocurvature models and isocurvature content of the best fit
general isocurvature models.]{Left panel: the contours of 68\%,
95\%, 99\% likelihood content in the $(\omega_\BAR, h)$ plane for
purely adiabatic models (shadows of green, smaller contours) and
for mixed models (red to yellow, large contours). The likelihoods
are obtained by maximizing the nuisance parameters, and the
initial conditions correlation matrix $\bs{M}$ for mixed (\ie
general isocurvature) models. For mixed models, the lowest
$\chi^2$ corresponds to even higher values of $\omega_\BAR$ and
$h$ than those shown in the plot. Right panel: the isocurvature
content $\ga$ defined in (\ref{eq:isoc_content}) of the best fit
mixed model as function of the parameters $(\omega_\BAR, h)$. A
larger value for $\ga$ indicates a predominance of the
isocurvature modes on the adiabatic one.} \label{fig:genic_like}
\end{figure}

For both models we determine the likelihood functions of the
cosmological parameters $\omega_\BAR$ and $h$ by maximizing the
initial conditions correlation matrix and the nuisance parameters.
The result is shown in the left panel of \FIG{fig:genic_like}
where the likelihood contours in the $(\omega_\BAR, h)$ plane are
indicated for purely adiabatic and for mixed (general
isocurvature) models. It is remarkable the extent to which the
innermost $1\si$ contour opens up, once we allow for isocurvature
components. Strangely, the least likely region is the upper left
corner which contains the value of $\omega_\BAR=0.019\pm 0.02$
inferred from BBN \citep{Burles:2000zk} and the Hubble space
telescope key project value for the Hubble parameter
\citep{Freedman:2000cf} of $h = 0.72\pm 0.08$. Moreover, there is
absolutely no upper limit for $\omega_\BAR$ within the range
investigated here! This is explained by the fact that the
strongest features of a high baryon density universe, the
asymmetry between even and odd acoustic peaks and the shift of the
peak position due to the change in the sound velocity, can be
fully compensated by an admixture of isocurvature modes (see left
panel of \FIG{fig:genic_bestfit}). A very high baryon density can
therefore easily be accommodated into this framework. However, for
high $\omega_\BAR$ and low $h$, it is difficult to find a good fit
because there is not enough power in the secondary peak region due
to the early integrated Sachs-Wolfe effect boosting the first
peak.

We define the isocurvature content of a mixed model as
 \be \label{eq:isoc_content}
 \ga \equiv \dfrac{M_{22} + M_{33} + M_{44}}{\tr M} \eqcomma
 \ee
where $M_{11}$ denotes the adiabatic mode amplitude.  The
isocurvature content of the model shown in the left panel of
\FIG{fig:genic_bestfit} is only $\ga = 0.12$, while for the
parameter choice in the right panel one has $\ga = 0.69$. Hence,
if the cosmological parameters are close to those chosen in the
left panel, we can conclude that the cosmic perturbations are
predominantly adiabatic.

In the right panel of \FIG{fig:genic_like} we plot the
isocurvature content, $\ga$, of the best fit model obtained by
minimizing $\chi^2$ by variation of the initial conditions for
given values of the cosmological parameters. Clearly, the further
away we move from the region of parameter space well fitted by the
purely adiabatic model, the higher the isocurvature contribution
needed to fit the data becomes.

 The main non-adiabatic component of our best fits is
the neutrino entropy mode. This was to be expected, since this
mode and its correlator with the adiabatic mode can shift the peak
positions and can substantially add or subtract power from the
second peak \citep{Bucher:1999re}. A crucial point is, therefore,
to know whether such a mode can appear in a realistic structure
formation scenario. It is known that for interacting species the
non-adiabatic part of the perturbations tends to decay with time.
Therefore, the generation of a neutrino entropy component can only
occur after neutrino decoupling, that is at $T \lesssim 1
\UUNIT{MeV}{}$ (see \citealp{Gordon:2003hw} for a discussion). A
neutrino isocurvature perturbation could also be due to a fourth
species of sterile neutrinos which may have decoupled very early
in the history of the Universe.
 The same remark also applies of course to the CDM isocurvature mode.
Note that the energy density of this fourth neutrino type cannot
be very high, in order not to contradict the light element
abundances, but there is nothing which prevents (at least in
principle) the presence of large perturbations in this component.

\subsection{How important is the assumption of adiabaticity?}

We have shown that in allowing for isocurvature perturbations, one
can fit very well pre-WMAP CMB data with cosmological parameters
which differ considerably from the ones preferred by adiabatic
perturbations alone. More importantly, allowing for generic
initial conditions, the ranges of cosmological parameters which
can fit the CMB anisotropy data open up to an extent to become
nearly meaningless. On the other hand, assuming measurements of
cosmological parameters from other methods like direct
measurements of the Hubble parameter which yield $h\sim 0.65$ and
BBN which implies $\omega_\BAR \sim 0.02$, we can use the CMB to
limit the isocurvature contribution in the initial conditions (or
other unconventional features) and thereby learn something about
the very early universe, \ie, the inflationary phase which has
generated these initial conditions. For cosmological parameters in
the range preferred by other CMB independent measurements
($\Omega_\Lambda \sim 0.7$, $\Omega_\MAT \sim 0.3$, $h \sim 0.65$,
$\omega_\BAR \sim 0.02$) the isocurvature contribution in the
initial conditions has to be relatively modest ($\ga \lesssim
0.3$).  We have also checked explicitly that, given these
cosmological parameters, a purely isocurvature model, \ie one with
$M_{11}=0$, cannot fit the data.

Finally, and most importantly, our work shows the danger of
calling parameter estimation by CMB anisotropy experiments a
``parameter measurement'' since the results depend so sensitively
(and quite unexpectedly) on the underlying model assumptions. We
rather consider CMB anisotropies as an excellent tool to test
model assumptions or consistency.  In the light of these findings,
non-CMB measurements of cosmological parameters acquire even more
importance. In short, CMB is the ideal tool to investigate the
{\it primordial parameters} for cosmic structure formation (\ie
the initial conditions), while there are many other possibilities
to constrain {\it cosmological parameters} ($\Omega_X$, $h$, etc),
which we have to use in order to obtain good limits for possible
isocurvature perturbations.

As shown in \cite{Bucher:2000hy} and discussed in
\SEC{chap:genic;sec:future}, CMB temperature anisotropies alone,
even if measured with optimal precision limited by cosmic
variance, do not allow the degeneracy between cosmological
parameters and initial conditions to be removed. Polarization
measurements represent an additional non-trivial means to lift
this degeneracy and might constrain the contribution of the
isocurvature modes to about $10$\% accuracy \citep{Bucher:2000hy}.
The main reason for this is that polarization is mostly sensitive
to the quadrupole of the photon distribution rather than the
photon density perturbation, these two quantities depending in a
different way on the initial conditions. In the same vein, using
the normalization of the matter power spectrum (provided it can be
measured accurately) also helps to break some of the degeneracies
induced by the isocurvature modes, as we show in the next section.

\clearpage
\section{The cosmological constant problem}
\label{chap:genic;sec:lambda}

Ever since the beginning of modern cosmology, one of the most
enigmatic ingredients has been the cosmological constant.
\cite{Einstein:1917} introduced it to find static cosmological
solutions (which are, however, unstable). Later, when the
expansion of the Universe had been established, he reportedly
called it his ``greatest blunder''. In relativistic quantum field
theory, for symmetry reasons the vacuum energy momentum tensor is
of the form $\ep g_{\mu \nu}$ for some constant energy density
$\ep$. The quantity $\La = 8 \pi G \ep$ can be interpreted as a
cosmological constant. Typical values of $\ep$ expected from
particle physics come, for example, from the super-symmetry
breaking scale which is expected to be of the order of $\ep \gsim
1 \UUNIT{TeV}{4}$ leading to $\La \gsim 1.7 \times 10^{-26}
\UUNIT{GeV}{2}$, and corresponding to $\Om_\La \gsim 10^{58}$.
Recall that for the density parameter $\Om_\La \equiv \ep /
\rho_\CRIT = \La / (8 \pi G \rho_\CRIT)$, where $\rho_\CRIT = 8.1
\times 10^{-47} \,h^2 \UUNIT{GeV}{4}$ is the critical density and
the fudge factor $h$ is defined by $H_0 = 100 \,h \UUNIT{km}{}
\UUNIT{s}{-1} \UUNIT{Mpc}{-1}$, lying in the interval $0.5
\lesssim h \lesssim 0.8$. $H_0$ is the Hubble parameter today.

Such a result is clearly in contradiction with kinematical
observations of the expansion of the universe, which tell us that
the value of $\Om_\TOT$, the density parameter for the total
matter-energy content of the universe, is of the order of unity, $
\calo({\Om_\TOT}) \sim 1$. For a long time, this apparent
contradiction has been accepted by most cosmologists and particle
physicists, convinced that there must be some deep, not yet
understood reason that vacuum energy --- which is not felt by
gauge-interactions --- does not affect the gravitational field
either, and hence we measure effectively $\La = 0$. This slightly
unsatisfactory situation became really disturbing in 1998, as two
groups, which had measured luminosity distances to type Ia
supernovae, independently announced that the expansion of the
universe is accelerated in the way expected in a universe
dominated by a cosmological
constant~\citep{Riess:1998cb,Perlmutter:1998np}. More recent
measurements, which extend to higher redshift, seem to strengthen
this conclusion \citep{Tonry:2003zg,Riess:2004nr}, obtaining
values of the order $\calo(\Om_\MAT) \sim \calo(\OLa) \sim 1$ and
cannot be explained by any sensible high energy physics model.
Tracking scalar fields or
quintessence~\citep{Ratra:1988rm,Wetterich:1988fm} and other
similar ideas~\citep{Ferreira:1997au} have been introduced in
order to mitigate the smallness problem --- \ie, the fact that
$\ep \sim 10^{- 46} \UUNIT{GeV}{4}$. However, none of those is
completely successful and really convincing at the moment, see
\cite{Straumann:2002he,Sahni:2004ai} for reviews.

\subsection{Does structure formation need a cosmological constant?}

After the supernovae~Ia results, cosmologists have found many
other data-sets which also require a non-vanishing cosmological
constant. The most prominent fact is that CMB anisotropies
indicate a flat universe, $\Om_\TOT = \Om_\MAT +\OLa = 1$, while
measurements of clustering of matter, \eg, the galaxy power
spectrum, require $\Ga \equiv h \Om_\MAT \simeq 0.2$. But also CMB
data alone, with some reasonable prior on the Hubble parameter,
point to $\OLa > 0$ at high significance \citep{Spergel:2003cb}.

This cosmological constant problem is probably the greatest enigma
in present cosmology. The supernova results are therefore under
detailed scrutiny, and there has been a significant amount of work
aiming at finding an alternative explanation for the data, see \eg
\cite{Meszaros:2002np,Blanchard:2003du,Alam:2004jy}. Cosmological
observations are usually very sensitive to systematic errors which
are often very difficult to discover. Therefore, in cosmology an
observational result is usually accepted by the scientific
community only if several independent data-sets lead to the same
conclusion. But this seems to be exactly the case for the
cosmological constant.

It is therefore imperative to investigate in detail whether
present structure formation data does require a cosmological
constant, by asking whether enlarging the space of models for
structure formation does mitigate the cosmological constant
problem. There are several ways to enlarge the model space, \eg
one may allow for features in the primordial power spectrum, like
a kink~\citep{Barriga:2000nk}. Here we study the cosmological
constant problem in relation to the initial conditions for the
cosmological perturbations.

In a first step we discuss once more the usual results obtained
assuming purely adiabatic models and we investigate the extent to
which pre-WMAP CMB data alone or combined with large-scale
structure measurements require $\OLa \neq 0$ in a flat universe,
presenting the findings published in \cite{Trotta:2002iz}. We
shall first proceed with the usual Bayesian analysis, but we also
discuss the results which are obtained in a frequentist approach.
We find that even if $\Om_\La = 0$ is outside the high likelihood
region in a Bayesian approach this is no longer the case from the
frequentist point of view. In other words the probability that a
model with vanishing $\Om_\La$ leads to the present-day observed
CMB and large-scale structure data is not exceedingly small.

We then study how the results are modified if we allow for general
isocurvature contributions to the initial conditions. In this
first study of the matter power spectrum from general isocurvature
modes we discover that a COBE-normalized matter power spectrum
reproduces the observed amplitude only if it is highly dominated
by the adiabatic component. Hence the isocurvature modes cannot
contribute significantly to the matter power spectrum and do not
lead to a degeneracy in the initial conditions for the matter
power spectrum when combined with CMB data.

\subsection{CMB and large scale structure data analysis}

The pre-WMAP CMB measurements, from
BOOMERanG~\citep{Netterfield:2001yq}, MAXIMA~\citep{Lee:2001yp},
DASI~\citep{Halverson:2001yy},
VSA~\citep{Scott:2002th,Taylor:2002ti}, CBI~\citep{Pearson:2002tr}
and Archeops~\citep{Benoit:2002mk} are in very good agreement up
to the third peak in the angular temperature power spectrum of CMB
anisotropies, $\ell \sim 1000$. In our analysis we therefore use
the COBE data~\citep{Smoot:1992td,Bennett:1994gg} in the
decorrelated compilation of \cite{Tegmark:1997jr} (7 points
excluding the quadrupole) for the $\ell$ region $3 \leq \ell \leq
20$ and the BOOMERanG data to cover the higher $\ell$ part of the
spectrum (19 points in the range $100 \leq \ell \leq 1000$). Since
Archeops has the smallest error bars in the region of the first
acoustic peak, we also include this data-set (16 points in the
range $15 \leq \ell \leq 350$). Including any of the other
mentioned data does not influence our results significantly. The
BOOMERanG and Archeops absolute calibration errors ($10\%$ and
$7\%$ at $1\si$, respectively) as well as the uncertainty of the
BOOMERanG beam size are included as additional Gaussian nuisance
parameters, and are maximized over.  We make use of the Archeops
window functions available from the \cite{Archeops:Website}, while
for BOOMERanG a top-hat window is assumed. For the matter power
spectrum, we use the galaxy-galaxy power spectrum from the 2dF
data which is obtained from the redshift of about $10^5$
galaxies~\citep{Tegmark:2001jh}. We include only the 22
decorrelated points in the linear regime, \ie, in the range $0.017
\leq k \leq 0.314 \quad [h \UUNIT{Mpc}{-1}]$, and the window
functions of \cite{Tegmark:2001jh} which can be found at
\cite{Tegmark:Website}.

Our grid of models is restricted to flat universes and we assume
purely scalar perturbations. Since the goal here is more to make a
conceptual point than to consider the most generic model, we fix
the baryon density to the BBN preferred value $\Om_\BAR h^2 \equiv
\om_\BAR = 0.020$~\citep{Burles:2000zk} and we investigate the
following 3-dimensional grid in the space of cosmological
parameters:
\begin{alignat}{3}
  0.35  & < h                         &< \quad & 1.00 \step{0.025} \notag\eqcomma \\
  0.00  &< \OLa                       &< \quad & 0.95 \step{0.05}
  \eqcomma\\
  0.80  &< n_\SCAL                     &< \quad & 1.20 \step{0.05}
  \notag
  \eqcomma
 \end{alignat}
where $n_\SCAL$ is the scalar spectral index, which again we take
to be the same for all modes,  and the numbers in parenthesis give
the step size we use. The total matter content $\Om_\MAT \equiv
\Om_\CDM + \Om_\BAR$ is $\Om_\MAT = 1 - \OLa$, and $\Om_\CDM$
indicates the cold dark matter contribution. For all models the
optical depth of reionization is $\tau = 0$ and we have three
families of massless neutrinos. For each grid point we compute the
ten CMB and matter power spectra, one for each independent set of
initial conditions, as explained in \SEC{chap:params;sec:ic}. The
initial condition correlation matrix $\bs{M}$ is parameterized
using the ten dimensional hypercube parameters presented on page
\pageref{sec:hypercube_params}.

For a given initial conditions correlation matrix $\bs{M}$ and
spectral index $n_\SCAL$, we quantify the isocurvature
contribution to the CMB temperature anisotropy by the
phenomenological parameter $\beta$ defined as
\begin{equation} \label{eq:isoc_content_beta}
\beta \equiv \frac{\displaystyle
                   \sum_{\scriptscriptstyle X = \CI, \NIV, \NID}
                   \left<(\ell(\ell+1))C_{\ell}^{\scriptscriptstyle (X,X)}
                   \right>_\ell}
                  {\displaystyle \sum_{\scriptscriptstyle
                   Y = \AD, \CI, \NIV, \NID}
                   \left< \ell (\ell + 1) C_\ell^{\scriptscriptstyle (Y,Y)}
                   \right>_\ell} ,
\end{equation}
where the average $\left<\cdot\right>$ is taken in the $\ell$
range of interest, in our case $3 \leq \ell \leq 1000$, and where
$C_{\ell}^{\scriptscriptstyle (X,X)}$ stands for the
auto-correlator of the CMB anisotropies with initial conditions
$X$. This quantity measures the average power of the adiabatic and
isocurvature modes over the full multipole range, and therefore it
gives a more phenomenological description of the isocurvature
contribution than the parameter $\gamma$ used in the previous
section, and defined in \rrp{eq:isoc_content}.

As highlighted in \SEC{chap:data;sec:Bayesian}, the correct
interpretation of Bayesian statistics is in terms of most likely
regions in parameter space, while the frequentist approach is
required in order to obtain exclusion intervals for the
parameters. In order to answer the question of whether the CMB and
large scale structure data exclude with a given confidence the
value $\OLa = 0$, we use the frequentist statistics, and compare
the result with the usual Bayesian approach.

\subsection{Adiabatic perturbations}

We first fit CMB data only ($N = 42$) by maximizing $M = 7$
parameters, \ie, the three nuisance parameters, $n_\SCAL$, $h$,
$\OLa$ and the overall amplitude of the adiabatic spectrum, and we
find (Bayesian likelihood intervals on $\OLa$ alone):
\begin{equation}
\OLa = 0.80 \lims{-0.35}{+0.10} \mbox{ at $2 \si$ $\quad$ and
$\quad$}
            \lims{-0.80}{+0.12}
 \mbox{ at $3 \si$}.
\end{equation}
The asymmetry in the intervals arises because the value of $\OLa$
for our maximum likelihood (ML) model is relatively large. One
could achieve a better precision in determining the ML value of
$\OLa$ by using a finer grid and varying $\omega_\BAR$ as well,
which has extensively been done in the literature and is not the
scope of this work. Moreover, the position of the acoustic peaks
in CMB anisotropies is mostly sensitive to the age of the universe
at recombination, which depends only on $\Om_\MAT h^2$, and to the
angular diameter distance, which depends on $\Om_\MAT$, $\OLa$ and
the curvature of the universe. When the universe is flat, the
angular diameter distance is weakly dependent on the relative
amounts of $\Om_\MAT$ and $\OLa$ as long as $\OLa$ is not too
large, see \SEC{chap:params;sec:acoustic} and
\FIGPAG{fig:shift_param2D}. Hence, one can achieve a sufficiently
low value of $\Om_\MAT h^2$ either via a large cosmological
constant or via a very low Hubble parameter, $h \lsim 0.45$.

\begin{figure}[tb]
\centering
\includegraphics[width=\twofigswidth]{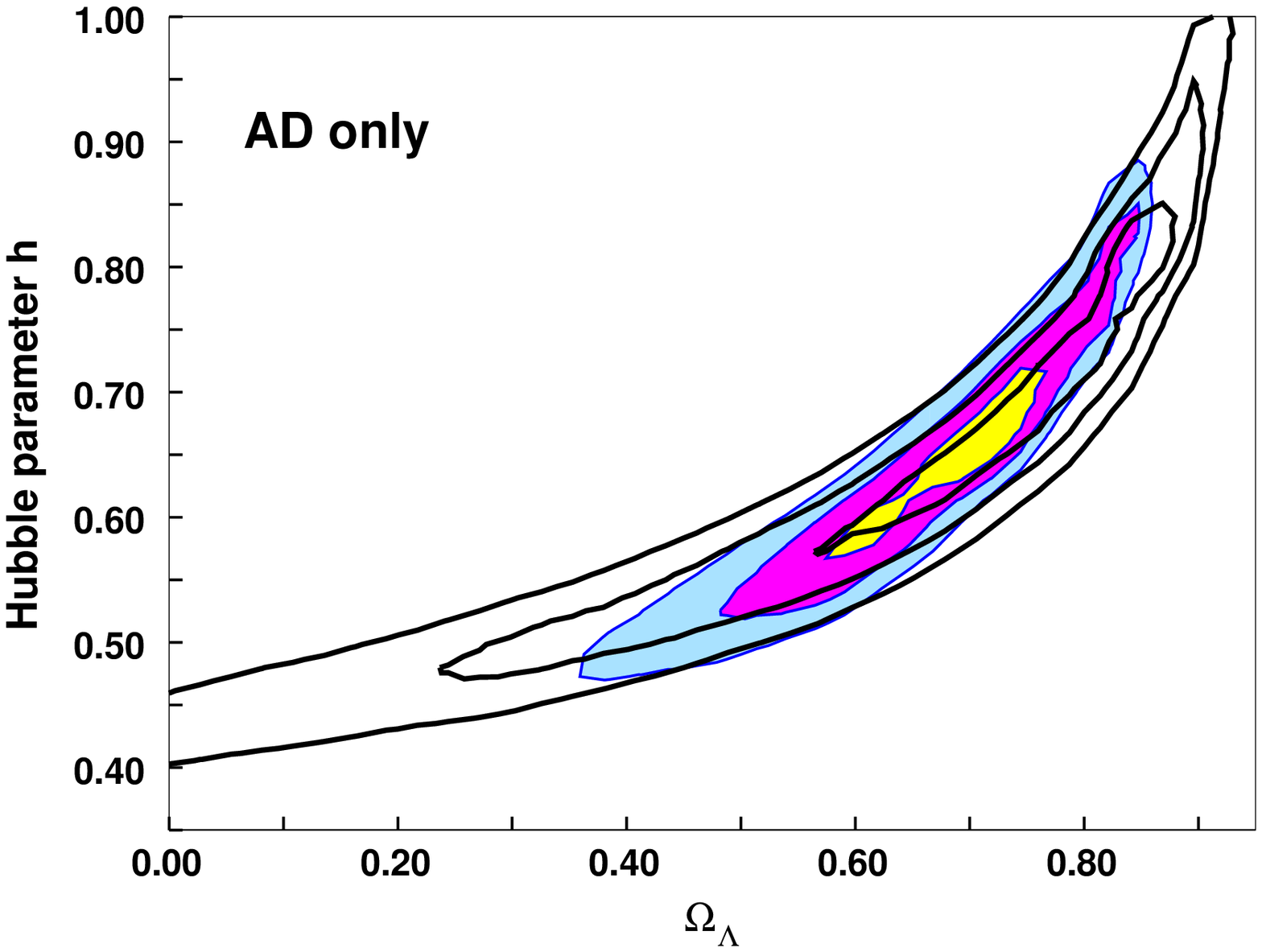}\hfill%
\includegraphics[width=\twofigswidth]{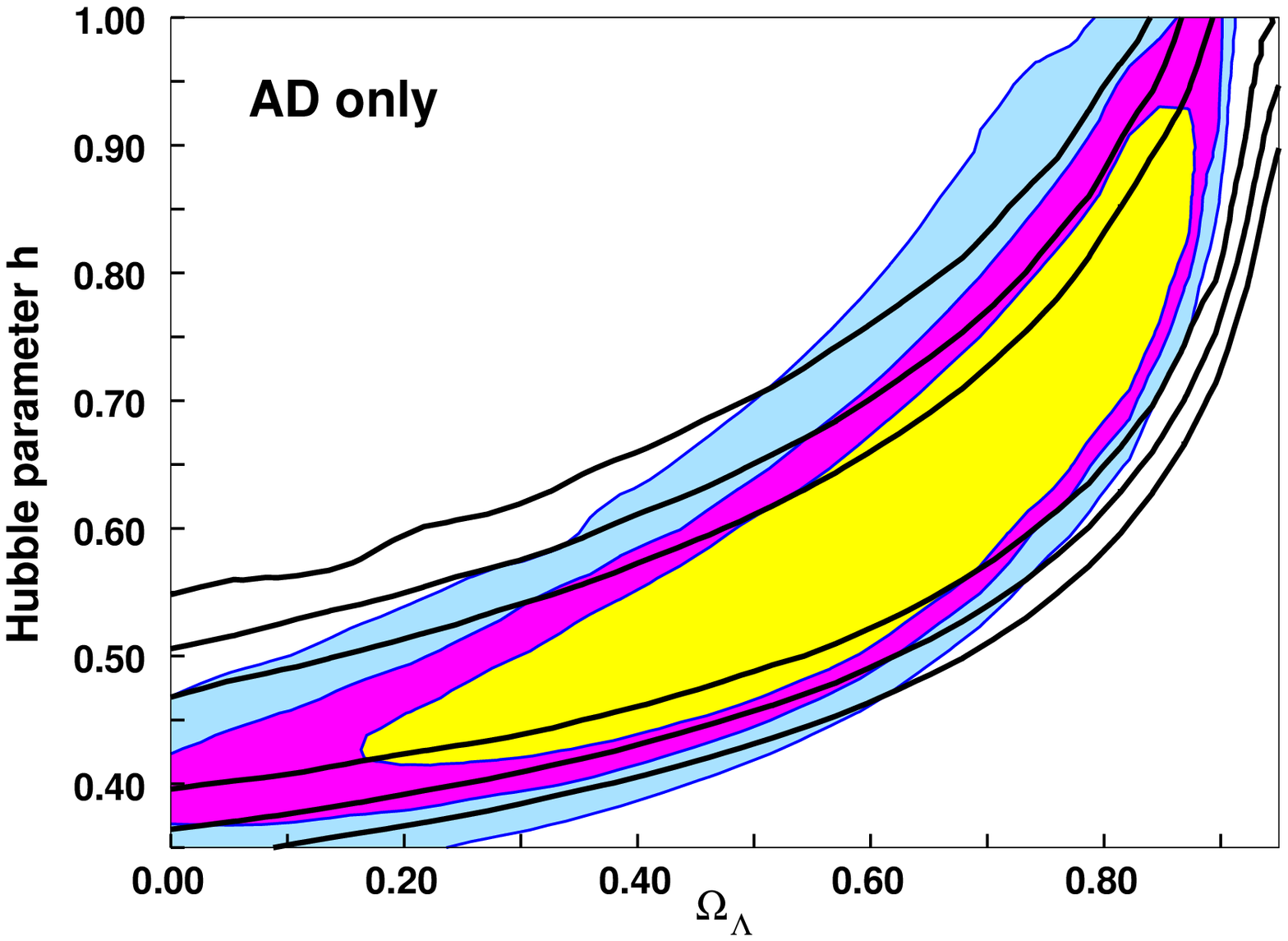}\hfill
\caption[Bayesian and frequentist likelihood contours in the
$(\OLa, h)$ plane.]{Joint likelihood contours (Bayesian, left
panel) and confidence contours (frequentist, right panel), with
CMB only (solid lines,$1 \si$, $2 \si$, $3 \si$ contours) and
CMB+2dF (filled) for purely adiabatic initial conditions. In the
right panel, the number of effective degrees of freedom is $F_\EFF
= 31$ for CMB alone $F_\EFF = 50$ for CMB+2dF.}
\label{fig:genic_ad_only}
\end{figure}

We now include the matter power spectrum $P_\MAT$, assuming
$P_\MAT = b^2 P_\GAL$, where $P_\GAL$ is the observed galaxy power
spectrum and $b$ some unknown bias factor (assumed to be scale
independent), over which we maximize. Inclusion of this data in
the analysis breaks the $\OLa$, $h$ degeneracy, since $P_\MAT$ is
mainly sensitive to the shape parameter $\Gamma \equiv \Om_\MAT
h$. We therefore obtain significantly tighter overall likelihood
intervals for $\OLa$:
\begin{equation}
\OLa = 0.70 \lims{-0.17}{+0.13} \mbox{ at $2 \si$ $\quad$ and
$\quad$}
            \lims{-0.27}{+0.15}
\mbox{ at $3 \si$} \eqdot
\end{equation}
We plot joint likelihood contours (Bayesian) for $\OLa$, $h$ with
purely adiabatic initial conditions in the left panel of
\FIG{fig:genic_ad_only}.
From the Bayesian analysis, one concludes that CMB and 2dF
together require a non-zero cosmological constant at very high
significance, more than $7 \si$ for the points in our grid! Note
that the ML point has a reduced chi-square $\hat{\chi}^2_{F = 56}
= 0.59$, significantly less than unity.

The frequentist analysis, however, excludes a much smaller region
of parameter space, \CF the right panel of
\FIG{fig:genic_ad_only}. The frequentist contours must be drawn
for the effective number of degrees of freedom, \ie, using the
number of effectively independent data points. We can therefore
roughly take into account a $10\%$ correlation, which is the
maximum correlation between data points given in
\cite{Netterfield:2001yq,Benoit:2002mk}, by replacing $F$ by the
effective number of degrees of freedom, $F_\EFF = 0.9 N - M$, and
rounding to the next larger integer (to be conservative). One
could argue that the BOOMERanG and Archeops data points are not
completely independent, since BOOMERanG observed a portion of the
same sky patch as measured by Archeops. This possible correlation
is difficult to quantify, but should not be too important since
the sky portion observed by Archeops is a factor of 10 larger than
BOOMERanG's and therefore we ignore it here. The right panel of
\FIG{fig:genic_ad_only} is drawn with $F_\EFF = 31$ for CMB alone
and $F_\EFF = 50$ for CMB+2dF, but we have checked that our
results do not change much if we use a $5\%$ correlation.

\begin{figure}[!tb]
\centering
\includegraphics[width=\twofigswidth]{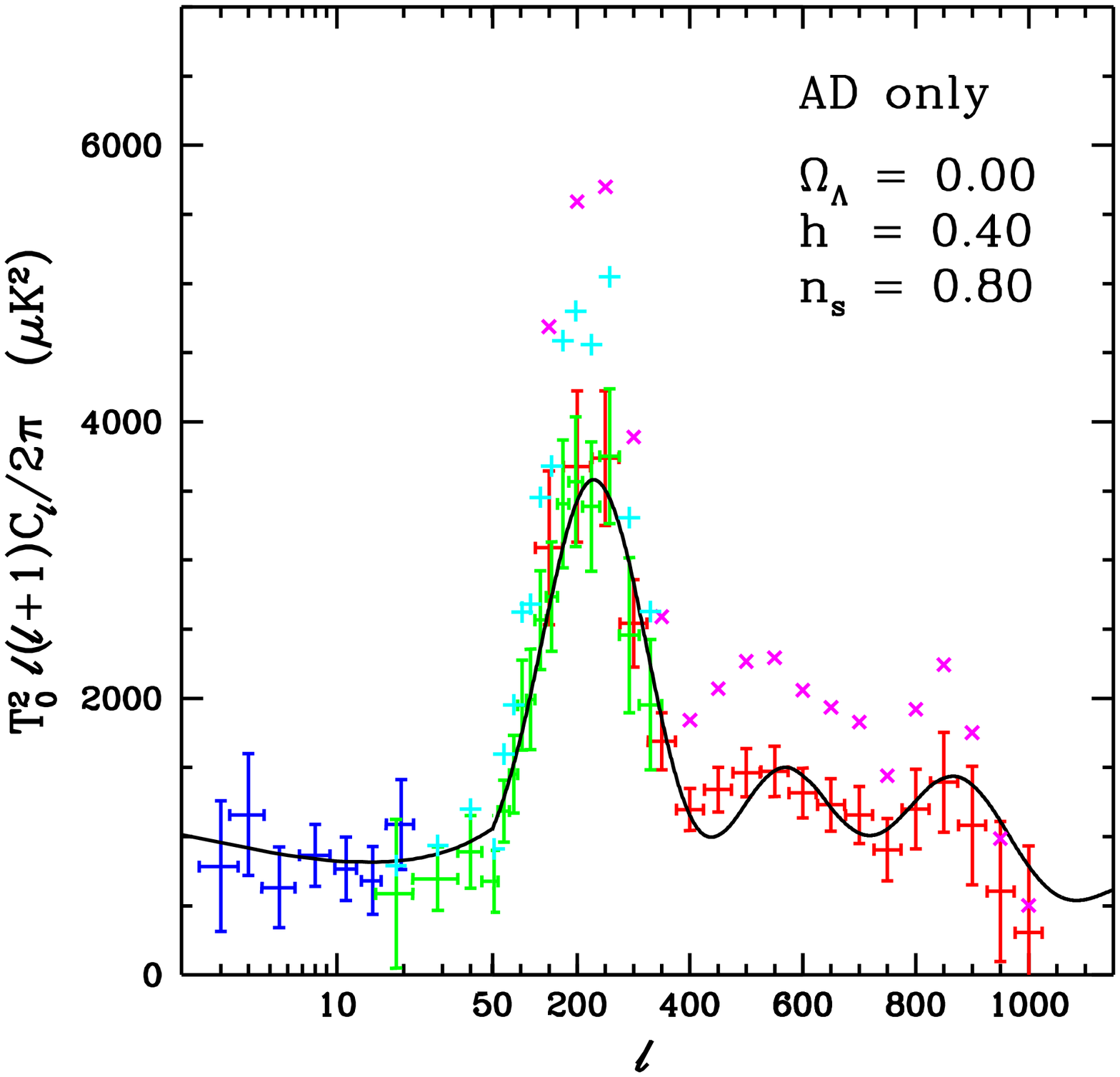}\hfill%
\includegraphics[width=\twofigswidth]{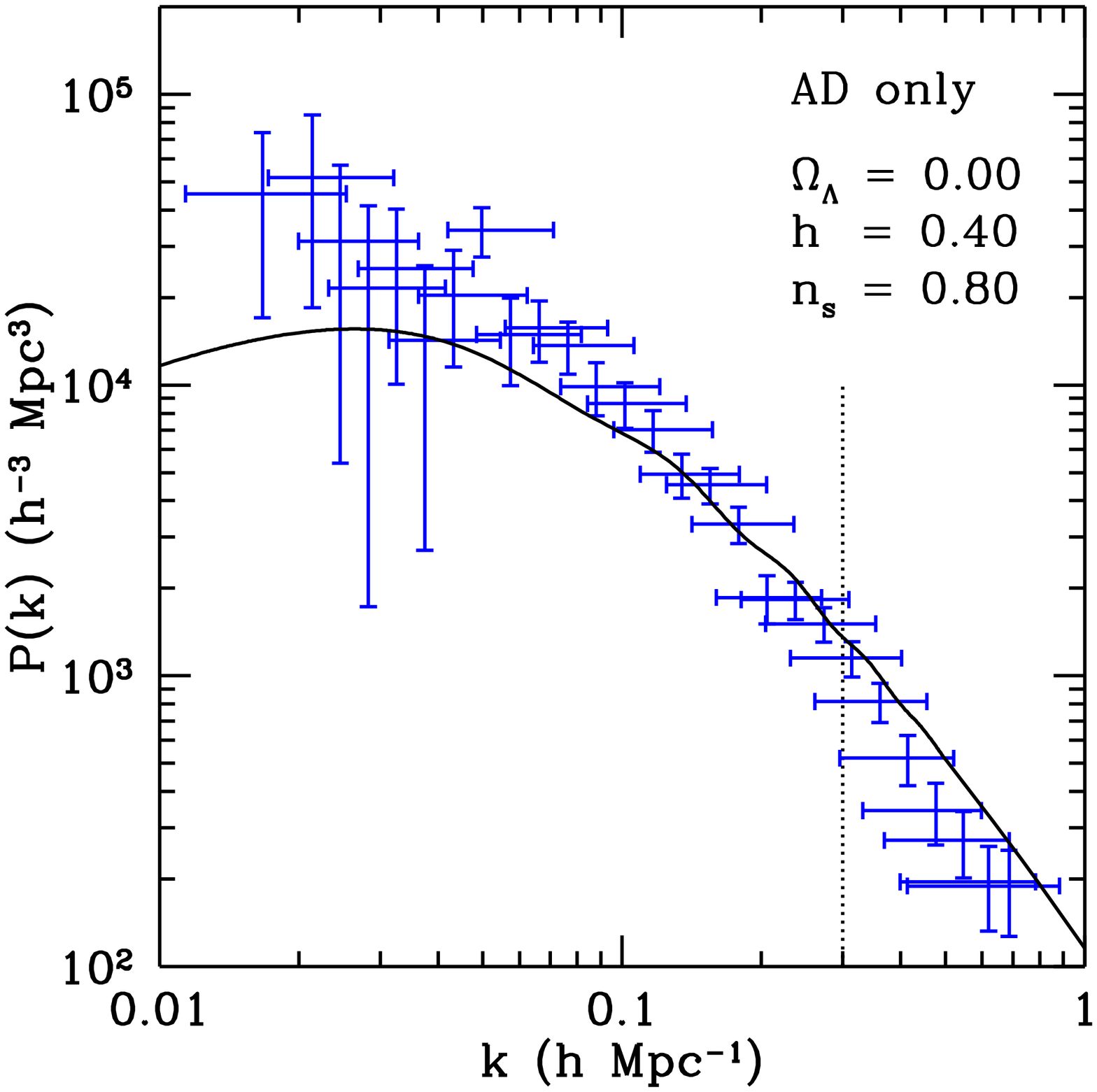}\hfill
\caption[Best fit of CMB and 2dF data compatible with $\OLa=0$ for
purely adiabatic models.]{Best fit with $\OLa = 0$ and purely
adiabatic initial conditions, compatible with CMB and 2dF data
within $2\si$ confidence level (frequentist). In the right panel,
only the 2dF data points left of the vertical, dotted line
--- \ie, in the linear region --- have been included in the analysis. Note the low
CMB first acoustic peak in the left panel due to the joint effect
of the red spectral index and of the absence of early ISW effect.
In this fit, the calibration of BOOMERanG (red/dark gray
errorbars) and Archeops (green/light gray errorbars) has been
reduced by $34\%$ and $26\%$, respectively. To appreciate the
difference, we plot the non recalibrated value of the BOOMERanG
and Archeops data points as diagonal/magenta crosses and
vertical/light blue crosses, respectively. Even though the fit is
``by eye'' very good, it seems highly unlikely that the
calibration error is so large.} \label{fig:genic_AD_OL0}
\end{figure}

It is interesting to note that there are regions in the left panel
which are excluded with a certain confidence by CMB data alone but
are no longer excluded at the same confidence when we include the
2dF data. In other words, it would seem that taking into account
more data and therefore more knowledge about the universe, does
not systematically exclude more models, \ie, the CMB+2dF contours
are not always contained in the CMB alone contours. This apparent
contradiction vanishes when one realizes that the confidence
limits on, \eg, $\OLa$ alone in the frequentist approach are just
the projection of the confidence contours of the right panel on
the $\OLa$ axis. One can readily verify in the right panel that
the confidence limits for the combined data-set are always smaller
than the ones for CMB data alone. There are points with $\OLa = 0$
and $h \simeq 0.40$ which are still compatible within $2 \si$ with
both 2dF and CMB data, at the price of pushing somewhat the other
parameters. In the best fit with $\OLa = 0$ shown in
\FIG{fig:genic_AD_OL0}, one has to live with a red spectral index
$n_\SCAL = 0.80$. Furthermore, the calibration of the BOOMERanG
and Archeops data points is reduced in this fit by $34\%$ and
$26\%$, respectively, \ie, more than 3 times the quoted $1 \si$
systematic error.

In both cases, it is clear that one can exploit the $\OLa$, $h$
degeneracy to fit CMB data alone with a model having $\OLa = 0$.
For a flat universe like the one we are considering, one has then
to use a much smaller value of the Hubble parameter than the one
indicated by other measurements, most notably the HST Key
Project~\citep{Freedman:2000cf}, which gives $h=0.72 \pm 0.08$.
The 2dF data are mainly sensitive to the shape parameter $\Gamma
\sim 0.2$, hence 2dF with $\Om_\MAT = 1.0$ would require an even
lower value of $h$ which is not compatible with CMB. Therefore
inclusion of 2dF data tends to exclude any flat model without a
cosmological constant. Summing up, for purely adiabatic initial
conditions the Bayesian approach gives very strong support to
$\OLa \neq 0$; in the more conservative frequentist point of view,
while $\OLa \neq 0$ cannot be excluded with very high confidence,
the combination of 2dF and pre-WMAP CMB data start to be
incompatible with a flat universe with vanishing cosmological
constant. These conclusions are in qualitative agreement with
previous works using comparable data
\citep{Netterfield:2001yq,Pryke:2001yz,Lewis:2002ah,Wang:2001gy,Durrer:2001jz,Rubino-Martin:2002rc,Benoit:2002mm}.
In the next section we investigate the stability of these well
known results with respect to inclusion of non-adiabatic initial
conditions.

\subsection{Mixed adiabatic and isocurvature perturbations}

We now enlarge the space of models by including all possible
isocurvature modes with arbitrary correlations among themselves
and the adiabatic mode as described in the previous section, but
with the restriction that all modes have the same spectral index.
We first consider CMB data only and maximize over initial
conditions. The number of parameters increases by nine and the
number of degrees of freedom decreases correspondingly with
respect to the purely adiabatic case considered above.

\begin{figure}[tb]
\centering
\includegraphics[width=\twofigswidth]{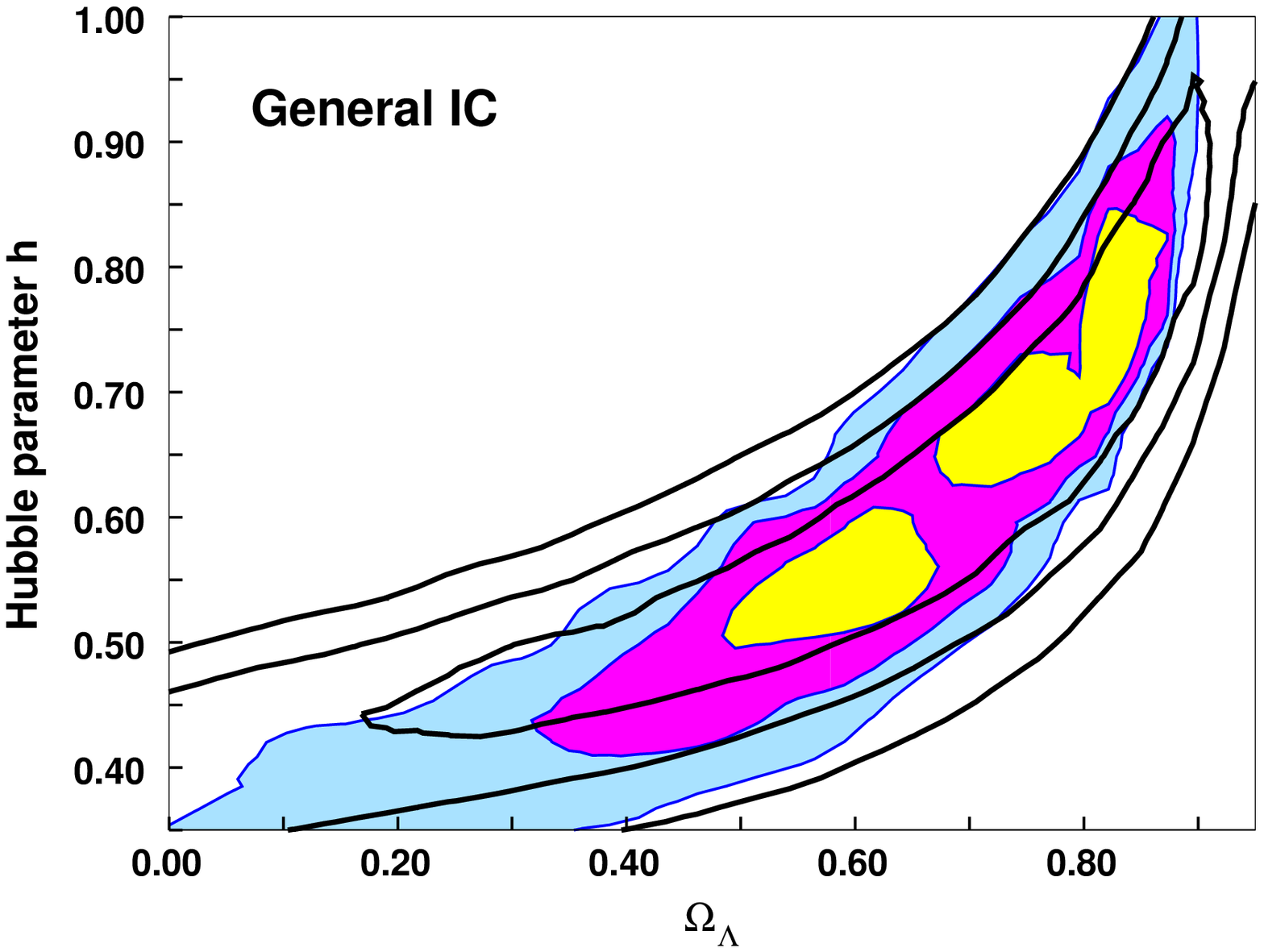}\hfill%
\includegraphics[width=\twofigswidth]{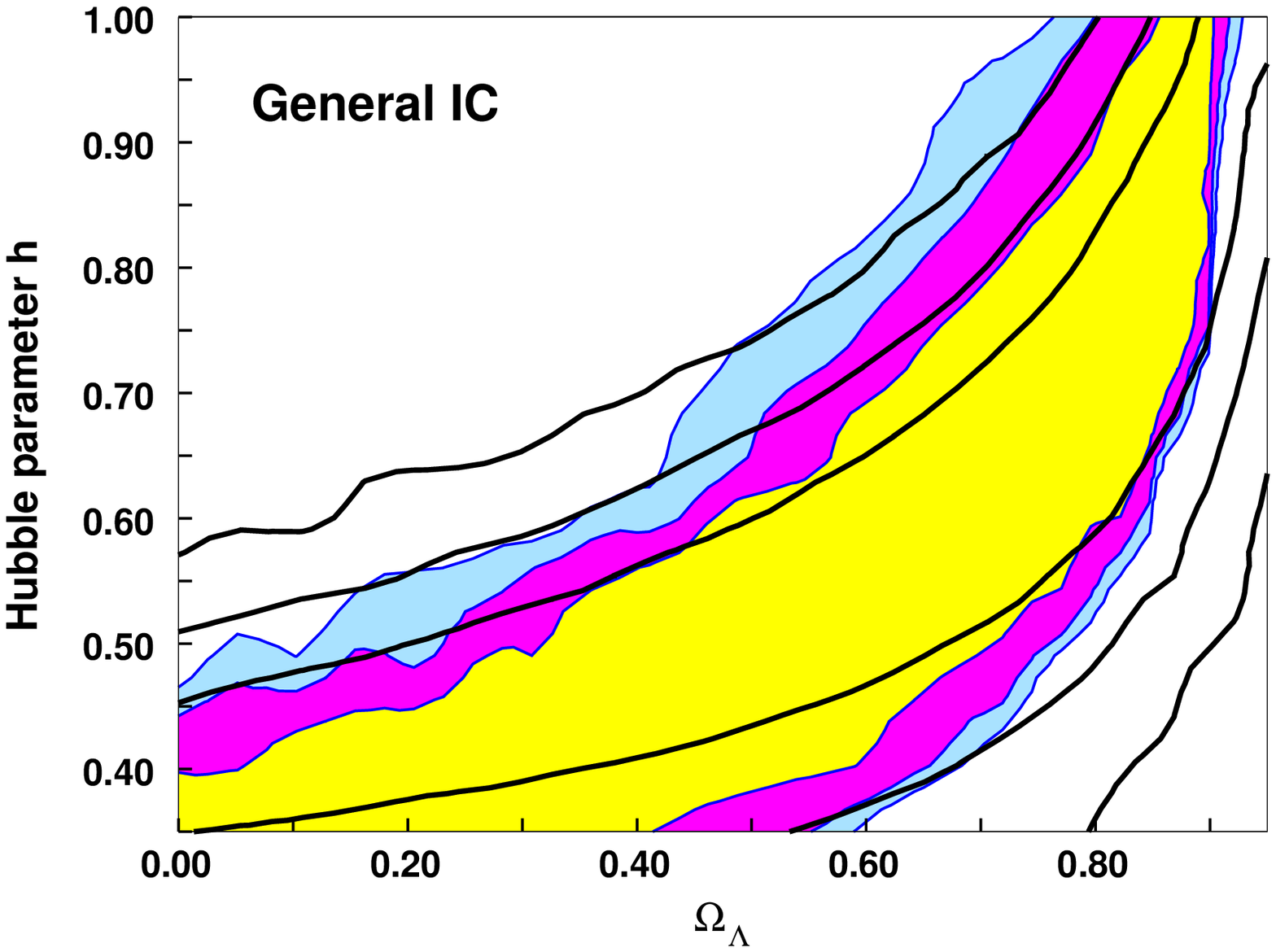}\hfill
\caption[Bayesian and frequentist likelihood contours in the
$(\OLa, h)$ plane for general isocurvature models.]{Joint
likelihood contours (Bayesian, left panel) and confidence contours
(frequentist, right panel), with CMB only (solid lines) and
CMB+2dF (filled) after maximization over general isocurvature
initial conditions. The likelihood/probability content is $1 \si$,
$2 \si$, $3 \si$, from the center to the outside. The disconnected
$1\si$ region in the left panel is an artificial feature due to
the grid resolution. In the right panel, the number of effective
degrees of freedom is $F_\EFF = 22$ for CMB alone $F_\EFF = 41$
for CMB+2dF.} \label{fig:genic_mixed}
\end{figure}

\begin{figure}[tb]
\centering
\includegraphics[width=\onefigwidth]{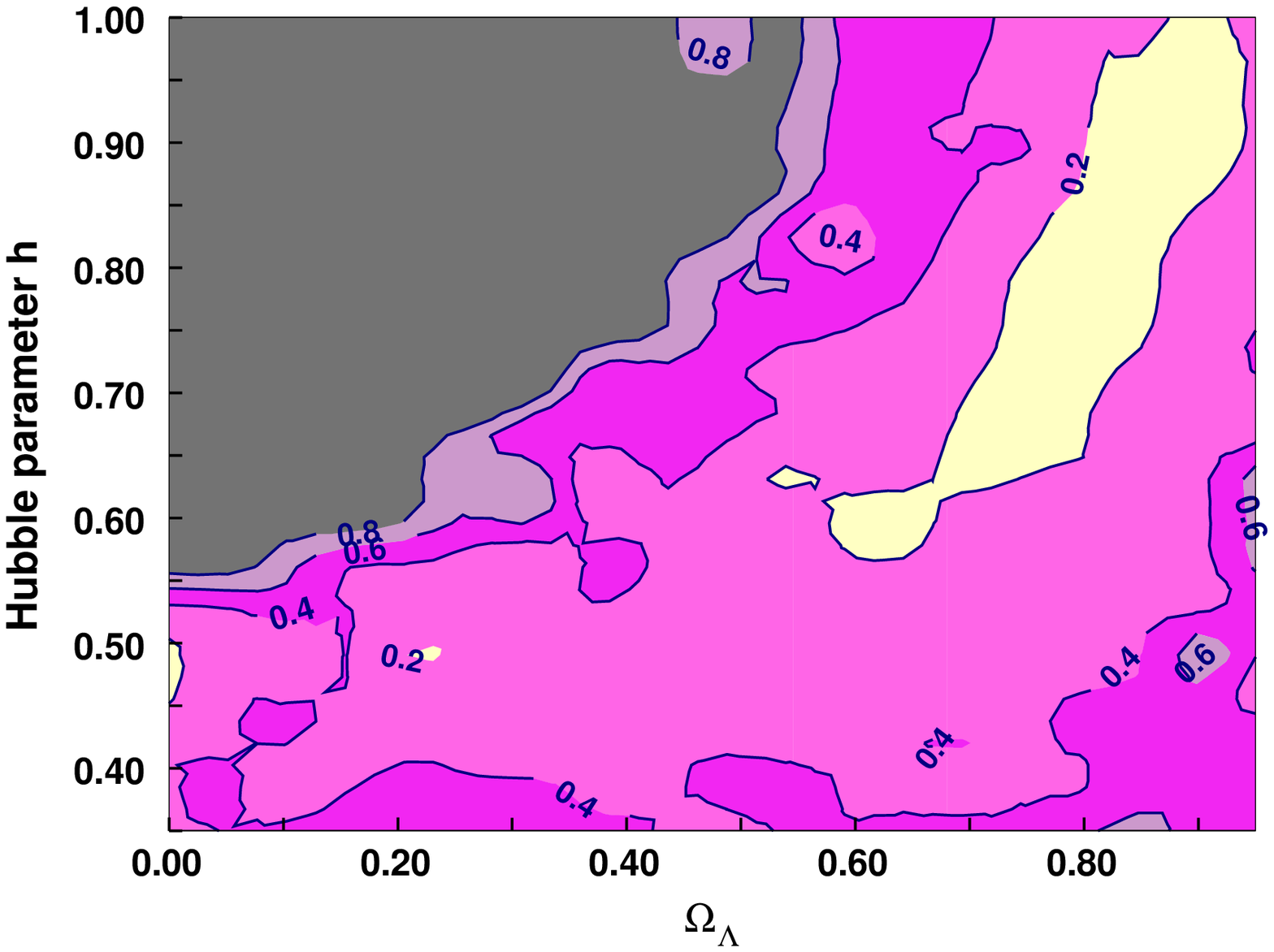}
\caption[Isocurvature content of the best fit
models.]{Isocurvature content $0.0 \leq \beta \leq 1.0$ of best
fit models with CMB and 2dF data. The contours are for $\beta =
0.20, 0.40, 0.60, 0.80$ from the center to the outside.}
\label{fig:genic_beta}
\end{figure}

Likelihood (Bayesian, left panel of \FIG{fig:genic_mixed}) and
confidence (frequentist, right panel of \FIG{fig:genic_mixed})
contours widen up somewhat along the degeneracy line. The
enlargement is less dramatic than in the case of the baryon
density presented in \SEC{chap:genic;sec:precision}. This is
partially due to our prior of flatness which reduces the space of
models to those which are almost degenerate in the angular
diameter distance. Most of our models have the first acoustic peak
of the adiabatic mode already in the region preferred by
experiments, hence in most of the fits, isocurvature modes play a
modest role, especially in the parameter regions with large
$\OLa$, $h$ (\CF~ \FIG{fig:genic_beta} and the discussion below).
Nevertheless, because of the $\OLa$, $h$ degeneracy, even a modest
widening of the contours along the degeneracy line results in an
important enlargement of the likelihood limits. The ML point does
not depart very much from the purely adiabatic case, but now we
cannot constrain $\OLa$ at more than $1\sigma$ (Bayesian, CMB
only):
\begin{equation}
\OLa = 0.85 \lims{-0.35}{+0.05} \mbox{ at $1 \si$ \eqcomma }
\end{equation}
and no limits for $0.0 \leq \OLa \leq 0.95$ at higher confidence.

\begin{figure}[tb]
\centering
\includegraphics[width=\twofigswidth]{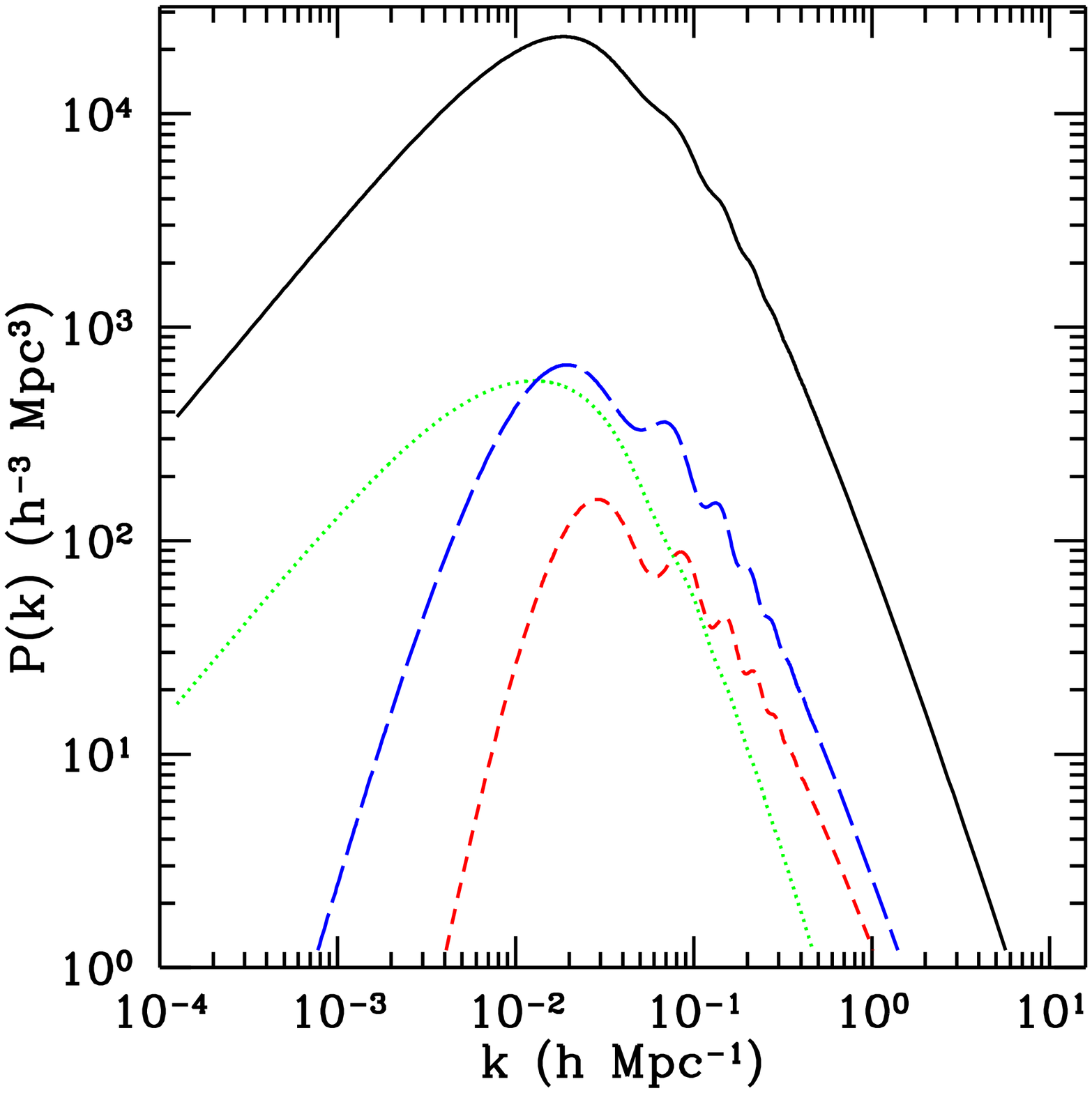}\hfill%
\includegraphics[width=\twofigswidth]{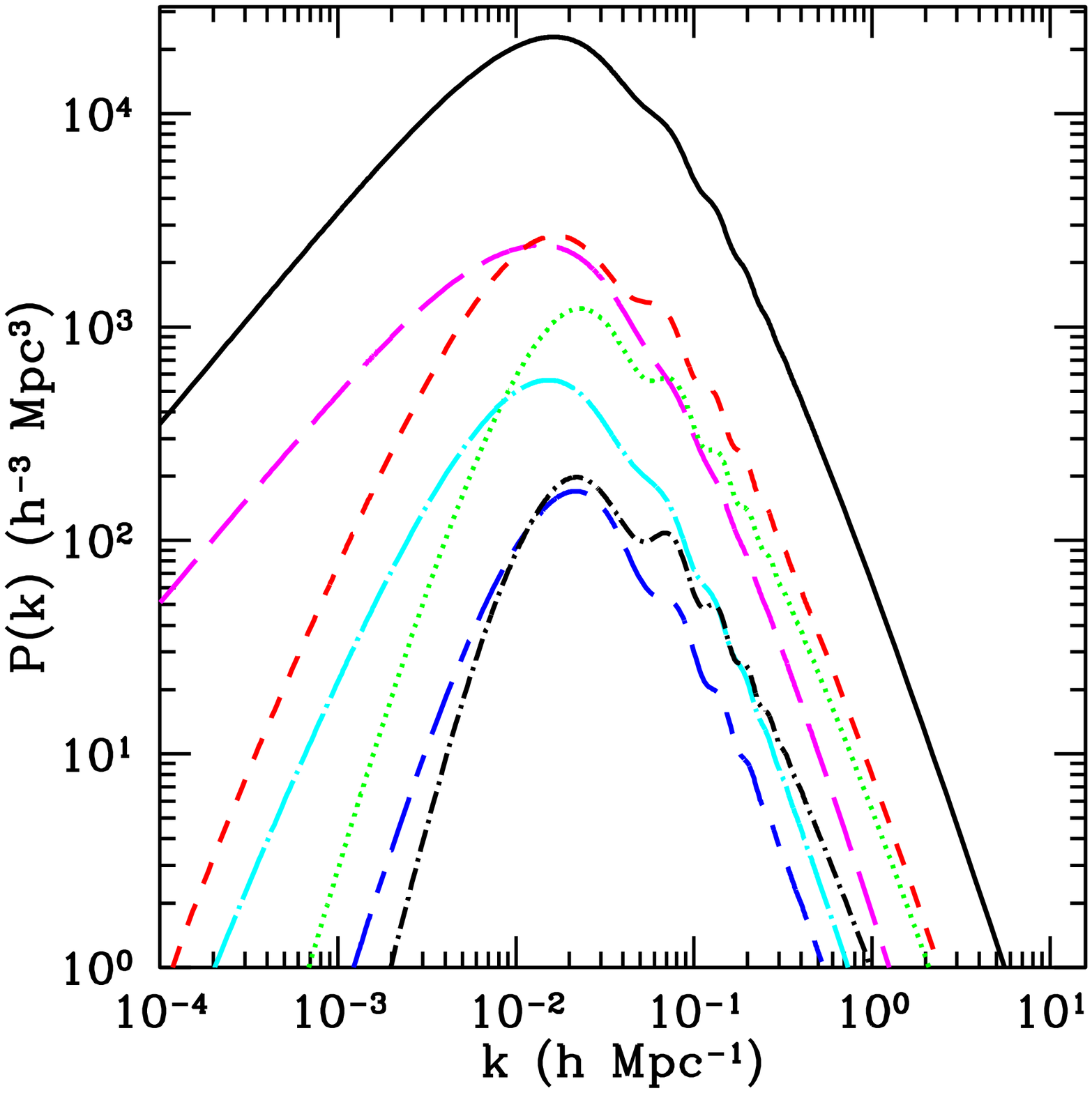}
\caption[Dark matter power spectra for adiabatic and isocurvaure
initial conditions.]{Dark matter power spectra of the different
auto- (left panel) and cross-correlators (right panel) for a
concordance model with $\OLa = 0.70$, $h = 0.65$, $n_\SCAL = 1.0$,
$\omega_\BAR = 0.020$, with the corresponding CMB power spectrum
COBE-normalized. The color and line style codes are as follows:
 in the left panel, adiabatic (AD): solid/black line; CDM isocurvature (CI):
dotted/green line; neutrino density (ND): short-dashed/red line;
neutrino velocity (NV): long-dashed/blue line;
 in the right panel, AD: solid/black line
(for comparison), $\left<\AD,\CI \right>$: long-dashed/magenta
line, $\left<\AD,\NID \right>$: dotted/green line,
$\left<\AD,\NIV\right>$: short-dashed/red line,
$\left<\CI,\NID\right>$: dot-short dashed/blue line,
$\left<\CI,\NIV\right>$: dot-long dashed/light-blue line, and
$\left<\NID,\NIV\right>$: dot-short dashed/black line. The
adiabatic mode is by far dominant over all others.}
\label{fig:genic_PS}
\end{figure}

In \FIG{fig:genic_PS} we plot the dark matter power spectra of the
different auto- (left panel) and cross-correlators (right panel)
for a concordance model. The norm of each pure mode (AD, CI, ND,
NV) is chosen such that the corresponding CMB power spectrum is
COBE-normalized. The cross-correlators are normalized according to
totally correlated spectra, \ie
\begin{equation}
M_{(\rm{X},\rm{Y})} = \sqrt{M_{\rm{X}} M_{\rm{Y}}/2} \eqcomma
\end{equation}
where $M_{(\rm{X},\rm{Y})}$ denotes the norm of the
cross-correlator between the modes $X$,$Y$ and $M_{\rm X}$ the
norm of the pure mode $X$. A crucial result is that the
COBE-normalized amplitude of the adiabatic matter power spectrum
is nearly two orders of magnitude larger than the isocurvature
contribution. The main reason for this is the amplitude of the
Sachs-Wolfe plateau which is about $\frac{1}{3} \Phi$ for
adiabatic perturbations and $2 \Phi$ for isocurvature
perturbations, where $\Phi$ is the gravitational potential at last
scattering, see \rr{eq:dT_AD} and \rrp{eq:dT_ISO}. This difference
of a factor of about $36$ in the power spectrum on large scales is
clearly visible in the comparison of $P_\AD$ and $P_\CI$ (the
difference increases at smaller scales). The case of the neutrino
modes is even worse since they start with vanishing dark matter
perturbations. That the CDM isocurvature matter power spectrum is
much lower than the adiabatic one has been known for some time
(see \eg \citealp{Stompor:1996py,Pierpaoli:1999zj}). However, it
was not recognized before that the same holds true for the
neutrino isocurvature matter power spectra as well, and -- more
importantly -- that this leads to a way to break the strong
degeneracy among initial conditions which is present in the CMB
power spectrum alone.

\begin{figure}[tb]
\centering
\includegraphics[width=\onefigwidth]{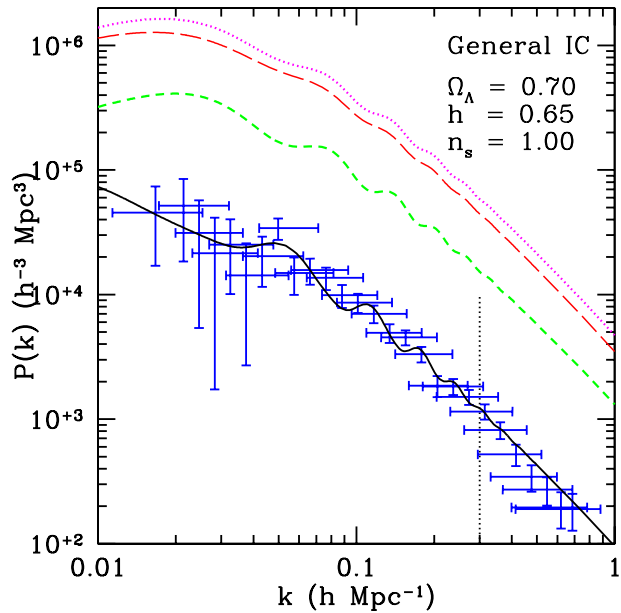}
\caption[Concordance model fit with general isocurvature initial
conditions and 2dF data only.]{Concordance model fit with general
isocurvature initial conditions and 2dF data only. The total
spectrum (solid/black) is the result of a large cancellation of
the purely adiabatic part (long-dashed/red) by the large, negative
sum of the various correlators (dotted/magenta, plotted in
absolute value). The short-dashed/green curve is the sum of the
three pure isocurvature modes, CI, ND and NV. Note that the
resulting total spectrum is less than one tenth of the purely
adiabatic part.} \label{fig:genic_ps_only}
\end{figure}

In an analysis with general initial conditions including the 2dF
data only we obtain very broad likelihood and confidence contours
which exclude only the lower right corner of the $(\OLa, h)$
plane. In contrast to the CMB power spectrum, the matter power
spectrum can be fitted with extremely high adiabatic and
isocurvature contributions, which are then typically cancelled by
large anti-correlations between the spectra. This behavior is
exemplified for a model with general isocurvature initial
conditions and $\OLa = 0.70$, $h = 0.65$, $n_\SCAL = 1.0$ in
\FIG{fig:genic_ps_only}. The best fits with 2dF data only are
dominated by large isocurvature cross-correlations. Clearly, the
resulting CMB power spectrum is highly inconsistent with the COBE
data. Hence such ``bizarre'' possibilities are immediately ruled
out once we include CMB data. Conversely, moderate isocurvature
contributions can help fitting the CMB data, and do not influence
the matter power spectrum, which is completely dominated by the
adiabatic mode alone.

Combining CMB and 2dF data we find now (Bayesian, mixed
isocurvature models):
\begin{equation}
\OLa = 0.65 \lims{-0.25}{+0.22} \mbox{ at $2 \si$ $\quad$ and
$\quad$}
            \lims{-0.48}{+0.25}
\mbox{ at 3$\si$} \eqdot
\end{equation}
The likelihood limits are larger than for the purely adiabatic
case but it is interesting that the Bayesian analysis still
excludes $\OLa = 0$ at more than $3 \si$ even with general initial
conditions, for the class of models considered here. Because of
the above explained reason, the widening of the limits is not as
drastic as one might fear. Therefore, combination of CMB and LSS
measurements turn out to be an ideal tool to constrain the
isocurvature contribution to the initial conditions.

\begin{figure}[tb]
\centering
\includegraphics[width=\twofigswidth]{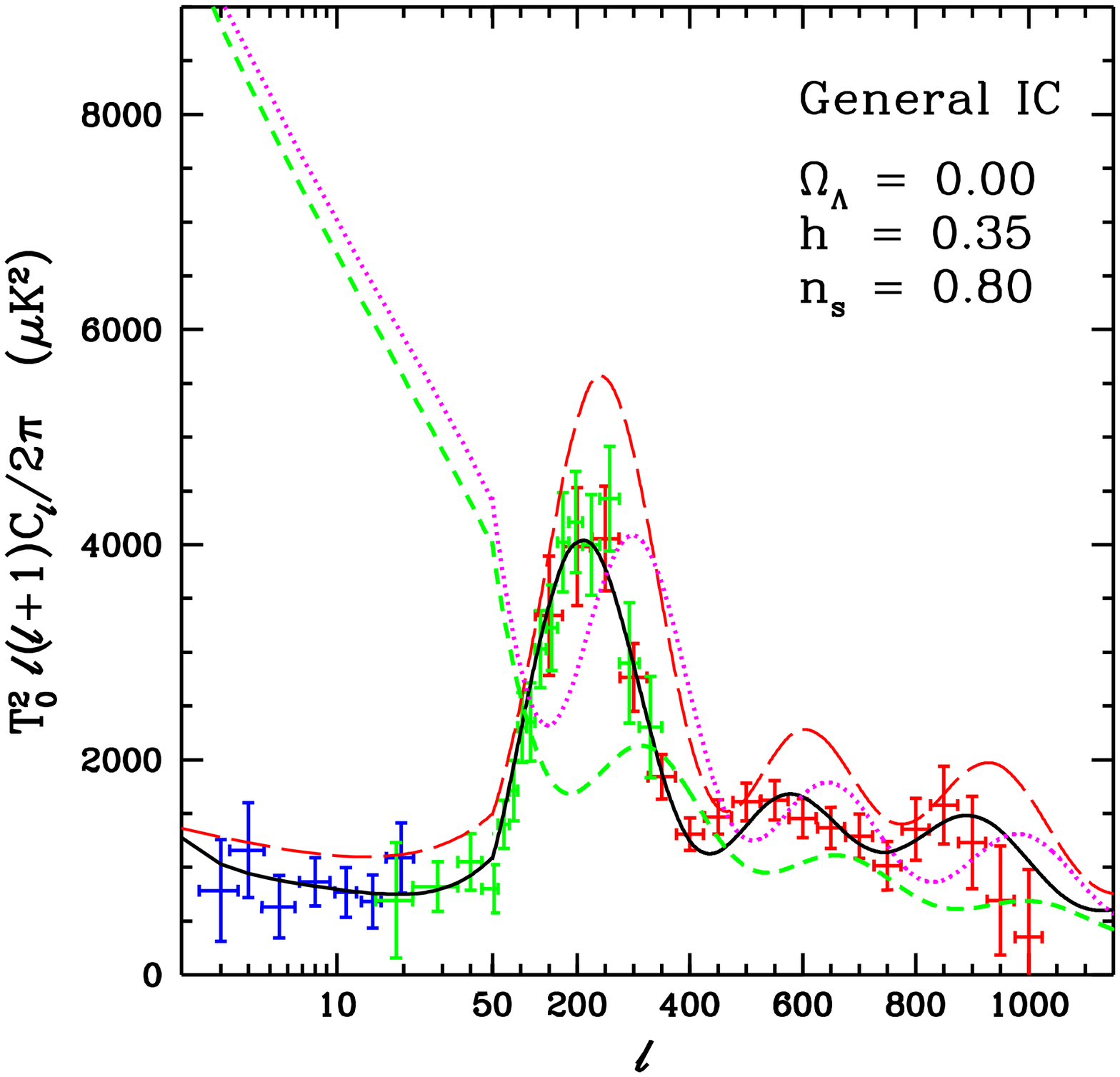}\hfill%
\includegraphics[width=\twofigswidth]{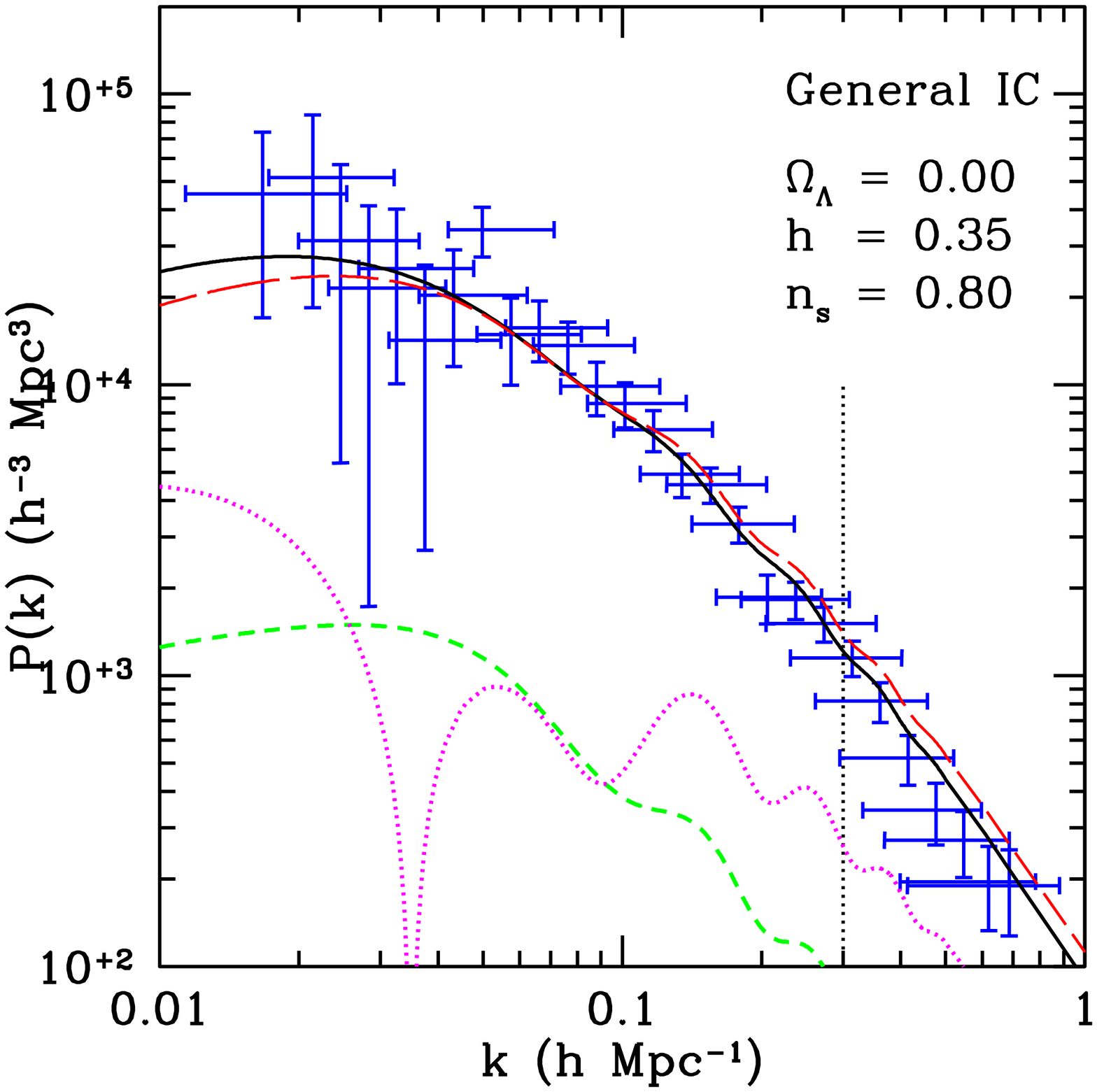}\hfill
 \caption[Best fit with mixed isocurvature models and
$\OLa = 0$.]{Best fit with general isocurvature models and $\OLa =
0$. As for the purely adiabatic case, even with general initial
conditions the absence of the cosmological constant suppresses in
an important way the height of the first peak. In both panels we
plot the best total spectrum (solid/black), the purely adiabatic
contribution (long-dashed/red), the sum of the pure isocurvature
modes (short-dashed/green) and the sum of the correlators
(dotted/magenta, multiplied by $-1$ in the left panel and in
absolute value in the right panel). The matter power spectrum is
completely dominated by the adiabatic mode, while the correlators
play an important role in cancelling unwanted contributions in the
CMB power spectrum at the level of the first peak and especially
in the COBE region. For this model we have an isocurvature content
$\beta = 0.39$, while the BOOMERanG and Archeops calibrations are
reduced by $28\%$ and $12\%$, respectively. The color codes for
the error-bars are the same as in \FIG{fig:genic_AD_OL0}.}
\label{fig:genic_AM_LambdaZero}
\end{figure}

From the frequentist point of view, one notices that the region in
the $\OLa, h$ plane which is incompatible with data at more than
$3\si$ is nearly independent on the choice of initial conditions
(compare the right panels of \FIG{fig:genic_ad_only} and
\FIG{fig:genic_mixed}). Enlarging the space of initial conditions
seemingly does not have a relevant benefit on fitting CMB and 2dF
data with or without a cosmological constant. The reason for this
is that the (COBE-normalized) matter power spectrum is dominated
by its adiabatic component and therefore the requirement $\Om_\MAT
h \sim 0.2$ remains valid.  In \FIG{fig:genic_AM_LambdaZero} we
plot the best fit model with general initial conditions and $\OLa
= 0$. We summarize our likelihood and confidence intervals on
$\OLa$ (this parameter only) in \TAB{table:genic_Lambda}.

In \FIG{fig:genic_beta} we plot the isocurvature contribution to
the best fit models with CMB and 2dF in terms of the parameter
$\beta$ defined in (\ref{eq:isoc_content_beta}). The best fit with
$\OLa = 0$ has an isocurvature contribution of about $40\%$. We
can put a constraint on the maximal isocurvature contribution
allowed by combining this plot with the exclusion plot obtained
with the frequentist approach, \FIG{fig:genic_mixed} right panel.
The result is that frequentist statistics limits the isocurvature
content $\beta$ to be
 \be
 \beta \lsim 0.4 \quad \text{($2\si$ c.l.).}
 \ee
\begin{table}[!tb]
\centering
\begin{tabularx}{\linewidth}{|X|c|ccc|ccc|cc|}
\hline \hline
 \multicolumn{10}{|c|}{Purely adiabatic}  \\
 \hline
 \multicolumn{2}{|c}{ }        & \multicolumn{3}{|c}{Bayesian\footnotemark[1]} \rule{0pt}{4ex}
                              & \multicolumn{3}{|c}{Frequentist\footnotemark[2]} \rule{0pt}{4ex} &  &\\
 Data-sets  & $\quad \OLa$ \quad & $1\si$ & $2\si$  & $3\si$ & $1\si$ & $2\si$
          & $3\si$ & $F$ & $\chi^2/F$\\
\hline CMB \rule{0pt}{4ex}
          & $0.80$& $\lims{-0.08}{+0.08}$
                  & $\lims{-0.35}{+0.10}$
                  & $\lims{\quad -}{+0.12}$
                  & $ < 0.93 $
                  & $ - $
                  & $ - $
                  & $35 $    & $0.58$\\
CMB +2dF \rule{0pt}{4ex}
          & $0.70$& $\lims{-0.05}{+0.05}$
                  & $\lims{-0.17}{+0.13}$
                  & $\lims{-0.27}{+0.15}$
                  & $ {\OLa}\lims{>0.15}{<0.90} $
                  & $ \quad < 0.92 $
                  & $\quad < 0.92 $
                  & $56 $    & $0.59$\\
\hline
\multicolumn{10}{|c|}{General isocurvature}  \\
\hline CMB \rule{0pt}{4ex}
          & $0.85$& $\lims{-0.35}{+0.05}$
                  & $ - $
                  & $ - $
                  & $ - $
                  & $ - $
                  & $ - $
                  &  $26 $   & $0.74$\\
CMB+2dF \rule{0pt}{4ex}
          & $0.65$& $\lims{-0.10}{+0.15}$
                  & $\lims{-0.25}{+0.22}$
                  & $\lims{-0.48}{+0.25}$
                  & $ < 0.90 $
                  & $ < 0.92 $
                  & $\quad < 0.95 $
                  &  $47 $   & $0.67$\\
\hline \hline
\end{tabularx}

\begin{tabularx}{\linewidth}{X}
$^1${~Likelihood interval.}\\
$^2${~Region not excluded by data with given confidence.}
\end{tabularx}
\caption[Likelihood (Bayesian) and confidence (frequentist)
intervals for $\OLa$ alone.]{Likelihood (Bayesian) and confidence
(frequentist) intervals for $\OLa$ alone (all other parameters
maximized). A bar, $-$, indicates that at the given
likelihood/confidence level the analysis cannot constraint $\OLa$
in the range $0.0 \leq \OLa \leq 0.95$. Where the quoted interval
is smaller than our grid resolution, an interpolation between
models has been used. \label{table:genic_Lambda}}
\end{table}

\subsection{Do isocurvature perturbations mitigate the $\La$ problem?}

There are three main conclusions we can draw from these results.
The first one is not new, but seems to be dangerously forgotten in
recent cosmological parameters estimation literature: namely that
likelihood contours cannot be used as ``exclusion plots''. The
latter are usually substantially wider, less stringent. A more
rigorous possibility are frequentist probabilities, which however
suffer from the dependence on the number of really independent
measurements which is often very difficult to come by.

Secondly, we have found that in COBE-normalized fluctuations, the
matter power spectrum has negligible isocurvature contributions
and is essentially given by the adiabatic mode. Hence the shape of
the observed matter power spectrum still requires $\Om_\MAT h
\simeq 0.2$, independent of the choice of initial conditions. Due
to this behavior, the condition $\Om = \OLa + \Om_\MAT = 1$
requires either a cosmological constant or a very small value for
the Hubble parameter, independently from the isocurvature
contribution to the initial conditions.

The third conclusion concerns the presence of a cosmological
constant from pre-WMAP CMB data combined with the 2dF matter power
spectrum: For flat models, a likelihood (Bayesian) analysis
strongly favors a non-vanishing cosmological constant. Even if we
allow for isocurvature contributions with arbitrary correlations,
a vanishing cosmological constant is still outside the $3 \si$
likelihood range. It is possible that there are open models, which
we did not consider here, in which the NV mode would be dominant,:
this because it presents a first acoustic peak at $\ell = 170$ in
flat models, which would be displaced to a larger multipole value,
as preferred by data, in an open Universe, thereby possibly giving
a good fit to CMB data and allow for the observed shape parameter
$\Ga$ with a reasonable value of $h$. This question remains to be
investigated in detail.

The situation changes considerably in the frequentist approach.
There, even for purely adiabatic models, $\OLa = 0$ is still
within $3 \si$ for a value of $h \le 0.48$ which is marginally
defendable.  The conclusion does not change very much when we
allow for generic initial conditions.

\clearpage
\section{Precision cosmology independent of initial conditions}
\label{chap:genic;sec:future}

\begin{figure}[!b]
\centering
\includegraphics[angle=270,width=\twofigswidth]{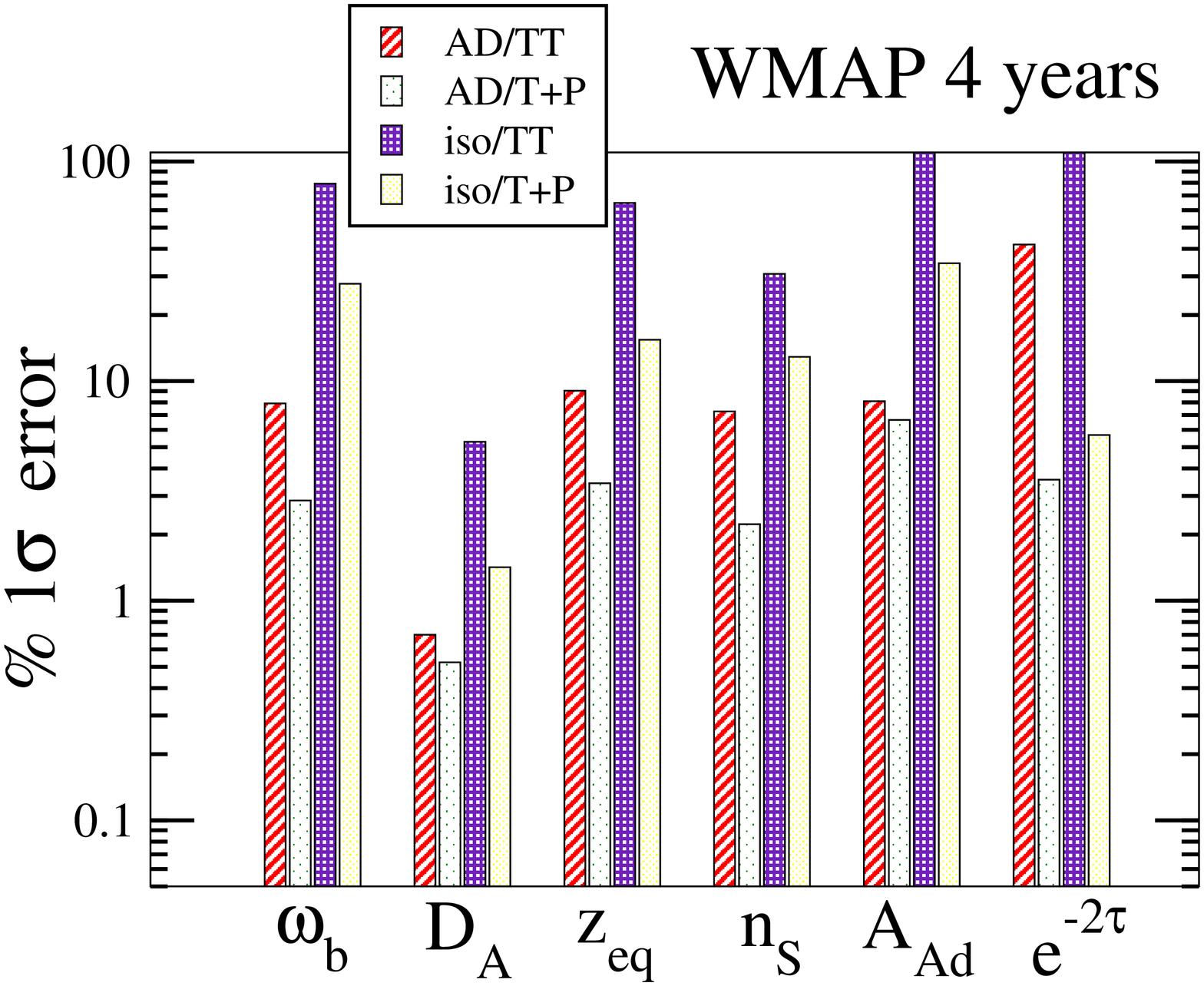}\hfill%
\includegraphics[angle=270,width=\twofigswidth]{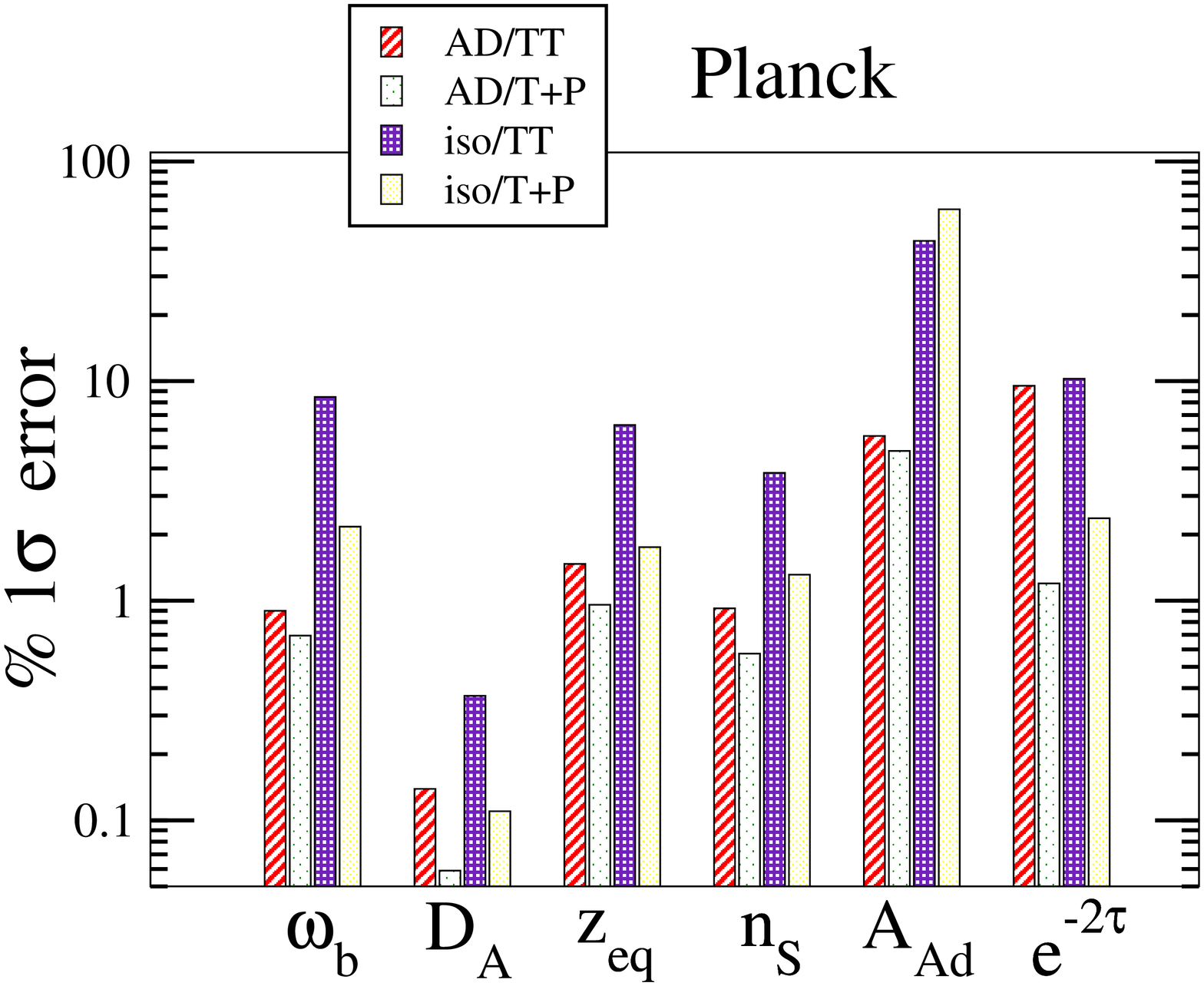}
\caption[Forecast for initial conditions independent determination
of normal parameters.]{Fisher matrix forecast for the percent
$1\sigma$ errors on six quantities which are well determined by
CMB alone with and without inclusion of general isocurvature
initial conditions. The left (right) panel is a forecast for WMAP
four year mission (Planck). From left to right, on the abscissa
axis: the baryon density, $\omega_b$, the angular diameter
distance $D_A$, the redshift of matter-radiation equality
$z_{\eq}$, the scalar spectral index $n_\SCAL$, the scalar
adiabatic amplitude $A_{\text{Ad}}$ and a function of the optical
depth to reionization, $\tau_\reion$. In the legend, ``AD'' means
that only adiabatic fluctuations were included, ``iso'' means that
general isocurvature modes were included and marginalized over.
``TT'' uses temperature information alone, ``T+P'' has
temperature, E-T correlation and E-polarization.}
\label{fig:genic_fma}
\end{figure}

As we have seen, it is difficult to simultaneously constrain both
the type of initial conditions and the cosmological parameters
using CMB alone. The future high accuracy measurements of CMB
polarization will help substantially in breaking degeneracies
between initial conditions. The degeneracies in the parameter
dependence of temperature and polarization are almost orthogonal,
and polarization can therefore lift ``flat directions'' in
parameter space.

To determine cosmological parameters independently on the initial
conditions, one includes general isocurvature modes, and then
marginalize over them. \cite{Bucher:2000kb,Bucher:2000hy}
considered forecasts for WMAP and Planck, and found that admitting
isocurvature modes would ruin the ability of WMAP to determine the
cosmological parameters with temperature information only. They
also highlighted that polarization measurements would be decisive
in assisting into the reconstruction of the cosmological
parameters when allowing for general isocurvature initial
conditions. Their results were obtained with a Fisher matrix
analysis on a cosmological parameter set which, according to
\cite{Kosowsky:2002zt}, leads to large overestimates of the
expected errors. We have reproduced their study
\citep{Trotta:2004}, using for the Fisher matrix forecast the
normal parameter set described in \SEC{chap:params:sec:normal} so
that we obtain forecasts not for the highly degenerate directions
defined by the cosmological parameters, but rather for orthogonal
combinations which are well measured by the CMB. Along these
directions, forecasts are much more reliable. The main features
are summarized in \FIG{fig:genic_fma}, where we plot the expected
$1\sigma$ error in percent for the six quantities which are
directly probed by the CMB with good accuracy (see figure
caption). We omit the energy density in the cosmological constant,
which is ill-determined with CMB alone because of the geometrical
degeneracy. We do not restrict our analysis to flat models, but
include spaces with non-zero curvature.

For WMAP the errors on normal parameters will increase roughly by
a factor ten with respect to the purely adiabatic scenario if one
marginalizes over general initial conditions, when temperature
information alone is considered (\CF first and third bar in the
left panel). When the full polarization information is included,
however, the errors will still be within approximately 10 to 30\%
even in the general isocurvature scenario. From the right panel,
we deduce that for the Planck experiment the worsening of the
errors will be much less if the high quality polarization
information is included. Roughly speaking, by including
isocurvature modes we expect errors which are larger than in the
adiabatic case by about a factor of two, but mostly still within
the few percent accuracy. These findings are in qualitative
agreement with \cite{Bucher:2000hy}, while providing a
quantitatively more reliable estimate of the expected accuracy.

This shows that the CMB alone will be able to provide high
precision cosmology even if the strong assumption of purely
adiabatic initial conditions will be relaxed. Combining CMB
results with other observation which independently constrain the
cosmological parameters, will enable us to fully open this window
to the mysterious epoch of the very early universe.

\chapter*{Publication list}
\newcommand{\PRL}{Phys.~Rev.~Lett.\ }
\newcommand{\PhysD}{Phys.~Rev.~D}
\newcommand{\MNRAS}{Mon.~Not.~R.~Astron.~Soc.\ }
\newcommand{\astro}[1]{{\texttt{\small  astro-ph/#1}}}
\newcommand{\rt}{{\bf R.~Trotta}}
\newcommand{\jrob}[4]{#1, {\it #2}, {\bf #3}, #4}

\begin{enumerate}
\item {\bf Trotta, R.} \& Durrer, R. (2004).
\newblock Testing the paradigm of adiabaticity.
\newblock In Ruffini et~al., editors, {\em Proceedings of the X Marcel Grossman
  Meeting, 20-26 July 2003, Rio de Janeiro}, astro-ph/0402032.
\newblock To appear.

\item Rocha, G.,{\bf Trotta, R.}, Martins, C., Melchiorri, A.,
Avelino, P., Bean, R., \&
  Viana, P. (2004).
\newblock Measuring $\alpha$ in the early {Universe: CMB} polarization,
  reionization and the {Fisher} matrix analysis.
\newblock {\em Mon. Not. Roy. Astron. Soc.}, 352:20--38, astro-ph/0309211.

\item{\bf Trotta, R.} \& Hansen, S.~H. (2004).
\newblock {Observing the helium abundance with CMB}.
\newblock {\em Phys. Rev.}, D69:023509, astro-ph/0306588.

\item Rocha, G.,{\bf Trotta, R.}, Martins, C., Melchiorri, A.,
Avelino, P., Bean, R., \&
  Viana, P. (2003).
\newblock New constraints on varying $\alpha$
\newblock {\em New Astronomy Reviews}, Proceedings of the 2nd CMBNET
Meeting, Oxford, 20-21 February 2003, Oxford, UK, 47: 863-869,
astro-ph/0309205.

\item{\bf Trotta, R.} (2003).
\newblock The cosmological constant and the
paradigm of adiabaticity.
\newblock {\em New Astronomy Reviews}, Proceedings of the 2nd CMBNET
Meeting, Oxford, 20-21 February 2003, Oxford, UK, 47: 769-774,
astro-ph/0304525.

\item Martins, C. J. A.~P., Melchiorri, A., Rocha, G.,{\bf Trotta,
R.}, Avelino, P., \&
  Viana, P. (2004).
\newblock {WMAP} constraints on varying $\alpha$ and the promise of
  reionization.
\newblock {\em Phys. Lett. B}, 585:29, astro-ph/0302295.

\item{\bf Trotta, R.}, Riazuelo, A., \& Durrer, R. (2003).
\newblock {The cosmological constant and general isocurvature initial
  conditions}.
\newblock {\em Phys. Rev.}, D67:063520, astro-ph/0211600.

\item Martins, C. J. A.~P., Melchiorri, A.,{\bf Trotta, R.}, Bean,
R., Rocha, G., Avelino,
  P., \& Viana, P. (2002).
\newblock Measuring $\alpha$ in the early universe: {CMB} temperature,
  large-scale structure and {Fisher} matrix analysis.
\newblock {\em Phys. Rev.}, D66:023505, astro-ph/0203149.

\item Bowen, R., Hansen, S.~H., Melchiorri, A., Silk, J., \& {\bf
Trotta, R.} (2002).
\newblock The impact of an extra background of relativistic particles on the
  cosmological parameters derived from microwave background anisotropies.
\newblock {\em Mon. Not. Roy. Astron. Soc.}, 334:760, astro-ph/0110636.

\item{\bf Trotta, R.}, Riazuelo, A., \& Durrer, R. (2001).
\newblock Reproducing cosmic microwave background anisotropies with mixed
  isocurvature perturbations.
\newblock {\em Phys. Rev. Lett.}, 87:231301, astro-ph/0104017.

\end{enumerate}

\bibliographystyle{../Bibstyle/hapalike-r}

\fancyhf{}
 \fancyhead[LE]{\bf\thepage}
 \fancyhead[CE]{}
 \fancyhead[RE]{\bf Bibliography}
 \fancyhead[LO]{\bf Bibliography}
 \fancyhead[CO]{}
 \fancyhead[RO]{\bf\thepage}





\end{document}